\documentclass[notitlepage]{disithesis}
\usepackage{times}
\usepackage{epsfig}
\usepackage{amsthm}
\usepackage{amscd}
\usepackage{pictexwd,dcpic}
\usepackage{algorithm}
\usepackage{algorithmic}
\usepackage{algorithm}
\usepackage{listings}             \usepackage[multidot]{grffile}
\usepackage[outdir=./]{epstopdf}
\usepackage{hyperref}
\usepackage{makeidx}
\usepackage{graphicx}
\usepackage{mathtools}
\usepackage{graphicx}
\usepackage{hyperref}
\usepackage{pldoc}

 \makeindex
\hypersetup{
	colorlinks   = true, urlcolor     = blue, linkcolor    = blue, citecolor   = red }

\def \real {\rm I\kern-.2em R}
\def \natural {\rm I\kern-.15em N}
\def \integer {\rm Z\kern-.2em Z}
\def \lsx {[\kern-.1em [}
\def \rsx {]\kern-.1em ]}

\newcommand{\margine}[1]{}  \newcommand{\ignore}[1]{}

			\newcommand{\vettore}[1]{{\bf{#1}}}

\newcommand{\abs}[1]{{\|{#1}\|}}
\newcommand{\norm}[1]{{\|{#1}\|}}
\newcommand{\interior}[1]{\stackrel{\circ}{#1}}
\newcommand{\mnsgc}[1]{\widehat{#1}}

\def\NOTA #1  {}
\def\JOURNAL #1  {}
\def\NOTACOMP #1  {}

\def\PAO #1 {} 

\def \real {{\rm I\kern-.2em R}}
\def \espace {{\rm I\kern-.2em E}}
\def \int {{\rm I\kern-.2em N}}
\def \bool {{\rm I\kern-.2em B}}
\def \cnum {{\rm I\kern-.5em C}}
\def \cell {{\rm I\kern-.2em \Gamma}}

\def\Box {\fbox{\rule{0ex}{0.5ex}\rule{0.5ex}{0ex}}}  

            \def\MESH  {\Gamma}

\def\Seq    {{\tt Seq}}

\newtheorem{definition}{Definition}[section]
\newtheorem{opt-definition}[definition]{Definition~*}
\newtheorem{property}{Property}[section]
\newtheorem{opt-property}[property]{Property~*}
\newtheorem{remark}[property]{Remark}
\newtheorem{opt-conjecture}[property]{Conjecture~*}
\newtheorem{theorem}{Theorem}[section]
\newtheorem{lemma}{Lemma}[section]
\newtheorem{mexample}{Example}[section]
\newenvironment{example}{\begin{mexample}\rm}{\hfill\Box\end{mexample}}
\newenvironment{short-example}{\begin{mexample}\rm}{\hfill\Box\end{mexample}}
\newenvironment{data}[2]
{\begin{xdata}[{#2}]\label{#1}\mbox{}\hrule\end{xdata}}
{\hrule\mbox{}\\}
\newtheorem{xdata}{Data Structure}[section]
\newenvironment{algo}[2]
{\begin{xalgo}[{#2}]\label{#1}\mbox{}\hrule\end{xalgo}}
{\hrule\mbox{}\\}
\newtheorem{xalgo}{Algorithm}[section]

  \newcommand{\trel}[1]{S_{{#1}}}      \newcommand{\decom}{\nabla\cdot}   \newcommand{\join}{\bullet} \newcommand{\bnd}[1]{\partial#1}  \newcommand{\carrier}[1]{\Delta(#1)} \newcommand{\lk}[1]{lk(#1)} \newcommand{\xlk}[2]{lk({#2},{#1})} \newcommand{\str}[1]{\star{#1}} \newcommand{\xstr}[2]{star({#2},{#1})} \newcommand{\clstar}[1]{\overline{\star{#1}}} \newcommand{\xclstar}[2]{\overline{\xstr{#1}{#2}}} \newcommand{\closure}[1]{\overline{#1}}  \newcommand{\funct}[3]{{#1}:{#2}\rightarrow {#3}}                          \def\ord{\mbox{\rm dim}}

                               \def\topo #1 {{\cal T}( #1 )}

\def\n    {{\rm n}}

\def\T    {{\rm T}}

\def\asc    {{abstract simplicial complex}}      
\def\asm    {{abstract simplicial map}}

\def\trix    {{\Upsilon}}

\def\verteq    {\stimap}      
\def\Verteq    {\Stimap}      
\def\tsimeq    {simplex}      
\def\natu    {essential}

\def\redo    {reduced}      
\def\cdecs  {{initial-quasi-manifolds}}
\def\cdec  {{initial-quasi-manifold}}
\def\Cdec  {{Initial-Quasi-Manifold}}
\def\cano    {{ standard}}      
\def\Cano    {{ Standard}}      
\def\ra    {{ regularly adjacent}}

\def\NRA  {\cal NM}       
\def\qm    {{quasi-manifold}}      
\def\wrt    {{with respect to}}      
\def\occup    {{storage requirement}}      
\def\Qm    {{Quasi-manifold}}      
\def\Quot    {{Quotient}}      
\def\quot    {{quotient}}      
\def\sglinst {{simplex gluing instruction}}
\def\Sglinst {{Simplex Gluing instruction}}
\def\gl {{gluing}}
\def\inst {{instruction}}
\def\insts {{instructions}}
\def\iqm    {{initial-quasi-manifold}}      
\def\Iqm    {{Initial-quasi-manifold}}

\def\eqt {{equating}}
\def\Eqt {{Equating}}

\def\stimap {{stitching}}
\def\Stimap {{Stitching}}

\def\dec {{decomposition}}
\def\Dec {{Decomposition}}
\def\gw {{generalized winged}}
\def\Gw {{Generalized Winged}}
\def\cdecrep {{extended winged}}
\def\Cdecrep {{Extended Winged}}
\def\NMWDS {{Non-manifold Winged}}
\def\nmwds {{non-manifold winged}}
\def\ewds {{EWDS}}
\def\ids {{implicit data structure}}
\def\CC {{\rm CC}}
\def\FTT {{\rm FTT}}
\def\FVV {{\rm FVV}}
\def\SIZE {{\rm SIZE}}
\def\TBase {{\rm TBase}}
\def\TBaseAddr {{\rm TBaseAddr}}
\def\TAddr {{\rm TVAddr}}
\def\IIBND {{\rm Bnd}}
\def\IITAddr {{\rm BndAddr}}
\def\VBase {{\rm VBase}}
\def\TTPP {{\rm TT''}}
\def\TVPP {{\rm TV''}}
\def\TVXP {{\rm TVI''}}
\def\TTP {{\rm TT'}}
\def\TVP {{\rm TV'}}
\def\VTS {$\mbox{\rm VT}^\star$}

\def\glb {{\wedge}}
\def\lub {{\vee}}

\def\gteq {\geq}

\def\lt {<}
\def\gt {>}
\def\poeq {\cong}        
\def\Adj    {{\cal A}}

\def\second    {{\prime\prime}}      
\def\third    {{\prime\prime\prime}}      
      
\newcommand{\dual}[1]{{\breve{#1}}}
\newcommand{\brk}[1]{{<\!#1\!\!>}}
\newcommand{\ualg}[2]{{{<}{#1},{#2}{>}}}

\newcommand{\binomial}[2]{{#1 \choose #2}}
\newcommand{\factorial}[1]{{#1!}}
\newcommand{\ceiling}[1]{\left\lceil{#1}\right\rceil}
\newcommand{\floor}[1]{\left\lfloor{#1}\right\rfloor}
\newcommand{\ordcomp}[1]{\Theta(#1)}
\newcommand{\lesscomp}[1]{O(#1)}
\newcommand{\morecomp}[1]{\Omega(#1)}

\newcommand{\starring}[3]{#1\stackrel{#2}{\longrightarrow}#3}

\newcommand{\send}[2]{#1\mapsto #2}
\newcommand{\canon}[1]{\decom{#1}}
\newcommand{\tr}[1]{{\rm trie}(#1)}
\newcommand{\FT}[1]{V^{#1}T}
\def\split0 {\nabla}
\newcommand{\xsplit}[2]{{#2}\split0 {#1}}
\def\splitsimp0 {\nabla}
\newcommand{\xsplitsimp}[2]{{#2}\splitsimp0 {#1}}
\def\splitstar0 {{\ge^{\nabla}}^{\star}}
\newcommand{\xsplitstar}[2]{{#2}\splitstar0 {#1}}
\def\splitmap {\sigma_{\nabla}}
\def\domain {{\rm domain}}
\def\TopSimplex {{\rm TopSimplex}}
\def\MinT {{\rm MinT}}
\def\MaxT {{\rm MaxT}}
\def\Vertex {{\rm Vertex}}
\def\MinV {{\rm MinV}}
\def\MaxV {{\rm MaxV}}

\def\INDEX {{\rm INDEX}}
\def\SIMPLEX {{\rm SIMPLEX}}

\newcommand{\nerve}[1]{{\cal N}({#1})}

\def\streq {\sim} 
\def\simpeq {\cong}  \def\pwleq {\approx}  \def\equivert {\approx} 
 
\def\equatsimp {\leftrightarrow} 
\newcommand{\closeq}[1]{{#1}^\star} \def\morphle {\prec} 
\def\quotle {\le} 
\def\refine {\preceq} 
\newcommand{\notation}[1]{{#1}\glossary{Notation #1}} \newcommand{\emi}[1]{{#1}\index{#1}}
\newcommand{\emd}[1]{{\bf #1}\index{#1}} \newcommand{\ems}[1]{{\em #1}\index{#1}} \newcommand{\ema}[2]{{#1} { #2}\index{#1!#2}\index{#2!#1}}
\newcommand{\emas}[2]{{\em #1} {\em #2}\index{#1!#2}\index{#2!#1}} \newcommand{\emad}[2]{{\bf #1} {\bf #2}\index{#1!#2}\index{#2!#1}} \def\dc {<} 
 
\def\sumeq {+} 
\def\proeq {\cdot} 
\def\prodec {\downarrow} 
\def\sumdec {\uparrow}

\def\iff {if and only if } 
\def\siff {iff} 
\def\cal {\mathcal} \def\latt {Appendix \ref{sec:thlatt}}
\def\optt {Appendix \ref{sec:opt}}

\def\CComp   {\MESH}          \def\AComp   {\Omega}          \def\topAComp   {\AComp^\top}                              \def\toptheta   {\theta^\top}

\def\Rtop   {R^\top}

\newcommand{\newred}[1]{{{#1}}}

\begin{document}

\title{Decomposition and Modeling \\in the Non--Manifold Domain}
\author{Franco Morando
\\
 Department of Computer and Information Sciences,\\ Universit\`a di Genova,\\
 Via Dodecaneso 35, 16146 Genova - Italy\\
 {\tt morando@disi.unige.it}\\
\mbox{ }\\
\mbox{ }\\
\mbox{ }\\
\mbox{ }\\
}
\submityear{2003}
\submitmonth{January}
\technumber{02}
\maketitle
\begin{addresspage}
{\bf
Dottorato di Ricerca in Informatica\\
Dipartimento di Informatica e Scienze dell'Informazione\\
Universit\`a degli Studi di Genova\\[2ex]}

DISI, Univ. di Genova\\
via Dodecaneso 35\\
I-16146 Genova, Italy\\
{\tt http://www.disi.unige.it/}\\[2ex]

{\bf
Ph.D. Thesis in Computer Science}\\[2ex]

Submitted by Franco Morando
\\
DISI, Univ. di Genova\\
{\tt morando@disi.unige.it}\\[2ex]
Date of submission:
December 2002\\[2ex]
Title:
{Decomposition and Modeling in the Non­-Manifold Domain}\\[2ex]
Advisor: Enrico Puppo\\
DISI, Univ. di Genova\\
{\tt puppo@disi.unige.it}\\[2ex]
Supervisor: Leila De Floriani\\
DISI, Univ. di Genova\\
{\tt deflo@disi.unige.it}\\[2ex]
\end{addresspage}
\dedication{Ai miei genitori, a mia moglie, alle mie figlie}
\begin{acknowledgements}
I wish to thank my supervisor Prof.\ Leila De Floriani for her suggestions and my advisor Prof.\ Enrico Puppo for helpful discussions. I wish to thank 
also  the external reviewers
Prof.\ Pascal Lienhardt and Prof.\ Alberto Paoluzzi for their insightful advice.
I also want to thank to all wonderful people at DISI  the 
Department of Computer and Information Science of the 
University of Genova for their contributions 
to the work presented here.
\end{acknowledgements}
\tableofcontents

\begin{abstract}
The problem of decomposing non­-manifold object  has already been studied in
solid modeling. However, the few proposed solutions 
are limited to the  problem of decomposing solids described through their
boundaries. 
In this thesis we study the problem of decomposing an 
arbitrary 
non-manifold simplicial complex into more regular components.
A formal notion  of decomposition is developed using combinatorial topology.
The proposed decomposition is unique, for a given complex, 
and is computable for complexes of any dimension.  
A decomposition algorithm is proposed. This algorithm
splits the input complex into a set of connected components
in a time proportional to the size of the input. 
The algorithm splits non-manifold surfaces into manifold components.
In three or higher dimensions a decomposition into manifold parts is not 
always possible.
Thus, in higher dimensions, we decompose a non-manifold into a decidable super class of 
manifolds, that we  call, {\em initial-quasi-manifolds}.
Initial-quasi-manifolds are then  carefully characterized  and a 
definition of this class, in term of local topological properties,
is established. 
        
We also defined a two-layered data 
structure, the {\em \cdecrep} data structure. This data structure is a
dimension independent data structure
conceived to model non-manifolds through their decomposition into
\cdec parts.
Our two layered data structure describes the structure of 
the decomposition and each component.separately.
Each decomposition component, in our description, is encoded using an 
extended version of 
the {\em winged representation} \cite{PaoAl93}. 
In the second layer we encode the connectivity structure of the decomposition.
We analyze the space requirements of the \cdecrep\ data structure and 
give algorithms to build and navigate it.
Finally, we discuss time requirements for the computation of
topological relations and show that for surfaces and tetrahedralizations 
embedded in $\real^3$ all topological relations can be extracted in optimal 
time.

This approach offers a compact, dimension independent, 
representation for non­-manifolds that can be useful whenever the modeled 
object has few non­-manifold singularities.
\end{abstract}
 \chapter{Introduction}
{
In point set topology a closed {\em manifold} object  
is a subset of the Euclidean space for which
the neighborhood of each internal point is locally equivalent to an open ball.
An objects that do not fulfill this property at one or more points
is what is usually  called a {\em non-manifold} object.
Manifolds deserved and continue to deserve a lot of theoretical investigation
from topology. Non-manifolds are less studied and therefore less
known objects. The main reason for this is that non-manifold surfaces seems highly
unstructured and, therefore, it seems that there is a  little chance to find meaningful
theoretical results for them.
Nevertheless, non-manifolds tend to populate the computer graphics field.
}

Geometric meshes with polygonal cells are widely used representations of three-dimensional objects.
Meshes are ubiquitous within several applicative domains including: 
CAD, Computer Graphics, virtual reality, scientific data visualization  
and finite element analysis.
As devices  for three-dimensional object reconstruction become more and more common \cite{BerRus00}, non-manifold
meshes are likely to become relevant in most applications dealing with three-dimensional objects.
For instance, 
{as reported in \cite{Gui99}, in a database of 300 meshes used for MPEG-4 core experiments,
mainly obtained from the Web, more than
half of the models
were represented by  non-manifold meshes.
}

The problem of representing and manipulating meshes with 
non-manifold
topology has been studied in solid modeling, mainly by the end of the '80s (see, e.g.,
\cite{Gur90,RosCon90,Wei88a}), because of its
relevance in CAD/CAM applications.
As a consequence, presently, 
there exist a few non-manifold modelers that represent 3D objects
by a mix of  
wireframe, 2D surfaces and 3D solids.

In non-manifold modelers, non-manifold objects are described 
through meshes with non-manifold features that are usually
encoded directly in an underlying non-manifold data structure.
Motivations for using non-manifold modeling and non-manifold
data structures have been pointed out by
several authors \cite{ChaAnd95,Gur90,RosCon90,Wei88a}. For instance, Boolean operators are closed in the {\em r-set} domain, that is a subset of  the non-manifold domain.
Sweeping or offset operations may generate parts of different
dimensionality, non-manifold topology is required in different
product
development phases, such as conceptual design, analysis or manufacturing
\cite{ChaAnd95,SriAl95}.

Non-manifolds  support the representation of  complex objects  made of 
parts of different {\em dimensionality}. Closed surfaces are used  to represent three-dimensional parts
(enclosed volumes), open surfaces  are used to represent two-dimensional parts, lines  are used to represent
one-dimensional parts, and points  are used to represent zero-dimensional parts.
The general idea is that
some parts of an object must be represented by a lower dimensional object
when seen at a sufficiently high level of abstraction.
Using non-manifolds, each part of an object can be represented by a geometric complex of
the proper dimensionality and
characterized by some geometrical and topological shape features.
Different parts are then glued together to form a non-manifold complex.

The  superior expressive power of {\em non-manifolds} is  also
established in a number of papers on surface simplification (see e.g., \cite{ElSVar99,SanVar98,GarHec98,PopHop97,RosBor93,Ronfard96,Sch97} ).
These papers show that, if we want
an {\em intelligible} simplified model below a certain size,
our simplification must modify the topology of the original mesh 
and create a non-manifold mesh.
{
\begin{figure}[h]
\begin{center}
\framebox{
\parbox[c][0.37\textwidth]{0.28\textwidth}{
\begin{center}
\vfill
\epsfig{file=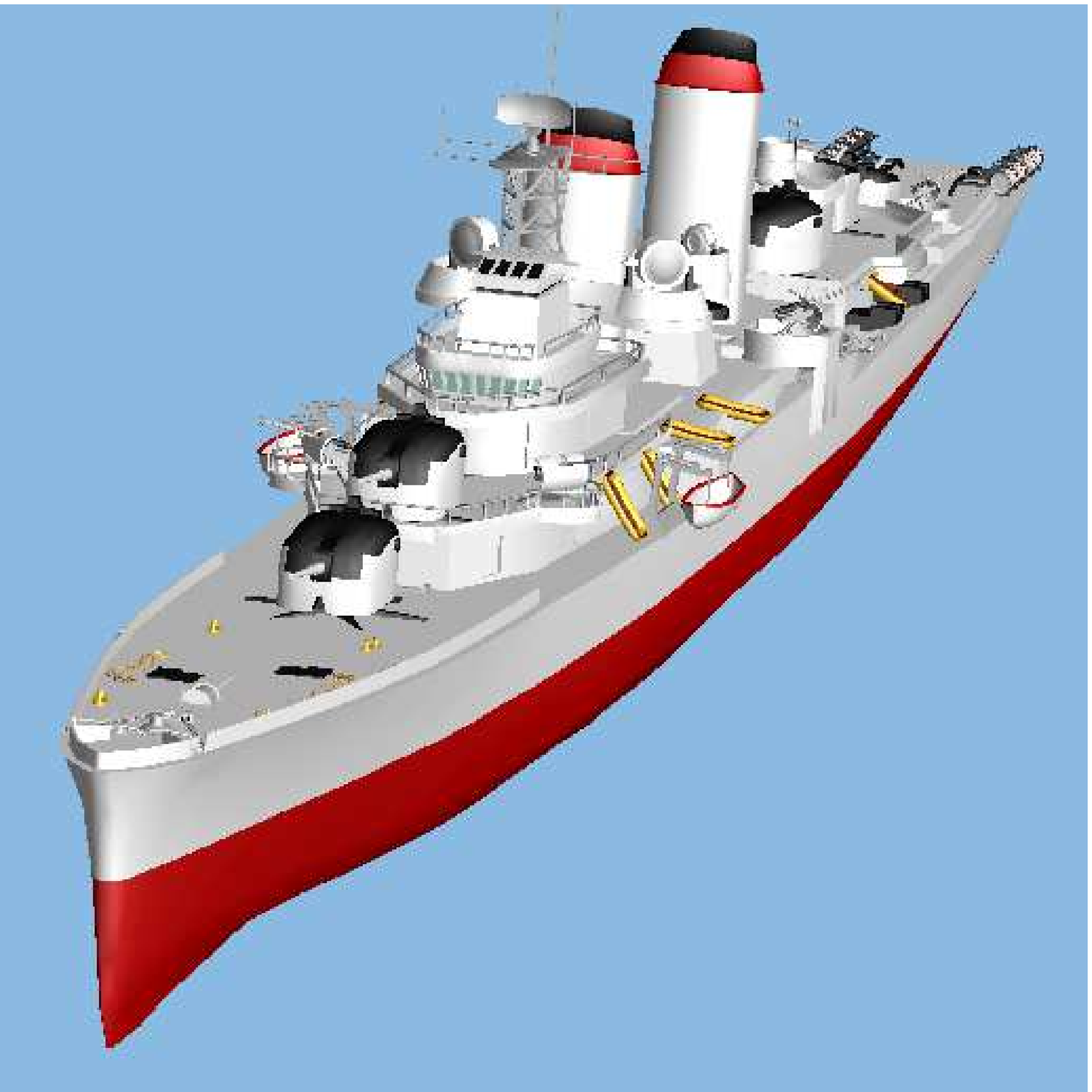,width=0.28\textwidth}
\vfill
(a)\end{center}
}
}
\framebox{
\parbox[c][0.37\textwidth]{0.28\textwidth}{
\begin{center}
\vfill
\epsfig{file=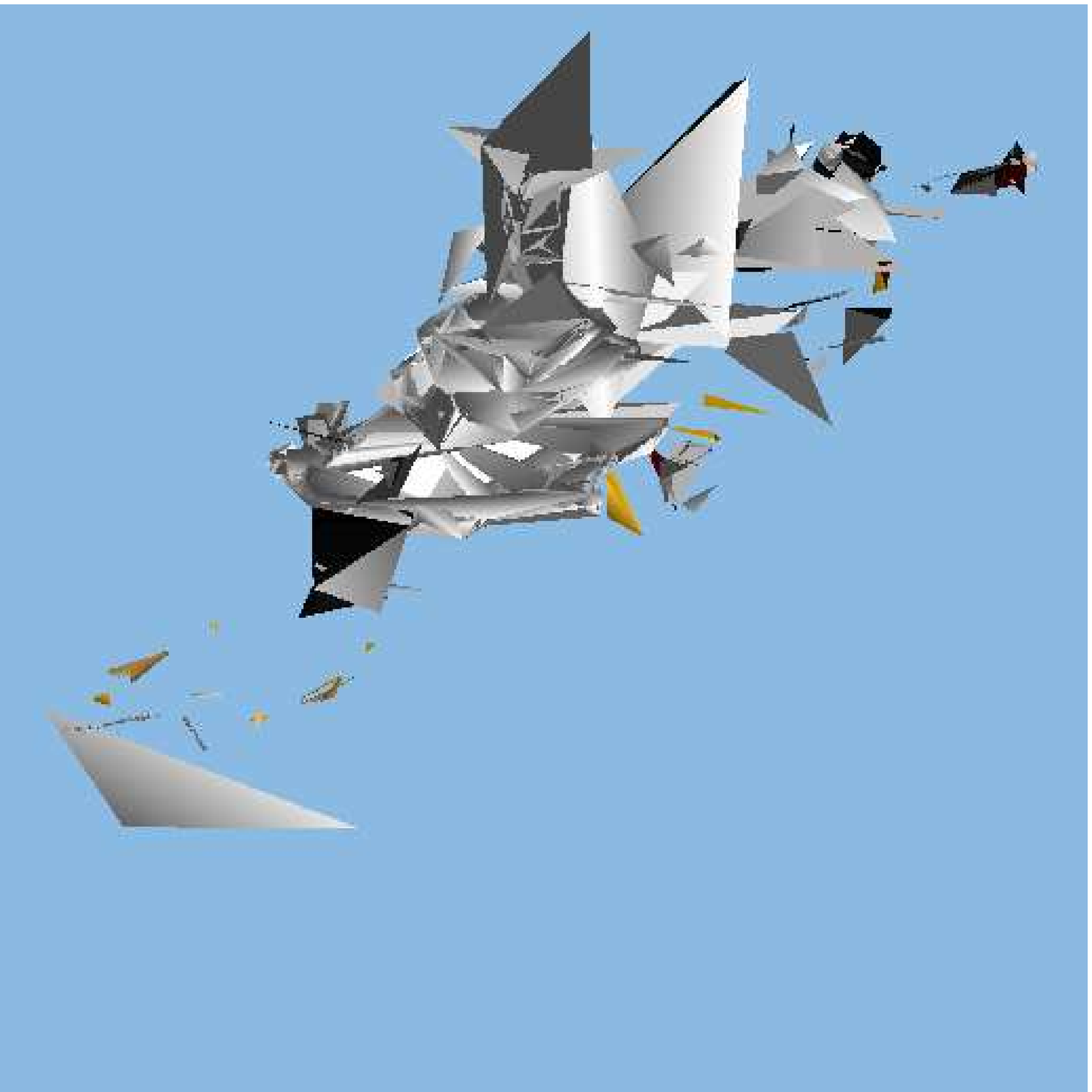,width=0.28\textwidth}
\vfill
(b)\end{center}
}
}
\framebox{
\parbox[c][0.37\textwidth]{0.28\textwidth}{
\begin{center}
\epsfig{file=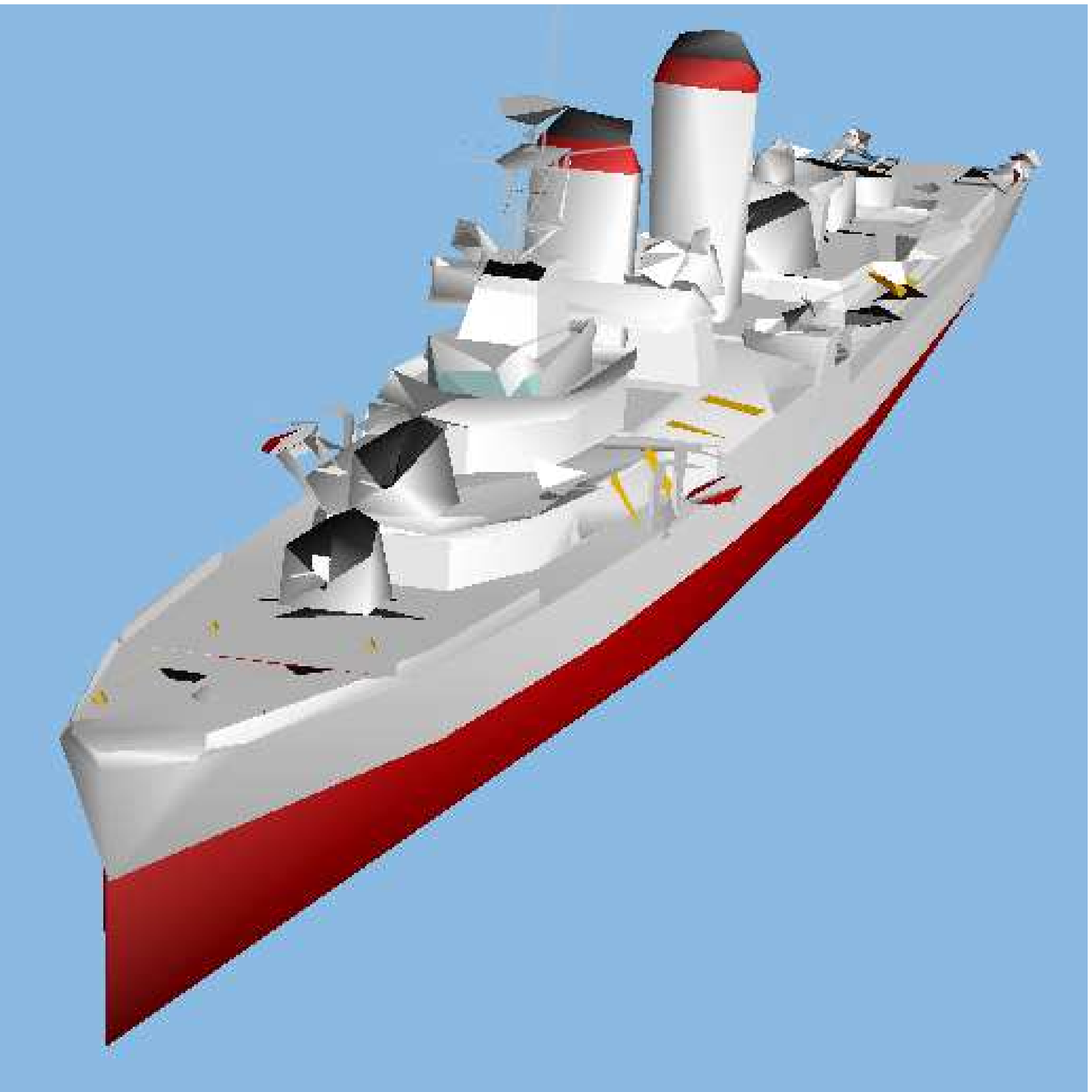,width=0.28\textwidth}
\vfill(c)
\end{center}
}
}
\end{center}
\caption{The  model in Figure (a) (from \cite{PopHop97}) has:  117, distinct,
manifold connected  components, 167744 triangles and 83799 vertices.
Figure (b) shows a simplified model for (a) with 117 connected  components, 
1154 vertices and 2522 triangles. 
This model is obtained from  (a) by a simplification process that
does not modify the (manifold) topology of the original 117 components.
Figure (c) shows simplified model for (a) that  has  only 56 connected  
components, 1517 triangles,
89 dangling edges and 5 isolated vertices. Model in (c) is a non-manifold
version of  model (a) and it is a much more intelligible version of
(a) and even  more compact than (b).}
\label{fig:hoppesav}
\end{figure}
}

As the figure \ref{fig:hoppesav} shows,  non-manifolds are relevant in simplification but there are other applications where singularities are essential.
For instance, one can
model the semantic content of an image
with an object of mixed dimensionality (e.g., see \cite{Kov89}).
Recently non-manifold   models become 
important to provide input to {\em model databases} \cite{SenBoy95}. In this context non-manifolds are used for 3D shape recognition and classification.
Indeed, a detailed and non simplified manifold   mesh is not structured enough to be
used directly for such purposes.
A manifold  mesh describes the shape of an object as a whole.
On the other hand a manifold cannot provide explicit
information neither on the subdivision of an object into parts, nor on its
morphological features.

Finally,  singularities may arise
as an undesired side-effects. This happens, for instance, 
in  features  extraction from images or  in 3D reconstruction.
Non-manifold singularities appear also as a byproduct of coarse 
discretization.

In summary, non-manifold objects are relevant in a number of computer graphic applications. On the other hand, non-manifolds, probably for their apparent unstructured nature,
are less studied than manifolds and few characterizations
of particular classes of non-manifolds (e.g., r-sets and pseudomanifolds) exist .

\section{Motivation of the Thesis}
{
As we have seen, in several applicative domains, 
non-manifold  are  essential elements.
In spite of this fact, non-manifold features are often neither 
detected nor modeled correctly.
We believe that this is a consequence of the fact that a 
mathematical framework specialized for non-manifoldness is missing.
Thus,  few approaches to non-manifold modeling exist.
Furthermore, they are limited to surfaces \cite{Gur90,LeeLee01,Wei86}. 
Most of approaches for modeling volumetric data (i.e. tetrahedralizations) are limited to the 
manifold domain \cite{DobLas87,LopTav97}.

Another problem is that existing data structures for boundary 
representations of non-manifold solids 
\cite{Gur90,LeeLee01,Wei86} are quite space-consuming. This is a consequence
of the fact that these modeling approaches implicitly 
assume that non-manifoldness can occur very often in the model. The resulting data structures are 
designed to accommodate a singularity everywhere in the modeled object. 
Thus, storage costs 
do not scale with the number of non-manifold singularities. 
On the other hand, much more compact 
data structures for subdivided 2-manifolds and
3-manifolds  do exist \cite{Bau72,DobLas87,GuiSto85,LopTav97,Man83}. 

{
One of the conjectures at the basis of this work
is that it could be possible,
for a wide class of objects, to provide a more compact representation.
This seems  possible by modeling a complex  through its decomposition.
}
We 
expected to obtain compact non-manifold modeling  by
breaking a non-manifold mesh into (possibly) manifold parts  and by coding both object parts and assembly  separately.

The major problem in taking advantage of this idea
lies in the fact that a decomposition of a non-manifold object
is not easily available.
We believe that the decomposition concept, in general,  is not clearly 
defined, too.
Some attempts in this direction are limited to surfaces
\cite{Des92,FaRa92,Gui99,Gui98,RosCad99}.
}

{In general, all existing proposals develop  a
decomposition approach that partition a complex into maximal manifold or
pseudomanifold connected components. 
This 
requirement about maximal components is quite ''natural'' since, otherwise
the collection of all top simplices in the original complex, each considered as separate 
component, would be a dumb solution to the decomposition problem. 
Unfortunately, already for surfaces, it easy to spot examples where 
several non equivalent, non trivial, decompositions 
exist (see the example in Figures \ref{fig:moebius1} and \ref{fig:moebius2}). 
This point is not sufficiently considered 
in existing approaches (with 
the notable exception of \cite{RosCad99}). 
}

{Furthermore, decomposing into {\em manifolds} seems to be a theoretically
hard problem.
As a  consequence of some 
classical results in combinatorial topology \cite{Mar58,Vol74}, 
there could not exist a decomposition algorithm,
for $d\ge 6$, 
that splits a generic $d$-complex into maximal manifold parts.
Such a decomposition problem is actually 
equivalent to the recognition problem for d-manifolds. This problem is settled for d = 4 \cite{Tho94}, it is still an open 
problem for d = 5, and is known to be unsolvable for $d\ge 6$ \cite{Vol74}. 
}
{
\begin{figure}[h]
\begin{center}
\framebox{
\parbox[c][0.43\textwidth]{0.30\textwidth}{
\begin{center}
\vfill
\psfig{file=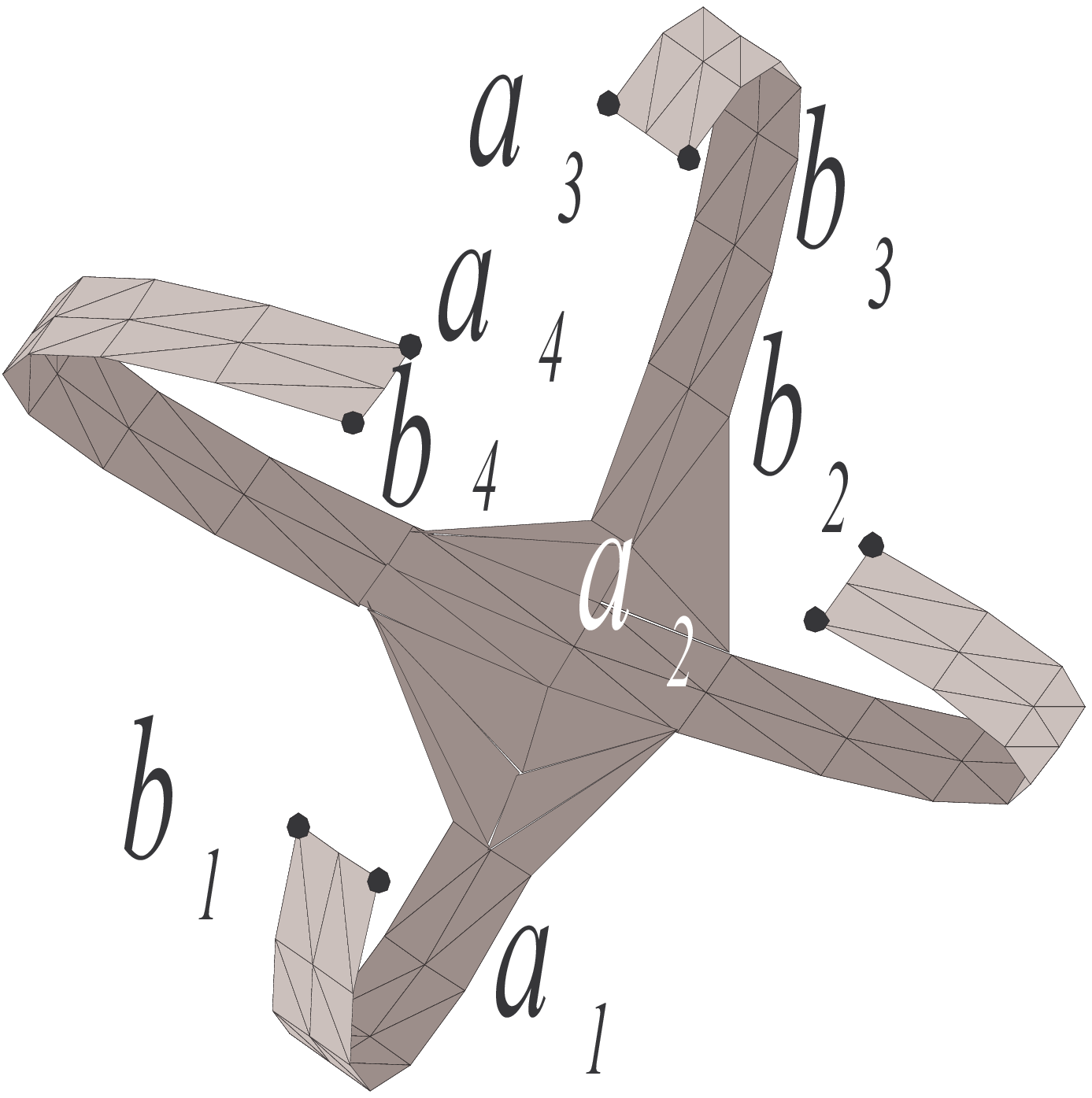,width=0.30\textwidth}
\vfill
{$\cal M$}\\
\vfill
(a)\end{center}
}
}
\framebox{
\parbox[c][0.43\textwidth]{0.50\textwidth}{
\begin{center}
\vfill
\psfig{file=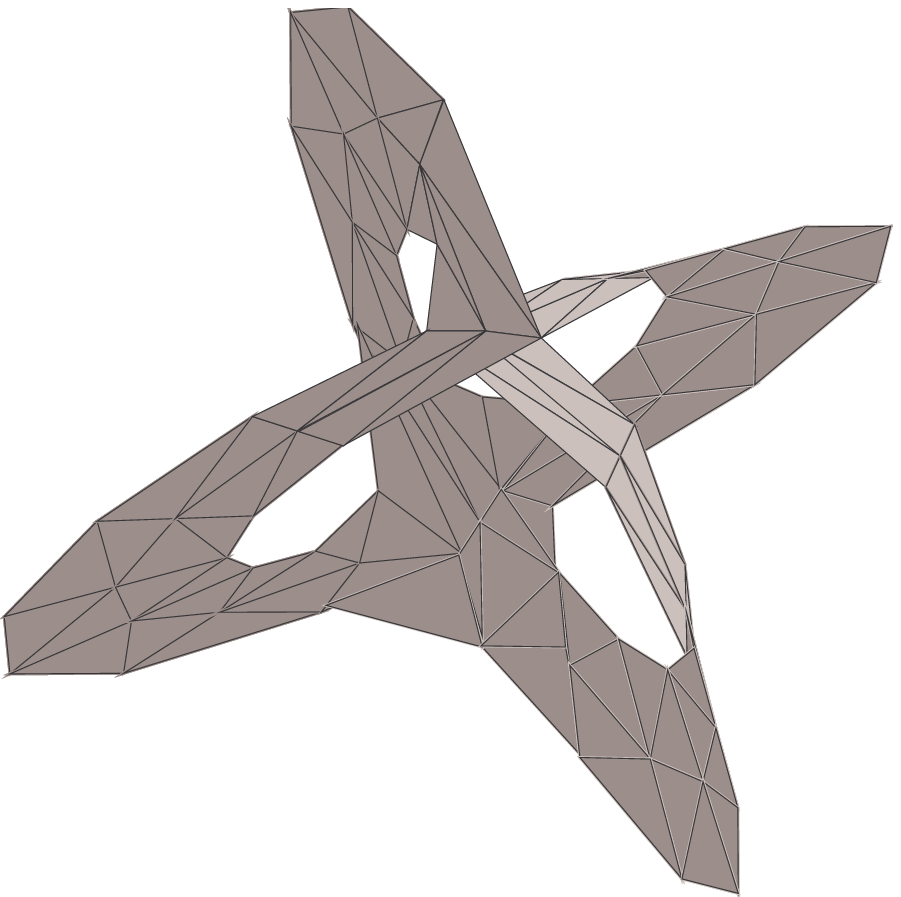,width=0.30\textwidth}
\vfill
{$\{a_1\equivert a_2\equivert a_3\equivert a_4, b_1\equivert b_2\equivert b_3\equivert b_4\}$}\\
\vfill
(b)\end{center}
}
}
\end{center}
\caption{
The Figure  \ref{fig:moebius1}b is a non-pseudomanifold
complex obtained from the sheet  of Figure \ref{fig:moebius1}a identifying
all segments of the form $a_i b_i$, i.e. ''applying''
equations 
$\{a_1\equivert a_2\equivert a_3\equivert a_4, b_1\equivert b_2\equivert b_3\equivert b_4\}$.
Below each complex in Figures \ref{fig:moebius1} and \ref{fig:moebius2} 
is reported the set of equations that tells how  to {\em stitch} together
the sheet $\cal M$ of Figure \ref{fig:moebius1}a to build the complex in  
Figure \ref{fig:moebius1}b.
}
\label{fig:moebius1}
\end{figure}
}
{
\begin{figure}[h]
\begin{center}
\framebox{
\parbox[c][0.40\textwidth]{0.40\textwidth}{
\begin{center}
\vfill
\psfig{file=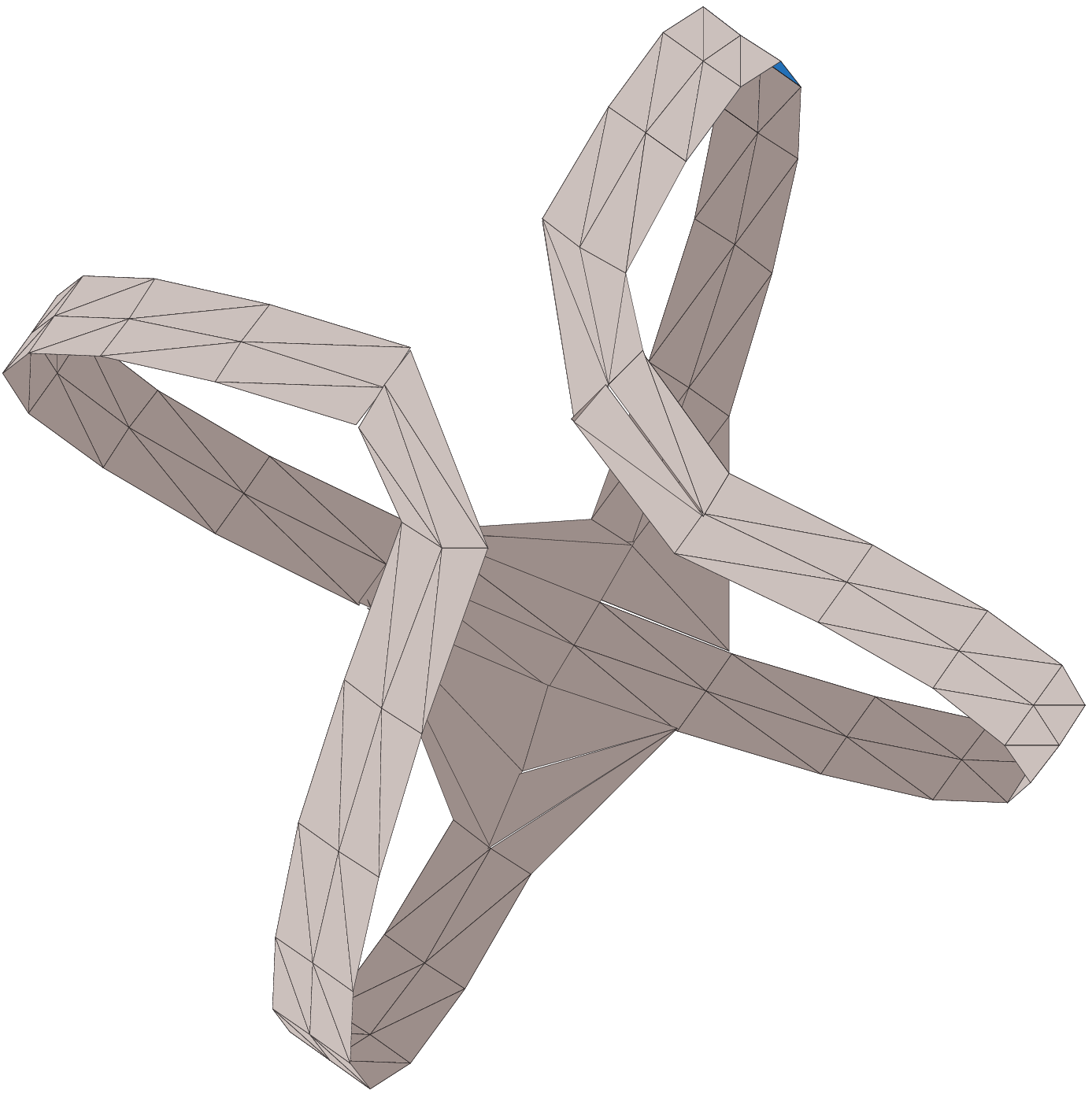,width=0.30\textwidth}
\vfill
{${\{a_1{\equivert}a_4,b_1{\equivert}b_4,a_2{\equivert}a_3,b_2{\equivert}b_3\}}$}\\
\vfill (a)
\end{center}
}
}
\framebox{
\parbox[c][0.40\textwidth]{0.40\textwidth}{
\begin{center}
\psfig{file=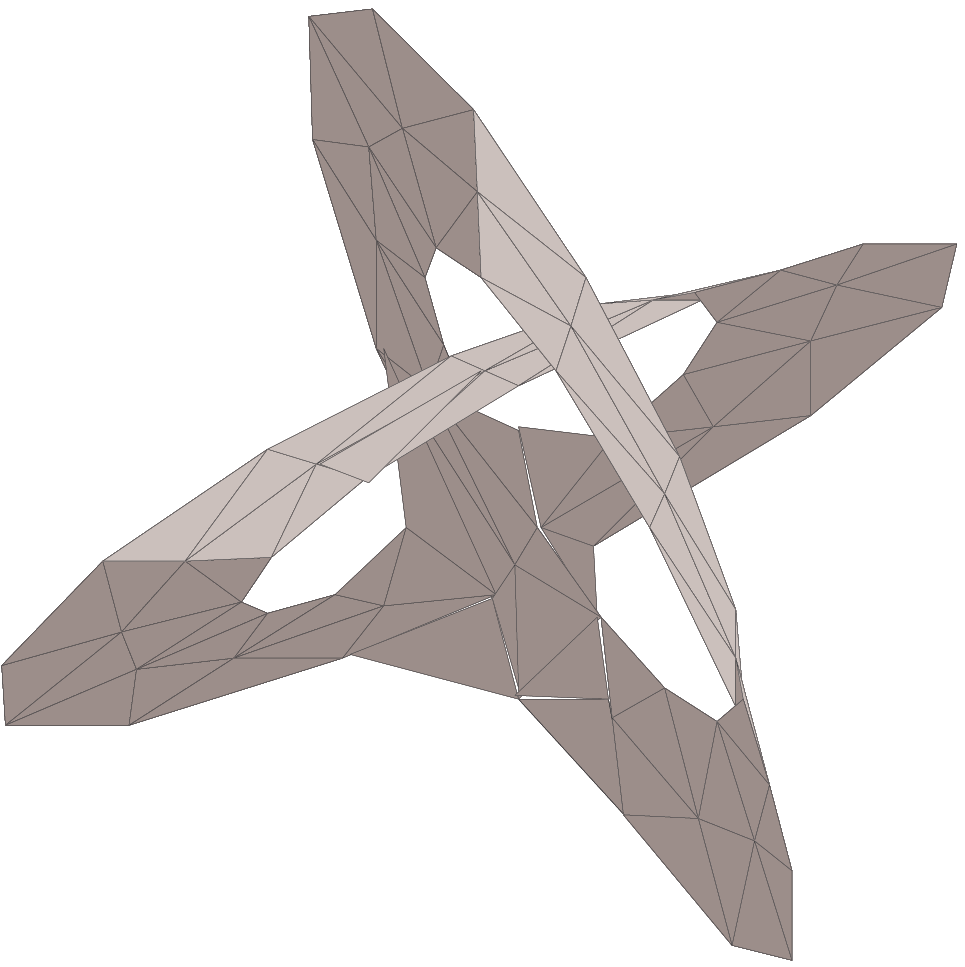,width=0.30\textwidth}
\vfill
{${\{a_1{\equivert}a_3,b_1{\equivert}b_3,a_2{\equivert}a_4,b_2{\equivert}b_4\}}$}\\
\vfill (b)
\end{center}
}
}
\end{center}
\caption{
The two complexes in Figures  \ref{fig:moebius2}a and \ref{fig:moebius2}b
are two different  decompositions
of the complex in Figures  \ref{fig:moebius1}b.
Both such decompositions consist of one
connected component and can be obtained from complex of 
Figure  \ref{fig:moebius1}b with a minimal number of cuts.
Nevertheless, these two optimal decompositions are non homeomorphic.
The complex in \ref{fig:moebius2}a is orientable while the complex in
\ref{fig:moebius2}b is not.
}
\label{fig:moebius2}
\end{figure}
}
\section{Goal of the Research}
The goal of this thesis is to study the non-manifold
domain through decomposition and to develop a non-manifold
modeling approach based on this decomposition. 

\subsection{Decomposition}
A possible approach for decomposing a non-manifold object is to cut it at those elements
(vertices, edges, faces, etc.) where non-manifold singularities occur.
The result of such a decomposition should be a collection  of
singularity-free components.
Different components should be linked together at geometric elements
where singularities occur.
Figure \ref{fig:joints} depicts an example of a non-manifold object
and of one of its possible decompositions.
{
	\begin{figure}[h]
		\begin{center}
			\framebox{
				\parbox[c][0.40\textwidth]{0.30\textwidth}{
					\vfill
					\psfig{file=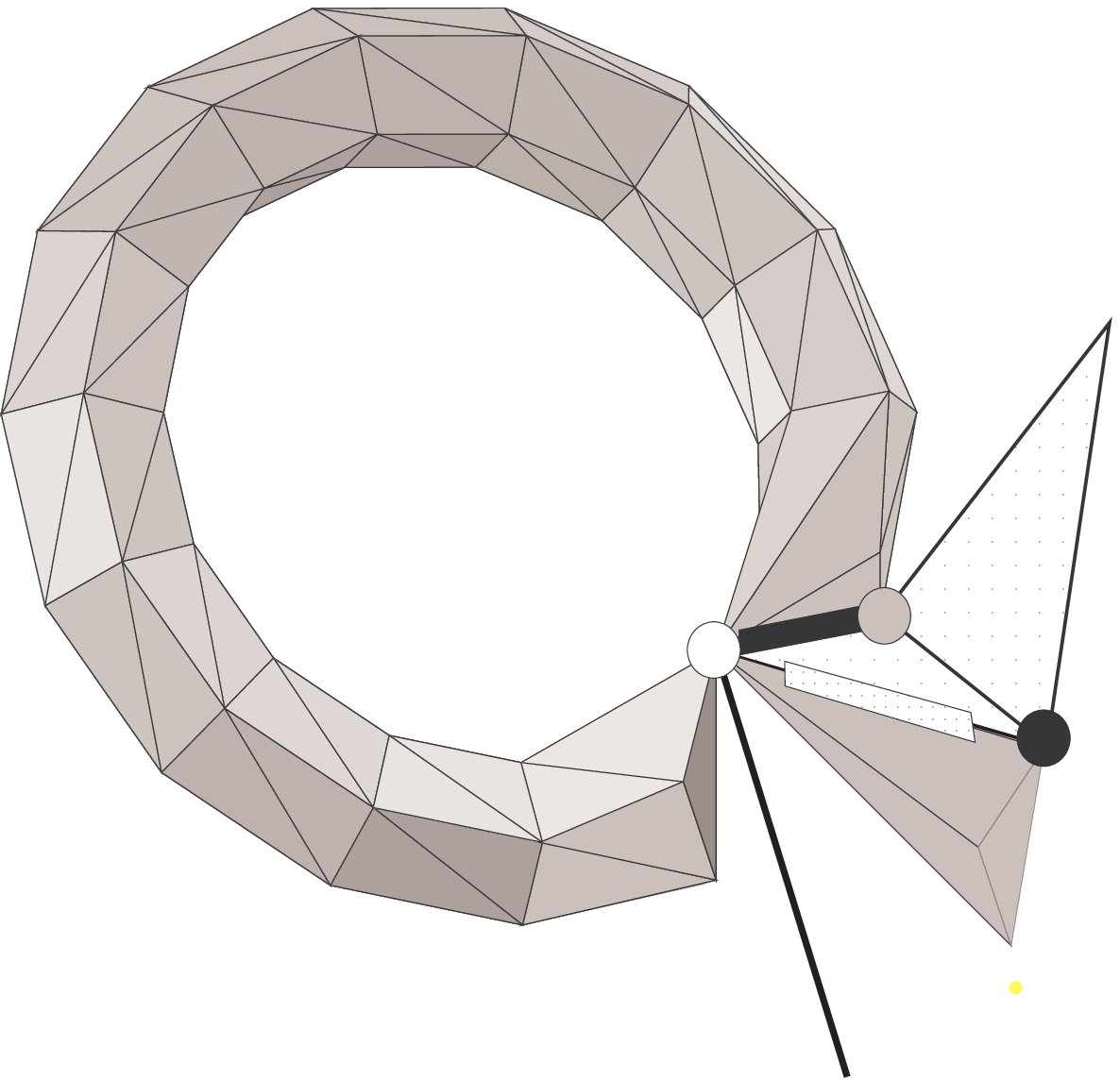,width=0.30\textwidth}
					\vfill
					\begin{center}
						(a)
					\end{center}
				}
			}
			\framebox{
				\parbox[c][0.40\textwidth]{0.30\textwidth}{
					\vfill
					\psfig{file=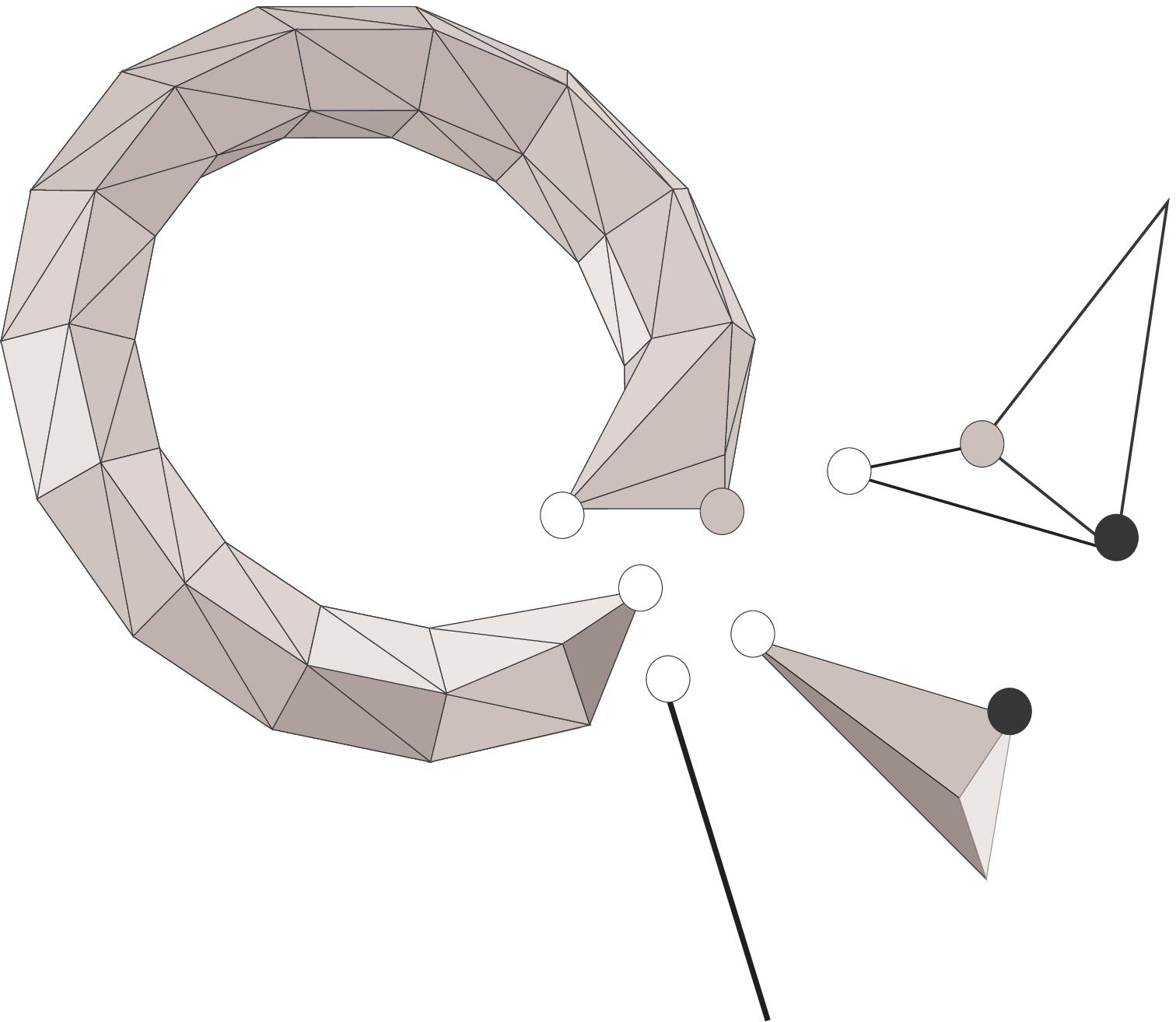,width=0.30\textwidth}
					\vfill
					\begin{center}(b)\end{center}
				}
			}
		\end{center}
		\caption{An example (a) of a non-manifold object (described by a three-dimensional simplicial complex made of tetrahedra, triangles and edges)
			with a dangling edge (A) and a
			dangling surface formed by two triangles (B) and (C)
			and its decomposition (b) into ''simpler'' components.}
		\label{fig:joints}
	\end{figure}
}

{
However, this definition poses some problems. In 
two or higher dimensions,
we have the above mentioned problem of non-uniqueness of
the decomposition (see Figure \ref{fig:moebius2}).
In three or higher dimensions 
a decomposition into manifold components
may need to introduce artificial cuts through certain objects.
Figure \ref{fig:pinchedpie}a shows an example of such an object: 
this complex consists of fourteen tetrahedra forming a
fan around point $p$. This object is a non-manifold object and point
$p$ is a singularity.
In order to eliminate the singularity, we necessarily have to cut the
object through a manifold face, like the triangle $pqr$
(see Figure \ref{fig:pinchedpie}b). This, again, can be done in several ways.
}
{
\begin{figure}[h]
\begin{center}
\framebox{
\parbox[c][0.40\textwidth]{0.30\textwidth}{
\vfill
\psfig{file=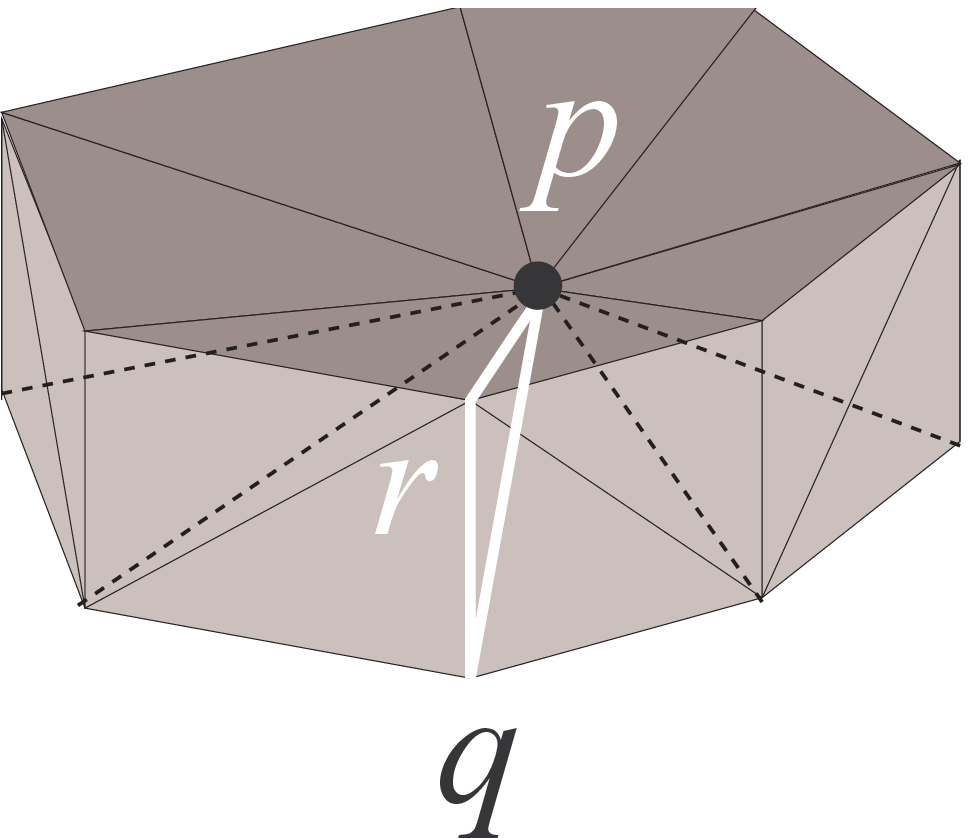,width=0.30\textwidth}
\vfill
\begin{center}(a)\end{center}
}
}
\framebox{
\parbox[c][0.40\textwidth]{0.30\textwidth}{
\vfill
\psfig{file=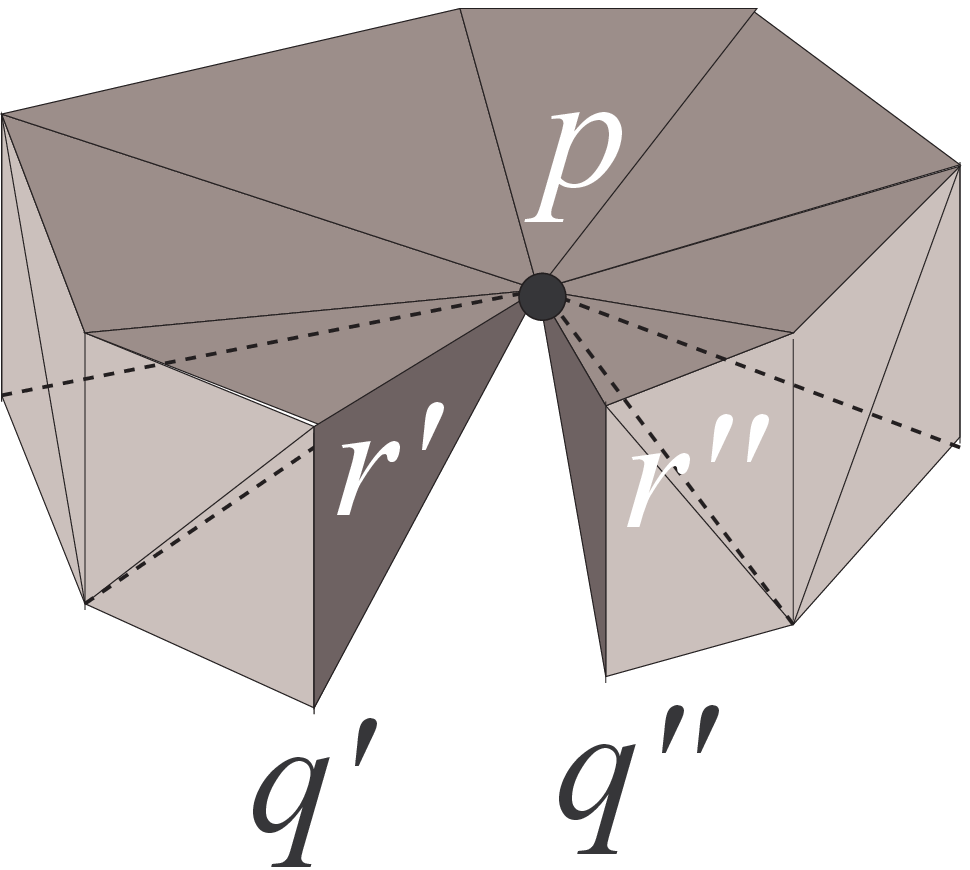,width=0.30\textwidth}
\vfill
\begin{center}(b)\end{center}
}
}
\end{center}
\caption{A non-manifold complex made up of of fourteen tetrahedra
(a) that can be decomposed into a $3$-manifold (b)
 by a cut at the non singular (thick) triangle $pqr$.}
\label{fig:pinchedpie}
\end{figure}
}

From examples like this we became aware that a  decomposition problem 
actually exists and started to look for a theoretical solution.
Thus, our first goal was to define a notion of decomposition 
that identifies a 
unique decomposition even if  several  decompositions
exist. A second issue was to characterize those complexes,
like the one of Figure \ref{fig:pinchedpie}, that appear as 
{\em unbreakable}. 
As a consequence, we expected to devise a decomposition algorithm
that aims to a unique solution and thus does not  need to use
any heuristic based optimization process as in \cite{RosCad99}.
Furthermore, we assumed that this subject could be investigated with a
dimension independent approach that should characterize, in a uniform
framework, unbreakable $3$-complexes, 
like the one in Figure \ref{fig:pinchedpie}. 

To study the decomposition problem we considered 
a description of non-manifold objects by using 
{\em \asc es} as basic modeling tools. 
In this way it is possible to study singularities from a purely
combinatorial point of view. 
To this aim we adopted the  framework
of combinatorial topology \cite{Gla70,Hud69} as basic
mathematical tool. Moreover, we felt that combinatorial models could be the necessary basis for designing effective data structures in solid modeling.

A second issue was the study of the  non-manifold domain through the
decomposition  process. We expect that it could be possible to
give a topological characterization of the class of complexes that do not split
nicely under decomposition (e.g., the complex of Figure \ref{fig:pinchedpie}).
In turn this will yield  a characterization of the topological
properties of the parts produced by the decomposition process.

Finally, 
in order to define a unique decomposition, we expected that 
the  set of all possible decompositions can be 
equipped with an order relation of the type ''{\em more decomposed than}''
such that the set of decompositions will be  both a
partially ordered set (poset) and a lattice.

\subsection{Modeling through Decomposition}
The first part of this thesis deals with the achievement of previously mentioned goals, i.e. the definition of a dimension independent decomposition algorithm. This is the key enabling factor for the 
definition of a dimension independent data structure. The second part of this thesis deals with a  particular approach to  
non-manifold modeling 
through complex decomposition.
The basic idea behind this kind of modeling is
that the information contained in a solid model comes from two rather  {\em independent} sources of information
that are the {\em structure} of the object and the description of 
its {\em parts}. 
The main goal of this second part of the thesis is to show that 
a compact representation for non-manifolds can be devised representing
structure and parts separately.
Parts are usually more regular (i.e. {\em manifolds})
and very compact modeling schemes are known for manifolds. 
Our hope was to show that modeling separately the decomposition structure and
its parts would lead to a compact modeling approach.
We expected to use less space than approaches that  use data structures that can encode non-manifold singularities everywhere. 
Second we expect that it would have been easy to extend this approach 
beyond the realm of surfaces to a generic $d$-complex.  

We expected to attain these goals using
a layered data structure that exploits the outcomes of the decomposition
process. 
In this (foreseen) two-layer data structure the upper layer should be used 
to encode the structure of the decomposition. 
On the other hand, the lower level will be used to 
encode the  components of the decomposition. 
Our initial assumption was that the 
theoretical results obtained for the decomposition process should
support this claim. Furthermore, we expected that, 
in the decomposition process, just singular (i.e. non-manifold) 
vertices are duplicated across different components.
We will call here, and in all the thesis, these vertices the
{\em splitting} vertices. 
The solution seemed to be feasible and especially elegant since the
large amount of theoretical  work behind the decomposition process allows us 
to describe non-manifoldness just keeping track of splitting vertices.

Thus, the upper layer actually is a thin layer, just encoding splitting vertices. The hard part of 
the modeling effort goes in  the definition of 
a data structure that models the components that comes out of our decomposition.
As the example of Figure \ref{fig:pinchedpie} shows, in 
three or higher dimensions there are non-manifold complexes that appear with no
assembly structure. They are inherently {\em unbreakable} 
and must be modeled as a single component.
Thus, the part, we concentrate our research on, was to match 
the decomposition outcomes with  a modeling approach that models
correctly the class of  components arising from the 
decomposition.

{
Finally, another (self-imposed) commitment 
was the need for a detailed {\em theoretical} analysis of the two 
layer data structure.
In this theoretical analysis our goal was to evaluate  both space and time 
requirements for the proposed data structure.
More in particular the goal of this evaluation is detailed in the following
checklists.

The checklist for the evaluation of space requirements was basically
made up of the following tasks:
\begin{itemize}
\item evaluate space necessary to model decomposition components;
\item evaluate space requirements 
when our approach is used to model 
$2$-manifolds, $3$-manifolds or $d$-manifolds. Compare this with existing 
approaches for manifolds. 
\item evaluate space needed to model the connecting structure of the decomposition;
\item evaluate space needed to model a non-manifold through its
decomposition;
\item compare our space requirements with existing approaches 
for non-manifold modeling;
\item  evaluate the critical ratio between manifoldness/non-manifoldness
that makes our approach more compact than others  in non-manifold 
modeling.
\end{itemize}
Another evaluation criterion has been  time complexity for
the extraction of topological relations 
(e.g., given a vertex, find all top simplices incident into 
that vertex).  
The checklist for this evaluation can be divided into two checklists.
The first is for the performance of the data structure for 
components. This checklist is the following:
\begin{itemize}
\item evaluate time needed to extract topological relations within a 
component; 
\item find under which conditions topological relations can be extracted in a time that is linear with respect to the size of the output; 
\item evaluate the influence of dimensionality on time complexity and find 
conditions, if any, 
in which topological relations cannot be extracted in a time that 
is linear with respect to the size of the output; 
\end{itemize}
Another checklist is used to evaluate time requirements for the 
overall two-layered data structure. \begin{itemize}
\item
evaluate the time necessary 
to build the data structure using the outcomes of the decomposition
process
\item evaluate the time complexity for extracting topological relations in a 
complex modeled through its decomposition;
\item find under which conditions topological relations can be extracted in a time that is linear with respect to the size of the output; 
\item evaluate the influence of dimensionality on time complexity and find 
conditions, if any, 
in which topological relations cannot be extracted in a time that 
is linear with respect to the size of the output; 
\item find a relation between the extent of non-manifoldness in the 
model and the time required to extract topological relations
\end{itemize}
Finally, we aimed, with the highest priority,  at meeting, 
the following requirements:
\begin{itemize}
\item  our approach must be more compact than existing approaches
for non-manifolds whenever the extent of  non-manifold situations is
substantially negligible in the combinatorial structure.
\item 
when the extent of the non-manifold situations is negligible, the 
space requirements must be comparable with that of most common 
approaches for manifold modeling;
\item 
the extraction of all topological relations should be performed
in a time that is linear with respect to the size of the output.
This should happen at least  in most relevant
applicative domains. In particular,
 extraction in linear time should be guaranteed
for non-manifold complexes of dimension two and  three emdeddable
in the Euclidean space. 
\end{itemize}
\section{Contribution of the Thesis}
This thesis studies, from a mathematical point of view, 
the problem of decomposing  non-manifolds in any arbitrary
dimension
and presents a dimension independent data structure for 
non-manifold modeling through complex decomposition.

The work in this thesis starts from a 
precise, mathematical statement of the decomposition problem. 
Based on this we give a dimension-independent notion of a 
{\em \cano} decomposition.
The problem of non-uniqueness of the
decomposition is discussed and  settled by defining a criterion to select 
the most general decomposition among all possible options. 
Existence and uniqueness of this decomposition  is mathematically established
and an effective algorithm to compute this {\em \cano} decomposition is proposed.
The topological properties of components in our decomposition 
are studied and precisely characterized.

We have  developed a framework for object 
decomposition that captures, 
through a systematic approach, all possible decompositions of
an input complex $\AComp$. Obviously there are several decompositions of $\AComp$. They are somehow {\em intermediate} between $\AComp$
itself, and the complex $\AComp^{\top}$ formed by the totally disconnected 
collection of all top simplices in $\AComp$.
We show that such decompositions form a {\em lattice} in which the top is the 
complex $\AComp^{\top}$ and the bottom is the complex $\AComp$.
Transitions between an element and its immediate successors, in this lattice, occurs through 
{\em stitching} a pairs of vertices.

In this lattice we define the {\em \cano\ decomposition} of
$\AComp$ (denoted by $\canon{\AComp}$) as that complex that is obtained  
from $\AComp^{\top}$ by {\em gluing}
all top $h$-simplices putting {\em glue} just on $(h-1)$-faces 
that are manifold faces in $\AComp$. We give a mathematical, dimension
independent, formulation of intuitive concepts such as:
{\em cutting}, {\em stitching vertices} and {\em gluing faces}. 
Next, we have 
proven that the {\em \cano\ decomposition} is unique and
that it is the most general 
decomposition that can be obtained by cutting the original complex
only at non-manifold faces. 
The connected components of the \cano\ 
decomposition are thus complexes, like the one of Figure \ref{fig:pinchedpie},
from which singularities cannot be
eliminated by cutting the complex at manifold faces.
We call such a complex an {\em initial-quasi-manifold} 
and we develop a 
characterization of \cdec\ complexes in term of 
local topological properties of the complex.

\Cdec\ complexes are studied and compared with the (few) 
existing classes of non-manifolds. 
In particular, we have proven that \cdec s are manifold 
in dimension two, i.e. the class of 2-\cdec and 2-manifolds coincide. 
In dimension $d=3$ or
higher d-\cdec are neither manifolds nor pseudomanifolds i.e. there are \cdec that are neither manifolds nor pseudomanifolds. Quasi-manifolds, introduced by \cite{Lie94}  are a proper subset of \cdec s, being
the set  of \cdec\ that are also pseudomanifolds.
A  rather counter intuitive finding of this analysis is that
there exist non-pseudomanifold $3$-complexes  (although not imbeddable in
$\real^3$) that can be generated by 
gluing together tetrahedra at triangles 
where just two tetrahedra glue at time. 
In other words a non-pseudomanifold adjacency, where three tetrahedra share the same triangular face, can be induced
using the (usual) manifold glue (i.e. manifold adjacency) on
triangles.

The \Cdec s, unlike manifolds, are a decidable class of 
complexes in any dimension.  The \cano\ decomposition itself can be computed 
in linear time with respect to.  the size of the complex $\AComp$.
An algorithm to compute the \cano\ decomposition
$\canon{\AComp}$  is proposed and we have shown that the 
output of this algorithm is sufficient to build a two-layered data
structure for $\AComp$.   

{
Using the results of the decomposition investigation we defined a two-layered data structure
that we called the {\em \nmwds\ representation}.
The \nmwds\ representation represents a 
non-manifold complex using its \cano\ decomposition.
First each component is encoded 
using an extension of the {\em Winged Representation} \cite{PaoAl93}.
We called this extension the {\em \cdecrep\ representation}.
Second, in an upper layer, we  encode
instructions necessary to stitch \iqm\ components together.
}

The \nmwds\ representation is designed to be extremely compact and yet  
to support retrieval of all topological relations  in  time 
linear with respect to the size of the output.
This time performance is achieved for $2$-manifolds and for 
$3$-manifolds embeddable in $\real^3$.
The  proposed data structure is more compact than existing data structures for
non-manifold surface modeling. In particular, the proposed data structure is fairly good for
 objects made up of few {\em nearly} manifold parts 
tied together with (a not-so-large number of) non-manifold joints.

\section{Thesis Outline}
This thesis consists of nine chapters  plus an Appendix.

{{\bf Chapter \ref{ch:related}}, provides an overview of the 
state of the art.
We review modeling approaches for non-manifolds,
for $3$-manifolds and the few dimension-independent modeling approaches.
The reviewed modeling approaches are presented in a uniform framework and
space requirements for each approach is evaluated.
In the second part of this chapter we review papers on non-manifold
surface decomposition.
Finally, a certain number of classic results in combinatorial topology are
presented in order to give an account of the known theoretical problems 
one can meet when going in higher dimension.
}

In {\bf Chapter \ref{sec:cellular}}, we introduce some basic
notions from combinatorial topology.
In this chapter we added also some 
results from point set topology.
This material,
although helpful to understand combinatorial concepts, is
actually unnecessary to develop our results.   
This optional material
is reported in this chapter with a starred header (e.g., {\bf Definition~*}).

However, we will use these geometric concepts both 
in examples and in our quotations from classical 
handbooks in combinatorial topology (mainly \cite{Gla70,Hud69}).
In other words, we need geometric concepts in order
to state classic results in combinatorial topology in their original
form.

At the end of this chapter
we will introduce, in Section \ref{sec:nerve}, the three 
not-so-standard concepts of: {\em nerve, pasting and \quot\ space},  
The  {\em nerve} concept is needed for the definition of the  {\em \quot s} of an \asc\ $\AComp^\prime$ modulo an equivalence 
relation $R$
(denoted by $\AComp^\prime/R$).
\Quot s, in turn, are  crucial in the definition of the {\em decomposition}
concept that will come in Chapter \ref{ch:decomp}.

In {\bf Chapter \ref{ch:quotlat}},
we first present the relation between abstract simplicial maps and \quot s.
We show that the set of all \quot s
of a given \asc\ $\AComp^\prime$ form a lattice that we called the
{\em \quot\ lattice}. 
The \quot\ lattice is isomorphic to a
well known lattice $\Pi_n$ called the {\em partition lattice}.
Mathematical properties of this lattice are  given in \latt 
that gives a short introduction to the notions from Lattice
Theory needed in this thesis. 
However, in the first part of this chapter, relevant properties
of lattices are 
summarized and restated, in an intuitive form,  
using a language closer to the subject of this thesis.  

Lattices, in the context of this thesis, will be used as  the
structure in which we order the decompositions of a given complex
(we anticipate that decompositions are a sublattice of the 
\quot\ lattice that will be is introduced in Chapter
\ref{ch:decomp}).
There is a clear benefit from
organizing decompositions into a  lattice.
In fact, in this way 
we grant a least upper bound for any arbitrary set 
of decompositions. 
This will be a key issue to define a unique decomposition.

Another key idea in the development of this thesis is the fact that we
can manipulate \quot s $\AComp^\prime/R$ using the set of
equations $E$ that defines $R$.
In particular we are interested in the fact
that some topological properties of a \quot\ $\AComp/E$
can be restated in terms of syntactic properties of the set $E$.
The manipulation of these syntactic objects give us a useful tool
to treat topological problems. 
Since these equations identify two
vertices together, we will call them  {\em \verteq\ equations}.
By the end of this chapter, we therefore introduce  { \verteq\ equations} and give 
the relation between a set of equations $E$ and  the 
{\em \quot\ lattice}.

In {\bf Chapter \ref{ch:decomp}},
we define the conditions 
that make {$\AComp^\prime$} 
a decomposition of the complex $\AComp$ obtained as the \quot\ {$\AComp^\prime/E$}.
Intuitively, a complex $\AComp^\prime$ is a decomposition of $\AComp$ if 
we can obtain $\AComp$
pasting together pieces of $\AComp^\prime$. 
Furthermore, we expect that nothing shrinks passing from 
$\AComp^\prime$ to  $\AComp$.
Following this idea, in this chapter, we define the notion of decomposition 
and define a sublattice of a specific \quot\ lattice that we 
called the  {\em \dec\ lattice}.
This lattice  contains an isomorphic copy for any \dec\ 
of a given complex $\AComp$. 
On top of the \dec\ lattice we have the totally exploded version
$\topAComp$ of $\AComp$. This is the complex consisting of all top
simplices in $\AComp$, each one considered as a distinct connected component.
At the bottom of the \dec\ lattice we have
(an isomorphic copy of) the complex $\AComp$.
We can walk on the \dec\ lattice from  $\topAComp$ to
$\AComp$ adding equations whose
basic effect is to stitch together two Vertices 
that belongs to two distinct simplices.

In {\bf Chapter \ref{sec:classify}}, we present  a more abstract view
of the \dec\ lattice. This view brings us closer to the 
solution of the decomposition problem.
In the previous  chapter we have studied the \dec\ lattice 
for a complex $\AComp$. We have seen that 
we can walk on the \dec\ lattice adding equations. Each equation has the 
effect of stitching together two Vertices that 
belongs to two distinct simplices.
This view of the \dec\ lattice is too fine-grained to be useful in this context.
In this chapter we take a different  look to the \dec\ lattice.
We imagine that we do not have the option to stitch just two vertices at 
time but we are forced to glue together two top simplices  $\theta_1$ and 
$\theta_2$ by gluing together 
all Vertices that  $\theta_1$ and $\theta_2$
have in common in $\AComp$. 
We will call this move a {\em \sglinst}.
Obviously,  \verteq\ equations provides a more, fine grained, view of the 
\dec\ lattice. In turn a \sglinst\ is, basically, a macro expression for
a set of \verteq\ equations. 

Thus, in this chapter, we introduce \sglinst s and define 
the subset of \dec s generated by a set of \sglinst s ${\cal E}$
(usually denoted by {$\topAComp/{\cal E}$}).
A discussion on the structure of the set of decompositions 
{$\topAComp/{\cal E}$} closes this chapter. 
In particular we show that 
not all \dec\ can be generated as a  \quot\ of the form 
{$\topAComp/{\cal E}$}.
Furthermore we show that 
the set of decompositions of the form {$\topAComp/{\cal E}$} is 
not a sublattice of the \quot\ lattice. 
Nevertheless, the (fewer) complexes of the form  {$\topAComp/{\cal E}$} are 
sufficient to treat the decomposition problem. This is a consequence of 
two fundamental lemmas stated at the beginning of 
Chapter \ref{ch:stdec}. 

}

In {\bf Chapter \ref{cc:toposgi}}, 
we study topological  properties
of the decomposition {$\topAComp/{{\cal E}}$}
studying the  syntactic properties of the set {$\cal E$}.
It is possible to relate the topological properties of 
{$\topAComp/{{\cal E}}$} with syntactic properties of the set
of \gl\ \inst\ {${\cal E}$}.
We first  consider the usual topological properties 
defined in Chapter \ref{sec:cellular}  such as regularity, connectivity,
pseudomanifoldness and manifoldness. 

Next we will consider 
{\em \Qm s}
\cite{Lie94} and a superset of \qm s we called {\em \Cdec s}.
In this chapter \qm s are defined in terms of syntactical  properties 
of the generating set of \sglinst s {${\cal E}$}. This definition is 
proven to be equivalent to the definition given by \cite{Lie94}.

Then \cdec s are defined in term of syntactical  properties 
of the generating set ${\cal E}$, too.
Next we prove that initial-quasi-manifolds can be defined in terms 
of local properties each vertex must have. 
Indeed we have found that, in an initial-quasi-manifold, the  star of each vertex
has a constant peculiar structure.  In fact, every couple of top d-simplices in a star must be connected 
with a path of d-simplices, each linked to the other via a (d-1)-manifold (non singular) joint.

This local property is sufficient to   prove that initial-quasi-manifold d-complexes are a proper superset 
of d-manifolds for $d\ge 3$. 
They coincide with manifolds for d = 2. 
They are a decidable set of d-complexes for any d. 
Finally we give an example of an initial-quasi-manifold 
tetrahedralization that is not  pseudomanifold. 
Such a tetrahedralization, however, cannot be embedded in $\real^3$ . 

In {\bf Chapter \ref{ch:stdec}}, 
we first prove two results that enables us  to use just \sglinst s in order
to treat the decomposition problem.
We prove that sets of \sglinst s are sufficient to label 
{\em every path} from any decomposition {$\AComp/{\equivert}$}
down to $\AComp$.
Thus we restrict our attention to transformations induced
by sets of \sglinst s ${\cal E}$ and study the relation between
syntactic properties of the set ${\cal E}$ and topological 
properties of the transformation from $\AComp^\prime$ to
$\AComp^\prime/{\cal E}$.

Next we define the class of decompositions we are interested in.
In particular we are interested in
decompositions that split only at non manifold simplices.
We will introduce in this chapter the class of ''interesting''
decompositions that 
we called {\em \natu} decompositions.
Then, we define the {\em \cano\ decomposition} as the
the least upper bound of the set of \natu\ decompositions.
Due to lattice structure, this complex exists and is unique.
We prove that such a least upper bound is still an \natu\ decomposition.
Several properties of the \cano\ decomposition are given, then.
In particular, we prove that the connected  components
of the \cano\ decomposition are \cdec s.

Next, we present an algorithm that transforms a 
complex into its standard decomposition by a sequence of local operations modifying just simplices 
which are incident at a vertex. Each local operation is computed using local information about 
the star of the vertex (i.e., the set of simplices incident to a vertex). 
Finally we prove that this computation can be done  in  $\lesscomp{t\log{t}}$,
where $t$ is the number of top simplices in the original complex.

In {\bf Chapter \ref{ch:nmmdl}}, we define
a two layer data structure,
we called the {\em \nmwds\ representation}.
The \nmwds\ representation represents a 
non-manifold complex using its decomposition.

Each component is encoded 
using an extension of the {\em winged representation} \cite{PaoAl93}.
This extension, that we called the {\em \cdecrep\ representation}, is
carefully presented
and its space 
requirement are assessed. We give 
algorithms to construct the \cdecrep\ data structure using the results of the 
decomposition process. Next we develop algorithms to extract 
topological relations in a single component.
The complexity of these operations is then analyzed.  

In a second step, we define a data structures that encodes
the information
necessary to stitch components together.
This completes the definition of the \nmwds\ data structure.
Algorithms to build this data structure are proposed and their time 
complexity is evaluated.
Next, we develop algorithms to extract 
topological relations in the non-manifold complex.
Finally, space requirements of the \nmwds\  representation
are compared  with space
requirements of other modeling approaches. The 
conditions that make this structure more suitable than others are
discussed.

In {\bf Chapter \ref{ch:concl}} we  briefly summarize the results 
the results of this thesis.open problems.

In {\bf \latt}, we resume basic notions of Lattice Theory and introduce 
the {\em partition lattice}.
In {\bf Appendix \ref{sec:opt}}, we describe the (rather tedious)  details of a space optimization for the \Cdecrep\ Data Structure. 
In {\bf Appendix \ref{sec:prolog}}, we describe a Prolog program that checks the correctness of Example \ref{app:example}.
 \chapter{State of the Art}
\label{sec:related}
\label{ch:related}
\section{Introduction}
In this thesis we develop a  decomposition procedure for a
generic simplicial complex. This decomposition procedure cuts
the simplicial complexes only at non-manifold singularities.
This decomposition is used to build a two layer data structure.
In the lower layer we represent decomposition components.
In the upper layer we tie together decomposition components.
Whenever the  decomposed complex is a $2$-complex we have that
decomposition components are manifold surfaces. 
Thus, this two layer approach, gives a data structure whose \occup\  might be
similar to the \occup s of standard data structures for manifold modeling.
This could happen whenever the degree of non-manifoldness is  low.
The \occup\ then
scales up with the degree of non-manifoldness 
in the decomposed complex. 

We found that the subject of this thesis, 
with a careful choice of the theoretical framework, 
can be developed with a dimension independent formulation
and thus we developed a dimension independent approach.

Existing related literature for this kind of study is 
surely the literature on manifold and non-manifold modeling.
For this reason, in the first part of this  chapter,
{we review modeling approaches for manifold surfaces, 
for $3$-manifolds, for non-manifolds and 
the few dimension-independent modeling approaches.
The reviewed modeling approaches are presented in a uniform framework and
space requirements for each approach is evaluated.
In the second part of this chapter we review papers on non-manifold
decomposition.
Finally, a certain number of classic results in combinatorial topology are
presented in order to give an account of the known theoretical problems
one can meet in a dimension independent formulation.

This chapter is organized as follows.
In Section \ref{sec:maninomani}, we discuss a basic problem  in the
relation between approaches for
manifold and non-manifold modeling then we revise modeling approaches
for manifold surfaces (Section \ref{sec:manisurf}) 
and for $3$-manifolds (Section \ref{sec:manitetra}). 
Since we are developing a
dimension independent approach,  in Section \ref{sec:dimindip}
we insert a  review of dimension independent modeling approaches.
Next we revise approaches to model cellular subdivisions of
non-manifolds (Section \ref{sec:nonmanisol}). 
This analysis shows that classic 
approaches for non-manifold modeling are space inefficient when compared
with approaches for manifold modeling. 

Next, in Section \ref{sec:decompart}, we review papers on decomposition of non-manifold models.
Finally, in Section \ref{sec:teorat},
we give a rationale for the purely combinatorial
framework we developed in this thesis and show that
this is necessary if one wants to develop a dimension independent
study of decompositions. \section{Manifold and non-manifold Modeling}
\label{sec:maninomani}
None of the existing modeling approaches for non-manifolds is completely
satisfactory. The few approaches that can represent the full
domain of non-manifold cellular subdivision of non-manifolds 
(e.g. Weiler's Radial Edge \cite{Wei86}) are definitely
space inefficient over the manifold domain.
The classical data structures for manifold surfaces
\cite{Bau72,GuiSto85,PreMul78}
outperform existing data structures for non-manifolds when
the latter are used to model manifolds.
This is shown, for instance, by the quantitative analysis developed in
\cite{LeeLee01} where several
non-manifold modeling  schemes are compared with the two classical 
data structures for boundary representation of manifold objects.
i.e. the Winged--Edge (WE) \cite{Bau72} and the Half Edge
\cite{Man83}
(see also \cite{Lie91} for another comparison).
Some of the results of comparisons in \cite{LeeLee01}
are summarized in Table \ref{tab:haratio}:
{
\begin{table}
{
\begin{center}
\begin{tabular}{|c|l|c|l|}\hline\hline
\multicolumn{2}{|c|}{Modeling Data Structure} & Ratio to WE & Representation Domain \\ \hline
\multicolumn{2}{|c|}{Winged Edge} & 1 & cellular 2-manifolds \\ \hline
1& Radial Edge & 4.4  & cellular 2-complexes \\ \hline
2& Partial Edge & 2.1  & cellular 2-complexes \\ \hline
3& Half Edge & 1.2   & cellular 2-manifolds \\ \hline
\end{tabular}
\end{center}
}\caption{Storage costs normalized with respect to winged-edge storage
requirements. Data are from Table 4 in \cite{LeeLee01}}
\label{tab:haratio}
\end{table}
}
The analysis in \cite{LeeLee01} shows that the radial--edge data
structure encodes manifold surfaces taking more than four times the space 
required by the winged--edge.

{
All data structures for non-manifold modeling have high
storage requirements if compared with data structures for manifold modeling.
None of the classic data structures for non-manifold modeling have storage 
requirements that scales with the degree of non-manifoldness in the modeled
object. In other words these data structures seems extremely space consuming
when they are used to encode manifolds or ''nearly'' manifold complexes.   
This situation, far from being satisfactory, is one  of the starting points
of this thesis.  To fully understand this problem,
in Sections \ref{sec:manisurf} and \ref{sec:manitetra} we review 
classical results for manifold modeling. This will 
provide a benchmark against which we will compare the data
structure for modeling the decomposition components we will describe in
Section \ref{sec:ewrep}.  

In section \ref{sec:dimindip} we present four 
dimension--independent modeling
approaches: the cell-tuple \cite{Bri93}, the selective geometric complexes
\cite{RosCon90}, the n-G-maps \cite{Lie91} and the winged representation \cite{PaoAl93}.
These provides another set of benchmarks for 
the data structure designed in this thesis.
Furthermore, at least for n-G-maps, some results in this thesis,
mainly  Property \ref{sec:quasi} can be
quite useful in the study of this modeling approaches, 
while the results in Chapter \ref{ch:nmmdl} builds upon an extended 
version of the Winged Representation  \cite{PaoBer93}  and extends it  to 
a dimension independent approach for the non-manifold domain.

In section \ref{sec:nonmanisol} we present three modeling approaches that can model
cellular subdivisions of non-manifolds realizable in $\real^3$. These approaches are reviewed
and presented stressing the fact that they all can be understood as
small variations around the original scheme presented in the radial-edge data
structure.    
These provides a set of benchmarks for storage requirements against which we
will compare our data structure.
 
In section \ref{sec:conclsta} we will resume the shortcomings of this
review and discuss the relation of the reviewed material with the  
results of this thesis.
}

\subsection{Data structure for encoding cellular decompositions of manifold
surfaces} \label{sec:manisurf}
In this section we review major approaches to represent $2$-manifolds.
We start presenting  classic data structure for $2$-manifolds 
(Winged--Edge \cite{Bau72}, DCEL \cite{PreMul78},
Half-Edge \cite{Man83,Wei85}).
Next we analyze structures based on
the Incidence Graph \cite{Edel87, Woo85}. 
In the following sections 
we will present data structures for 
$3$-manifolds (the Facet Edge \cite{DobLas87} and the Handle-Face \cite{LopTav97}
data structures).

For each data structure, we will  give an expression for space requirements 
\wrt\  the number of geometric entities in the model.
To this aim, in the following, we will denote with $v$ the number of vertices,
with $e$ the number of edges
and with with $f$ the number of top faces. Similarly, we will use pairs of letters $V$,$E$,$F$ (e.g. VE)  
to denote relations between elements of the model.
We will say that element $x$ of type $X$ and element
$y$ of type $Y$ are in $XY$ relation if one is face of the other. 
If $x_1$ and $x_2$ are both of the same dimension 
(i.e. they are of type $X$), and they share a proper face of 
maximun dimension, then we will say that they are in the
$XX$ relation.  
Thus two faces are in a $FF$ relation if they
share an edge. Two egdes are in a $EE$ relation if they share a vertex.
Any $XX$ relation  is called an {\em adjacency} relation 
while any $XY$ relation, for $X\neq Y$ is called an {\em incidence} relation. 
In general, we will call the $XY$ and the $XX$
relations {\em topological relations}.
Finally, we will  denote with $XY^*$ a function that  is a 
subset of the $XY$ relation. Thus the $VE^*$ relation gives an 
edge incident into a given vertex.

We will call the {\em extraction} of  an $XY$ relation the retrieval of
all $Y$ elements that are in an $XY$ relation with a given $X$ element
$x$. Thus, for instance, to  extract the $VE$ relation we have to find all
the edges that are incident into a given vertex $v$. 
All data structures listed in this section supports the 
extraction  of all topological relations  in a time that 
is linear \wrt\  the size of the output. 
\subsubsection{The \ema{winged--edge}{Data Structure}}
\label{sec:winged}
{
The winged­-edge data structure \cite{Bau72} 
represents each edge of a manifold surface using eight references that 
points eight cells that are incident to an edge $e$. 
With reference to Figure \ref{fig:winged} we have that the eight 
references relative to the thick edge $e$ are: 
two references (PVT, NVT) for incident vertices (encoding the EV relation), 
two  references (PFACE, NFACE) for incident
faces (encoding the EF relation) 
and four references (PCW, PCCW, NCW and NCCW) to the incident edges 
that share with $e$ the same faces and the same vertices. These four references
represents a subset of the EE relation.

For a given edge we choose arbitrarily the first extreme vertex 
PVT and the second extreme vertex NVT thus assigning an orientation to
$e$ from PVT to NVT.
Face PFACE is the face on the left of someone traveling on the oriented edge
standing outside of  the surface.
A simple convention is at the basis of names for the four references
PCW, PCCW, NCW and NCCW.
We have that in the above names, CW stands for clockwise, 
CCW stands for counter-clockwise,
N stands for next and P stands for previous. 
We judge clockwise and counter-clockwise rotations by standing outside 
the surface.
Note that this definition implies that we assume we
are modeling an orientable surface.
Thus the reference   PCW, 
stored for a certain edge $e$, references  the
previous edge, in clockwise order,   around the source vertex PVT. 
The four edges PCW, PCCW, NCW and NCCW are  the 
so called {\em wings} of the thick oriented edge $e$ in Figure \ref{fig:winged}.
\begin{figure}[h]
\centerline{\fbox{\psfig{file=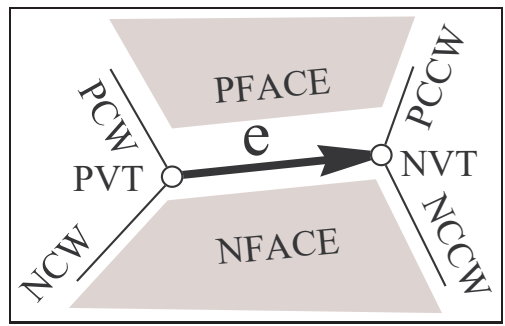}}}
\caption{A geometric realization of an adge in a surface modeled by a 
winged--edge
data structure. Edges and vertices are labeled with references relative 
to the thick oriented edge $e$} 
\label{fig:winged}
\end{figure}

It can be proven that this data structure models orientable 
2­-manifolds subdivided into cell complexes \cite{Wei85}.
To extract all topological relations we need to introduce  
a reference to an incident edge for both vertices and faces 
(i.e., the $VE^*$ and the $FE^*$ relation). 
If we want to retrieve all edges around a face  in a
given, clockwise (CW) or counterclockwise (CCW), orientation we must check that 
the edge we are considering has an orientation coherent with the given 
orientation.
This can be checked by a pair of lookup  into PFACE and NFACE.
These lookups must be repeated for each extracted edge around a face $f$. 
A similar remark holds for the problem of 
retrieving all edges around a vertex  in a
given (clockwise or counterclockwise) orientation.
In this case a double lookup to  PVT and  NVT is needed for each
extracted edge.

A double lookup may also be used if we want to
extend the WE to a non-orientable surface. 
If the modeled surface might be non orientable
then  we cannot assume that the pair of wings
incident  at PVT (NVT) are  labeled as PCW and NCW
(PCCW and NCCW) using   a counter-clockwise 
rotation order judged standing {\em outside} 
the surface.
For a non orientable surface labels PCW and NCW
(PCCW and NCCW) will be assigned using {\em some}
rotational order around  PVT (NVT).
The only constraint is that PCW,edge $e$ and PCCW must bound face PFACE and NCW,$e$,NCCW must bound NFACE. 
When, stating from $e$, 
we extract a new edge $e\prime$,
a first double lookup is needed to find the
orientation of $e\prime$. This first lookup 
will decide whether the wertex $v$, shared by $e$ 
and $e\prime$, is either $e\prime.PVT$ 
or $e\prime.NVT$.
A second double lookup into the pair of wings 
incident to  the vertex $v$ 
(recall $v$ is shared by $e$ and  $e\prime$),
will decide whether $e$ and $e\prime$ use 
coherent rotational orientation for ordering
wings around $v$.

Taking into account the \occup\ of $VE^*$ and $FE^*$ 
we have that the \occup\ for this data structure is of $8e+v+f$.  
It is easy to see that pointers PCW, PCCW, NCW and NCCW organize
edges around a vertex into a doubly--linked circular list. 
Therefore all topological relations can be computed in optimal time.
Variants are possible where either vertices (PVT,NVT) or the facets
(PFACE,NFACE) can be omitted losing only some of the traversal 
capabilities.

\subsubsection{The \ema{quad--edge}{Data Structure} }
The quad-edge \cite{GuiSto85} use the same  data structure of the winged--edge
but organize the four edge pointers (PCW, PCCW, NCW and NCCW) in a different 
way. We reported these four pointers for an edge $e$ in Figure \ref{fig:quad}.
\begin{figure}[h]
\centerline{\fbox{\psfig{file=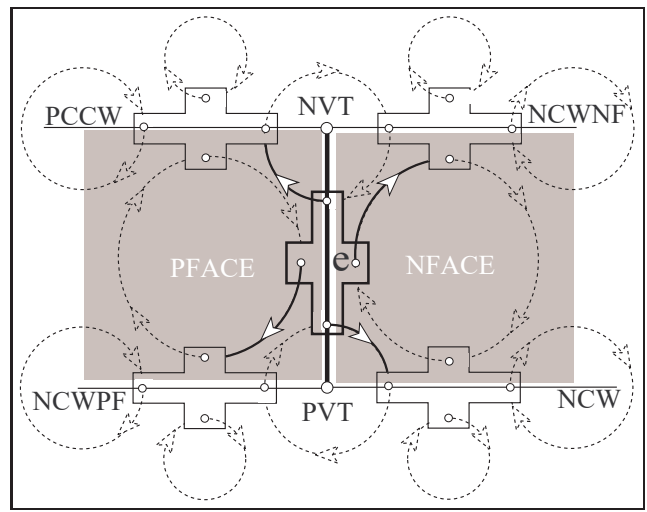,width=0.5\textwidth}}}
\caption{A geometric realization of an adge in a surface modeled by a 
quad--edge
data structure. Local clockwise orientation is assumed.
Edges and vertices are labeled with references relative 
to the thick oriented edge $e$ }
\label{fig:quad}
\end{figure}
The two data structures differ in the way they define the 
references they use. First we have that PVT, NVT and PFACE and NFACE 
are defined as, respectively, the two incident vertices and the two 
incident faces to edge $e$ (thick in Figure \ref{fig:quad}).
To explain the names of references we first assume that  there exist
a local coherent orientation around vertices and for loops delimiting faces.
In Figure \ref{fig:quad} a CW orientation is chosen.
This orientation  induces a cyclic ordering of edges around each 
vertex. References NPVT and NNVT store the next vertex, after
$e$ in the ciclic ordering of edges respectively around vertex PVT and NVT.
Similarly, references NPF and NNF store the next vertex, after
$e$ in the ciclic ordering of edges respectively around face PFACE and NFACE.
The definition of the quad--edge
data structure do not assume  the total orientation
of the surface to be encoded.
Note that if we reverse the local orientation we will store the same 
four reference in a different order. It can be proved that with a 
pair of look-ups we can decide if the orientation of each edge
among NPVT, NNVT, NPF and NVF is the same of orientation of $e$ or not.
Thus, this data structure can encode non--orientable surfaces and
has the same storage requirements of the winged-edge.

In the following we will analyze more compact alternatives to the 
winged--edge based on the deletion of two of the ''wings''.
However, if one wants to support situations where curved edges and  
faces with one or two edges are allowed, then all four edge pointers must remain. 
Otherwise, the traversal around a vertex or around a facet 
is no longer uniquely defined \cite{Wei85}. 
We start with a winged-­edge data structure where the wings PCCW and 
NCCW are omitted. This is the so called Doubly Connected Edge List (DCEL). 
}
\subsubsection{The \ema{DCEL}{Data Structure} }
{
The DCEL (Doubly Connected Edge List) data structure \cite{PreMul78} 
can represent orientable surfaces and
assumes that all edges receive an orientation. 
Then the DCEL represents each oriented edge of the surface using 
six references: 
two references (PVT, NVT) for incident vertices (i.e. the EV relation), 
two  references (PFACE, NFACE) for incident
faces (i.e. the EF relation) 
and the two references (PCW and NCCW) as defined in
section \ref{sec:winged}.
{
\begin{figure}[h]
\centerline{\fbox{\psfig{file=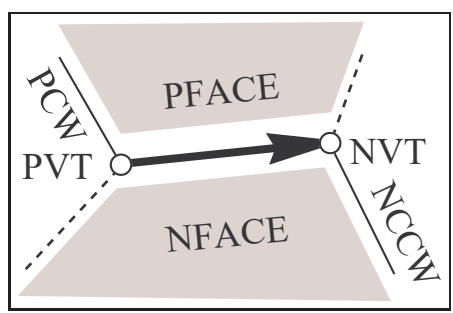}}}
\caption{A geometric realization of an edge in a surface modeled by a DCEL 
data structure.} 
\label{fig:dcel}
\end{figure}
}
These two references represents a portion of the EE relation.
It is easy to see that edges around a vertex are 
linked in a simply linked circular list. 
The  next element in this list is the next edge around a vertex in 
CCW order.
With these  relations  we can retrieve all topological relation 
in optimal time. The only limitation is that the $FE$ and  the $VE$ 
relations are extracted in a particular order i.e., the edges around
a vertex are returned in  CCW order and edges around a face are
returned in CW order.

Again, to extract all topological relations we need to model both vertices
and oriented  faces  with a reference to an incident edge 
(i.e. $VE^*$ and 
$FE^*$ relation). 
Taking into account the  $VE^*$ and the $TE^*$ relations the \occup\ 
for this data structure is equal to $6e+v+f$.  

\subsubsection{The \ema{Half-Edge}{Data Structure} }
\label{sec:he}
With the term Half--Edge we denote a number of data structures that split the 
winged-edge representation 
i.e., the eight  references:  PVT, NVT, PFACE, NFACE, PCW, PCCW, NCW and NCCW
into two similar nodes, called 
{\em half-edges}. Total information is preserved because each
half-edge points to the other half using a mutual reference called OTHERH.
Two options are described in \cite{Wei85} as the {\em face-edge}
data structure (FES) and the {\em vertex-edge}  data structure (VES).
The FES keeps in each half-edge (see Figure \ref{fig:halfedge}b) 
the four references to NFACE, PVT, NCW, NCCW 
from the winged--edge data structure (see Figure \ref{fig:halfedge}a).
The VES keeps the four references
to PFACE, PVT, NCW, PCW (see Figure \ref{fig:halfedge}c)
in the half-edge description.
{
\begin{figure}[h]
\begin{minipage}{0.29\textwidth}
\fbox{\psfig{file=wingedok.eps,width=\textwidth}}
\begin{center}(a)\end{center}
\end{minipage}
\hfill
\begin{minipage}{0.29\textwidth}
\fbox{\psfig{file=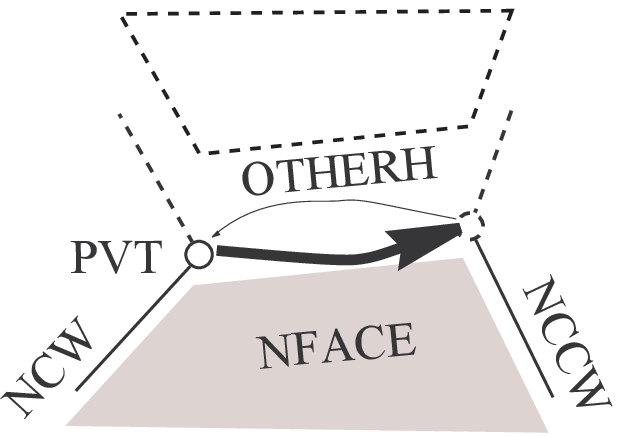,width=\textwidth}}
\begin{center}(b)\end{center}
\end{minipage}
\hfill
\begin{minipage}{0.29\textwidth}
\fbox{\psfig{file=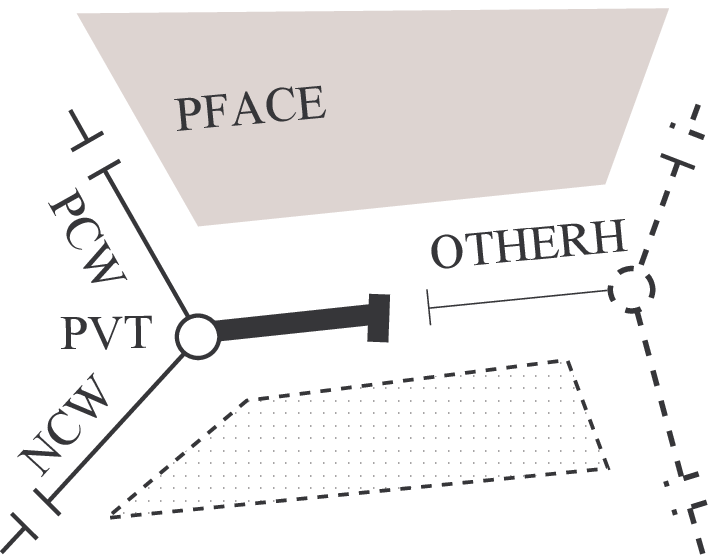,width=\textwidth}}
\begin{center}(c)\end{center}
\end{minipage}
\caption{A geometric realization of an edge in a surface modeled by a
winged--edge (a) and one (the thick one) 
of the two possible half-edges 
in the face--edge structure (FES) (b) 
and in the vertex--edge VES structure (c)}
\label{fig:halfedge}
\end{figure}
}
It is easy to see that these references link all edges incident to 
a vertex in a doubly--linked list.
To extract all topological relations we need to model both vertices
and faces  with a reference to an incident half-edge 
(i.e. $VE^*$ and $FE^*$ relation). 
Thus the \occup\ for these two variant of the half-edge
is equal to $10e+f+v$.

By storing the  two additional pointers in the half--edge 
data structure we can extract edges bounding a given face either
in CW and in CCW order without any need of doing a double 
lookup  into PFACE and NFACE as needed with the winged-edge structure. 
For this reason, the  reference to NFACE in the FES 
and the reference to PFACE in the VES half-edge 
are unnecessary whenever one is not interested in the EF relation.
With a similar argument one can delete the PVT references if 
the EV relation is not needed.  
Next, one can exploit the ideas used in the DCEL approach and reduce
the \occup\ by deleting a pointer in each half-edge \cite{Ket99}.
For the FES one just need to keep, in the half-edge, the
three references
to NFACE, PVT, NCCW (see Figure \ref{fig:dcelhalfedge}a).
For the VES one just need to keep in the half-edge the three 
references
to PFACE, PVT,  PCW (see Figure \ref{fig:dcelhalfedge}b).
{
\begin{figure}
\begin{center}
\begin{minipage}{0.29\textwidth}
\fbox{\psfig{file=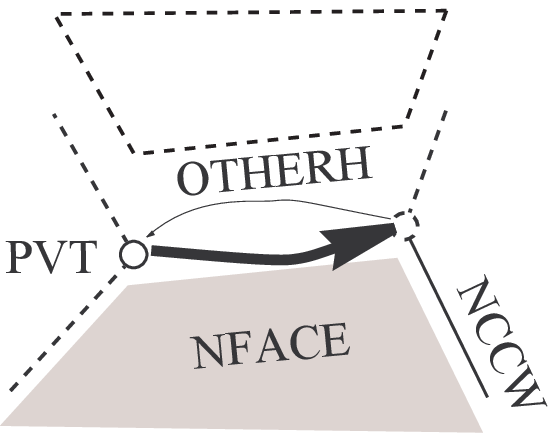,width=\textwidth}}
\begin{center}(a)\end{center}
\end{minipage}
\begin{minipage}{0.29\textwidth}
\fbox{\psfig{file=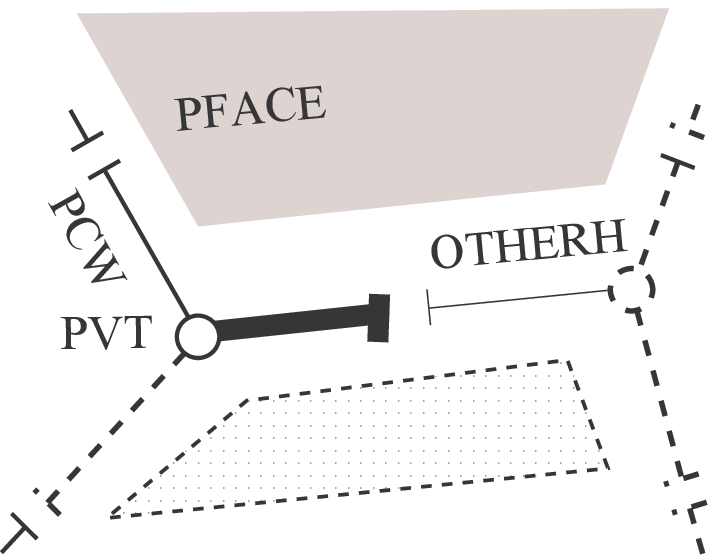,width=\textwidth}}
\begin{center}(b)\end{center}
\end{minipage}
\end{center}
\caption{A geometric realization of an edge in a surface modeled by 
a reduced half-edges following 
the FE-structure half-edge (a) and the
VE-structure half-edge (b)}
\label{fig:dcelhalfedge}
\end{figure}
}
The \occup\ for these schemes is  therefore equal to $8e+v+f$.
In this way, the edges around a vertex are linked in a simply linked circular list. 
With these schemes $FE$ and  $VE$ 
relations are extracted in a fixed order.
In particular the edges around
a vertex are returned in  CCW order and the edges around a face are
returned in CW order. 
Note that again we have no need to reference
to PFACE or NFACE or PVT to extract VE  and FE relations.
If we take this option and delete these references we obtain an
\occup\ equal to $4e+v+f$.

\subsubsection{Comparison} In conclusion we have five data structures:
the winged--edge (WE), the quad--edge, the DCEL and the FES and VES variants 
to the  half-edge (HE) data structures.
These data structures
offer a range of solutions for the representation of a manifold
surface with different memory requirements. With small differences they
all supports optimal 
extraction of topological relations. The different options are 
summarized in Table \ref{tab:mani}.
\begin{table}
{
\begin{center}
\begin{tabular}{|c|c|l|}\hline\hline
\multicolumn{2}{|c|}{Space} & Relations Modeled \\ \hline
1&$f$&${\rm FE}^*$ relation\\ \hline
2&$v$&${\rm VE}^*$ relation\\ \hline
3&$2e$&EV relation\\ \hline
4&$2e$&EF relation\\ \hline
5&$2e$&VE/$e$ and FE/$e$ relation in fixed order (reduced HE)\\ \hline
6&$4e$&VE/$e$ and EV relation in fixed  order with double lookup (DCEL)\\ \hline
7&$4e$&FE/$e$ and EF relation in  fixed order with double lookup (DCEL)\\ \hline
8&$6e$&EV, EF, VE/$e$ and FE/$e$ relation in fixed order with double lookup (DCEL)\\ \hline
9&$6e$&VE/$e$ and FE/$e$ relation in CW and CCW order (HE)\\ \hline
10&$6e$&VE/$e$ and EV relation in CW and CCW order with double lookup (WE)\\ \hline
11&$6e$&FE/$e$ and EF relation in CW and CCW order  with double lookup (WE)
\\ \hline
12&$8e$&EV, EF, VE/$e$ and FE/$e$ relation as in $10$ and $11$ (WE)
\\ \hline
\multicolumn{2}{|c|}{$f+v+6e$} & DCEL (FE and  VE in fixed order) (1+2+8)
\\ \hline
\multicolumn{2}{|c|}{$f+v+8e$} & WE (FE and  VE in CW and CCW order with double lookup) (1+2+12) \\ \hline
\multicolumn{2}{|c|}{$f+v+10e$} & HE (FE and  VE in CW and CCW order without double lookup) (1+2+9+3+4) \\ \hline
\multicolumn{2}{|c|}{$8e$} & Symmetric Data Structure   \\ \hline
\end{tabular}
\end{center}
}
\caption{Storage requirements for different data structures for 2-manifold 
modeling}
\label{tab:mani}
\end{table}
The most basic solution is a reduced half-edge where all references,
but those between half-edges, are deleted \cite{Ket99}.
This takes $4e$ and
supports the extraction of VE (FE) relation in optimal
time  provided that a starting edge $e$ incident to the given vertex
(face) is known. We denote these relations with VE/$e$ and FE/$e$.
In this case edges are returned in a fixed order.
If this is not acceptable, the space saving supported by a DCEL-like
optimization is not possible and this raises the \occup\ to $6e$.
To provide this starting edge we add $f$ references to compute ${\rm FE}^*$ 
and $v$ references to compute ${\rm VE}^*$.
To provide the EV relation  $2e$ references must be added. We can add
this either to a reduced half-edge and pay a total of  $6e$, 
or to a non-reduced half-edge and pay $8e$.
In both cases, we can merge together the two half-edges and save
$2e$ deleting the OTHERH references. This merge implies,
during the  extraction of the VE relation,
an additional double lookup for each edge visited.
Similar remarks holds for the extraction of the  EF relation.

\subsubsection{Incidence Graph and the Symmetric Data Structure}
A quite straightforward representation
scheme for any cell complex can be obtained by using
modeling approaches for graphs.
Indeed, we can have a node for each
cell and an edge for every adjacency relation.
This is  the idea that is behind 
{\em  incidence graph} \cite{Edel87}.

For $2$-dimensional complexes this, possibly, implies to store the
six relations VE, VF, EV, EF, FE, FV.
Obviously this
scheme is redundant, since, for example,
the vertices
adjacent to face, can be detected using
adjacency between vertices and edges
together with adjacency between edges and faces.
A simplified incidence graph, called the 
{\em symmetric} data structure is proposed in \cite{Woo85}.
In this scheme
redundancy is limited by representing only relations between cells
whose dimension differ of just one unit. For 2-dimensional
complexes we will just represent adjacency between vertices and edges
and between faces and edges. Thus only the four relations 
EV, VE, EF and FE are represented.
It easy to see that the EV relation takes $2e$ references to be encoded and 
$2e$ references are necessary to encode the EF relation in a closed 
$2$-manifold.
It is easy to see that the same space is needed to encode the VE and FE 
relations.
Thus the symmetric data structure encodes a closed $2$-manifold with 
$8e$ references. As Table \ref{tab:mani} shows 
this is the most compact solution to manifold modeling.
}
\subsection{Data Structures for encoding three-manifolds}
\label{sec:manitetra}
In this section, we  revise most important approaches for the 
representation of $3$-manifolds represented through cellular decompositions.
We first present the  
Facet Edge \cite{DobLas87} data structure (FES) and the 
Handle-Face \cite{LopTav97} data structure. Then we present an extension of 
the symmetric data structure for $3$-manifolds represented 
through simplicial complexes \cite{BruDeFPup89,Def99}.
{
\subsubsection{The \ema{Facet-Edge}{Data Structure} }
\label{sec:fe}
The facet-edge \cite{DobLas87} scheme
has been developed 
conceived to represent a cellular 
subdivision of the 3-sphere through its $2$-skeleton (i.e. the set of
all $2$-faces of cells). 
By this approach one can represent
cell $2$-complexes where an arbitrary number of faces 
(called here {\em facets})
are incident to an edge.
Facet-edge is actually an extension of the quad-edge scheme.
However, the relation with the quad-edge will be discussed further on. 

In this approach we have multiple
representations for each edge plus an algebra of
operators.
The main idea is that for each oriented
edge $e$   and for each oriented $2$-face $f$ we have a pair 
{$\brk{f,e}$} called the {\em facet-edge} pair.

The facet-edge pair {$\brk{f,e}$} contains  two orientated object: 
$f$ and $e$. 
The orientation of the
face $f$ is given by a cyclic ordering of edges of $f$. 
The spin orientation of the edge $e$ 
is induced by the orientation of edge $e$ and induce  a cyclic ordering 
on the set of facets incident to edge $e$.
For a given unoriented face and a given unoriented incident edge, four
possible facet-edge pairs are possible. 
An algebra of operations is given
to switch between facet-edge pairs  in order to traverse the
data structure.
The operators in this algebra allow retrieving, for each
facet-edge pair {$a=\brk{f,e}$}, the following entities
(see Figure \ref{fig:facetedge}):
\begin{itemize}
\item  the next edge on the cycle of faces that bounds
the oriented face $f$. This edge is denoted as $a\cdot Enext$;
\item  the next face  in the oriented  sequence of  faces
around edge  $e$. This face is denoted as $a\cdot Fnext$;
\end{itemize}
This approach introduces also two operators to change the orientation 
of the facet-edge pair. These are $a\cdot Spin$, that reverses 
the order of rotation around the edge, and $a\cdot Clock$, 
that reverses both
the order of rotation around the edge and within the face.

The effect of these two operations is resumed in the {\em handcuff}
diagram adapted from \cite{DobLas87}.
In this diagram  a facet--edge pair $a=\brk{e,f}$ is represended by two
oriented cycles. A circle is placed around oriented edge $e$.
The  orientetion of this first circle must be
CCW when judged by someone whose feet to head orientation is 
that of edge $e$. Thus, to change the spin orientation around an oriented 
edge $e$ we simply have to pass from $e$ to $-e$.
The second circle is placed on the face $f$ and its 
orientation must be that of the loop boundary of $f$.
{
\begin{figure}[h]
\begin{center}
\fbox{
\parbox[c][9.5cm]{0.40\textwidth}{
\vfill
\psfig{file=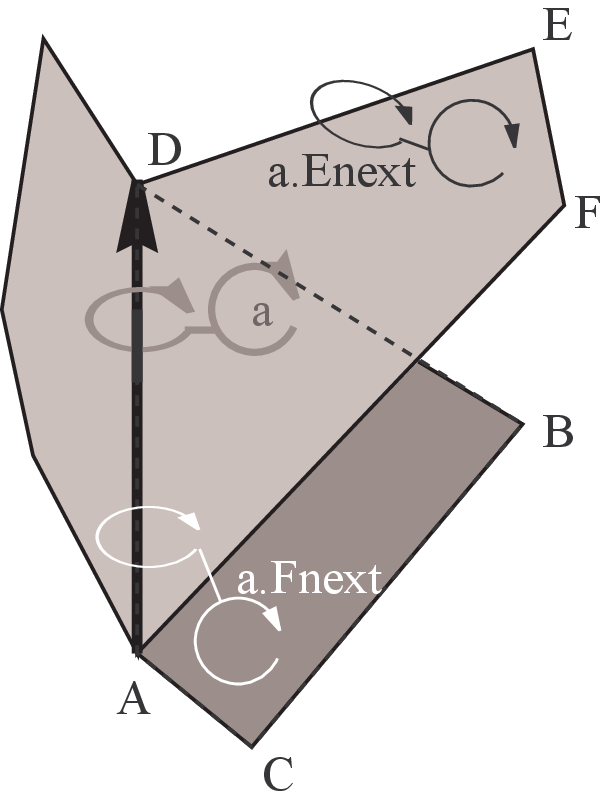,width=0.40\textwidth}
\vfill
\begin{center}(a)\end{center}
}
}
\fbox{
\parbox[c][9.5cm]{0.40\textwidth}{
\vfill
\psfig{file=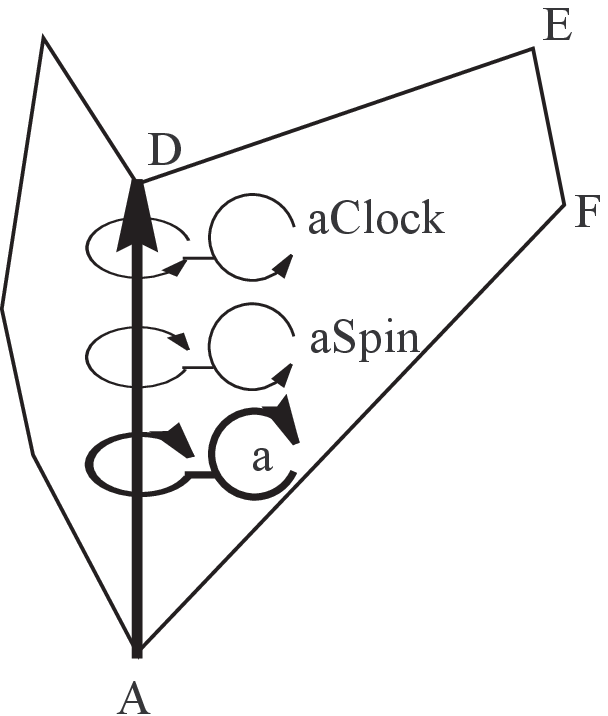,width=0.40\textwidth}
\vfill
\begin{center}(b)\end{center}
}
}
\end{center}
\caption{The operators Clock, Spin, ENext and FNext applied to the facet edge
$\brk{AD,ADFE}$}
\label{fig:facetedge}
\end{figure}
}
{
Thus, in Figure \ref{fig:facetedge}a 
the facet-edge {$a=\brk{AD,ADEF}$},
associated with the 
directed edge  $AD$  and with the oriented face $ADEF$, 
is represented by the two thick, 
dark gray, oriented circles.
In Figure \ref{fig:facetedge}a 
we report also the two facet-edges that results from 
the application of the two operators 
ENext and FNext to the facet-edge {$a=\brk{AD,ADEF}$}. 
The facet-edge $a\cdot Fnext=\brk{AD,BCAD}$ 
is represented by the two white oriented cycles in Figure
\ref{fig:facetedge}a.
The facet-edge $a\cdot Enext=\brk{DE,ADEF}$ 
is represented by the two black oriented cycles in Figure
\ref{fig:facetedge}a.
}
{
In Figure \ref{fig:facetedge}b we report the facet-edges that results from 
the application of the two operators 
Clock, Spin, 
The facet-edge $a\cdot Spin=\brk{AD,DAFE}$ 
is represented by the first two oriented cycles in Figure
\ref{fig:facetedge}b.
The facet-edge $a\cdot Clock=\brk{DA,DAFE}$ 
is represented by the next two oriented cycles in Figure
\ref{fig:facetedge}b.
}

Starting from the facet-edge {$\brk{f,e}$}
and composing these four operations we can obtain: 
\begin{itemize}
\item  the previous edge in the cycle of faces that bound
the oriented  face $f$;
\item  the previous face  in the oriented  sequence of
faces around edge  $e$.
\end{itemize}
With these operations we can retrieve a cycle of edges for every
face and a cycle of faces adjacent to a certain edge.

The data structure used for this representation is made up of a collection
of arrays storing four pointers.
For a given unoriented face and a given unoriented incident edge 
we recall that four possible facet-edge pairs are possible. 
All of them are represented by the so called  {\em facet-edge node}.  
In the data structure presented in \cite{DobLas87} the facet-edge node for 
{$\brk{f,e}$} is represented by four references.
Thus the internal data structure is similar to that of the quad--edge
(see Figure \ref{fig:quad}).
{
\begin{figure}[h]
\centerline{\fbox{\psfig{file=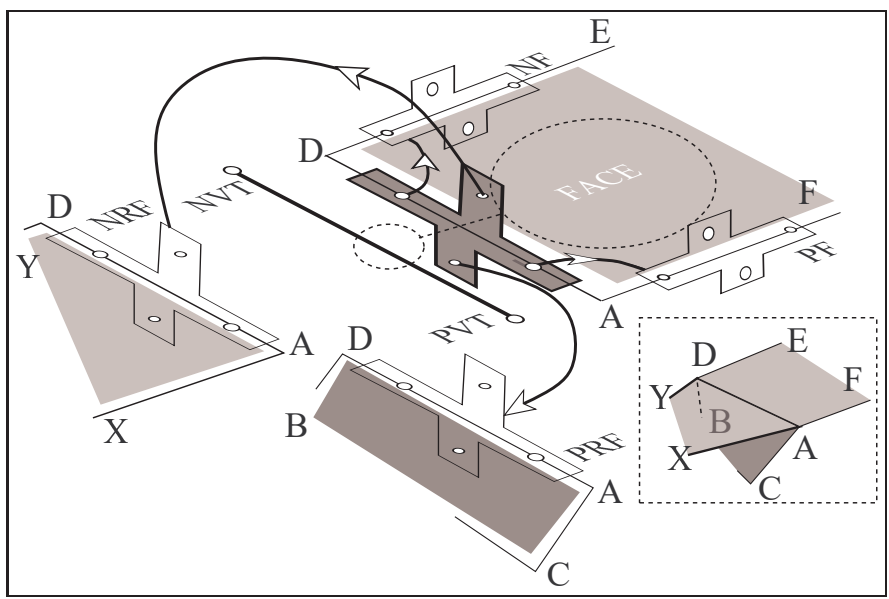,width=0.7\textwidth}}}
\caption{A fragment of the facet-edge
data structure for three $2$-cells sharing the edge $AD$}
\label{fig:quadfacet}
\end{figure}
}
In Figure \ref{fig:quadfacet} we report 
a fragment of the facet-edge
data structure for three $2$-cells sharing the edge $AD$.
This complex is reported in the dashed box in the  lower right corner of 
Figure \ref{fig:quadfacet}.
In this figure we depict 
the facet-edge node (drawn in dark gray  with thick border) 
for the for four facet-edges associated with  
edge $AD$ and face ADEF also denoted by {$FACE$}.  
In the same drawing we report 
the four facet-edge nodes referenced by this facet-edge node.
These four facet-edge nodes are:
\begin{itemize}
\item  the facet-edge node for the next and the previous facet-edge in the
cycle of edges for the face {$FACE$}. 
These are denoted by $NF$ and $PF$ 
in Figure \ref{fig:quadfacet}.
\item  the facet-edge node for the next and the previous facet-edge  
in the cycle  of faces around edge  $AD$.
These are denoted by $NRF$ and $PRF$ in Figure \ref{fig:quadfacet}.
\end{itemize}
Thus, in this data structure, face-nodes form a doubly-linked lists around
each edge and a doubly linked list for each face. 
Note that in the original data structure presented in \cite{DobLas87}
references to the face FACE and to
vertices PVT and NVT are not mentioned. 

The facet-edge is represented by a record called
{\em facet-edge reference} that contains a reference to a facet-edge node
plus three bits to encode the possible orientation of the edge and of 
the facet.
It can be proven \cite{DobLas87}
that is possible to implement all operations $Fnext$, $Enext$, $Spin$ and 
$Clock$ by  transforming facet-edge references.
Without entering the details of the implementation of these operators 
we simply note that
operations $Enext$ and $Fnext$ will change the referenced face-node while
operations $Spin$ and $Clock$ do not change the referenced face-node but
simply alters the bits in the facet-edge reference.

The \occup\ of this data structure can be easily evaluated for a  
simplicial subdivision.
For a simplicial subdivision we spend 12 
references for each triangle in the $2$-skeleton of the modeled
$3$-complex. Thus the \occup\ of this data structure is of $12f$ for a 
simplicial complex whose $2$-skeleton has $f$ triangles.

The original paper \cite{DobLas87} does not present algorithms 
to extract topological relations nor it introduces entities to model 
explicitly 
vertices, 2-cells and 3-cells. Given a facet-edge pair {$\brk{f,e}$}
it is easy to see that, in a cell subdivision, 
we can compute in linear time both the EF
relation for an edge $e$ and the FE relation for a face $f$.    
Given a facet-edge incident to a given polyhedral cell it is possible
to extract all  facet-edges incident to that polyhedral cell.
It is easy to see  that for a simplicial subdivision
the TE relation (recall that TE in this case stands for top-3-cell to edge)
can be computed in linear time whenever a facet-edge
incident to a given top 3-cell is available.
}
\subsubsection{\ema{Handle Face}{Data Structure}}
\label{sec:handleface}
The Handle-Face data structure \cite{LopTav97}
is designed to represent 3-manifolds
described by cell complexes. Each cell is represented  
through its boundary that, in turn, is represented through
the reduced FES data structure (see Section \ref{sec:fe} and Figure \ref{fig:dcelhalfedge}a).
A complete FES data  structure is introduced for each $3$-cell.
elements to model vertices edges and faces are duplicated for each 
$3$-cell. 

The data structure introduce four  basic topological  entities
(see Figure \ref{fig:handleface}): Vertices (V), 
Edges (E),  Faces (F) and Surfaces (S). Surfaces bounds $3$-cells.
{
\begin{figure}[h]
\centerline{\fbox{\psfig{file=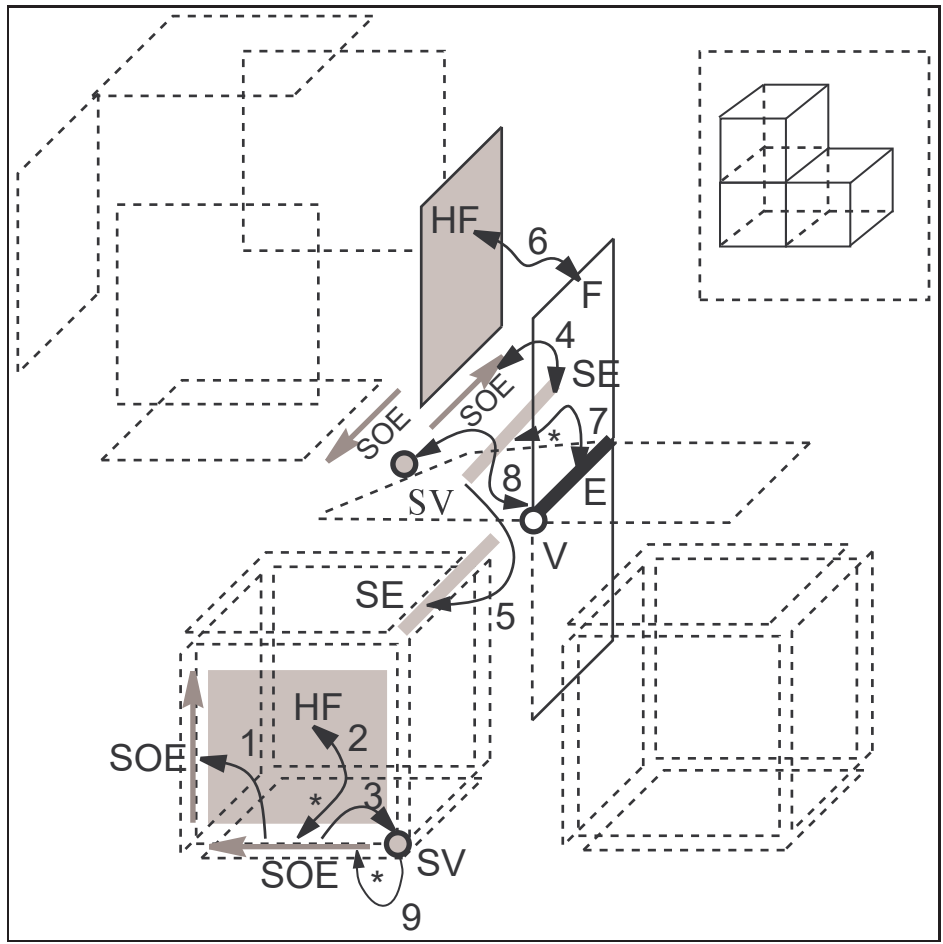}}}
\caption{Relations among objects in the handle-face data structure for
the complex made up of three cubes (fremed in the top right corner).
Arrows from objects X to object Y   means that every instance of
object X  must store a reference to the appropriate Y object.
}
\label{fig:handleface}
\end{figure}
}
For each vertex a distinct topological entity, 
called the {\em surface vertex} (SV), is introduced for each $3$-cell incident
to a given vertex.
For each edge a distinct topological entity, 
called the {\em surface edge} (SE), is introduced for each $3$-cell incident
to a given edge.
For each face a distinct topological entity, 
called the {\em half-face} (HF), is introduced for each $3$-cell incident
to a given face.
Note that up to two HF are introduced for each face.
For each SE  two oriented edges are introduced, called
{\em  surface oriented  edge}  (SOE). 

The topological entities: SOE, SV, HF models each surface using the
scheme of the FES. In this framework the SOE entity plays the role of the
half-edge. The only difference is that the OTHERH pointer 
(see Figure \ref{fig:dcelhalfedge}a) is not 
introduced. Instead of OTHERH  the {\em surface edge} (SE) 
node references two half-edges and is referenced by these
two half-edges (we labeled this  reference 
with $4$ in Figure  \ref{fig:handleface}). 
This double reference between SOE an SE plays the role of 
the pointer OTHERH.

Using  the integer labels  from Figure \ref{fig:handleface} we now list all 
references for all topological elements in the handle-face data structure.
The star $*$ on one tip of a relation from X to Y denotes the ${\rm XY}^{*}$
relation (i.e. we store just one Y element in relation with X).

Following  the FES scheme, in the handle-face data
structure each SOE reference (1) the next SOE in the loop
of edges that bounds an HF. The SOE reference also the HF (2) and the 
SV at its origin (3). 
A SV stores a reference to one (9) of the incident SOE.
The SE are tied together in a cycle of references (5)  of SE
modeling the set of $2$-faces around an edge in a $3$-complex.
The three basic topological  entities: Vertices (V), 
Edges (E) and Faces (F) reference the corresponding lower level 
entities i.e., each face reference two HF and is referenced by two HF (6).
Each HF stores a reference to one (relation 2 in the direction of
the arrow with $*$) of  the incident SOE.
Each edge node is referenced (7)
by each SE and points a single SOE (we recall that  the star $*$ on 
one tip of relation 7 denotes a partial relation).   
Finally the vertex node  V is referenced by each SV and reference (8)
all the SV. 

For each cell with $e$ edges this structure consumes $4e$ references
to encode the SOEs and $4e$ references for the SEs (note that
there are two arrows on the relations labeled with  4).
To evaluate storage requirements in term of number of
faces, edges and vertices  we must make some assumption on 
the $3$-complex we are modeling.  
We assume that we are interested in a  model with simplicial cells and we
evaluate the space needed to encode a tetrahedron
In this case the cell
has six edges and thus the tetrahedron boundary takes 48 references.
References (4) to the F node require  eight more references.
References (8) to the V node require other  eight references.
References (7) to the E node require other  six references and 
we neglect the partial reference from E to one of its SE.
References (relation 2, the arrow labeled with $*$) to a SOE incident to an HF takes other four
references.
References (9) to a SOE incident to an SV takes other four
references.
Thus in this data structure we use 78 references to model a tetrahedron.
The resulting data strucutre is extremely verbose and its time efficency is 
not investigated in \cite{LopTav97}.  However, it is easy to
see that all topological relations can be recovered in optimal time.

The paper \cite{LopTav97} shows that 
the handle-face
data structure supports a certain type of editing operations
(the so called Morse operators)
performed by attaching handle-bodies to an existing manifold.
\subsubsection{The \ema{Three-dimensional Symmetric}{Structure}}
\label{sec:simm3d}

In \cite{Def99} is studied the problem of representing
simplicial decomposition for the the class of
$3$-manifolds with boundary. To this aim one can model
directly the four topological entities: tetrahedra (T), triangles (F), 
edges (E) and vertices (V) and store some of the relations between
these entities.  
The Three-dimensional Symmetric Structure (TSS)  stores
relations TF, FT, FE and EV and stores, for each vertex 
an incident edge (i.e. the  {${\rm VE}^*$} relation) 
and for each edge an incident face (i.e. the  ${\rm EF}^*$ relation).
Since each tetrahedron has four triangular
faces the TF  relations can be stored
using $4t$ references.
Since each triangular face has three edges
the FE  relations can be stored
using $3f$ references.
Since each edge has two incident vertices
the EV  relations can be stored
using $2e$ references.
The partial relations {${\rm VE}^*$}
and {${\rm EF}^*$} takes respectively $v$ and $e$ references.
Thus this extension of the simmetric structure to $3$-manifolds takes 
$8t+3f+3e+v$  references to encode a simplicial $3$-complex.

It can be proven \cite{Def99} that the EF relation can be recovered 
in time proportional to the size of the output. This can be done
by recovering first a face incident to a given edge $e$ 
with the {${\rm EF}^*$}
relation. Then using the FT and the TF relations we find all
tetrahedra incident to edge $e$ and with relation FE we retrieve the faces
incident to $e$. 

Using the algorithm for the EF extraction and the 
{${\rm VE}^*$} relation we can build the VE relation combining
the EF, FE and EV relations. In both methods more elements than what is 
needed are visited but it can be proved that the amount of unnecessary 
visits has an upper  bound that is  linear \wrt\ the size of the output.
Thus the EV can be extracted in linear time, as well.
Combining the stored relations and the EF and VE relations it can be
proved \cite{Bruzzone90} that  all topological relations can be extracted.
It can be proved, using results in Section  \ref{sec:facenum}, that actually all
topological relations can be computed in a time  
that is  linear \wrt\ the size of the output.

\subsubsection{Comparison for $3$-manifold modeling} 
The following table summarizes space requirements for the 
Facet Edge \cite{DobLas87} data structure,
the Handle-Face \cite{LopTav97} data structure and for the
extension of the symmetric data structure to $3$-manifolds (TSS).
To compare this with the TSS we assume that we have to model
a simplicial $3$-complex.
{
\begin{center}
\begin{tabular}{|c|c|}\hline\hline
{Space Requirement} & Data Structure \\ \hline
$78t$& Handle--Face\\ \hline
$12f$& Facet-Edge (vertices and tetrahedra not explicitly modeled) \\ \hline
$t+18f$& Facet-Edge (vertices and tetrahedra explicitly modeled) \\ \hline
$8t+3f+3e+v$& TSS \\ \hline
\end{tabular}
\end{center}
}

To compare the storage requirements of the three approaches we
use the relations $f\le 4t$, $e\le 6t$, $v\le 4t$, $6e\le 4f$, and $4v\le 4f$.
Note that we reach the upper bound (i.e.,  $f=4t$, $e=6t$, $v=4t$, $6e=4f$
and $4v=4f$)
for a simplicial $3$-complex where each simplex is a distinct 
connected component (we called such a complex a 
{\em totally exploded} complex).
These relations holds for all $3$-complexes that can be assembled, 
starting from a totally exploded tetrahedralization, \gl\ two
triangles together. It can be proven (see Property \ref{pro:iqmrel} Parts \ref{pro:mdcdec} and \ref{pro:d1conn} )  that all $3$-manifolds can be assembled in this way.

Note that $6e\le 4f$ and $4v\le 4f$ holds in a totally exploded
tetrahedralization and there is no way to glue two triangles together
without identifying at least a pair of points and two pairs of edges.
Thus, every time we glue together two triangles, we decrease $f$ of one
unit we decrease $v$ of {\em at least} one  units and we decrease $e$ of 
at least two units.
Thus $6e\le 4f$ and $4v\le 4f$ must hold in every complex
assembled in this way and in particular it must hold in a $3$-manifold.

Using $f\le 4t$, $e\le 6t$ and  $v\le 4t$
we have $t+18f\le 73t$ and $8t+3f+3e+v\le 42t$. This shows
that the handle-face data structure is the most expensive data structure
and its storage cost is more than 1.86 times the storage cost of the TSS.
Next using $6e\le 4f$ and $4v\le 4f$ and the fact that. in a $3$-manifold
$4t\le 2f$ we have that storage requirement for the TSS i.e.
$8t+3f+3e+v$ can be written as $t+7t+3f+3e+v\le t+3.5f+3f+2f+f=t+9.5f$.
This proves that the TSS data structure is more compact than the 
facet--edge data structure even if we consider the original version of
the facet--edge using just $12f$ pointers.
The TSS saves $8.5f$ references over the facet-edge. Using
$4t\le 2f$ we have that this saving represent   at least the 46\%
of the storage requirements of the facet-edge. Thus the facet--edge
require more than 1.85 times the storage used by TSS.
The above analysis is summarized in the following table that
shows lower bounds for space requirements normalized vs. space
requirements for the TSS.
{
\begin{center}
\begin{tabular}{|c|c|}\hline\hline
{Normalized Space Requirement} & Data Structure \\ \hline
$\ge 1.86$& Handle--Face \\ \hline
$\ge 1.85$& Facet-Edge  \\ \hline
$1$& TSS \\ \hline
\end{tabular}
\end{center}
}

\subsection{Dimension independent data strucutres for encoding cell complexes}
\label{sec:dimindip}
In this section we report four approaches to dimension independent 
modeling: the cell-tuple approach \cite{Bri93}, 
the Winged Representation \cite{PaoAl93}, the n-G-maps \cite{Lie91}
and the Selective Geometric Complexes (SGC) \cite{RosCon90}.
One thing to note is that all these approaches can model $d$-manifolds
but surely, for each one of these approaches, the representation
domain is larger than the class of $d$-manifolds. 
In general we have 
that non recognizability of $d$-manifolds for 
$d\ge 6$ (see Property \ref{teo:halting}) implies that it is not possible
to have a  dimension independent representation whose applicative
domain is exactly the class of $d$-manifolds.
Any algorithm that will encode a generic $d$-complex into 
such a representation will act as a decision procedure for the 
class of manifolds.
None of the above approaches, with the exception of SGCs, can model 
completely the non-manifold domain. 
The SGC can model the full generality of 
the non-manifold domain. We also mention that  in \cite{LieElt93}, 
is presented an extension
of n-G-Maps that also models the whole non-manifold domain.

As we anticipated, the modeling approach  in this class have a stronger 
relations with the results of this thesis. 
In Chapter \ref{ch:nmmdl} we extend
the winged representation to the non-manifold domain.
Another  byproduct of  the results in this thesis is the exact definition
of the representation domain for the winged representation that happens 
to be the set of \qm\ that, in turn, is 
the representation domain of n-G-maps. 
{
\subsubsection{\emi{Selective Geometric Complexes} (SGC)}
{SGC \cite{RosCon90} is a modeling scheme
based on a notion of cell complex similar to CW complexes.
A {\em regular finite CW complex} for a metrizable topological space $X$ 
(see for instance \S 7.3 in   \cite{Jan94})
is a collection $\Gamma$ of subsets  of $X$, called {\em cells}, 
such that:
\begin{itemize}
\item for each cell $c$ there exist an integer $k\ge 0$ such that
$c$ is homeomorphic to the open $k$-ball $B^k=\{x\in\real^k|\|x\|<1\}$ and
the closure of $c$ is homeomorphic to a closed $k$-ball; 
\item $\Gamma$ is a partition of $X$; 
\item the boundary of each $k$-cell is homeomorphic to the $(k-1)$-sphere
$S^{k-1}$ and  can be expressed as the union of cells in
$\Gamma$.
\end{itemize}
A $k$-cell is a cell homeomorphic to $B^k$.

The cell concept in CW complexes is devised to attack topological problems,
whereas SGC cells have been tailored to the needs of geometric modeling.
As we quote form \cite{RosCon90}, {\em the differences lies almost 
exclusively in the concept of what constitutes the fundamental entity:
the cell}.
{
SGCs are a compromise between
simplicial complexes  and CW complexes.
CW complexes are, by far, too abstract.
In fact, every solid, homeomorphic to a closed $3$-ball, can be
expressed  by the same CW complex.
For representing such a solid it is  sufficient
a CW complex, whose
combinatorial structure reduces to the triplet $\{p,S^2-p,B^3\}$
(being $p$ any point of $S^2$).
We have that cell  $c_0=p$ is a point, i.e., a 0-cell.
The cell  $c_2=S^2-p$ is homeomorphic to $B^2$ i.e., a 2-cell.
The last cell $c_3=B^3$ is a a 3-cell.
The three cells are organized so that
$\bnd c_3= c_2+c_0$;  and $\bnd c_2=c_0$.

On the other hand, simplicial complexes are
too detailed. Indeed, infinitely many simplicial
representations are possible, for a given topological space.
Even if, we just consider complexes with a minimal number of 
simplices, we are left
with a non-unique representations. For a cubic surface, for
example, we have $2^6$ different,
simplicial complexes with twelve triangles.

For this cubic surface,
for example, SGC
provides the quite ''natural'' representation with: 6
2-cells,  12 1-cells and 8 0-cells.
However, SGC are, still, fairly abstract. Indeed,
with SGC, we can express, with a finite complex,
some unbounded domains.
SGC can also code: non-manifolds, open set,
domains with missing internal points and
non regular simplicial complexes.
So, SGC stands midway between CW and simplicial
complexes, supporting natural and  unique  representations.
The expressive power of SGC cells is quite broad since they can
encode cell complexes  with open and closed cells and
with cells  with internal vertices and edges.

SGC cells are defined using concepts from the theory of algebraic 
varieties and stratification. 
An {\em algebraic variety} \cite{Whi57} is any closed subspace of the Euclidean 
space $\real^d$ that is the locus of common real zeros of a finite set of
real polynomials in $d$ variables. 
A variety that cannot be decomposed 
as the union some other varieties is called an {\em irreducible
algebraic variety}.
An irreducible algebraic variety $V$  can still be partitioned, using
differential properties, into a regular part $R$ and
a singular part $S$.
It can be proved that the set of
connected components of $R=V-S$ must be a finite set of open manifolds that
are sub-manifolds of $V$ \cite{Boo86,War71}. Each of these sub-manifolds is called an {\em extent} of $V$.

Furthermore, it can be proven that the set of singular points $S$ 
is both a closed set and a variety. 
Thus $S$ will have its extent too. These extents will be considered 
as extents of $V$. too.
The theory of stratification \cite{Col75,Joh83,Wal74} guarantees that $V$ can be decomposed into the disjoint union of
a set of connected open sub-manifolds each sub-manifold being included
into   an extent of $V$. This set is called a 
{\em manifold decomposition} of $V$. 
A manifold decomposition {${\cal M}$} has a cellular structure, i.e., 
the boundary of an element in a manifold decomposition  of $V$
(denoted by {${\cal M}$})
can be expressed as the disjoint union of a finite set of elements 
in  {${\cal M}$}.

With this theoretical foundation SGC defines a cell $c$ as any connected open
subset of an extent of an algebraic variety. 
We will denote with $c.E$ the extent in which cell $c$  is contained.

A SGC {${\cal C}$} is any collection of disjoint cells $c_i$ such that the 
boundary of each cell is the disjoint union of a finite set of 
elements in the SGC {${\cal C}$}.
Furthermore, for all cells $c_i$ on the boundary of a cell $c$,
cells $c_i$ in a SGC are constrained
to stay either on the interior  of the extent  $c.E$
(i.e. $c_i\subset c.E$) or completely outside the extent $c.E$
(i.e., $c_i\cap {c.E}=\emptyset$). 

This definition of SGC cells guarantees that we can
always find a common refinement of two intersecting
cells from two different SGC.
The intersection of two algebraic
varieties yields an algebraic variety of lower
dimension.
From these intersections we can select new
boundary  cells, that cut the intersecting cells
of higher dimension.
This cut defines the common refinement we need.
Obviously, this approach
assumes that we are able to computationally intersect
algebraic varieties without too many problems.

Cells in SGCs are defined so that we can
always find a SCG complex that describes the intersection of two
cells.
This choice is the  key factor that allows SCG to support a rich set of operations 
including 
Boolean operations, boundary and interior operation and
regularized boolean operations.
{

The encoding of adjacency relations in SGCs is 
done using as a  complete incidence graph \cite{Edel87}. 
In fact each cell $c$ bears a reference
to the cells that are on the boundary of $c$ and to cells that have
$c$ in their boundary.
This seems quite space consuming if compared with data structures we 
previously reviewed. 
However it is clearly unfair to compare the 
combinatorial structure of an SGC complex against, let say for instance,
the combinatorial structure of a simplicial subdivision. 
In SGC complexes are modeled with a cell complex using quite complex cells.
Thus, in  SGC, the structure of the cell complex, represented 
via an incidence graph, can be regarded as an upper layer that ties together
quite complex cells. In this we see a relationship between SGC
and our two layer data structure devised in Chapter \ref{ch:nmmdl}.

SGC cells are then grouped together using a generic cellular structure 
encoded using the incidence graph. As a result,  
with SGC,  we can  handle any kind of non-manifold complexes,
including non-regular complexes with dangling
edges and isolated points. 
}

SGCs supports uniqueness of representation.
Different SGC representations are possible for a 
given topological space. However SGCs representation can be ordered
so that, redundant SGC representations are recognized
and just one, minimal, CSG representation
is maintained. 
It is possible to define a {\em simplification}
procedure that compresses redundant SGCs to such a minimal
representation. 
SGC simplification, actually induces a poset structure
over SGC complexes. To see how this takes place we
need, again, to introduce some definitions.
}

{
We will say that two cells of the same extent
can be {\em joined} \siff\ they can be merged into one  deleting their
common external boundary.
We will say that a SGC $C$ can be {\em simplified}
\siff\ there are two or more cells in $C$ that can
be joined. 
If a  new complex $C'$ results from a set
of  simplifications performed on $C$, then,
we will say that $C$ will be a {\em refinement}
of $C'$ and we will write $C'<C$.
For instance in Figure \ref{fig:sgc} we have that 
the complex in (a) is a refinement of the complex in (b).
{
\begin{figure}[h]
\centerline{
\fbox{
   \epsfig{file=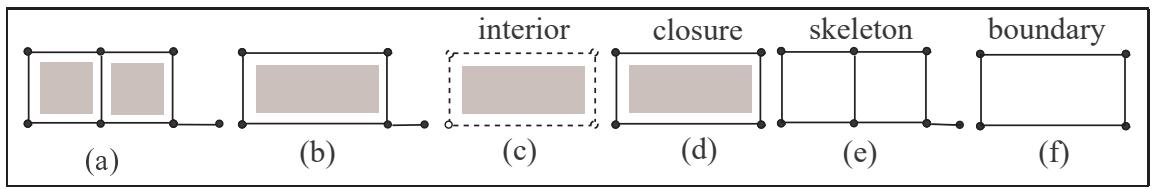, width=.9\textwidth}
}
 }
\caption{Operations in SGC:
A non minimal SGC (a), then its minimal SGC (b), then
its regularized version  (d) obtained taking the
closure of the interior of (b).
Topological operations in SGC:
SGC (c) is the interior of (b);
(d) is the closure of (c);
(f) is the boundary of (d);
(e) is is the result of a boundary operation
on the non minimal SGC in (a).}
\label{fig:sgc}
\end{figure}
}
The notion of refinement defines a partial
order over the class of SGC.
In the poset ordered  by the refinement relation $<$ 
for every SGC $C$, there is a  {\em lower} minimal element
$\mnsgc{C}$, such that $\mnsgc{C}<C$.
Such a minimal element $\mnsgc{C}$ can, actually, be computed with a
simplification algorithm. If $A$ and $B$ represent
tha same topological space, then,  it can be proved that,
$\mnsgc{A}=\mnsgc{B}$. Hence, we can say that SGC provides
a unique representation for a given topological space that
can be subdivided as a SGC. 
For instance in Figure \ref{fig:sgc} we have that the SGC in (b) is the
minimal representation for the complex in (a). 
From a minimal  SGC representation of a topological space
$CC$, we can extract a SGC representation
for the interior $\interior{CC}$ and the boundary $\bnd{CC}$ of  $CC$.
A SGC complex for the interior $\interior{CC}$ of $CC$
is obtained, selecting all the cells
of maximal dimension in the SGC
for $CC$ (in Figure \ref{fig:sgc} in (c) we have the interior of 
the SGC in (a)).

The boundary  of 
the interior $\interior{CC}$ can be obtained selecting the
cells of $CC$, that are on the boundary of cells
in $\interior{CC}$ (in  Figure \ref{fig:sgc} in (f) we have
the boundary of the SGC in (c)).
We note that, obviously, we obtain wrong results if we
apply those definitions to a non-minimal SGC.
For example,  Figure \ref{fig:sgc} (e), is what
remains if we take the boundary  of the interior of the  SGC in (a).
Recall that the SGC in (a) is a non minimal, version of
the SGC in (b).
The complex in (e) is usually called the 1-skeleton of (a).

We have seen that we can, easily, compute
$\bnd\interior{CC}$.  However, in general,  we are
not able to  compute $\bnd{CC}$.
However, if $CC$ is regular, then
$CC=\overline{\interior{CC}}$ and  then
$\bnd{CC}=\bnd\interior{CC}$ ($\overline{A}$ is the topological closure of $A$).
If $CC$ is not regular, but it is closed, then
we can extract a {\em regularized} version
of $CC$, taking  $\overline{\interior{CC}}=
\interior{CC}\cup\bnd{\interior{CC}}$
(the SGC in  Figure \ref{fig:sgc} (d) is the regularized version of the 
complex in (a)).
}
{
\begin{figure}[h]
\centerline{
\fbox{
   \epsfig{file=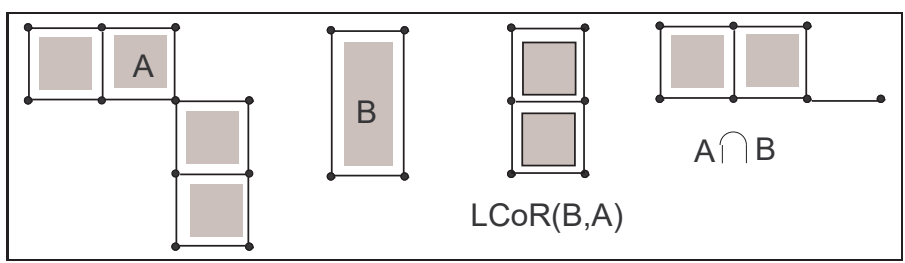, width=.7\textwidth}
}
 }
\caption{On the left two SGC $A$ and $B$. In the middle 
the LCoR(B,A). On the right, non-minimal SGC for
$A\cap B$.
}
\label{fig:sgccup}
\end{figure}
}

{
A nice feature of SGC is that the refinement partial order
$<$ always admits a least compatible refinement ($LCoR$) for any pair of elements.
We say that $A$ and $B$ are compatible, 
if their common part, $A\cap B$ is
represented by the same complex both in $A$
and in $B$.
For instance,  in Figure \ref{fig:sgccup} we have that
$A$ and $B$ are not compatible.
If we take two SGCs $A$ and $B$ 
there always exists
a pair of minimal compatible SGCs $A'$ and $B'$ such that $A<A'$ and $B<B'$. 
The pair of complexes $\{A',B'\}$
will be 
denoted by $Subdivide(A,B)$.

In \cite{RosCon90} is presented an algorithm that compute $Subdivide(A,B)$.
This is based on the joint computation of 
the least refinement of $A$ that can be made compatible \wrt\ $B$
(we will denote this complex with $LCoR(A,B)$) and of
the least refinement of $B$ that can be made compatible \wrt\ $A$
(i.e. $LCoR(B,A)$)

We note that, in general, $LCoR(A,B)\neq LCoR(B,A)$.
Referring to Figure \ref{fig:sgccup} we have that:
$LCoR(A,B)=A$ and $LCoR(B,A)$ is the SGC in the middle 
of Figure  \ref{fig:sgccup}.
Furthermore, $LCoR(A,B)$ is not compatible
\wrt\ $B$, but it is compatible with
$LCoR(B,A)$. With this notation, we can, finally, describe
$Subdivide(A,B)$ 
as $Subdivide(A,B)=\brk{LCoR(A,I),LCoR(B,I)}$ where
$I=LCoR(A,B)\cap LCoR(B,A)$ is the intersection.
Then referring to Figure \ref{fig:sgccup} we have that
$Subdivide(A,B)=\brk{A,LCoR(B,A)}$.

When the SGCs for $Subdivide(A,B)$ is available,
then  we can extract,  from $Subdivide(A,B)$, a SGC  for
both $A\cup B$ and $A\cap B$.
For example, in  Figure \ref{fig:sgccup} we
compute  $A\cap B$ taking cells, that are
both in $LCoR(A,B)$ (i.e. A) and $LCoR(B,A)$ and obtain the complex
on the right of Figure \ref{fig:sgccup}.
Then, referring back to Figure \ref{fig:sgc} (a)
we simplify the  SGC (a) for $A\cap B$ and obtain
\ref{fig:sgc} (b), that is a minimal SGC.
Steps in Figures \ref{fig:sgc} (c) and \ref{fig:sgc} (d) leads
to the regularized version of $A\cap B$.
In this way, SGC supports both
Boolean operations and regularized Boolean operations.

{
We note that, to compute $Subdivide(A,B)$,
we need to check   every pair of cells in $A$ and $B$
for intersection. This might require 
$\ordcomp{\abs{A}\abs{B}}$ intersections between pairs of
algebraic varieties. 
In \cite{RosCon90} a complexity analysis for this operation is not reported.
}

In conclusion, SGC supports both
Boolean and regularized Boolean operations.
It is possible to define a {\em simplification}
procedure that compresses redundant SGCs to its minimal
representation SGC \wrt\  the ordering induced by the
simplification notion.
A SGC usually has an extremely compact, combinatorial structure,
while more complex geometric information might be used to code cells.
Therefore, the adoption of SGC representations shifts
the complexity towards the geometry of cells.

}
}
}
{
\subsubsection{The \emi{Cell-Tuple}}
The cell-tuple offer a scheme to encode any  CW complex of a d-manifold.
Given a finite regular  CW complex  $\Gamma$ for a $d$-manifold ${\cal M}$
we define a {\em subdivided manifold}, as 
the pair $\{{\cal M},\Gamma\}$.
A {\em cell-tuple}  for the subdivided manifold {$\{{\cal M},\Gamma\}$}
is any sequence of $d+1$ cells  of $\Gamma$, $t=(c_0,\ldots,c_{d})$
such that $c_i$ is a cell of dimension $i$ and such that
$c_i$ is a face of $c_{i+1}$.

The cell tuple structure for a subdivided $d$-manifold {$\{{\cal M},\Gamma\}$}
is given by the set ${\cal T}$ of all cell-tuples for $\Gamma$ 
together with a set of
symmetric relations, denoted by ${\rm switch}_i$, such that, 
two cell-tuples $\tau$ and $\tau^\prime$ are in relation \wrt\  
${\rm switch}_i$ \iff\ they differ just
for the $i$-th cell. 
It is easy to see that for a top $d$-cell that  is a $d$-simplex we 
must introduce $\factorial{(d+1)}$  cell-tuples in ${\cal T}$.
{
\begin{figure}[h]
\centerline{\framebox{\psfig{file=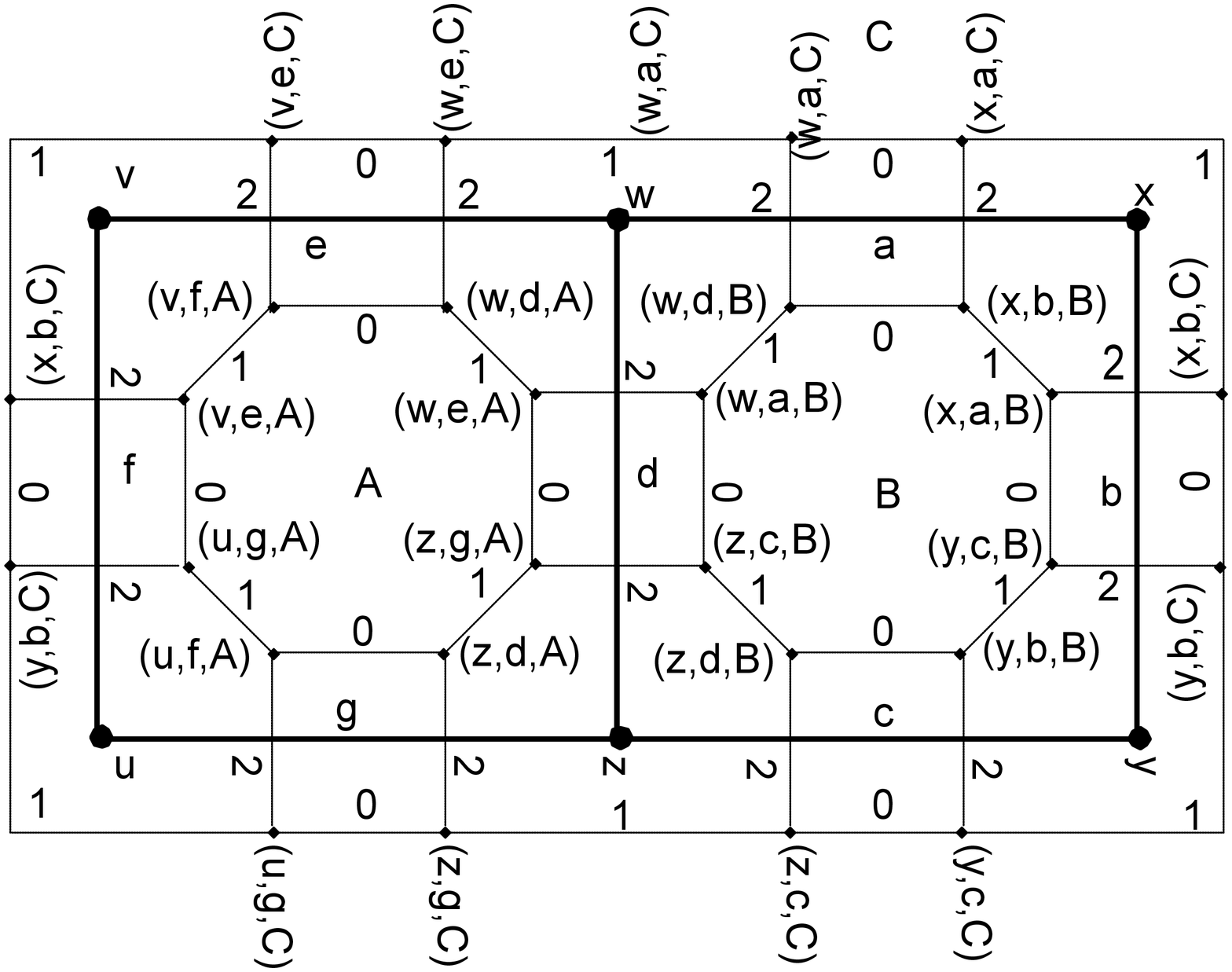,width=0.8\textwidth}}}
\caption{The graph of cell-tuples (with dashed arcs) 
for the complex made up of the 
the two square faces $A=uvwz$ and $B=wxyz$ and of the surrounding
unbound face $C$. Small dots labeled with triples are cell-tuples. Dashed arcs links two cell tuples that are in the ${\rm switch}_i$ relation being $i$ the arc label.}
\label{fig:celltuple}
\end{figure}
}
The cell-tuple structure  
can be represented by a graph whose nodes are the cell-tuples in 
{${\cal T}$} and such that there is an arc (labeled with $i$) in between
the two cell-tuples $\tau$ and $\tau^\prime$ whenever
$\tau$ and $\tau^\prime$ are in relation \wrt\  ${\rm switch}_i$.
An example of this graph representation is in Figure~\ref{fig:celltuple}.

It can be proven that the relations ${\rm switch}_i$ are actually functions,
i.e.  there is just one edge labeled with $i$ leaving from any node 
in the cell-tuple graph. 
Thus, all the nodes in this graph must be
of order $(d+1)$.  It can be proven \cite{Bri93} 
that there is a bijection between 
$i$-cells in the complex $\Gamma$ and cycles that do not contain arcs
labeled with $i$. An $i$-cell is incident to a $k$-cell \iff\ the associated
cycles share some vertices. Thus all topological relations can be extracted
in optimal time traversing the graph of cell-tuples.
{
\begin{figure}[h]
\begin{center}
\psfig{file=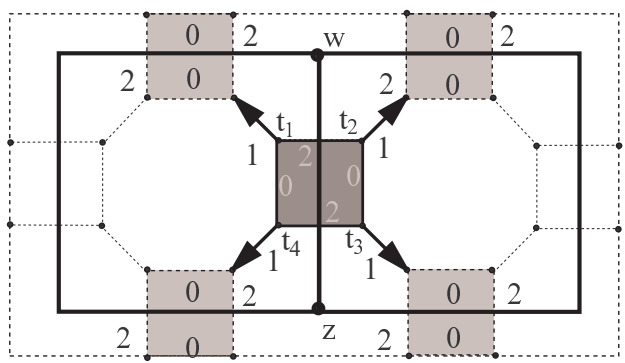,width=.5\textwidth}
\end{center}
\caption{Translation of the cell tuple of Figure  \ref{fig:celltuple}
into winged-edge and quad-edge}
\label{fig:celltuwe}
\end{figure}
}

A cell-tuple structure for a subdivided $2$-manifold
${\cal M}$ can be translated into a winged-edge data structure, 
a quad-edge data structure, a FES and a VES.
Examples of this translation for the complex of Figure \ref{fig:celltuple}
are given in Figure \ref{fig:celltuwe}.
A cell-tuple structure for a subdivided $3$-manifold
can be translated into a facet-edge data structure.
All these translations are better understood if we refer to the graph
representation of a cell tuple data structure. 
In this graph we will call a $\{x,y\}$-orbit (or $xy$-orbit for short) 
a cycle whose edges are labeled only with $x$ and $y$. 
In a similar way  we define $x$-orbits and $xyz$-orbits.
With some abuse of notation, given an orbit $o$  
we will call the $o$ orbit also the corresponding orbit
in the algebraic cell tuple structure i.e. 
the set of cell tuples associated to graph vertices 
in the orbit.
With this notation we can define how to translate a cell-tuple graph into
various data structures.

To obtain a winged-edge data structure or a quad-edge data structure we
consider 02-orbits.   
Let $o$ be the (set of cell-tuples in a) 02-orbit (in dark gray in Figure \ref{fig:celltuwe}) given by
$o=\{t_1,t_2,t_3,t_4\}$ being, for instance, 
${\rm switch}_0(t_{1})=t_4$,
${\rm switch}_2(t_{4})=t_3$,
${\rm switch}_0(t_{3})=t_2$ and
${\rm switch}_2(t_{2})=t_1$.
For  each 02-orbit $o$ we introduce a winged-edge (or a quad-edge)
whose four wings (solid arrows in Figure \ref{fig:celltuwe}) 
points the other  four winged-edges
(in light gray in Figure \ref{fig:celltuwe}) introduced
for the four  02-orbits containing, respectively, 
${\rm switch}_1(t_1)$, ${\rm switch}_1(t_2)$, ${\rm switch}_1(t_3)$ and   ${\rm switch}_1(t_4)$. 
{
\begin{figure}[h]
\begin{center}
\begin{minipage}{0.48\textwidth}
\psfig{file=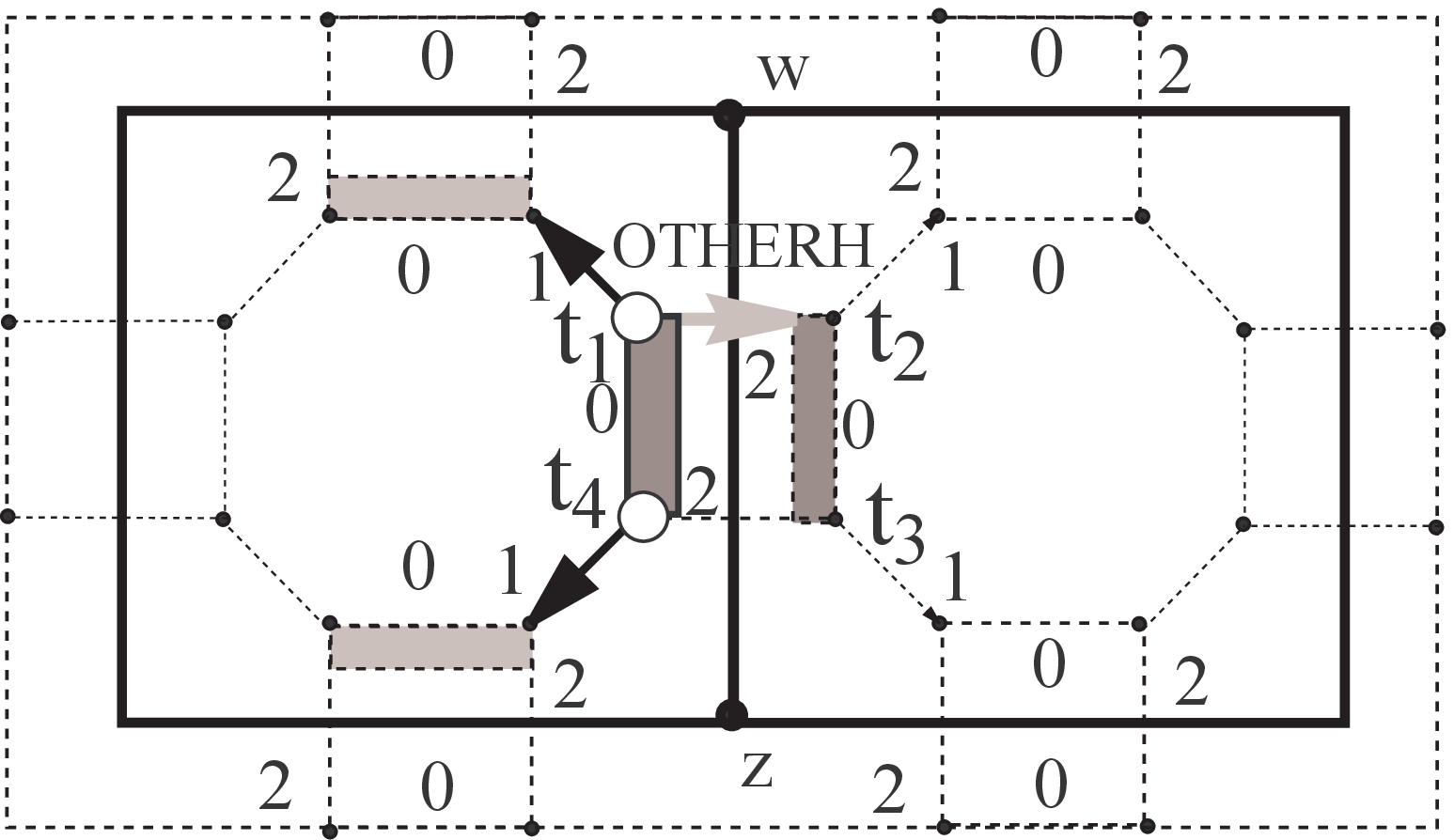,width=\textwidth}
\begin{center}(a)\end{center}
\end{minipage}
\hfill
\begin{minipage}{0.48\textwidth}
\psfig{file=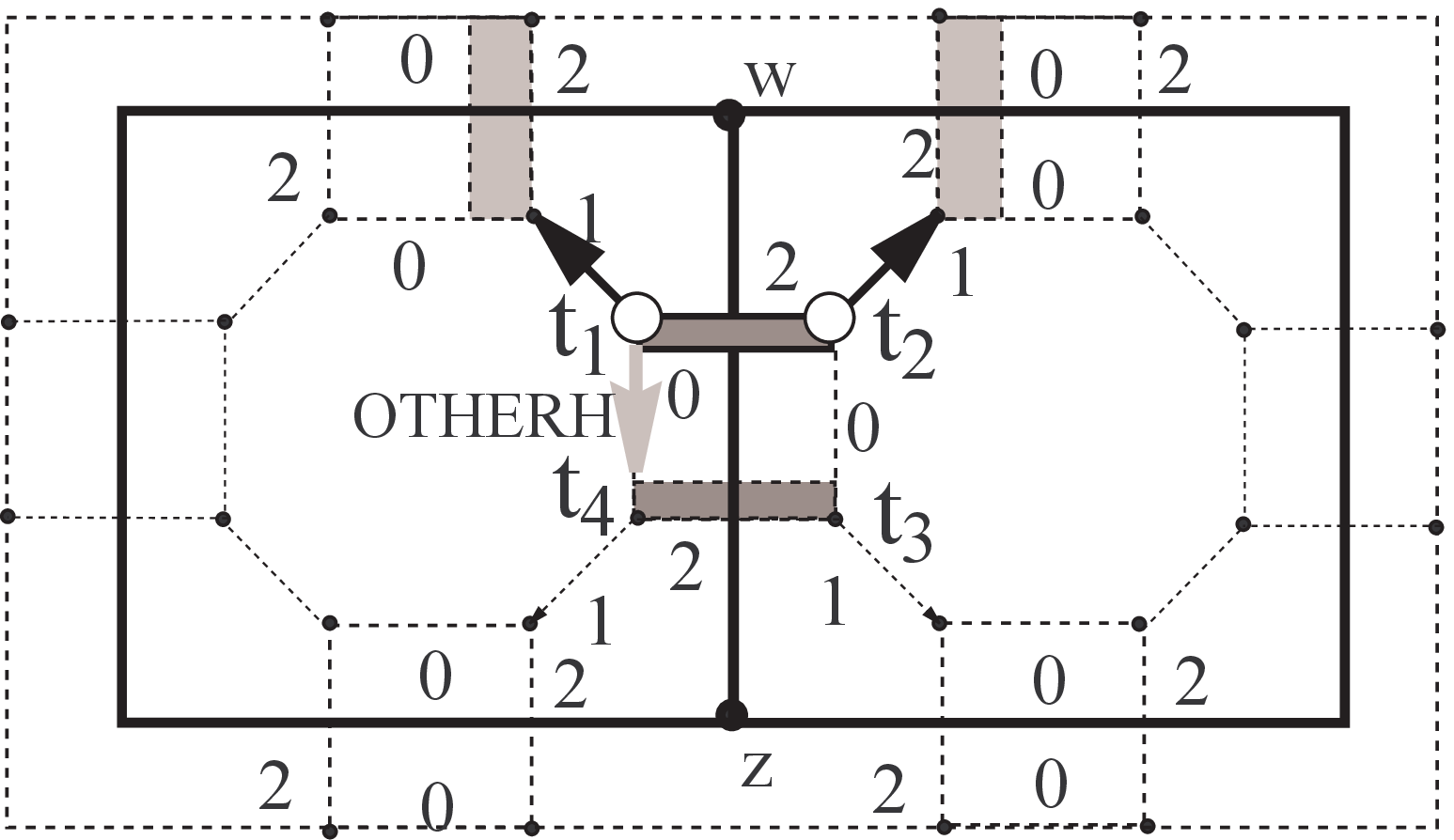,width=\textwidth}
\begin{center}(b)\end{center}
\end{minipage}
\end{center}
\caption{Translation of the cell tuple of Figure  \ref{fig:celltuple}
into  FES (a) and VES (b) data structures}
\label{fig:celltutr}
\end{figure}
}
To obtain a FES half edge, (see Section \ref{sec:he})  we consider 0-orbits   
Let $o$ be the 0-orbit (in dark gray in Figure \ref{fig:celltutr} (a)) given by
the pair $t_1$ and $t_4$ with 
${\rm switch}_0(t_{1})=t_4$. 
For  each 0-orbit $o$ we introduce a FE half-edge 
whose pointer OTHERH points the half-edge  introduced
for the  0-orbits containing the tuple pointed by ${\rm switch}_2$ (in dark gray in Figure \ref{fig:celltutr} (a)). In this case points $t_2$ since
${\rm switch}_2(t_1)=t_2$. 
In Figure \ref{fig:celltutr}(a) the two wings points with solid black arrows  the half-edges introduced for  the results of the mappings
${\rm switch}_1(t_1)$ and ${\rm switch}_1(t_4)$.

Similarly, to obtain a VES half edge (see Section \ref{sec:he}), we consider 2-orbits   
Let $o$ be the 2-orbit (in dark gray in Figure \ref{fig:celltutr}b) given by
$o=\{t_1,t_2\}$ . This is a 2-orbit since 
${\rm switch}_2(t_{1})=t_2$. 
For  each 2-orbit $o$ we introduce a VE half-edge 
whose pointer OTHERH points to the half-edge  introduced
for the  2-orbit containing  $t_3$ and $t_4$ (in dark  gray in Figure \ref{fig:celltutr}(b)).
Therefore OTHERH is translated using ${\rm switch}_0$ 
indeed ${\rm switch}_0(t_1)=t_4$.  
The two wings (solid black arrows in Figure \ref{fig:celltutr} (b)) 
points the half-edges introduced for the orbits we can reach thru the maps
${\rm switch}_1(t_1)$ and ${\rm switch}_1(t_2)$.

More complex is the relation between the facet-edge  (see Section \ref{sec:fe})
data structure and the cell-tuple. To understand this relation we must 
consider the fragment of the facet-edge structure reported in Figure 
\ref{fig:quadfacet}. In Figure {\ref{fig:celltufacet} (a)}  we draw the cell-tuple graph for  the 
fragment of complex of Figure \ref{fig:quadfacet}.
{
\begin{figure}[h]
\begin{center}
\fbox{
\parbox[c][7.5cm]{0.46\textwidth}{
\vfill
\psfig{file=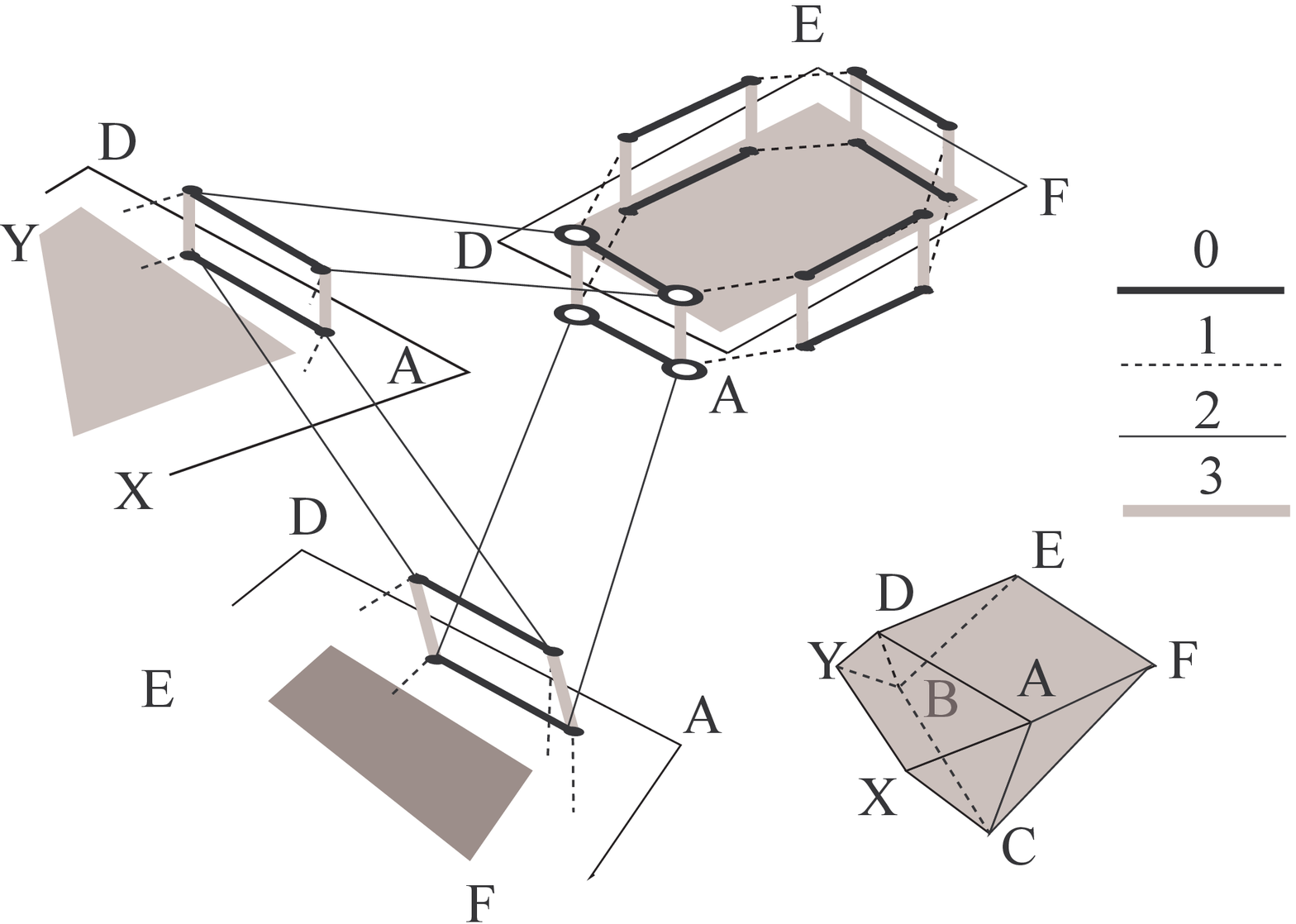,width=.46\textwidth}
\vfill
\begin{center}(a)\end{center}
}
}
\fbox{
\parbox[c][7.5cm]{0.46\textwidth}{
\vfill
\psfig{file=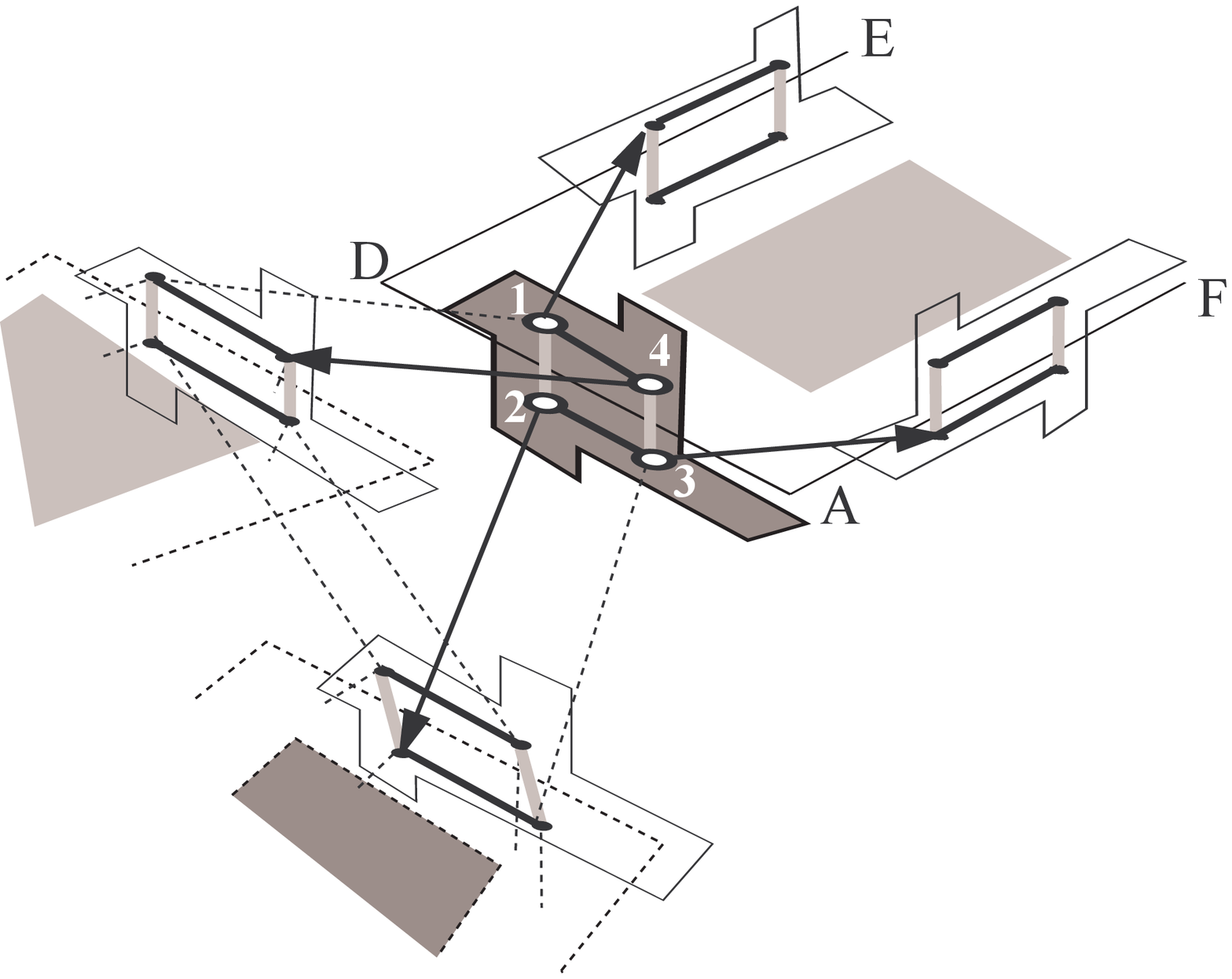,width=.46\textwidth}
\vfill
\begin{center}(b)\end{center}
}
}
\end{center}
\caption{Translation of the cell-tuple data structure for the complex in
the lower right corner of (a) into the corresponding
facet-edge data structure in (b) and also in Figure   \ref{fig:quadfacet}}
\label{fig:celltufacet}
\end{figure}
}
In the graph fragment of Figure \ref{fig:celltufacet}(a) we report the cell-tuples containing the face DAFE
(the rectangle on the right) and the cell-tuples containing the edge DA. 
Note that each face is encoded as if it was surrounded by a outer space. Thus for each edge we take four cell tuples.
Note that in this graph of cell tuples edges are drawn with different styles instead of
using labels. A translation for these styles 
is reported on the left of
Figure \ref{fig:celltufacet}(a).  
We can translate this graph into the facet-edge data 
structure associating a facet-edge to each cell-tuple.
The four large white spots in Figure {\ref{fig:celltufacet}(a)} are 
translated into the four facet-edges that can be associated to the
four variants of the facet edge pair $\brk{AD,DAFE}$. 
However, in the
facet-edge data structure facet-edge are not represented  directly.
In fact, the basic entity in the facet-edge data structure  is the 
facet-edge node. Therefore, to translate the cell-tuple data structure for a 
subdivided $3$-manifold into a facet-edge data structure  we must introduce
a facet-edge node for each 03-orbit in the cell-tuple graph.
This is what is represented in Figure {\ref{fig:celltufacet} (b)}.
A facet-edge node is associated  with the
03-orbit (in dark gray in Figure {\ref{fig:celltufacet}(b)}) given by
$o=\{t_1,t_2,t_3,t_4\}$ (in Figure \ref{fig:celltufacet}(b) the cell-tuple $t_i$, for $i$ equal to
1,2,3 and 4,
is the white blob labeled with $i$).
We have, for $i=2,4$,
${\rm switch}_3(t_{i-1})=t_i$ and ${\rm switch}_0(t_{i})=t_{(i+1) mod 4}$.
For  each 03-orbit  we introduce a facet-edge node 
whose four references  (solid arrows in Figure {\ref{fig:celltufacet} (b)})
points the other  four facet-edge nodes introduced
for the four  03-orbits containing, respectively, 
${\rm switch}_1(t_1)$, ${\rm switch}_2(t_2)$, ${\rm switch}_1(t_3)$ and   ${\rm switch}_2(t_4)$. 
The final facet-edge data structure is the one shown in Figure 
\ref{fig:quadfacet}. This completes the description of the translation of the  Cell Tuple into the  Facet Edge. 

The domain of representation of cell-tuples is restricted to
subdivided $d$-manifolds but
it can be proven \cite{Lie91} that the representation domain of cell-tuple
is wider than this. We  discuss this issue in the next section showing the
relation between cell-tuples and n-G-maps.

The cell-tuple representation is extremely verbose. If we encode a simplicial
subdivision of a $d$-manifold  we must store, for each $d$-simplex, 
{$\factorial{(d+1)}$} tuples each containing $(d+1)$ elements.
If we represent the cell tuple
data structure through its graph we have to accomodate 
$\factorial{(d+1)}$ nodes for each $d$-simplex.  
Each graph node  is of order $(d+1)$.  
We can implement nodes as an array of $(d+1)$  references to 
other $(d+1)$ nodes. 
In both cases we have to use {$(d+1)\factorial{(d+1)}$} reference for 
each $d$-simplex.
}

{
\subsubsection{\emi{n-G-maps}}
\label{sec:ngmap}
The n-G-maps \cite{Lie91}
is an implicit cell model, like the cell-tuple, but
n-G-maps have
an expressive power higher \wrt\  cell-tuples
since they can describe a subclass of pseudomanifolds,
called {\em cellular quasi-manifolds}.
An n-G-map model is described by a set of {\em paste} relationship  
$\alpha_i$
between primitive elements, called {\em darts}.
For a $d$-dimensional space $d+1$
paste relationships are necessary.
Relationships must satisfy certain
constraint that enforce coherence in the representation.
The three constraints are:
\begin{itemize}
\item
each $\alpha_i$ must be an involution, i.e., $\alpha_i^2=\alpha_i$;
\item
$\alpha_i$, for $0\le i<n$, must not have fixed points;
\item
$\alpha_i\alpha_{i+2+k}$ must be an involution for all $0\le i<i+2+k\le n$
\end{itemize}
Whenever $\alpha_n$ is also an involution the n-G-map is called {\em closed}.

{
When these constraints are satisfied, we can  look at
kernel relationships as relations that paste together darts.
Pasting together darts leads to the modeled object.
{
{
\begin{figure}[h]
\centerline{\fbox{\psfig{file=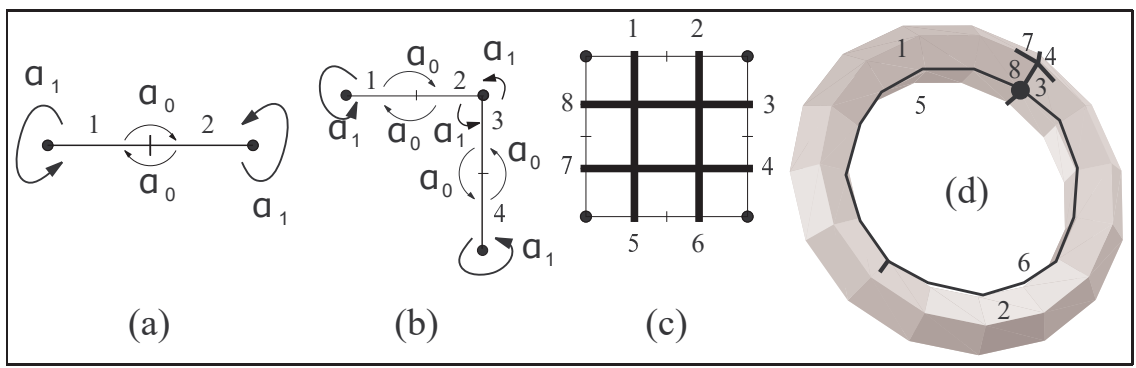, width=.9\textwidth}}}
\caption{Examples of n-G-maps}
\label{fig:ngmap}
\end{figure}
}
In Figure  \ref{fig:ngmap} we find examples of complexes modeled with
n-G-maps.
From left to right we have an example of a 1-G-map formed by two darts
labeled $1$ and $2$. In the complex of Figure \ref{fig:ngmap}a 
involution $\alpha_0$ is defined completely
by the equation $\alpha_0(1)=2$ while involution $\alpha_1$ is the identity. 
Therefore the n-G-map of Figure \ref{fig:ngmap}a  is an example of a
non-closed n-G-map.
The complex in Figure  \ref{fig:ngmap}b is an example of a 
non-closed  1-G-map defined by
$\alpha_0(1)=2$,
$\alpha_0(3)=4$,
$\alpha_1(1)=1$,
$\alpha_1(3)=2$ and
$\alpha_1(4)=4$.
The complex in Figure \ref{fig:ngmap}c, if we do not consider the four
thick black stripes, is the 1-G-map for the perimeter of a square.
This is
obtained using four darts and the involutions: 
$\alpha_0(1)=2$,
$\alpha_0(3)=4$,
$\alpha_0(5)=6$,
$\alpha_0(7)=8$,
$\alpha_1(1)=8$,
$\alpha_1(2)=3$,
$\alpha_1(4)=6$ and
$\alpha_1(5)=7$.
Note that arrow labels are omitted in Figure \ref{fig:ngmap}c.
This  defines a closed 1-G-map. We can transform this into the toroidal 
surface of Figure \ref{fig:ngmap}c by stitching together darts: 
1 and 5, 2 and 6, 3 and 8, 4 and 7.
In general a surface is modeled by a 2-G-map that can be obtained
from a 1-G-map by 
introducing the involution $\alpha_2$
In this case the involution we need to build the toroidal surface is defined by:
$\alpha_2(1)=5$,
$\alpha_2(2)=6$,
$\alpha_2(3)=8$ and
$\alpha_2(4)=7$.
}
This is denoted in Figure \ref{fig:ngmap}c 
(following the graphical conventions in \cite{Lie91}) 
by the four thick black stripes.

After these examples we can say that, in general,
in n-G-maps the modeled object is built pasting together darts
with $\alpha_i$ pasting relations.
Relation 
$\alpha_0$ connects darts to form edges. For this reason the
dart is usually graphically represented as a  VE half-edge. 
Then, by connecting
darts (with $\alpha_1$) we connect adjacent edges to form both
cycles  and open paths.
Cycles   will be used to model  2-cells.
Then we connect darts in adjacent 2-cells (with $\alpha_2$)
to form a surface with or without boundaries.
Surfaces without boundary will model a 3-cells.
Going on, (with $\alpha_3$) we can connect 3-cells
and build a $3$-complex.
}

The representation domain of n-G-maps is the set of {\em Cellular \Qm}
\cite{Lie94}. This class of complexes is especially relevant in this
thesis and therefore we will give a deeper insight into this definition. 
Cellular \qm\ are pseudomanifolds whose cellular decomposition can be 
triangulated into a simplicial set that is a simplicial \qm. To introduce
simplicial \qm\ we first need to introduce {\em numbered simplicial sets}. 

A numbered simplicial $d$-dimensional set is a simplicial  $d$-complex 
whose vertices are labeled with integers from $0$ to $d$. 
In each $h$-simplex of a numbered simplicial $d$-dimensional set 
the $h+1$ labels for vertices must be distinct. 
Not all simplicial complexes can become a numbered simplicial complexes
adding a labeling to vertices.
For instance, the boundary of the tetrahedron is an example of a
$2$-complex that cannot be labeled as a  numbered simplicial complex. 
Once that we have assigned labels 0,1,2 to three vertices of the tetrahedron,
there is no vay to assign a label from 0 to 2 to the fourth vertex
of the tetrahedron.
Whatever will be our labeling choice we will not have  
distinct labels for all triangles on the boundary of the tetrahedron.
This proves that the boundary of a tetrahedron is not a numbered  simplicial complex.
However it can be proven that is always possible to find a such a labeling for
the  barycentric subdivision of a given simplicial 
complex so that  the subdivision becomes a numbered simplicial complex.

A simplicial \qm\ $d$-complex is a numbered simplicial $d$-complex that 
can be obtained from a collection of disjoint $d$-simplices identifying
$(d-1)$ faces in such a way that at most two $d$-simplices share a $(d-1)$-face.
Cellular \qm\ are defined as pseudomanifolds that have a triangulation that
is a simplicial \qm.

Note that according to the original definition the boundary of 
the tetrahedron is a simplicial complex that is a cellular \qm\ but 
{\em it is not} a numbered simplicial \qm.

In this thesis we will consider only simplicial subdivision, therefore
we will use the term \qm\ to denote a simplicial complex that is a cellular
\qm. It is easy to show that 
all and alone the cellular \qm\ that are simplicial
can be obtained from a collection of disjoint $d$-simplices identifying
$(d-1)$ faces in a way that at most two $d$-simplices share a $(d-1)$-face.

{
For surfaces (i.e. 2-manifolds) it has been proven \cite{Gri76}
that all subdivision of $2$-manifolds can be expressed 
by gathering 
cells homeomorphic to a disk and by identifying edges. Thus 
we can say that 2-G-maps can express all and alone 
surfaces and \qm\ $2$-complexes coincide with $2$-manifolds. 
In higher dimension we can always find a bijection
between classes of topological spaces and  sub-classes of n-G-maps.
However, in general, it is unknown  which is class of topological 
spaces that can be expressed by gathering 
$d$-cells homeomorphic to a $d$-ball and by identifying cell $(d-1)$-faces. 
}

Even if the n-G-maps  modeling cannot express the whole
non-manifold domain,
n-G-maps are the most expressive scheme among implicit cell
representations.
Their expressive power  is beyond that of cell-tuples.
A representation based on {\em chains} of n-G-Maps \cite{LieElt93}
can be used to represent arbitrary cell complexes with a mix
of open and closed cells.
In this scheme, we have a 2-level
hierarchy where both chains and  cells must be represented.
A straightforward
implementation  of n-G-Map chains must implement explicitly both
chains and n-G-maps, the latter being used to represent cells.

There is a tight relation between cell-tuples and n-G-maps as abstract
combinatorial objects. 
According to \cite{Bri93} one can translate a cell-tuple into an
n-G-map by introducing a dart $d(t_i)$ for each cell tuple $t_i$
and by defining  the involution $\alpha_k$ as
$\alpha_k(d(t_i))=d({\rm switch}_k(t_i))$.
A minor difference between n-G-maps and cell-tuples is in the way they
handle boundaries. The cell-tuple approach assumes that the modeled 
$d$-complex in surrounded by a special $d$-cell $c_\infty$.
On the other hand n-G-maps bend darts
into  fixpoints for the involution $\alpha_n$ to express boundaries.
To cope with this distinction one have to pre-process
a cell-tuple representation before translating it into n-G-maps and
redefine ${\rm switch}_n(t)$ as  ${\rm switch}_n(t)=t$
whenever the 
cell-tuple ${\rm switch}_n(t)$ contains $c_\infty$.

On the other hand, following \cite{Lie94b},  one 
can find examples of 2-G-maps
for a certain subdivision of a subdivided $2$-manifold that cannot 
be translated directly into a cell-tuple.
Let consider for instance the triangle $A=xyz$ in Figure \ref{fig:nocelltu} (a)
with an internal point $t$ and an internal edge $d$.
This correspond to the 2-G-map in Figure \ref{fig:nocelltu} (b) where darts
1 and 4 and darts 2 and 3 are tied together by the involution $\alpha_2$
(thick black stripes). This cannot be expressed as a cell-tuple
structure introducing a cell-tuple for each dart.
If we attempt to do this we end up in a cell-tuple structure
containing the three tuples $(A,b,x)$,$(A,c,x)$ and $(A,d,x)$.
This is not allowed in the cell-tuple scheme.
Indeed, in this case, we have 
that relation ${\rm switch}_1$ is not a function.
In fact,  there are two possible results when ${\rm switch}_1$ is 
applied to each of these three tuples above.
{
\begin{figure}[h]
\centerline{\fbox{\psfig{file=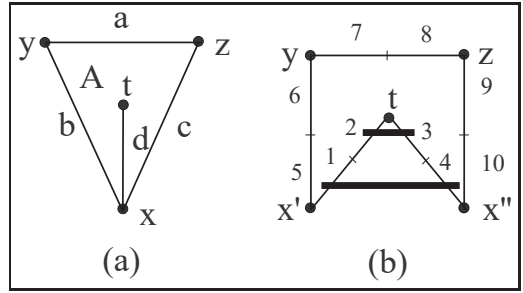, width=.4\textwidth}}}
\caption{An example of a 2-G-maps that cannot be translated directly into a 
cell-tuple}
\label{fig:nocelltu}
\end{figure}
}

An implementation of n-G-maps is proposed \cite{Ber89,Lie89b} using
$d+1$ pointers for each dart in a $d$-dimensional complex. 
In a simplicial subdivision each
$d$-simplex will require $\factorial{(d+1)}$ darts and thus
$(d+1)\factorial{(d+1)}$ references. 
}

{
\subsubsection{The \ema{Winged}{Representation}}
\label{sec:wingedpao}
The winged representation \cite{PaoAl93} is a dimension independent
modeling approach  for simplicial subdivisions.
The domain of the winged  representation is the subclass of the
topological sub-spaces of $\real^n$ 
that can be represented by a uniformly dimensional simplicial
subdivision such that a $(d-1)$-simplex is adjacent at most to 
two $d$-simplices.
The original Winged Representation, 
for a uniformly dimensional $d$-complex $\AComp$, is a pair 
$(\AComp^{[d]},\Adj)$ where
{$\AComp^{[d]}$} is the subset of the $d$-complex $\AComp$ made up of
all  $d$-simplices in $\AComp$ and $\Adj$ is an
{\em adjacency function} that associates each $d$-simplex with the
$(d+1)$-tuple of $d$-simplices that are adjacent to it.
Both vertices and $d$-simplices are represented through indexes and
$d$-simplices in  {$\AComp^{[d]}$} are represented as (d+1)-tuples 
of vertex indexes. Thus the data structure for 
$\AComp^{[d]}$ will be a function that gives,
for each top simplex index $t$, the $(d+1)$ indexes of the vertices in $t$.
This is often called the TV relation (Top simplex to Vertex).
Similarly the data structure for the adjacency function $\Adj$  
will be a function that gives,
for each top simplex index $t$, the $(d+1)$ indexes of the top simplices
adjacent to $t$.
For simplices that are not adjacent exactly to $(d+1)$ $d$-simplices,
the special symbol $\bot$ is used, at some places, in the corresponding
$(d+1)$ tuple to mean ''no adjacency''.
This is often called the TT relation (Top simplex to Top simplex).
For this reason the winged representation is sometimes called  the TV,TT
data structure or {\em indexed data structure with adjacency}.

The original work introducing the name ''winged representation'' 
\cite{PaoAl93} defines a functional language 
that allows to manipulate polyhedra that are geometric realizations
of winged representations. A rich set of operations is provided
including: boundary operator, extrusion operators, editing 
through the application of  simplicial maps and set theoretic operations.
In particular the algorithm for boundary extraction \cite{FerPao91} 
shows that one can effectively navigate the simplicial complex 
using relations $\Adj$.
In spite of this flexibility
the winged representation is extremely compact since  it uses $6f$ references
to encode a manifold surface and $8t$ references to encode a 
tetrahedralization. 
In general, $2(d+1)$ references are needed
for each top $d$-simplex in the modeled $d$-complex.
In \optt, we give an optimization procedure that can encode
implicitly some information in this representation saving 
$v$ references.

The problem of extracting all topological relations 
out of this representation for $d$-manifolds is briefly reported  in the following.
A first problem is to extract, for a given  $0\le m\le d$, all $m$-simplices 
incident to a given vertex $v$.
The  extraction of these, vertex based, 
topological relations can be performed adding $v$ references to encode 
the partial relation ${\rm VT}^{*}$
that gives, for each vertex, an incident $d$-simplex. 
The optimization in \optt,  encodes
implicitly, this relation, as well.

In order to extract, for given  $0<i< m\le d$, all $m$-simplices 
incident to a given $i$-simplex $\gamma$,
can introduce similar partial relations (denoted by $\FT{i}$)
that gives, for each $i$-simplex, for $0<i<d$, 
an incident $d$-simplex.

In this thesis (see Section  \ref{sec:trie}) we designed 
a {\em trie} \cite{Fre60} based data structure
that can encode all these $\FT{m}$ partial relations using less than  
three references for each $m$-simplex for $0\le m\le (d-1)$ 
and less than two references for each $(d-1)$-simplex. 
(See Property \ref{pro:vmtsize}).

Thus, 
we can encode  the winged representation and all these partial relations 
$\FT{i}$ for our simplicial
complex using $6f+2e+3v$ references for surfaces and 
$8t+2f+3e+3v$ references for  tetrahedralizations.
Using the optimization in \optt\ this reduces to, respectively,
$6f+2e+v$ references for surfaces and 
$8t+2f+3e+v$ references for  tetrahedralizations.

For a simplicial subdivision of a  $2$-manifold we have $3f\le 2e$ and thus
we have an \occup\ smaller than $6e+v$. For a closed manifold 
homeomorphic to a sphere we have $v=(6+e)/3$ and thus the \occup\ 
for this kind of surface an optimized winged representation takes less
than $6.3e+2$, thus being more compact than all other solutions so
far presented. We recall that the best \occup\ for $2$-manifolds, up
to now, was $8e$ provided by the symmetric data structure (see Table 
\ref{tab:mani}).

For a $3$-manifolds
the winged representation
is more compact than the symmetric structure for $3$-manifolds (recall
that in Section \ref{sec:simm3d} we have shown that
the symmetric structure takes $8t+3f+3e+v$). Thus, also for three manifolds,
an optimized winged structure is the most compact solution for a 
tetrahedralization.

It can be proven (see Properties \ref{pro:compl}, \ref{pro:complnm}, 
\ref{pro:optext}, \ref{pro:corrSnh} and \ref{pro:nmopt})
that, for a $d$-manifold, for $d=2$ and $d=3$,
all topological relations can be extracted in linear time from the 
winged representation.

In this thesis we will study the winged representation in order to 
extend it to an efficient data structure to encode the components of 
our decomposition. Thus  several results in this thesis are linked with
this  data structure. A brief discussion of these links is 
reported at the end of this  chapter.
}
\subsection{Non-Manifold Modeling Data Structures}
\label{sec:nonmanisol}
In this section we review some modeling approaches that have been devised
to model cellular subdivisions of solid objects with  non-manifold situations.
We first note that all these approaches assume that at most
two $3$-cells in a $3$-complex shares the same $2$-face.
Thus, the intended modeling domain coincides with the subset of $3$-polyhedra 
embeddable in $\real^3$.
In related literature we encounter different subclasses of this modeling 
domain.
The term \ems{r-set} \cite{req80} is used
to denote the set of solids that can be expressed as cellular complexes where
all cells have dimension $3$.
An r-set has the property that it 
coincides with the closure of its interior.

A proper subclass of r-sets is the class of
\emas{manifold}{solids} that is simply the class of solids bounded by a
geometric realizations, in $\real^3$,
of a closed orientable $2$-manifold surface.
Manifold solids  can be built with a finite number of finitary operations
called {\em Euler operators} \cite{req80,Hop89} 

The class of r-sets that are not collections of manifold solids is called
the set of \emas{non-manifold}{solids}.
Within this class five types of non-manifold situations can occur
(sometimes called {\em special notches}) The are shown in Figure \ref{fig:nonmanisolid} from (a) to (e).
{
\begin{figure}[h]
\centerline{\fbox{\psfig{file=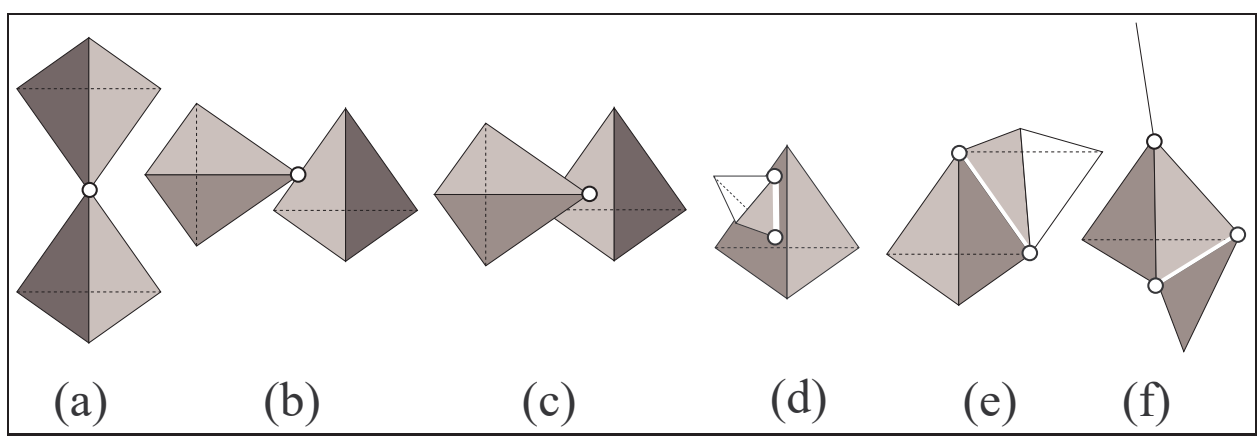, width=\textwidth}}}
\caption{Examples of r-sets that are  non-manifold solids 
with five different types of special notches (from a to e)
and of pseudomanifold solids (from a to c). In (f) we have a non-manifold
solid with dangling edges and dangling faces. 
In white, in each solid, we have the set of points that are not manifold
points according to the topological definition of manifoldness.}
\label{fig:nonmanisolid}
\end{figure}
}
The subclass of non-manifold solids where conditions of Figures
\ref{fig:nonmanisolid} (d) and \ref{fig:nonmanisolid} (e) 
do not occur is called the class of \emas{pseudomanifold}{solids}.
Note that special notches are non-manifold situations also in the boundary
of the solids. Furthermore  the boundary of a pseudomanifold solid is a
pseudomanifold surface.

Finally we have that the class  of  $3$-polyhedra imbeddable in $\real^3$ is
more general than the class of r-sets since these solids might
have {\em dangling} edges and dangling faces. 
In Figure \ref{fig:nonmanisolid}(f) we have a complex with a dangling edge 
and a dangling triangle.
All the data structures listed in this section can model this, more
general, class of solids.
We will refer this super-class of non-manifolds as a 
\emas{realizable}{non-manifold} (in $\real^3$) while the
term non-manifold will be reserved to the topological concept.
In this thesis we treat this larger class of ''non-manifolds'' and we
do not assume embedding in $\real^3$ unless explicitly stated.

We note that data structures listed in this section
cannot model all non-manifold 
$3$-complexes. For instance the complex (embeddable in $\real^4$),
consisting of three tetrahedra sharing a triangle, cannot be modeled with
approaches listed in this section.

A second point to note is that many of the  reviewed approaches present 
both a concrete data structure and a set of {\em operators} 
to stepwise build the associated concrete data structure. 
For instance Weiler's work \cite{Wei88b} defines a set of 
operators called NMT that support the construction of a realizable 
non-manifold.
Other approaches defines a set of operators 
that can modify  a non-manifold solid preserving some sort of 
invariant. This invariant is usually presented as an extension of 
the Euler-Poincare formula for the non-manifold domain. 
The related operators are therefore called {\em generalized Euler operators}. 
In \cite{Masu92,YamKim95,LeeLee01}, for instance, different extension
of the Euler-Poincare formula are presented and related 
sets of Euler operators are introduced. 
This approach, follows  the work of 
Mantyla  \cite{Man84} that shows that the Euler operators 
are both complete and sound for the class of two-manifolds.
In particular the strong result in Weiler's   work is the fact that he gave
an inversion algorithm for the Euler operators \cite{Wei88b} producing a
sequence of Euler operations to build a given two-manifold.

The problem 
of creating a non-manifold data structure, when no sequence of operations is available, 
is partially addressed in \cite{Mcoutofcore}. In this paper is devised an
out-of-core implementation of an algorithm that
takes a (possibly huge) unstructured list of triangles in a STL input 
and builds a, possibly un-coherent, non-manifold data structure called LEDS.
However, if a daemon properly present triangles in the 
STL file, in a 
certain order, then the algorithm can output a LEDS that is topologically 
coherent with the non-manifold solid  it describes.  
In a companion paper \cite{mcm02b} this  requirement is mitigated  with
a {\em slicing} algorithm. This algorithm
requires to have all triangle normals pointing out of the
solid they bound. With this information a coherent topology is 
obtained.

In this thesis we describe the abstract notion of the decomposition of 
a  simplicial $d$-complex and design algorithms to build and
navigate a compact data structure that can represent generic, 
non-manifold, simplicial $d$-complexes using the result of the decomposition
process. In this framework the definition of construction operators
for this data structure is bypassed. 
Therefore the related literature on non-manifold operators, is neglected in this chapter.  
In this we follow the approach 
in \cite{Mcoutofcore} and we operate on an unordered set of simplices.
However we do not assume a particular ordering for our input and yet
we produce a coherent data structure from which we can extract all
topological relations.

A short discussion on the possible relation between the decomposition 
problem and non-manifold Euler operators is discussed at the end of this 
chapter.  

The three data structures  we are going to review in this section about non-manifold data structures are:
the  Radial--Edge data structure (RES) \cite{Wei86},
the Tri-Ciclic Cusp (TCC) \cite{Gur90} and 
the Partial Entity data structure (PES) \cite{LeeLee01} 
The RES data structure is historically the first proposal for cellular 
subdivisions of realizable  non-manifolds.
The similar data structures \cite{Gur90,YamKim95,LeeLee01,McM00} are
basically more compact revision  of the radial--edge data structure.
Related paper points out some pitfalls of the RES and propose some 
variation to correct it.
{
\subsubsection{\emi{Radial Edge}}
The Radial Edge Structure (RES) \cite{Wei86} 
is a modeling approach
where a realizable non-manifold is represented by a set of basic
elements called {\em uses}. Each use encodes a specific pair of 
instances of topological elements
(i.e. vertices, faces, edges, etc. ) in some adjaceny relation.
For instance, the edge use is the encoding of an edge-face pair within
the ternary relation among edges. faces and cells.

There are seven topological elements that are considered in the
RES. These are: the region, the shell, the face, 
the  loop, the edge and the vertex:
We report here the origial description of these seven RES  
topological entities.
Between parenthesis we report the shorthands used in Figure 
\ref{fig:radial} to denote the RES topological entities.
{
\begin{figure}[hpt]
\centerline{
\fbox{
   \epsfig{file=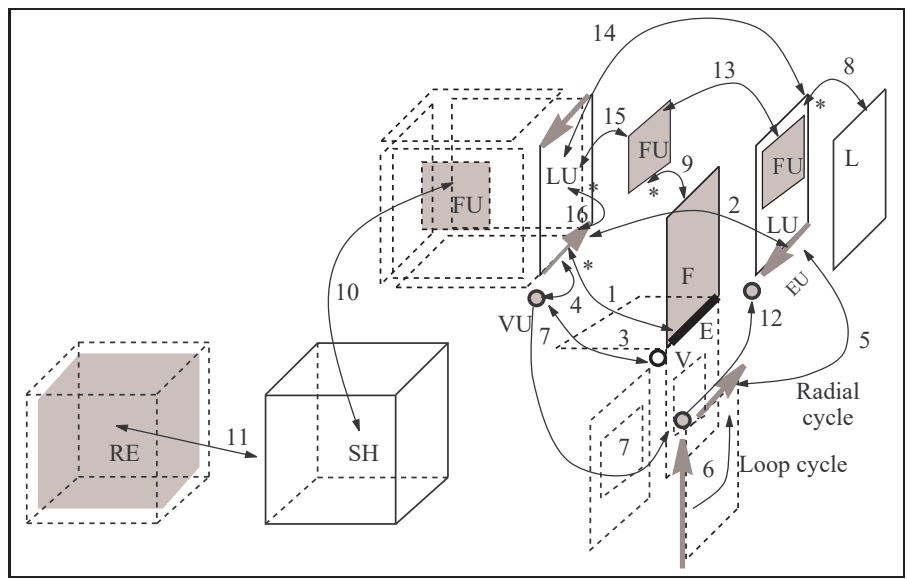,width=.7\textwidth}
}
}
\caption{Topological entities and uses in a radial--edge data structure 
together with the mutual references as in BRL-CAD \cite{brlcad} 
implementation}
\label{fig:radial}
\end{figure}
}

A {\em model} is a single three-dimensional topological modeling space, 
consisting of one or more distinct  regions of space. A model is not strictly a topological element as such, but acts as a repository for all topological elements contained in a geometric model

A {\em region} (RE) is a volume of space. 
There is always at least one in a model. Only one region in a model 
may have infinite extent; 
all others have a finite extent, 
and when more than one region exists in a model, all regions have a boundary.

A {\em shell} (SH) is an oriented boundary surface of a region. 
A single region may have more than one shell, as in the case  
of a solid object with a void contained within it. 
A shell may consist of a connected set of faces which form a closed volume or may be an open set of adjacent faces, a wireframe, or a combination of these, or even a single point.

A {\em face} (F)  is a bounded portion of a shell. 
It is orientable, though not oriented, as two region boundaries 
(shells) may use different sides of the same face. 
Thus only the use of a face by a shell is oriented. 
Strictly speaking, a face consists of the piece of surface it covers, 
but does not include its boundaries.

A {\em loop} (L) is a connected boundary of a single face. 
A face may have one or more loops, for example  a polygon would require 
one loop and a face with a hole in it would require two loops. 
Loops normally consist of an alternating sequence of edges and vertices 
in an open circuit, but may consist of only a single vertex. 
Loops are also orientable but not oriented, 
as they bound a face which may be used by up to two different shells. 
Thus, it is the use of a loop that is oriented (this will be introduced later
as the {\em loop use}).

An {\em edge} (E) is a portion of a loop boundary between two vertices. 
Topologically, an edge is a boundary curve segment which may serve as part 
of a loop boundary for one or more faces which meet at that edge. 
Every edge is bounded by a vertex at each end (possibly the same one). 
An edge is orientable, though not oriented; 
it is the use of an edge which is oriented
(this will be introduced later
as the {\em edge use}).

A {\em vertex} (V) is a topologically unique point in space, that is, 
no two vertices may exist at the same geometric location. 
Single vertices may also serve as boundaries of faces and as complete shell 
boundaries.

The usage in a shell of the four topological entities: faces, loops edges and
vertices is explicitly represented in the RES 
through objects called {\em uses}. 
A use is instantiated for each occurrence of the corresponding 
topological entity in a 
particular shell.  Thus the RES introduces four types of
uses objects.

A {\em face--use} (FU) is one of the two uses (sides) of a face. 
Face-uses, the use of a face by a shell, are oriented with respect 
to the face geometry.

A {\em loop--use} (LU) is one of the uses of a loop associated with one of the 
two uses of a face. It is oriented with respect to the associated face use.

An {\em edge--use} (EU) is an oriented boundary curve segment on a loop-use of 
a face-use and represents the use of an edge by that loop-use, 
or if a wireframe edge, by endpoint vertices. 
Orientation is specified with respect to edge geometry. 
There may be many uses of a single edge in a model, but there will always 
be an even number of edge-uses. 
A wireframe edge produces two edge uses, one for each end of the edge.

A {\em vertex--use} (VU) is a structure representing the adjacency use of a 
vertex by an edge as an edge point, 
by a loop in the case of a single vertex loop, or by a shell in the case of a single vertex shell.

A FU is bound by a set of LU. 
For a simply-connected face we have just one LU. For multiply-connected  
faces we have multiple LUs and one of them encloses the others. 
We call the enclosing LU a bounding LU.
In Figure \ref{fig:twozone}(a) face BDEF is a simply connected face while
face bounded by ABCD is multiply connected. Cycle ABCD enclose cycle GHIJ.

Regions, i.e.  $3$-cells, are defined by  shells of face-uses.
Face-uses comes in pairs for each face.
Obviously there is a bijection between FUs and bounding LUs.
The orientation of the bounding LUs for a certain FU $fu$ 
is a positive sense of rotation determined standing on $fu$ outside 
of the region bounded by the shell to which $fu$ belongs.

The radial--edge is an explicit cell scheme since
each topological entity is represented by a set of uses.
In fact we have:
\begin{itemize}
\item
A region, i.e. a $3$-cell
is bounded by a set of {\em shells}.
\item A shell is defined by a set of FUs and EUs.
EUs at this level models dangling edges.
Alternatively, a shell can be an isolated VU.
\item A face is represented by a pair of FUs, one for each
orientation;
\item A loop is represented by a pair of LUs, one for each
orientation; One or more loop-uses  bound a face use. More that one loop
uses is needed for faces that are not simply connected.
An oriented loop, i.e. a LU, is defined by a set of EU.
Alternatively an  oriented loop can be an isolated VU.
\item An edge is represented by a set of EUs, two EUs are
introduced for each
face  adjacent to an edge.
The two EUs correspond to the two possible
orientations of the face;
\item A vertex is represented by a set of VU.
There is a vertex use for each face use incident to that vertex.
Multiple vertex uses are linked together in a unique list. 
\end{itemize}

We note that
the choice of linking all vertex uses in a unique list implies that we have
the same sort of list of vertex uses for the two situations in 
Figure \ref{fig:twozone} (b).
In this sense the radial--edge do not models correctly all the
pseudomanifold boundaries.
{
\begin{figure}[h]
\fbox{
\parbox[c][5cm]{0.25\textwidth}{
\vfill
\psfig{file=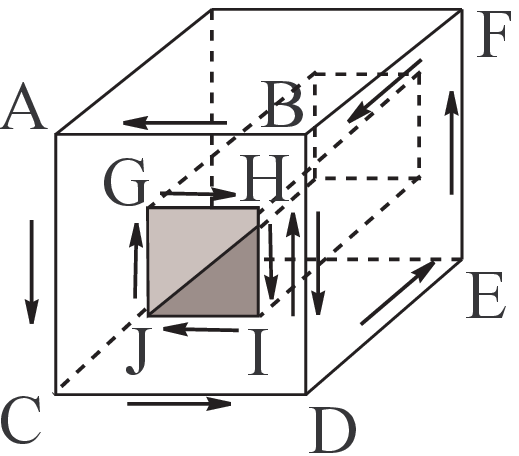,width=0.25\textwidth}
\begin{center}(a)\end{center}
}
}
\hfill
\fbox{
\parbox[c][5cm]{0.7\textwidth}{
\vfill
\epsfig{file=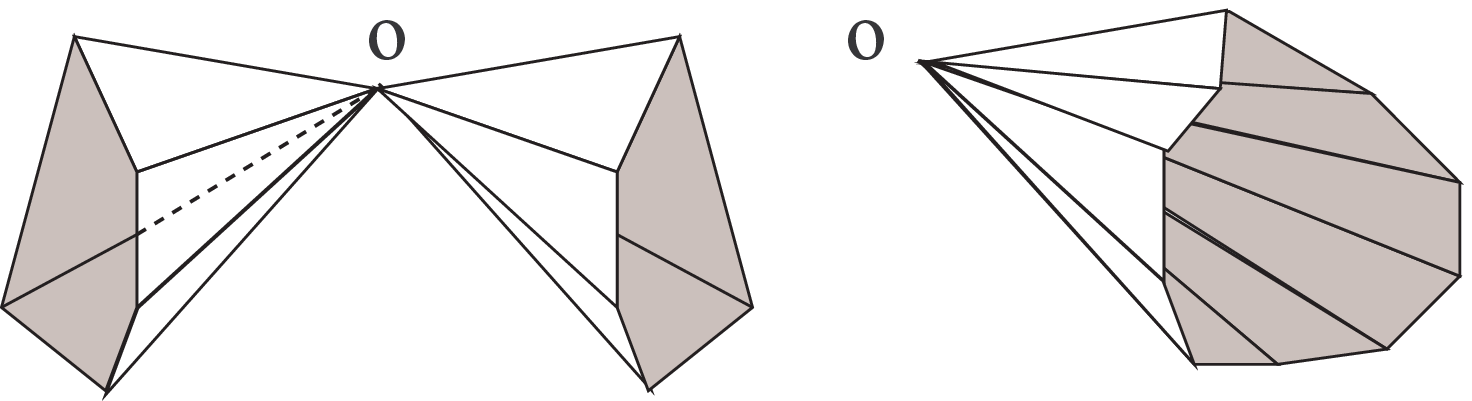,width=.7\textwidth}
\begin{center}(b)\end{center}
}
}
\caption{Orientation of loops in the RES (a) and a pair of different 
fans of ten triangles (b). 
Both fans implies that vertex $o$ has twenty vertex uses in its
radial--edge representation. In both situations the twenty uses are linked 
in a unique list.}
\label{fig:twozone}
\end{figure}
}
This problem is properly adressed by the tri-ciclic cusp data structure
\cite{Gur90} that introduce disks around vertices to group vertex uses
(see Section \ref{sec:triciclic}).

The radial--edge data structure has several 
slightly different implementations.
Some of them are part of commercial packages (e.g. Smlib \cite{smlib}).
Weiler itself at Autodesk revised the original data structure  to take
care of the above mentioned problem of isolated non-manifold vertices \cite{Wei96}. 
In this review we found convenient to describe the 
radial--edge data structure implemented in the  NMG package in
BRL-CAD \cite{brlcad}
(as described in \cite{But91}).
Thus, in Figure \ref{fig:radial} we reported all arrows for mutual references
inserted in nodes that implements the radial--edge topological entities. 
Arrow  tips labeled with a star denotes partial relations.
For a partial relation  just one reference is stored
even if the elements to be referenced are many.
Other elements will be recovered using alternative paths in this diagram. 

Arrows labeled with integers in Figures \ref{fig:radial} and \ref{fig:twoeu}
denotes relation between objects implementing topological entities. 
{
We will describe briefly the meaning of all these relations:

EUs for an edge are linked pairwise (by relation 2) and
are  organized into two 
cycles of edge uses. More in particular:
{
\begin{figure}[h]
\fbox{
\parbox[c][8cm]{0.49\textwidth}{
\psfig{file=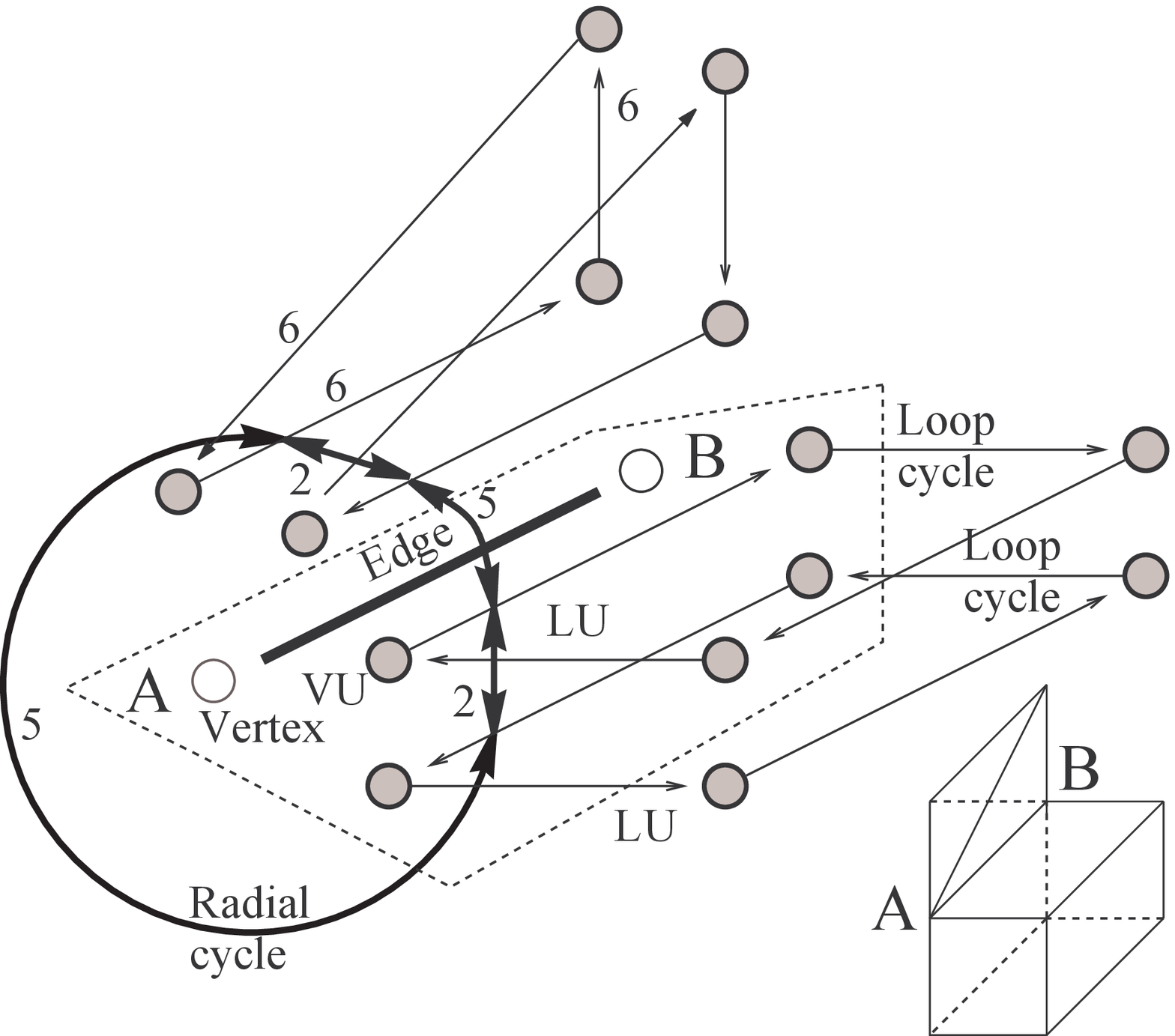,width=.49\textwidth}
\begin{center}(a)\end{center}
}
}
\hfill
\fbox{
\parbox[c][8cm]{0.44\textwidth}{
\vfill
\psfig{file=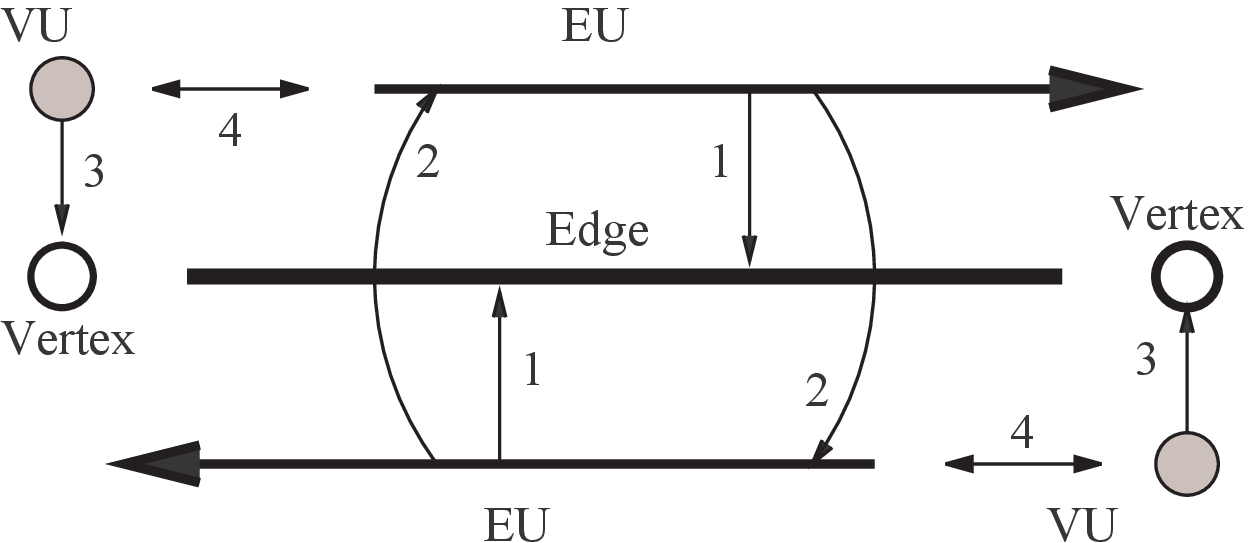,width=0.44\textwidth}
\vfill
\begin{center}(b)\end{center}
}
}
\caption{An example of radial and loop cycles in the fragment of 
the radial--edge around edge AB 
for the complex in the right lower corner of figure  (a) and an example (b) 
of relations for  a  pair of EU associated to the use of an edge
by a certain face. Figure (b) is the detail of the framed portion of 
figure (a)
(both adapted from \cite{But91}) 
}
\label{fig:twoeu}
\end{figure}
}\begin{itemize}
\item EUs are then organized into two  cycles:
One for all EUs
participating to a LU (this is what is called the {\em loop cycle} (6))
and one for all EUs of a given edge.
This is what we called the {\em radial cycle} (i.e. 2-5 cycle) 
(see Figure \ref{fig:twoeu} (a)).
\item A binary
relationship (2) is defined for the pair of  EUs for the
same edge that corresponds to the two possible
orientations of an adjacent face (see Figure \ref{fig:twoeu} (b));
\end{itemize}
The uses reference their parent in the RES hierarchy, thus
LU reference L (8), FU reference F (9), EU reference E (1), 
The inverse relations for the above three relations
are all stored as partial relations.
A VU reference (3) its parent  V and the inverse relation is 
stored completely since all VU for a certain vertex are linked in a list (7).
FUs comes in pairs for each face and corresponding pairs of FUs are 
related by relation 13.
Similarly LUs comes in pairs, two for each loop, and corresponding 
pairs of LUs are related by relation 14.
A single  bounding LU correspond to each FU in the bijection 15. 
For non-simply connected faces several LU are linked in a list
(not shown in the diagram of Figure  \ref{fig:radial}) and
the FU points to the head of this list.
A LU reference (16) one of its EUs and each EU reference (16) 
the LU to which it belongs.
All FUs reference (10) the shell they belongs to and
a shell reference one of these FU, others are linked in a list 
(not shown in the diagram of Figure  \ref{fig:radial}).
All shells reference (11) the region they bound to and
a region reference one of these shells, others are linked in a list
(not shown in the diagram of Figure  \ref{fig:radial}).
}

{
The evaluation of space requirements for this representation is 
possible if we assume something about the nature of the cellular
subdivision of the model. 
Following statistical assumptions, in \cite{LeeLee01}, a \occup\ of
$4.41$ times the space needed by the winged--edge is reported in
\cite{LeeLee01}.
This is evaluated assuming that the RES is used to encode
the boundary of a single shell manifold solid.

{We evaluate here the number of references needed to encode a 
simplicial subdivision of a regular 3-complex. 
To this aim we first list all the instances of radial--edge elements
we need  to encode a tetrahedron and report 
between parenthesis the number of 
references required by each instance. This is evaluated as the number of 
arrows coming out of the corresponding object in Figure \ref{fig:radial}.    
Thus, to model a tetrahedral cell we need  one region (1), one shell (2), a list of
four FU (4 references for the list and 3  references for each FU), 
four LU (4). For each triangle
we need three EU (7) and three VU (3). This sums to 155 references.
Note that we omitted space needed for faces (F), loops (L),  edges (E) and 
vertices (V) that will add  one reference for each of these entities. 
If one is forced to encode a tetrahedralization with this scheme
the \occup\ will be 155t+2f+e+v.  We count two references for each 
face since we have to introduce one loop for each face.
}
}
}
\subsubsection{The \ema{Tri-cyclic cusp}{Data Structure}} 
\label{sec:triciclic}
This data structure \cite{Gur90} is similar to the RES. 
Variations are introduced to correct a RES limitation.
Some topological elements are added, some elements in the RES are 
renamed and some 
names of the RES are used to denote  a different element with
some awkward overlapping of names. In Figure \ref{fig:tri}
we report a fragment of a  tri--cyclic cusp data structure 
for the complex in the lower left corner of this figure.
{
\begin{figure}[hpt]
\centerline{
\fbox{
   \epsfig{file=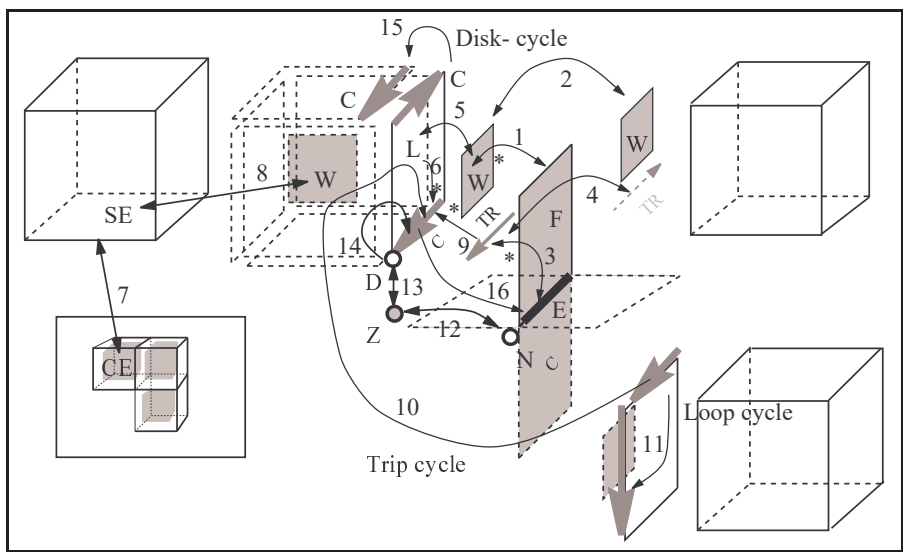,width=.7\textwidth}
}
 }
\caption{Elements in a fragment of a  tri--cyclic cusp data structure for the complex in the lower left corner.
Arrows represent a relation between topological entities described in
 \cite{Gur90}. }
\label{fig:tri}
\end{figure}
}
In the following description of the tri-ciclic data structure 
we report in  parenthesis the labels used in Figure \ref{fig:tri}
to denote topological elements in the data structure.

A first thing to note is that, 
the Loop (L) of the tri-ciclic data structure
actually plays the role of the loop-use in the RES while, 
in the tri-ciclic data structure, 
the loop node of the RES is omitted.

{
A cell (CE) in this cellular subdivision is delimited by one or more 
{\em seals} (SE) each seal being an orientable manifold surface.
In this data structure each face
is represented by two distinct
element one for each oriented face.
Oriented faces are   called {\em walls} (W). Walls corresponds 
to face-uses in the RES. 
Thus each seal is represented 
by a collection of walls with coherent orientation. 
More than one seal is used to model cells with cavities.
}

This data-structure  is an improvement
with respect to the radial--edge since it orders edges around a vertex 
into separate lists whenever several distinct fans are incident to 
a vertex.
Each separate fan is called a {\em disk} and the circular list of 
edges that belongs to a fan is called a {\em disk cycle}.   
Disks partition the 3D space around a vertex into {\em zones}.
{
\begin{figure}[hpt]
\centerline{
\fbox{
   \epsfig{file=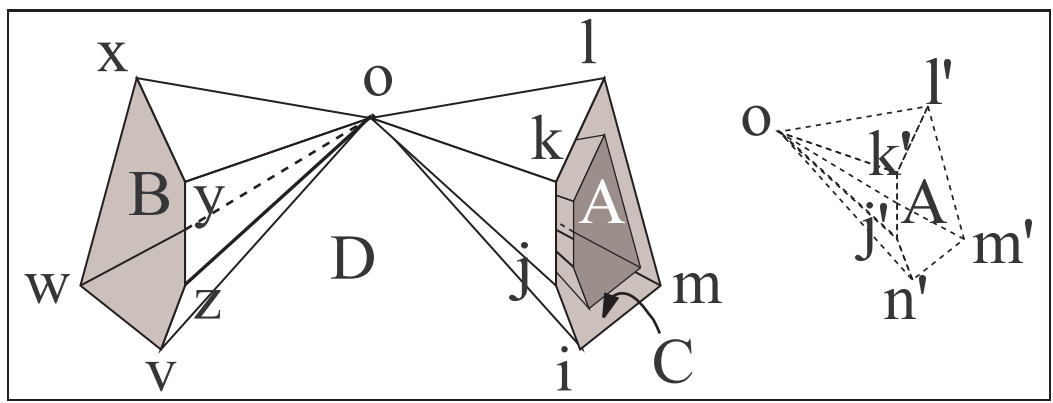}
}
 }
\caption{A example illustrating the {\em zone} concept in 
the tri--cyclic cusp representation}
\label{fig:zones}
\end{figure}
}
See for instance the situation of vertex $o$ in Figure
\ref{fig:zones}. In this figure let us consider the three cones 
from $o$ to the perimeter of the three 
polygons $xyzvw$, $ijklm$ and $i'j'k'l'm'$. 
In the situation of Figure \ref{fig:zones}
let us imagine that we have six disks, 
two coincident disks for each cone. 
Two disks are associated with each $1$-connected component 
in a vertex star.
The three components (cones) partition 
the neighborhood of vertex $o$ into three finite zones, denoted by
$A$, $B$, $C$, and an unbound zone $D$. Each zone is bound by two
cones. Thus we can partition  the six disks into four
disjoint pairs that are the boundaries for the four zones.

In the tri-ciclic
data structure each of the triangles in Figure \ref{fig:zones}  
is represented by two distinct
oriented faces (i.e. {\em walls}) that correspond 
to face-uses in the RES. 
Edges delimiting each triangle are 
duplicated for each wall. These duplicated edges are called 
{\em cusps} (C). Cusps plays the role of edge-uses in the RES.
Thus, in the tri-ciclic data structure, 
each side of each cone is actually represented by a  distinct
structure given by a list of cusps. This list of cusps is called a
{\em disk cycle}. 
{
Each cycle defines the
topological entity we called the {\em disk}.
To model this topological entity a specific topological element called
disk (D) is introduced in the data structure. 

As we have seen in Figure \ref{fig:zones} disks 
partition the space around a vertex into {\em zones}.
To model this topological entity another specific topological element 
called zone (Z) is introduced in the data structure. 
Each zone is delimited by one or more disks and 
note that we can have zones delimited by more than two disks.
An unbound zone might be delimited by just one disk.
The entity representing the zone reference these disks and 
the vertex at the apex of each disk.
For each pair of zones sharing
a boundary we have two disks, one for each side of  the common 
boundary.  Therefore each disk delimits just one {\em and not two} zones.
}

Thus in the tri-ciclic cusp data structure a vertex is modeled
by a {\em node} (N) that references a list of zones (Z) 
each of which is 
delimited (and reference) two disk cycles represented through
disks (D).

A {\em cusp} (C) is used to model 
a number of different topological entities.
Cusps can represent dangling edges or isolated vertices.
However, usually a cusp models an edge adjacent to  a wall. 
In this case the cusp participate
to the above mentioned disk cycle. Cusps 
are inserted into other two cycles. A second cycle,
called the loop cycle links cusps that delimit a wall.
A third cycle, 
called the {\em trip cycle}. links all of cusps that have the 
same orientation and are associated to the same edge.    
The name of the representation comes from the fact that
there are three cyclic relations between cusps:
the {\em disk cycle}, the {\em loops cycle} and the {\em trip cycle}.
The loop cycle links cusps that delimit a wall while
adjacency between walls is defined grouping adjacent edges (i.e cusps) into a
cycle  called  the trip cycle. Finally all cusps adjacent to the same vertex 
are grouped into cycles that defines vertex disks.

Finally we analyze the relations between the topological
elements in this data structure using integer labels 
of Figure \ref{fig:tri}. A cell in a tri-ciclic data structure reference one or more 
delimiting {\em seals} (7). Each seal reference (8) the 
collection of walls with coherent orientation that defines the seal. 

A face reference (through the partial relation 1) two {\em walls} (W) representing the two possible 
orientations for a face. Each wall must reference its parent face
(1) and its opposite wall (2).

Edges reference (through the partial relation 3) two {\em trips} (TR) representing the two possible 
orientations for an edge. 
Each trip must reference its parent edge (3) and its opposite trip
(4). The trip reference (9) one of the cusps in the associated trip 
cycle (10). Each cusp reference the corresponding edge (16).

A wall is bounded by one or more loop cycles (11).
Thus a wall must reference (5) one or more loops. Each
loop reference (5) the wall it belongs to. A loop reference
(6) one of the cusps in the associated loop cycle.

A {\em node} (N) (i.e. a vertex)  reference (12) 
the set of zones its neighborhood is divided into.
A zone reference (12) its parent node and its disks (13).
Each disk reference (13) the zone it bounds. 
Disks reference (14) one of the cusps in the associated disk cycle (15).

{
The evaluation of space requirements for this representation
heavily depends  on the kind of cellular subdivision adopted
{We evaluate here the number of references needed to encode a 
a simplicial subdivision of a tetrahedron.  
To this aim we list all the instances of tri--ciclic elements
we need  to encode the tetrahedron and report 
between parenthesis the number of 
references required by each instance. This is evaluated as the number of 
arrows coming out of each object in Figure \ref{fig:tri}.    
To model a tetrahedral cell we need  
one cell (1), one seal (1), a list of
four triangular walls (4 references for the list and 3
references for each W), four loops (L) (2). 
For each triangular wall we need three cusps (4). Finally each vertex of the 
tetrahedron must be associated with a disk. Thus we need four disks (2)
and four zones (2). One more pointer is needed for each
zone to build a  list of zones. 
This sums to 94 references.
Note that we omit space needed for faces (F),   edges (E) and 
nodes (N) that will add  one reference for each of these entites. 
If one is forced to encode a tetrahedralization with this scheme
the \occup\ will be 94t+f+e+v.  
}
}
\subsubsection{The \ema{Partial Entity}{Data Structure} }
The partial entity data structure (PES) \cite{LeeLee01}  
is a reduced version of the RES obtained
by neglecting loop-uses. 
Thus, for each face, we have  a single loop (L).
In the face-uses, that here are called {\em partial faces} (PF), is stored a
flag that allows to compare the orientation of the loop with the
orientation of the partial face. 
In Figure  \ref{fig:partial} we list all entites in the PES. Large
characters are used to denote  entites that
differ from those in the RES.
{
\begin{figure}[hpt]
\centerline{
\fbox{
   \epsfig{file=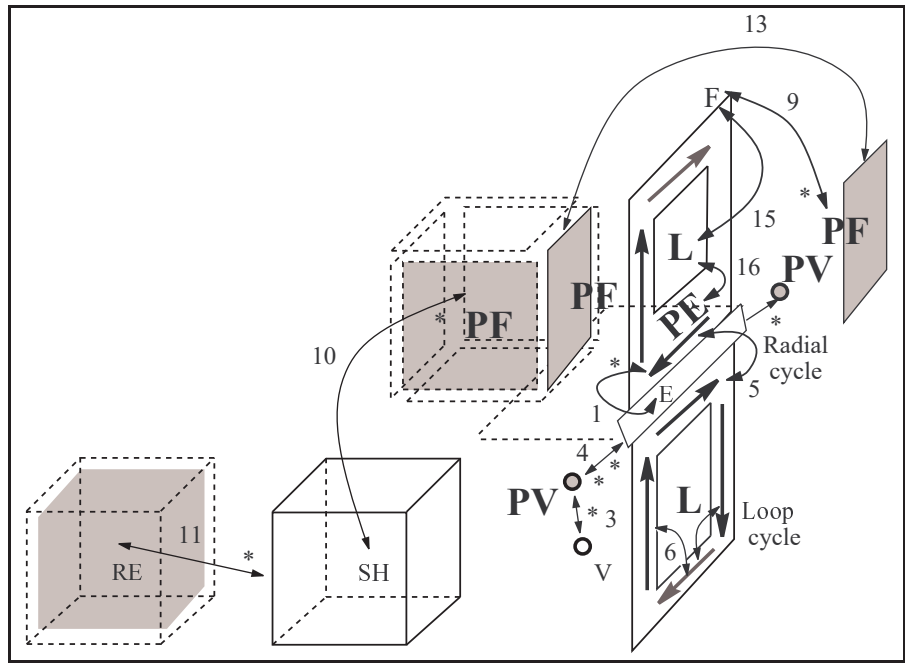, width=0.7\textwidth}
}
}
\caption{Topological entities in the partial entity data structure}
\label{fig:partial}
\end{figure}
}

Loops are obtained   as {\em doubly-linked} cycles of 
entities called {\em partial edges}. Partial edges plays the role of 
edge uses in RES  but, \wrt\  them, we have just a partial edge 
for each face adjacent to a given edge. On the contrary, 
in the RES, we have to introduce an edge use for each 
(boundary of a) cell adjacent to a given edge.

{
We will describe briefly the meaning of all the relations in
Figure \ref{fig:partial}.
We use non-contiguous integers to label relations. 
This is to keep labeling used for RES whenever possible.
We have used labels 1,3,4,5,6,9,10,11,13,15,16 to highlight
the fact that the corresponding relations in the RES have 
a similar meaning.
 
A region (i.e. a cell) is bounded by one or more cells. 
All shells reference (11) the region they bound to and
a region reference one of these shells, others are linked in a list
(not shown in the diagram of Figure  \ref{fig:partial}).
All PFs reference (10) the shell they belongs to and
a shell reference one of these PF, others are linked in a list 
(not shown in the diagram of Figure  \ref{fig:partial}).

PEs for an edge are  organized into two cycles implemented 
with doubly-linked lists.
One cycle is for all PEs
participating to a loop (this is what is called the {\em loop cycle} (6))
and one for all PEs of a given edge.
This is called the {\em radial cycle} (i.e. 5 cycle).

Partial elements  reference their non-partial counterpart
in the PES hierarchy, 
thus PE reference E (1), PF reference F (9), PV reference V (3).
The inverse relations for the above three relations
are all stored as partial relations.

PFs comes in pairs, two for each face, and corresponding pairs of PEs are 
related by relation 13.
A single  bounding L correspond to each F in the bijection 15. 
For non-simply connected faces several L are linked in a list
(not shown in the diagram of Figure  \ref{fig:partial}) and
the F points to the head of this list.
A loop L reference (16) one of its PEs and each PE reference (16) 
the loop it belongs to.
A PV points an incident edge E and E points  the two  PV one
for each endpoint (4).   
}

{
The evaluation of space requirements for this representation
heavily depends  on the kind of cellular subdivision adopted.
Following statistical assumptions the PES
is assigned in \cite{LeeLee01} a \occup\ of
$2.17$ times the space needed by the winged--edge. This is evaluated
assuming that we have to encode
the boundary of a manifold solid enclosed by a single shell.
Under these assumptions the RES is more expensive \wrt\ PES
by a factor 2.03. 
}

{Alternatively to this analysis we evaluate here the number of 
references needed to encode 
a simplicial subdivision of a tetrahedron. 
To this aim we list all the instances of PES classes
needed  to encode a tetrahedron and report 
between parenthesis the number of 
references required by each instance. This is evaluated as the number of 
arrows coming out of each object in Figure \ref{fig:partial}.    
To model a tetrahedral cell we need  
one cell (1), one seal (1), a list of
four triangular PF (4 references for the list and 2
references for each PF), four loops (L) (2). 
For each triangular PF we need three PE (5). Finally each vertex of the 
tetrahedron must be associated to a PV. Thus we need four PV (2).
One more pointer is needed for each
PV to build a  list of PV. 
Note that references for each loop and PEs are
17 per face and must be counted just once for each face. 
In this way this
structure saves a lot of space \wrt\  the RES.
The remaining references
sums to 27 references for each tetrahedron.
Note that we omit space needed for faces (F),   edges (E) and 
vertices (V) that will add $2f+2e+v$   references. If one is forced to encode a tetrahedralization with this scheme
the \occup\ will be 27t+19f+2e+v.  
If we assume a manifold tetrahedralization we can say that 
$4t\le 2f$. If we forget boundaries  and assume  
and we get an occupation of 65 references for each tetrahedron that is less 
than half of the space needed by the RES.
}
\subsection{Conclusions}
We have reviewed several approaches to model realizable non-manifolds 
and for each approach we have computed the \occup\ required to
encode a simplicial subdivision with $t$ tetrahedra, $f$ faces,
$e$ edges and $v$ vertices. The results of this analysis are 
summarized in Table \ref{tab:nonmani}.
{
\begin{table}
{
\begin{center}
\begin{tabular}{|c|c|l|}\hline\hline
\multicolumn{2}{|c|}{Modeling Data Structure} & \occup\ \\ \hline
1& Radial Edge & $155t+2f+e+v$  \\ \hline
2& Tri-Cyclic Cusps & $94t+f+e+v$  \\ \hline
3& Partial Entity & $27t+19f+2e+v$  \\ \hline
\end{tabular}
\end{center}
}\caption{Storage cost for non-manifold data structures}
\label{tab:nonmani}
\end{table}
}
We already reported a  ratio of 2.01  between  \occup\ of
RES and PES following results in \cite{LeeLee01}.
It can be proven that the \occup\ of the  Tri-Cyclic Cusp data
structure is intermediate between \occup\ of RES and PES.
Thus the \occup\ of PES is the best in this class and still
its \occup, according to \cite{LeeLee01}, is at least twice the \occup\
needed by the standard data structures for manifold surfaces (e.g.
the  winged-edge).

\section{Decomposition of non-manifold surfaces and solids}
\label{sec:decompart}
{
The idea of representing polyhedra  through its decomposition  
is already present in literature since 1984.
The original formulation of the problem is contained
in papers about {\em notch cutting} \cite{Cha84}.
However in this area of research the emphasis was on the decomposition
of a generic solid polyhedron ${\cal P}$ into convex polyhedra. 
The non-manifold
edges and vertices in the boundary of the polyhedron ${\cal P}$
were called {\em special notches} and their removal is performed
using a geometric decision procedure \cite{Baj91}.
This can lead to a decomposition based on the geometry that might
not be satisfactory in general. 

Depending on the surface normal we use,
the two boxes sharing an edge in Figure \ref{fig:mcmains} (a) 
might be decomposed into two
boxes (Figure \ref{fig:mcmains} (b)) or into a single volume obtained making the two boxes
communicate through the common edge split in two  (Figure \ref{fig:mcmains} (c)).
This is clearly pointed out in \cite{mcm02b} from which we  quote the examples in Figure \ref{fig:mcmains}
\begin{figure}[h]
\begin{center}
\parbox[c][7.5cm]{0.46\textwidth}{
\vfill
\psfig{file=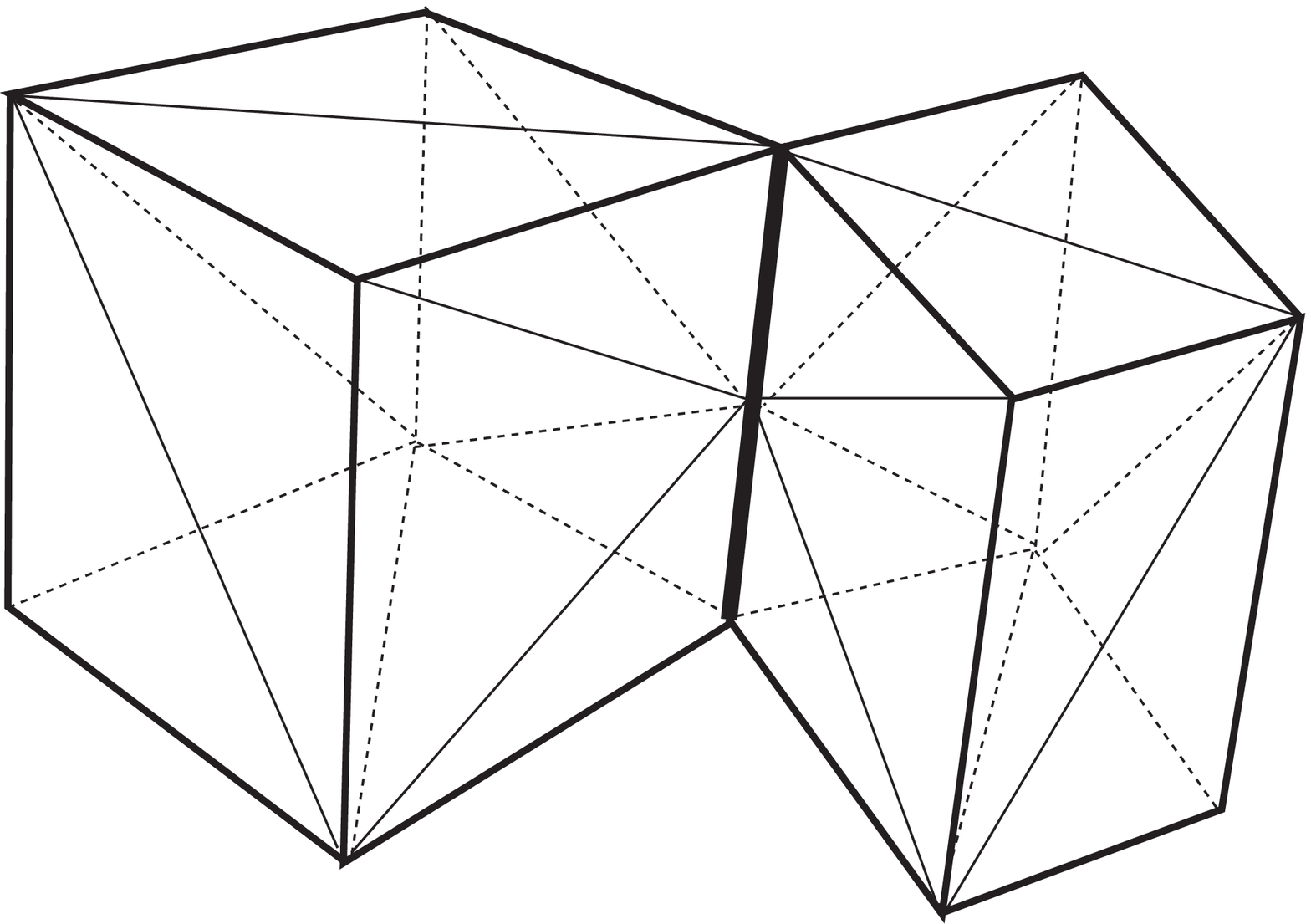,width=.46\textwidth}
\vfill
\begin{center}(a)\end{center}
}
\parbox[c][7.5cm]{0.46\textwidth}{
\vfill
\psfig{file=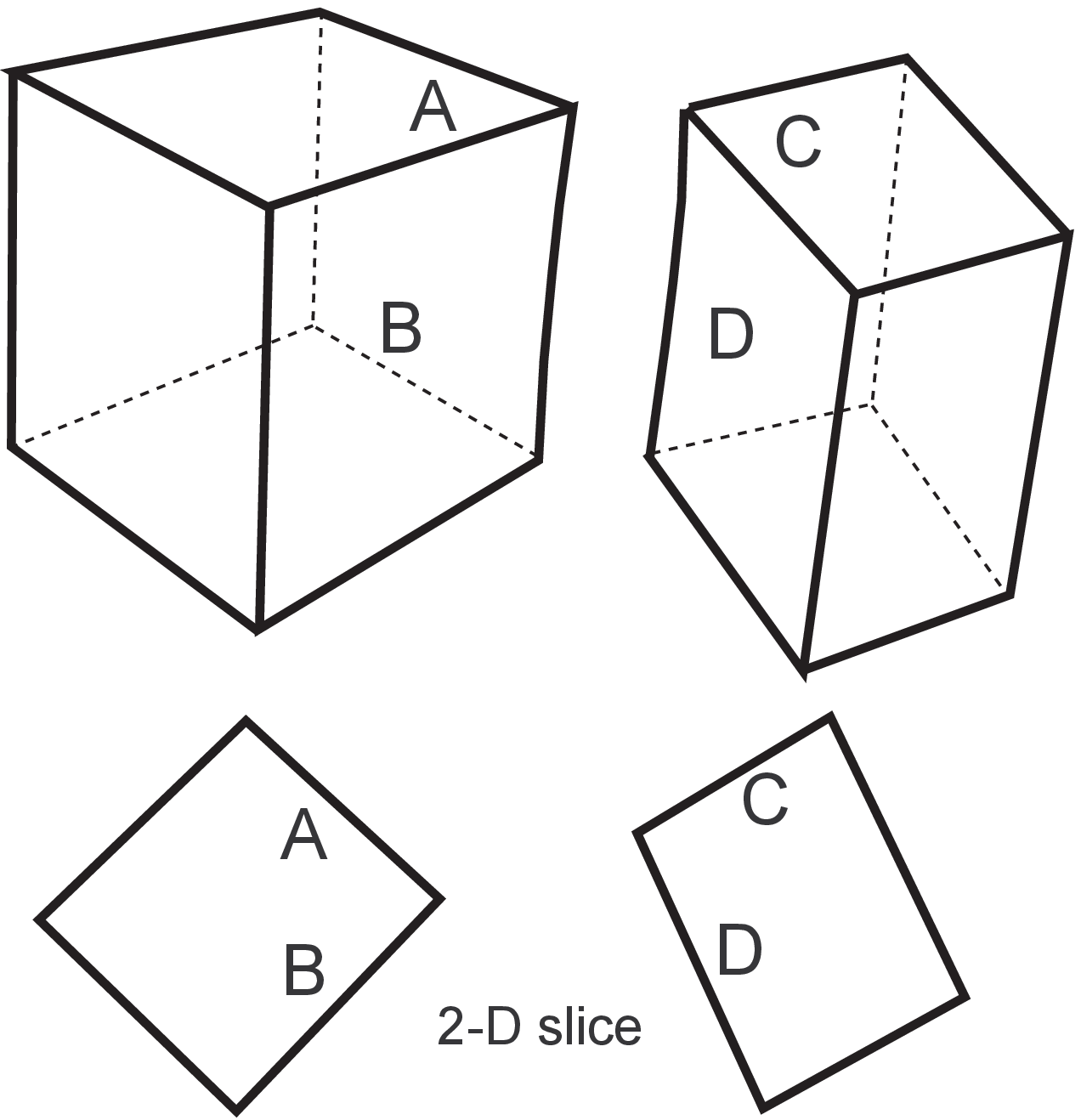,width=.46\textwidth}
\vfill
\begin{center}(b)\end{center}
}
\parbox[c][7.5cm]{0.46\textwidth}{
\vfill
\psfig{file=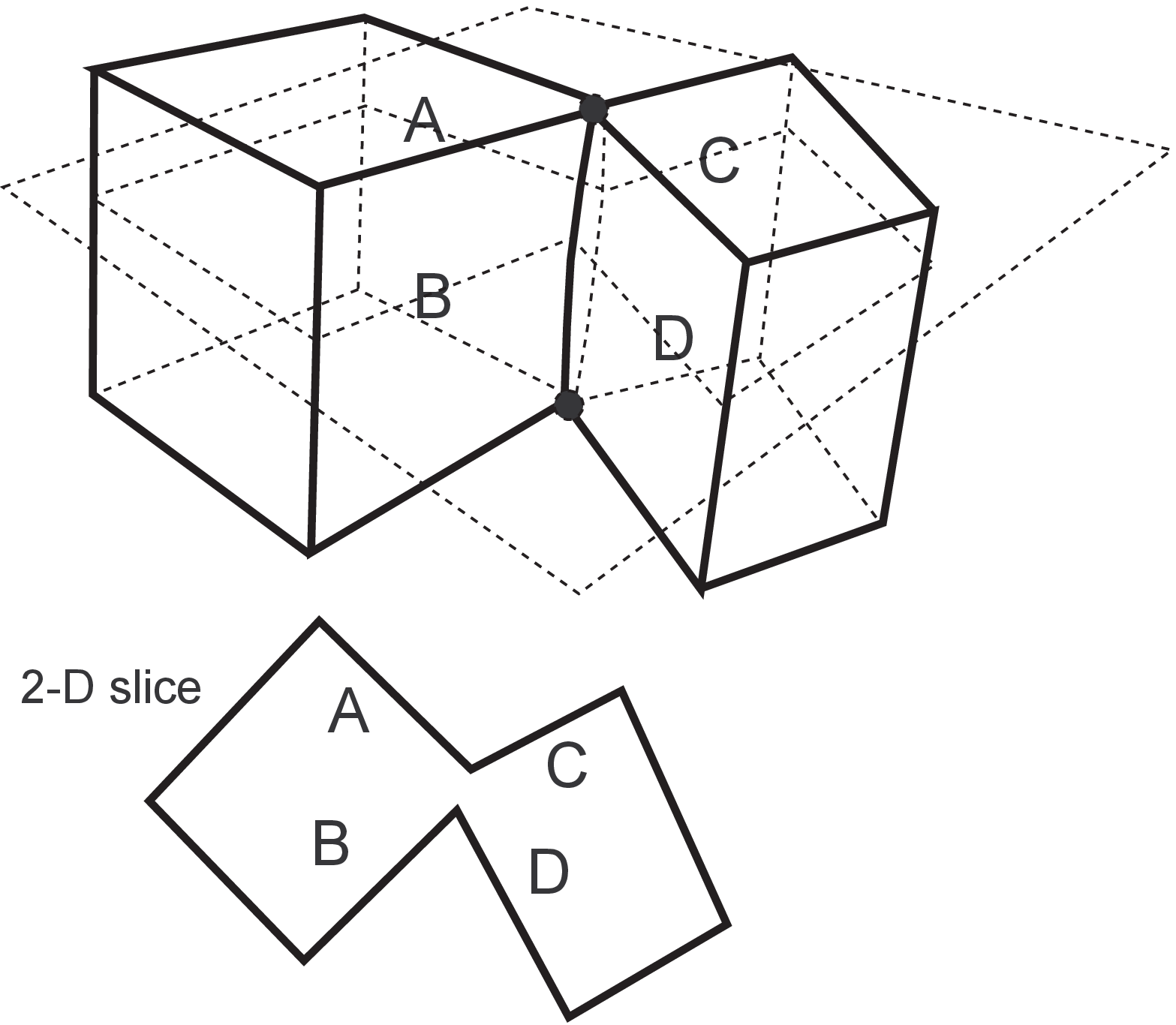,width=.46\textwidth}
\vfill
\begin{center}(c)\end{center}
}
\end{center}
\caption{Ambiguities in decomposition of non-manifold solids. 
A non-manifold solid (a) 
and two possible patterns for the underlying non-manifold data structure (b)
and (c) (adapted from \cite{McM00})}
\label{fig:mcmains}
\end{figure}
}

Nevertheless, several topological modeling approaches assumes that a 
decomposition procedure 
for $2$-complexes is available and propose a modeling approach based  
on  the decomposition of the boundary of the solid to be modeled.

\subsection{\ema{TCD}{Two-manifold Cell Decomposition Graph} }
In \cite{FaRa92} the decomposition of \cite{Baj91} is taken as 
starting point. Thus a decomposition of the original 
non-manifold solid is assumed. This decomposition is assumed to provide
manifold  components by duplicating edges and vertices at special
notches. 

Each component of the decomposition is taken then as 
a separate cell. Thus cells are the interior of
orientable closed $2$-manifolds. These cells are represented through
their boundary that in turn is modeled  with a 
standard cellular decomposition for 2-manifolds.
Edges and vertices that are copies of special notches are then grouped 
together using an hypergraph called the two-manifold cell decomposition
graph (TCD). 
In the original paper access primitives to the representation are not 
presented, nor it is detailed how to build the proposed data structure using 
decomposition results. The paper contained a claim that a set of two Euler 
operators and their inverses is sufficient to implement needed primitives
for a non-manifold solid modeler.

\subsection{The \ema{non-manifold spine}{representation} }
In \cite{Des92} a non-manifold solid  is proven to be equal  
to the limit (\wrt\ a certain metric) of a sequence of manifold solids. 
This theoretical result is used to prove that extending the usual set of Euler
operators one can define non-manifold solids through 
these sequences. Obviously the  extended operators act on a  sequence 
of manifold solids and
returns a sequence of  manifold solids. 
However, it can be proven that both these 
sequences and the operators admit a finite representation.
To prove this result this paper presents a data structure for 
non-manifold solids.
In this data structure the non-manifold solid is represented through a set 
of manifold solids together with  the set of all non-manifold points.
In a non-manifold solid this set is a graph called the {\em non-manifold spine}.
In the manifold solids are inserted finite combinatorial objects
called  {\em  infinitesimal faces} whose semantics is the sequence
of faces needed to approximate the non-manifold r-set.
Infinitesimal faces behaves as ordinary faces within approximating 
manifold solids. Infinitesimal faces 
are involved in an adjacency relations with the  elements of the 
manifold spine.   These adjacency are collectively stored within an 
hash table. We will call the {\em   manifold spine representation} the
combination of these four components:
the manifold spine, the infinitesimal faces, the set of manifold solids
with  infinitesimal faces and the encoding of the adjacency 
relations between the spine and the infinitesimal faces.  
Completeness of the set of extended Euler operators is proven
in two steps. First it is proven that the manifold spine representation
can model all approximating sequences of manifold solids.
Next it is proven that 
introduced operators are designed to  build all valid instances of this
representation.
In this paper 
it is not detailed how to build the proposed decomposition of the non-manifold
solid. However it is proven that it is possible to build  the
proposed  data structure with the extended Euler operators introduced by this
scheme.
Higher level primitives, like a sweep operator, are implemented using this
set of Euler operators.  However, is still responsibility of the user  to give the correct sequence of Euler operations that can
build a certain non-manifold solid.

\subsection{\emi{Cutting and Stitching}}
{
Cutting and stitching is presented in a pair of works that
do not address  directly the non-manifold modeling problem.
On the contrary at least one of them \cite{Gui98} tries to avoid
non-manifoldness 
by converting a  non-manifold surface into a manifold surface.
The work consider a  more general class of non-manifolds since it do not
assume to work with solids and do not use the notion of interior. 
No user supplied modeling is assumed and the topological input data are 
assumed to be available in a raw list of faces. 
The conversion is performed in a two step process. In a first step
the original complex is decomposed. Two decomposition algorithms are
presented.
The algorithms 
presented in this thesis can be considered as a dimension independent 
extension of these algorithms to non regular complexes of arbitrary dimension.
Further discussion on this relation is presented in the last section of 
this chapter.
The cutting algorithm proposed in \cite{Gui98} is shown to have complexity
{\em proportional to the number of {\rm (non-manifold)} marked edges times
times the largest number of corners in a vertex star}.

In a second step the result of cutting is reprocessed in order to
{\em stitch} together edges that were cut in the previous step. 
This action is called
a {\em stitching}. This second step must produce a manifold 
surface. A particular manifold stitch is produced using greedy strategies.
Two criteria are proposed. One is to attempts to maximize the number
of edges that  stitches together 
(called a {\em maximal length} stitch). Another algorithm 
attempts to maximize the number of  vertices that stitches together.
(called a {\em maximal size} stitch).
Two strategies are presented. The first strategy
stitches only edges or vertices that were
previously cut by the cutting algorithm. In a second strategy this 
constraint is removed and stitching is promoted using geometric
proximity.

The two algorithms are used to convert a non-manifold
surface  into a manifold surface that can be  used as input to  
simplification algorithms for manifolds. The result of the conversion is also
fed to a compression algorithm for manifolds \cite{TauRos98}
based on {\em  topological surgery}.
The compressed  surface will have the same geometric realization 
of the original non-manifold complex but will retain the combinatorial
structure of the decomposed complex.
A later work \cite{Gui99} extends
the topological surgery compression scheme to handle 
directly the non-manifold surface and deliver a compressed version that
preserves the combinatorial structure of the non-manifold surface.
Non-manifold compression is performed
by cutting first along non-manifold points following the cutting process
in \cite{Gui99}.
Then parts are compressed using topological surgery. 
The non-manifold structure is  compressed 
adding the encoding for the inversion of the  cutting process through 
a set of stitching instructions.
Note  that cutting in \cite{Gui98} is limited to regular $2$-complexes.
The proposed cutting algorithm is shown to have complexity
{\em proportional to the number of {\rm (non-manifold)} marked edges times
the largest number of marked edges times the largest number 
of corners in a vertex star}.
It is easy to exhibit simplicial subdivisions  of non-manifold surfaces 
where using  this statement we can  predict 
a processing time that is  quadratic \wrt\ the number of faces.
To obtain our dimension independent extension algorithm we simply restated 
the algorithms presented in \cite{Gui98} in a recursive style.
The result of cutting is referred to be {\em the manifold with  the maximum 
number of vertices and the maximum number of  components that cuts through the
singular edges and vertices and nowhere else}.
}
\subsection{\emi{Matchmaker}}
Finally, Matchmaker is an algorithm, presented in \cite{RosCad99},
that is especially conceived to decompose
the boundary of r-sets (or even a 2-complex) into a set of manifold surfaces.
This approach, although limited to non-manifold 2-complexes, attempts to
minimize the number of vertex replications introduced by the decomposition
process. In a situation like that of Figure
\ref{fig:match} (a) 
\begin{figure}[h]
	\begin{center}
\parbox[c][7.5cm]{0.46\textwidth}{
			\vfill
			\psfig{file=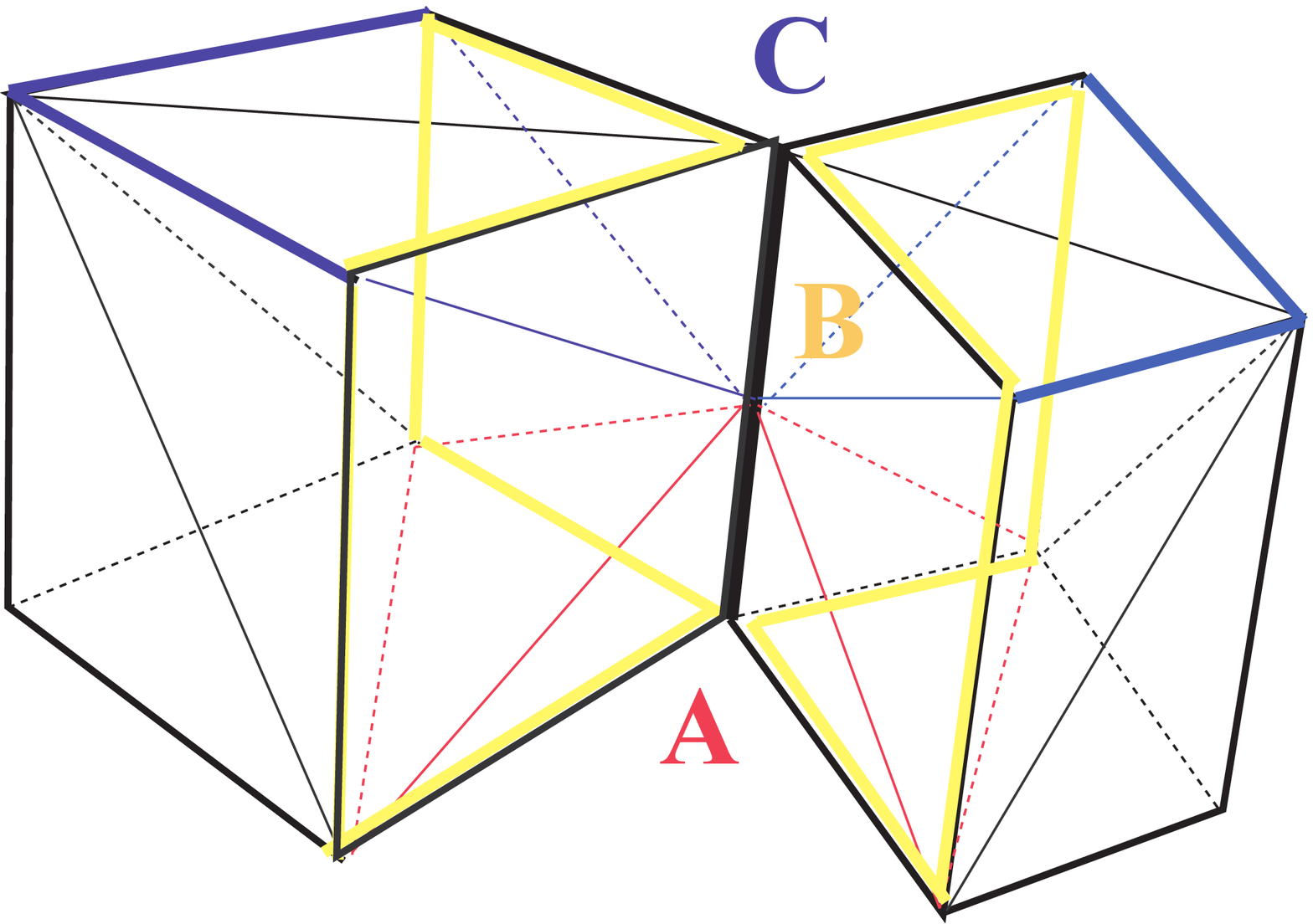,width=.46\textwidth}
			\vfill
			\begin{center}(a)\end{center}
}\mbox{ }\mbox{ }\mbox{ }
\parbox[c][7.5cm]{0.46\textwidth}{
			\vfill
			\psfig{file=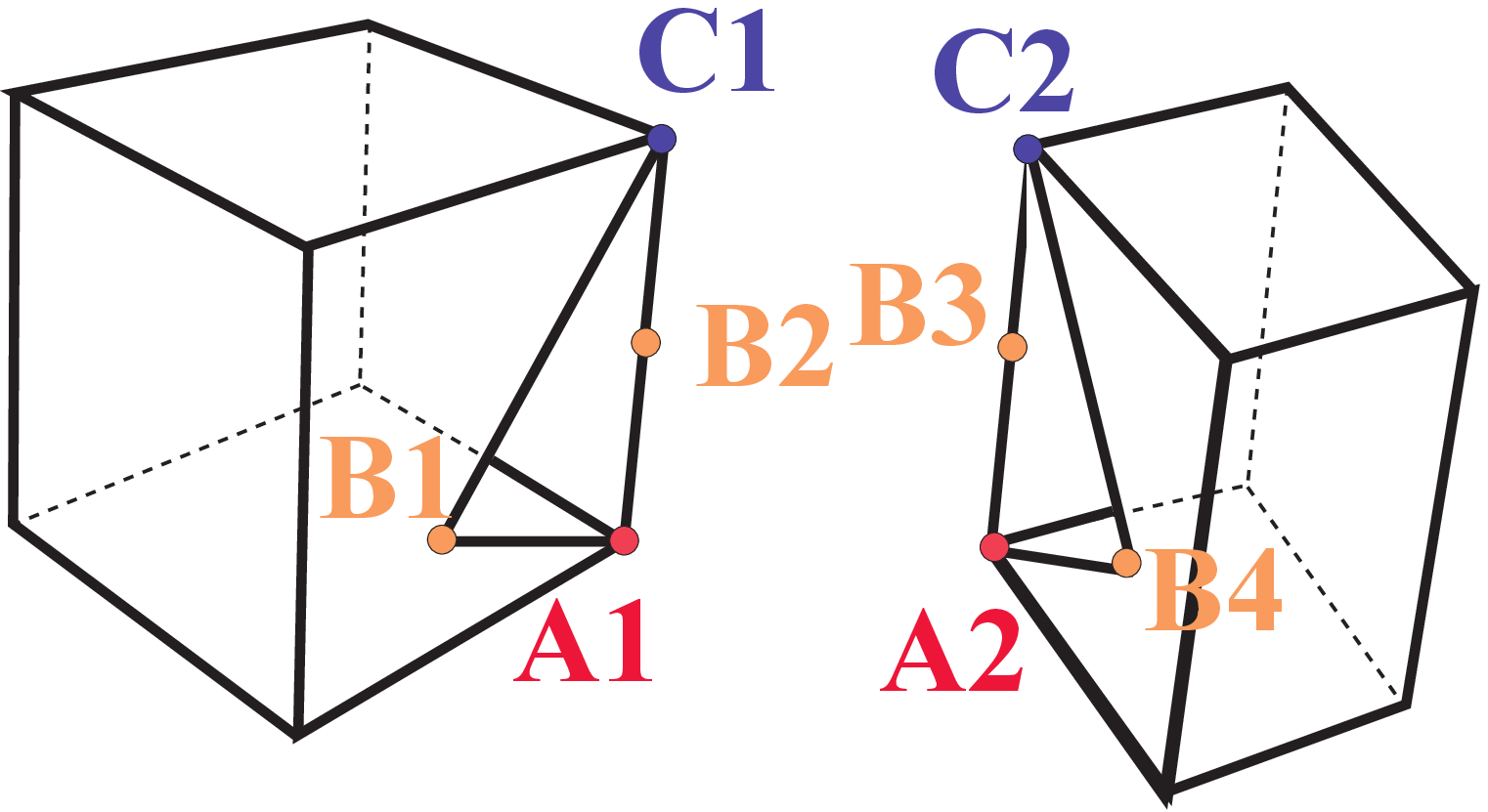,width=.46\textwidth}
			\vfill
			\begin{center}(b)\end{center}
}
\parbox[c][7.5cm]{0.46\textwidth}{
			\vfill
			\psfig{file=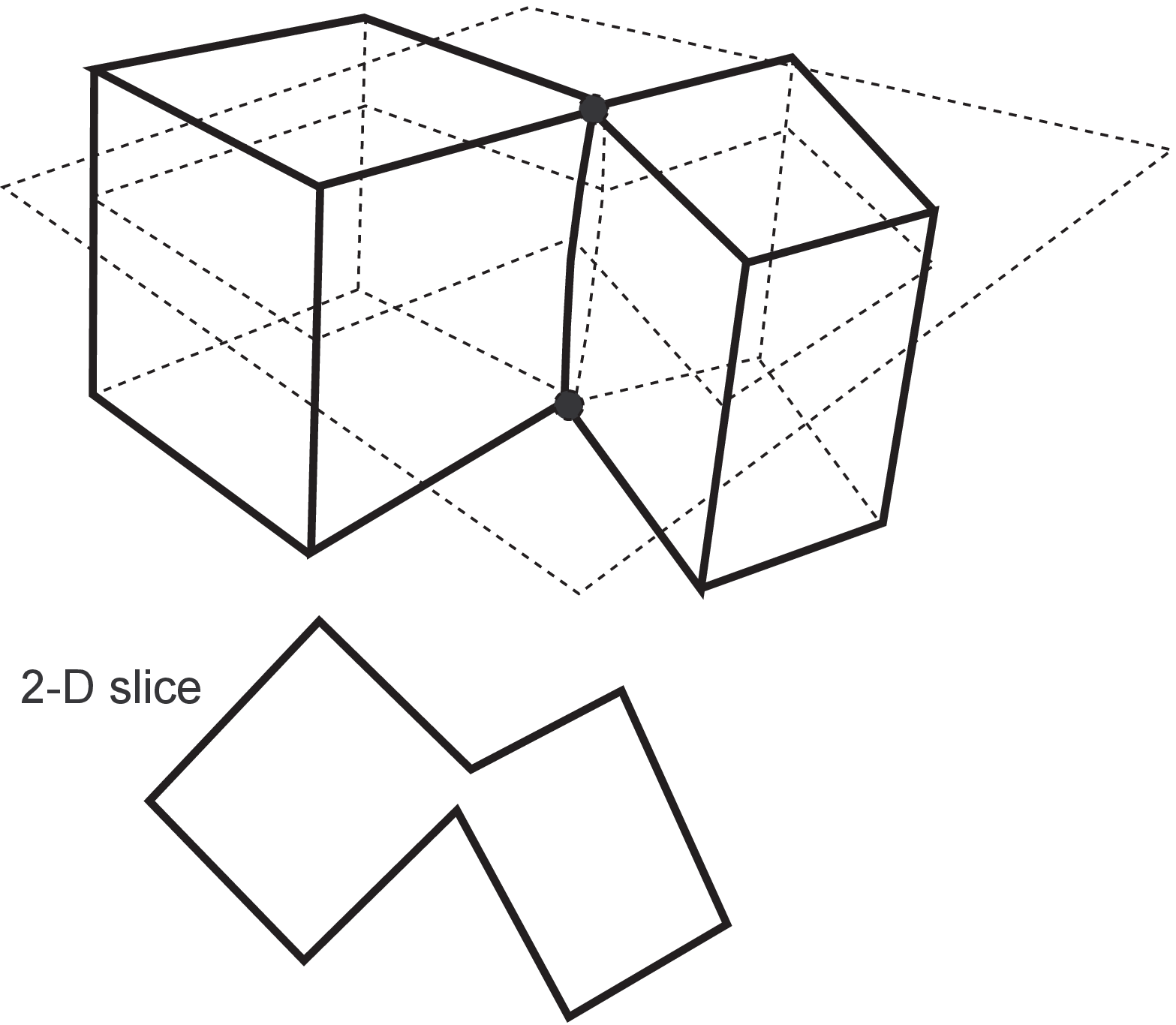,width=.46\textwidth}
			\vfill
			\begin{center}(c)\end{center}
		}
\mbox{ }\mbox{ }\mbox{ }
\parbox[c][7.5cm]{0.46\textwidth}{
			\vfill
			\psfig{file=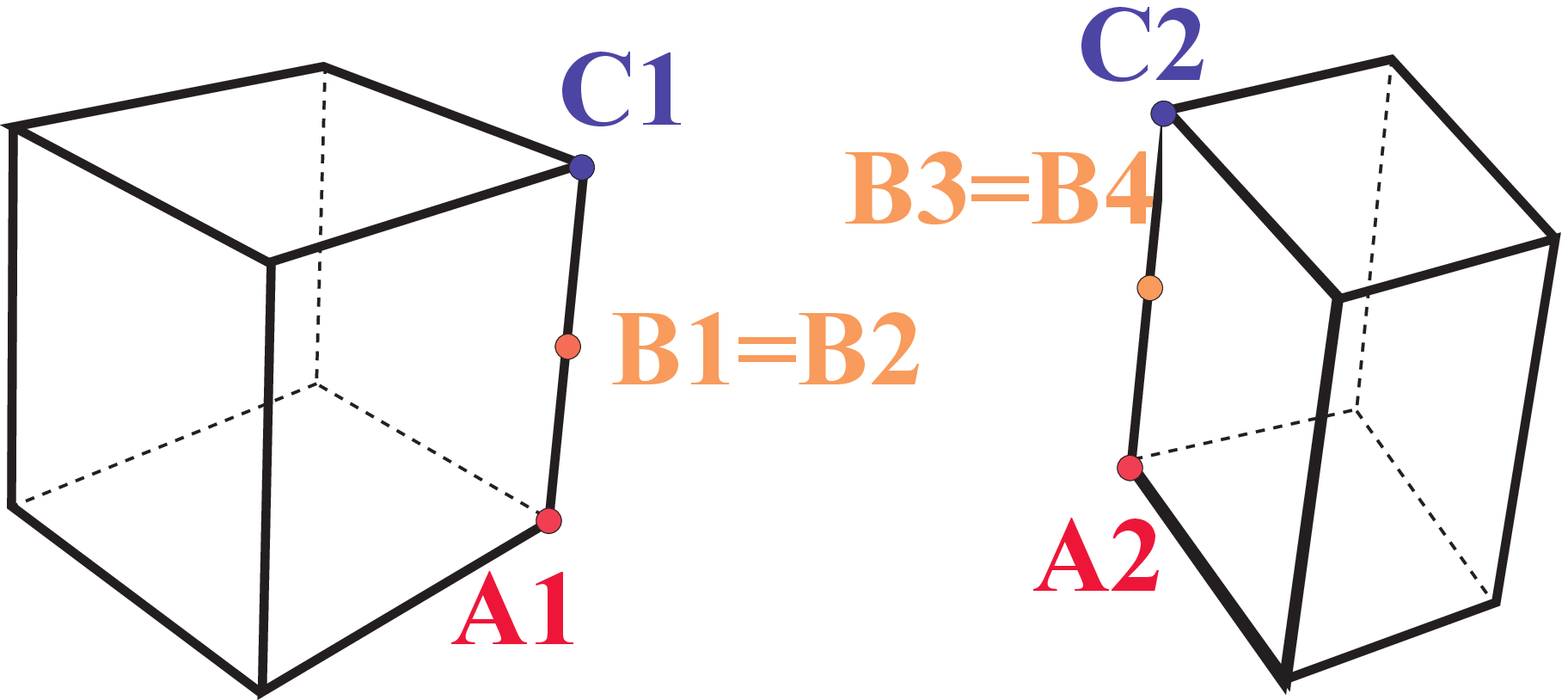,width=.46\textwidth}
			\vfill
			\begin{center}(d)\end{center}
		}
\end{center}
	\caption{Matchmaker and the decomposition of non-manifold solids. 
		A non-manifold solid (a) 
		and three possible decompositions: Matchmaker (c) the Canonical Decomposition (in this thesis) (b)
		and another option as in \cite{mcm02b}(d) }
	\label{fig:match}
\end{figure}
Matchmaker turns the original solid (a) into a manifold model as in (c) without introducing vertices duplication as in (d). The resulting model is manifold because of the geometric embedding. In the embedding Matchmaker assumes that edges can be duplicated introducing a couple of  infinitesimally curved edges. 
The two cubes becomes a single volume obtained making the two boxes
communicate through the common edge split in two.
The authors calls this kind of surface {\em edge manifold}. As they say: {\em if these edges were bent by an infinitely small amount in the	appropriate direction, the resulting shape would either be manifold or would have only isolated non-manifold vertices. We say that the resulting model is edge-manifold.} The algorithm introduces duplication for  isolated non-manifold vertices and obtain a manifold embedding for the original model. Going on with this idea the algorithm also corrects self intersections in the geometric model. The whole approach is different from the one presented in this thesis. Indeed, with Matchmaker, in the combinatorial structure of the decomposed model, the common edge in Figure  \ref{fig:match} (a) remains adjacent to four faces and manifoldness is obtained in the embedding by curving edges. A second distinction is about the type of algorithm. With Matchmaker the output depends on some greedy choices that {\em  may result in vertex
replications that could be avoided}.  Indeed we try to devise a decomposition that is the {\em  most general} among possible decompositions.  As a consequence we introduce more vertex replication. In the case of Figure \ref{fig:match} (a) our decomposition will go recursively to decompose the colored 1-complexes of Figure  \ref{fig:match} (a) and count their connected components. Thus for A we decompose the graph in red and count two connected components. The same happens for C in blue. The decomposition for B accounts for four components (the four Cs in yellow).
Thus we introduce two copies for A and C, and four copies for B.
The resulting decomposition is in (b) and is the {\em most  general} since both cases in (c)  and (d) can be obtained by further {\em stitching}. Yet is not too general {\em cutting} only at non-manifold edges. We felt that all the details about how to define italicized words require the framework of combinatorial topology.

\section{A rationale for a combinatorial approach}
\label{sec:teorat}
In this thesis we basically present a dimension independent
decomposition scheme for non-manifolds and a related data structure.
In this section we give motivation for the role we reserved,
in this thesis, to 
combinatorial topology (a.k.a. piecewise linear topology or PL-topology).
To this aim we report basic negative results that
characterize the incomplete relation  between combinatorial and point
set topology in higher dimension. This incomplete relation 
must be taken into account when choosing a theoretical framework
for a  dimension independent formulation. 

\subsection{Motivations}

For the particular study we carried out in this thesis we felt uneasy
to develop this study in a  geometric framework and,
as we have reported, 
it is not essentially  mandatory to embed the theoretical framework 
in $\real^k$.
Furthermore all results in this  thesis do not mention the particular 
embedding. This is the reason why we found nice and profitable
to have a completely combinatorial presentation.
Finally, in the context of this thesis, the value of
dimension independent results obtained in the geometric
settlement (i.e. $\real^k$) is questionable. The value of these results
is based on the shaking ground of the
incomplete relationship between point set and combinatorial topology.
To show that a combinatorial approach is essentially due
we will outline,  in this section, classic negative results about the
relation  between point set and combinatorial topology.

However, before starting this review of related results in 
combinatorial topology, we  try to explain  why they are related
with this work. We try to do this   with an example.

With the following example we want to show that there are reasonable
clues that it is not feasible to 
investigate  the subject of this thesis in $\real^k$ and
then  translate back theoretical results from $\real^k$ into the
language of cellular decompositions.

To build such an example we take one of 
the main results of this thesis and see
a possible similar result formulated in the language of point set
topology. In particular, in this example, we assume that we have derived 
(somehow) this 
''continuous'' formulation and enlight the difficulties that are behind a
translation  of the continuous result into a result for cellular 
(or simplicial) subdivisions.

One of the central results in this thesis is that there  exist  a unique
maximal decomposition (denoted by $\canon{\AComp}$)
of a given simplicial $d$-complex
$\AComp$ obtained cutting $\AComp$ only at non-manifold simplices. 
Thus the continuous analogue of this result might state that, 
{\em for a given 
polyhedron $P$ in $\real^k$, there is a  unique, up to homeomorphism,
maximally decomposed polyhedron $\canon{P}$ that can be mapped into $P$ by a 
continuous function  
such that the common  image of two distinct points in $\canon{P}$ 
is always a point that has not a neighborhood homeomorphic
to $\real^k$}. By {\em maximally decomposed} here we mean that there 
is not another
polyhedron $P^\prime$ with the above characteristics and such that
$\canon{P}$ is the continuous image of $P^\prime$.

Let us forget the problem of producing a correct ''discrete'' theorem that 
exploits {\em completely} this (hypothetical) result and let us just mention
the fact that this continuous version seems a quite hard result to prove.
Surely a simple consequence  of such a result is that, 
given an abstract simplicial complex $\AComp$, ve must define the
discrete decomposition   $\canon{\AComp}$ as any triangulation
of the decomposition {$\canon{P(\AComp)}$} of 
a geometric realization {$P(\AComp)$} of $\AComp$. 
This definition is actually
faulty in higher dimension  since an arbitrary triangulation of a 
manifold need not to be a combinatorial manifold. 
This is one of the classic  negative results \cite{Edw75} we will
mention in the review of the following  section.
Thus, not all points that are manifold points in
{$\canon{P(\AComp)}$} must be manifold vertices in $\canon{\AComp}$ 
and consequently
the class of non-manifold singularities that are removed in 
$\canon{\AComp}$ is not clearly identified.
Thus, we found interesting to review, in this section, 
all mathematical results
that characterize the incomplete relation between ''continuous'' topological
concepts  and their  ''discrete'' analogue.
 
\subsection{Topological and Combinatorial Manifolds}
Manifolds are usually introduced by a rather geometric  definition that
actually gives the notion of  {\em topological} $d$-manifolds.
A topological $d$-manifolds is a 
topological  subspace of some
Euclidean space {\em such that every point has an open  neighborhood that is homeomorphic to  $\real^d$}. 
However, often a theoretical model closer to the  computerized
representation is needed. Therefore, one it is forced  to
modify this definition and 
introduce an analogous combinatorial  definition that views
manifolds as a collection of related discrete entities
(e.g.  $d$-simplices).

A first step in this direction   is to define a slightly different
class of objects called {\em triangulated manifolds}.
Triangulated manifolds are defined as topological manifolds that
are also  {\em geometric simplicial complexes}  i.e.
collection of discrete entities (triangles or tetrahedra) glued together.
A topological manifold is {\em triangulable} iff it is equivalent 
(i.e. {\em  homeomorphic}) to a triangulated manifold. 
Triangulable $d$-manifolds are a {\em proper} subclass of topological
$d$-manifolds. In fact there are manifolds that do  not have
a  triangulated equivalent. This is already true for $d=4$
(see \cite{AkbCar90} that combines results by  Freedman  \cite{FreeLuo89,FreeQui90} 
and classical Casson results \cite{Lut00}  Pg. 5).

If we restrict our attention to the subclass of geometric simplicial complex 
it can be proved (see \cite{Dey99} p. 2 for instance) 
that a simplicial complex is a triangulated  $d$-manifold 
iff the {\em star} of simplices around each vertex (see Definition \ref{par:linkdef}) is homeomorphic to $\real^d$.
Equivalently, we can say 
that a simplicial complex is a $d$-manifold 
iff the {\em link} of each vertex (i.e. , in manifolds, the boundary of the vertex  star)
is homeomorphic to the $(d-1)$-sphere.
Therefore, to recognize $d$-manifolds we  must be able to recognize the  equivalence
of a simplicial $(d-1)$-complex with a  $(d-1)$-sphere.

To have a {\em combinatorial} analogue  of manifoldneess definition
one must turn from vertex link equivalence based on
homeomorphism to PL-equivalence based on 
{\em piecewise linear} homeomorphism (PL-homeomorphism).
PL-equivalence proves to be truly combinatorial (see \cite{Veg97} p. 520)
and at least semi-decidable.
In fact, two $d$-dimensional simplicial complexes are PL-homeomorphic
iff they are {\em stellar equivalent}.
and simplicial complexes are stellar equivalent if they can be
transformed into a common refinement by a finite sequence of
discrete operations, called {\em starring} operations.
Each starring operation is a finite operation 
that modifies the combinatorial structure of the complex.
Therefore, stellar equivalence is at least a semi-decidable relation
(see \cite{Lic99} for a survey on stellar and bistellar equivalence).

The   adoption of PL-equivalence, also called {\em combinatorial equivalence}
leads to a first definition of the closed
{\em combinatorial $d$-manifold}, that is: 
{\em A closed  combinatorial $d$-manifold is a triangulated manifold
where every point has a link that is piece-wise linearly homeomorphic 
(PL-homeomorphic) to the $(d-1)$-sphere}.

This definition of combinatorial equivalence is recognized as the best
discrete analogue of the topological definition of homeomorphism and 
still the analogy is known to be unsatisfactory.
In fact, in general, it is false that
{\em two $d$-complexes that are homeomorphic must be PL-homeomorphic 
(PL-equivalent)}. 
The italicized assumption is called Steinitz's {\em  Hauptvermutung} and 
it was the basic assumption behind combinatorial topology.
This assumption clearly legitimate the claim that we can study
some properties of topological spaces w.l.o.g. by studying properties 
of complexes. 

{This assumption is known to be true for $d$-complexes for $d\le 2$ 
\cite{Pap43}
and for $d=3$  \cite{Moi52} and is known to be false in general.
In fact Milnor \cite{Mil61} provided the first example of a pair of
non-manifold $7$-complexes that have homeomorphic geometric realizations and 
yet that are not PL-equivalent.
Besides Milnor counterexample for $d=7$ the Hauptvermutung
is an open problem for $d\ge 4$.
}

Obviously PL-homeomorphism is semi-decidable via stellar equivalence,
however, this do not implies that the definition we have 
given for  $d$-manifolds is still satisfactory.
In fact, we can use semi-decidable stellar equivalence to see if the link 
of each vertex is PL-homeomorphic to a {\em particular} triangulation 
of the  $(d-1)$-sphere.  
Thus to validate a $d$-manifold we must select a particular reference 
simplicial complex for the $(d-1)$-sphere and then check every vertex link. 
Thus, in the end the best discrete analogue of the
definition of the closed topological manifold is the following:
{\em A closed combinatorial $d$-manifold is a triangulated manifold
where every point has a link that is piece-wise linearly homeomorphic 
(PL-homeomorphic) to the $(d-1)$-simplex}   

However the results of this link checking should not depend on the particular
triangulation selected for the $(d-1)$-sphere. 
This is possible iff all triangulations of the $d$-sphere are PL-equivalent. 
The correctness of this  assumption is actually  
an open problem for $4$-spheres.   
On the other hand it has been proved that, for  $d\ne 4$,  
the standard $d$-simplex is a unique reference structure for
all triangulations of $d$-sphere that are {\em also} a combinatorial manifold 
(see \cite{Moi52} for $d\le 3$ and \cite{KirSie77} for $d\ge 5$). 
However, already for $d=5$, it has been discovered an
example of  a triangulable $5$-sphere that is triangulable
into a complex that is not a combinatorial manifold in the sense of the 
definition above \cite{Edw75}.

Thus, as far as we are concerned in this thesis, for $d\ge 4$ it is 
not that obvious tha we can translate a  results on topological manifoldness
to  a similar result in term of combinatorial manifoldness. 
We feel that this suggest to develop a dimension
independent study of this kind with an approach that makes little or
no reference to concepts from point set topology.
Even if we stay in the combinatorial domain, still we have to comply
with  some other limitations when we are developing a dimension 
independent approach. We briefly report them here since an easy 
consequence of these results is that a decomposition
into manifold parts is not always possible in higher dimension
(see  Property \ref{pro:nondecdec}).

\subsection{Manifold limitations in higher dimension \label{sec:higherdim}}
We have seen that the Hauptvermutung is false and thus
combinatorial equivalence is less powerful
that topological equivalence.
However, even if we accept to
work with combinatorial equivalence still we have that for $d\ge 4$ 
combinatorial equivalence is only semi-decidable. This is a consequence
of a result by Markov \cite{Mar58}. This result states that there 
exist $4$-complex $\AComp_0$ such that it is impossible to decide, 
for any other complex $\AComp$, 
if $\AComp_0$ and $\AComp$ are combinatorially equivalent.

Following Markov approach Novikov proved \cite{Vol74} that  there exist a
combinatorial $d$-manifold $M_d$, for any $d\ge 5$, s.t. we cannot
decide whether $M_d$ is a $d$-sphere or not.
Novikov, following a technique that was already in Markov paper
\cite{Mar58}, proves this
result  (as reported in \cite{Vol74}) showing that
the problem of recognizing a $d$-sphere for $d\ge 5$ implies the problem
of recognizing a trivial group in a finite sequence of finitely
generated groups. It is known \cite{Adj57} that this problem, in turn,
implies  the Halting Problem.
The relation between the recognizability of the
$d$-sphere and  the halting problem is unfolded, for 
instance, in \cite{Nab96}.
These results, in turn, impair the possibility of recognizing combinatorial 
$d$-manifolds for $d\ge 6$ (see Theorem \ref{teo:halting}).
An easy consequence of this is that a decomposition
into manifold parts is not always possible in higher dimension
(see  Property \ref{pro:nondecdec}).

\subsection{Summary}
In the end, the best founded  definition for
combinatorial $d$-manifolds requires the link of every vertex to be
{\em PL-homeomorphic to the boundary standard $d$-simplex}.
This definition is truly combinatorial since PL-homeomorphism is at least
semi-decidable via stellar equivalence.
Combinatorial $d$-manifolds then are a proper subclass of triangulated
$d$-manifolds already for $d=5$.
In fact there exist  a triangulable $5$-sphere 
that is not triangulable as a combinatorial manifold \cite{Edw75}.

This short review of the fundamental problems in the definition
of $d$-manifolds proves that the above definition of
combinatorial manifold is, up to now, the best combinatorial
analogue of the local euclidean condition required by topological
manifolds.

Nevertheless, some problems remains and
we can say that things works completely fine only in dimension lower than three.
For $d\le 3$ all topological $d$-manifolds are triangulable
\cite{Rad26,Moi52} homeomorphism and PL-homeomorphism are equivalent
(i.e. the Hauptvermutung is true),  the $d$-sphere has a unique structure
(i.e. the  boundary of the standard $(d+1)$-simplex) and,
for $d\le 3$, the $d$-sphere is algorithmically recognizable \cite{Tho94}.
However the classification of three manifolds is still an open problem.

For $d=4$ there exist a
$4$-manifold not triangulable and it is unknown if the
$4$-sphere has a unique structure or whether it is recognizable.
For $d\ge 4$ 
the Hauptvermutung is an open problem and 
both homeomorphism and PL-homeomorphism between $d$-manifolds are not decidable.
For $d\ge 5$ the $d$-spheres that are triangulable as
combinatorial $d$-manifolds has a  unique structure, however already for $d=5$, 
we find a triangulable $5$-sphere that 
is not triangulable as a combinatorial manifold \cite{Edw75}.
For  $d\ge 4$ $d$-manifolds can not be classified \cite{Mar58} and for $d\ge 5$ the  the $d$-sphere  is not recognizable \cite{Vol74}.

All these limitations, especially those in the  classification of  $d$-manifolds, 
impair the possibility of using them as building blocks in the decomposition of 
non-manifolds.
Furthermore, for these negative results it is not possible to build a
satisfactory relation between results derived in $\real^k$, using
point set topology definitions and their combinatorial counterpart.
For this reason a purely combinatorial approach has been adopted in this thesis.

\section{Conclusions}
\label{sec:conclsta}
In this section we collect some notes about the relation between the results 
in this thesis and reviewed    related works.

\subsection{Representation domain for cell complexes}
A first relation that might be considered is the relation of the results in this thesis  with several modeling approaches presented in Section
\ref{sec:nonmanisol}. If we omit approaches for $2$-manifolds all
other approaches pretend to model cell complexes where only two $3$-cells 
are incident to any given $2$-cells.
{In general cells are defined by  shells of face uses or similar entities.
Face uses comes in pairs for each face, Therefore
there is no way to
model a non-pseudomanifold $3$-complex made up, for instance, by three
tetrahedra sharing the same triangle.
A similar limitation  is present also in dimension independent modeling
schemes.
}{
Modeling schemes artificially constrain their domain of representation
to some well known topological class 
but 
the problem of characterizing the domain of the representation is still
open. 
More precisely we 
cannot say what is exactly the subclass of the
topological subspace of $\real^n$ 
that can be represented by a uniformly dimensional cellular
subdivision such that a $(d-1)$-cell is always adjacent at most to 
two $d$-cells. 

In this thesis we will study this domain for simplicial subdivisions.  
We will
show that the representation domain of the winged representation is,
the set of \Qm s defined by Lienhardt \cite{Lie94} (see Property  \ref{pro:quasi} and Definition \ref{def:quasi}). Next we will give a characterization of 
this class of complexes in term of local topological properties. 
It is well konwn that \qm\ surfaces are ordinary $2$-manifold surfaces and
in dimension three of higher \qm\ are a proper superset of manifolds.

In this thesis, in order to study the decomposition problem,
a different point of view 
is investigated assuming an {\em operational}
view of the winged representation. In fact we can give the winged 
representation an operational interpretation considering the 
adjacency function $\Adj$ as a set of \gl\ \inst s that tells how to 
stitch together $d$-simplices \gl\ them at $(d-1)$-faces. 
Obviously we constrain \gl\ \inst s not to glue together more than
two $d$-simplices at time.
However note that a \gl\ between two $d$-simplices might
induce \gl s between other $d$-simplices.
Note that this view is exactly what we have in mind when we want to
find out what is represented by a given instance of a certain 
modeling data structure stored by some program in some computer
memory.

Surprisingly enough the representation domain for this operational 
view is not the same of the original representation domain
for the winged representation.
We show that by this \gl\ process we generate a class of objects,
we called \Cdec, that are a proper superset of \Qm.
We will give a characterization of this class of complexes in term of 
local topological properties
(see Property  \ref{pro:quasi} and Definition \ref{def:quasi}). 
The most surprising fact coming from this analysis is that 
non-pseudomanifold complexes can be built with this operational view.
Thus a non-pseudomanifold $3$-complex can be stored in most
data structures for non-manifold solids
even if this kind of  non-pseudomanifoldness is not directly represented
by these data structures.  
In other words we found out that  a set of 
\gl\ \inst s that do not to glue together more than
two $3$-simplices at time might generate a tetrahedralization where
three tetrahedra {\em must} be incident to the same,  non-pseudomanifold
triangle  (see Example \ref{app:example}).
}  

\subsection{Extension of the winged representation to the non-manifold domain}
In section \ref{sec:nonmanisol} we presented three modeling approaches that can model
cellular subdivisions of realizable non-manifolds. These approaches are reviewed
and presented stressing the fact that they all can be understood as
small variations around the original scheme present in the radial-edge data
structure.   On the other hand, in Chapter \ref{ch:nmmdl} 
we define a data structure that represents a strong departure 
from this model. 
{
In this we see a strong relation with the winged representation 
(see Section \ref{sec:wingedpao}) because
our data structure is essentially
an extended version  of the winged representation that is capable of 
handling complexes that are \cdec.
With this data structure at hand, 
we devised algorithms to extract all topological relations 
(See algorithms \ref{algo:s0h}, \ref{algo:snm},
\ref{algo:Snmnm} and \ref{algo:Snm}).
Extraction algorithms use auxiliary data structures that we 
introduced to model
non-manifoldness.
However, these data structure vanishes for $d$-manifolds and, 
thus, in the manifold domain, 
our algorithms reduce to algorithms presented for the winged data structure.
}

\subsection{Decomposition and non-manifold cell complex construction}
{
It is also interesting to note that the partial  entity data structure (PES)
pretend to treat vertices much like we do in our decomposition. 
In fact a vertex is modeled
as a list of {\em partial vertices} and a partial vertex in introduced
for each manifold surface incident at that vertex. Quoting from 
\cite{LeeLee01} we can say that:
{\em ''In a non-manifold model, a vertex can be adjacent to an arbitrary
number of two-manifold surfaces... 
The readers may imagine that the p-vertices for a vertex
are formed by splitting the vertex into as many pieces as the
adjacent surfaces.''}
However this paper do not  give any enlightenment on how to do this {\em splitting}
and actually, already for non-manifold surfaces, 
several, non  homeomorphic, options exists 
(recall the example of Figure \ref{fig:moebius1}).  
On the other hand the paper \cite{LeeLee01} presents extended Euler 
operations  that 
can be implemented using the PES and reports that a solid modeler
has been built on top of these operations. 
In other words the (PES) modeling approach actually {\em assumes} 
that the user can imagine the decomposition of the non-manifold he wants
to model 
and offer him a solid modeler.

The definition of the \cano\ decomposition, we will give in Chapter
\ref{ch:stdec}, can support a different approach to data strucutres such as PES.
Consider that decomposition process works on an unstructured presentation
of the complex to be decomposed. A 
byproduct of   the decomposition algorithm \ref{algo:decomp} 
is the list of the partial vertices for
what we called a {\em splitting vertex} (note that we prefer to
use the term {\em vertex copies} for this kind of partial vertices).
Thus a particular application of our approach can be the definition and
the construction of a PES for a given simplicial complex, whenever
raw data are available.

This idea is, for instance, at the basis of the work in \cite{McM00}.
The work in \cite{McM00} propose a RES-like  data structure and
gives algorithm to build such data structure starting from raw STL data. 
There is a basic difference between the two approaches. 
We develop a notion of a unique most general 
decomposition 
while the conversion in \cite{mcm02b}
chooses some rules of thumb based
on geometric conventions and produce a, 
not so general,
decomposition in order to rebuild cells that are meaningful solids.
However meaningfulness of the result is based on a particular heuristics and
on the assumption that the STL raw input data contains correct 
information on the orientation of surfaces bounding cells.
}

\subsection{Decomposition, SGC and the two layered approach}
{
Finally 
we stress the fact that 
the data structure devised in Chapter \ref{ch:nmmdl} of this thesis 
is related with the idea behind  SGC.
SGC proposes a cell complex
whose cells are chosen following known results about manifold
decompositions of algebraic varieties in the theory of 
stratification. SGC essentially propose a two level data structure.
In the upper layer
SGC models the combinatorial structure of this decomposition 
using the incidence graph.
The decomposition components are then modeled as
geometric objects (i.e. extents of algebraic varieties).
In analogy with SGC, we developed a two layer approach based
on a decomposition of the object to be modeled.
With respect to SGC we developed 
a combinatorial  approach to study the decomposition process
for the combinatorial object.
We used  combinatorial structures to model
both the decomposition structure 
and the decomposition components.
On the other hand, SGC uses theory of  stratification  to
postulate a decomposition of the topological space to
be modeled.
Then SGC models just the upper
layer with a combinatorial approach using the incidence graph.
and leaves the encoding  of components of the cellular decomposition in the domain of geometry. 
}

\chapter{Background \label{sec:cellular}}
\section{Introduction}
The subject developed in this work lies within a purely
combinatorial framework. 
In this section, we summarize some background from Combinatorial 
Topology.
We refer to
\cite{Gla70} Pg. 7--48  and  \cite{Hud69} Chap.  1
for a thorough treatment.

We will use geometric concepts and $\real^n$ both 
in examples and in quotations from classical handbooks in
combinatorial topology (mainly \cite{Hud69,Gla70}).
We note that any reference to geometry in
these standard handbooks of Combinatorial Topology, is actually  unnecessary both in the general settlement and in this work. 
We feel that, in general, many notions in Combinatorial Topology 
deserve, and can have, a more ''combinatorial'' presentation.
This  is clearly pointed out by Lickorish in \cite{Lic99}. Lickorish, speaking 
about the traditional definition of {\em combinatorial $n$-manifold}
(see Definition \ref{def:combman}) argues: 
''{\em \ldots this traditional definition in not exactly ''combinatorial''.
The result described later do show it to be equivalent to other formulation
with a stronger claim to this epithet}''.
As clearly pointed out in a survey  by R. Klette 
(see \cite{Kle00} \S 4)  
this idea of  abstraction from geometrical aspects is already present
in the work of Tucker \cite{Tuc33} and in the work of 
Reidemeister \cite{Red38}. 

Following this idea we kept the presentation as ''combinatorial'' as
possible and singled out optional reference to geometry with star like
headers (e.g. {\bf Definition~*}).
This material is  surely helpful to understand combinatorial concepts, but,
from a strictly formal point of view, the introduction  of
these  geometric concepts is not mandatory neither to develop the general 
theory  nor to develop our results (see the already cited work of Lickorish
\cite{Lic99} for discussion of this issue and for a ''purely combinatorial''
presentation of some results in Combinatorial Topology) . 

At the end of this chapter, after having introduced basic concepts we introduce in Section \ref{sec:nerve} the three 
not-so-standard concepts of: {\em Nerve, Pasting and \Quot\ Space},  
The  {\em Nerve} concept is needed for the definition of {\em \Quot s}
that in turn is  crucial for the definition of the decomposition concept that
will come in the next chapter. 

We like to warn the reader  that not all background concepts used in this
thesis are contained in this section. 
This thesis, and the  material presented in this section, sometimes 
reference concepts from Lattice Theory. 
The related Lattice Theory background is presented 
in \latt. 
Thus, the \latt\ contains also a few background concepts used in this thesis. 
They are collected in a separate appendix because these concepts do not
belong to the usual mainstream  in Computer Grahpics.
For this reason, too, our exposition try to use  results from Lattice Theory
with a certain appeal to intuition supporting concepts with examples from
this applicative domain. 
However careful references to \latt\ are inserted whenever  needed.

\section{Simplicial Complexes\label{sec:asc}}
We start this section by introducing  
{\em abstract simplicial complexes} and  then we develop some 
geometric concepts to  give examples of {\em geometric 
realizations} of an abstract simplicial complex.

\begin{definition}[Abstract Simplicial Complexes \cite{Spa66} Pg. 108 or \cite{Ago76} Pg. 47]
\label{def:asc}
Let $V$ be a finite set of elements that we call  \emd{vertices}.
An \emd{abstract simplicial complex} $\AComp$ with vertices in $V$ is a subset  of the set of
(non empty) parts of $V$
such that:
\begin{itemize}
\item
for every vertex  $v\in V$ we have that $\{v\}\in \AComp$;
\item
if $\gamma \subset V$  is an element of  $\AComp$, then every
subset
of $\gamma$ is also an element of $\AComp$.
\end{itemize}
\end{definition}
Each element of $\AComp$ is called an \emas{abstract}{simplex}.
In the literature, usually the term {\em abstract} is omitted \cite{Spa66}.
Sometimes (as in \cite{Veg97}) the term {\em comcnatorial} is used instead of
{\em abstract}.

If $\xi$ is a subset of an abstract simplex $\gamma$ then, by the above
definition, $\xi$ must be an abstract simplex. In this case, $\xi$ is called
a \ems{face} of $\gamma$ (written $\xi\le\gamma$).
The face relation is a {\em partial order} over abstract simplices in $\AComp$
and the pair $\ualg{\AComp}{\le}$ is a {\em poset} 
(see \latt\ Section \ref{sec:poset}).
We will say that $\xi$ is a \emas{proper}{face} of $\gamma$
(written $\xi<\gamma$) \iff\ $\xi\le\gamma$ and $\xi\ne\gamma$.
In the following we will mainly use abstract simplicial
complexes and therefore, sometimes, we will use the terms {\em complex} 
or \emas{abstract}{complex} to denote an abstract
{\em simplicial} complex.

To each abstract simplex $\gamma \in \AComp$ we associate an integer  $dim(\gamma)$, called
\ems{dimension} of $\gamma$. The dimension of $\gamma$ is defined by $\ord(\gamma)=|\gamma|-1$, where $|\gamma|$ is the
number of vertices in $\gamma$.
A complex $\AComp$  is called {\em $d$-dimensional} or a
{\em $d$-complex} if
$\max_{\gamma\in\AComp}(\ord(\gamma))=d$.
A simplex of dimension $s$ is called an  $s$-simplex.
Each $d$-simplex of a $d$-complex $\AComp$ is called a {\em maximal  simplex} of $\AComp$.

The set of all cells of dimension smaller or equal to $m$ is called the
{\em $m$-skeleton} of $\AComp$ (denoted by $\AComp^m$). It is easy to see
that $\AComp^m$ is a subcomplex of $\AComp$. 
We will use the notation  {$\AComp^{[m]}$} to denote the set of all 
$m$-simplices in $\AComp$ (i.e. {$\AComp^{[m]}=\AComp^{m}-\AComp^{(m-1)}$}.

\subsection{Embedding and Geometric Examples}
In the following we will actually develop the subject of this work,
using only abstract simplicial complexes,
with no reference to the possible geometry of the complex.
Geometry and $\real^3$ will come into play in our examples.
We use geometric examples since we believe that geometric complexes
provide an intuitive representation of combinatorial  concepts.
In this view, in this paragraph, we
introduce basic notions for {\em geometric} simplicial complexes
mainly from \cite{Veg97}.

In this paragraph we first use the notion of {\em homeomorphism}.
We will seldom use this concept in the mainstream development of this work
(this will be referenced mainly in examples and  optional material).
Therefore we do not report full definitions here. We 
assume that  the reader is familiar with this  notion and with 
the related notion of {\em topological space} and of
{\em metric space}. We refer the reader to  \cite{Kel55} 
for a general 
reference and possibly to \cite{KorKor68} Pag. 386-388 for a short but 
accurate reminder on this subject.

In the following example we will restrict our attention 
to the well known topological space $\real^n$. By $\real^n$ we denote
the cartesian product  of $n$ copies of
$\real$ equipped with the {\em Euclidean} topology 
induced by the standard Euclidean metric $\|\vettore{x}\|$. 
While putting forward this assumption, we 
note that  sometimes $\real^n$ is not comfortable enough 
to develop more sophisticated examples (e.g. Freedman' s counterexamples \cite{FreeLuo89} \cite{FreeQui90}).
Nevertheless, our restriction to models and examples in $\real^n$ is 
perfectly legal since necessary exceptions arise only in some 
related results that are not essential for developing the 
subject of this work.

\begin{opt-definition}[Geometric Simplex \cite{Veg97} Pg. 519]
\label{def:geomsimp}
Let us consider the Cartesian product $\real^n$,
equipped with the standard Euclidean topology, and let 
$A=\{\vettore{p}_i|1\le k\le (d+1)\}$ be a set of $d+1$
{\em affinely independent} points in $\real^n$ with $(0\leq d\leq n)$.
A \emad{geometric}{$d$-dimensional simplex} $\sigma$ in $\real^n$
is the closed set that is the locus of points  $\vettore{p}$ such that 
$\vettore{p}=\Sigma_{i=1}^{d+1}{\lambda_i\vettore{p}_i}$
with {$\lambda_i$} a set of positive coefficients such that
$\Sigma_{i=1}^{d+1}{\lambda_i}=1$.
\end{opt-definition}
In the situation of the above definition we will say that $\vettore{p}$ is
the convex combination of the  points $\vettore{p}_i$.
The coefficients $\lambda_i$ are called the \emas{barycentric}{coordinates}
of {$\vettore{p}$} w.r.t. the set of points in $A$.
It can be proved (see \cite{EngSve92} Pg. 10, for instance) that 
barycentric coordinates of a point  {$\vettore{p}$}, w.r.t. a certain set  
of points $A$, are uniquely determined.
We will also say that the set  $A$ {\em spans} the geometric simplex $\sigma$.
Any geometric simplex spanned by a proper subset of $A$ will be called
a  {\em face} of $\sigma$.
A $0$-dimensional geometric simplex will be called a  {\em point} 
or a {\em geometric vertex}.
A $1$-dimensional geometric simplex will be called an {\em edge}.
A $2$-dimensional geometric simplex will be called a {\em triangle}.
A $3$-dimensional geometric simplex will be called a {\em tetrahedron}.
The union of a set of (possibly non disjoint) geometric simplices in
$\real^n$ is a subset of $\real^n$ that is  called  a 
{\em geometric} \ems{polyhedron} (or polyhedron for short).

We note that, in a polyhedron, two geometric simplices 
can share internal points. 
If we forbid this we obtain the notion of
{\em geometric simplicial complex}. Geometric simplicial complexes are
the geometric counterpart of  abstract simplicial complexes.
\begin{opt-definition}[Geometric Simplicial Complex]
A \emas{geometric}{simplicial complex} $K$ in $\real^n$ is a finite set
of geometric simplices  in $\real^n$ such that:
\begin{enumerate}
\item all faces of
a simplex $\sigma\in K$ are in $K$;
\item for any pair,  $\sigma$ and $\sigma^\prime$,
of geometric simplices in $K$
their intersection is either empty or $\sigma\cap\sigma^\prime$ is
a face of both $\sigma$ and $\sigma^\prime$.
\end{enumerate}
\end{opt-definition}
Any geometric  simplicial complex  $K$ implicitly defines a
topological compact subspace  of $\real^n$ (denoted by \notation{$|K|$}) 
given by the union
of all simplices $\sigma$ in $K$ (i.e. $|K|=\cup_{\sigma\in K}{\sigma}$).
We will call such a space the  geometric polyhedron 
associated with the geometric simplicial complex $K$.
It can be proven (see \cite{Hud69} Corollary 1.7 Pg. 14)
that for any geometric polyhedron $P$ there exist 
a geometric simplicial complex $K$ 
such that $P=|K|$ (i.e.   $P$ is the geometric polyhedron 
associated with the geometric simplicial complex $K$).
For this reason in the following we will always denote geometric polyhedra
with $|K|$. 

Abstract simplicial complexes provide an abstract view over the set of  
polyhedra. On the other hand a polyhedron provides a 
\emas{geometric}{realization} for an abstract simplicial complex.
Therefore, geometric realizations will be used
to give graphical examples of abstract simplicial complexes.
\begin{opt-definition}[Geometric Realization]
\label{def:georel}
Let $\AComp$ be an abstract simplicial $d$-complex with vertices in $V$.
We will say that a geometric simplicial complex $K$ in $\real^n$
is a {\em geometric realization} (or \ems{embedding})
of $\AComp$  in $\real^n$
\iff\ there is a injective mapping
$\funct{\vettore{p}(v)}{V}{\real^n}$ such that:
\begin{enumerate}
\item for each vertex $v\in V$ the point $\vettore{p}(v)$ is a  
geometric vertex  of $K$;
\item for any abstract simplex $\gamma=\{v_i\}$
in $\AComp$ the set  $\{\vettore{p}(v_i)\}$ spans  a
geometric simplex in $K$.
\end{enumerate}
\end{opt-definition}
It can be proven (see \cite{Gla70} Pg. 5) that
any abstract simplicial $d$-complex admits  geometric realization in
$\real^n$ for $n\ge(2d+1)$.
Even if many geometric realization of $\AComp$ are possible in $\real^n$
it can be  proven (see \cite{Veg97} Pg. 522) that the associated polyhedra 
are all homeomorphic in $\real^n$.

The polyhedron associated with a geometric realization of an abstract
simplicial complex $\AComp$ will be called a \ems{carrier} for $\AComp$.
The class of polyhedra that are carriers of an embedding of
$\AComp$ will be denoted by \notation{$\carrier{\AComp}$}.

Sometimes it is possible to find an embedding of a $d$-complex $\AComp$
into $\real^n$ for $n<(2d+1)$ (see \cite{Veg97} Pg. 529 for a survey on
embeddability). In particular, in our examples we will always use
complexes embeddable into $\real^3$.

Now let us come to drawings that are presented in the running  examples in this work
(see for instance the polyhedron in Figure \ref{fig:es3complab}a).
Our sample drawings
usually represent a
{\em labeled geometric simplicial complex} (see \cite{Ago76} Pg. 48)
i.e. a  polyhedron in $\real^3$ with some labeling for
vertices.
In presenting our drawings we  assume
that, by looking at the labeling at vertices,
the reader is able to perceive the drawing of the polyhedron in $\real^3$
as a geometric simplicial complex and therefore she (or he) can ''see'' the
abstract simplicial complex behind its geometric realization.
According to this assumption we will often refer to drawings in figures
as to ''{\em the (abstract) simplicial complex in figure}'' as a shortcut for
''{\em the (abstract) simplicial complex the polyhedron in figure is a
realization of}''.

\begin{figure}
\begin{minipage}{0.49\textwidth}
\fbox{\psfig{file=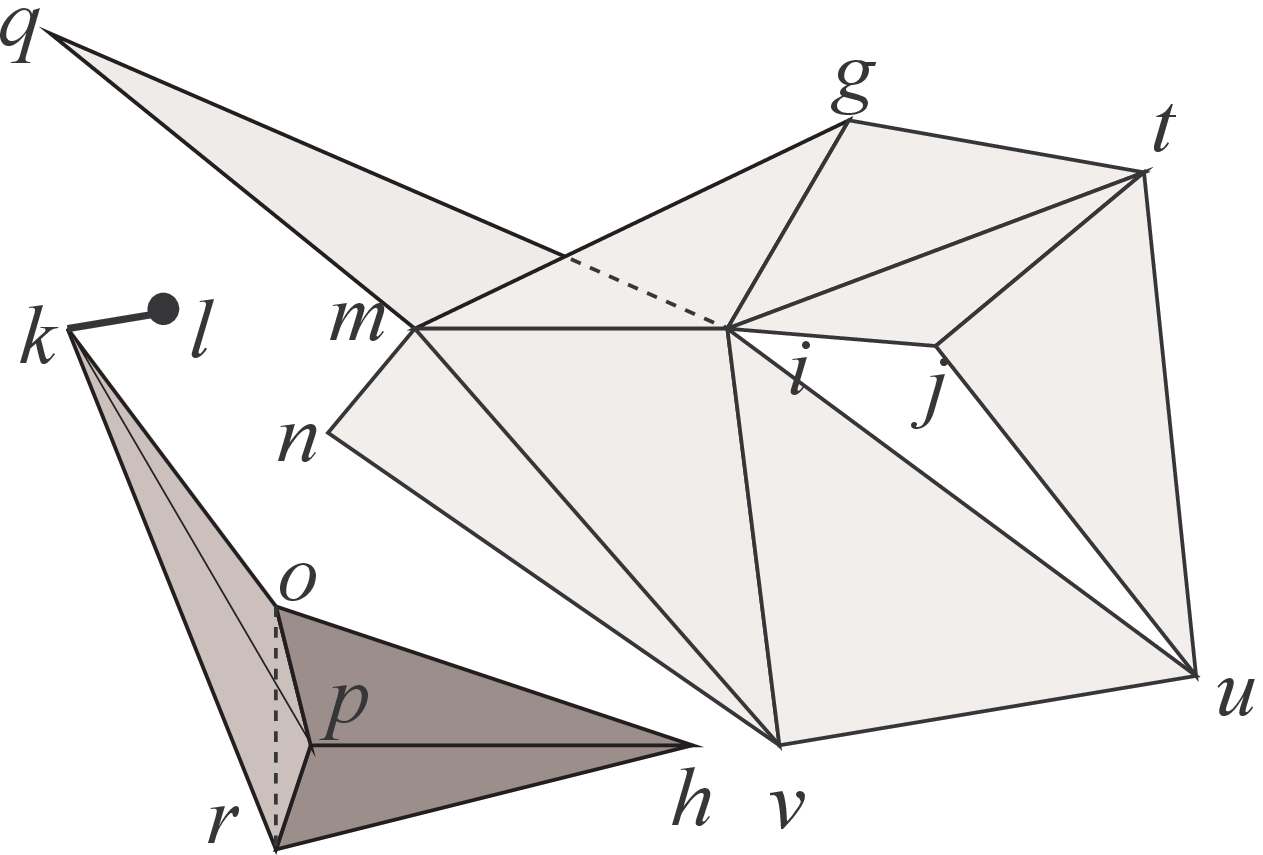,width=\textwidth}}
\begin{center}(a)\end{center}
\end{minipage}
\hfill
\begin{minipage}{0.49\textwidth}
\fbox{\psfig{file=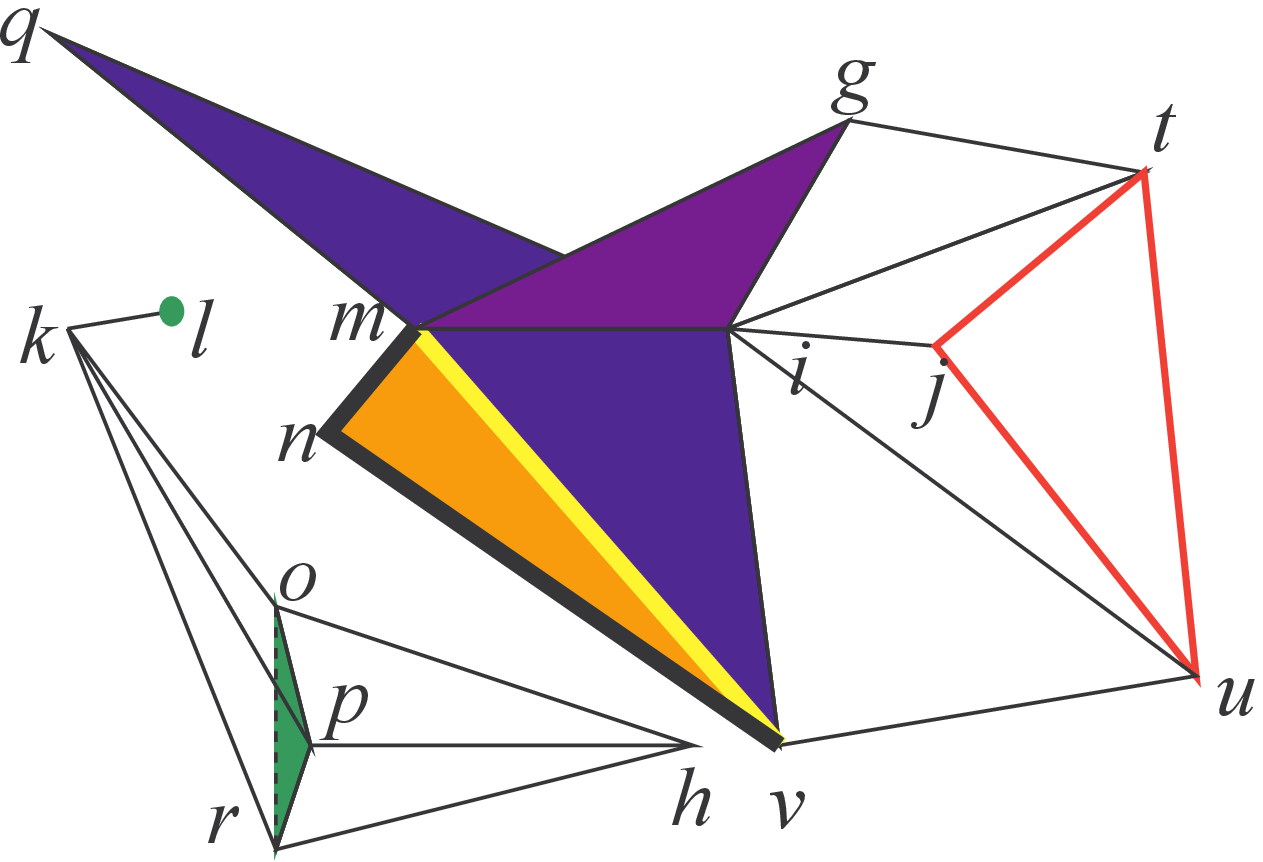,width=\textwidth}}
\begin{center}(b)\end{center}
\end{minipage}
\caption{An example of a geometric realization of an abstract simplicial
complex (a) and examples (b) of some combinatorial concepts
(See \S\ \ref{par:linkdef}).}
\label{fig:es3complab}
\end{figure}
\begin{example}
In Figure \ref{fig:es3complab}a we depict a polyhedron of $\real^3$ that
is a {\em geometric simplicial complex}. In turn this geometric
simplicial complex can be seen as  the geometric
realization of an abstract simplicial complex. Let us call  $\AComp$
this abstract simplicial complex.
According to the conventions outlined above, we will say,
for instance,  that
{\em  the abstract simplicial complex $\AComp$ in Figure
\ref{fig:es3complab}a is a $3$-complex}.

In this view, the finite set of points of $\real^3$ $\{g,h,i,j,k,\ldots\},$ in
Figure \ref{fig:es3complab}a, stands
for the set of Vertices  of the abstract simplicial complex $\AComp$.
Similarly, in examples, we will talk about points,
edges, triangles and tetrahedra to mean the corresponding simplices in
the abstract simplicial complex $\AComp$, and so,
for instance, the segments
$kr$, $kp$, $ko$, $kl$,
stands, respectively, for the $1$-simplices
$\{k,r\}$, $\{k,p\}$, $\{k,o\}$, $\{k,l\}$ of $\AComp$.
Similarly the $mnv$ triangle in the geometric complex  in Figure \ref{fig:es3complab}
stands for the $2$-simplex $\{m,n,v\}$ in the abstract simplicial
complex $\AComp$.
The presence of two tetrahedra in Figure \ref{fig:es3complab}a:
$kopr$ (in light gray) and $hrop$ (in dark gray) means that in $\AComp$
we have just two $3$-simplices $\{k,o,p,r\}$ and $\{h,r,o,p\}$.
Finally  note that a simplex is a face of another simplex
in $\AComp$ \iff\ this face relation holds in the geometric example.
For instance, in Figure \ref{fig:es3complab}a we have that, 
in $\real^3$, the triangle
$rop$ is face of both $kopr$ and $hrop$. 
At a combinatorial level we will have that the simplex  $\{r,o,p\}$
in $\AComp$ is a face for both $\{k,o,p,r\}$ and $\{h,r,o,p\}$.
\end{example}

In the following, we will mainly introduce definitions
and properties for {\em abstract} simplicial complexes.  
Therefore, whenever there is no ambiguity, we will
freely use the terms {\em simplex} and {\em complex} to mean
{\em abstract simplex} and  {\em abstract simplicial complex}.
On the other hand the adjective  {\em geometric} will be assumed
(and therefore omitted) when dealing with geometric examples.
\subsection{Boundary, Star, Link, Subcomplexes and Closures\label{par:linkdef}}

The \ems{boundary} \notation{$\bnd{\gamma}$} of a simplex  $\gamma$ is defined to be the set
of all proper faces of $\gamma$.
Similarly, the \ems{coboundary} or \ems{star}  of a simplex $\gamma$
is defined as $\star{\gamma}=\{\xi\in \AComp\ |\ \gamma\subseteq\xi\}$.
Cells $\xi$ in \notation{$\star{\gamma}$} are called \ems{cofaces} of $\gamma$.
Any simplex $\gamma$ such that
$\star{\gamma}=\{\gamma\}$ is called a \ems{top simplex} of
$\AComp$.
Two distinct simplices are said to be \ems{incident} \iff\
one of them is a face of the other.

The \ems{link} of a simplex $\gamma$, denoted by $\lk{\gamma}$, is the set of
all faces of cofaces of $\gamma$, that are not incident to $\gamma$.

The star and the link of a simplex $\gamma$ are 
defined by referring to the
surrounding complex $\AComp$.
We can emphasize this reference by using
$\xstr{\AComp}{\gamma}$, instead of $\str{\gamma}$ and
$\xlk{\AComp}{\gamma}$, instead of $\lk{\gamma}$.
\begin{example}
Considering the complex $\AComp$ in 
Figure \ref{fig:es3complab}b we have that the boundary of
the triangle $jtu$ is
$\bnd{jtu}=\{j,t,u,jt,ju,tu\}$ (in red) and
the star (coboundary) of simplex $mi$ is given by
$\star{mi}=\{mi,mig,miq,mij \}$ (in violet blue).
The link of point $k$ can be found considering the simplices contaning $k$, i.e.  $\star{k}=\{k,kl,ko,kp,kr,kop,kpr,kor,kopr\}$ and taking the set of faces not containing the  point $k$ , i.e. the triangle $opr$, vertex $l$ together with
all triangle faces and hence $\lk{k}=\{l,o,p,r,op,pr,ro,opr \}$ 
(in green).
Finally we note that the top simplices in $\AComp$ are eleven:
the eight grey triangles on the right of Figure \ref{fig:es3complab}a and
the two tetrahedra $kopr$, $hopr$ and the edge $kl$.
\end{example}

A subset $\Gamma$ of $\AComp$ is \ems{closed}  (in $\AComp$)
if, for every simplex
$\gamma\in\Gamma$ we have that $\bnd{\gamma}\subset\Gamma$ i.e. 
all simplices of the  boundary
of $\gamma$ are also simplices of $\Gamma$.
Note that a closed subset of $\AComp$ is always a complex.

A subset of simplices $\AComp^{\prime}\subset \AComp$ is a complex
on its own \iff\ $\AComp^\prime$ is a closed set of simplices.
In this case $\AComp^{\prime}$ is
said to be a \ems{subcomplex} of $\AComp$.
In general, given a set of simplices $\Gamma$, the closure 
\notation{$\closure{\Gamma}$} of $\Gamma$
is the smallest subcomplex whose simplices include those in $\Gamma$.
In particular, for a simplex $\gamma$, we will use the notation
$\closure{\gamma}$ as a shortcut for $\closure{\{\gamma\}}$.

\begin{short-example}
Consider, for instance, the following three subsets of the
abstract simplicial complex $\AComp$ in Figure \ref{fig:es3complab}b:
$A=\{mnv\}$ (in orange); $B=\{mnv,mn,nv\}$ (in black and orange);
$C=\{m,n,v,mn,nv,vm,mnv\}$ (in black, orange and yellow).
Both  subsets $A$ and $B$ are  not closed in the
original complex $\AComp$, while $C$ is closed in the original complex $\AComp$ and $C=\closure{A}=\closure{B}$.
\end{short-example}

Note that the star of a simplex is very rarely 
a subcomplex while the link of a simplex
is always a subcomplex.
With these definitions, we can write
that, for simplex $\gamma$,
$\lk{\gamma}=\clstar{\gamma}-\star{\closure{\gamma}}$.
Again, we can emphasize the dependence of the link on the complex $\AComp$
by writing the identity $\lk{\gamma}=\clstar{\gamma}-\star{\closure{\gamma}}$
as
$\xlk{\AComp}{\gamma}=\xclstar{\AComp}{\gamma}-\xstr{\AComp}{\closure{\gamma}}$.

\begin{figure}
\begin{minipage}{0.30\textwidth}
\fbox{\psfig{file=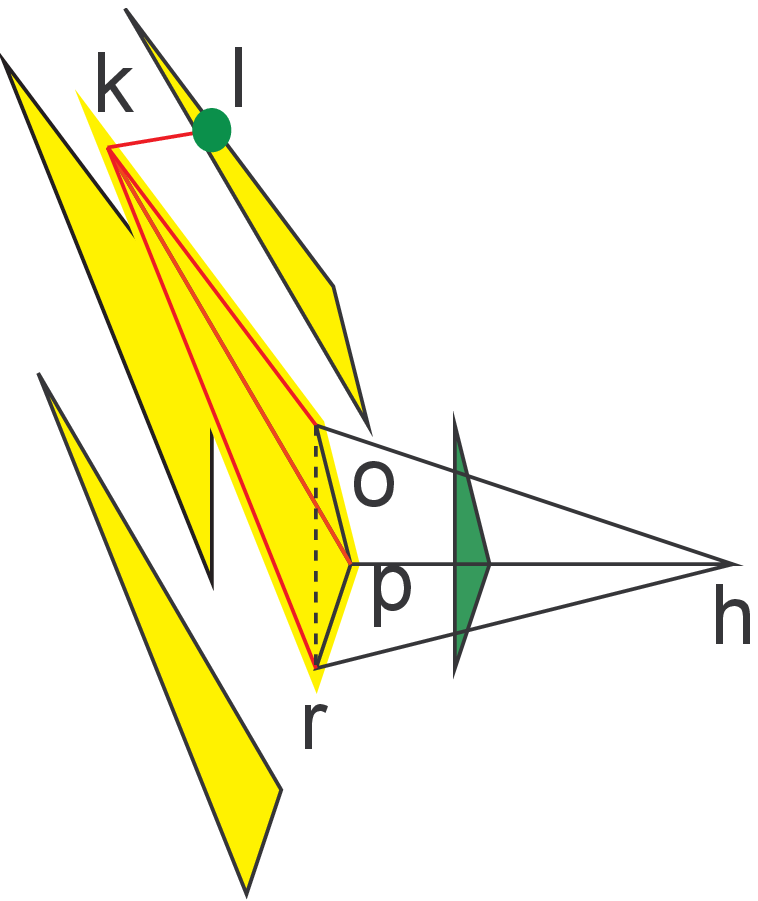,width=\textwidth}}
\begin{center}(a)\end{center}
\end{minipage}
\hfill
\begin{minipage}{0.30\textwidth}
\fbox{\psfig{file=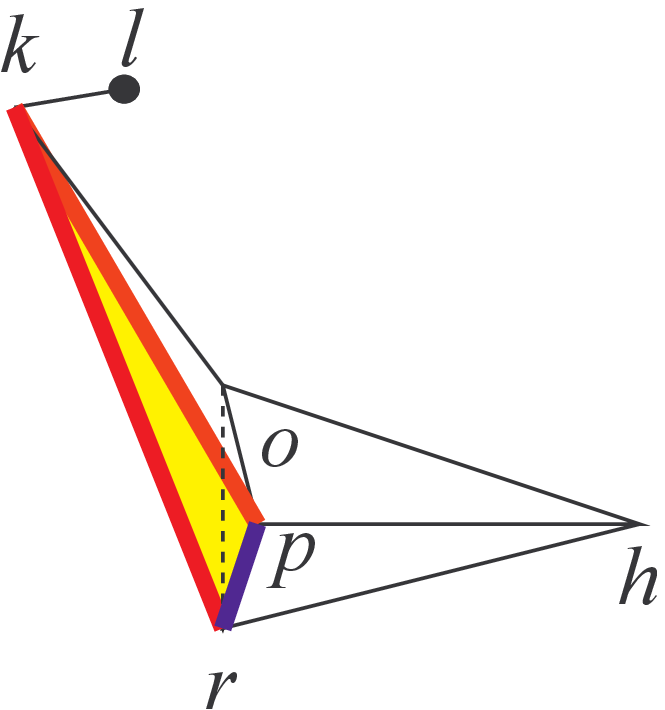,width=\textwidth}}
\begin{center}(b)\end{center}
\end{minipage}
\caption{Examples of some combinatorial identities in a $3$-complex
(See \S\ \ref{par:linkdef}).}
\label{fig:es3complcd}
\end{figure}
\begin{example}
In the complex of Figure \ref{fig:es3complcd}a we already noted that $\lk{k}=\{l, o, p, r, op, pr, ro, opr\}$ (in  green).
It is easy to see that $\star{k}=\{k,kl,ko,kp,kr,kop,kpr,kor,kopr\}$
(in yellow, green  and red in Figure  \ref{fig:es3complcd}a) and
$\clstar{k}=\star{k}\cup\{l,o,p,r,op,pr,ro,opr\}$.  Therefore
$\lk{k}=\clstar{k}-\star{k}$ being $k=\closure{k}$.
Note again that $\star{k}$ is not a closed  subcomplex.

Let us consider Figure \ref{fig:es3complcd}b and the subcomplex
$\Delta=\closure{kpr}=\{kpr,kp,kr,pr,k,p,r\}$ ($\Delta$ 
is in red, blue and yellow
in Figure \ref{fig:es3complcd}b).
It is easy to see, by definition,  that $\xlk{\Delta}{k}$ is given by the
blue segment $rp$ in Figure \ref{fig:es3complcd}b.
Let us verify this using the identity
$\xlk{\AComp}{\gamma}=
\xclstar{\AComp}{\gamma}-\xstr{\AComp}{\closure{\gamma}}$ with
$\AComp=\Delta$ and $\gamma=\{k\}$. Note that where this is not
ambiguous, we use $k$ as a shortcut for $\{k\}$.
It is easy to see that $\xstr{\Delta}{k}=\{k,kp,kr,kpr\}$ (in red and yellow) and
taking the closure  $\xclstar{\Delta}{k}$ in $\Delta$ we have
$\xclstar{\Delta}{k}=\xstr{\Delta}{k}\cup\{p,r,pr\}$.  
Therefore, being $k=\closure{k}$,
we rewrite the identity
$\xlk{\Delta}{\gamma}=\xclstar{\Delta}{\gamma}-\xstr{\Delta}{\closure{\gamma}}$ as
$\xlk{\Delta}{k}=\xclstar{\Delta}{k}-\xstr{\Delta}{k}=\{p,r,pr\}$ and  find the blue
segment $pr$ plus all its faces.
Note that $\xlk{\AComp}{k}$ is the subcomplex in  green in
Figure \ref{fig:es3complcd}a and, therefore, $\xlk{\Delta}{k}\neq\xlk{\AComp}{k}$
\end{example}

Let $\Phi$ and $\Gamma$ be   two set of simplices 
such that for every $\phi\in\Phi$ and $\gamma\in\Gamma$ we have
$\phi\cap\gamma=\emptyset$.
We define (following for instance \cite{Spa66} Pg. 109) 
the join $\Phi\join\Gamma$ as  the set of
simplices
$\Phi\cup\Gamma\cup
(\cup_{\phi\in\Phi,\gamma\in\Gamma}\{\phi\cup\gamma\})$.
We note that the  join operator $\join$ is both commutative and
associative and, for any set of simplices $\Gamma$, we have
$\Gamma\join\emptyset=\emptyset\join\Gamma=\emptyset$.

It is easy to prove that if 
$\Phi$ and $\Gamma$ are two abstract simplicial complexes, then
the join $\Phi\join\Gamma$ is an abstract simplicial complex.
In particular for a simplex $\gamma$ we write
$\gamma\join\Phi$ as shorthand for $\closure{\{\gamma\}}\join\Phi$.
If $\phi$ and $\gamma$
are two disjoint simplices  we  denote with $\phi\join\gamma$ the
complex $\phi\join\gamma=
\closure{\{\phi\}}\join\closure{\{\gamma\}}=
\closure{\phi\cup\gamma}$.
If  $\gamma$ is made up of  single vertex $w$ we will call
the join $\{w\}\join\Phi$ the \ems{cone} from $w$ to $\Phi$.

Where this is not ambiguous we will use the notation 
$\phi\join\gamma$ also  to denote the {\em simplex} 
$\phi\cup\gamma$.
With this notation we
have that the link of $\gamma$ in a certain complex $\AComp$ is given by
\newred{
$\xlk{\AComp}{\gamma}=\{\phi\in\AComp|\gamma\join\phi\subset\AComp\}$
}
Similarly it is easy to verify that
\newred{
$\clstar{\gamma}=\closure{\gamma}\join\lk{\gamma}$.
}

\begin{example}
As an example let us consider an application of join between  complexes
to compute the closed star of vertex $k$ in Figure \ref{fig:es3complcd}a
using formula $\clstar{\gamma}=\closure{\gamma}\join\lk{\gamma}$.
We already know that $\clstar{k}$ is given by all the colored parts in
Figure \ref{fig:es3complcd}a.
In the complex of Figure \ref{fig:es3complcd}a
we have that $\lk{k}=\{l,o,p,r,op,pr,ro,opr\}$
(in  green in Figure \ref{fig:es3complcd}a).
The identity $\clstar{\gamma}=\closure{\gamma}\join\lk{\gamma}$.
with $\gamma=\{k\}$ becomes
$\clstar{k}=k\join\lk{k}$, where we used $k$ as a shortcut for $\{k\}$ and
for $\{k\}=\closure{\{k\}}$.
To join  $k$ with its link we  just have to join $k$ with $l$ and $opr$.
By definition of join we have that
$\{k\}\join\{l\}=\closure{kl}$ (i.e. the $kl$   edge).
Similarly, joining $k$ with triangle $opr$
we get the tetrahedron $kopr$
(in red, green and yellow in Figure \ref{fig:es3complcd}a).
This shows that joining $k$ with its link we obtain the closed star of $k$.
\end{example}

\subsection{\label{par:conndef}Adjacency, Paths and Connected Complexes}
Two simplices are called \ems{$s$-adjacent} if they share an
$s$-face; in particular, two
$p$-simplices are said to be \ems{adjacent} if they are
$(p-1)$-adjacent.

A {\em $h$-path} is a sequence of simplices
$(\gamma_i)_{i=0}^k$ such that two successive simplices $\gamma_{i-1}$ $\gamma_{i}$
are $h$-adjacent.
Two simplices $\gamma$ and $\gamma^{\prime}$ are
{\em $h$-connected}, \siff\  there exist a $h$-path $(\gamma_i)_{i=0}^k$
such that $\gamma$ is a face of $\gamma_0$ and $\gamma^\prime$ is a face of $\gamma_k$.
A subset $\AComp^{\prime}$ of a complex $\AComp$ is called {\em $h$-connected}
\siff\ every pair of its vertices are $h$-connected.
Any maximum $h$-connected subcomplex of a complex $\AComp$
is called a {\em $h$-connected component} of $\AComp$.
Usually (see \cite{Ago76} Pg. 40) when one forget to mention $h$ and talks 
about ''connectivity'' means $0$-connectivity.

\begin{short-example}
According to the definition above, the pair of tetrahedra $kopr$
and $hopr$ in the complex of Figure \ref{fig:es3compleg}a
are both $0$-adjacent, $1$-adjacent and
$2$-adjacent. Hence they  also represent a pair of adjacent $3$-simplices.
It is easy to show that the complex of Figure \ref{fig:es3compleg}a
has two connected components one of
which is formed by the three simplices $kl$, $kopr$, and $hopr$
(with all their faces)
and the other which is formed by the eight gray triangles on the right.
For any pair of simplices in each component there exist a path 
in between.
For instance, the two simplices $l$ and $h$ are
connected by the thick gray path $(lk,kr,rh)$.
Simplices $gm$ and $nv$ are $0$-connected via the black thick $0$-path 
$(gm,mn,nv)$. 
The $2$-simplices $gmi$ and $nvm$ are $1$-connected via the three dark gray 
triangles that belongs to the $1$-path $(gmi,mvi,nvm)$. 
With similar remarks one can show that the subcomplex made up of the eight
gray triangles 
is a $1$-connected $2$-complex.
\end{short-example}

\begin{figure}
\begin{minipage}{0.49\textwidth}
\fbox{\psfig{file=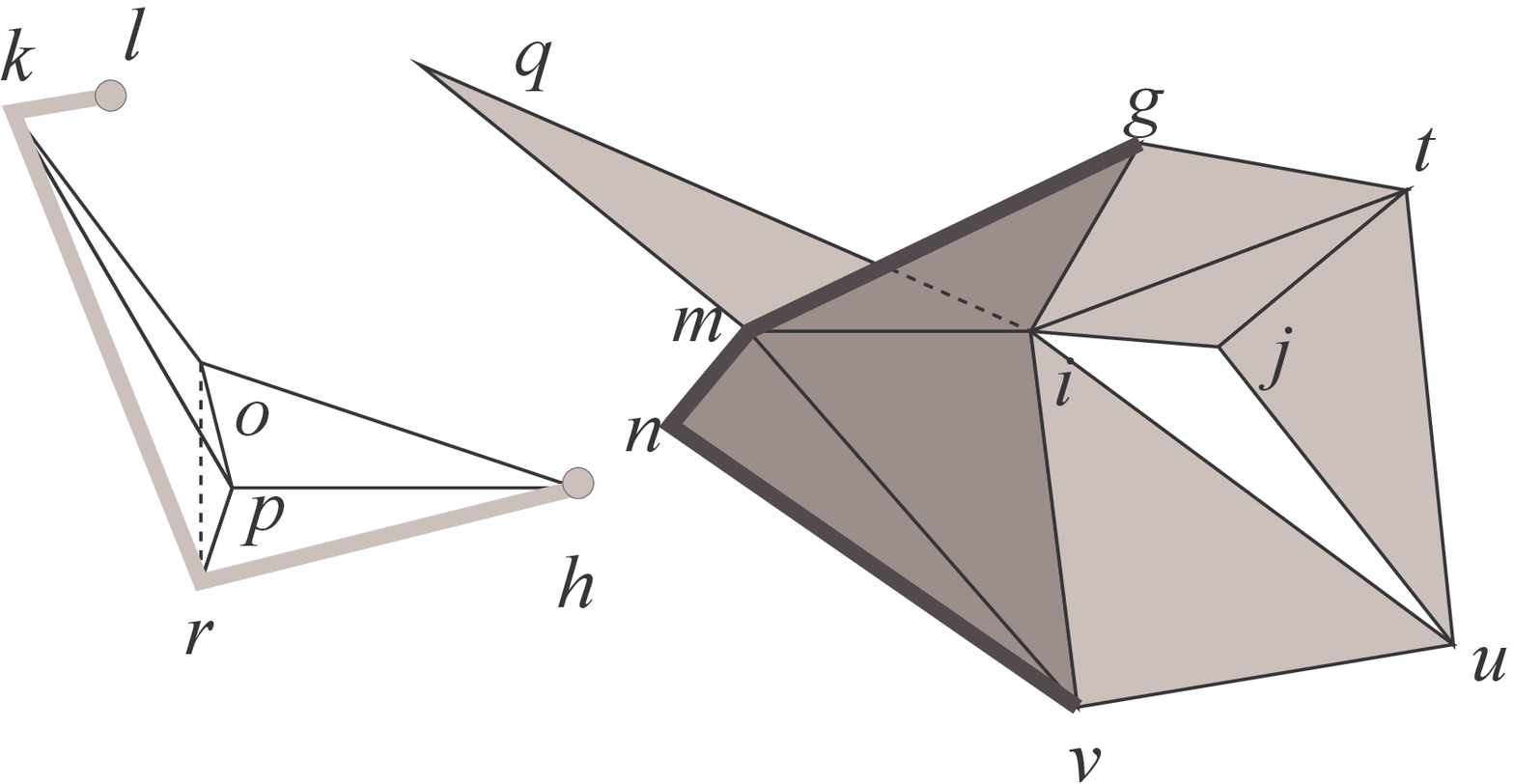,width=\textwidth}}
\begin{center}a\end{center}
\end{minipage}
\hfill
\begin{minipage}{0.49\textwidth}
\fbox{\psfig{file=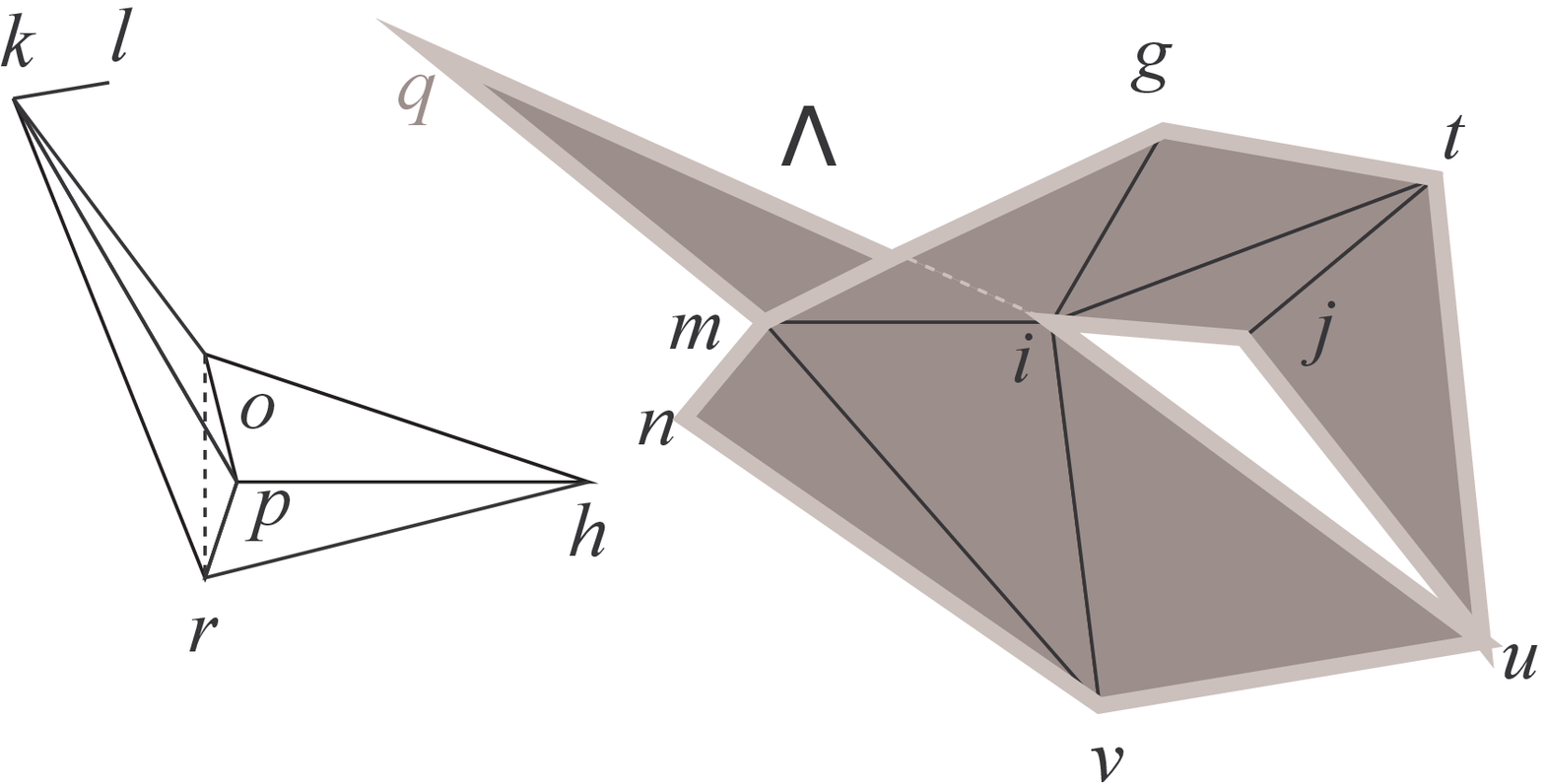,width=\textwidth}}
\begin{center}b\end{center}
\end{minipage}
\caption{Examples of paths in connected components (a) 
(See \S\ \ref{par:conndef}) and of regular and non-regular 
complexes (See \S\ \ref{par:reg})(b)
.}
\label{fig:es3compleg}
\end{figure}

\subsection{Regular Complexes\label{par:reg}}
A $d$-complex where all top simplices are maximal
(i.e. of dimension $d$) is called \ems{regular}
or {\em uniformly $d$-dimensional.}

Let $\AComp$ be a regular complex, and $\AComp^{d-1}_b$ be the
collection of all its
$(d-1)$-simplices
having only one incident $d$-simplex.
The subcomplex $\partial\AComp=\closure{\AComp^{d-1}_b}$
is called the \ems{boundary} of
$\AComp$,
while the collection of simplices
$\AComp-\partial\AComp$ is called the set of
\emas{internal}{simplices} of
$\AComp$.

\begin{short-example}
The complex of Figure \ref{fig:es3compleg}b  is a $3$-complex
whose maximal simplices are the two tetrahedra $kopr$ and $hopr$.
The complex is not regular since the top simplices are
not only the two tetrahedra $kopr$ and $hopr$ 
but also the edge $kl$  and the the eight gray  triangles on the right,
The whole complex is not regular while
there are three regular subcomplexes that are
the closure of $\{kopr,hopr\}$,  the edge $kl$ and the complex $\Lambda$
made up of the eight gray triangles on the right.
The boundary of $\Lambda$ is given by the eleven thick gray segments
in Figure \ref{fig:es3compleg}b.
\end{short-example}

\section{Abstract Simplicial Maps\label{par:maps}}
\subsection{Introduction}
In this paragraph we introduce {\em abstract simplicial maps} and 
show that they provide a categorical structure
for abstract simplicial complexes. 
In this section, we follow mainly the approach in \cite{Ago76,EngSve92,Spa66}
and define (abstract) simplicial maps as applications between
(abstract) simplicial complexes. Other approaches 
\cite{Gla70,Hud69,Veg97} define simplicial maps as applications 
between geometric simplicial complexes.
We retain both approaches here in order to state some results in  their 
original form. In order to make a clear distinction between these two
options we will call {\em abstract simplicial maps} those between abstract
simplicial complexes and {\em geometric simplicial maps} those between 
geometric simplicial complexes.

\subsection{Abstract Simplicial Maps}
Let $\cal V$ be a set of symbols, that we call the {\em universe of vertices.}
W.l.o.g., from now on we will  assume that all complexes have their
vertices in $\cal V$.
A \emas{vertex}{map} is any map $f$ between vertices,
i.e., $\funct{f}{\cal V}{\cal V}$.
Moreover, given a vertex map $f$, we will use notation
$f[\send{a}{x},\send{b}{y},\ldots]$
as a shorthand for the map $f'$ that differs from $f$ only at vertices
specified between brackets, i.e., $f'(a)=x$, $f'(b)=y$, and so on.
If $W$ is a set of vertices we will use $[\send{W}{x}]$ as a
shortcut for $[\send{w}{x}|w\in W]$.
We will denote by $\funct{\Lambda}{\cal V}{\cal V}$ the identity vertex map and
we will use the notation $[\send{a}{x},\send{b}{y},\ldots]$ as a shortcut for
$\Lambda[\send{a}{x},\send{b}{y},\ldots]$

\begin{definition}[Abstract Simplicial Map]
\label{def:abssimap}
Given a pair of abstract simplicial complexes
$\AComp$ and $\AComp^\prime$ we will say that a vertex map
$f$ defines (or induces) a \emd{simplex map} from $\AComp^\prime$  to
$\AComp$ if for every simplex $\gamma\in\AComp^\prime$
the set of Vertices $f(\gamma)=\{f(v)\ |\ v\in\gamma\}$
is a simplex in $\AComp$.
We will use the term  \emad{abstract}{simplicial map} to denote both
vertex maps and simplex maps.
\end{definition}
Obviously an abstract simplicial maps  preserve the face relation i.e. if $\gamma\le\beta$ then $f(\gamma)\le f(\beta)$.
We can extend \asm\ from simplices  to complexes by defining
$f(\AComp')=\{f(\gamma)\ |\ \gamma\in\AComp'\}$.
It is easy to see that if $\AComp'$ is a complex, 
then $\AComp=f(\AComp')$ is a complex, too.
In the following we will use $f^{-1}(v)$ to denote the 
inverse image along $\funct{f}{\cal V}{\cal V}$. of the set $\{v\}$.
Similarly   will use $f^{-1}(\gamma)$ to denote the 
inverse image along the simplex map induced by $f$ of the set $\{\gamma\}$.
Note that, with this notation $f^{-1}(v)$ denotes a set of vertices, while
$f^{-1}(\{v\})$ denotes a set of simplices.

In the following we will often use the same symbol (e.g., $f$)
to denote both the vertex map and the
simplex map . Similarly we will use the same
symbol (e.g., $f^{-1}$) to denote the two  inverses.
However, note  that by no means $f^{-1}(\gamma)$
can be recovered  using only the knowledge of all sets $f^{-1}(v)$
for all $v\in\gamma$.
\begin{figure}[h]
\begin{center}
\framebox{
\parbox[c][0.26\textwidth]{0.45\textwidth}{
\psfig{file=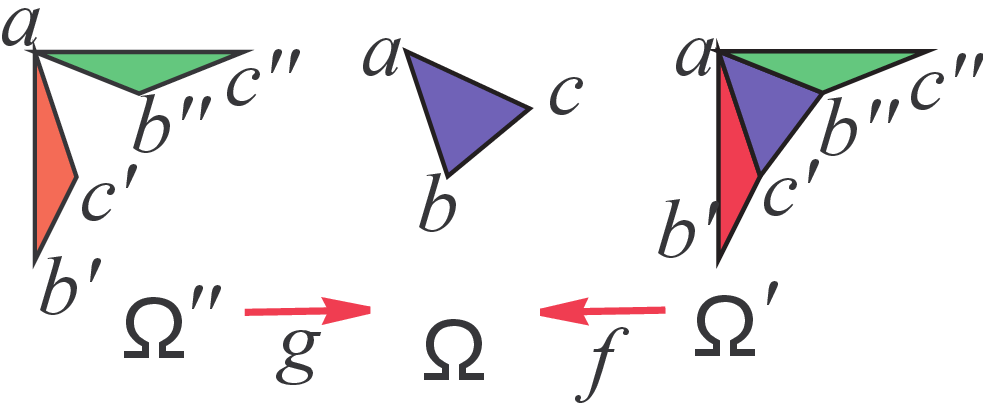,width=0.45\textwidth} \begin{center}(a)\end{center}
}
}
\framebox{
\parbox[c][0.26\textwidth]{0.45\textwidth}{
\psfig{file=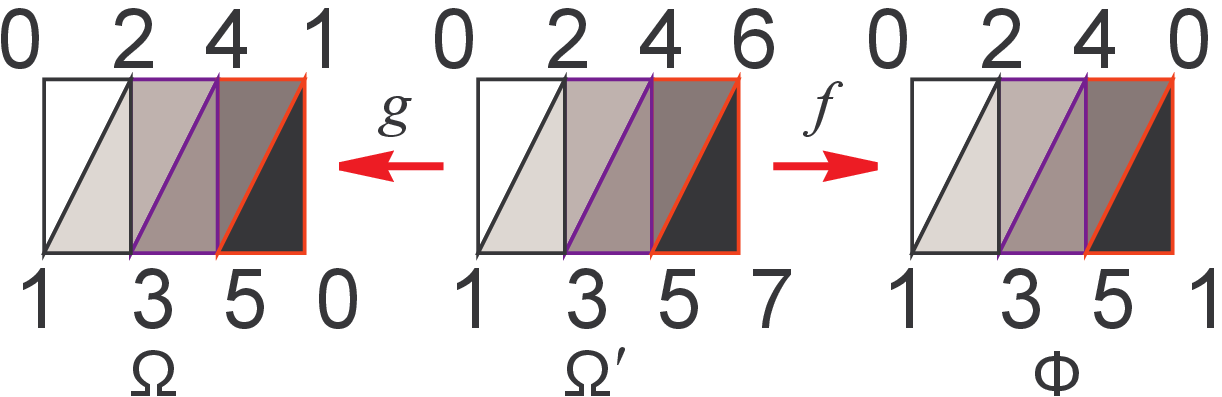,width=0.45\textwidth}
\begin{center}(b)\end{center}
}
}
\end{center}
\caption{Examples of:
two distinct abstract simplicial maps with 
$f^{-1}(v)=g^{-1}(v)$ for all $v$ (a); 
two distinct vertex map with
the same simplex stitching on top simplices (b)
}
\label{fig:simpsti}
\end{figure}
\begin{short-example}
Consider for instance the three complexes $\AComp$, $\AComp^\prime$ and
$\AComp^\second$ in Figure \ref{fig:simpsti}a. 
We have that $f(\AComp^\prime)=\AComp$ and
$g(\AComp^\second)=\AComp$. For all Vertices $v$ in $\AComp$ we
have $f^{-1}(v)=g^{-1}(v)$ and yet the two maps $f$ and $g$ are
different. In fact we have $f^{-1}(a)=g^{-1}(a)=\{a\}$;
$f^{-1}(b)=g^{-1}(b)=\{b^\prime,b^\second\}$ and
$f^{-1}(c)=g^{-1}(c)=\{c^\prime,c^\second\}$.
Nevertheless $f^{-1}(abc)=\AComp^\prime$ and
$g^{-1}(abc)=\AComp^\second$.
\end{short-example}
Similarly note that the knowledge of the restriction of a
simplex map $f$
to top simplices is not sufficient to define $f$ completely.

\begin{short-example}
In Figure \ref{fig:simpsti}b we have an example of two
abstract simplicial map with the same simplex map on triangles (and on edges) 
and distinct vertex maps. In fact  the vertex maps
$f=[\send{6}{0},\send{7}{1}]$ and
$g=[\send{6}{1},\send{7}{0}]$ induces two  simplex stitchings
$\funct{f}{\AComp^\prime}{\Phi}$ and
$\funct{g}{\AComp^\prime}{\AComp}$ such that, for any $x$, $y$ and $z$
$f(xyz)=g(xyz)$ and $f(xy)=g(xy)$.
However, note that $f$ and $g$ are not strictly the same simplex
map since their codomain $\Phi$ and $\AComp$ are not
isomorphic. This can easily be proven (see also the comment on Figure \ref{fig:mapsgen}a) by noticing that $\Phi$ is orientable while
$\AComp$ it is not.
Furthermore, obviously, $f$ and $g$ differ on $0$-simplices.
\end{short-example}
\subsection{Isomorphism}
A particular class of simplicial maps is given by maps
induced by a consistent renaming of Vertices. Such a kind 
of mapping is called an \ems{isomorphism}.
\begin{definition}[Isomorphism]
An \emd{isomorphism} is a bijective \asm.
\end{definition}
A standard characterization of isomorphic complexes is therefore
the following.
\begin{definition}[Isomorphic Complexes (see \cite{Lic99} Pg.  301)]
Two abstract simplicial complexes $\AComp$ and $\AComp^\prime$ are
\emd{isomorphic} (written $\AComp\simpeq\AComp^\prime$) 
\iff\ there exist a bijection between their
vertices that induces a bijection between their simplices.
\end{definition}
Indeed, it is easy to prove (see \cite{EngSve92} Prop. 2.5.2)
that a simplicial map induced by a bijection between Vertices is an 
isomorphism; that the composition of two isomorphisms is an 
isomorphism and that the inverse map of an isomorphism is an isomorphism, too.

The following property gives another characterization of isomorphic 
complexes.
\begin{property}
\label{pro:iso}
Two abstract simplicial complexes
$\AComp$ and $\AComp^\prime$ are
{\em isomorphic} \iff\ there exist
two abstract simplicial maps $f$ and $g$ s.t.
$f(\AComp^\prime)=\AComp$ and $g(\AComp)=\AComp^\prime$
\begin{proof}
First note that if  $\AComp$ and $\AComp^\prime$ are
isomorphic there exist a bijection $h$ between vertices
that is also a bijection between simplices therefore
we have an invertible simplicial map whose inverse is
again a simplex map therefore taking  $f=h$ and $g=h^{-1}$
we prove that for an isomorphic pair of complexes
$\AComp$ and $\AComp^\prime$
there exist
two abstract simplicial maps $f$ and $g$ s.t.
$f(\AComp^\prime)=\AComp$ and $g(\AComp)=\AComp^\prime$.

Conversely if such an $f$ and a $g$ exist, then both
the vertex map $f$ and the induced simplex map  $f$ must be
surjective  since it spans all vertices and all simplices in
$\AComp$. Therefore the number of vertices (simplices)
in $\AComp$ is smaller or equal to the number of
vertices (simplices)  in $\AComp^\prime$.
Using the same argument for $g$ we conclude that also
$g$ must be surjective and hence the number of vertices
(simplices) in $\AComp$ must be equal to the number of
vertices (simplices)  in $\AComp^\prime$.
Since no vertex can disappear, $f$ must be a bijection
between vertices. Since $f$ is an abstract simplicial map
the simplex map $f$ must be also a bijection between
simplices. Note that not necessarily $f=g^{-1}$.
\end{proof}
\end{property}

It is easy to see that the set
of abstract simplicial  complexes can be equipped with a
categorical structure using abstract simplicial maps as
morphisms. This provides 
the set of abstract simplicial complexes with a
preorder (see Section \ref{sec:poset} in \latt\ for the definition of 
preorder) 
\begin{property}
\label{pro:mapposet}
The set of abstract simplicial complexes becomes a
category using  abstract simplicial maps as morphisms.
Furthermore, the set of abstract simplicial complexes has a preorder
$\morphle$.  This preorder is completely  defined by asking
that $\AComp\morphle\AComp^\prime$ \iff\ there exist an abstract
simplicial map $f$ s.t. $f(\AComp^\prime)=\AComp$.
\begin{proof}
The categorical structure 
(see, for instance, \cite{Spa66} Pg. 14 and Pg. 110) 
comes form the fact that
the  functional composition of two abstract simplicial maps is still
an abstract simplicial map.
Since  $\Lambda(\AComp)=\AComp$
we have that $\AComp\morphle\AComp$ (i.e. $\morphle$ is reflexive and
the identity is a morphism).
Transitivity again  comes easily from the fact that
the composition of two abstract simplicial maps is still
an abstract simplicial map.
\end{proof}
\end{property}
Note that to have a preorder we just need a
a reflexive and transitive relation.
In fact, in general
$\AComp\morphle\AComp^\prime$ and $\AComp^\prime\morphle\AComp$
do not implies $\AComp^\prime=\AComp$. 
However,  by Property \ref{pro:iso}, this implies
that the two abstract simplicial complexes
$\AComp$ and $\AComp^\prime$ are isomorphic.

Now we can give similar definitions for maps between geometric 
simplicial complexes. 
Following this idea we will report the definitions for 
{\em geometric simplicial maps} and {\em geometric simplicial equivalence}.
Indeed note that a   geometric simplicial map is actually a map between
a pair of polyhedra $P$ and $P^\prime$. Such a map can be a geometric simplicial map or not
with reference to a  particular pair of geometric simplicial complexes 
$K$ and $K^\prime$. These geometric
simplicial complexes $K$ and $K^\prime$  must be chosen 
such that $P=|K|$ and $P^\prime=|K^\prime|$. 
This is expressed by the following definition.

\begin{opt-definition}[Geometric Simplicial Map (See \cite{Veg97} Pg. 520)]
A continuous  map $\funct{\phi}{|K|}{|L|}$ 
is a \emad{geometric}{simplicial map} w.r.t. the geometric simplicial
complexes $K$ and $L$ \iff:
\begin{enumerate}
\item for every set of geometric vertices 
$A=\{\vettore{p}_i\}$ that spans a geometric simplex $\sigma_A$ 
in $K$, the set $\phi(A)$ spans a geometric simplex in $L$.
\item for every geometric vertex $\vettore{p}\in\sigma_A$ 
if $\lambda_i$ are such that
$\vettore{p}=\Sigma_i{\lambda_i\vettore{p}_i}$ then we have 
$\phi(\vettore{p})=\Sigma_i{\lambda_i \phi(\vettore{p}_i)}$
\end{enumerate}
\end{opt-definition}
The second condition in the above definition implies that a geometric
simplicial map $\funct{\phi}{|K|}{|L|}$ is completely determined by 
its restriction to  the set of  geometric vertices of $|K|$.
  
It is easy to show (see \cite{Spa66} Pg. 113) that there is a category of
polyhedra and of geometric simplicial maps.
There is an obvious equivalence between this category
and the category of abstract simplicial complexes and 
abstract simplicial maps. 
In fact let $\AComp$ and $\AComp^\prime$ be two abstract
simplicial complexes with vertices in $V$ and $V^\prime$. 
Let us consider two injective mappings 
$\funct{\vettore{p}}{V}{\real^n}$ and
$\funct{\vettore{p}^\prime}{V^\prime}{\real^m}$ and let 
$K$ and $K^\prime$ be the geometric realization of
$\AComp$ and $\AComp^\prime$ defined by $\vettore{p}$  and 
$\vettore{p}^\prime$.
Now, for any abstract  simplicial map 
$\funct{f}{\AComp^\prime}{\AComp}$, there is 
a  geometric simplicial map
$\funct{\phi}{|K^\prime|}{|K|}$ given by   
$\phi(\vettore{x})=\vettore{p}(f({\vettore{p}^\prime}^{-1}(\vettore{x})))$.
Conversely, any geometric  simplicial map 
$\funct{\phi}{|K^\prime|}{|K|}$ defines an  abstract simplicial map
$\funct{f}{\AComp^\prime}{\AComp}$ given by   
$f(v)=\vettore{p}^{-1}(\phi({\vettore{p}^\prime}(v)))$.
With the above naming, the correspondence between  $f$ and $\phi$ is expressed
by the following commutative  diagram.
\[
\begin{CD}
\AComp @>f>> \AComp^\prime\\
@V\vettore{p}VV @VV{\vettore{p}^\prime}V \\
|K|@>\phi>>|K^\prime|
\end{CD}
\]
Note that $f$ and $\phi$ are well defined being
$\vettore{p}^{-1}$ and  ${\vettore{p}^\prime}^{-1}$ defined over
points (i.e. geometric $0$-simplices) of $K$ and $K^\prime$.
Very often the geometric simplicial map $\phi$ corresponding to
an abstract simplicial map $f$ is denoted by $|f|$.
It is easy to show (see \cite{EngSve92} Pg. 104) that for any abstract
simplicial map $f$ and $g$ we have that $|f|$ is continuos w.r.t.
the standard Euclidean topology over $\real^n$. Furthermore
we have that $|fg|=|f||g|$ and that $f$ is an isomorphism \iff\
$|f|$ is an homeomorphism (see \cite{Ago76} Pg. 48).
A more general approach, that do not assume $|K|$ to be  imbedded
into $\real^n$, is possible but is outside the scope of this work
(see \cite{Spa66} Pg. 110-114).

The notion of isomorphism is categorical and hence created and preserved
when passing from the category of \asc\ to the
equivalent  category of geometric simplicial complexes. 
The created isomorphism induce an isomorphism relation between pairs of
geometric simplicial complexes. This is usually called 
{\em geometric simplicial equivalence}.
The following definition gives the standard characterization for this
equivalence.

\begin{opt-definition}[Geometric Simplicial Equivalence (See \cite{Veg97} Pg. 520)\label{def:simpleq}]
Two geometric simplicial complexes $K$ and $L$  are \emad{simplicially} 
{equivalent} or \emad{simplicially}{isomorphic} \iff\ there are two geometric simplicial
maps $\funct{\phi}{|K|}{|L|}$ and $\funct{\psi}{|L|}{|K|}$ such
that $\phi(|K|)=|L|$ and $\psi(|L|)=|K|$.
In this case we will write $K\simpeq L$.
\end{opt-definition}
For sake of simplicity, here and in the following,
we omit the adjective {\em geometric}
and talk about \emas{simplicial}{equivalence} to mean
{\em geometric simplicial equivalence}.
On the other hand, the term {\em isomorphism}
will be reserved to the corresponding relation between abstract simplicial
complexes.

The equivalence between the category of \asc\ an the
category of geometric simplicial complexes implies that 
two abstract simplicial complexes are isomorphic \iff\ they
have simplicially equivalent geometric realizations. 
In particular, geometric simplicial complexes
that are geometric realizations of a complex $\AComp$ are all 
simplicially equivalent. Therefore all the polyhedra in
$\carrier{\AComp}$ must be homeomorphic.

Given two geometric realizations $K$ and $K^\prime$ 
of two abstract  simplicial complexes
$\AComp$ and $\AComp^\prime$, we have seen that there is a one to one
correspondence between abstract simplicial maps, between $\AComp$ and
$\AComp^\prime$, and geometric simplicial maps w.r.t. $K$ and 
$K^\prime$. We have already stressed that, in examples, we use geometric 
simplicial complexes (e.g. $K$ and $K^\prime$) and 
perspective  drawings of   polyhedra
in $\real^3$ (e.g. $|K|$ and $|K^\prime|$) as a handy presentation 
for the corresponding abstract simplicial complexes 
($\AComp$ and $\AComp^\prime$). 
Similarly, in examples, 
we will use geometric simplicial maps (e.g.$\funct{|g|}{|K^\prime|}{|K|}$) 
as a handy presentation for the associated abstract simplicial maps (i.e. $\funct{g}{\AComp^\prime}{\AComp}$). 
With this assumption, in examples, we will omit the distinction between 
$g$ and $|g|$ and talk freely  about the abstract simplicial map  $g$ with no
further reference to the  
geometric simplicial map $|g|$  actually presented in drawings.
\begin{figure}
\begin{center}
\begin{minipage}{0.69\textwidth}
\fbox{\psfig{file=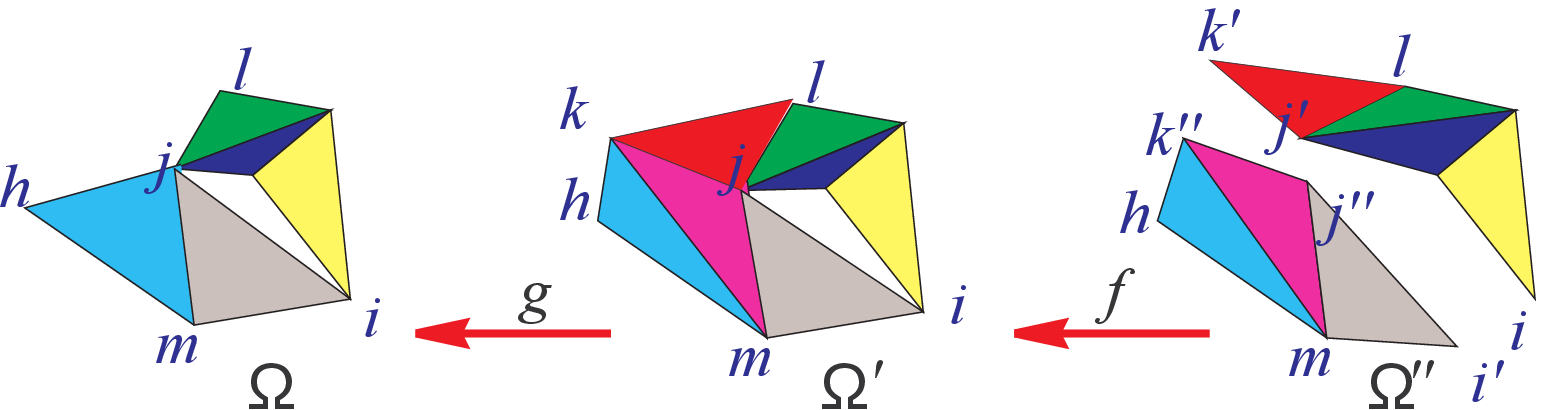,width=\textwidth}}
\begin{center}(a)\end{center}
\fbox{\psfig{file=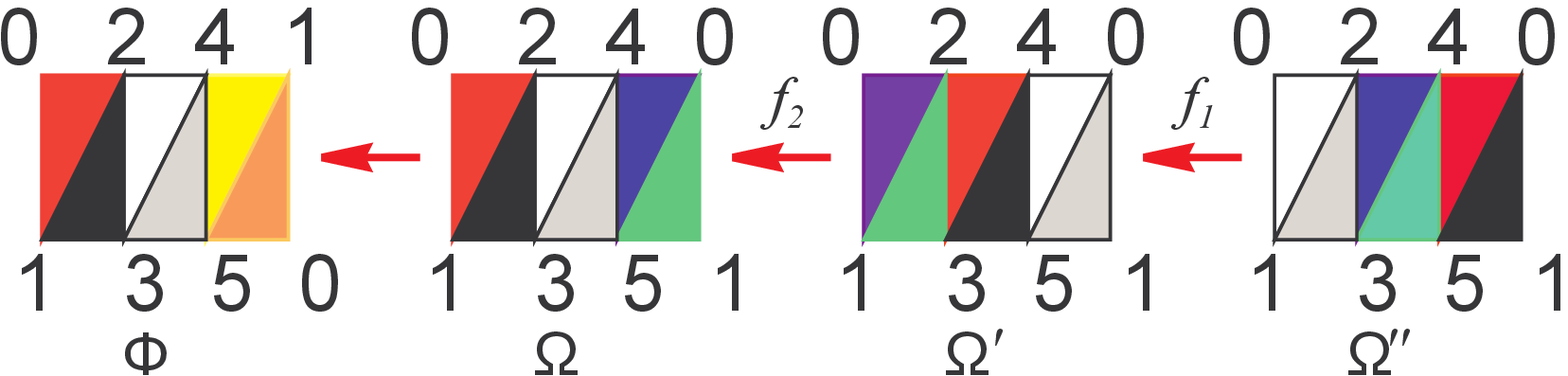,width=\textwidth}} \begin{center}(b)\end{center}
\end{minipage}
\hfill
\framebox{
\begin{minipage}{0.13\textwidth}
\psfig{file=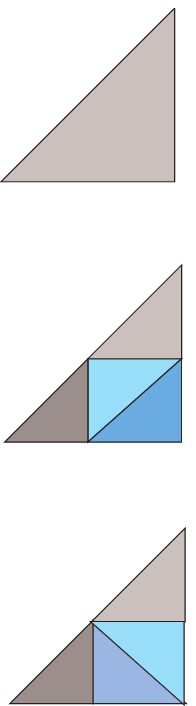,width=\textwidth} \begin{center}(c)\end{center}
\end{minipage}
}
\end{center}
\caption{
Examples of: 
abstract simplicial maps (a);
two isomorphisms:$f_1$ and $f_2$ and a complex $\Phi$ non isomorphic with $\AComp$ (b); three complexes that are homeomorphic but not isomorphic (c) (See Example  \ref{exa:mapsgen} )}
\label{fig:mapsgen}
\end{figure}
\begin{example}
\label{exa:mapsgen}
See now, for example, abstract simplicial maps
$f$ and $g$ (represented through geometric simplicial maps)
in Figure \ref{fig:mapsgen}a. 
We have that 
$f=[\send{k^\prime}{k},\send{k^\second}{k},
\send{j^\prime}{j},\send{j^\second}{j}]$,
and $g=[\send{k}{j}]$. Colors in figures denotes the mapping 
between triangles i.e., triangles $k^\prime j^\prime l$ in $\AComp^\second$ 
and $kjl$ in $\AComp^\prime$ receive the same color (red) 
because $f(k^\prime j^\prime l)=kjl$.
Similarly, there is no red triangle in $\AComp$ because
$g$ collapses triangle $kjl$ to edge $jl$ and $kjm$ to edge $jm$.
In Figure \ref{fig:mapsgen}b we have three examples of isomorphic
complexes $\AComp$, $\AComp^\prime$ and $\AComp^\second$.  
We can see that they are isomorphic by considering the fact
that they are linked by two simplicial maps: 
$\funct{f_1}{\AComp^\second}{\AComp^\prime}$ and 
$\funct{f_2}{\AComp^\prime}{\AComp}$ induced by
the vertex map $f(x)=x+4\bmod 6$. 
This vertex map $f$ is a bijection of the set of vertices $\{0,1,2,3,4,5\}$
onto itself.
It is easy to check that 
that $f$ extends to a bijection between simplices. 
This can be done by checking exhaustively that the six triangles in
$\AComp^\second$ (or in $\AComp^\prime$) are 
mapped by $f_1$ (or by $f_2$) into other six triangles in $\AComp^\prime$
(or in $\AComp$). In Figure \ref{fig:mapsgen}b
we colored triangles so that a triangle and its the image via $f_1$ (or $f_2$)
receive the same color. For instance  triangle $015$ in
$\AComp^\second$ and triangle $453$ in $\AComp^\prime$ are
both in plain black because $f_1(015)=453$. 
 
The leftmost complex $\Phi$ is an example of a complex that is not isomorphic
to  $\AComp$. 
In fact, any geometric realization of $\Phi$ must be homeomorphic 
to the Moebius strip and this cannot be homeomorphic to
a geometric realizations of
$\AComp$. Any geometric realization of $\AComp$ must be
homeomorphic to a plain
orientable strip i.e., a sphere with two holes.
The fact that $\AComp$ and $\Phi$ are not isomorphic can be verified 
by checking exhaustively that any vertex map associated 
to a permutation of $\{0,1,2,3,4,5\}$ cannot be extended to 
a simplex map.
If we try, for instance, to extend the identity on
$\{0,1,2,3,4,5\}$
from a vertex map to a simplex map we fail because
triangle $045$ is not mapped to any triangle
(i.e., $045$ does not exist in $\Phi$). 
Similarly, if we consider vertex map $[\send{2}{3},\send{3}{2}]$ 
we have that triangle $345$ is mapped to $245$ that does not exist
as a triangle in $\Phi$.  
An exhaustive analysis of the $6!$ permutations of
$\{0,1,2,3,4,5\}$ reveals that the two abstract simplicial complexes
$\Phi$ and $\AComp$ are not isomorphic.

However note that, in general,
geometric realizations of two non-isomorphic complexes can still be 
homeomorphic.
Let us consider, for instance, the 
two complexes in Figure \ref{fig:mapsgen}c. They have homeomorphic
polyhedra but they are not the same abstract complex. However we can
observe that they are related by some sort of {\em subdivision}
process. 
\end{example}
The remark at the end of the previous example
is the basis of the definition of a 
combinatorial  analogue of homeomorphism. 
We will discuss this subject in the next paragraph in order to give  
a combinatorial definition of manifolds and non-manifolds. 
The first step in this direction is the definition of 
{\em stellar equivalence}

\section{Stellar Equivalence\label{par:star}}
We like to start this section by pointing out that 
definitions and results outlined in this 
paragraph account for a theoretical development that started 
in the 1920's and 1930's by M.H. A. Newman \cite{New26} and 
J. W. Alexander \cite{Ale30}. 
The concept of {\em stellar equivalence}
brings into the combinatorial framework a concept somehow equivalent 
to that of homeomorphism. 
We define stellar equivalence here and 
discuss the relation between stellar equivalence and
homeomorphism.  
Next, we report limitations on decidability of  stellar equivalence.
In Section \ref{par:mani} we will  use  stellar equivalence to define 
{\em combinatorial manifolds} and {\em non-manifolds}. 

\begin{definition}[Starring (See \cite{Hud69} Pg. 8)]
Given a $d$-simplex $\gamma\in\AComp$ and a point $w$ s.t.  $\{w\}\notin\AComp$
we define the operation of \ems{starring} the simplex
$\gamma$ at the point $w$ as the operation that transforms
the complex $\AComp$ into the new complex $\AComp^\prime$
obtained from $\AComp$ with the following steps:
\begin{itemize}
\item delete the (open) star $\str{\gamma}$;
\item for each simplex in $\phi\in\str{\gamma}$ and
for  each vertex  $v\in\gamma$ add
the simplex $\{w\}+\phi-\{v\}$ plus all its faces.
\end{itemize}
In symbols we have that $\AComp^\prime=(\AComp-\str{\gamma})
\cup(\cup_{\phi\in\str{\gamma},v\in\gamma}\closure{\{w\}+\phi-\{v\}})$
\end{definition}
It is easy to prove that the added subcomplex is given by
$\{w\}\join\bnd{\gamma}\join\lk{\gamma}$ i.e.,
$\AComp^\prime=(\AComp-\str{\gamma})\cup\{w\}\join\bnd{\gamma}\join\lk{\gamma}$.
We will say that $\AComp^\prime$ is obtained from $\AComp$ by the
{\em elementary stellar subdivision} 
of the  simplex $\gamma$ at point $w$  and
we will use the notation
$\starring{\AComp}{(\gamma,w)}{\AComp^\prime}$ to denote this.
A \emas{stellar}{subdivision} 
of $\AComp$
is any abstract simplicial complex $\AComp^\prime$ obtained by a sequence
of elementary stellar subdivisions on $\AComp$.

Similarly when complex $\AComp^\prime$ is given and a vertex $w$ in
$\AComp^\prime$ is
such that there exist $\gamma$ and $\AComp$ for which
$\starring{\AComp}{(\gamma,w)}{\AComp^\prime}$ we will say that we can
\ems{weld} $\AComp^\prime$ at $w$ yielding $\gamma$. We will express this
with the notation $\starring{\AComp^\prime}{{(\gamma,w)}^{-1}}{\AComp}$.
We will say that $\AComp$ is obtained from $\AComp^\prime$ by the
\emas{stellar}{weld} at point $w$  yielding the  simplex $\gamma$.

The term \emas{starring}{operation} will be used to denote both
stellar weld and elementary stellar subdivision.
Two abstract simplicial complexes $\AComp$ and $\AComp^\prime$ will be
\emas{stellar}{equivalent} (denoted by $\AComp \streq \AComp^\prime$)
\iff\ there exists a finite sequence of
starring operations and simplicial isomorphisms that transforms $\AComp$
into $\AComp^\prime$.

A theorem by Neumann \cite{New31} proves that restricting starring
operations to 1-simplices we still obtain the stellar equivalence
defined above.

\begin{figure}
\begin{minipage}{0.49\textwidth} 
\fbox{ \psfig{file=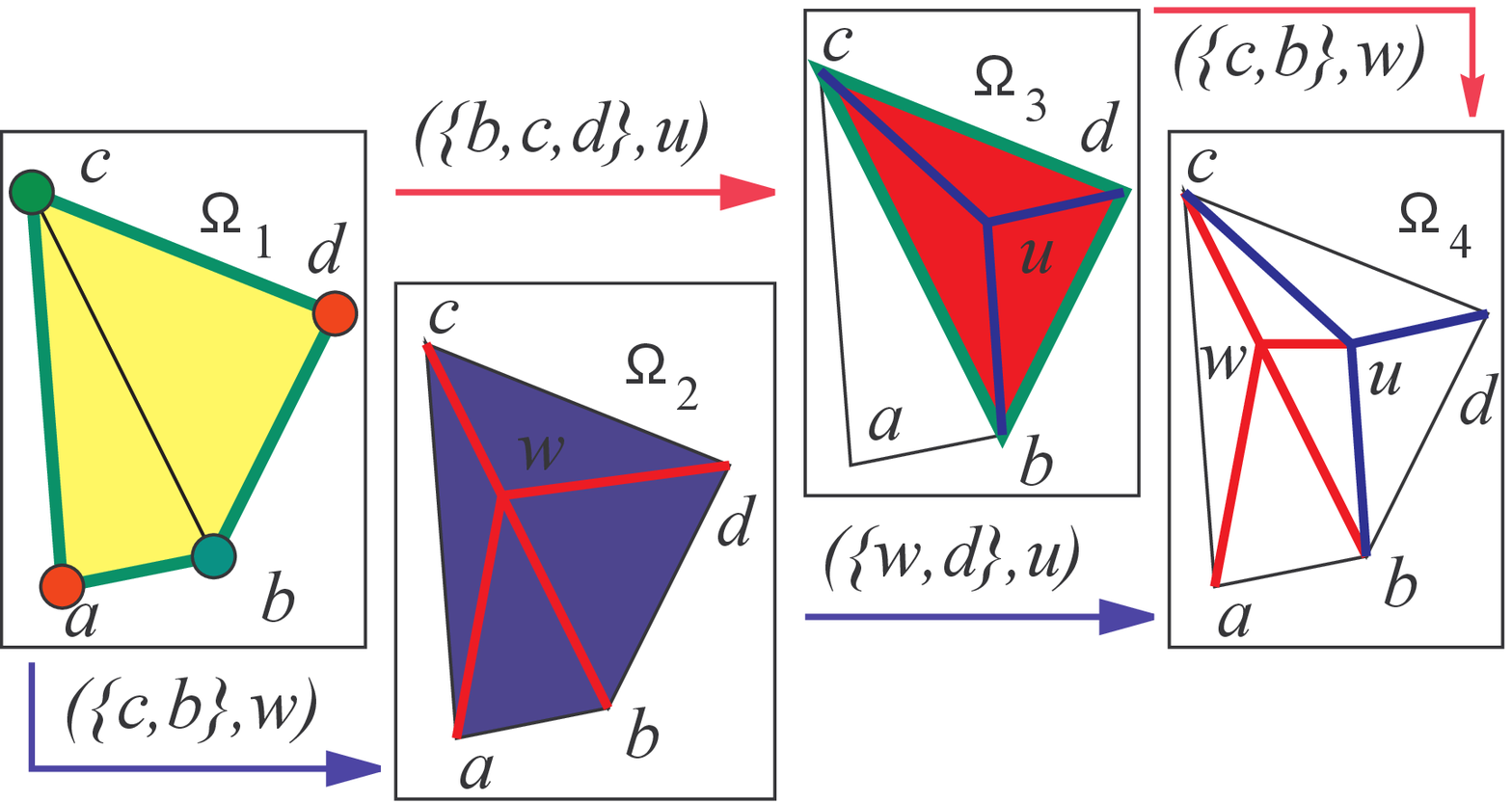,width=\textwidth} }
\begin{center}a\end{center}
\end{minipage}
\hfill
\begin{minipage}{0.47\textwidth} 
\fbox{ \psfig{file=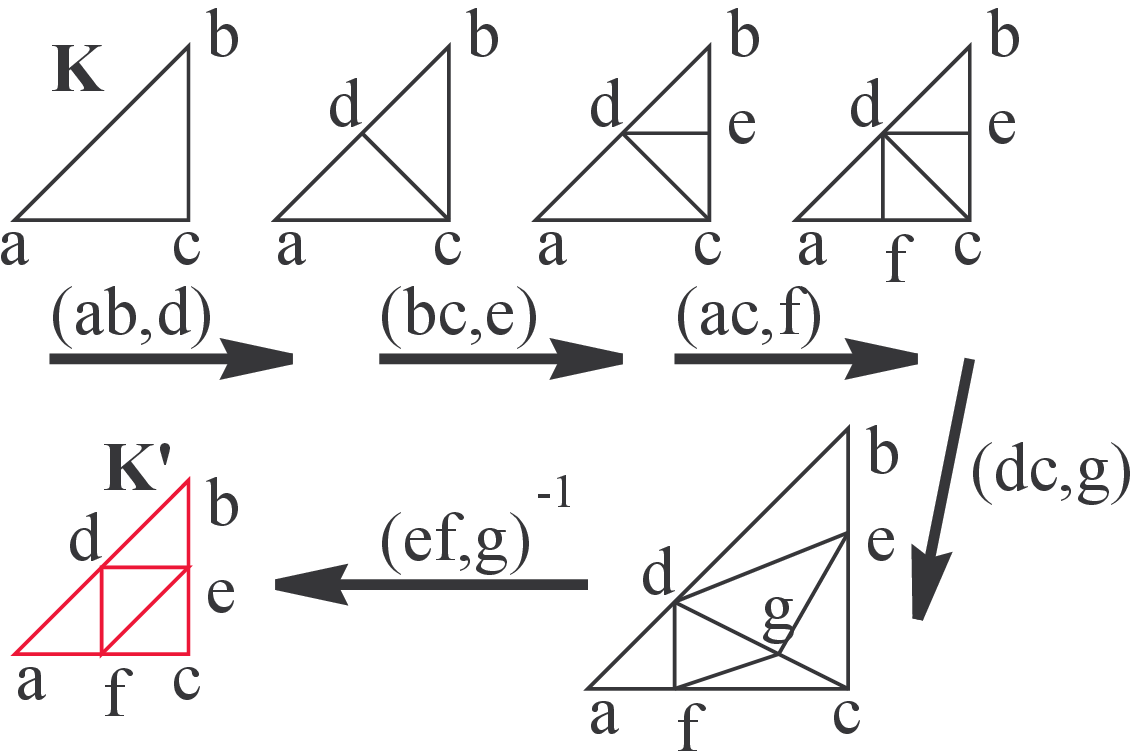,width=\textwidth} }
\begin{center}b\end{center}
\end{minipage}
\caption{Starring (a) and stellar equivalence (b) (See Example \ref{exapar:star} for (a) and \ref{exapar:star2} for (b) )}
\label{fig:starring}
\end{figure}
\begin{example}
\label{exapar:star}
As an example of application of this definition
consider, for instance, the abstract simplicial  complex $\AComp_1$ in Figure
\ref{fig:starring}a.
Complexes $\AComp_2$ and  $\AComp_3$ are stellar subdivisions of complex $\AComp_1$
obtained by stellar  subdivision of, respectively, edge $\{b,c\}$ at $w$
and of triangle $\{b,c,d\}$ at $u$ i.e., in symbols,
$\starring{\AComp_1}{(\{b,c\},w)}{\AComp_2}$  and
$\starring{\AComp_1}{(\{b,c,d\},u)}{\AComp_3}$.

Let us consider in detail the first stellar subdivision along the blue path
i.e., $\starring{\AComp_1}{(\{b,c\},w)}{\AComp_2}$.
To subdivide  simplex $\{b,c\}$ at $w$ we
first delete the yellow triangles  $\{a,b,c\}$, $\{b,c,d\}$ and edge $\{b,c\}$
that are the (open) star $\str{bc}$.
We leave in place the boundary of this star 
$\bnd{(\str{bc})}=\closure{\{ca,ab,bd,dc\}}$
(i.e. the thick green lines).
Then, according to definition we must add
$\{w\}\join\bnd{\{b,c\}}\join\lk{\{b,c\}}$.
Since  $\bnd{\{b,c\}}=\{\{b\},\{c\}\}$
(in green in Figure \ref{fig:starring}a) and
$\lk{\{b,c\}}=\{\{a\},\{d\}\}$ (in red in Figure \ref{fig:starring}a)
we add $\{w\}\join\{\{b\},\{c\}\}\join\{\{a\},\{d\}\}$.
The join $\{\{b\},\{c\}\}\join\{\{a\},\{d\}\}$ is given
by the four segments: 
$ca$,$ab$,$bd$,$dc$ and therefore we must add the cones from $w$ to these four segment.
This gives  the four blue triangles in  $\AComp_2$.
We must add these four triangles
together with their faces.
By taking faces
of the blue triangles we also add segments  $ca$, $ab$, $bd$ and $dc$.
Note that these segments were not deleted since they are not in the open
star of $bc$.

Now let us consider in detail the first stellar subdivision along the red path
$\starring{\AComp_1}{(\{b,c,d\},u)}{\AComp_3}$.
The starring subdivision of $\{b,c,d\}$  at $u$ is obtained
by deleting  $\{b,c,d\}=\str{\{b,c,d\}}$.
We leave in place the boundary of this star 
$\bnd{(\str{bcd})}=\closure{\{cb,bd,dc\}}$
(i.e. the thick green lines).
After this deletion we  add the cone from $u$
to the boundary of triangle $\{b,c,d\}$, being  $\lk{\{b,c,d\}}=\emptyset$.
Since $\bnd{\{b,c,d\}}$ is given by the three segments $bc$, $cd$ and $db$
the simplices to be added are obtained
by taking the cone from $u$ to these three segments. In
this way we obtain the three
red triangles of  $\AComp_3$ in Figure \ref{fig:starring}a.
These triangles are added together with their faces.
However note that segments $bd$, $dc$ and $cb$, that must be added now, 
were not deleted removing the open star of
triangle $\{b,c,d\}$.

Let us see a very simple example of a stellar equivalence.
We want to show that $\AComp_2$ and $\AComp_3$ are stellar
equivalent (i.e. . $\AComp_2\streq \AComp_3$).
We follow blue and red path to $\AComp_4$.
The blue path is completed  by the  stellar subdivision of edge $\{w,d\}$ at $u$
in $\AComp_2$. The red path is completed  by the  stellar subdivision 
of edge $\{b,c\}$ at $w$ in $\AComp_3$.
Both paths join at $\AComp_4$.  In symbols we have that:
$\starring{\AComp_2}{(\{w,d\},u)}{\AComp_4}$  and
$\starring{\AComp_3}{(\{b,c\},w)}{\AComp_4}$.
The way in which we obtained complex $\AComp_4$ proves that complexes
$\AComp_2$ and $\AComp_3$ are stellar
equivalent. 

Note that if, by any chance, we perform these two last stellar subdivisions
by introducing two new Vertices $u^\prime$
and $w^\prime$, instead of $u$ and $w$ (i.e. if we perform the 
stellar subdivisions given by 
$\starring{\AComp_2}{(\{w,d\},u^\prime)}{\AComp_4^\prime}$  and
$\starring{\AComp_3}{(\{b,c\},w^\prime)}{\AComp_4^\second}$),
we still obtain
stellar equivalence between $\AComp_2$ and $\AComp_3$
since the resulting complexes $\AComp_4^\prime$ and $\AComp_4^\second$,
although distinct, are isomorphic.
In fact the renaming of Vertices $[\send{u^\prime}{u}]$ and
$[\send{w^\prime}{w}]$ defines two isomorphisms that sends,
respectively  ${\AComp_4^\prime}$ and ${\AComp_4^\second}$ into $\AComp_4$.
\end{example}

Next we recall here some basic results on stellar equivalence
for future reference.
A first property of stellar equivalence is that
joining stellar equivalent complexes we obtain
stellar equivalent complexes.
\begin{property}[See \cite{Lic99} Pg. 303]
Let us consider two pairs,  $\AComp_1$, $\AComp_1^\prime$
and  $\AComp_2$, $\AComp_2^\prime$,
of  abstract simplicial complexes 
such that  $\AComp_1\streq\AComp_1^\prime$ and
$\AComp_2\streq\AComp_2^\prime$ and such that  
$\AComp_1\join\AComp_2$ and $\AComp_1^\prime\join\AComp_2^\prime$
are well defined complexes,
then we have
$\AComp_1\join\AComp_2\streq\AComp_1^\prime\join\AComp_2^\prime$
\end{property}
In particular the cone from a vertex  to two stellar equivalent complexes
gives stellar equivalent complexes.

The next step in this overview  will be the presentation of an
equivalence between  geometric simplicial complexes that
mimics stellar equivalence between abstract simplicial complexes.  
Such an equivalence, called {\em piece-wise linear} equivalence,
will be used in Section \ref{par:manidef} to give the (classical)
definition of {\em combinatorial} manifolds.
To develop this notion we first need to introduce the notion
of {\em subdivision} relation between geometric simplicial complexes.
\begin{opt-definition}[Subdivision (\cite{Gla70} Pg. 7)]
A geometric simplicial complex $K^\prime$ is a  \emd{subdivision} of another
geometric simplicial complex $K$ \iff\ $|K|=|K^\prime|$ and for  every 
simplex $\sigma^\prime\in K^\prime$ there exist a simplex 
$\sigma\in K$ such that $\sigma^\prime\subset\sigma$.
\end{opt-definition}
Note that, in the condition $\sigma^\prime\subset\sigma$, the
two simplices $\sigma$ and $\sigma^\prime$ are {\em geometric} simplices,
i.e. subsets of $\real^n$.
\begin{short-example}
\label{exapar:star2}
It is easy to show that if $\AComp^\prime$ is a stellar subdivision
of $\AComp$ then there exist two geometric simplicial complexes $K$ and 
$K^\prime$ that are geometric realization of, respectively, $\AComp$ and
$\AComp^\prime$ such that $K^\prime$ is a subdivision of $K$.
The converse is not always true.
In fact not all subdivisions are stellar subdivisions.
This is easy to see considering the two geometric complexes $K$
and $K^\prime$ of Figure 
\ref{fig:starring}b. The geometric complex 
$K^\prime$ (in red)
is a subdivision of the complex $K$ (i.e. the single  triangle $abc$).
The geometric complex $K^\prime$ can not be the geometric
realization of a stellar subdivision of $abc$. 
We can prove this by contradiction.
Let assume that such a stellar subdivision exist.
First we note that every stellar subdivision
introduce a new vertex. 
Therefore just three elementary
stellar subdivision must produce the red complex $K^\prime$ out of $abc$ 
introducing $d$, $e$ and $f$.
If we consider all the possible moves it is easy to see that
every elementary stellar subdivision
that introduces a vertex on a boundary edge creates 
a vertex of order $3$. 
See for instance the moves in Figure 
\ref{fig:starring}b.
Therefore, the last elementary subdivision of this possible stellar subdivision from $K$ to $K^\prime$ must introduce 
a vertex of order $3$.
The contradiction comes from the fact that all added
Vertices, i.e. $d$, $e$ and $f$ are of order $4$.
Therefore $K^\prime$ is not a stellar {\em subdivision} of
$K$.
\end{short-example}

The notion of subdivision seems to be more general than the notion 
of stellar subdivision. However 
stellar subdivision admits a combinatorial definition while generic
subdivisions do not.
Indeed, we can define stellar subdivisions for an abstract simplcial complex
$\AComp$ with no reference to its geometric (polyhedral)  
realization. Stellar subdivisions are not the only possible option for
a combinatorial notion of subdivision.
Another purely combinatorial  definition of  a subdivision is the so
called {\em barycentric subdivision}. Barycentric 
subdivisions can be defined directly on the   
abstract simplicial complex with no reference to a geometric
realization. We do not report  this notion here
(see \cite{Gla70} Pg, 7) simply because it can be proved
that any barycentric subdivision is a stellar subdivision while 
the converse it is trivially false.
In general the problem of whether stellar subdivision can mimics
completely generic subdivisions can be presented as the claim:
\begin{opt-conjecture}
Let $\AComp$ and $\AComp^\prime$ be two abstract simplicial complexes
and let $K$ and $K^\prime$ be two particular geometric realization 
for $\AComp$ and $\AComp^\prime$, respectively. If $|K|=|K^\prime|$
then there exist a common stellar subdivision
for $\AComp$ and $\AComp^\prime$
\end{opt-conjecture}
The above claim is a classic conjecture (see \cite{Hud69} Pg. 14)
that is still unsolved.
A purely combinatorial formulation  of (a slighty more general version
of) the above claim,  presented in  \cite{Lic99} (see \cite{Lic99} Pg. 311),
is the following:
\begin{opt-conjecture}
Let $\AComp$ and $\AComp^\prime$ be two abstract simplicial complexes
If $\AComp$ and $\AComp^\prime$ are stellar equivalent
then there exist a common (up to isomorphism) stellar {\em subdivision}
for $\AComp$ and $\AComp^\prime$.
\end{opt-conjecture}
Since the real power of stellar subdivisions is still an open problem, 
in the following, we must
consider the full stellar equivalence i.e. the equivalence
generated by
both subdivisions and welds.
Stellar equivalence is surely more powerful than equivalence 
based on stellar subdivisions.
In the following, we first give an example of stellar equivalence and 
then consider the relation between
stellar equivalence and another equivalence
based on subdivisons (called  {\em piecewise linear equivalence}).
\begin{short-example}
The sequence of starring operations
in Figure \ref{fig:starring}b shows that $K^\prime$ can be the 
geometric realization of a complex that is stellar equivalent to
$K$. This equivalence is established by four subdivisions and a final weld.
\end{short-example} 
Indeed, as we will see in the following, stellar equivalence and equivalence
based on subdivisions are actually the same equivalence. 
In order  to see this, we must carefully define equivalence based on
subdivisions. This will bring in the ideas of {\em piecewise linear map}
and {\em piecewise linear equivalence}.
These  notions will guide us  to the definition of an equivalence
that is the best combinatorial analogue of homeomorphism for polyhedra.

\begin{opt-definition}[Piecewise Linear Simplicial Map (See \cite{Gla70} Pg. 13)]
A continuous  map $\funct{\phi}{|K|}{|L|}$ 
is a \emad{piecewise linear (p.l.)}{simplicial map} from  the geometric 
simplicial complex $K$ to the geometric simplicial complex $L$ \iff\ 
$K$ and $L$ have two subdivisions $K^\prime$, $L^\prime$ such that $\phi$ 
is a geometric
simplicial map from $K^\prime$ to $L^\prime$.
\end{opt-definition}
In the situation of the above definition, if $K^\prime$ and $L^\prime$ are
simplicially equivalent, then the p.l. simplicial map $\phi$ is a homeomorphism
called a \emas{piecewise linear}{homeomorphism}.
Directly from the definition, it is easy to see that
all simplicial maps are p.l. maps and all
simplicial homeomorphisms are p.l. homeomorphisms.
Geometric simplicial complexes  with p.l. simplicial maps constitutes a 
category (see \cite{Gla70} Theorem I.6). In particular, in this category,
we can define an equivalence between geometric simplicial complexes.
This will result in the notion of PL-equivalence.

\begin{opt-definition}[Piecewise Linear Equivalence (PL-equivalence \cite{Gla70} Pg.13)]
Two geometric simplicial complexes $K$ and $L$ are PL-equivalent \iff\ 
they have two simplicially equivalent subdivisions.
In this case we will write $K\pwleq L$.
\end{opt-definition}
Directly from the definition above, we have that two geometric simplicial complexes
that are simplicially equivalent must be PL-equivalent.
Therefore, all geometric simplicial complexes in $\carrier{\AComp}$ are
PL-equivalent.
We will say that two abstract simplicial  complexes are PL-equivalent
\iff\ their carriers contains PL-equivalent polyhedra. 

Therefore, we have that PL-equivalence induce an equivalence upon 
abstract simplicial complexes. This equivalence, although defined using
(geometric) subdivisions, must admit a combinatorial
definition. This definition is actually given by the notion of 
stellar equivalence. This is stated by the following theorem that is 
a central result in combinatorial topology.
\begin{theorem}[Equivalence of Stellar and PL-theory (See \cite{Gla70} Pg. 41 or \cite{Lic99} Pg. 311)]
\label{theo:starpwl}
Let $\AComp$ amd $\AComp^\prime$ be two abstract simplicial complexes
with geometric realization $K$ and $K^\prime$. We have that {$\AComp\streq \AComp^\prime$} \iff\ $K\pwleq K^\prime$.
\end{theorem}
On the ground of the previous result,
in the following, we will use the term \emas{combinatorial}{equivalence}
to denote both PL-equivalence and stellar equivalence.
Combinatorial equivalence implies homeomorphism between 
polyhedra that are geometric realizations of combinatorially equivalent
complexes. Therefore we have that if $\AComp\streq\AComp^\prime$
then $\carrier{\AComp}$ and $\carrier{\AComp^\prime}$ must contain
homeomorphic polyhedra.
The converse is the well known \ems{Hauptvermutung}, 
which states that two
abstract simplicial complexes that have  homeomorphic geometric
realization are combinatorially equivalent.
This property is known to be true for $d$-complexes for $d\le 2$ 
\cite{Pap43}
and for $d=3$  \cite{Moi52} and it is known to be false in general.
In fact, Milnor \cite{Mil61} provided the first example of a pair of
non-manifold $7$-complexes
that have homeomorphic geometric realizations, and yet
are not combinatorially  equivalent.

Combinatorial equivalence is not equivalent to homeomorphism 
for $d$-complexes for $d\ge 7$. However, already for $d\ge 4$, 
combinatorial equivalence is only semi-decidable. This is a consequence
of a result by Markov \cite{Mar58}. This result states that there 
exist a $4$-complex $\AComp_0$ such that it is impossible to decide, 
for any other complex $\AComp$, 
if $\AComp_0\streq\AComp$.

We end this section by quoting
a result that we report for forthcoming reference.
\begin{property}
Let $\AComp$ and $\AComp^\prime$ be two stellar
equivalent abstract simplicial complexes 
(i.e. {$\AComp\streq \AComp^\prime$}). 
Let $w$ be a vertex that remains unchanged across the starring operations
that bring $\AComp$ to $\AComp^\prime$, 
then the links {$\xlk{\AComp}{w}$}  and 
{$\xlk{\AComp^\prime}{w}$} are stellar equivalent.
\begin{proof}
See Corollary 1.15 in  \cite{Hud69} Pg. 23 and Theorem \ref{theo:starpwl}.
\end{proof}
\end{property}

\section{\label{par:manidef}Manifoldness}
Having defined  combinatorial equivalence, we can step into the 
definition of {\em combinatorial manifolds}. Before doing that, we define a
superclass of manifolds called {\em pseudomanifolds} and 
introduce the first notion of {\em non-manifoldness}.  
Next we will introduce {\em topological} and {\em combinatorial} 
manifolds and discuss their relation between them.

\subsection{Pseudomanifolds}
Let $\gamma$ be a $(d-1)$-simplex in a $d$-complex $\AComp$.
We say that $\gamma$ is a \emas{manifold}{$(d-1)$-simplex}
\iff\ the closed star of $\gamma$ is a regular $d$-complex
containing just one or at most two 
$d$-simplices incident at $\gamma$.

The two  $d$-simplices $\gamma$ and $\gamma^{\prime}$ are said to be
\emas{manifold}{connected}, in a abstract simplicial complex $\AComp$, 
\iff\ there exist a sequence of $d$-simplices (called a \ems{manifold path})
$(\gamma_i)_{i=0}^k$ with $\gamma=\gamma_0$, $\gamma^\prime=\gamma^k$
and such that two successive complexes
$\gamma_{i-1}$, $\gamma_{i}$, $i=1,\ldots, k$ are adjiacent via a manifold
$(d-1)$-simplex
(i.e. $\gamma_{i-1}\cap\gamma_{i}$ is a manifold $(d-1)$-simplex in $\AComp$).
Note that we do not ask $\AComp$ to be regular.
Proceeding as with $h$-connectivity we can define 
{\em manifold-connectivity} and 
{\em manifold-connected} components. 

Having defined manifold $(d-1)$-simplices we can step into the definition
of {\em pseudomanifolds}.
\begin{definition}[Pseudomanifold  see \cite{Spa66} Pg. 150]
A regular $(d-1)$-connected $d$-complex where all 
$(d-1)$-simplices are manifold is called a
\emad{combinatorial}{pseudomanifold} (possibly with boundary).
\end{definition}

\begin{figure}
\begin{minipage}{0.49\textwidth}
\fbox{\psfig{file=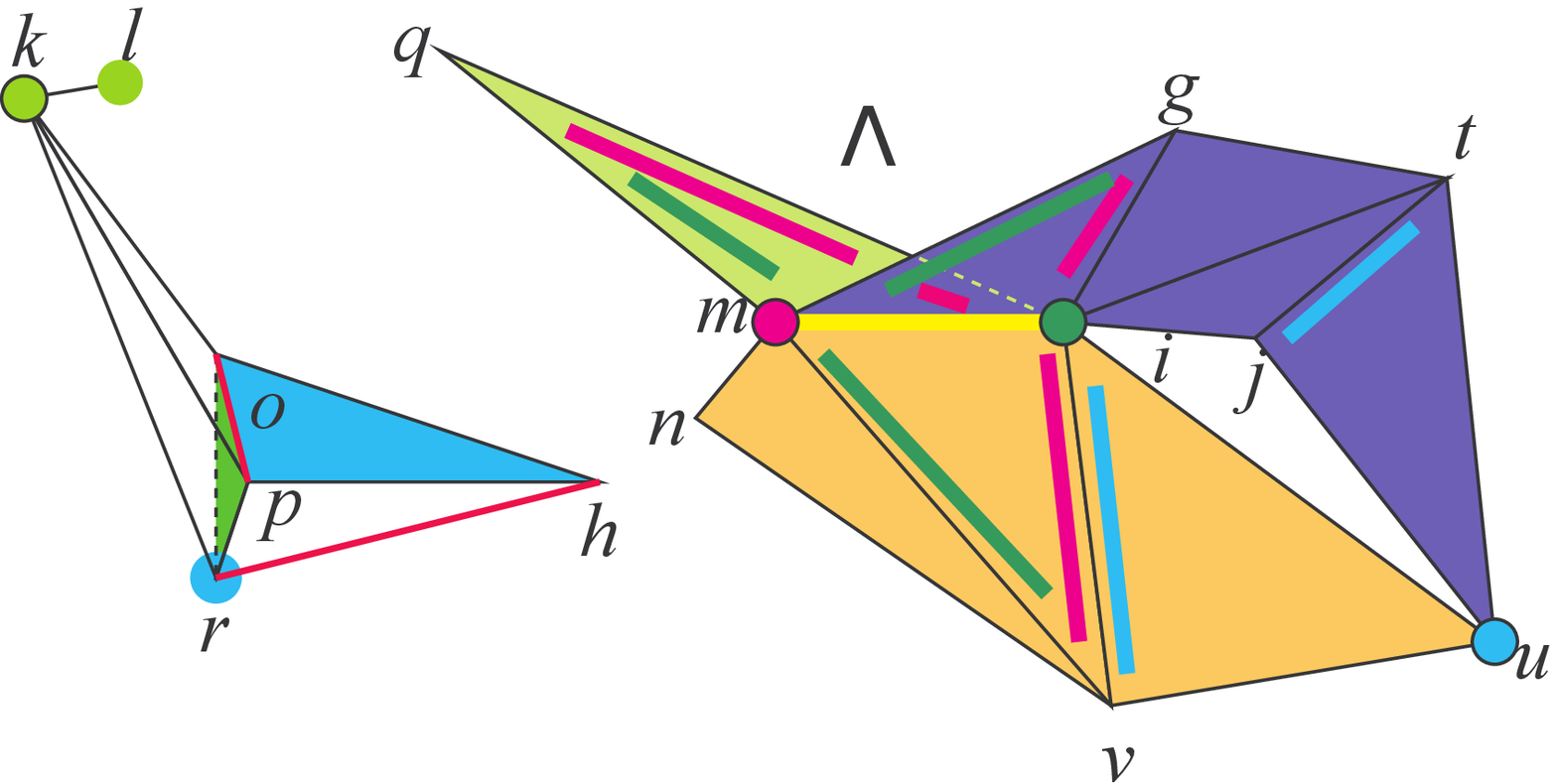,width=\textwidth}}
\begin{center}(a)\end{center}
\end{minipage}
\hfill
\begin{minipage}{0.49\textwidth}
\fbox{\psfig{file=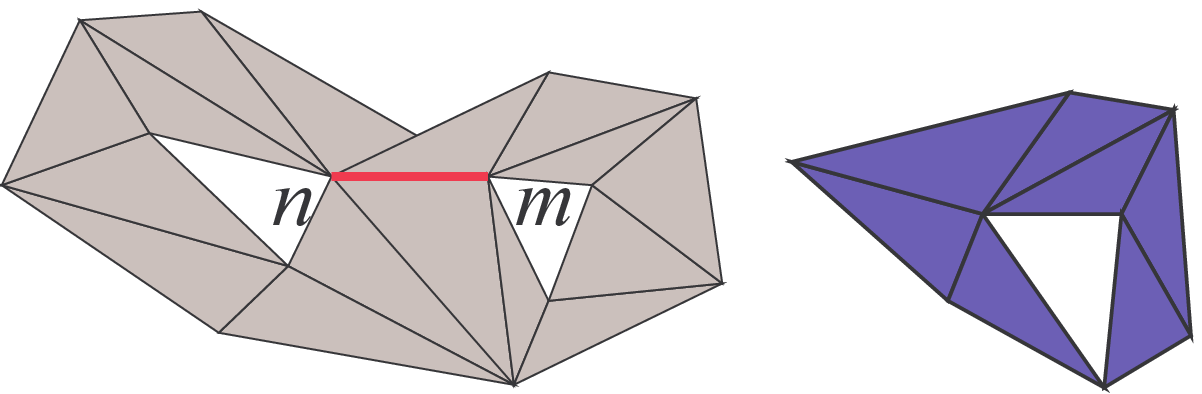,width=\textwidth}}
\begin{center}(b)\end{center}
\end{minipage}
\caption{Manifold and non-manifold simplices in a $2$-complex 
(See Examples \ref{expar:manidef}) and \ref{expar:reg})
.}
\label{fig:es3complfh}
\end{figure}

\begin{short-example}
\label{expar:manidef}
Consider, for instance, the subcomplex $\Lambda$ in Figure 
\ref{fig:es3complfh}a.
All $1$-simplices (segments) in the regular $2$-complex
$\Lambda$ are manifold $1$-simplices but the
segment $mi$ in yellow. This is not a manifold $1$-simplex.
The complex $\Lambda$ is neither a pseudomanifold nor a manifold-connected
component. It is easy to see that in $\Lambda$ we have three manifold
connected components these are respectively the triangle $qmi$,
the three orange triangles (i.e. $nmv$,$miv$ and $viu$) and 
the four violet triangles (i.e. $mig$, $git$, $itj$ and $tju$).

The gray subcomplex in Figure \ref{fig:es3complfh}b is an example of a 
manifold connected complex that is not a pseudomanifold since
the red segment $nm$ is not a manifold $1$-simplex. Finally  the
blue complex in Figure  \ref{fig:es3complfh}b is an example of
a $2$-pseudomanifold.
\end{short-example}

\subsection{Topological and Combinatorial Manifolds\label{par:mani}}
In this paragraph we will introduce both \emas{topological}{manifolds} 
and \ems{combinatorial manifolds} and discuss the distinction between
these two definitions.

The definition of {\em combinatorial manifolds}, being
based on stellar equivalence, is a
truly combinatorial  definition. 
By a {\em truly combinatorial}
definition we mean a definition
such that there exist a semi-decision algorithm that 
stops when it has realized that
the input encodes a combinatorial manifold.

We note that combinatorial manifolds are the only 
known analogue of topological manifolds that admit 
a semi-decidable definition.
This fact, although foundational for geometric modeling and computer
graphics, receives little attention 
in handbooks of combinatorial topology (with the notable exception of 
\cite{Gla70}). This sort of forgetfulness
is clearly  pointed out in the survey in \cite{Lic99} 
that has been  a precious help for writing this short introduction to this subject.

Note that combinatorial manifoldness is just semi-decidable and
is not decidable.
In fact there is no  {\em decision procedure} that stops
and says if the input was a combinatorial $d$-manifold or not.
Such an algorithm is impossible to build for $d\ge 6$. 
This is an easy consequence of a result by S.Novikov  
\cite{Vol74}.

In the  standard Euclidean topological space $\real^d$ 
we define the {\em standard closed unit $d$-\ems{ball}} as the set 
${\cal B}^d=\{\vettore{x}\in\real^d|\|\vettore{x}\|\le 1\}$ 
where $\|\vettore{x}\|$ is
the standard Euclidean metric over the Cartesian product $\real^d$.
The standard  closed unit $d$-ball in $\real^d$
will be denoted by ${\cal B}^d$.

With this definition we can introduce topological manifolds:
\begin{opt-definition}[Topological $d$-manifold \cite{Gla70}]
A topological connected $d$-manifold 
is a  topological connected metric space 
where each point has a  neighborhood homeomorphic 
to $\real^d$ or to the closed unit ball
${\cal B}^d$ in $\real^d$.
\end{opt-definition}
An alternative equivalent definition is the following: 
\begin{opt-definition}[Topological $d$-manifold \cite{Veg97} Pag. 518]
A topological connected $d$-manifold 
is a  topological connected metric space 
where each point has a  neighborhood homeomorphic 
to the standard open unit $d$-ball {$\stackrel{\circ}{{\cal B}^d}$}
or to the closed unit half $d$-ball ${\cal B}^d_{+}$.
\end{opt-definition}
{
We denote with {$\stackrel{\circ}{{\cal B}^d}$}
the {\em standard open unit $d$-\ems{ball}} i.e. the set 
$\stackrel{\circ}{{\cal B}^d}=
\{\vettore{x}\in\real^d|\,\|\vettore{x}\|<1\}$ 
Similarly we denote with ${\cal B}^d_{+}$
the {\em standard closed unit half $d$-\ems{ball}} i.e. the set 
${\cal B}^d_{+}=\{\vettore{x}\in\real^d|\,\|\vettore{x}\|<1\,\,\mbox{\rm and}\,\,
x_i\ge 0\,\,\mbox{for}\,\,i=1,\ldots,d\}$ 
}

A topological $d$-manifold that is a geometric polyhedron is usually 
called a  \emas{triangulable}{manifold}.
From this definition, we have that a topological $d$-manifold is triangulable
\iff\ it is the carrier of some abstract simplicial complex.
Note that there are examples \cite{AkbCar90}
of topological $4$-manifolds that are not triangulable.
It is unknown if there exist non triangulable  topological 
$d$-manifolds for $d>4$.

In the following we will use the term  \emas{standard}{$d$-simplex} to
denote the abstract simplicial complex $\Delta^d$ obtained as the set of
parts of a set of
$d+1$ vertices.
It is easy to see that the complex  $\Delta^d$ is defined up to isomorphism.

A \emas{combinatorial}{$d$-ball} is any abstract simplicial complex $B^d$
that is combinatorially equivalent to  $\Delta^d$.
A \emas{combinatorial}{$d$-sphere} is any simplicial complex $S^d$
that is combinatorially equivalent to  $\bnd{\Delta^{d+1}}$.
Note that a combinatorial $0$-ball is any singleton $\{\{v\}\}$ for any
vertex $v$, while a combinatorial $0$-sphere, being
combinatorially equivalent to  $\bnd{\Delta^{1}}$ is any couple
$\{\{v\},\{w\}\}$ for any pair of distinct vertices  $v$  and $w$.

\begin{definition}[Combinatorial Manifold (see \cite{Gla70} Pg. 19 )]
\label{def:combman}
Let $v$ be a vertex in a regular $d$-complex $\AComp$.
We say that $v$ is a \emad{manifold}{vertex} \iff\ its link $\lk{v}$ is a 
$(d-1)$-complex that is combinatorially equivalent either to:
\begin{enumerate}
\item
the boundary of a $d$-dimensional simplex (i.e., a $(d-1)$-sphere) 
if $v$ is an internal vertex, or to
\item
a $(d-1)$-simplex (i.e., a $(d-1)$-ball) if $v$ is a boundary vertex.
\end{enumerate}
Otherwise, we say that $v$ is a \emad{non-manifold}{vertex} in $\AComp$.

A regular $d$-complex where all vertices are manifold
is called a \emad{combinatorial}{$d$-manifold} (possibly with boundary).
\end{definition}
It is easy to prove that a combinatorial $d$-manifold is also a 
combinatorial $d$-pseudomanifold.
All combinatorial $d$-manifolds are topological $d$-manifolds.
More precisely,
it is easy to see that the polyhedron associated to a geometric realization
of a combinatorial $d$-manifold is a topological $d$-manifold
(Remark Pg. 26 in \cite{Hud69}).
Therefore, all polyhedra in the carrier of a combinatorial $d$-manifold
are topological manifolds.

However there are topological manifolds that are not contained in the 
carrier of a combinatorial manifold. 
In fact, not all topological $d$-manifolds are triangulable as a combinatorial
manifold.
It has been proved that all topological $1$-manifolds and 
$2$-manifolds are triangulable as
a combinatorial manifold
\cite{Rad26}. Also $3$-manifolds are known to be triangulable as 
combinatorial $3$-manifolds \cite{Moi52}.
Freedman (see \cite{FreeLuo89} \cite{FreeQui90})
constructed an example of a topological
$4$-manifold that cannot be triangulated as a combinatorial manifold.

We have seen in Section \ref{par:star} that combinatorial equivalence
is less powerful than topological equivalence (i.e. homeomorphism).
We recall that Milnor in \cite{Mil61} disproved the Hauptvermutung, 
by  providing the first counterexample in this sense.
Milnor counterexample is based on the
construction of  a pair of
$7$-complexes that have homeomorphic geometric realizations and yet
that are not combinatorially  equivalent.
Since these two complexes were non-manifold it is reasonable to ask
if the Hauptvermutung  can still hold within the class of  combinatorial
manifolds. This claim was called the \emas{manifold}{Hauptvermutung}.

The manifold Hauptvermutung
was solved negatively in the late 1960's 
\cite{KirSie69} \cite{KirSie77}.
In fact, for instance from the construction in \cite{Can79}, for any $d\ge 5$, 
one can build an abstract simplicial 
$d$-complex that is not a combinatorial $d$-manifold
and whose carrier is homeomorphic to the standard $d$-sphere.
Since the standard $d$-sphere can always be triangulated as
a combinatorial $d$-manifold this provides a counterexample
to the combinatorial Hauptvermutung, already for $d>5$, 
in the tiny realm of the triangulations of the $d$-sphere. 

Hence, topological manifolds are a proper super-class of combinatorial
manifolds and homeomorphism is stronger than combinatorial equivalence
even within the subclass of combinatorial manifolds.
Furthermore, combinatorial equivalence is just semi-decidable and not decidable
even within the subclass of combinatorial manifolds.
In fact, the recalled  result by Markov \cite{Mar58}, already 
in its original form, applies to $4$-manifolds. We recall that 
Markov result  states  that there exist a {\em combinatorial  $4$-manifold}
$\AComp_0$ such that it is impossible to decide, 
for any other {\em combinatorial manifold} $\AComp$, 
if $\AComp_0\streq\AComp$.

Another deep theoretical result, by S. Novikov \cite{Vol74}, shows that 
the problem of recognizing the $d$-sphere for $d\ge 5$ implies the problem
of recognizing a trivial group in a finite sequence of finitely
generated groups. It is known \cite{Adj57} that this problem, in turn,
implies  the Halting Problem 
(see \cite{Nab96} Pg. 1-4 for an introduction on the relation between
these problems).
In \cite{Nab96} the relation between the recognizability of the 
$d$-sphere and  the halting problem is explicitly stated.
Thus we report here this  latter''effective'' version of the 
Novikov result:
\begin{theorem}[Theorem 2.1 in \cite{Nab96}] 
\label{teo:halting}
There exist an algorithm which, for  any $d\ge 5$, any combinatorial
manifold  $M$ of dimension $d$, any given Turing machine $T$ and its 
input $w$ constructs  another combinatorial manifold  $R_T(w)$ such that
$M$ and $R_T(v)$ are combinatorially equivalent to \iff\ T halts on $w$. 
\end{theorem}
\begin{proof}Theorem \ref{teo:halting} is essentially an excerpt from
Theorem 2.1 in \cite{Nab96}.  restated using the 
the notations in this work.
\end{proof}
If we take $M=S^d$ and assume that we can recognize the $d$-sphere
then, by Theorem \ref{teo:halting}, we can decide the Halting Problem.

It is well known \cite{Nab96} that the  existence of an algorithm 
recognizing whether or not a $(d+1)$-complex is a combinatorial 
$(d+1)$-manifold 
is equivalent to the recognizability of the $d$-sphere. 
So, this decision problem is known to be
solvable for $d=1,2,3$ \cite{Tho94},
unsolvable for $d\ge 5$ and open for $d=4$.

\begin{example}
As an example of application of the above definitions consider, for instance, 
the complex of Figure \ref{fig:es3complfh}a.
According to our  definition, Vertices $u$,
$m$ and $i$ are non-manifold Vertices.
In fact, they are all boundary Vertices and none of the links
$\lk{u}$ (segments in pale blue),
$\lk{m}$ (three segments in violet), and
$\lk{i}$ (three segments in green) is combinatorially equivalent to  a  $1$-simplex.
This can be easily proven by observing that combinatorially
equivalent complexes must have homeomorphic geometric realizations and
neither the violet, nor the green nor, the pale blue $1$-complexes are homeomorphic
to a segment. 
In fact, the pale blue link is disconnected while, in the violet link,
the vertex $i$ is incident to three $1$-simplices. The same problem
exists at vertex $m$ in the green link.
\label{expar:reg}
\end{example}
In the following we will restrict our attention to combinatorial
manifolds. Therefore, in the following, very often, we will omit the term 
combinatorial and 
talk about manifolds and manifoldness to mean {\em combinatorial}
manifoldness.

\subsection{Non-manifold simplices}
We have already defined non-manifold $(d-1)$-simplices and non-manifold
vertices in a regular complex.
We now extend  this definition to any  $s$-simplices for any  $0\le s<d$ in an
arbitrary (possibly non regular) $d$-complex.
\begin{definition}
\label{def:simplnonmani}
Let $\gamma$ be a $s$-simplex in a $d$-complex $\AComp$ with $0\le s\le (d-1)$
we will say that $\gamma$ is a \emas{manifold}{$s$-simplex}  
\iff\ $\lk{\gamma}$ is
a regular $h$-complex that is
combinatorially equivalent either to
the $h$-sphere or to a $h$-ball, for some  $h\le d-s-1$
\end{definition}
Note that the above definition do not require $\AComp$ 
to be a regular complex.
Indeed, if $\AComp$ is  a regular $d$-complex we have $h=d-s-1$.
In this case, in a regular $d$-complex, for $s=0$ and $s=d-1$
the above definition  gives, respectively, the conditions for manifold
vertices and manifold $(d-1)$-faces in a regular complex.

An $s$-simplex that is not a manifold $s$-simplex will be called a
\emas{non-manifold}{$s$-simplex}.
In the following, we will see that in a combinatorial manifold all simplices
are manifold and hence its boundary is manifold, too.
On the other hand we have that a regular complex has non manifold 
simplices \iff\ it is not a manifold complex.
Furthermore we will see that manifoldness is  preserved by
combinatorial equivalence.  A simplex can be manifold even if all its
faces are non manifold, while the converse it is not true.
This is stated by the following properties.
\begin{property}
\label{pro:maninonmani}
\mbox{}
\begin{enumerate}
\item \label{pro:0top} If $\AComp$ is a combinatorial $d$-manifold then $\bnd{\AComp}$ is a combinatorial $(d-1)$-manifold without boundary.
\item \label{pro:1top}  If a combinatorial $d$-manifold $\AComp$ is combinatorially
equivalent to another complex   $\AComp^\prime$, then $\AComp^\prime$ is
a combinatorial $d$-manifold, too.
\item \label{pro:2top}
In a combinatorial manifold all simplices are manifold simplices.
\item \label{pro:3top} All faces of a non-manifold simplex are non manifold simplices.
\end{enumerate}
\begin{proof}
Property \ref{pro:0top} is proven in \cite{Gla70} Pg. 21.
Properties \ref{pro:1top} and  \ref{pro:2top} are proven in
\cite{Lic99} (see Lemma 3.2 p. 304).
The property  \ref{pro:3top} can be proven as follows.
Let us assume that there exist a manifold face of a non manifold simplex
$\gamma$
and derive a contradiction. 
Figure \ref{fig:proofm}a depicts 
a situation coherent with the assumption that $\gamma$ (in violet blue)
is a non manifold simplex.
Let $\zeta$ (vertex in orange) be a manifold face of the non manifold simplex 
$\gamma$ and
let $\omega$ (vertex in green) be such that $\omega\cap\zeta=\emptyset$ and
$\gamma=\zeta\cup\omega$, then we have that 
$\closure{\{\gamma\}}=\zeta\join\omega$.
Now, it is easy to prove, (see for instance Lemma in \cite{Gla70} Pg. 20) that
$\lk{\gamma}=\xlk{\xlk{\AComp}{\zeta}}{\omega}$ (${\xlk{\AComp}{\zeta}}$ are the three edges in orange and  $\lk{\gamma}$ are the three vertices in violet). Since, by hypothesis,
$\zeta$ is a manifold
simplex, we have that $\xlk{\AComp}{\zeta}$ 
(the three thick orange segments) is
combinatorially equivalent either to a ball or a sphere and
hence, by Part \ref{pro:1top} in this property, 
$\xlk{\AComp}{\zeta}$ is a manifold complex. Now
$\omega$ is a simplex in $\xlk{\AComp}{\zeta}$ and hence, in this complex, the
link of  $\omega$ (the three violet blue thick dots), i.e.  
$\xlk{\lk{\zeta}}{\omega}$, must be
combinatorially equivalent either to a sphere or to a ball.
Therefore $\lk{\gamma}$ is combinatorially equivalent either
to a sphere or to a
ball (and in the figure the 0-sphere, i.e. two vertices fails to exist). This cannot be true being $\gamma$ a non-manifold simplex.
\end{proof}
\end{property}
\begin{figure}[h]
\begin{minipage}{0.25\textwidth}
\fbox{\psfig{file=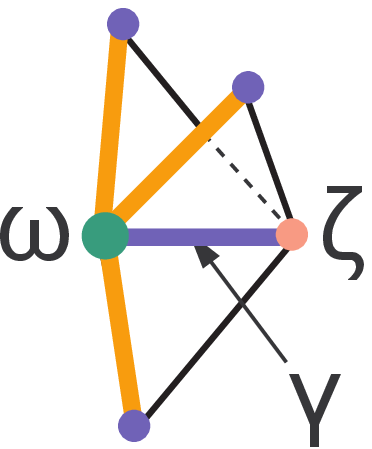,width=\textwidth}}
\begin{center}(a)\end{center}
\end{minipage}
\hfill
\begin{minipage}{0.70\textwidth}
\fbox{\psfig{file=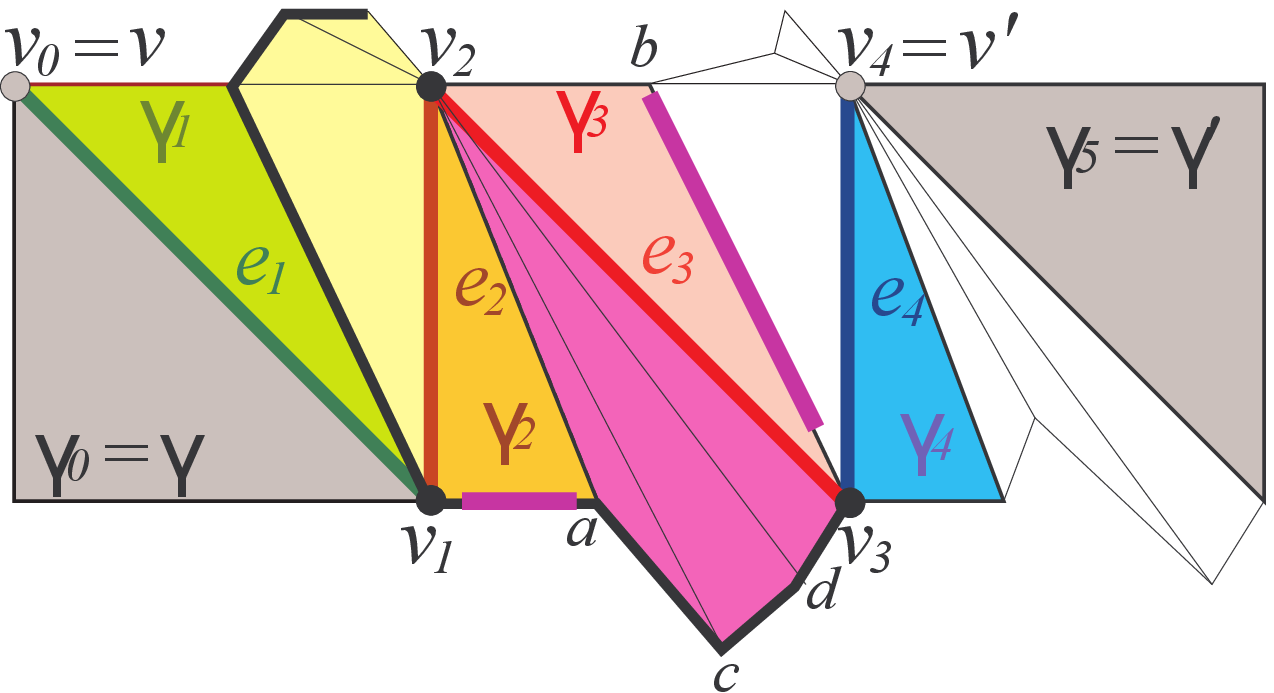,width=\textwidth}}
\begin{center}(b)\end{center}
\end{minipage}
\caption{Proof of Property \ref{pro:maninonmani} Part \ref{pro:3top} (a)
and of Property \ref{pro:manid1conn} (b)}
\label{fig:proofm}
\end{figure}
Using the property above, it is easy to  prove that a connected 
combinatorial  $d$-manifold is $(d-1)$-connected. 
This is expressed by the property below that  will be useful in the
following.	
\begin{property} 
\label{pro:manid1conn}
A combinatorial $d$-manifold
is connected if and only if it is $(d-1)$-manifold connected.
\begin{proof}
We have just to prove that 
a combinatorial $d$-manifold
is connected if and only if it is $(d-1)$-connected.
In fact, a $(d-1)$-face in a combinatorial $d$-manifold can have
up to two cofaces. So, in this context, a $(d-1)$-connected
component is also a  manifold-connected component. 

The easiest way to prove
that a combinatorial connected $d$-manifold.
is $(d-1)$-connected is  by induction on $d$.
If $d=1$ the property is obvious. In fact ,
a complex $\AComp$ it is   connected if and only if is $0$-connected
(see Section \ref{par:conndef}).
Let us assume that property holds for connected $h$-manifolds, for $h<d$,
and let us prove that it holds for connected $d$-manifolds, too.
In Figure \ref{fig:proofm}b we show 
a situation, for $d=2$, which is coherent with the notations chosen in 
this proof.
Let $\gamma$ and $\gamma^{\prime}$ be two  $d$-simplices in $\AComp$
(in gray).
Let $v$ and $v^\prime$ be two Vertices in $\AComp$ 
(the two gray blobs in figure)
such that
$v\in\gamma$ and  $v^\prime\in\gamma^{\prime}$. 
Being $\AComp$ connected, there exist a path 
$(v_{i})_{i=0}^{n}$ in $\AComp$, 
from $v=v_0$ to $v^\prime=v_n$ ($n=4$ in the figure) and
such that, for every two consecutive Vertices $v_{i-1}$ and 
$v_{i}$, the pair {$\{v_{i-1},v_{i}\}$} is a $1$-simplex.
For each edge {$e_i=\{v_{i-1},v_{i}\}$} (in green, brown, red and blue)
let us select a $d$-coface 
of $e_i$. Let $\gamma_i$ be this coface (i.e. $e_i\subset\gamma_i$).
Let us consider the sequence of the  $n$ selected $d$-simplices
$(\gamma_i)_{i=1}^{n}$ (in green, brown, pink and pale blue).
Let us extend this sequence with $\gamma_0=\gamma$ 
and with $\gamma_{n+1}=\gamma^\prime$.
We will show that we can insert a $(d-1)$-path between each couple of 
consecutive $d$-simplices $\gamma_i$ and $\gamma_{i+1}$ in the sequence
$(\gamma_i)_{i=0}^{n+1}$. This will  create a $(d-1)$-path 
from $\gamma=\gamma_0$ and $\gamma^\prime=\gamma_{n+1}$.
To do this we note that for any $i=0,\ldots,n$ (we chose $i=2$ in figure), 
the two consecutive
$d$-simplices $\gamma_i$ and $\gamma_{i+1}$ share the vertex $v_i$.
Therefore, both $\gamma_i$ ($\gamma_2$ in brown) and $\gamma_{i+1}$ ($\gamma_3$ in pink) belong 
to $\str{v_i}$ ($\str{v_2}$ is made up of eight triangles 3 in yellow one in brown three in violet one in pink).
Being $\AComp$ a manifold, we have that $\lk{v_i}$ 
($\lk{v_2}$ is the thick black line plus violet $v_3b$)
is combinatorially 
equivalent either to the boundary of a $d$-dimensional simplex or
to a $(d-1)$-simplex. In both cases, by Property \ref{pro:maninonmani}
Part \ref{pro:1top},  we have
that $\lk{v_i}$ is a $(d-1)$-manifold  and hence, by
inductive hypothesis, $\lk{v_i}$ must be  $(d-2)$-connected. We have now that 
both $\gamma_i-\{v_i\}$ and $\gamma_{i+1}-\{v_i\}$ (edges $v_1a$, $v_3b$ 
in violet) 
belong to $\lk{v_i}$.
Therefore, there must be a $(d-2)$-path $(\gamma^{(i)}_k)$ in $\lk{v_i}$
from  $\gamma_i-\{v_i\}$ to $\gamma_{i+1}-\{v_i\}$
(the $0$-path $v_1a$, $ac$, $cd$, $dv_3$ and $v_3b$).
Hence, the sequence of complexes $(\{v_i\}\join\gamma^{(i)}_k)$ 
will be a $(d-1)$-path from  $\gamma_i$ to $\gamma_{i+1}$ (the $1$-path
made up of $\gamma_2$, $\gamma_3$ and the three  violet triangles in figure).
Therefore, we can build a $(d-1)$-path between any pair of 
$d$-simplices,
$\gamma=\gamma_0$ and $\gamma^\prime=\gamma_{n+1}$,
in the connected $d$-manifold $\AComp$.
\end{proof}
\end{property}

A non manifold simplex that is not the face of another non manifold simplex
is called a \emas{top}{non manifold simplex}.
\begin{remark}
\label{rem:manicof}
Note that a top non manifold simplex  needs not necessarily  to be a
top simplex.
Property \ref{pro:maninonmani} Part \ref{pro:3top} implies that a
manifold simplex cannot be face of a non-manifold simplex and, hence, all
cofaces of a manifold simplex must be manifold simplices.
\end{remark}

\begin{short-example}
Coming back to the example of Figure \ref{fig:es3complfh}a we have that
the triangle $poh$ in pale blue is a manifold $2$-simplex since its link
is the regular $0$-complex $r$ (in pale blue). This vertex is trivially
combinatorially equivalent to the $0$-ball.
Similarly the $1$-simplex $rh$ (in red) is a manifold $1$-simplex since its
link is the regular $1$-complex $op$ (in red). This is
combinatorially equivalent to the $1$-ball.
As a direct consequence of Property \ref{pro:maninonmani} Part \ref{pro:3top}
we have that cofaces of the manifold simplex $rh$ must be manifold simplices. 
To check this consider
$rh$ cofaces i.e.  triangles $rph$ and $roh$ and tetrahedron $rohp$. 
The cofaces of dimension $2$ has links
$\lk{rph}=o$ and. $\lk{roh}=p$. The links $o$ and $p$  trivially satisfy 
the manifold condition in Definition \ref{def:simplnonmani}.
Finally we have that the vertex $k$ (in green) is a non-manifold vertex
because its link (in green too)
is made up of vertex $l$ and of the triangle $rop$ and hence is not regular.
As further examples we can  look at non-manifold edge $mi$ we can apply
Property \ref{pro:maninonmani} Part \ref{pro:3top}
to see that  all faces of $mi$ (i.e. Vertices $m$ and $i$) are non manifold
Vertices.
Vertices $m$ and $j$ are examples of non-manifold simplices that
are not top non-manifold simplices.
In turn edge $mi$ is an example of a top non-manifold simplex that
is not a top simplex.
\end{short-example}

We end this section with a property that will be useful in 
Chapter \ref{ch:nmmdl}.
\begin{property}
\label{pro:ballcone}
Let $\AComp$ be a $d$-sphere  and
let $\gamma$ be a $n$-simplex whose vertices are not in $\AComp$. 
We have that $\gamma\join\AComp$ is
a  $(d+n+1)$-ball.
\end{property}
\begin{proof}
If $\AComp$ is  a $d$-sphere then  it  is easy
to see that  the cone  $\gamma\join\AComp$ is   stellar equivalent to the 
$(d+n+1)$-ball (see for instance \cite{Hud69} Lemma 1.13, Part 2, Pg. 22). 
This proves that 
when $\AComp$ is  a $d$-sphere then  $\gamma\join\AComp$ must be
a $(d+n+1)$-ball.
\end{proof}

\section{Nerve, Pasting  and \Quot\ space}
\label{sec:nerve}
In this section we introduce 
the notion of  {\em pasting}.
This notion is sometimes introduced as a tool for  the
topological  classification of
closed manifold surfaces
(see  for instance \cite{EngSve92} Pg. 98 and Pg. 222 and \cite{Spa66} Pg.108
and Pg. 152).
Indeed, this classification is usually obtained by
showing that each closed $2$-manifold can be constructed  
by  {\em pasting} a triangulated rectangular
sheet at its boundary. 
\NOTA{
In section \ref{sec:nerves}, 
we will show that the notion of pasting is quite general and can provide
an alternative view to the category of abstract simplicial complexes with
abstract simplicial maps. 
}
Indeed, the result of pasting a complex $\AComp$ onto itself is
intimately related with the {\em \quot} space for a  geometric
realizations of $\AComp$ (see \cite{EngSve92} Pg. 372). 
With this idea in mind we report here the basic
notions of  {\em covering}, {\em nerve} and of {\em pasting}.
\begin{definition}[Covering see \cite{Spa66} Pg. 152 \cite{EngSve92} Pg. 61]
\label{def:covering}
A \ems{covering} of a set $X$ is a collection of subsets  of $X$ 
whose union gives $X$.
\end{definition}
Given two coverings {${\cal W}$} and ${\cal U}$ we will say that 
{${\cal W}$}
is a \ems{refinement} of {${\cal U}$} (denoted by {${\cal W}\refine{\cal U}$})
if there is a function $\funct{\phi}{\cal W}{\cal U}$ such that
for each $W\in{\cal W}$ we have $W\subset\phi(W)$.
The function $\phi$ is called a \emas{canonical}{covering projection}
from {${\cal W}$} to {${\cal U}$}.
In this situation we will say that {${\cal U}$} is a \ems{coarsening} of 
{${\cal W}$}.
For  coverings of  a set $X$ it is easy to see that the relation $\refine$ is both
reflexive and transitive and it is antisymmetric. In the case $X$ is a
finite set, 
the set of coverings of finite set $X$ is a poset ordered by the refinement
relation ($\refine$).
A covering whose elements are pairwise disjoint is called a \ems{partition}.
Sets in a partition are called \emas{partition}{blocks}.

Given a covering {${\cal U}$} for a set $X$ we can associate an 
abstract simplicial complex $\nerve{\cal U}$
with the covering {${\cal U}$}. The complex $\nerve{\cal U}$ is called 
the {\em nerve} of the covering {${\cal U}$}. The informal idea behind the
concept of the nerve of a covering is better understood if referenced to a
less abstract settlement.  
In fact, let us consider
the particular case in which:
\begin{itemize}
\item $X$ is a subset of the standard Euclidean space $\real^n$ and $X$ is a polyhedron;
\item  ${\cal U}$. is a covering of $X$ such that
each element in {${\cal U}$} contains just one vertex from $X$. 
\end{itemize}
In this situation, it is easy to see that the  geometric realization of  
$\nerve{\cal U}$ can be the polyhedron $X$
For instance it is easy to see that if {${\cal U}$} is 
the covering induced by closed
faces in a Voronoi diagram, then the associated Delunay 
triangulation is a possible geometric realization for 
the nerve {${\nerve{\cal U}}$}.
With this idea in mind we can report the definition of nerve (see
\cite{Spa66} Pg. 109)
\begin{definition}[Nerve]
\label{def:nerve}
Given a covering {${\cal U}$} of finite set $X$ the \emd{nerve} of  
{${\cal U}$} (denoted by {${\nerve{\cal U}}$}) is 
the abstract simplicial complex such that:
\begin{itemize}
\item the set {${\cal U}$} is the set of Vertices of {${\nerve{\cal U}}$} and 
\item $\gamma=\{v_1,\ldots,v_n\}$ is a simplex in {${\nerve{\cal U}}$}
\iff\ $\cap_{i=1,n}{v_i}\neq\emptyset$
\end{itemize}
\end{definition}

\begin{figure}
\begin{minipage}{0.49\textwidth}
\fbox{\psfig{file=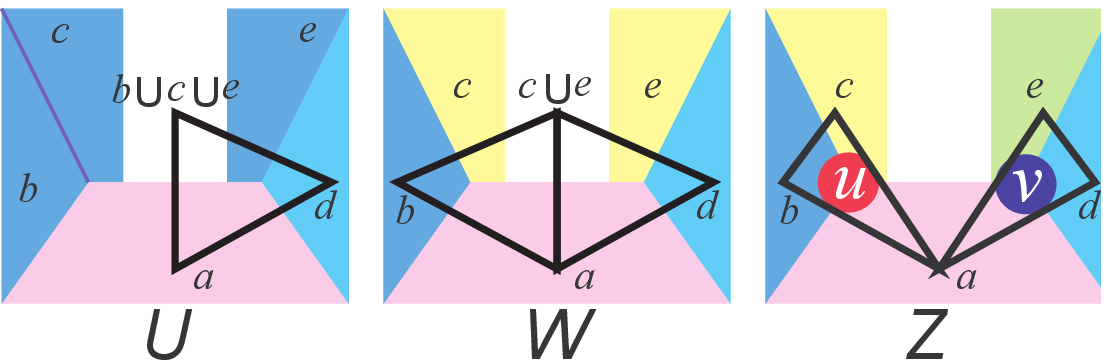,width=\textwidth}}
\begin{center}(a)\end{center}
\end{minipage}
\hfill
\begin{minipage}{0.49\textwidth}
\fbox{\psfig{file=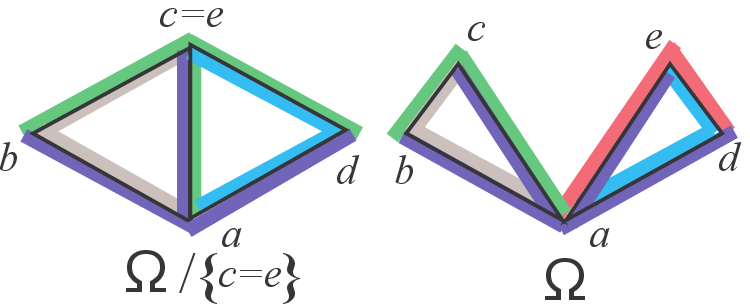,width=\textwidth}}
\begin{center}(b)\end{center}
\end{minipage}
\caption{An example of two coverings with
the associated nerve (a) and an example of a pasting}
\label{fig:nerve}
\end{figure}
\begin{example}
As an example of application of the above definitions consider 
the three coverings ${\cal U}$, ${\cal W}$ and {${\cal Z}$} for the 
U shaped domain in Figure \ref{fig:nerve}a.
Let us concentrate first on the rightmost covering {${\cal Z}$}.
The covering {${\cal Z}$} is made up of the five colored closed polygons 
$a$ (in pink), $b$ (in blue), $c$ (in yellow), $d$ (in pale blue),
and $e$ (in green). 
Each polygon contains its boundary.
It is easy to see that $a$, $b$ and $c$ has a non empty intersection in the
vertex $u$ (in red) and $a$, $d$ and $e$ has a non empty intersection in the
vertex $v$ (in blue). 
Therefore the nerve {$\nerve{\cal Z}$} must contain triangles 
$abc$ and $ade$.  It is easy to see that the nerve {$\nerve{\cal Z}$}
is {\em exactly} the $2$-complex made up of these two triangles. 
\end{example}

An abstract simplicial  complex $\AComp$ can always be seen as the nerve of 
a particular  covering of $\AComp$. This is expressed by the following
property (see Lemma 2.3.13 in \cite{EngSve92})
\begin{property}
\label{pro:basiccover}
Let be $\AComp$ an abstract simplicial complex with Vertices in $V$ and
let be {${\cal S}$} the covering of $\AComp$ given by the stars of Vertices
in $\AComp$ (i.e.  {${\cal S}=\{\str{v}|v\in V\}$}) then $\AComp$
is isomorphic to the nerve  {${\nerve{\cal S}}$}
\end{property}
\begin{proof} To prove  this property we note that
it can be proved (see Lemma 2.3.13 in \cite{EngSve92}) that  
$\gamma=\{v_1,\ldots,v_n\}$ is a simplex in $\AComp$ \iff\
$\cap_{i=1,n}{\str{v_i}}\neq\emptyset$.
This happens \iff\ $\{\str{v_1},\ldots,\str{v_n}\}$ is a simplex in 
{${\nerve{\cal S}}$}. Therefore the renaming of Vertices
$\funct{i}{V}{\cal S}$ that sends $v$ into $\str{v}$ is an isomorphism 
between  $\AComp$ and {${\nerve{\cal S}}$}.
\end{proof}
We have seen that each abstract simplicial complex can be obtained
as a nerve.
Similarly there is some relation between
the  refinement relation between  the 
corresponding coverings and
abstract simplicial maps across nerves. 
More precisely each refinement {${\cal W}$} of a  covering
{${\cal U}$} induce an abstract simplicial
map from  {$\nerve{{\cal W}}$} to {$\nerve{{\cal U}}$}.
This is expressed by the following property.
\begin{property}\label{pro:refmap}
Let  {${\cal W}$} and {${\cal U}$} be two coverings 
such that {${\cal W}$} is a refinement of {${\cal U}$} 
(i.e. {${\cal W}\refine{\cal U}$})
and let be $\funct{\phi}{\cal W}{\cal U}$ 
a canonical covering projection from {${\cal W}$} to {${\cal U}$}.
Then $\phi$ is a vertex map from Vertices of {$\nerve{{\cal W}}$} to
Vertices {$\nerve{{\cal U}}$} that induce an abstract simplicial map
between the two nerves.
\end{property}
\begin{proof}
See \cite{Spa66} Pg. 152
\end{proof}
Following the notation of Property \ref{pro:mapposet} we have that 
{${\cal W}\refine{\cal U}$} implies that
 $\nerve{{\cal U}}\morphle\nerve{{\cal W}}$.
\begin{example}
The covering {${\cal Z}$} in Figure \ref{fig:nerve}a is an example of 
a refinement of the covering {${\cal W}$}.
In fact the function $[\send{c}{(c\cup e)},\send{e}{(c\cup e)}]$
is the canonical covering projection $\funct{\phi}{\cal Z}{\cal W}$ 
such that, for each $Z\in{\cal Z}$ we have $Z\subset\phi(Z)$.
It is easy to see that $\phi$ induce an abstract simplicial map from
{$\nerve{\cal Z}$} to {$\nerve{\cal W}$}.  
This map is defined by the vertex 
map that sends $c$ 
into $\phi(c)=c\cup e$, and $e$ into $\phi(e)=c\cup e$.
Note that $c\cup e$ is a vertex in the nerve  {$\nerve{\cal W}$}.
Similarly, it is easy to see that 
the covering {${\cal W}$} is a refinement of  the covering {${\cal U}$}
with the canonical covering projection $\funct{\psi}{\cal W}{\cal U}$ 
given by $\psi=[\send{(c\cup e)}{(b\cup c\cup e)},\send{b}{(b\cup c\cup e)}]$.
\end{example}

A particular class of nerves is the class of {\em pastings} of a complex 
$\AComp$.
Pasting are usually introduced to classify closed $2$-manifolds 
(e.g. \cite{EngSve92}). In this context we use this concept for different goal,
namely to investigate
the lattice structure in the category of abstract simplicial complexes
and abstract simplicial map.
The informal idea behind pasting is quite straightforward.
Let us consider an abstract simplicial complex $\AComp$ with Vertices in $V$.
Given an equivalence relation $R$ on $V$  we
can  transform complex $\AComp$ into the new complex $\AComp/R$ 
by \ems{pasting} together two or more Vertices in $V$ according to $R$. 
In this context  the equivalence relation $R$ is used to 
specify which Vertices must be identified. 
We assume that the reader is familiar with the standard terminology for
equivalence classes. A short resume is provided in \latt\ 
Section \ref{sec:partposet}.

The formal definition of the complex $\AComp/R$
is given considering $\AComp/R$ as the nerve of a particular  
covering of $\AComp$ induced 
by $R$.
\begin{definition}[Pasting (see \cite{EngSve92} Pg. 222]
\label{def:quot}
Let $\AComp$ be an abstract simplicial complex with Vertices in $V$ and let
$R$ be  an equivalence relation on $V$. 
Let $[v]$ be the equivalence class for $v$ induced by $R$ and let 
{$R_{[v]}$} be the 
set of simplices in $\AComp$ given by the union of  open stars 
of Vertices in $[v]$ 
(i.e. {${R_{[v]}}=\cup_{w\in [v]}{\str{w}}$}).
In this situation the complex $\AComp/R$ is defined as  the nerve of 
the covering
${\cal R}=\{{R_{[v]}}|v\in V\}$.
\end{definition}
Note that, for a given equivalence relation $R$, the covering
${\cal R}=\{{R_{[v]}}|v\in V\}$ is a 
coarsening of the  covering
{${\cal S}=\{\str{v}|v\in V\}$} defined in Property \ref{pro:basiccover} (i.e.
${\cal S}\refine{\cal R}$). In particular,
if we denote with
$\Delta_V=\{(v,v)|v\in V\}$, the identity relation or the {\em diagonal} in
$V\times V$ we have that  ${\cal S}={\cal R}$ \iff\ $R=\Delta_V$.

Directly from the Definition \ref{def:quot} 
we have that $\AComp/\Delta_V$
is isomorphic to $\AComp$ 
(recall Property \ref{pro:basiccover} and the fact that $\AComp/\Delta_V=\nerve{\cal S}$).
Similarly  we have that $\AComp/(V\times V)$ is the complex made up
of a single isolated vertex. 
From Property \ref{pro:refmap} we have that, for any equivalence $R$ it holds 
$\AComp/(V\times V)\morphle\AComp/R\morphle\AComp/\Delta_V\simpeq\AComp$
The relation $R$ will be called the \emas{generating}{relation} or the \emas{divisor}{relation} for the \quot   $\AComp/R$. 

\begin{example}
As an example of pasting consider 
the complex $\AComp$ in Figure \ref{fig:nerve}b. We have that 
$\AComp$ is isomorphic to the nerve  {${\nerve{\cal S}}$} where
{${\cal S}$} is the covering of $\AComp$ given by the stars of Vertices
in $\AComp$. In the complex on the right of Figure  \ref{fig:nerve}b, 
we have depicted with the same color  the edges in each vertex star. 
For instance, the four blue thick edges belongs 
to the star of the vertex $a$.
The star of each vertex is made up of the two colored thick edges with 
the triangle between them.  The common vertex
completes this star.
For instance, the star $\str{a}$ is made up of the 
vertex $a$, of the four edges in blue together with the two triangles 
$abc$ and $ade$.
In the complex on the left of Figure \ref{fig:nerve}b we have 
depicted a coarsening {${\cal S}^\prime$} of the covering {${\cal S}$}. 
The coarser version {${\cal S}^\prime$} is obtained merging 
together the stars of $c$ (in green on the right) and of  $e$ (in red) into 
a single item $(\str{c}\cup\str{e})$ (in green on the left).
This new item is  made up of: the Vertices $c$ and $e$; 
the four green edges on the left, i.e.  $bc$, $ac$, $ae$ and $ed$;
the two triangles $abc$ and  $ade$.
It is easy to see, by exhaustive intersection 
of elements in {${\cal S}^\prime$}, that
the complex on the left is  isomorphic to the 
nerve of the covering {${\cal S}^\prime$}. We close this example by showing,
by Definition \ref{def:quot},
that the nerve $\nerve{{\cal S}^\prime}$ is isomorphic to the pasting 
$\AComp/R$, with $R$ is the equivalence  given by $R=\{(e,c),(c,e)\}$.
Indeed, using the notations of Definition \ref{def:quot}, we have 
$R_{[e]}=R_{[c]}=(\str{c}\cup\str{e})$ while 
$R_{[v]}=\str{v}$ for all other Vertices $v$. Therefore
{${\cal R}={{\cal S}^\prime}$} and thus
$\AComp/R=\nerve{{\cal S}^\prime}$.
\end{example}
As we anticipated in the introduction of this section, in the
following the complex $\AComp/R$ will be called the \ems{\quot} 
of $\AComp$ induced by the relation $R$.
We prefer {\em \quot} instead of {\em pasting} because the abstract simplicial
complex $\AComp/R$ is intimately related with the
{\em topological} quotient space of the geometric 
realizations of $\AComp$. To state this relation
formally we assume that the reader is 
familiar with the notion of quotient of a {\em topological space} $X$
(see \cite{EngSve92} Pg. 369 or \cite{Spa66} Pg. 5).
Given a topological space $X$ and an equivalence relation 
$R\subset X\times X$ we will denote with  $X/R$ the topological space that
is the quotient of $X$ induced by $R$. 
Assuming this notion  we can  state the following property.
\begin{opt-property}[\Quot\ Geometric Realization]
Let be $\AComp$ an abstract simplicial complex with Vertices in $V$ and let
be  $R$ an equivalence relation on $V$. 
Let be $K$ a geometric realization of $\AComp$.
In this situation it can be proved that there exist an equivalence relation
${\vettore{p}(R)}$ over $|K|\times|K|$ such that
any geometric realization of the 
abstract simplicial complex  $\AComp/R$
will be homeomorphic to the quotient space $|K|/{\vettore{p}(R)}$.
\end{opt-property}
\begin{proof} The proof is sketched in \cite{EngSve92} Pg. 372. 
Here  we just report the construction of the equivalence ${\vettore{p}(R)}$.
For each point ${\vettore{q}}$ in the
polyhedrom $|K|$ let be  $\{\vettore{q}_i|i=1,\ldots,d+1\}$ the
smallest geometric simplex that contains $\vettore{q}$ and let be 
$\lambda^{\vettore{q}}_i$ its barycentric coordinates (i.e.  
$\vettore{q}=\Sigma_{i=1}^{d+1}{\lambda_i^{\vettore{q}}\vettore{q}_i}$
(see Definition \ref{def:geomsimp}).
Let $\vettore{p}(v)$ be the geometric realization of the vertex $v\in V$
and let ${\vettore{p}([v])}$ be the  set of geometric realization of
Vertices in an equivalence class $[v]$ 
induced by equivalence $R$. 
Finally  we define $\lambda_{[v]}^{\vettore{q}}$ as the 
partial sum $\lambda_{[v]}^{\vettore{q}}=
\Sigma_{\vettore{q}_i\in{\vettore{p}([v])}}{\lambda_i^{\vettore{q}}}$.
With this notation
the equivalence   ${\vettore{p}(R)}$ is defined by requiring 
that $({\vettore{q}},{\vettore{r}})\in{\vettore{p}(R)}$ \iff\ 
$\lambda_{[v]}^{\vettore{q}}=\lambda_{[v]}^{\vettore{r}}$ for all 
the equivalence classes $[v]$ induced by  $R$.
It is easy to show that the realizations of equivalent Vertices in $R$ 
are equivalent w.r.t.  $\vettore{p}(R)$ (i.e.
$\vettore{p}(R) \supset \{(\vettore{p}(u),\vettore{p}(u))|(u,v)\in R\}$)
Similarly if $\vettore{q}$ and  $\vettore{r}$ are two points within the 
same simplex with barycentric coordinates
$\vettore{q}=\Sigma_{i=1}^{d+1}{\lambda^{\vettore{q}}_i\vettore{q}_i}$
and
$\vettore{r}=\Sigma_{i=1}^{d+1}{\lambda^{\vettore{r}}_i\vettore{r}_i}$
we will have that 
$\vettore{q}$ and  $\vettore{r}$ are equivalent in  $\vettore{p}(R)$ 
whenever, for all $i=1,2,\ldots,(d+1)$, points $\vettore{q}_i$ 
and $\vettore{r}_i$ are equivalent w.r.t. $\vettore{p}(R)$ 
(i.e. $({\vettore{q}_i},{\vettore{r}_i})\in{\vettore{p}(R)}$
for all $i=1,2,\ldots,(d+1)$ implies that 
$({\vettore{q}},{\vettore{r}})\in{\vettore{p}(R)}$).
\end{proof}
\begin{example}
Returning to the situation of Figure \ref{fig:nerve}b 
we have that equivalence
$R=\{(e,c),(c,e)\}$ induce an equivalence ${\vettore{p}(R)}$.
It is easy to see that ${\vettore{p}(R)}$ is
made up of all pairs of geometric Vertices of the form 
$({\vettore{q}},{\vettore{r}})$
and $({\vettore{r}},{\vettore{q}})$ such that:
point ${\vettore{q}}$ is in the   segment $ac$; 
point ${\vettore{r}}$ is in the  segment $ae$; points 
${\vettore{q}}$, ${\vettore{r}}$ have equal barycentric coordinates.
\end{example}

 \chapter{The \Quot\ Lattice}
\label{ch:quotlat}

\section{Introduction}
There is an obvious relation between abstract simplicial maps and \quot s.
Indeed a \quot\ $\AComp^\prime/R$ 
is the image of $\AComp^\prime$ via an appropriate abstract simplicial  map 
$f_R$.
Conversely for  each  \asm\  $\funct{f}{\AComp^\prime}{\AComp}$
there exist an appropriate equivalence relation $R_f$ such that
the image {$f(\AComp^\prime)$} is isomorphic to the \quot\ 
$\AComp^\prime/R_f$.
In this chapter we present this relation in detail 
(in Property \ref{pro:uniqverquot}) and show that the set of all \quot s
of a given \asc\ $\AComp^\prime$ form a lattice we called the
{\em \Quot\ Lattice}. 

The \Quot\ Lattice is isomorphic to a
well known lattice $\Pi_n$ called the {\em partition lattice}.
Mathematical properties of this lattice are  given in Appendix \ref{sec:pin}.
However, in the first part of this chapter, relevant properties are 
summarized and restated, in an intuitive form,  
using a language closer to the subject of this thesis.  

The \quot\ lattice, in the context of this thesis, will be used as  the
structure in which we order the decompositions of a given complex.
The relevant factor is that, being this structure a lattice, 
we can expect to have
a least upper bound for any arbitrary set of decompositions. 
This will be a key issue to define a unique decomposition.

Another key idea in the development of this thesis is the fact that we
can manipulate \quot s $\AComp^\prime/R$ using the set of
equations $E$ that defines $R$.
In particular, we are interested in the fact
that some topological properties can be restated in term  of
syntactical properties for equations.
The manipulation of these syntactical objects will give us an  alternative,
and sometimes fruitful, way to the treat topological problems. 
Since these equations identify two
vertices together we will call them  {\em \Verteq\ Equations}.
Thus, in this chapter, we finally introduce  { \Verteq\ Equations} and enlight 
the relation between sets of equations and  the {\em \Quot\ Lattice}.
Some  examples are provided to lead the reader along this path.  

\section{Maps and Equivalence}
We start this path with the property below that
detals the relation between abstract simplicial maps and \quot s
$\AComp^\prime/R$. 
In particular parts \ref{pro:quotmapunique} and  \ref{pro:quotmapclass}
further details the structure of the nerve $\AComp/R$ introduced by
Definition \ref{def:quot}.

\begin{property}
\label{pro:uniqverquot}
Let be $\AComp$ an abstract simplicial complex with Vertices in $V$ and
let be $R$ and equivalence relation on $V$.
In this situation the following properties holds:
\begin{enumerate}
\item \label{pro:quotmapunique} 
for any vertex $w$ in $\AComp$ there is a unique vertex $W$ in the 
nerve $\AComp/R$ such that $\str{w}\subset W$ (or $\{w\}\in W$);  
\item \label{pro:quotmapclass} 
A pair of  Vertices $u$ and $w$ in $\AComp$ are equivalent w.r.t. $R$ 
(i.e.  $(u,w)\in R$) \iff\ there is a unique vertex $W$ in the 
\quot\ $\AComp/R$ such that $\{u\}\in W$ and   $\{w\}\in W$;  
\item \label{pro:quotmap}
For any \quot\ $\AComp/R$ we have
$\AComp/R\morphle\AComp$ and $\AComp/R\simpeq\AComp$ \iff\ $R=\Delta_V$;
\item \label{pro:mapquot}
For any abstract simplicial complex $\AComp\morphle\AComp^\prime$
there exist an equivalence relation $R_f$ such that  
$\AComp^\prime/R_f\simpeq\AComp$;
\end{enumerate}
\end{property}
\begin{proof}
To prove Parts \ref{pro:quotmapunique} and  \ref{pro:quotmapclass}
we proceed as follows.
We recall from Definition \ref{def:quot}  that
the complex {$\AComp/R$} is defined as  the nerve of 
the covering ${\cal R}=\{{R_{[v]}}|v\in V\}$ 
with  {${R_{[v]}}=\cup_{w\in [v]}{\str{w}}$}.
Recall also that, in this formula, $[v]$ is the equivalence 
class for $v$ w.r.t $R$. From the definion of nerve 
(see Definition \ref{def:nerve}) we have that Vertices of {$\AComp/R$} are
the elements ${R_{[v]}}\in{\cal R}$. 
We have that $\{w\}\in\str{v}$ \iff\ $w=v$ and therefore 
$\{w\}\in{R_{[v]}}$ \iff\ $w\in[v]$. By transitivity we have
that $\str{w}\in{R_{[v]}}$ \iff\ $w\in[v]$.
We have that $(u,w)\in R$ \iff\ there is a unique equivalence class $[v]$ that
contains both  $u$ and $w$. This is  equivalent to ask to have a unique
vertex ${R_{[v]}}$ in {$\AComp/R$} such that {$\{u\}\in{R_{[v]}}$} and
{$\{w\}\in{R_{[v]}}$}. This proves Part \ref{pro:quotmapclass}.
To prove Part \ref{pro:quotmapunique} we recall that 
$\str{w}\in{R_{[v]}}$ \iff\ $w\in[v]$.
Since equivalence classes
are disjoint there must be just one vertex ${R_{[v]}}$
in  {$\AComp/R$} that contains $\str{w}$. This shows also that
there must be just one vertex ${R_{[v]}}$
in  {$\AComp/R$} that contains $\{w\}$.

To prove Part \ref{pro:quotmap} we proceed as follows.
We have seen (see remark after Definition \ref{def:quot})
that $\AComp/R\morphle\AComp/\Delta_V\simpeq\AComp$,
therefore we just have to prove that 
$\AComp/R\simpeq\AComp$ implies $R=\Delta_V$.
Let us assume that  $\AComp/R\simpeq\AComp$ for some  $R\neq\Delta_V$
and derive a contradiction.
Indeed if the equivalence $R$ is not empty the number of
equivalence classes for $R$ is lower that $|V|$ and the  covering
${\cal R}=\{{R_{[v]}}|v\in V\}$ is such that $|{\cal R}|<|V|$.
This leads to a contradiction since $\AComp/R=\nerve{{\cal R}}$ has less
Vertices than $\AComp$ and hence cannot be isomorphic to $\AComp$.

To Prove part \ref{pro:mapquot} we proceed as follows.
Let $f$ be the map such that $f(\AComp^\prime)=\AComp$. We
define $R_f$ 
to be the equivalence relation such that $(u,v)\in R_f$ \iff\ $f(u)=f(v)$.  
Let $[v]$ be  a generic equivalence class w.r.t. $R_f$ and 
let $i$ be the renaming of Vertices $\funct{i}{\AComp^\prime/R_f}{\AComp}$
defined by $i({R_{[v]}})=f(v)$. 
We want to show that $i$ is an isomorphism.
We first note that the definition of $i$ is sound since it do not depends 
on the vertex $v$. Indeed we have that $[u]=[v]$ \iff\ $f(u)=f(v)$.
Now we need to show that $i$ maps every simplex in {${\AComp^\prime/R_f}$}
into a simplex of $\AComp$.
By Part \ref{pro:quotmap} of this property, we know that there is an abstract simplicial map $g$
that  sends {$\AComp^\prime$} into {${\AComp^\prime/R_f}$}. This is given
by the map $g$ defined by $g(v)={R_{[v]}}$. Thus 
we have that $g(v)={R_{[v]}}$ and we already know that $i({R_{[v]}})=f(v)$. 
Therefore, for each simplex $\gamma\in{\AComp^\prime}$,
we can write $i(g(\gamma))=f(\gamma)$. 
Every simplex in {${\AComp^\prime/R_f}$}
must be of the form $g(\gamma)$, for some $\gamma\in{\AComp^\prime}$.
Hence, the renaming of Vertices $i$
maps every simplex $g(\gamma)$ in {${\AComp^\prime/R_f}$} into simplex 
$i(g(\gamma))=f(\gamma)$.
This proves that $i$  is an isomorphism and that $\AComp^\prime/R_f\simpeq\AComp$.
\end{proof}

In the situation of Property \ref{pro:uniqverquot} we will denote with
$f_R$ the abstract simplicial map $\funct{f_R}{\AComp}{\AComp/R}$ defined
as $f_R(v)=R_{[v]}$. For any simplex $\gamma\in\AComp$ we use the 
symbol {$\gamma/R$} to denote the simplex that is the  image  of $\gamma$
in $\AComp/R$ (i.e. $\gamma/R=f_R(\gamma)$). We will call
{$\gamma/R$} the \ems{pasted} version of $\gamma$ via relation $R$.
With some abuse of notation 
we will denote $\{v\}/R$ with $v/R$.
With this notation we can express Part \ref{pro:quotmapclass} of the
previous property  by saying that $u/R=v/R$ \iff\ $(u,v)\in R$.
Note that $v\in\gamma$ implies that  $v/R\in \gamma/R$ while the converse, 
in general, is false.

\section{\Quot\ Lattice}
{
The set of abstract simplicial complexes $\AComp$  with the preorder relation 
$\AComp\morphle\AComp^\prime$ is not a poset.
As we will see in the following we are interested in finding a 
least upper bound for certain sets of complexes. Therefore we actually need a poset and a  lattice structure.
A possible option to construct such a lattice 
is to identify isomorphic complexes and work with
{\em classes} of isomorphic complexes (instead of working with plain complexes).
In this way, the extension of relation $\morphle$ to classes of isomorphic
complexes becomes antisymmetric. Furthermore, it is easy to prove that
the poset for these classes of isomorphic complexes has a lattice structure.

Unfortunately  this choice, although theoretically elegant, leads to 
an approach that is of limited interest for applications.
In fact  we should not forget that \asc es are used  here
to capture the combinatorial
structure of a model that still has some geometric realization
(see Definition \ref{def:georel}).
Thus, the geometric realizations of two isomorphic complexes  
can be quite different
and we could not consider them as equivalent.
Thus we need to consider
isomorphic and non-identical \asc es  as distinct 
objects. 

In the following we will define a lattice that preserve this 
distinction thus satisfying this basic requirement from 
the applicative domain. }
In particular,  we will first  restrict our attention to \quot s of a
generic complex $\AComp^\prime$ and devise a lattice structure 
for the set of \quot s $\AComp^\prime/R$. 
Later on, in chapter \ref{ch:stdec},
the structure of complex $\AComp^\prime$ will be specialized to
have all the decompositions of a given complex $\AComp$ as \quot s of
a certain complex $\AComp^\prime$.
We will call the lattice of \quot s of $\AComp^\prime$
the  \emas{\Quot}{lattice} (for $\AComp^\prime$).
The \quot\  lattice is isomorphic to a well known lattice called the 
\emas{partition}{lattice}. The properties of this lattice, together with
the formal background necessary to state them are reported in  
\latt.
Here we will use these theoretical tools (i.e. Lattice Theory) in an intuitive
fashion. However note that, beyond the informal style of this presentation,
examples, properties and definitions are grounded in a sound theoretical framework.
We simply make our exposition more intuitive and less formal.
A full account on this formal background for $\Pi_n$  is given 
in Section \ref{sec:pin}.

The restriction to the lattice of \quot s actually do not impair generality.
In fact in the \quot\ lattice for $\AComp^\prime$ we have
representatives for all complexes that
can be obtained modifying $\AComp^\prime$.
Indeed, we have shown (See forthcoming Property \ref{pro:uniqverquot} Part \ref{pro:mapquot}) that for any complex 
$\AComp$  such that $\AComp\morphle\AComp^\prime$ there exist a \quot\ 
$\AComp^\prime/R$ isomorphic to $\AComp$.
Therefore in the following, in order to study the possible modifications of
$\AComp^\prime$ we will restrict our attention to the elements in the
\quot\ lattice for $\AComp^\prime$.
We will study the properties of elements of the form
$\AComp^\prime/R$ by studying the properties of the relation $R$.
This implies that whenever two  isomorphic objects
$\AComp/R_1$ and  $\AComp/R_2$ are  generated by 
distinct
quotients with distinct relations $R_1\neq R_2$.
will be treated as distinct objects.

In order to give here a short account on the {\em partition lattice}
(see \latt) we note that the set
of  equivalences on $V$
is a poset ordered by the standard
set inclusion
(i.e. $        R_1\le        R_2$ if and only if
$        R_1\subset        R_2$).
The poset of  equivalences on $V$
has a maximum (i.e., the relation $V\times V$)  
and a minimun (i.e., the identity relation $\Delta_V$),
being $\Delta_V\subset R\subset V\times V$. 
Then, if $R_1$ and $R_2$ are two equivalences
we define their sum ${R_1}\sumeq{R_2}$
as the smallest equivalence containing the two.
Similarly
we define their product ${R_1}\proeq{R_2}$
as the intersection ${R_1}\cap{R_2}$.
Note that the intersection of two equivalences is still an equivalence and
is the greatest equivalence contained in them.
With these two operations,
the poset of equivalences becomes the partition lattice $\Pi_n$.
In fact sum and product of two equivalence is still an equivalence.

This lattice induce a poset and a lattice over the set of \quot s.
We called this poset and this lattice respectively
the \emas{\quot}{poset} and the \emas{\quot}{lattice}.
\begin{definition}[\Quot\ poset]
\label{def:quotposet}
The set of \quot s of a given complex $\AComp^\prime$ is a poset with the
ordering relation $\quotle$ defined as $\AComp/R_2\quotle \AComp/R_1$
\iff\ $R_1\subset R_2$
\end{definition}
By the remark at the end of Property \ref{pro:uniqverquot} we have that the
mapping that sends $\AComp$ into $\AComp/R$ is injective and thus the
\quot\ poset is  anti-isomorphic to the partition lattice $\Pi_n$ 
where $n$ is the number of Vertices in $\AComp$.
We recall that two posets are anti-isomorphic \iff\ they are isomorphic
but we exchange the direction of the ordering passing from a poset to the
other (see \latt\ Section \ref{sec:poset})

The \quot\ poset has a greatest and
a least element. In fact, for any relation $R$ we have:
${\AComp^\prime/(V\times V)}\quotle \AComp^\prime/R\quotle\AComp^\prime/\Delta_V$.
Note that the complex $\AComp^\prime/\Delta_V$ is isomrphic to $\AComp^\prime$
(see remark at the end of Definition \ref{def:quot}) while 
${\AComp^\prime/(V\times V)}$ is the complex made up of  single isolated vertex.

The $\quotle$ ordering in the  \quot\  poset implies the 
preorder given
by the relation $\morphle$  (associated with 
abstract simplicial maps).  Indeed relations induced by non
identical isomorphisms are missing.
This connection between these two relations 
is  detailed in the property below.
\begin{property} \label{pro:equiquot} Let $\AComp^\prime$ be an abstract simplicial complex with  
Vertices in $V$ and let
$R_1$ and $R_2$ be two equivalence relations on $V$.
In this situation if $\AComp^\prime/R_2\quotle \AComp^\prime/R_1$
then $\AComp^\prime/R_2\morphle \AComp^\prime/R_1$
\end{property}
\begin{proof}
To Prove this property  we proceed as follows.
Since $\AComp^\prime/R_2\quotle \AComp^\prime/R_1$ we have that $R_1\subset R_2$.
Let $[v]_1$ and $[v]_2$ the equivalence classes of $v$ w.r.t.,
respectively, $R_1$ and $R_2$.
We have that $[v]_1\subset [v]_2$ for all $v\in V$ \iff\ $R_1\subset R_2$.
Since we have that $[v]_1\subset [v]_2$ for all $v\in V$
we can say that the covering ${\cal R}_1=\{{R_{[v]_1}}|v\in V\}$ 
is a refinement of  ${\cal R}_2=\{{R_{[v]_2}}|v\in V\}$.
Finally,  by Property \ref{pro:refmap}, we have that 
$\AComp^\prime/R_2\morphle \AComp^\prime/R_1$
\end{proof}{
It is easy to see that, 
the situation of Property \ref{pro:equiquot} (whenever $R_1\subset R_2$,
we have that Vertices in $\AComp^\prime/R_2$ are obtained as
union of Vertices from  $\AComp^\prime/R_2$, in
particular we have that ${R_{[v]_2}}=\cup_{u\in[v]_2}{R_{[u]_1}}$. 
}
We note that the converse is not true, in particular there exist
pairs of isomorphic \quot s $\AComp/R_2$ and $\AComp/R_1$ for which
$\AComp/R_2\morphle \AComp/R_1$ and $\AComp/R_1\morphle \AComp/R_2$
and neither $R_1\subset R_2$ nor $R_2\subset R_1$
(i.e. neither  $\AComp/R_2\quotle \AComp/R_1$ nor 
$\AComp/R_1\quotle \AComp/R_2$).
In the next example we present one of these situations.
{
\begin{figure}[h]
\begin{center}
\fbox{\psfig{file=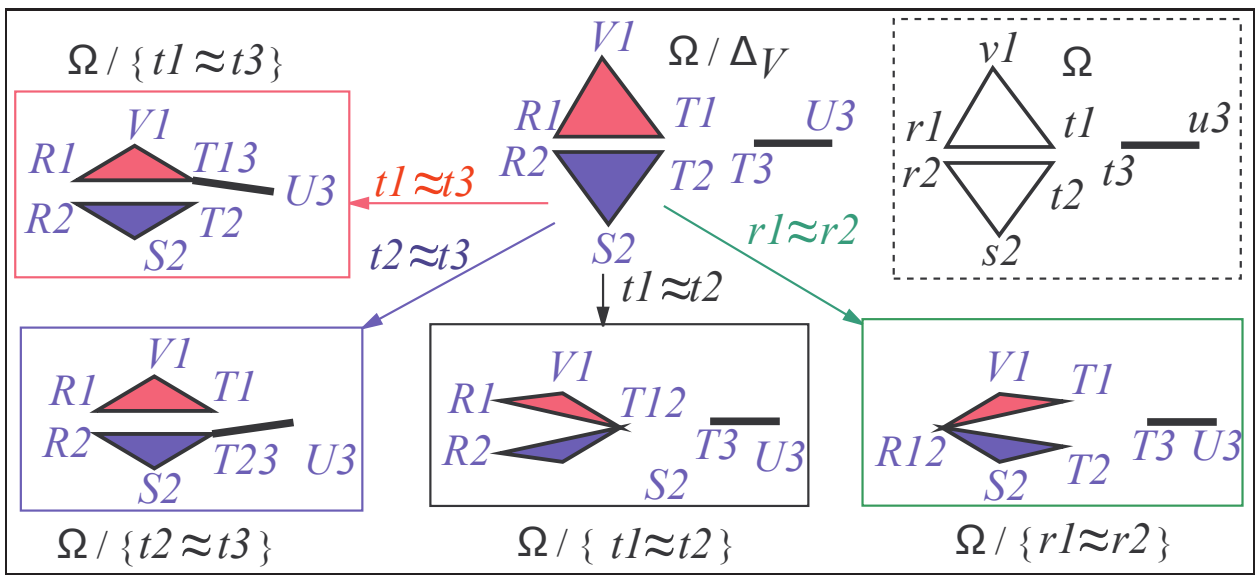,width=\textwidth}}
\end{center}
\caption{An example of a poset of \quot s}
\label{fig:latticequo1}
\end{figure}

\begin{example}
\label{ex:latticequo1}
As an example of application of the definitions above
consider Figure \ref{fig:latticequo1}.
In Figure \ref{fig:latticequo1} we sketched a portion of the poset
of \quot s $\AComp/R$.
The complex $\AComp$ (in the dashed frame on the top right)
is  the $2$-complex made up of the two triangles
$r1v1t1$, $r2v2t2$  and of the segment $t3u3$ .
We recall that Vertices in the \quot\ complex  $\AComp/R$ are collection of simplices 
and, for sake of clarity, we used for this collection a label of the form $Xn$ (i.e. V1 R1 etc.).
These labels are chosen following a certain convention.
Label $Vn$ is used for vertex $R_{[vn]}$ in the \quot\ $\AComp/\Delta_V$.
Note that vertex $Vn$ is a set of simplices from $\AComp$
(see Definition \ref{def:nerve} and \ref{def:quot}).
Label $Rxy$ is used for vertex $R_{[rx]}\cup R_{[ry]}$ and similarly
label $Txyz$ (in forthcoming examples) will be
used for  vertex $R_{[tx]}\cup R_{[ty]}\cup R_{[tz]}$.

With these assumption we can use the annotated Hasse diagram
of Figure  \ref{fig:latticequo1} to show the application of 
Property \ref{pro:uniqverquot} and 
of Property \ref{pro:equiquot}
(see Section \ref{sec:poset} in \latt\ for a definiton of Hasse diagram).
For our convenience we used arrows that are not provided by standard
Hasse diagrams.
Here and in the following we use the notation $\{t1\equivert t2\}$ 
to denote the smallest equivalence that contains $(t1,t2)$.

A first thing to note is this. If an arrow exist from the  $\AComp/R_1$ to 
$\AComp/R_2$ then  it must be $R_2\subset R_1$.
Consider for instance the complex $\AComp/\Delta_V$ and the
complex we labeled with {$\AComp/\{t1\equivert t2\}$}.
This complex  can be obtained
by stitching together the two triangles in {$\AComp/\Delta_V$}
at $t1$ and $t2$.
Looking at vertex labels in the \quot\ {$\AComp/\{t1\equivert t2\}$} we
can verify what stated in 
Property \ref{pro:uniqverquot} Part \ref{pro:quotmapclass}.
In fact all
Vertices in $\AComp/\Delta_V$ remains unchanged in
{$\AComp/\{t1\equivert t2\}$} but
Vertices labeled $T1$, $T2$ that maps to $T12$.

According to the notation introduced at the end of Property
\ref{pro:uniqverquot} we have that simplex $R2T23$ can be
denoted as  $r2t2/\{t2\equivert t3\}$.

In this figure each arrow denotes the application
of  the \asm\ foreseen by Property \ref{pro:equiquot}.
If an arrow is labeled with equation $vx\equivert vy$ 
it is easy ot see that the \asm\  is induced by
the vertex map $[\send{Vx,Vy}{Vxy}]$.
Arrows in this diagram represent the abstract simplicial maps
foreseen by  Property \ref{pro:equiquot} since
trivially, for any relation $R$, we have $\AComp/R\quotle\AComp/\Delta_V$.

Finally note that the four non trivial \quot s are pairwise isomorphic.
For instance \quot\ {$\AComp/\{t_1\equivert t3\}$} and 
{$\AComp/\{t_2\equivert t3\}$} are isomorphic and no
inclusion holds between the generating relations 
$\{t1\equivert t3\}$ and $\{t2\equivert t3\}$.
\end{example}

The \quot\ poset  becomes a lattice 
with the two operations $\sumdec$ (sum or join) and $\prodec$ (product or meet)
defined by:
$$(\AComp^\prime/{R_1})\prodec(\AComp^\prime/{R_2})=
\AComp^\prime/({{R_1}\sumeq{R_2}})\hspace{1cm}
(\AComp^\prime/{R_1})\sumdec(\AComp^\prime/{R_2})=
\AComp^\prime/({{R_1}\proeq{R_2}})$$
We will call this lattice the \emas{\quot}{lattice}.
This is anti-isomorphic w.r.t. the  partition lattice $\Pi_n$.

We found convenient to use arrows for lattice operators 
$\prodec$ and $\sumdec$ to support  intuition.
Indeed, according to the usual convention  in Hasse diagrams (see \latt\ Section \ref{sec:poset})
of having greater elements up we found  convenient to use
the symbol $\prodec$ for the  meet since  this is the greatest lower bound.
Similarly we choose the symbol
$\sumdec$ for the join because this is the least upper bound.
Note that in the \latt\ the basic results about Lattice Theory are reported using the more usual convention where the sum or join $\sumdec$ is  represented by the $\lub$ symbol.

We recall that $\AComp^\prime\morphle\AComp^\second$
if and only if  there exist a \asm\ $f$ such that $\AComp^\prime=f(\AComp^\second)$. 
Therefore, we defined the sense of  order $\morphle$ to be graphically coherent
with arrow for $f$. The \asm s goes from greater elements to smaller elements
w.r.t. $\morphle$ and $A\morphle B$ intuitively reads as 
{\em A is less exploded than B} or {\em B is more detailed than A}.

Note that passing from equivalences to complexes we have two corresponding
but opposite orders.
As a consequence
the sum of two  equivalences
${{        R_1}\sumeq{        R_2}}$ gives the product (meet)
among complexes $(\AComp/{        R_1})\prodec(\AComp/{        R_2})$.
Similarly the product of two  equivalences
${{        R_1}\proeq{        R_2}}$ gives the sum (join)
among complexes $(\AComp/{        R_1})\sumdec(\AComp/{        R_2})$.
Sometimes, in next sections, we will
prefer to use
$\AComp/({{        R_1}\sumeq{        R_2}})$ and
$\AComp/({{        R_1}\proeq{        R_2})}$,
instead of 
as $(\AComp/{        R_1})\sumdec(\AComp/{        R_2})$
and
$(\AComp/{        R_1})\prodec(\AComp/{        R_2})$
being usually  interested in the operations on the  equivalences that are 
behind the operations  $\sumdec$ and $\prodec$.
}
\begin{figure}[h]
\begin{center}
\fbox{\psfig{file=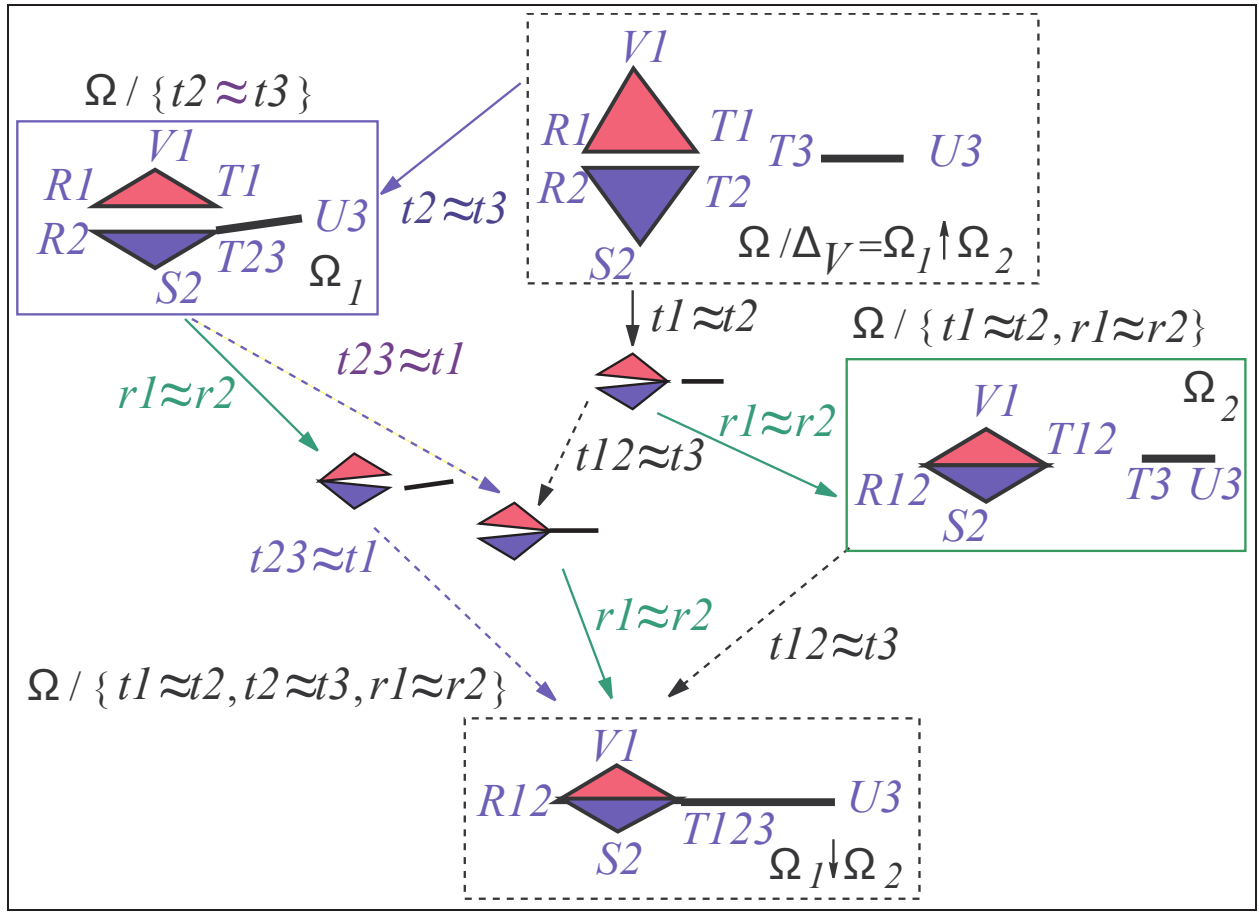,width=\textwidth}}
\end{center}
\caption{An example of portion of  a lattice of \quot s}
\label{fig:latticequo2}
\end{figure}
\begin{example}
\label{ex:latticequo2}
In Figure  \ref{fig:latticequo2} we present another portion of the poset of 
Figure \ref{fig:latticequo1}.
In this second figure we extend the form of arrow labels using 
also labels of the form $vxy\equivert vz$.
This means that the corresponding one hop \asm\ is induced by
the vertex map $[\send{Vx,Vy,Vz}{Vxyz}]$.
This second figure shows an instance of two
lattice operations, $\AComp_1\sumdec \AComp_2$ and 
$\AComp_1\prodec \AComp_2$, between the two \quot s 
$\AComp_1=\AComp/\{t2\equivert t3\}$ and 
$\AComp_2=\AComp/\{r1\equivert r2,1\equivert t2\}$.  
We note that we have just one hop from
$\AComp_1\sumdec \AComp_2$ to $\AComp_1$ and
just one hop from  $\AComp_2$ to $\AComp_1\prodec \AComp_2$.
This proves that these two complexes are, respectively, the least upper bound
and the greatest lower bound for the pair $\{\AComp_1,\AComp_2\}$.
In this figure we used the notation 
$\{t1\equivert t2, t2\equivert t3, r1\equivert r2\}$
to denote the smallest equivalence that equates $t1$ with $t2$ and $t3$ and
$r1$ with $r2$.
If we consider the two relations generating $\AComp_1$ and $\AComp_2$ 
(i.e. $\{t2\equivert t3\}$ and $\{t1\equivert t2, r1\equivert r2\}$),
then it is easy to verify that the least upper bound
and the greatest lower bound for the pair $\{\AComp_1,\AComp_2\}$ 
are given respectively
by the \quot\ of $\AComp$ 
with the intersection
(i.e. $\Delta_V=\{t2\equivert t3\}\cap\{t1\equivert t2, r1\equivert r2\}$) 
and with the 
sum  (i.e. $\{t1\equivert t2, t2\equivert t3, r1\equivert r2\}$) of 
the two equivalence relations generating $\AComp_1$ and $\AComp_2$. 
\end{example}

We have shown that the \quot\ lattice and the lattice of equivalences
are anti-isomorphic. It is quite easy to see that the set of 
simplices in an \asc\ $\AComp$, ordered by set inclusion, 
is a lattice, too.  This lattice  is usually called the 
{\em face lattice} $\AComp$. Lattice operations for this lattice
are usual set theoretic union (join) and intersection (meet). 
This explain why the name join is used for the operation
{$\gamma_1\join\gamma_2$}. However note that the join  
{$\gamma_1\join\gamma_2$}  from Combinatorial Topology  
is a partial version of the join in the   face  lattice.
In fact {$\gamma_1\join\gamma_2$} undefined whenever 
{$\gamma_1\cap\gamma_2\neq\emptyset$}.

The \quot\ operation is linear w.r.t. operations in the face lattice, 
More precisely the following identities holds:
\begin{property}
\label{pro:lattid}
Let $\gamma_1$ and $\gamma_2$ be two simplices in the \asc\ $\AComp$.
Let $V$ be the set of vertices in $\AComp$
and let $R$ be an equivalence relation on $V$.
In this situation the following identites holds:  
\begin{eqnarray}
\frac{\gamma_1}{R}\cup\frac{\gamma_2}{R}&=&\frac{\gamma_1\cup\gamma_2}{R}\\
\frac{\gamma_1}{R}\cap\frac{\gamma_2}{R}&=&\frac{\gamma_1\cap\gamma_2}{R}\\
\gamma_1\le\gamma_2&\Rightarrow&\frac{\gamma_1}{R}\le\frac{\gamma_2}{R}
\label{eg:min}
\end{eqnarray}
\end{property}

In this framework we are interested also 
in the extension of the \quot\ notation
(i.e. $\AComp/{\equivert}$) to the composition of two or more equivalences in
$V \times V$. Note that if $\equivert_1$ and $\equivert_2$ are two relations
in $V \times V$ the composition of two \quot s 
$(\AComp/{\equivert_1})/{\equivert_2}$ is undefined. Indeed, using the 
previous definitions, we can attach a meaning to  
$(\AComp/{\equivert_1})/{\equivert_2}$ \iff\ the relation $\equivert_2$ is 
a relation among Vertices of $\AComp/{\equivert_1}$.
This is not used at all and, usually, both
$\equivert_1$ and $\equivert_2$ are relations between the Vertices
of $\AComp$. In this latter situation we use the notation
$(\AComp/{\equivert_1})/{\equivert_2}$ 
after the following definitions.
\begin{definition}
\label{def:quotquot}
Let $\AComp$ be an abstract simplicial complex with Vertices in $V$ and
let $\gamma$ be a simplex in $\AComp$. Let  $\equivert_1$ and $\equivert_2$
be two relations in $V \times V$
(i.e. .$\equivert_1\subset V \times V$ and $\equivert_2\subset V \times V$).
In this situation we define:
\[
\begin{array}{ccccc}
(\AComp/{\equivert_1})/{\equivert_2}&=&
\AComp/({\equivert_1}\sumeq{\equivert_2})&=&
(\AComp/{\equivert_2})/{\equivert_1};
\\
(\gamma/{\equivert_1})/{\equivert_2}&=&
\gamma/({\equivert_1}\sumeq{\equivert_2})&=& 
(\gamma/{\equivert_2})/{\equivert_1};
\end{array}
\]
\end{definition}

\section{\Verteq\ Equations}{
We close this chapter with a discussion on  the relation between 
equivalence and sets of equations. We already used 
equations (e.g. $\{t2\equivert t3\}$) in examples. 
In this section we will study the relation between equations and the
structure of the \quot\ lattice.
Indeed the partition lattice, and so the lattice of equivalences and
the \quot\ lattice, belongs
to a special class of lattices called {\em geometric lattices}.
The basic property of a finite geometric lattice is that each element
in a geometric lattice
can be expressed as the join of a finite number of a set of generators
called {\em points}.
Thus, for the lattice of equivalence relations the 
{\em points} are those equivalence relations
that are generated by a single equation `(e.g. $\{t2\equivert t3\}$). 
Obviously, in this thesis, we do not call them points to avoid confusion with 
the Vertices in \asc.
We recall that an \asc\ $\AComp$, ordered with the 
face relations is a lattice called the {\em face lattice}.
The face lattice  is geometric lattice, too and 
$0$-simplices $\{v\}$ in $\AComp$ are the points for this lattice.
A short introduction to these concepts is in Section \ref{sec:pin}
of \latt.
}

In the following we will use 
expressions of the form {$u\equivert v$}, called \emas{\verteq}{equations},
to denote the smallest equivalence that contains the couple $(u,v)$. 
As a consequence {$u\equivert v$} and {$v\equivert u$} will denote the
same object.
We will say that the equivalence ${\equivert}\subset V\times V$ 
\ems{satisfies} the equation $u\equivert v$  if and only if 
$(u,v)\in{\equivert}$.
We will write $u\not\equivert v$ if and only if $(u,v)\not\in{\equivert}$.
In this case we will say that the equivalence $\equivert$ do not
satisfies the equation $u\equivert v$.
An equivalence $\equivert$ is usually  given by a set of equations
of the form $E=\{u_i\equivert v_i|i=1,\ldots k\}$. The equivalence
given by such a set of equations $E$, denoted
by $\equivert^E$, is  the
the smallest equivalence $\equivert^E$
such that $\equivert^E$ satisfies all the equations in $E$.
(i.e. $u_i\equivert^E v_i$ for $i=1,\ldots,k$).
We will say that $E$ denotes or generates $\equivert^E$

If  $E$ is a  set of equations that generates a
an equivalence, denoted by $\equivert^E$,
we will use $E$ as a shortcut for $\equivert^E$ in all the expressions
where this is not ambiguous. In particular we will
write $\equivert\sumeq \;E$ and
$\equivert\proeq\; E$ as a shortcut
for $\equivert\sumeq \equivert^E$  and $\equivert\proeq \equivert^E$.
Similarly we will use $\AComp/E$ as a shortcut for $\AComp/{\equivert^E}$
and $\gamma/E$ as a shortcut for $\gamma/\equivert^E$.

\subsection{Independent Equations}
We can generate an equivalence using several sets of equations. However
there exists sets of equations that are, in some sense, redundant.
The concept of {\em redundancy} is perfectly captured  by the  
specific notion of {\em independence} in geometric lattices. 
In this subsection we will adapt this notion to our
particular framework and report the related results (without proof).
Section \ref{sec:lattice} in the \latt\ and
in particular the material following Definition \ref{def:indip}
discuss  the general notion of
independence in geometric lattices and lists the related results.

A set of equations $E$ is \ems{redundant} if the equivalence $\equivert^E$
can be generated by a subset of $E$.
A non redundant set of equations is called  
a set of \emas{independent}{equations}.
All non redundant sets of equations that generates the same equivalence 
contains the same number of equations. This number is called 
the \ems{rank} index of the equivalence. If ${\equivert_2}$ is the immediate 
superior of $\equivert_1$ then there exist an equation $u\equivert v$
s.t. $u\not\equivert_1 v$ and 
${\equivert_1}+{\{u\equivert v\}}={\equivert_2}$.
In this case we will say that ${\{u\equivert v\}}$ {\em labels} the
one hop chain from
${\equivert_1}$ to ${\equivert_2}$ 
(or from $\AComp/{\equivert_2}$ to $\AComp/{\equivert_1}$).
Longer chains will be labeled by sequences  of independent equations.
Note that same chain can be labeled by different labels and the same label
can be used to label different chains in the \quot\ lattice.

In general some care must be taken not to overlook this analogy
between complexes, equivalences and equations.
Indeed there is not a one to one correspondence between set of equations and
equivalences.
So, for instance, $\equivert^{E_1}=\equivert^{E_2}+\equivert^E$ 
(that can also be written as $E_1=E_2+E$) do not implies that
$E_1= E_2\cup E$ 

{
\begin{figure}[h]
\begin{center}
\fbox{\psfig{file=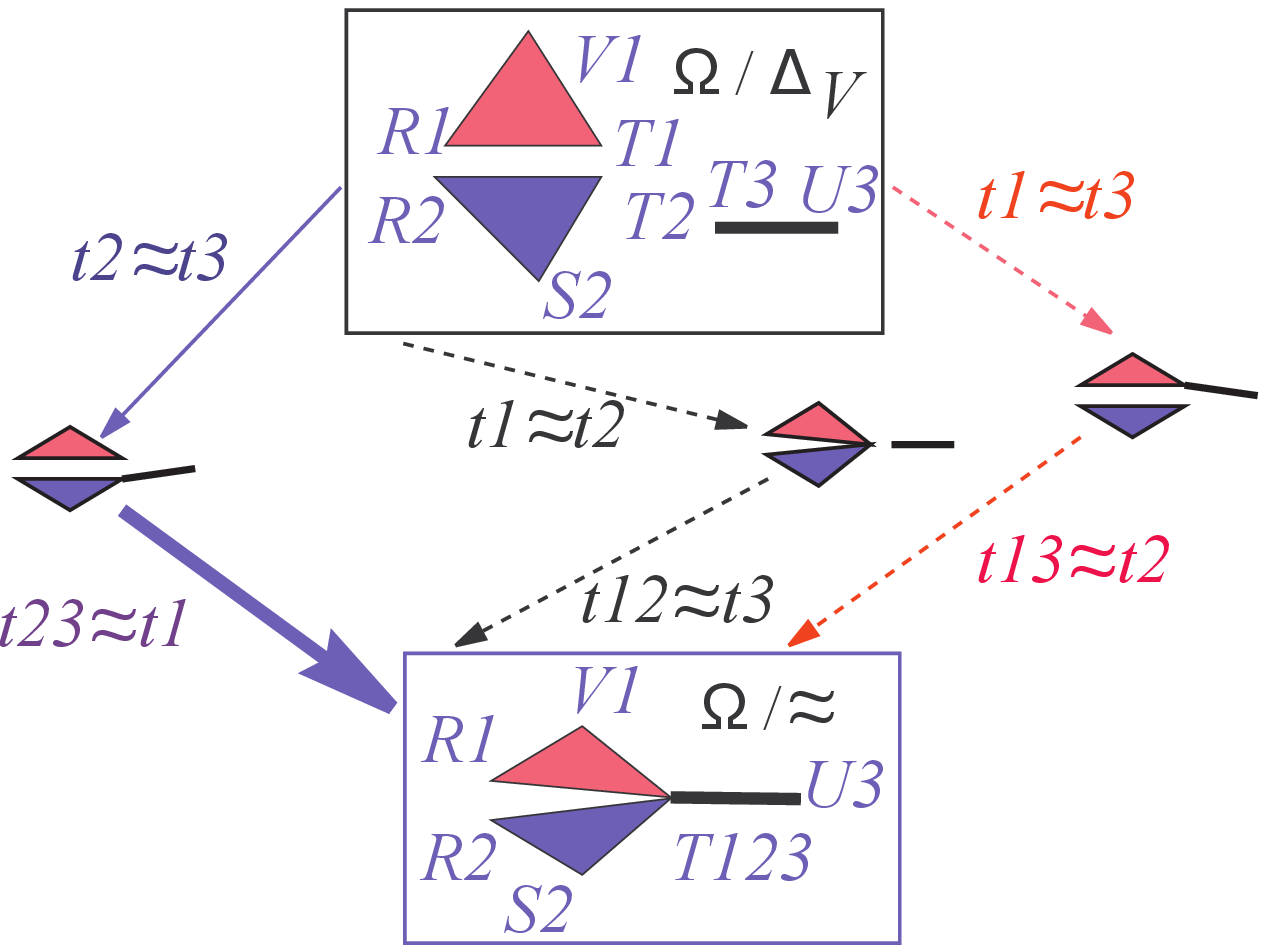,width=.7\textwidth}}
\end{center}
\caption{Different sets of equations can generate  equivalence $\equivert$
in $\AComp/{\equivert}$ 
(See Example  \ref{ex:equats}) 
}
\label{fig:lattport}
\end{figure}
\begin{example}
\label{ex:equats}
Note that different sets of equations can be associated
to the equivalence $\equivert$.  Let us consider
the portion of the \quot\ lattice in Figure \ref{fig:lattport}.
In this figure we use the label $vxy\equivert vz$ to say that the 
corresponding one hop path can be labeled either with $vx\equivert vz$ or by 
$vy\equivert vz$.
Looking at the three paths  (solid or dashed) in Figure \ref{fig:lattport}
we have that several pairs of equations can be used to generate the
equivalence $\equivert$. According to the labeling conventions
we can build  a set of equations for $\equivert$
by collecting equations that labels the one hop steps in 
the Hasse diagram of Figure  \ref{fig:lattport}.
We just have to collect labels on a path from $\AComp/\Delta_V$
down to $\AComp/{\equivert}$. Following the solid path we can take 
$E=\{t1\equivert t3, t3\equivert t2\}$
or
$E^\prime=\{t1\equivert t3, t1\equivert t2\}$ and
still have ${\equivert}={\equivert^E}={\equivert^{E^\prime}}$.
These two variants  are obtained 
since, from the second (thick) solid arrow,  we can get either equation 
$t3\equivert t2$ or equation $t1\equivert t2$.
The {\em same} sets  comes from the two dashed paths on the right of
Figure \ref{fig:lattport}.
Following this path  we always collect one of the two sets $E$ or $E^\prime$.
This fact is a consequence of a general property of geometric 
lattices called {\em semimodularity}. This property will be introduced 
in the following subsection.

Finally, as an example of application  of Definition \ref{def:quotquot},
we note that following the solid path we can write $\AComp/{\equivert}$
as $(\AComp/\{t1\equivert t3\})/\{t1\equivert t2\}$.
\end{example}
}

\subsection{Semimodularity}
Although, different labels are possible for a 
path, even for a one hop path (see, for instance, the solid
thick edge in Figure \ref{fig:lattport}),
nevertheless, we have that the size of the set of independent equations that labels a path
must be constant.  In general, let us consider two a starting point 
{$\AComp/{\equivert^{E_x}}$} and an ending point {$\AComp/{ \equivert^{E_y}}$} (or shortly
{$\equivert^{E_x}$} and {$\equivert^{E_y}$})
we can collect different labels traveling through different paths
from {$\equivert^{E_x}$} to {$\equivert^{E_y}$}.
However, we always have the option to collect the  {\em same} 
labels (i.e. the same unordered set of equations) traveling through  different paths from {$\equivert^{E_x}$} to 
{$\equivert^{E_y}$}. 
This  is a consequence of 
a general property for geometric lattices called {\em semimodularity}.
Semimodularity is formally introduced in
\latt\ Section  \ref{sec:pin}, Definition \ref{def:semi}. In the Appendix, together 
with semimodularity definition, we list relevant properties for semimodular latices.
The partition lattice  $\Pi_n$, being a geometric lattice, is semimodular.
Thus the \quot\ lattice, being anti-isomorphic w.r.t. the partition lattice,
is semimodular, too.

In this framework semimodularity gives us an interesting property for labels, 
(and for equations).
{In fact, in general, extending the construction of the diagram of
Example  \ref{ex:equats} to a larger 
portion of the \quot\ lattice one can prove that, 
given two equivalences $E_a$ and $E_b$ 
(note we use $E_a$ as a shorthand for $\equivert^{E_a}$) 
with a common  upper  bound $E_{x}$ and a  common lower bound $E_{y}$ 
we can use the same 
label $E_{xa}$ for the chain from $E_{x}$ to $E_{a}$ and for the chain from 
$E_{b}$ to $E_{y}$.
Similarly 
we can use the same 
label $E_{xb}$
for the chain from $E_{x}$ to $E_{b}$ and for the chain from $E_{a}$ to $E_{y}$.
With reference to the situation depicted in the following diagram we can say that semimodularity ensures that {\em parallel} arrows can receive the same labeling :

\edef\thinlines{\the\catcode`@ }\catcode`@ = 11
\let\@oldatcatcode = \thinlines

\def\smash@@{\relax \ifmmode\def\next{\mathpalette\mathsm@sh}\else\let\next\makesm@sh
  \fi\next}
\def\makesm@sh#1{\setbox\z@\hbox{#1}\finsm@sh}
\def\mathsm@sh#1#2{\setbox\z@\hbox{$\m@th#1{#2}$}\finsm@sh}
\def\finsm@sh{\ht\z@\z@ \dp\z@\z@ \box\z@}

\edef\@oldandcatcode{\the\catcode`& }\catcode`& = 11

\def\&whilenoop#1{}\def\&whiledim#1\do #2{\ifdim #1\relax#2\&iwhiledim{#1\relax#2}\fi}\def\&iwhiledim#1{\ifdim #1\let\&nextwhile=\&iwhiledim 
        \else\let\&nextwhile=\&whilenoop\fi\&nextwhile{#1}}

\newif\if&negarg
\newdimen\&wholewidth
\newdimen\&halfwidth

\font\tenln=line10

\def\thinlines{\let\&linefnt\tenln \let\&circlefnt\tencirc
  \&wholewidth\fontdimen8\tenln \&halfwidth .5\&wholewidth}\def\thicklines{\let\&linefnt\tenlnw \let\&circlefnt\tencircw
  \&wholewidth\fontdimen8\tenlnw \&halfwidth .5\&wholewidth}

\def\drawline(#1,#2)#3{\&xarg #1\relax \&yarg #2\relax \&linelen=#3\relax
  \ifnum\&xarg =0 \&vline \else \ifnum\&yarg =0 \&hline \else \&sline\fi\fi}

\def\&sline{\leavevmode
  \ifnum\&xarg< 0 \&negargtrue \&xarg -\&xarg \&yyarg -\&yarg
  \else \&negargfalse \&yyarg \&yarg \fi
  \ifnum \&yyarg >0 \&tempcnta\&yyarg \else \&tempcnta -\&yyarg \fi
  \ifnum\&tempcnta>6 \&badlinearg \&yyarg0 \fi
  \ifnum\&xarg>6 \&badlinearg \&xarg1 \fi
  \setbox\&linechar\hbox{\&linefnt\&getlinechar(\&xarg,\&yyarg)}\ifnum \&yyarg >0 \let\&upordown\raise \&clnht\z@
  \else\let\&upordown\lower \&clnht \ht\&linechar\fi
  \&clnwd=\wd\&linechar
  \&whiledim \&clnwd <\&linelen \do {\&upordown\&clnht\copy\&linechar
    \advance\&clnht \ht\&linechar
    \advance\&clnwd \wd\&linechar
  }\advance\&clnht -\ht\&linechar
  \advance\&clnwd -\wd\&linechar
  \&tempdima\&linelen\advance\&tempdima -\&clnwd
  \&tempdimb\&tempdima\advance\&tempdimb -\wd\&linechar
  \hskip\&tempdimb \multiply\&tempdima \@m
  \&tempcnta \&tempdima \&tempdima \wd\&linechar \divide\&tempcnta \&tempdima
  \&tempdima \ht\&linechar \multiply\&tempdima \&tempcnta
  \divide\&tempdima \@m
  \advance\&clnht \&tempdima
  \ifdim \&linelen <\wd\&linechar \hskip \wd\&linechar
  \else\&upordown\&clnht\copy\&linechar\fi}

\def\&hline{\vrule height \&halfwidth depth \&halfwidth width \&linelen}

\def\&getlinechar(#1,#2){\&tempcnta#1\relax\multiply\&tempcnta 8
  \advance\&tempcnta -9 \ifnum #2>0 \advance\&tempcnta #2\relax\else
  \advance\&tempcnta -#2\relax\advance\&tempcnta 64 \fi
  \char\&tempcnta}

\def\drawvector(#1,#2)#3{\&xarg #1\relax \&yarg #2\relax
  \&tempcnta \ifnum\&xarg<0 -\&xarg\else\&xarg\fi
  \ifnum\&tempcnta<5\relax \&linelen=#3\relax
    \ifnum\&xarg =0 \&vvector \else \ifnum\&yarg =0 \&hvector
    \else \&svector\fi\fi\else\&badlinearg\fi}

\def\&hvector{\ifnum\&xarg<0 \rlap{\&linefnt\&getlarrow(1,0)}\fi \&hline
  \ifnum\&xarg>0 \llap{\&linefnt\&getrarrow(1,0)}\fi}

\def\&vvector{\ifnum \&yarg <0 \&downvector \else \&upvector \fi}

\def\&svector{\&sline
  \&tempcnta\&yarg \ifnum\&tempcnta <0 \&tempcnta=-\&tempcnta\fi
  \ifnum\&tempcnta <5 
    \if&negarg\ifnum\&yarg>0                   \llap{\lower\ht\&linechar\hbox to\&linelen{\&linefnt
        \&getlarrow(\&xarg,\&yyarg)\hss}}\else \llap{\hbox to\&linelen{\&linefnt\&getlarrow(\&xarg,\&yyarg)\hss}}\fi
    \else\ifnum\&yarg>0                        \&tempdima\&linelen \multiply\&tempdima\&yarg
      \divide\&tempdima\&xarg \advance\&tempdima-\ht\&linechar
      \raise\&tempdima\llap{\&linefnt\&getrarrow(\&xarg,\&yyarg)}\else
      \&tempdima\&linelen \multiply\&tempdima-\&yarg \divide\&tempdima\&xarg
      \lower\&tempdima\llap{\&linefnt\&getrarrow(\&xarg,\&yyarg)}\fi\fi
  \else\&badlinearg\fi}

\def\&getlarrow(#1,#2){\ifnum #2 =\z@ \&tempcnta='33\else
\&tempcnta=#1\relax\multiply\&tempcnta \sixt@@n \advance\&tempcnta
-9 \&tempcntb=#2\relax\multiply\&tempcntb \tw@
\ifnum \&tempcntb >0 \advance\&tempcnta \&tempcntb\relax
\else\advance\&tempcnta -\&tempcntb\advance\&tempcnta 64
\fi\fi\char\&tempcnta}

\def\&getrarrow(#1,#2){\&tempcntb=#2\relax
\ifnum\&tempcntb < 0 \&tempcntb=-\&tempcntb\relax\fi
\ifcase \&tempcntb\relax \&tempcnta='55 \or 
\ifnum #1<3 \&tempcnta=#1\relax\multiply\&tempcnta
24 \advance\&tempcnta -6 \else \ifnum #1=3 \&tempcnta=49
\else\&tempcnta=58 \fi\fi\or 
\ifnum #1<3 \&tempcnta=#1\relax\multiply\&tempcnta
24 \advance\&tempcnta -3 \else \&tempcnta=51\fi\or 
\&tempcnta=#1\relax\multiply\&tempcnta
\sixt@@n \advance\&tempcnta -\tw@ \else
\&tempcnta=#1\relax\multiply\&tempcnta
\sixt@@n \advance\&tempcnta 7 \fi\ifnum #2<0 \advance\&tempcnta 64 \fi
\char\&tempcnta}

\def\&vline{\ifnum \&yarg <0 \&downline \else \&upline\fi}

\def\&upline{\hbox to \z@{\hskip -\&halfwidth \vrule width \&wholewidth
   height \&linelen depth \z@\hss}}

\def\&downline{\hbox to \z@{\hskip -\&halfwidth \vrule width \&wholewidth
   height \z@ depth \&linelen \hss}}

\def\&upvector{\&upline\setbox\&tempboxa\hbox{\&linefnt\char'66}\raise 
     \&linelen \hbox to\z@{\lower \ht\&tempboxa\box\&tempboxa\hss}}

\def\&downvector{\&downline\lower \&linelen
      \hbox to \z@{\&linefnt\char'77\hss}}

\def\&badlinearg{\errmessage{Bad \string\arrow\space argument.}}

\thinlines

\countdef\&xarg     0
\countdef\&yarg     2
\countdef\&yyarg    4
\countdef\&tempcnta 6
\countdef\&tempcntb 8

\dimendef\&linelen  0
\dimendef\&clnwd    2
\dimendef\&clnht    4
\dimendef\&tempdima 6
\dimendef\&tempdimb 8

\chardef\@arrbox    0
\chardef\&linechar  2
\chardef\&tempboxa  2

\let\lft^\let\rt_

\newif\if@pslope \def\@findslope(#1,#2){\ifnum#1>0
  \ifnum#2>0 \@pslopetrue \else\@pslopefalse\fi \else
  \ifnum#2>0 \@pslopefalse \else\@pslopetrue\fi\fi}

\def\generalsmap(#1,#2){\getm@rphposn(#1,#2)\plnmorph\futurelet\next\addm@rph}

\def\sline(#1,#2){\setbox\@arrbox=\hbox{\drawline(#1,#2){\sarrowlength}}\@findslope(#1,#2)\d@@blearrfalse\generalsmap(#1,#2)}\def\arrow(#1,#2){\setbox\@arrbox=\hbox{\drawvector(#1,#2){\sarrowlength}}\@findslope(#1,#2)\d@@blearrfalse\generalsmap(#1,#2)}

\newif\ifd@@blearr

\def\bisline(#1,#2){\@findslope(#1,#2)\if@pslope \let\@upordown\raise \else \let\@upordown\lower\fi
  \getch@nnel(#1,#2)\setbox\@arrbox=\hbox{\@upordown\@vchannel
    \rlap{\drawline(#1,#2){\sarrowlength}}\hskip\@hchannel\hbox{\drawline(#1,#2){\sarrowlength}}}\d@@blearrtrue\generalsmap(#1,#2)}\def\biarrow(#1,#2){\@findslope(#1,#2)\if@pslope \let\@upordown\raise \else \let\@upordown\lower\fi
  \getch@nnel(#1,#2)\setbox\@arrbox=\hbox{\@upordown\@vchannel
    \rlap{\drawvector(#1,#2){\sarrowlength}}\hskip\@hchannel\hbox{\drawvector(#1,#2){\sarrowlength}}}\d@@blearrtrue\generalsmap(#1,#2)}\def\adjarrow(#1,#2){\@findslope(#1,#2)\if@pslope \let\@upordown\raise \else \let\@upordown\lower\fi
  \getch@nnel(#1,#2)\setbox\@arrbox=\hbox{\@upordown\@vchannel
    \rlap{\drawvector(#1,#2){\sarrowlength}}\hskip\@hchannel\hbox{\drawvector(-#1,-#2){\sarrowlength}}}\d@@blearrtrue\generalsmap(#1,#2)}

\newif\ifrtm@rph
\def\@shiftmorph#1{\hbox{\setbox0=\hbox{$\scriptstyle#1$}\setbox1=\hbox{\hskip\@hm@rphshift\raise\@vm@rphshift\copy0}\wd1=\wd0 \ht1=\ht0 \dp1=\dp0 \box1}}\def\@hm@rphshift{\ifrtm@rph
  \ifdim\hmorphposnrt=\z@\hmorphposn\else\hmorphposnrt\fi \else
  \ifdim\hmorphposnlft=\z@\hmorphposn\else\hmorphposnlft\fi \fi}\def\@vm@rphshift{\ifrtm@rph
  \ifdim\vmorphposnrt=\z@\vmorphposn\else\vmorphposnrt\fi \else
  \ifdim\vmorphposnlft=\z@\vmorphposn\else\vmorphposnlft\fi \fi}

\def\addm@rph{\ifx\next\lft\let\temp=\lftmorph\else
  \ifx\next\rt\let\temp=\rtmorph\else\let\temp\relax\fi\fi \temp}

\def\plnmorph{\dimen1\wd\@arrbox \ifdim\dimen1<\z@ \dimen1-\dimen1\fi
  \vcenter{\box\@arrbox}}\def\lftmorph\lft#1{\rtm@rphfalse \setbox0=\@shiftmorph{#1}\if@pslope \let\@upordown\raise \else \let\@upordown\lower\fi
  \llap{\@upordown\@vmorphdflt\hbox to\dimen1{\hss \llap{\box0}\hss}\hskip\@hmorphdflt}\futurelet\next\addm@rph}\def\rtmorph\rt#1{\rtm@rphtrue \setbox0=\@shiftmorph{#1}\if@pslope \let\@upordown\lower \else \let\@upordown\raise\fi
  \llap{\@upordown\@vmorphdflt\hbox to\dimen1{\hss
    \rlap{\box0}\hss}\hskip-\@hmorphdflt}\futurelet\next\addm@rph}

\def\getm@rphposn(#1,#2){\ifd@@blearr \dimen@\morphdist \advance\dimen@ by
  .5\channelwidth \@getshift(#1,#2){\@hmorphdflt}{\@vmorphdflt}{\dimen@}\else
  \@getshift(#1,#2){\@hmorphdflt}{\@vmorphdflt}{\morphdist}\fi}

\def\getch@nnel(#1,#2){\ifdim\hchannel=\z@ \ifdim\vchannel=\z@
    \@getshift(#1,#2){\@hchannel}{\@vchannel}{\channelwidth}\else \@hchannel\hchannel \@vchannel\vchannel \fi
  \else \@hchannel\hchannel \@vchannel\vchannel \fi}

\def\@getshift(#1,#2)#3#4#5{\dimen@ #5\relax
  \&xarg #1\relax \&yarg #2\relax
  \ifnum\&xarg<0 \&xarg -\&xarg \fi
  \ifnum\&yarg<0 \&yarg -\&yarg \fi
  \ifnum\&xarg<\&yarg \&negargtrue \&yyarg\&xarg \&xarg\&yarg \&yarg\&yyarg\fi
  \ifcase\&xarg \or  \ifcase\&yarg    \dimen@i \z@ \dimen@ii \dimen@ \or \dimen@i .7071\dimen@ \dimen@ii .7071\dimen@ \fi \or
    \ifcase\&yarg    \or \dimen@i .4472\dimen@ \dimen@ii .8944\dimen@ \fi \or
    \ifcase\&yarg    \or \dimen@i .3162\dimen@ \dimen@ii .9486\dimen@ \or
      \dimen@i .5547\dimen@ \dimen@ii .8321\dimen@ \fi \or
    \ifcase\&yarg    \or \dimen@i .2425\dimen@ \dimen@ii .9701\dimen@ \or\or
      \dimen@i .6\dimen@ \dimen@ii .8\dimen@ \fi \or
    \ifcase\&yarg    \or \dimen@i .1961\dimen@ \dimen@ii .9801\dimen@ \or
      \dimen@i .3714\dimen@ \dimen@ii .9284\dimen@ \or
      \dimen@i .5144\dimen@ \dimen@ii .8575\dimen@ \or
      \dimen@i .6247\dimen@ \dimen@ii .7801\dimen@ \fi \or
    \ifcase\&yarg    \or \dimen@i .1645\dimen@ \dimen@ii .9864\dimen@ \or\or\or\or
      \dimen@i .6402\dimen@ \dimen@ii .7682\dimen@ \fi \fi
  \if&negarg \&tempdima\dimen@i \dimen@i\dimen@ii \dimen@ii\&tempdima\fi
  #3\dimen@i\relax #4\dimen@ii\relax }

\catcode`\&=4

\def\generalhmap{\futurelet\next\@generalhmap}\def\@generalhmap{\ifx\next^ \let\temp\generalhm@rph\else
  \ifx\next_ \let\temp\generalhm@rph\else \let\temp\m@kehmap\fi\fi \temp}\def\generalhm@rph#1#2{\ifx#1^
    \toks@=\expandafter{\the\toks@#1{\rtm@rphtrue\@shiftmorph{#2}}}\else
    \toks@=\expandafter{\the\toks@#1{\rtm@rphfalse\@shiftmorph{#2}}}\fi
  \generalhmap}\def\m@kehmap{\mathrel{\smash@@{\the\toks@}}}

\def\mapright{\toks@={\mathop{\vcenter{\smash@@{\drawrightarrow}}}\limits}\generalhmap}\def\mapleft{\toks@={\mathop{\vcenter{\smash@@{\drawleftarrow}}}\limits}\generalhmap}\def\bimapright{\toks@={\mathop{\vcenter{\smash@@{\drawbirightarrow}}}\limits}\generalhmap}\def\bimapleft{\toks@={\mathop{\vcenter{\smash@@{\drawbileftarrow}}}\limits}\generalhmap}\def\adjmapright{\toks@={\mathop{\vcenter{\smash@@{\drawadjrightarrow}}}\limits}\generalhmap}\def\adjmapleft{\toks@={\mathop{\vcenter{\smash@@{\drawadjleftarrow}}}\limits}\generalhmap}\def\hline{\toks@={\mathop{\vcenter{\smash@@{\drawhline}}}\limits}\generalhmap}\def\bihline{\toks@={\mathop{\vcenter{\smash@@{\drawbihline}}}\limits}\generalhmap}

\def\drawrightarrow{\hbox{\drawvector(1,0){\harrowlength}}}\def\drawleftarrow{\hbox{\drawvector(-1,0){\harrowlength}}}\def\drawbirightarrow{\hbox{\raise.5\channelwidth
  \hbox{\drawvector(1,0){\harrowlength}}\lower.5\channelwidth
  \llap{\drawvector(1,0){\harrowlength}}}}\def\drawbileftarrow{\hbox{\raise.5\channelwidth
  \hbox{\drawvector(-1,0){\harrowlength}}\lower.5\channelwidth
  \llap{\drawvector(-1,0){\harrowlength}}}}\def\drawadjrightarrow{\hbox{\raise.5\channelwidth
  \hbox{\drawvector(-1,0){\harrowlength}}\lower.5\channelwidth
  \llap{\drawvector(1,0){\harrowlength}}}}\def\drawadjleftarrow{\hbox{\raise.5\channelwidth
  \hbox{\drawvector(1,0){\harrowlength}}\lower.5\channelwidth
  \llap{\drawvector(-1,0){\harrowlength}}}}\def\drawhline{\hbox{\drawline(1,0){\harrowlength}}}\def\drawbihline{\hbox{\raise.5\channelwidth
  \hbox{\drawline(1,0){\harrowlength}}\lower.5\channelwidth
  \llap{\drawline(1,0){\harrowlength}}}}

\def\generalvmap{\futurelet\next\@generalvmap}\def\@generalvmap{\ifx\next\lft \let\temp\generalvm@rph\else
  \ifx\next\rt \let\temp\generalvm@rph\else \let\temp\m@kevmap\fi\fi \temp}\toksdef\toks@@=1
\def\generalvm@rph#1#2{\ifx#1\rt \toks@=\expandafter{\the\toks@
      \rlap{$\vcenter{\rtm@rphtrue\@shiftmorph{#2}}$}}\else \toks@@={\llap{$\vcenter{\rtm@rphfalse\@shiftmorph{#2}}$}}\toks@=\expandafter\expandafter\expandafter{\expandafter\the\expandafter
      \toks@@ \the\toks@}\fi \generalvmap}\def\m@kevmap{\the\toks@}

\def\mapdown{\toks@={\vcenter{\drawdownarrow}}\generalvmap}\def\mapup{\toks@={\vcenter{\drawuparrow}}\generalvmap}\def\bimapdown{\toks@={\vcenter{\drawbidownarrow}}\generalvmap}\def\bimapup{\toks@={\vcenter{\drawbiuparrow}}\generalvmap}\def\adjmapdown{\toks@={\vcenter{\drawadjdownarrow}}\generalvmap}\def\adjmapup{\toks@={\vcenter{\drawadjuparrow}}\generalvmap}\def\vline{\toks@={\vcenter{\drawvline}}\generalvmap}\def\bivline{\toks@={\vcenter{\drawbivline}}\generalvmap}

\def\drawdownarrow{\hbox to5pt{\hss\drawvector(0,-1){\varrowlength}\hss}}\def\drawuparrow{\hbox to5pt{\hss\drawvector(0,1){\varrowlength}\hss}}\def\drawbidownarrow{\hbox to5pt{\hss\hbox{\drawvector(0,-1){\varrowlength}}\hskip\channelwidth\hbox{\drawvector(0,-1){\varrowlength}}\hss}}\def\drawbiuparrow{\hbox to5pt{\hss\hbox{\drawvector(0,1){\varrowlength}}\hskip\channelwidth\hbox{\drawvector(0,1){\varrowlength}}\hss}}\def\drawadjdownarrow{\hbox to5pt{\hss\hbox{\drawvector(0,-1){\varrowlength}}\hskip\channelwidth\lower\varrowlength
  \hbox{\drawvector(0,1){\varrowlength}}\hss}}\def\drawadjuparrow{\hbox to5pt{\hss\hbox{\drawvector(0,1){\varrowlength}}\hskip\channelwidth\raise\varrowlength
  \hbox{\drawvector(0,-1){\varrowlength}}\hss}}\def\drawvline{\hbox to5pt{\hss\drawline(0,1){\varrowlength}\hss}}\def\drawbivline{\hbox to5pt{\hss\hbox{\drawline(0,1){\varrowlength}}\hskip\channelwidth\hbox{\drawline(0,1){\varrowlength}}\hss}}

\def\commdiag#1{\null\,
  \vcenter{\commdiagbaselines
  \m@th\ialign{\hfil$##$\hfil&&\hfil$\mkern4mu ##$\hfil\crcr
      \mathstrut\crcr\noalign{\kern-\baselineskip}
      #1\crcr\mathstrut\crcr\noalign{\kern-\baselineskip}}}\,}

\def\commdiagbaselines{\baselineskip15pt \lineskip3pt \lineskiplimit3pt }\def\gridcommdiag#1{\null\,
  \vcenter{\offinterlineskip
  \m@th\ialign{&\vbox to\vgrid{\vss
    \hbox to\hgrid{\hss\smash@@{$##$}\hss}}\crcr
      \mathstrut\crcr\noalign{\kern-\vgrid}
      #1\crcr\mathstrut\crcr\noalign{\kern-.5\vgrid}}}\,}

\newdimen\harrowlength \harrowlength=60pt
\newdimen\varrowlength \varrowlength=.618\harrowlength
\newdimen\sarrowlength \sarrowlength=\harrowlength

\newdimen\hmorphposn \hmorphposn=\z@
\newdimen\vmorphposn \vmorphposn=\z@
\newdimen\morphdist  \morphdist=4pt

\dimendef\@hmorphdflt 0       \dimendef\@vmorphdflt 2       

\newdimen\hmorphposnrt  \hmorphposnrt=\z@
\newdimen\hmorphposnlft \hmorphposnlft=\z@
\newdimen\vmorphposnrt  \vmorphposnrt=\z@
\newdimen\vmorphposnlft \vmorphposnlft=\z@
\let\hmorphposnup=\hmorphposnrt
\let\hmorphposndn=\hmorphposnlft
\let\vmorphposnup=\vmorphposnrt
\let\vmorphposndn=\vmorphposnlft

\newdimen\hgrid \hgrid=15pt
\newdimen\vgrid \vgrid=15pt

\newdimen\hchannel  \hchannel=0pt
\newdimen\vchannel  \vchannel=0pt
\newdimen\channelwidth \channelwidth=3pt

\dimendef\@hchannel 0         \dimendef\@vchannel 2         

\catcode`& = \@oldandcatcode
\catcode`@ = \@oldatcatcode

 {
	\varrowlength=50pt
	\harrowlength=50pt
	\sarrowlength=.30\harrowlength
	$$
	{
		\commdiag{
			&\mbox{\large $E_{x}$}\cr
			&\arrow(-1,-1)\lft{\mbox{\large $E_{xa}$}}\quad
			\arrow(1,-1)\rt{\mbox{\large $E_{xb}$}}\cr
			\mbox{\large $E_a$}&& \mbox{\large $E_b$} \cr
			&\arrow(1,-1)\lft{\mbox{\large $E_{xb}$}}\quad
			\arrow(-1,-1)\rt{\mbox{\large $E_{xa}$}}\cr
			&\mbox{\large $E_{y}$}
		}
	}
	$$
}

\begin{example}
\label{ex:lattconf}
As an example of application of semimodularity  consider
the poset in  Figure \ref{fig:lattconf}. This is 
generated taking all the one hop steps that leaves from $\AComp/\Delta_V$ 
and its successors.
Arrows received different colors and were depicted dashed or not. A table at the
bottom  of the figure gives labels for all the arrows types.
We can use colors in this diagram to verify that two paths with a common
start and common end can be labeled by the same set of equations.
First note that dashed arrows can be labeled with a couple of equations.
For instance dashed black arrows can be labeled with equation 
$t1\equivert t3$ or with equation $t2\equivert t3$.
Solid arrows can be labeled with only one equation. 
Coloring for solid arrows is chosen so that  red, blue
and black dashed arrows can be labeled only  with an equation
that labels solid arrows of the other two colors in the triple red blue and black.
Green do not participate to this scheme.
For instance dashed black arrows can be labeled with equations that
are on solid red and blue arrows
(i.e. $t1\equivert t3$ and $t2\equivert t3$).   

To check semimodularity in this tiny example we have to select two paths with a common start and a common end. 
Next we  label them and see if 
semimodularity holds. That's to say, we compare the  two paths to see if they can be 
labeled with the same set of equations.
So, for instance, if we compare two paths and in one we have a black dashed
arrow then, in the other path, we must have a red or blue solid arrow.
We can say that a black
dashed arrow can be {\em balanced} by a red or blue solid arrow.
Similarly we will say that a red
dashed arrow can be balanced by a black or blue solid arrow.
Finally a black or red solid arrow will balance  a blue dashed arrow.
Green solid arrows can only be balanced by another green solid arrow.
With these remarks in mind we can travel, for instance,   from
$\AComp/\Delta_V$ down to 
$\AComp/\{t1\equivert t2,t2\equivert t3,r1\equivert r2\}$ 
and see if the colors we collect balance.
Note that others starting and ending points are possible, too.
For this top to bottom travel, for instance, we can go down via the leftmost path i.e. red arrow, next 
dashed red and finally green arrow.
Another  path, that balances this, is, for instance, the rightmost i.e. 
green, black and dashed black.
Indeed, in the two paths we have two greens that balance each other, then red balance dashed black and dashed black balance with solid red.
Note that in this drawing things are organized so that  parallel arrows balances so one can easily see that semimodularity holds. 
\end{example}
{
\begin{figure}[h]
\fbox{\psfig{file=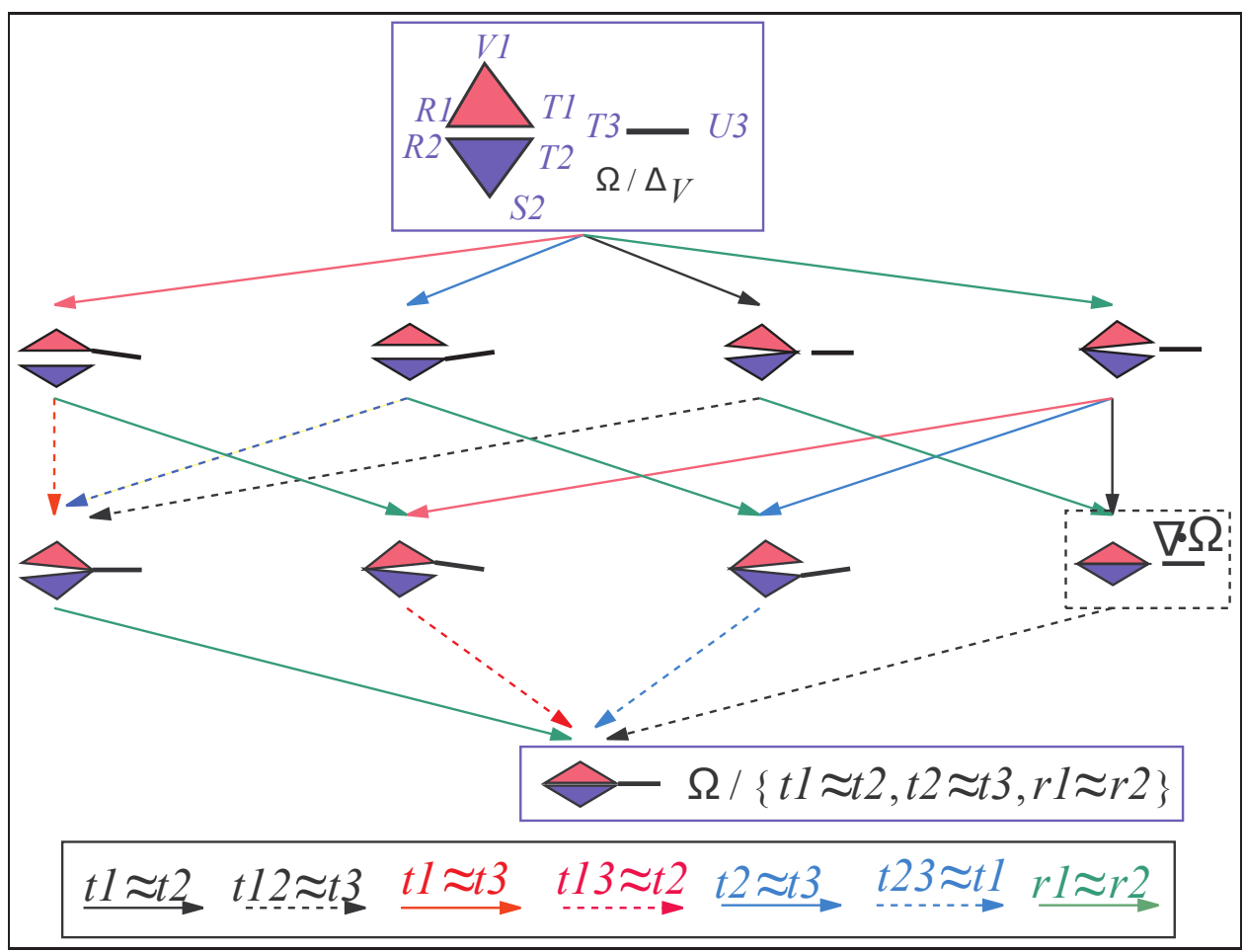,width=\textwidth}}
\caption{Semimodularity in the \quot\ lattice 
(See Example  \ref{ex:lattconf}). The complex $\canon{\AComp}$ is the 
{\em \cano\ decomposition} of the complex at the bottom of the lattice.
This complex is framed for 
future reference (see Chapter \ref{ch:stdec})}
\label{fig:lattconf}
\end{figure}
}
{
Closing this section we like to stress the fact that the main consequence of 
semimodularity is that \verteq\ equations
actually provides a powerful abstraction to denote transformation between
simplices.
This is especially true if we compare them against equivalences.
Indeed an equivalence is a {\em global} object whose form heavily
depends on (the set of vertices of) the complex it applies to.
Sets of  equations, on the contrary, do not depend on a 
particular complex and  can be used to label paths between different
complexes.
Each equation actually defines some sort of {\em rewrite rule} 
whose effect is local and  independent from the
global context in which it is applied. 
Thus \verteq\ equations  are the right {\em basic} tool 
to  denote  a transformation between complexes. 
Indeed,  as we will see,  \verteq\ equations are  
{\em too basic}
for our needs because they impose a too fine grained view on the 
\quot\ lattice.
To correct this problem, in Chapter \ref{sec:classify} 
we will introduce {\em \sglinst s} as a means to group together 
\verteq\ equations into more complex transformations. 
}

 \chapter{Decomposition Lattice}
\label{ch:decomp}
\section{Introduction}
{
Obviously not all the \quot s\ {$\AComp=\AComp^\prime/E$} of a given complex 
$\AComp^\prime$ are such that {$\AComp^\prime$} is a  decompositions of 
$\AComp$. For instance, by equation 
$E=\{a\equivert b\}$ we can readily collapse a triangle $abc$ into one of 
its edges $ac$ and an edge is not the decomposition of a triangle. Thus, in this chapter, we will characterize the set of
equivalences $R$ that makes {$\AComp^\prime$} a a decomposition of the
\quot\ {$\AComp=\AComp^\prime/R$}.
Intuitively, a complex $\AComp^\prime$ is a decomposition of $\AComp$ if 
pasting together pieces of $\AComp^\prime$ we can obtain $\AComp$.
Furthermore, we expect that nothing shrinks or collapse passing from 
$\AComp^\prime$ to $\AComp$.
More precisely intuition require to have a dimension preserving bijection
between top simplices in $\AComp^\prime$ and  
$\AComp$.
In this chapter we define the notion of decomposition 
and identify a sublattice of a particular \quot\ lattice that we 
called the  {\em decomposition lattice}.
This lattice  contains an isomorphic copy for any \dec\ 
of a given complex $\AComp$. 
}

\section{Decompositions}
We first define the notion of \dec\ using \asm s. In the next section
we will restrict our attention to a particular class of decompositions
that form a lattice and that contains an isomorphic representative for
each decomposition of a given complex.  
We recall that, by Property \ref{pro:uniqverquot}
Part \ref{pro:quotmap}, there is  an \asm\  associated with each 
equivalence $\equivert$.
\begin{definition}[Decomposition]
\label{def:decomposition}
An \asc\ $\AComp^\prime$ is a \emd{decomposition} of $\AComp$ \iff\
${\AComp^\prime/{\equivert}}\simpeq\AComp$ and the \asm\ from
$\AComp^\prime$ to $\AComp^\prime/{\equivert}$
is a dimension preserving map that induces a  bijection between top simplices.
\end{definition}
In the following we will introduce the lattice of  the
possible decomposition of $\AComp$ called the {\em \dec\ lattice}
of $\AComp$.
The  \dec\ lattice is a sublattice of the   \quot\ lattice for 
a particular complex $\topAComp$ that we will obtained as
the total decomposition of $\AComp$.
We called this complex the {\em totally exploded} decomposition of $\AComp$.
In the rest of this thesis we will consider only decomposition 
in the \dec\ lattice.
This restriction do not impairs the generality  of the approach. 
In fact, we will show that (see Property \ref{pro:declatt}), 
for any decomposition $\AComp^\second$,
there exists at least a  \dec\ in the \dec\ lattice that is isomorphic to $\AComp^\second$.

\section{The totally exploded decomposition $\topAComp$}
We start the construction of the {\em \dec\ lattice} with the 
definition of the {\em totally exploded} decomposition of a complex $\AComp$.
This will be denoted by  $\topAComp$.
We can intuitively define the {\em totally exploded}
decomposition of $\AComp$ by saying that 
$\topAComp$ is the decomposition of $\AComp$ where
each  top simplex in $\topAComp$ is a distinct connected component.
This requirement
completely defines $\topAComp$ up to isomorphism.
It is easy to see that the top simplices in $\topAComp$ must be the same,
for number and dimension, as those in $\AComp$.  
For instance in Figure \ref{fig:topdef}a complex  $\topAComp$ (on the
right) is the totally exploded version of complex $\AComp$
(on the left).

{
The intuitive definition of $\topAComp$  do not identify clearly 
the complex $\topAComp$. Indeed countably many different
isomorphic options exist for the choice of $\topAComp$. 
To make both theory and proofs straightforward 
we chose a particular naming for Vertices in  $\topAComp$.
Let  $\Theta$ be the set of top simplices in $\AComp$,
then, for each top simplex
$\theta=\{u,v,w,\ldots\}$, we will place in $\topAComp$ the simplex
$\toptheta=\{u_{\theta},v_{\theta},w_{\theta},\ldots\}$.
The above  conventions leads to the following definition.
\begin{definition}[Totally Exploded Decomposition]
\label{def:top}
Let be $\Theta$ the set of
top simplices in $\AComp$.
We will define the \emad{totally exploded}{decomposition} of a
complex $\AComp$, denoted by $\topAComp$, as the complex whose set
of top simplices is  $\Theta^\top=\{\toptheta |\theta\in \Theta\}$
with $\toptheta=\{v_{\theta}|v\in\theta\}$.  
\end{definition}
We will denote with $V^{\top}$ the set of Vertices in $\topAComp$.
It is easy to see that there is a distinct vertex $v_{\theta}$ in  $V^{\top}$
for each vertex $v\in V$ and for each top simplex $\theta$ in the star 
of $v$. Hence we have
$V^{\top}=\{v_{\theta}|v\in V\,\, {\rm and}\,\, \theta\in\xstr{\AComp}{v}\cap\Theta\}$.
}
{
\begin{figure}
\begin{minipage}{0.49\textwidth}
\fbox{\psfig{file=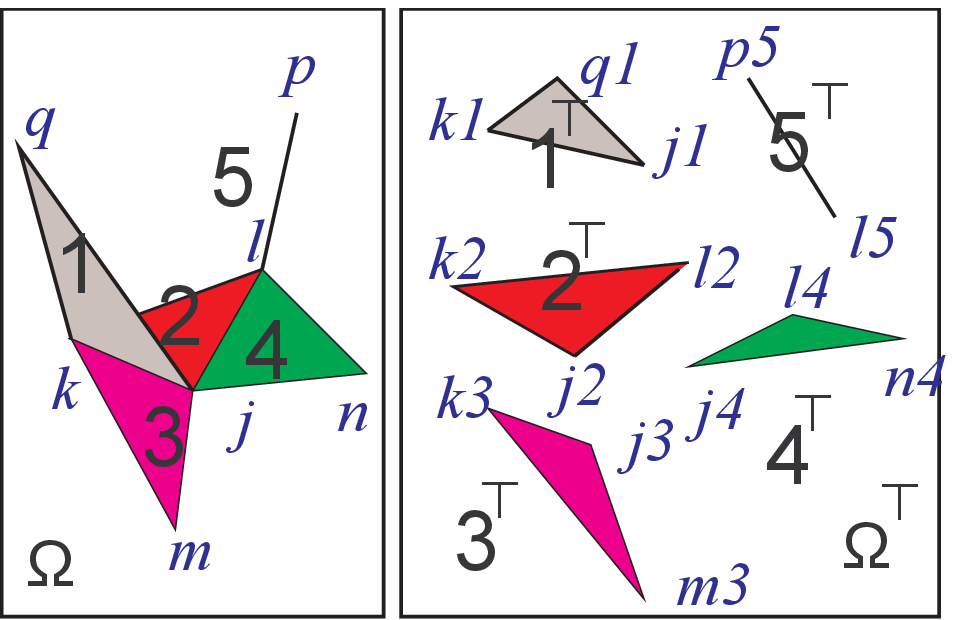,width=\textwidth}}
\begin{center}(a)\end{center}
\end{minipage}
\hfill
\begin{minipage}{0.49\textwidth}
\fbox{\psfig{file=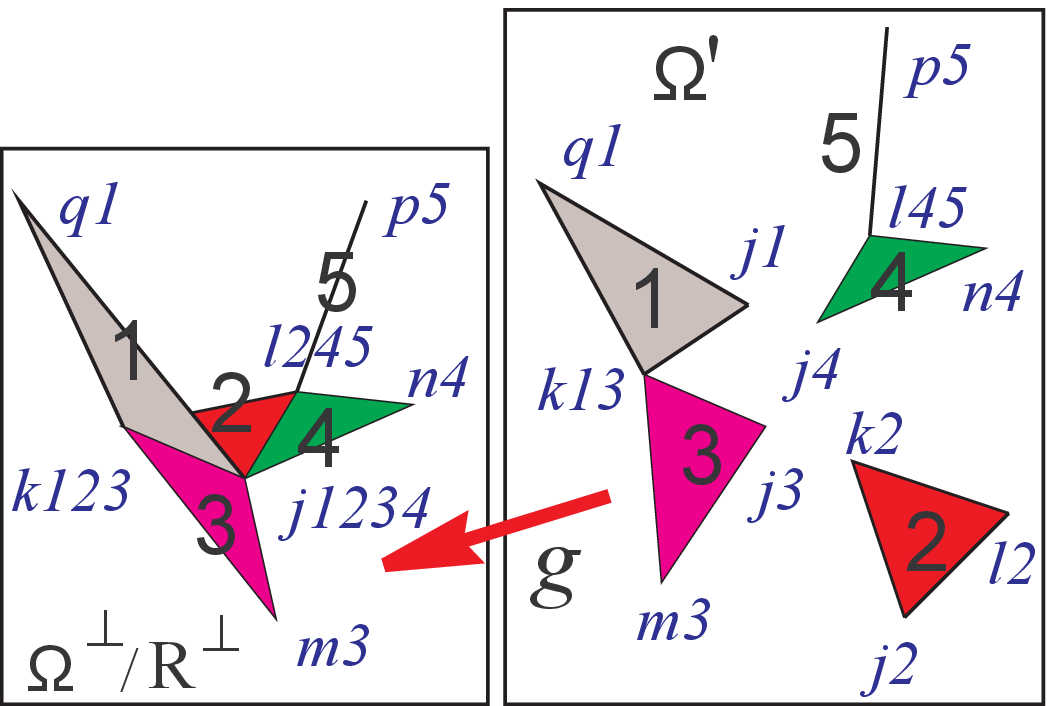,width=\textwidth}}
\begin{center}(b)\end{center}
\end{minipage}
\caption{Example of totally exploded decomposition (a) (see Definition \ref{def:top})
and of $\topAComp/\Rtop$ (b)  (see Property \ref{pro:topacomp}).}
\label{fig:topdef}
\end{figure}

\begin{example}
Consider for instance the complexes in Figure \ref{fig:topdef}a.
To enhance readability we have labeled top simplices in $\AComp$
with integers. We used  the label $1$, for instance,
as a shortcut for the simplex $\theta_1=\{k,q,j\}$
(we recall that in an abstract simplicial complex a $d$-simplex
is a set of $d+1$ Vertices).
Similarly, we have labeled  Vertices in $\AComp^\top$
with alphanumeric strings of the form $vx$ where $v$ is a vertex
of $\AComp$ and $x$ is the number that labels a simplex.
The label $vx$ stands for  the vertex  name $v_{\theta_x}$.
Note that, for instance, label $k1$ stands for $k_{\theta_1}$ that,
in turn
must be unfolded as $k_{\{k,q,j\}}$. For obvious reasons in
examples we will use shortcuts as $k1$ instead of this heavy notation.
Similarly the label $1^\top$ is a shortcut  for the simplex
$\theta_1^\top=\{k,q,j\}^\top=\{k1,q1,j1\}$. Unfolding shortcuts we can write
$\theta_1^\top=\{k_{\theta_1},q_{\theta_1},j_{\theta_1}\}$
that in turn, by unfolding
$1$ becomes $\{k_{\{k,q,j\}},q_{\{k,q,j\}},j_{\{k,q,j\}}\}$.
With these conventions we have that the complex   $\AComp^\top$
on the right of  Figure \ref{fig:topdef}a is the
totally exploded version of   $\AComp$ with Vertices chosen
according to the conventions described above.
\end{example}
}

We can {\em stitch} together top simplices from $\topAComp$ and
obtain a range of decomposition of $\AComp$ reaching, in the
end, $\AComp$ itself. This range of decomposition will be our
\dec\ lattice. A first step towards the definition of this lattice is the
following property that sets top and bottom elements for the
\dec\ lattice.
\begin{property}
\label{pro:topacomp}
The complex $\topAComp$ is a \dec\ of $\AComp$ and
there exist an equivalence   
$\Rtop$ such that {${\topAComp/\Rtop}\simpeq\AComp$}. The equivalence $\Rtop$ is unique up to isomorphism.
\end{property}
\begin{proof}
In the situation of the definition \ref{def:top} the vertex map 
$\funct{f}{\topAComp}{\AComp}$ defined by $f(v_{\theta})=v$ induce a
dimension preserving  \asm\ that is
a bijection between top simplices that  maps  simplex 
$\toptheta$ into $\theta$ and $f(\topAComp)=\AComp$. 
The map $f$ is uniquely identified by these conditions.
We recall that, by Property \ref{pro:uniqverquot}
Part \ref{pro:quotmap}, there is  a unique \asm\ associated with each 
equivalence.
Let {$R_{\top}$} be the equivalence
associated with the \asm\ $f$.
By Part \ref{pro:mapquot} of Property \ref{pro:uniqverquot} 
we have that {${\topAComp/\Rtop}\simpeq\AComp$}.
Therefore there exist an isomorphism $i$ from {${\topAComp/\Rtop}$} to
{$\AComp$}.  Composing $i$ with $f$ we obtain a dimension preserving
\asm\ between $\topAComp$ and $\AComp$. This proves that $\topAComp$ is
a \dec.
\end{proof}
As an example of application of these concepts consider the following 
example. 
\begin{example}
As an example of the construction of $\topAComp/\Rtop$
consider, for instance, the complex  $\AComp/\Rtop$ on the left of Figure
\ref{fig:topdef}b. It is easy to see that  the complexes
$\AComp$ in Figure \ref{fig:topdef}a and $\AComp^\bot$ are 
isomorphic through the isomorphism $i=[
\send{j1234}{j},
\send{k123}{k},
\send{l245}{l},
\send{m3}{m},
\send{n4}{n},
\send{p5}{p}]$

As an example of \dec\ consider,
for instance, the complex  $\AComp^\prime$ on the right of Figure
\ref{fig:topdef}b. The complex $\AComp^\prime$ is a
\dec\ for $\AComp$.
In fact it is easy to verify that the \asm\ induced by the vertex map
$g=[
\send{\{j1,j2,j3,j4\}}{j1234}
\send{\{k13,k2\}}{k123},
\send{\{l45,l2\}}{l245}
]$ is a dimension preserving \asm\ and we have that
$g(\AComp^\prime)=\AComp/\Rtop\simpeq\AComp$. Therefore $\AComp^\prime$
is a \dec\ for $\AComp$.
\end{example}

{
\section{The \Dec\ lattice}
\label{sec:declatt}
A first thing to note is that the  poset generated by all
\quot s of $\topAComp$ is larger than the lattice of \dec s.
Indeed the quotient lattice for  $\topAComp$ contains also complexes obtained collapsing some vertices.
Consider for instance a complex made up of two adjacent segments
$\AComp=\closure{\{ab,bc\}}$. We have that 
$\topAComp=\closure{\{a1b1,b2c2\}}$ is the
totally exploded version of $\AComp$. In this situation  the totally
exploded decomposition is the only non trivial decomposition for $\AComp$.
Still the
\quot\ $\topAComp/\{a1\equivert c2\}$ is a \quot\ of
$\topAComp$ that must not be in the  lattice of \dec s.

With  the results of Property  \ref{pro:topacomp} it is easy to delimit 
the \dec\ lattice as what is between the totally exploded
decomposition and $\AComp$.
With this idea in mind we are ready to define the \dec\ lattice.
\begin{definition}[\Dec\ Lattice]
Let be $\AComp$ a complex and let be $\Rtop$ be the equivalence such that
{${\topAComp/\Rtop}\simpeq\AComp$} 
then we define the \emad{\Dec}{Lattice} as the sublattice of the lattice of \quot s
of $\topAComp$  given 
by the closed interval 
$[{\topAComp/{R_{\top}}},{\topAComp/\Delta_{V^\top}}]$
where {$\Delta_{V^\top}$} is the identity relation over the set of 
vertices of $\topAComp$.
\label{dec:lat-2019}
\end{definition}
We note that the above definition relies on the existence and uniqueness of $\Rtop$ that is guaranteed by Property \ref{pro:topacomp}.

The \dec\ lattice is the   sublattice of the lattice of \quot s
of $\topAComp$ that is  anti-isomorphic
to the sublattice of the partiton  lattice given 
by the closed interval $[{\Delta_{V^\top}},R_{\top}]$,
(see in \latt\ the discussion after Example \ref{ex:hasse} for related definitions)
Whenever this is not ambiguous we will use $\topAComp$ to denote 
the top element in the \dec\ lattice i.e., the \quot\ 
$\topAComp/{\Delta_{V^\top}}$
Similarly we use $v_{\{\theta\}}$ for the  vertex 
$v_{\{\theta\}}/{\Delta_{V^\top}}$.
{
\begin{figure}[h]
\fbox{\psfig{file=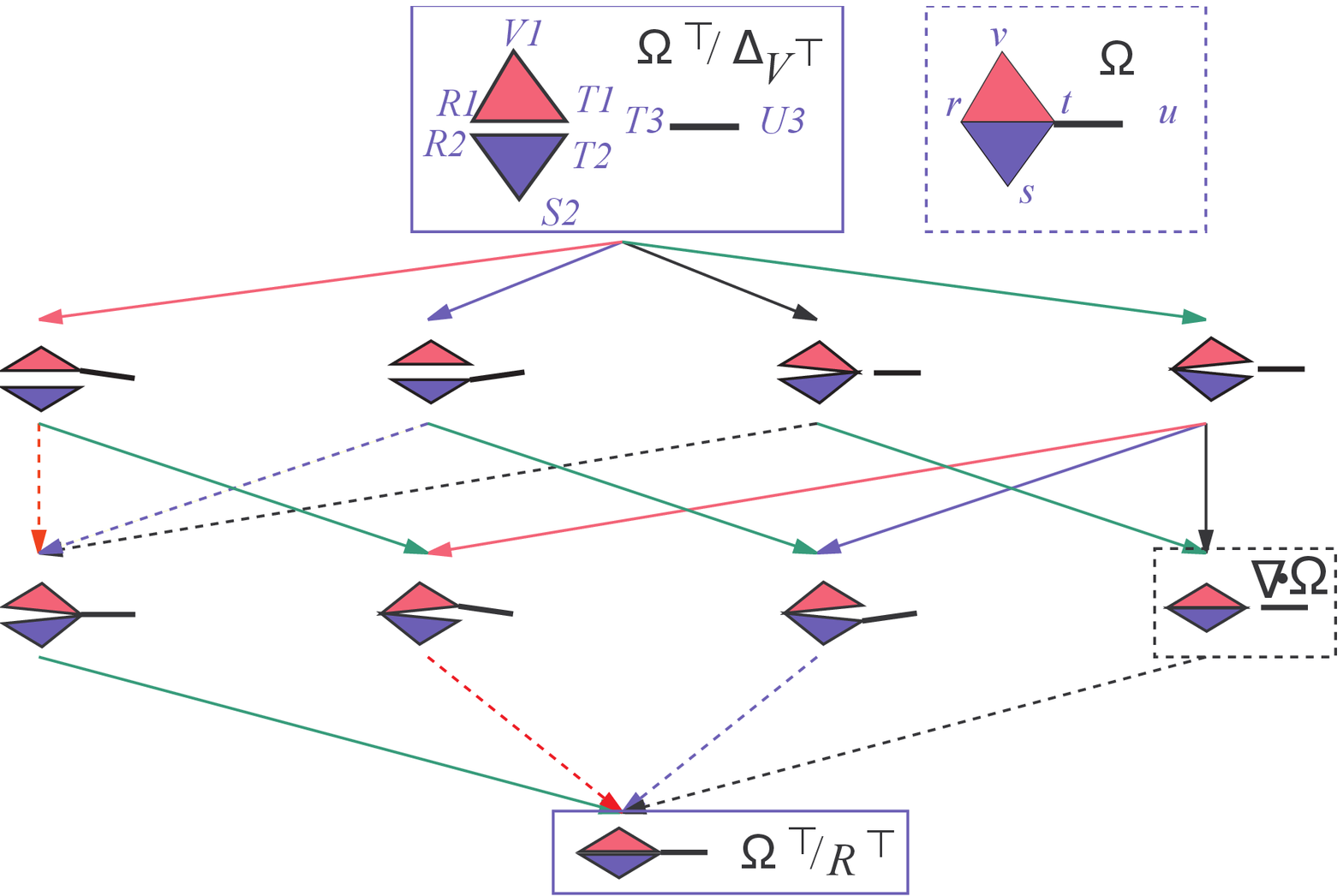,width=\textwidth}}
\caption{The \dec\ lattice 
(See also Example  \ref{ex:lattconf}) for the complex $\AComp$.
}
\label{fig:declatt}
\end{figure}
}

In Figure \ref{fig:declatt} we present  an example of a \dec\ lattice.
This lattice is isomorphic to the the lattice of Figure \ref{fig:lattconf}. The reader must understand that these two figures actually contains two {\em different} lattices. 
In Figure \ref{fig:lattconf}, in the previous chapter, the top element was a quotient of $\AComp$ and in this first example we have chosen for $\AComp$ a complex made up of three connected components, i.e. two triangles and one edge.  In Figure\ref{fig:declatt} we take for $\AComp$ the complex made up of one connected component, i.e. the two triangles and the edge are stitched together. Then, at the top, in this figure,  we have (a quotient of) $\topAComp$ that is, again,  a complex made up of three connected components,  two triangles and one edge. 

It is easy to prove that the \dec\ lattice is actually the 
lattice of {\em all and alone} the decompositions. 
This is expressed formally by the following two  properties.
\begin{property} 
\label{pro:alldec}
All \quot s in the decomposition lattice are decompositions.
\end{property}
\begin{proof}
Indeed, for a given element in the \dec\ lattice 
$\AComp^\prime=\topAComp/{\equivert}$ we have 
$\Delta_{V^\top}\le{\equivert}\le\Rtop$. Thus we
always find two \asm s
one from {$\topAComp/\Delta_{V^\top}$} to {$\AComp^\prime$} and
the other from {$\AComp^\prime$} to {$\topAComp/{R_{\top}}$}
(see Properties \ref{pro:equiquot} and \ref{pro:mapposet}).
Since there is a dimension preserving bijection between top simplices in
{$\topAComp/\Delta_{V^\top}$} and {$\topAComp/{R_{\top}}$} there must be a 
dimension preserving \asm\ from {$\AComp^\prime$} to 
{$\topAComp/{R_{\top}}$}. This will induce a bijection between top simplices 
in {$\AComp^\prime$} and {$\topAComp/{R_{\top}}$}. 
By Definition \ref{def:decomposition} this proves that {$\AComp^\prime$} is a decomposition of {$\topAComp/{R_{\top}}$}. 
\end{proof}
In general there is a dimension preserving \asm\ between any
ordered pair of \dec s in the \dec\ lattice (by ordered pair 
we mean two \dec s such that
$\topAComp/{\equivert_1}\morphle \topAComp/{\equivert_2}$).

In the following we will study decompositions by studying the properties of
the decomposition lattice. This is perfectly legal since all decompositions
are in the decomposition lattice up to isomorphism. This is stated by the
following property. 
\begin{property}
\label{pro:declatt}
Each decomposition has an isomorphic copy in the \dec\ lattice.
\end{property}
\begin{proof}
	\[
	\begin{CD}
	\AComp^\prime\simpeq(\AComp^\prime)^\top/(R^\prime)^\top @>f>>   \AComp\simpeq\topAComp/\Rtop\\
	@A(R^\prime)^\top Ah^\prime A 
\\
	(\AComp^\prime)^\top @<i<<\AComp^\top
	\end{CD}
	\]
Let $\AComp^\prime$ be a decomposition for $\AComp$. 
Being $\AComp^\prime$  a decomposition there must be  a dimension preserving \asm\ $f$ from $\AComp^\prime$ to 
$\AComp\simpeq\topAComp/\Rtop$.
The situation of this proof is summarized in the communtative diagram above.
This \asm\ $f$ must  be also  a bijection between top 
simplices in $\AComp$ and $\AComp^\prime$.
Hence, there must be a dimension
preserving bijection between top simplices in the  totally exploded
versions of $\AComp$ and in the totally exploded version of $\AComp^\prime$. Therefore there must be an
isomorphism between these two totally exploded versions. 
Let us denote with $i$ this isomorphism between {$(\AComp^\prime)^\top$} and $\AComp^\top$.
By Property \ref{pro:topacomp}
there will be an equivalence $(R^\prime)^\top$ and an associated  \asm\ $h^\prime$ from  {$(\AComp^\prime)^\top$} to $\AComp^\prime\simpeq(\AComp^\prime)^\top/(R^\prime)^\top$.
Similarly we introduce the \asm\   $h$ and equivalence $R^\top$ in the commutative diagram above.
If we compose this isomorphism with the \asm\ $h^\prime$ we obtain a dimension
preserving  \asm\ $g$ from $\topAComp$ to {$\AComp^\prime$} therefore $\AComp^\prime\morphle\topAComp$.
Hence, by Property \ref{pro:uniqverquot} Part  \ref{pro:mapquot}.
$\topAComp/R_g\simpeq\AComp^\prime$ being $R_g$ the
equivalence  associated with $g$.
This proves that $\topAComp/R_g$ is an element of the \quot\ lattice.
We have to show that is in the \dec\ lattice.
We have 
$\topAComp/\Rtop\simpeq\AComp\morphle\AComp^\prime\simpeq\topAComp/R_g$.
Thus $R_g\le\Rtop$ and therefore 
$\topAComp/R_g$ is an element of the \dec\ lattice and 
this complexes the proof.
\end{proof}
Hence in the following we will restrict our attention to the \dec\ lattice 
and assume that all equivalences we consider  are within 
$[{\Delta_{V^\top}},R_{\top}]$.

\section{\Eqt\ Simplices}
The elements in the \dec\ lattice of a 
complex $\AComp$ are generated by stitching together 
vertices from the totally exploded \dec\ of $\AComp$.  
 Vertices that are glued together must not belong to the same 
top simplex.  Thus, the \dec\ lattice must be anti-isomorphic to the poset  
of partitions of vertices in the totally exploded decomposition of
$\AComp$.
In the partition {$V^\top/{R_{\top}}$}
(i.e. the partition of $V^\top$ induced by {$R_{\top}$}) there will be a   
block $\pi_{w}$ for each vertex $w\in V$. The block $\pi_{w}$ will be
the subsets of $V^{\top}$ given by
{$\{w_{\theta}|\theta\in\xstr{\AComp}{w}\cap\Theta\}$}. 
Directly from the definition of pasting and from properties of the partition 
lattice $\Pi_n$ we have that:
\begin{property}[Structure of the \Dec\ Lattice]
\mbox{}
\label{pro:uniqver}
\begin{enumerate}
\item
Each element of the \dec\ lattice is a complex given by the nerve of a
refinement of the covering {$\{R_{[v]}|[v]\in(V^\top/{R_{\top}})\}$}; 
\item the poset of coverings that generate the \dec\ lattice,
ordered by the refinement relation, is 
isomorphic with the sublattice 
{$[V^\top/R_{\top},{V^\top/{\Delta^{V^\top}}}]$} 
of the partitions of $V^\top$;
\item in particular we pass from the immediate
superior {$V^\top/\equivert_1$} to its immediate inferior {$V^\top/\equivert_2$} in {$[V^\top/R_{\top},{V^\top/{\Delta^{V^\top}}}]$}
by uniting together two blocks in {$V^\top/\equivert_1$}. 
\end{enumerate}
\end{property}
}
{
By collapsing several pairs of vertices we can equate simplices that are distinct
in the totally exploded version of $\AComp$.
Let be $\equivert\in[{\Delta_{V^\top}},R_{\top}]$ 
an equivalence on the Vertices of 
$\topAComp$ and let {$\AComp^\prime$} be a  decomposition in the \dec\ 
lattice.
If two distinct simplices $\gamma^\prime_1$ and $\gamma^\prime_2$ in 
$\AComp^\prime$ have a common pasted simplex via $\equivert$
(i.e. $\gamma^\prime_1/{\equivert}=\gamma=\gamma^\prime_2/{\equivert}$)
we will say that $\gamma^\prime_1$ and $\gamma^\prime_2$ are \emad{\eqt}{simplices} for 
$\equivert$. In this situation we will say and that $\gamma$ is a
\emas{splitting}{simplex} that splits 
into $\gamma^\prime_1$ and $\gamma^\prime_2$ when undoing $\equivert$.
We will say that  $\gamma^\prime_1$ and $\gamma^\prime_2$ are two \emas{simplex}{copies} 
for $\gamma$ under $\equivert$. 
Note that, by hypothesis, {$\AComp^\prime/{\equivert}$} is still a 
decomposition.
We note that all the non-common faces of the simplex copies
(i.e. faces in
$(\cup_i{\bnd{\gamma^\prime_i}})-\cap_i{\bnd{\gamma^\prime_i}}$)
must be \eqt\ simplices for $\equivert$.

We will say that a simplex copy $\gamma^\prime$
is a \emas{manifold}{\eqt\ simplex}  
(resp. \emas{non manifold}{\eqt\ simplex}) w.r.t. $\equivert$
if the corresponding splitting simplex
(i.e. $\gamma^\prime/{\equivert}$) 
is a manifold (resp. non manifold) simplex.

If an \eqt\ simplex $\gamma^\prime\in\AComp^\prime$, for an equivalence
$\equivert$, is not a face of another \eqt\ simplex we will call 
$\gamma^\prime$ a \emas{top}{\eqt\ simplex} for $\equivert$.
Not all the simplex copies of a certain simplex need to be
a top \eqt\ simplex if one is.
Whenever the simplex $\gamma$ is a vertex we will talk about
\emas{vertex}{copies} and \emas{splitting}{vertex}
}

 \chapter{\Sglinst s}
\label{sec:classify}
\section{Introduction}
{
In the previous  chapter we have studied the lattice of \dec\ of
a complex $\AComp$. We have seen that this is anti-isomorphic to a
closed interval  sublattice of the partition lattice.
On top of the \dec\ lattice we have the totally exploded version
$\topAComp$ of
$\AComp$ and we can walk on the \dec\ lattice adding equations whose
basic effect is to glue together two vertexes of  $\topAComp$.
In this section we take a different  look at the \dec\ lattice.
We imagine that we do not have the option to glue single pairs of vertexes each time but
we are forced to glue together two top simplexes  $\theta^\top_1$ and 
$\theta^\top_2$ gluing together all vertexes they have in common in 
$\AComp$. We will call this move a {\em \sglinst} (or a \gl\ \inst\ for short).
Next, in Section \ref{sec:glinstlatt} we will show that \gl\ \inst s define 
another lattice that is a proper subset of the \dec\ lattice.
This will complete a three level hierarchy of lattices that we devised
to study the \dec\ complex. 

In this hierarchy, at the lowliest level,
we have the lattice of \quot s of $\topAComp$.
This \quot\ lattice spans between the totally exploded \dec\ of $\AComp$ 
and the
single vertex resulting from the total collapse of $\AComp$ into a point.
The lattice of \quot s of $\topAComp$ 
offers the finest granularity and the maximum extension 
providing a representative for  all possible modifications of
$\AComp$. 
We believe that several problems in computer graphics can be
modeled within the framework provided by this lattice. For instance the 
{\em Vertex Tree} in \cite{Luebke97} can be described easily as
an implementation of the closed interval 
$[\topAComp/(V^{\top}\times V^{\top}),\topAComp/\Rtop]$ in the
lattice of \quot s of $\topAComp$.

Then we introduced  another  lattice is more specific  for the
decomposition problem.
Indeed, in the previous chapter we have introduced the \dec\ lattice as a closed 
interval within the lattice of \quot\ of $\AComp$.
In this section, we will introduce  
the lattice of \dec\ generated by sets of \gl\ \inst s.
This latter  is a sublattice of the \dec\  lattice and
represents an abstraction on it.
This abstraction is central in the development of this thesis,
In fact,  Chapter \ref{cc:toposgi}, will be devoted to the
study of the topological properties of elements in this lattice.
In particular we will detail topological properties that are relevant for 
the \dec\ problem. 
In Chapter \ref{ch:stdec}, in particular in Lemma \ref{lemma:couple},
we will prove that this lattice actually represents the right abstraction 
to master the \dec\ problem and thus we build, in this lattice, our
definition of {\em \cano\ decomposition}. 
}

\section{\Sglinst s}
\label{sec:gluinst}
{We introduce \gl\ \inst s to have  an handy way to denote the 
set of \verteq\
equations needed to {\em completely
stitch together} top simplexes $\theta_1$ and
$\theta_2$. To understand what we mean
by ''{\em completely stitch together}''
we first recall that there is a bijection between top
simplexes in a complex $\AComp$ and top simplexes in any decomposition of
the complex $\AComp$.
Let us consider a pair of  top simplexes $\theta_1$ and $\theta_2$ in 
$\AComp$ incident at  the non-empty simplex
$\gamma$ (i.e. $\gamma=\theta_1\cap\theta_2$).
We might find  that, in a decomposition {$\topAComp/{\equivert}$}, the 
corresponding  two  top simplexes 
{$\theta_1^\top/{\equivert}$} and
{$\theta_2^\top/{\equivert}$} might share a simplex
$\gamma^\prime=
{\theta_1^\top/{\equivert}}\cap{\theta_2^\top/{\equivert}}$ whose dimension
is smaller than $\ord{(\gamma)}$.
For instance, let be $\theta_1$ and $\theta_2$ two tetrahedra in a 
$3$-complex $\AComp$ and let   $\gamma$ be 
their common simplex (i.e.  $\gamma=\theta_1\cap\theta_2$).
For instance $\gamma$ could be a triangle.
Now there are decomposition {$\topAComp/{\equivert}$}
where top tetrahedra 
{$\theta_1^\top/{\equivert}$} and
{$\theta_2^\top/{\equivert}$} 
do not share a full triangle but simply an edge or a tip.
In this case we  will say that $\theta_1$ and $\theta_2$
(or {$\theta_1^\top/{\equivert}$} and
{$\theta_2^\top/{\equivert}$}) 
{\em do not} completely stitch together.
On the other hand,  we will say that, in the decomposition {$\topAComp/{\equivert}$},
top simplexes $\theta_1$ and $\theta_2$ 
{\em completely} stitch together
if and only if top simplexes 
{$\theta_1^\top/{\equivert}$} and
{$\theta_2^\top/{\equivert}$} 
intersect at  a simplex 
with the same dimension of $\gamma=\theta_1\cap\theta_2$. 
In the case of our example of the two tetrahedra sharing a triangle in  $\AComp$ we will say that they completely 
stitch together in all decompositions $\topAComp/{\equivert}$
where they share a triangle.
}

A \sglinst\ (or simply a \emas{\gl}{\inst} for short) is
a pair, $g=\{\theta_1,\theta_2\}\subset \Theta$, 
of top simplexes in the set of top simplexes $\Theta$ of $\AComp$,
The \gl\ \inst\ $g$ will be usually written as
{$\theta_1\equatsimp\theta_2$} (or $\theta_2\equatsimp\theta_1$).
If $\gamma$ is the common simplex, i.e.
$\gamma=\theta_1\cap\theta_2$, to highlight the role of the common
simplex $\gamma$, we will use $g_\gamma$, 
instead of plain $g$, to denote the \gl\ \inst\ made up of
a pair of top simplexes that intersect at $\gamma=\theta_1\cap\theta_2$.

Simplex \gl\ \inst s are synctatic objects that denotes an equivalence 
on vertexes of $\topAComp$.
With  this idea in mind 
we define the set of \emas{equations}{associated} with
the \gl\ \inst\ {$\theta_1\equatsimp\theta_2$}.
as the set
$\{v_{\{\theta_1\}}\equivert v_{\{\theta_2\}}|v\in\theta_1\cap\theta_2\}$.
We will say that {${\theta_1\equatsimp\theta_2}$} denotes the equivalence
defined by this set of \verteq\ equations.
Whenever we need to put more emphasis on the denoted object we 
will use the symbol  {$\equivert^{\theta_1\equatsimp\theta_2}$} to
talk about the denoted equivalence. In all other cases, when this is not
ambiguous, we will use the notation {${\theta_1\equatsimp\theta_2}$}
to denote: the \sglinst, the associated set of \verteq\ equations and
the equivalence {$\equivert^{\theta_1\equatsimp\theta_2}$}.

{
We will say that the equivalence ${\equivert}$ {\em satisfies}\ 
instruction  {${\theta_1\equatsimp\theta_2}$} whenever
the equivalence
{$\equivert^{\theta_1\equatsimp\theta_2}$},
is contained in the 
equivalence ${\equivert}$ (i.e. $\equivert^{\theta_1\equatsimp\theta_2}\subset{\equivert}$)).
}            

{Given a \gl\ \inst\ {$\theta_1\equatsimp\theta_2$} 
we will define the \emas{\gl\ \inst}{order}  as
$\dim{(\theta_1\cap\theta_2)}$.
Note that the order of
$\theta_1\equatsimp\theta_2$ is 
one unit less the number  of \verteq\ equations associated with 
the \gl\ \inst.

Note that we do not ask the two top simplexes $\theta_1$ and $\theta_2$
to be incident.
However, if the two simplexes $\theta_1$ and $\theta_2$ are disjoint, we have
that the associated  set of equations is empty.
In this case we will call the \gl\ \inst\ 
$\theta_1\equatsimp\theta_2$ {\em empty} or {\em void}.
By convention we assign order $-1$ to empty \inst s.

\section{Sets of \gl\ \inst s}
Given a set ${\cal E}=\{g_i\}$ of \gl\ \inst s 
we associate a set of 
\verteq\ equations to {${\cal E}$} by taking all the
\verteq\ equations associated with each ${g_i}$.
We will say that {${\cal E}$} is associated 
(or denotes) this set of \verteq\ equations. 
We will use the symbol {$\equivert^{\cal E}$} 
(or ${\cal E}$ as a shortcut) to denote the 
equivalence induced by the set of \verteq\ equations
associated with  {${\cal E}$}.
We extend this notation to the empty set by taking the identity relation
for {$\equivert^{\emptyset}$}. 
If {$\equivert^{\cal E}\subset{\equivert}$} we will say that 
equivalence {$\equivert$} {\em satisfies} the set of \gl\ \inst s
${\cal E}$. This happens if and only if  {$\equivert$} 
satisfy all \verteq\ equations in {${\cal E}$}.
}
{Thus we will write {$\topAComp/{\cal E}$} and $\gamma/{\cal E}$ to
denote both {$\topAComp/{\equivert^{\cal E}}$} and $\gamma/{\equivert^{\cal E}}$.
In particular we will say that the set of \gl\ \inst s  ${\cal E}$
{\em generates} the \quot\ {$\topAComp/{{\cal E}}$}.
Furthermore, according to  Definition \ref{def:quotquot} it is perfectly legal
to write {$\AComp^\prime/{\cal E}$} for a \dec\ 
$\AComp^\prime=\topAComp/{\cal E}^\prime$. 
In fact this
unfolds to {$\AComp^\prime/{\cal E}=
(\topAComp/{\cal E}^\prime)/{\cal E}=
(\topAComp/{\equivert^{{\cal E}^\prime}})/{\equivert^{\cal E}}$}
and Definition \ref{def:quotquot}  this becomes
{$\topAComp/({\equivert^{{\cal E}^\prime}}+{\equivert^{\cal E}})$}
{
\begin{figure}[h]
\begin{center}
\begin{minipage}{0.24\textwidth}
\fbox{\psfig{file=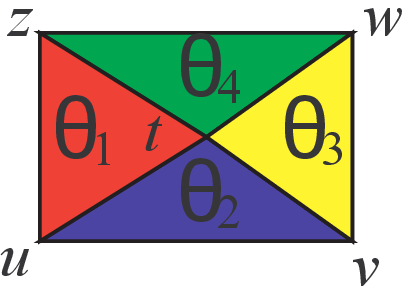,width=\textwidth}}
\begin{center}(a)\end{center}
\end{minipage}
\mbox{  }
\mbox{  }
\begin{minipage}{0.48\textwidth}
\fbox{\psfig{file=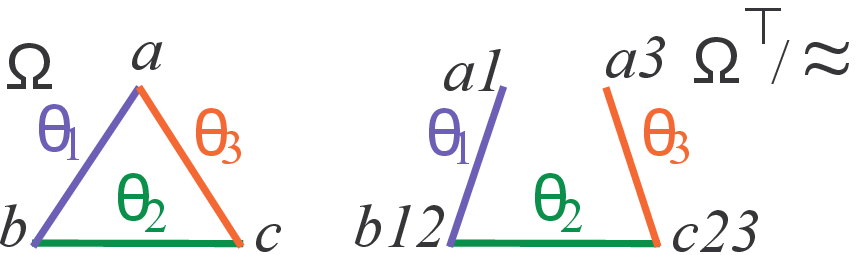,width=\textwidth}}
\begin{center}(b)\end{center}
\end{minipage}
\end{center}
\caption{
An example of a  complex with  redundant \gl\ \inst s (a) and
non transitivity for $\equatsimp$ (b)
}
\label{fig:tsimpab}
\end{figure}
}
{
\begin{example}
The {$\AComp/{\cal E}$} notation suggest an 
easy similarity between \quot s of the form
$\AComp/E$,  defined by  a set of {\em \verteq} equations $E$,
and \quot s of the form {${\AComp/{\cal E}}$},
defined by  a set of {\em \gl} \inst s {${\cal E}$}.
Undoubtly strong relation between these two families exists.
However, some care must be taken in extending 
concepts for sets of {\em \verteq} equations
to sets of {\em \gl} \inst s.
A first flaw is in the concept of {\em independent} set of equations.
There are quite obvious examples of non redundant sets of \gl\ \inst\
whose associated set of {\em \verteq} equations {\em is not} independent.

See for instance the complex of Figure \ref{fig:tsimpab}a.
Consider the set of \gl\ \inst s 
${\cal E}=
\{{\theta_1\equatsimp\theta_2},
{\theta_2\equatsimp\theta_3},{\theta_3\equatsimp\theta_4},
{\theta_4\equatsimp\theta_1}\}$.
The corresponding pairs of \verteq\ equations, a pair for each \gl\ 
\inst s, above are:
$\{{u1\equivert u2},{t1\equivert t2}\}$,
$\{{v2\equivert v3},{t2\equivert t3}\}$,
$\{{w3\equivert w4},{t3\equivert t4}\}$ and
$\{{z4\equivert z1},{t4\equivert t1}\}$.
The set of \verteq\ equations {${\cal E}$} is a set of non redundant 
\inst s. No \inst\ in the set {${\cal E}$} can be deleted without 
decomposing the generated complex.
For instance we cannot delete \gl\ \inst\  {$\theta_4\equatsimp\theta_1$}.
In fact, in this case, we have that equation ${z4\equivert z1}$ 
is not satisfied by the equivalence 
induced by the first three \gl\ \inst. However, the set of eight equations
associated with {${\cal E}$} is not a set of independent \verteq\ equations. 
In fact ${t4\equivert t1}$ is already satisfied by the equivalence 
induced by the six \verteq\ equations in the  first three pairs 
corresponding to the set of \gl\ \inst s:.
$\{{\theta_1\equatsimp\theta_2},
{\theta_2\equatsimp\theta_3},{\theta_3\equatsimp\theta_4}
\}$.
\end{example}
}
}
\section{The lattice of \quot s modulo \gl\ \inst s}
\label{sec:glinstlatt}
{In this  section we will show that sets of  \gl\ \inst s define 
another lattice that is a proper subset of the \dec\ lattice.
This will complete the three level hierarchy of lattices that we devised
to study the \dec\ complex. 

We will see in this section
that the lattice of \dec\ generated by sets of \gl\ \inst\
is a {\em point lattice} that is a proper subset of the  
\dec\ lattice.  In fact, in some cases,
not all elements of the \dec\ lattice can be 
generated by sets of \gl\ \inst. 
}

We will denote with $\closeq{\cal E}$ the set of all non
empty \gl\ \inst s satisfied by {$\equivert^{\cal E}$}.
Directly from the definition we have that 
$\equivert^{\closeq{\cal E}}=\equivert^{{\cal E}}$ and 
$\topAComp/\closeq{\cal E}=\topAComp/{\cal E}$. 
If a set of \gl\ \insts\ is such that $\closeq{\cal E}={\cal E}$
we will say that set {${\cal E}$} is {\em closed}.
Note that in a closed set of \gl\ \insts\ 
the symbol 
{${\theta_1\equatsimp\theta_2}$} do not denotes a transitive relation.
Indeed we may have complexes $\AComp$ with some decomposition
$\topAComp/{\equivert}$ for which we have that $\equivert$ satisfy both  
$\theta_1\equatsimp\theta_2$ and
$\theta_2\equatsimp\theta_3$ and yet we might find that 
$\equivert$ do not satisfies $\theta_1\equatsimp\theta_3$.
This might happens even if equivalence $\equivert$ is defined by a
closed   set  of \gl\ \inst s.

\begin{example}
The simplest example in this sense is given by the $1$-complex  $\AComp$ of Figure \ref{fig:tsimpab}b.
Complex $\AComp$ is  given by the three segments $\theta_1=ab$, $\theta_2=bc$ and 
$\theta_3=ca$.
Consider now the decomposition $\topAComp/{\equivert}$ where edges 
$\theta_1$ and  $\theta_3$ do not stitch together at $a$.
We have that equivalence $\equivert$ satisfy \gl\ \inst s
$\theta_1\equatsimp\theta_2$ and
$\theta_2\equatsimp\theta_3$. However equivalence
$\equivert$ do not satisfy $a1\equivert a3$ and therefore do not
satisfy \gl\ \insts\  $\theta_1\equatsimp\theta_3$.
Note that the set  of \verteq\ equations associated with the 
set of two \gl\ \inst s.
${\cal E}=\{\theta_1\equatsimp\theta_2, \theta_2\equatsimp\theta_3\}$
is exactly made up by the 
two \verteq\ equations $b1\equivert b2$ and 
$c2\equivert c3$. These two equations defines the equivalence $\equivert$.
No other \gl\ \inst\ is satisfied by 
$\equivert{=}\equivert^{\cal E}$ and  therefore the pair of \gl\ \insts\  ${\cal E}$ is closed.
\end{example}
Closed sets of \gl\ \inst s form a poset that is ordered by set inclusion.
To prove this we note that 
the intersection of two closed sets of \gl\ \inst s 
is still a closed set of \gl\ \inst s. Thus the set of closed \gl\ \inst s has
a \ems{closure\ }{property} (see \latt\  Section \ref{sec:lattice}).
It is easy to see why the set {$\closeq{{\cal E}_1}\cap\closeq{{\cal E}_2}$} is closed. If we need to add a gluing instruction to close it this must be both in {$\closeq{{\cal E}_1}$} and in {$\closeq{{\cal E}_2}$} since they are both closed.  So it must be in the intersection.
In general  a set with the closure property is a complete lattice ordered
by set inclusion. Whenever the closure property holds
it can be proven that 
lattice operations are
{$\closeq{{\cal E}_1}\cap\closeq{{\cal E}_2}$} and
{$\closeq{(\closeq{{\cal E}_1}\cup\closeq{{\cal E}_2})}$} giving, 
respectively,
the g.l.b. and the l.u.b. of
{$\closeq{{\cal E}_1}$} and {$\closeq{{\cal E}_2}$}.
Thus, we have that closed sets of \gl\ \inst s
form a lattice. 
By definition,
this lattice is a \ems{point\ }{lattice} (see \latt\ Definition \ref{def:pointlatt}). In fact all elements in this lattice
are generated joining (i.e. summing)  
a basic set of elements called {\em points} or
{\em atoms}.
The  points for this lattice are singletons  of the 
form $\{\theta_1\equatsimp\theta_2\}$ containing  
a single \gl\ \inst.  

{The  mapping that sends each set of \gl\ \inst s
{${\cal E}$} into the \dec\   {$\topAComp/{\cal E}$},
although not injective in general, becomes 
injective if restricted to closed sets of \gl\ \inst s.
This mapping sends the 
lattice of closed sets of  \gl\ \inst s into the set of
\dec\   generated by \gl\ \inst s. It is easy two see that
this mapping is \ems{antitone} (i.e. it reverses ordering,
 see \latt\ Section \ref{sec:poset}). Thus the set of  
\dec s   generated by \gl\ \inst s must be a lattice anti-isomorphic to the
lattice of closed sets of \gl\ \inst s.
Thus the set of
\dec s of the form {$\topAComp/{\cal E}$} is a lattice. 
Unfortunately this lattice, in general, is
not a sublattice of the \dec\ lattice.
Indeed it is quite easy  to build two closed sets of \gl\ \inst s 
{$\closeq{{\cal E}_1}$}  and {$\closeq{{\cal E}_2}$} such that the \dec\
$\topAComp/({\equivert^{\closeq{{\cal E}_1}}}\proeq\equivert^{\closeq{{{\cal E}_2}}})$ {\em is not} the l.u.b.  of
{$\topAComp/\closeq{{\cal E}_1}$}  and {$\topAComp/\closeq{{\cal E}_2}$} given by
$\topAComp/({\closeq{{\cal E}_1}}\cup\closeq{{{\cal E}_2}})$.
Therefore 
$\topAComp/({\equivert^{\closeq{{\cal E}_1}}}\proeq\equivert^{\closeq{{{\cal E}_2}}})$ {\em is not} a
\dec\ of the form {$\topAComp/{\cal E}$}.

\begin{figure}[h]
\begin{center}
\fbox{\psfig{file=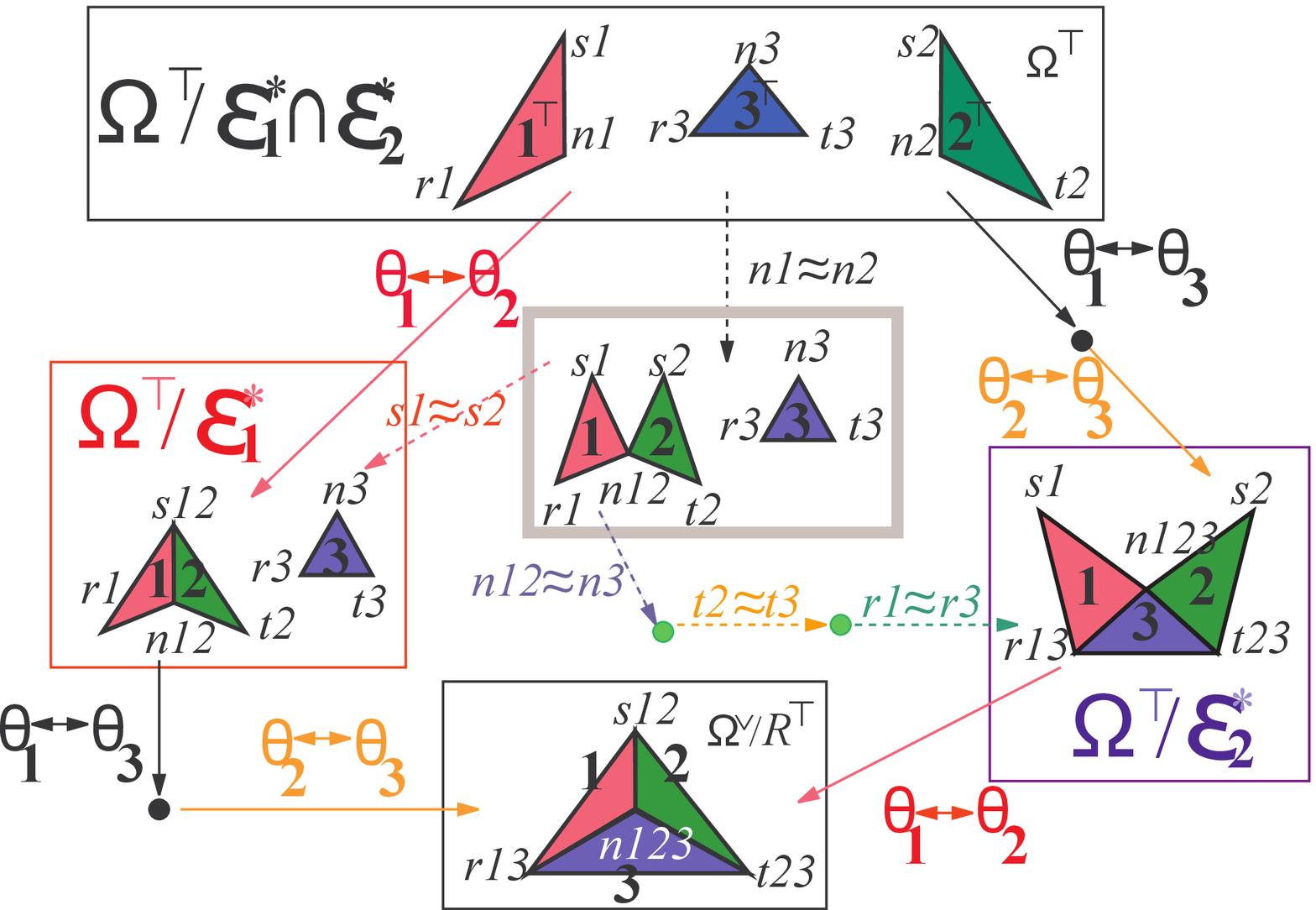,width=0.8\textwidth}}
\end{center}
\caption{A lattice where  
{${\topAComp/(\equivert^{{\cal E}_1}\proeq{\equivert^{{\cal E}_2}})}$} is not 
a \dec\ generated by a set of \gl\ \inst s}
\label{fig:nonind}
\end{figure}
\begin{example}
\label{ex:nonlatt}
Consider for instance the lattice in Figure \ref{fig:nonind}.
We have two disjoint sets  of \gl\ \inst s
{$\closeq{{\cal E}_1}=\{\theta_1\equatsimp\theta_2\}$} and {$\closeq{{\cal E}_2}=\{\theta_1\equatsimp\theta_3,\theta_2\equatsimp\theta_3\}$}
that generates the two decompositions {$\topAComp/\closeq{{\cal E}_1}$} and {$\topAComp/\closeq{{\cal E}_2}$} It is easy to verify that the intersection of the corresponding equivalences 
{$\equivert^{\closeq{{\cal E}_1}}$} and {$\equivert^{\closeq{{\cal E}_2}}$} 
is not empty.
This intersection is the equivalence defined by equation {$n1\equivert n2$}.

We have 
that the \dec\ {$\topAComp/{\{{n1\equivert n2}\}}$} is the least upper 
bound in the \dec\ lattice   
for the pair of complexes
{$\topAComp/\closeq{{\cal E}_1}$} and {$\topAComp/\closeq{{\cal E}_2}$}.
Similarly equivalence {$\{{n1\equivert n2}\}$} is the greatest lower bound
{${{\equivert^{\closeq{{\cal E}_1}}}\proeq{\equivert^{\closeq{{\cal E}_2}}}}$}.
Equivalence  {${\{{n1\equivert n2}\}}$} is not an equivalence of the
form ${\equivert^{{\cal E}}}$. Indeed there is not a  set of
\gl\ \inst s   ${\cal E}$ such that
{${\equivert^{{\cal E}}}={\{{n1\equivert n2}\}}$}.
\end{example}

{Furthermore, equivalence generated by  
sets of \gl\ \inst form a lattice that is not {\em semimodular}.
This can be seen in the example of Figure \ref{fig:nonsemim}
\begin{figure}[h]
\begin{center}
\fbox{\psfig{file=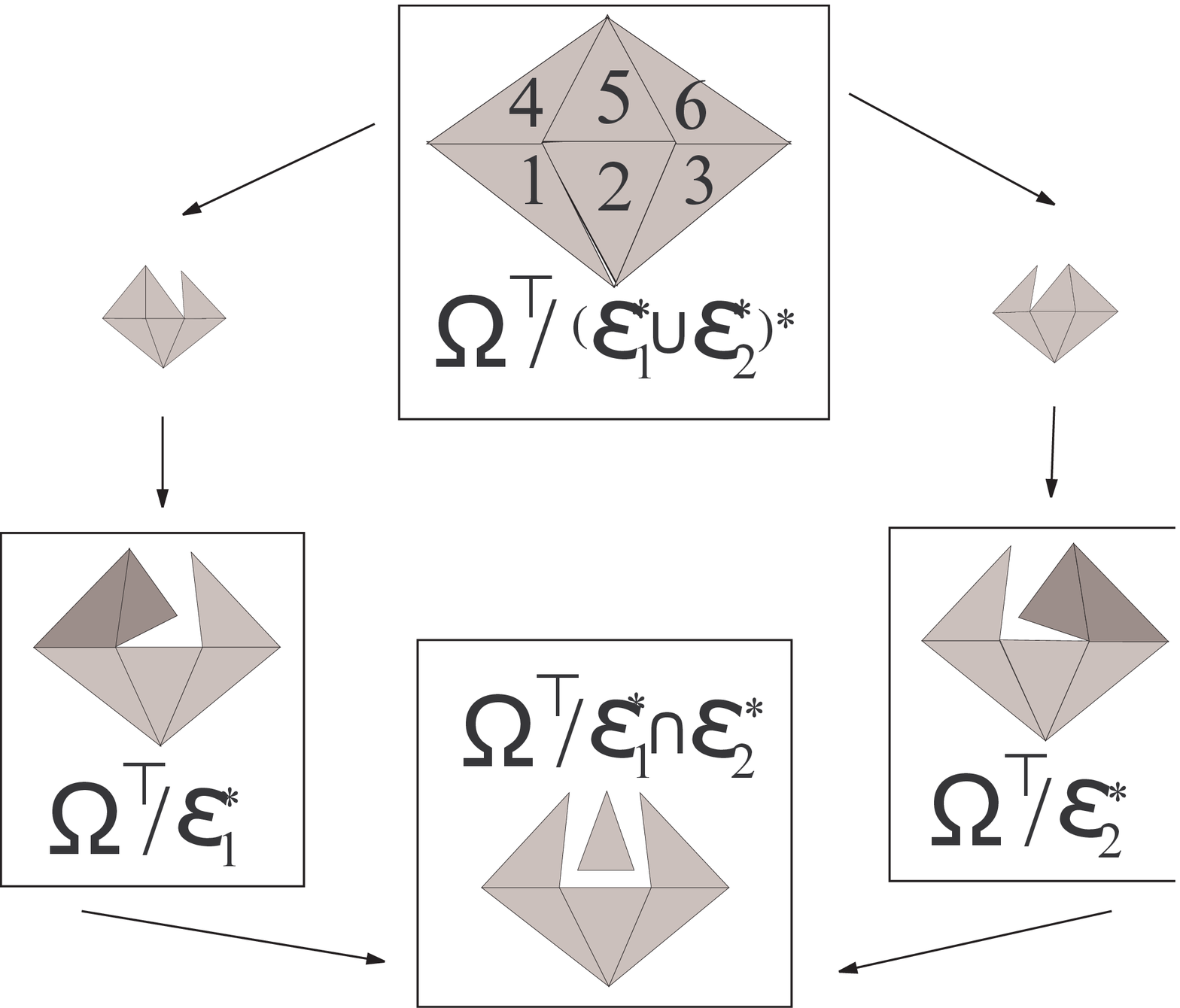,width=0.4\textwidth}}
\end{center}
\caption{A lattice where the two elements  
{${\topAComp/\closeq{{\cal E}_1}}$} and 
{${\topAComp/\closeq{{\cal E}_2}}$} 
are both immediate superior to 
{$\topAComp/{\closeq{{{\cal E}_1}}\cap{\closeq{{\cal E}_2}}}$} 
and there is not a
a common immediate superior for
{${\topAComp/\closeq{{\cal E}_1}}$} and 
{${\topAComp/\closeq{{\cal E}_2}}$} 
}
\label{fig:nonsemim}
\end{figure}
\begin{example}
We recall that 
a lattice is semimodular \iff\ whenever two elements has a 
common immediate inferior they also have a common immediate superior
(see \latt\ Definition \ref{def:semi}). In the lattice of 
Figure \ref{fig:nonsemim} we have
$${{\closeq{{\cal E}_1}\cap\closeq{{\cal E}_2}}=\{
{\theta_4\equatsimp \theta_1},
{\theta_1\equatsimp \theta_2},
{\theta_2\equatsimp \theta_3},
{\theta_3\equatsimp \theta_6}\}
}
$$. 
Then adding ${\theta_4\equatsimp \theta_5}$ (i.e. glue the two gray 
triangles on the left) we get ${\closeq{\cal E}_1}$
Adding ${\theta_5\equatsimp \theta_6}$ (i.e. glue the two gray 
triangles on the right) we get $\closeq{{\cal E}_2}$. The set 
$\closeq{\closeq{{\cal E}_1}\cap\closeq{{\cal E}_2}}$ is the l.u.b. for 
the pair {$\closeq{{\cal E}_1}$}  and {$\closeq{{\cal E}_2}$}. Unfortunately
this element is not the immediate superior for 
neither {$\closeq{{\cal E}_1}$}  nor {$\closeq{{\cal E}_2}$}. 
This is due to the 
presence of the two small unframed complexes in  
Figure \ref{fig:nonsemim}. This proves that the proposed lattice is 
not semimodular. 
\end{example}
}
The lack of semimodularity impair the possibility of having a grade for 
\dec\ based on the size of the set of \gl\ \inst s necessary to build them. 

\begin{figure}[h]
\begin{center}
\mbox{\psfig{file=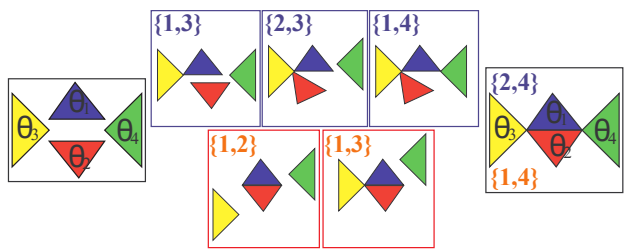,width=0.5\textwidth}}
\end{center}
\caption{Two ways of forming the complex on the right with four (upper path) 
and three (lower path) \gl\ \inst s. In both paths each \gl\ \inst\ always add
independent \verteq\ equations}
\label{fig:nondistmin}
\end{figure}
{
\begin{example}
Consider the situation of Figure \ref{fig:nondistmin}.
We can go from the complex on the extreme left to the  complex on the 
extreme right following two  paths. The upper path 
is made up of four hops.
Collecting \gl\ \inst s
that labels each hop we get a set of four \gl\ \inst s.
In each frame we report the \gl\ \inst\ that is necessary to reach the complex
at the end  of each hop. 
We use the shortcut $\{i,j\}$ for the \inst\ $\theta_i\equatsimp\theta_j$.
With this convention we collect, along the upper
path,  the four \gl\ \inst s:    
${\theta_1\equatsimp\theta_3}$,
${\theta_2\equatsimp\theta_3}$,
${\theta_1\equatsimp\theta_4}$ and
${\theta_2\equatsimp\theta_4}$.
Similarly, lower in Figure \ref{fig:nondistmin}, we have a path
made up of {\em just three} hops.
Along this path we collect just  {\em three} \inst s:
${\theta_1\equatsimp\theta_2}$,
${\theta_1\equatsimp\theta_3}$ and
${\theta_1\equatsimp\theta_4}$.
Both sets of \gl\ \inst s are associated with a set of four
independent \verteq\ equations.
Indeed the complex on the extreme left is of grade four in the \dec\
lattice. 
\end{example}
This example shows that a grade based on the number of \gl\ \inst\ is not 
possible and this is a major consequence of the lack of
semimodularity.
}

We close this section  with a very simple  example that shows the subset of 
the lattice of \dec\ in  Figure \ref{fig:lattconf}} obtained taking 
complexes generated by \gl\ \inst s.
\begin{figure}[h]
\begin{center}
\fbox{\psfig{file=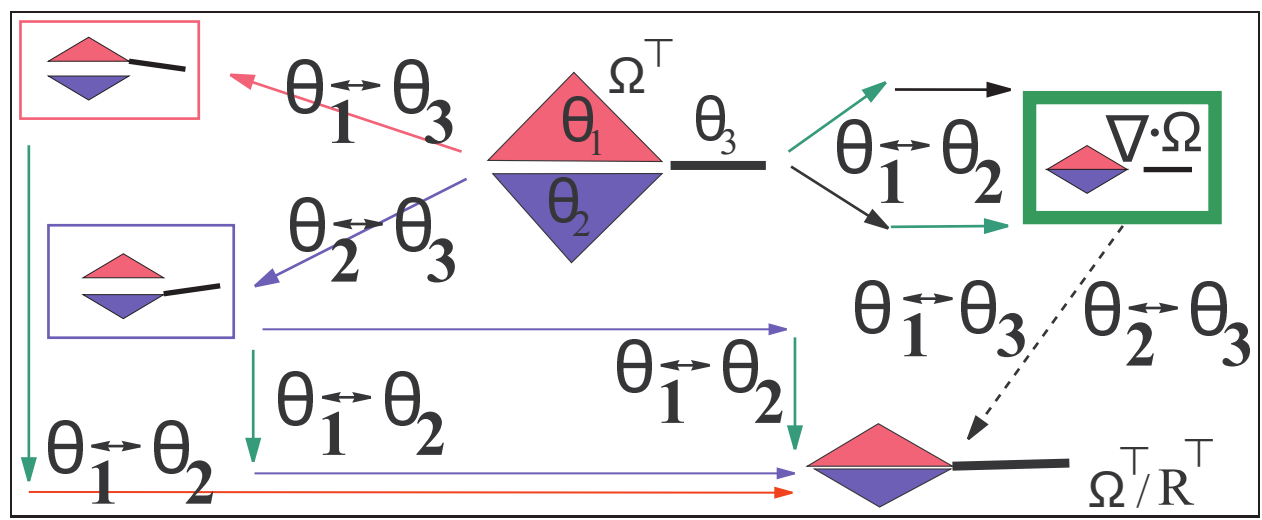,width=\textwidth}}
\end{center}
\caption{A \redo\  lattice out of lattice of Figure \ref{fig:lattconf}}
\label{fig:redlat}
\end{figure}
Figure \ref{fig:redlat} is a reduced version of Figure  
\ref{fig:lattconf} obtained by deleting 
decomposition that are not of the form {$\topAComp/{\cal E}$}.
Arrows are reported with the colors they have in Figure \ref{fig:lattconf}. 
Colored arrows denote  \verteq\ equations that's why path for a \gl\ \inst\ corresponds to more than one arrow. 
Groups of colored arrows are labeled with the \gl\ \inst\ that
transforms the complex at the origin of the arrow
into the complex at the tip of the arrow.
So, for instance, let us start at complex with the thick  frame. 
This is {$\topAComp/{\{\theta_1\equatsimp \theta_2\}}$}. 
By following the black dashed 
line we must add ${\theta_1\equatsimp \theta_3}$ or 
{${\theta_2\equatsimp \theta_3}$} and reach (the isomorphic copy of)
$\AComp$ at the bottom.
The bottom is generated either by the 
set $\{\theta_1\equatsimp \theta_2,\theta_1\equatsimp \theta_3\}$
or by the  set $\{\theta_1\equatsimp \theta_2,\theta_2\equatsimp \theta_3\}$.

 \chapter{\Sglinst s and Topological properties}
\label{cc:toposgi}
\section{Introduction}
{
Given a \dec\ generated by a set of \gl\ \inst\   {${\cal E}$},
it is possible to connect the topological properties of 
{$\topAComp/{{\cal E}}$} with properties of the set
of \gl\ \inst\ {${\cal E}$}.
Indeed we will develop an analysis of
complexes generated by a set of \gl\ \inst s {${\cal E}$}.
This analysis gives some topological  properties
for the decomposition {$\topAComp/{{\cal E}}$}
on the ground of properties of \gl\ \inst s in {${\cal E}$}.

Note that some of the properties for  {${\cal E}$}
are defined considering the relation between 
\inst s in the set {${\cal E}$} and the complex $\AComp$.
Properties for the set of \inst s {${\cal E}$} {\em do not}
refer to the generated complex {$\topAComp/{{\cal E}}$} and note that,
in general,  
{$\topAComp/{{\cal E}}$} {\em is not} {$\topAComp/{R_{\top}\simpeq\AComp}$}.
This formulation is due since we are interested in
the set of all complexes of the form {$\topAComp/{{\cal E}}$} that are
decompositions of $\AComp$. 

However, we note that
topologic properties of complex $\AComp$ can be discussed by 
considering the set of \gl\ \inst s that builds $\AComp$,
(i.e. the set of \gl\ \inst s\ {${\cal E}$} such that 
$\topAComp/{{\cal E}}=\topAComp/{R_{\top}\simpeq\AComp}$.
Thus,  results in this chapter
applies to all modeling approaches that builds a simplicial complex
\gl\ together top simplices using  operators that can be modeled by
\sglinst s. 

In the following we will
derive relations between the structure of
a set of \gl\ \inst s ${\cal E}$
and the topological properties of the complex $\topAComp/{{\cal E}}$.
We will first  consider the usual topological properties 
defined in Chapter \ref{sec:cellular}  such as regularity, connectivity,
pseudomanifoldness and manifoldness. Next we will consider \Qm\ 
\cite{Lie94} and a superset of \qm we called \cdec.
The latter being particularly relevant for the forthcoming study of
decompositions. 
}
\section{Regularity, Connectivity Pseudomanifoldness}
In this section we give a characterization of topological properties 
for $\topAComp/{\cal E}$ in term of 
property for the set of \gl\ \inst s ${\cal E}$.
\subsection{Regularity}
We start with regularity as defined in Section \ref{par:reg}.
We will say that a \gl\ \inst\ {${\theta_1\equatsimp\theta_2}$} 
is {\em regular} if $\ord{(\theta_1)}=\ord{(\theta_2)}$.
For a regular \inst\ {${\theta_1\equatsimp\theta_2}$}
we define the {\em dimension}
of the \inst\ as the dimension $\ord{(\theta_1)}=\ord{(\theta_2)}$.
Note that the dimension of a regular \inst\ must not be confused
with the {\em order} of an \inst\ given by
$\ord{(\theta_1\cap\theta_2)}$.

It is easy to show that we can generate a complex whose connected components
are regular if and only if we use a set of  regular \gl\ \inst s.
\begin{property}
\label{pro:regueqregu}
The connected components of\ \  {$\topAComp/{\cal E}$} are regular if and only if
all \inst s in {${\cal E}$} are regular
\begin{proof}
First let us prove that if {$\topAComp/{\cal E}$} has regular connected
components, then all \inst s in {${\cal E}$} must be regular.
If {${\theta_1\equatsimp\theta_2}$} is an \inst\ in {${\cal E}$} then
the two top simplices $\theta^\top_1/{\cal E}$ and $\theta^\top_2/{\cal E}$ 
share some simplex in {$\topAComp/{\cal E}$}. Therefore the two top 
simplices {$\theta^\top_1/{\cal E}$} and $\theta^\top_2/{\cal E}$ belongs to 
the same connected component. 
Since each connected component is regular
we have 
$\ord{({\theta^\top_1/{\cal E}})}=\ord{({\theta^\top_2/{\cal E}})}$. 
It is easy to see that we have, for $i=1,2$,
$\ord{({\theta^\top_i/{\cal E}})}=\ord{({\theta^\top_i})}=\ord{({\theta_i})}$. 
Thus 
$\ord{({\theta^\top_1/{\cal E}})}=\ord{({\theta^\top_2/{\cal E}})}$
gives
$\ord{({\theta_1})}=\ord{({\theta_2})}$
This proves that \inst\
{${\theta_1\equatsimp\theta_2}$} must be regular.

Conversely  let be {${\cal E}$} a set of regular \gl\ \inst s.
and let be $E$
the set of independent \verteq\ equations in the set of
\verteq\ equations associated with {${\cal E}$}.
For each equation {$v_{\theta_1}\equivert v_{\theta_2}$} in $E$
we have that $\ord{(\theta_1)}=\ord{(\theta_2)}$. 
In this case we will say that equation 
{$v_{\theta_1}\equivert v_{\theta_2}$} has dimension
$h=\ord{(\theta_1)}=\ord{(\theta_2)}$. 
We will prove that if  \verteq\ equations in $E$ has this property then the 
connected  components of the generated complex {$\topAComp/E$} 
are regular.
In particular the connected components of dimension $h$ are 
given by {$\topAComp_h/E_h$} where $\topAComp_h$ is the subcomplex of
$\topAComp$ of all $h$-simplices and $E_h$ is the subset of equations
of $E$ of dimension $h$.
We will prove this by induction on the number $|E|$ of independent
\verteq\ equations. 

For $|E|=0$ must be
{$E=\emptyset$} and $\topAComp/\emptyset\simeq\topAComp$. Obviously,
connected components in $\AComp^\top$ are regular. In fact, in 
$\AComp^\top$, each top simplex is a connected component on its own.

Now, for $|E|>0$ let us consider $E^\prime=E-\{v_{\theta_1}\equivert v_{\theta_2}\}$
By inductive hypothesis we have that {$\topAComp/{E^\prime}$} has
regular connected components. 
We have that, for $i=1,2$. 
{$v_{\theta_i}/{E^\prime}\in{\theta^\top_i}/{E^\prime}$}
and $\ord{(\theta^\top_1/{E^\prime})}=\ord{(\theta_1)}=\ord{(\theta_2)}=
\ord{(\theta^\top_2/{E^\prime})}=h$. So, by inductive hypothesis,
{$v_{\theta_1}/{E^\prime}$} and {$v_{\theta_2}/{E^\prime}$}
belongs to two regular  connected components
of  {$\topAComp/{E^\prime}$} of the dimension $h$ within
{$\topAComp_h/E^\prime_h$}.
Now adding
{$v_{\theta_1}\equivert v_{\theta_2}$} we map  
{$\topAComp/{E^\prime}$} 
into {$\topAComp/E$}. 
and {$v_{\theta_1}/{E^\prime}$} collapse with 
{$v_{\theta_2}/{E^\prime}$}. This possibly merges two
regular  connected components of dimension $h$.
Therefore {$\topAComp/E$}
will have regular connected components, too.
\end{proof}
\end {property}

The complex generated by a set of  regular \inst s {${\cal E}$} 
has regular connected components. From the proof of the 
previous property we have that all connected components
of dimension $h$ are generated by the subset of regular \inst s
of dimension $h$. This is stated by the following property.
\begin{property}
\label{pro:partregular}
Let be {${\cal E}$} a set of regular \gl\ \inst s for a 
$d$-complex $\AComp$.
For all $0\le h\le d$  let be 
${\cal E}_h$ the subset of \inst s of dimension $h$ and let be
$\topAComp_h$ the subcomplex  of\ \  $\topAComp$ containing all top 
simplices of\ \  $\topAComp$ of dimension $h$. 
In this situation the set of connected components of dimension $h$ 
is the subcomplex of\ \  {${\topAComp/{\cal E}}$}
given by $\topAComp_h/{{\cal E}_h}$. 
\end{property}

\subsection{Connectivity}
Another topological property that admits an easy characterization in 
term of \gl\ \insts\ is $h$-connectivity (see Section \ref{par:conndef}). 
It is easy to see that if {${\theta_1\equatsimp\theta_2}$} and
{${\theta_2\equatsimp\theta_3}$} are two \gl\ \insts\  in ${\cal E}$ then
$\theta_1$ and $\theta_3$ are at least $k$-connected where $k$ is the 
minimum order between that of the two \insts.
In general it can be proved that,
if we 
apply a set {${\cal E}_k$} of \gl\ \insts\ of order smaller or equal than $k$, we obtain several
sets of  top simplices bundled in $k$-connected components.
More precisely the following property holds:
\begin{property}
\label{pro:conn}
Let be $\AComp$ a $d$-complex with top simplices in $\Theta$ and 
let {${\cal E}$} be a set of \gl\ \inst s. For for any $k<d$ 
let {${\cal E}_k$} be the subset 
of  \gl\ \inst s  in {${\cal E}$} of order  smaller or equal to $k$.
Let $R_k$ be the smallest equivalence on $\Theta^\top$ that  contains the
relation
$\{(\theta^\top_1,\theta^\top_2)|{\theta_1\equatsimp\theta_2}\in{{\cal E}_k}\}$.
In this situation each block in the partition of top simplices 
$\Theta^\top/R_k$ gives a set of  top simplices
{$\topAComp/{\cal E}$} that are $k$-connected.
If  {${\cal E}$} is  a closed set of \gl\ \inst s then 
{$\Theta^\top/R_k$}
 gives the partition of $\Theta^\top$ induced by the $k$-connected 
components of\ \ {$\topAComp/{\cal E}$}.
\end{property}
\begin{proof}
For each pair of top simplices $\theta^\top_a$ and $\theta^\top_b$ 
in a block of partition {$\Theta^\top/R_k$} we can find
a sequence {$(\theta^\top_i)_{i=0}^{n}$} of top simplices 
that describe a $k$-path in {$\topAComp/{\cal E}$} 
(i.e. {$(\theta^\top_i/{\cal E})_{i=0}^{n}$} )
between  $\theta^\top_a/{\cal E}$ and $\theta^\top_b/{\cal E}$.
Indeed, being  $\theta^\top_a R_k \theta^\top_b$, we can select
the $\theta^\top_i$ such that $\theta^\top_a=\theta^\top_0$, $\theta^\top_n=\theta^\top_b$ and
such that {$\theta^\top_i R_k\theta^\top_{i+1}$}
(i.e. {$\theta_i\equatsimp\theta_{i+1}$} is in {${\cal E}_k$}).
Thus, $\theta^\top_i/{\cal E}$ and $\theta^\top_{i+1}/{\cal E}$ share a $k$-face
in  {$\topAComp/{\cal E}$}. This proves that every block in {$\Theta^\top/R_k$} is $k$-connected in {$\topAComp/{\cal E}$}.

Conversely if  $\theta^\top_a/{\cal E}$ and $\theta^\top_b/{\cal E}$ are
$k$-connected in {$\topAComp/{\cal E}$} we can select a $k$-path .
{$(\theta^\top_i/{\cal E})$} made up of top simplices in between
$\theta^\top_a/{\cal E}$ and $\theta^\top_b/{\cal E}$.
Thus, $\theta^\top_i/{\cal E}$ and $\theta^\top_{i+1}/{\cal E}$ must share
at least a $k$-face. Thus {$\theta_i\equatsimp\theta_{i+1}$} must
be a \gl\ \inst\ of order greater or equal to $k$ and must be 
satisfied by {${\cal E}$}. Being the set {${\cal E}$} closed we must have 
that {$\theta_i\equatsimp\theta_{i+1}$} is in {${\cal E}$} and thus
{$\theta^\top_i R_k\theta^\top_{i+1}$}. Thus $\theta^\top_a$ and $\theta^\top_b$ 
must be in the same block of the partition {$\Theta^\top/R_k$}.
\end{proof}
For a complex generated by a set of \gl\ \insts\ of order greater or equal 
to $h$ we have that {$\Theta^\top/R_k$} do not change for all $k\le h$.
Thus, this complex must have connected components that are at least 
$h$-connected.
However note that is easy to find examples of complexes generated by
set of \gl\ \insts\ of order $0$ that have non trivial $1$-connected 
components. See for instance the upper path for the construction  
of the rightmost complex in Figure  \ref{fig:nondistmin}.
This path contains the application of four \gl\ \inst s of order $0$
(i.e. 
${\theta_1\equatsimp\theta_3}$,
${\theta_2\equatsimp\theta_3}$,
${\theta_1\equatsimp\theta_4}$ and
${\theta_2\equatsimp\theta_4}$.) and yet the
generated complex has non
trivial $1$-connected components.
This do not contrasts with the second part of Property \ref{pro:conn}
since this set of four \gl\ \inst s of order $0$ is not a closed set of
\gl\ \inst s.
On the other hand,
if a complex is $h$-connected and is not $(h+1)$-connected, not
necessarily it can be 
generated by a set of \gl\ \insts\  of order lower than $h$. 
This is an obvious consequence of the fact that $h$-connectivity need
to hold all across the complex.
\begin{example}\label{ex:caramella}
Figure \ref{fig:caramella}a, shows an obvious example 
of a $0$-connected complexes that is not $1$-connected and yet can only
be generated by sets of \gl\ \insts\ that must
contain the \inst\  ${\theta_1\equatsimp\theta_2}$. 
This is an \inst\ of order $1$.
\end{example}
{
\begin{figure}[h]
\begin{center}
\framebox{
\parbox[c][0.45\textwidth]{0.30\textwidth}{
\psfig{file=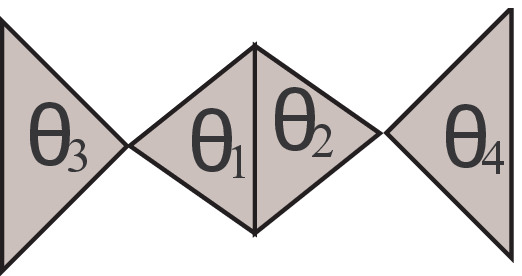,width=0.30\textwidth}
\begin{center}(a)\end{center}
}
}
\framebox{
\parbox[c][0.45\textwidth]{0.45\textwidth}{
\psfig{file=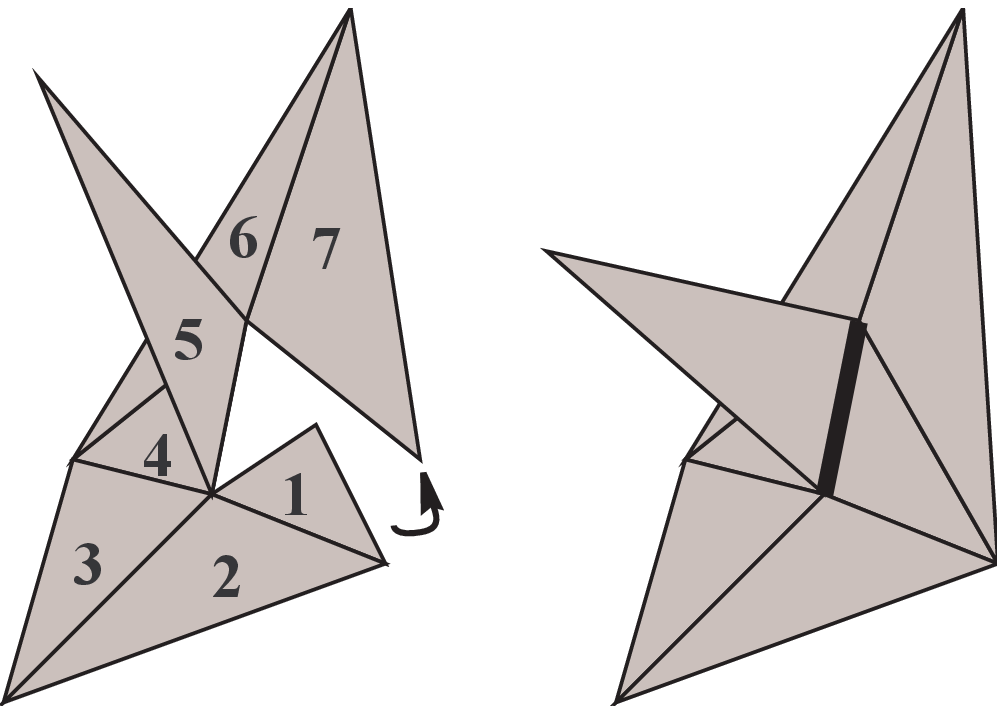,width=0.45\textwidth}
\begin{center}(b)\end{center}
}
}
\end{center}
\caption{An example (a) of a $0$-connected non $1$-connected complex that cannot
be generated using only instructions of order $0$ 
(see Example \ref{ex:caramella}). 
In (b) we have an example of a 
non-pseudomanifold complex generated by a non closed set set of 
pseudomanifold \inst s (see Example \ref{ex:nonpmnoncl})}
\label{fig:caramella}
\end{figure}
}

\subsection{Pseudomanifoldness}
\label{def:manifoldinst}
Next we analyze the relation between a set of \inst s {${\cal E}$} and
the pseudomanifoldness of the generated complex.
A non void \gl\ \inst\ $\theta_1\equatsimp\theta_2$
will be called a manifold (resp. non manifold) \inst\ (w.r.t $\AComp$) 
iff $\theta_1\cap\theta_2$ is a manifold 
(resp. non manifold) simplex
in $\AComp$.
In the following, unless otherwise stated, we will assume
that  manifold and non-manifold  \inst s are referred to $\AComp$.
Therefore we will simply say that  an \inst\   
$\theta_1\equatsimp\theta_2$ is a manifold  (non-manifold) \inst\
to mean that 
$\theta_1\equatsimp\theta_2$ is a manifold  (non-manifold) \inst\
w.r.t $\AComp$.

A first thing to note about pseudomanifoldness is that, in a $d$-complex,
we can not  have two \inst s  
that shares the same $(d-1)$-simplex whenever one of them is a manifold \inst.
This is stated by the following property.
\begin{property}
\label{pro:nopair}
Let be {${\cal E}$} the set of \gl\ \inst s that generate the
$d$-complex \ \ {$\topAComp/{\cal E}$}.
Let {$g_\gamma=\theta_1\equatsimp\theta_2$} be an \inst\ of order $(d-1)$ 
in {${\cal E}$} and 
let $\gamma=\theta_1\cap\theta_2$ be the common simplex of dimension $(d-1)$.
If  {$g_\gamma$} is a manifold \inst\ w.r.t. {$\topAComp/{\cal E}$}  
then, no other  pair $g^\prime_\gamma$, with common simplex\ \
$\gamma$, can exist in {${\cal E}$}. 
\begin{proof}
The common manifold simplex $\gamma$ must be of dimension $(d-1)$. 
Since $\gamma$ is a manifold simplex at most two $d$  
simplices can share the $(d-1)$ face $\gamma$. Therefore $g_\gamma$ is 
the unique couple of two simplices sharing $\gamma$.
\end{proof}
\end{property}
A set of \gl\ \inst s that pairwise satisfy the hypothesis of Property
of \ref{pro:nopair} is called a {\em pseudomanifold} set of \gl\ \inst s. Note that we assume that instructions are of order $(d-1)$. 
\begin{definition}[Pseudomanifold set of \gl\ \inst s]
\label{def:pminst}
A set of \gl\ \inst s {${{\cal E}}$} is called a 
set of \emas{pseudomanifold}{\gl}  \inst s if and only if
do not exist in {${\cal E}$} two \inst s 
of order $(d-1)$ that shares the same $(d-1)$-simplex.
\end{definition}
By Property \ref{pro:nopair}
a set of manifold \inst s is also a pseudomanifold set of \inst s.
Note that in a pseudomanifold set of \inst s not necessarily all
\inst s are manifold (see Example \ref{ex:nonpmnoncl}
for a pseudomanifold set of \inst s with a non-manifold \inst s).

An easy consequence of the Property \ref{pro:nopair} is that a closed sets 
of  \gl\ \inst s {$\closeq{\cal E}$} can generate a pseudomanifold
if and only if the subset of  \gl\  \inst s of order $(d-1)$ in
{$\closeq{\cal E}$} is a pseudomanifold set of \inst s.
This characterization of pseudomanifoldness is stated by the property below.
\begin{property}
\label{pro:eqpseudo}
Let ${\closeq{\cal E}}$ be a closed set of  \gl\ \inst s. A $(d-1)$-connected $d$-complex {$\topAComp/{\closeq{\cal E}}$} is a 
pseudomanifold $d$-complex if and only if ${\closeq{\cal E}}$ is a
pseudomanifold set of regular \inst s.
\end{property}
\begin{proof}
Let us first assume that the $d$-complex 
{$\topAComp/{\closeq{\cal E}}$} is a pseudomanifold and
let us prove that ${\closeq{\cal E}}$ ia a pseudomanifold  set of regular 
\inst s.
If the $d$-complex {$\topAComp/{\closeq{\cal E}}$} is a pseudomanifold the 
the complex {$\topAComp/{\closeq{\cal E}}$} is regular and
any $(d-1)$-simplex $\gamma$ in {$\topAComp/{\closeq{\cal E}}$} 
is  a  manifold $(d-1)$-simplex.
By Property \ref{pro:regueqregu} we have that the
set {${\closeq{\cal E}}$} must be a set of 
regular \inst s.
Furthermore any \inst s  of order $(d-1)$ in ${\closeq{\cal E}}$ 
must be a manifold \inst\ because all $(d-1)$-simplices are manifold in a pseudomanifold. By applying Property \ref{pro:nopair}
we get the uniqueness of pairs sharing a certain $(d-1)$-simplex. 
Therefore we have proven that {${\closeq{\cal E}}$} is a
pseudomanifold  set of regular \inst s.

Conversely if {${\closeq{\cal E}}$} is a closed set of regular \inst s
then, by Property \ref{pro:regueqregu}, we have that 
the $d$-complex {$\topAComp/{\closeq{\cal E}}$}
is regular.
Now,  let be {$\gamma^\top/{\closeq{\cal E}}$} a 
$(d-1)$-simplex in {$\topAComp/{\closeq{\cal E}}$}.
Let us assume that more than two $d$-simplices meet at {$\gamma^\top/{\closeq{\cal E}}$} and derive a contradiction.
If more than two simplices are incident to {$\gamma^\top/{\closeq{\cal E}}$} 
let us select three 
$d$-simplices $\theta_1^\top/{\closeq{\cal E}}$, 
$\theta_2^\top/{\closeq{\cal E}}$ and $\theta_3^\top/{\closeq{\cal E}}$
incident at {$\gamma^\top/{\closeq{\cal E}}$}.
Then let us consider the two  \inst s  $\theta_1\equatsimp\theta_2$
and $\theta_1\equatsimp\theta_3$. Since 
the $d$-simplices $\theta_1^\top/{\closeq{\cal E}}$, 
$\theta_2^\top/{\closeq{\cal E}}$ and $\theta_3^\top/{\closeq{\cal E}}$
share a $(d-1)$-simplex 
the two  \inst s  $\theta_1\equatsimp\theta_2$
and $\theta_1\equatsimp\theta_3$ must be satisfied by the equivalence
generated by ${\closeq{\cal E}}$.
Thus \inst s $\theta_1\equatsimp\theta_2$
and $\theta_1\equatsimp\theta_3$ must be 
in  {${\closeq{\cal E}}$}  because the set {${\closeq{\cal E}}$} is a closed
set  of \inst s.  This is against the hypothesis 
that do not exist two \inst s
of order $(d-1)$ in {$\closeq{\cal E}$}
such that the  two \inst s  shares the same $(d-1)$-simplex.
Therefore at most two $d$ simplices meet at a generic
$(d-1)$ simplex {$\gamma^\top/{\closeq{\cal E}}$}. 
Therefore, being   {$\topAComp/{\closeq{\cal E}}$} 
$(d-1)$-connected by hypothesis,  this proves that {$\topAComp/{\closeq{\cal E}}$} is a pseudomenifold.
\end{proof}

Note that the proof of the above property builds essentially on the closure of
the set of generating \inst s {${\closeq{\cal E}}$}.
Indeed, it is quite easy to find examples of non closed pseudomanifold
sets of regular 
\inst s {${{\cal E}}$} 
that   generates a non-pseudomanifold $d$-complex
The construction of
such an example is already possible for $d=2$ and rests on the possibility
that some \inst\ can be ''implicit'' within a non closed set of \gl\
\inst s {${{\cal E}}$}. We present this situation in the following example. 
\begin{example}\label{ex:nonpmnoncl}
The non-pseudomanifold $2$-complex on the right of Figure 
\ref{fig:caramella}b can be generated by the non 
closed set of \gl\  $1$-\inst s {$${\cal E}=\{
{\theta_1\equatsimp\theta_2},
{\theta_2\equatsimp\theta_3},
{\theta_3\equatsimp\theta_4}, 
{\theta_4\equatsimp\theta_5},
{\theta_4\equatsimp\theta_6},
{\theta_6\equatsimp\theta_7},
{\theta_7\equatsimp\theta_1}
\}$$}
This set of \gl\
\inst s induce the
\inst\ {$\theta_1\equatsimp\theta_5$} that is  not  present in the
non-closed generating set {${\cal E}$}. It is easy to see that
this example is not a counterexample to Property \ref{pro:eqpseudo}.
In fact, to close set {${\cal E}$} we must add \inst\ 
{$\theta_1\equatsimp\theta_5$} to the set {${\cal E}$}.
This instruction {\em clashes} with \inst\ {$\theta_4\equatsimp\theta_5$}
since \inst s {$\theta_4\equatsimp\theta_5$}
and {$\theta_1\equatsimp\theta_5$} share the black thick non-manifold edge
in Figure \ref{fig:caramella}b.
Adding the \inst\ {$\theta_1\equatsimp\theta_5$}
to the set  {${\cal E}$} we violate condition in
Definition \ref{def:pminst}  and obtain a non-pseudomanifold set of 
\gl\ \inst.
\end{example}

As we have seen in  Property  \ref{pro:eqpseudo} 
any closed pseudomanifold set of regular \inst s will
generate a pseudomanifold. The following property offer 
an alternative formulation of this fact.  
\begin{property}
\label{pro:cpmm}
In a closed pseudomanifold set of regular \inst s  {${\closeq{\cal E}}$} all
\inst s  of order $(d-1)$ are manifold w.r.t {$\topAComp/{\closeq{\cal E}}$}.
\begin{proof}
Let be {$\theta_1\equatsimp\theta_2$} an \inst\ of order $(d-1)$ with 
$\gamma={\theta_1\cap\theta_2}$. We have to prove that the $(d-1)$-simplex
$\gamma$ is a manifold simplex. Let us assume that 
$\gamma$ is not a manifold simplex and derive a contradicion.
If $\gamma$ is not a manifold simplex then more than two $d$-simplexs are 
incident to $\gamma$ let us select three 
$d$-simplices $\theta_1$, $\theta_2$ and $\theta_3$  incident at $\gamma$.
Then let us consider the two  \gl\ \inst s $\theta_1\equatsimp\theta_2$
and $\theta_1\equatsimp\theta_3$.
These two \gl\ \inst s  must be 
in  {${\closeq{\cal E}}$}  because the set {${\closeq{\cal E}}$} is a closed
set  of \gl\ \inst s.  This is against the hypothesis 
that {${\closeq{\cal E}}$} is a pseudomanifold set.
\end{proof}
\end{property}

\section{Manifoldness}
Next we want to study the relation between sets of \gl\ \inst s 
and manifoldness. This study will lead to the introduction of two 
classes of non manifold complexes. We called these
classes {\em \Qm} (after \cite{Lie94}) and
{\em \cdecs}.
This will show that it is possible to define several degrees
of non-manifoldness. Usually in leterature we just find two classes
of non-manifold complexes, namely regular complexes and pseudomanifolds.
In the following we will study the notion of \qm\ and
show that \qm\ can be generated by a particular
class of sets of \gl\ \inst s.
Later on, in order to characterize the connected components of our decomposition scheme
we will introduce a superset of \qm\ we called 
{\em \cdec\ } complexes. 

All these notions comes out quite naturally if one attempts to relate
manifoldness in the complex   {$\topAComp/{{\cal E}}$} with some property
for the set ${{\cal E}}$. A first step in this direction is the following
property  that gives an obvious relation between manifold complexes and 
sets of manifold \gl\ \inst s.
\begin{property}
If {$\topAComp/{{\cal E}}$} is a combinatorial manifold then all \inst s
in {${{\cal E}}$} must be manifold w.r.t {$\topAComp/{{\cal E}}$}.
\begin{proof}
If {${\theta_1\equatsimp\theta_2}$} is an \inst\ in {${\cal E}$} then
the two top simplices $\theta_1/{\cal E}$ and $\theta_2/{\cal E}$ 
share some simplex in
{$\topAComp/{\cal E}$}. This common simplex is a manifold simplex being
{$\topAComp/{{\cal E}}$} a manifold. Therefore
{${\theta_1\equatsimp\theta_2}$} is a manifold \inst.
\end{proof}
\end {property}
Note that the converse is not true. 
In fact, even if  all \inst s in {${\closeq{\cal E}}$} are manifold, it 
is still possible that  
{$\topAComp/{\closeq{\cal E}}$} is not  a combinatorial manifold.
\begin{example}
\label{ex:moebnonmani}
The simplest example of such a complex is given by the cone to 
the  triangulation of the Moebius strip (see Figure \ref{fig:nonmani}).
\begin{figure}[h]
\begin{center}
\fbox{\psfig{file=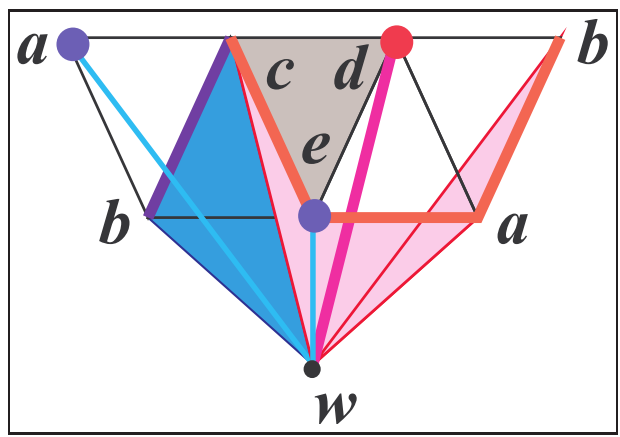,width=.48\textwidth}}
\end{center}
\caption{A non-manifold $3$-complex where all tetrahedra meet at 
manifold simplices}
\label{fig:nonmani}
\end{figure}
Let us consider the (unfolded) Moebius  strip in Figure \ref{fig:nonmani}. This 2-complex is made up
of five triangles ($abc$,$bce$,$ced$,$aed$ and $abd$).
Next, let us build the cone from $w$ to this triangulation
The first thing to note is that vertex $w$ is a non-manifold vertex. In fact,
the link of $w$ is a Moebius strip.
However, every pair of incident tetrahedra
in this cone shares  a triangle where just two tetrahedra meet.
This can be seen
looking at the triangulation of the Moebius strip in Figure \ref{fig:nonmani}.
In this triangulation every triangle  incident to an edge  with another shares  an edge where just two triangles meet.
For instance 
the triangle  $cde$ share 
edge $ce$ with triangle $bce$ and
edge $de$ with triangle $ade$.
When we take the cone from $w$, we have that
the common (red) vertex between two triangles
becomes a common edge between two tetrahedra 
(e.g. the  thick edge in violet). 
Similarly the common (blue) edge becomes a triangle 
(e.g. the triangle in pale blue).
It is easy to see that any pair of 
tetrahedra in this complex intersect at a manifold simplex. 
To show this, we first note that the link of the intersection
of two triangles in the Moebius strip is the link of a simplex within a 
manifold (indeed the Moebius strip is a manifold).
In particular, when the two triangles shares an edge
this link is made up  of a couple of points,
(e.g. the blue blobs in figure).
When the two triangles shares a single vertex
this link is a path made up  of  three consecutive 
edges (e.g. the the red thick lines in figure).
When we take the cone from $w$ 
we have that the cone to
the intersection of two triangles becomes the intersection
of two tetrahedra. The cone to the corresponding link becomes the link of the intersection of the two 
tetrahedra (e.g. the two pale blue edges at $w$ and the
three pale red triangles). 
From the form of these links 
we can say that any pair of 
tetrahedra in this complex intersect at a manifold simplex. 
For this reason, any \gl\ \inst\ {${\theta_1\equatsimp\theta_2}$}
have a manifold common simplex {$\theta={\theta_1\cap\theta_2}$}.
Therefore the
closed  set of \inst s that describes this complex is a set of manifold 
\inst s even if the $3$-complex is not a $3$-manifold at point $w$.
\end{example}
So, it is not possible to say that a closed set of manifold \inst s will
always generate a manifold.
Indeed, as  we will see in the following, it can be proven that
a  closed set of manifold \inst s  generates a {\em \qm}.
However, not all \qm\ can be generated by closed sets of manifold \inst s.
Thus, we point out that, the exact characterization of the set of complexes
generated by closed sets of manifold \gl\ \inst s is a problem left open
by this thesis.

To go on with our study on manifoldness we start by studying 
some properties of manifold \inst s.
A first fact is that manifold \inst s are  regular.
\begin{property}
\label{pro:maniregu}
Manifold \inst s are regular \inst s.
A manifold \inst\  cannot be made up of 
two top simplices of different dimensionality. 
\begin{proof}
If {${\theta_1\equatsimp\theta_2}$} is a manifold \inst,
then the two top simplices $\theta_1$ and $\theta_2$ 
share a simplex 
{$\gamma={\theta_1\cap\theta_2}$} that must be a manifold
simplex in the $d$-complex $\AComp$.
By Definition \ref{def:simplnonmani} the link $\lk{\gamma}$ must be an  
$h$-complex (whith $h=d-\dim{\gamma}-1$) that is combinatorially 
equivalent  either to the $h$-sphere or
to the  $h$-ball. In both cases, by Property \ref{pro:maninonmani}
Part \ref{pro:1top},  we have
that $\lk{\gamma}$ is an $h$-manifold and hence 
$\lk{\gamma}$ must be a regular $h$-complex.
Now $\theta_1-\gamma$ and $\theta_2-\gamma$ are two simplices in
the regular $h$-complex $\lk{\gamma}$. Hence $\theta_1-\gamma$ and 
$\theta_2-\gamma$ must have the same dimension $h$ and this implies
that both $\theta_1$ and $\theta_2$ have the same dimension.
Therefore \inst\ {${\theta_1\equatsimp\theta_2}$} is a regular
\inst.
\end{proof}
\end{property}
Manifoldness embodies some notion of regularity. Indeed the location
of top simplices around a manifold simplex follows a certain pattern.
Therefore no surprise if some of the \gl\ \inst s that stitch
top simplices around a 
manifold simplex are redundant. Indeed
all manifold \inst s $\theta_1\equatsimp\theta_2$ of order
smaller than the maximum 
(note that the maximum is $\dim{(\theta_i)}-1$) can be neglected.
This is one of the consequences of the following  property that
relate redundant \inst s with connectivity
(see Definition \ref{par:conndef}) in the generated complex. 
\begin{property}
\label{pro:delnond}
Let be $\theta_1$ and $\theta_2$ two top $d$-simplices in a $d$-complex $\AComp$.
Let be  {${\theta_1\equatsimp\theta_2}$} a \gl\ \inst\ 
and let be {$\gamma=\theta_1\cap\theta_2$}.
Then the following facts holds:
\begin{enumerate}
\item
\label{pro:delnod1}
If $\str{\gamma}$ is $(d-1)$-connected 
then there exist a set  of regular \gl\ \inst s 
{${\cal E}_\gamma$}, of 
order $(d-1)$, such that {${\theta_1\equatsimp\theta_2}$}
is satisfied by $\equivert^{{\cal E}_\gamma}$.
\item
\label{pro:delnodpm}
If $\str{\gamma}$ is $(d-1)$-manifold-connected then the set 
{${\cal E}_\gamma$} will be a pseudomanifold set of regular
\inst s.
\end{enumerate}
\begin{proof}
Not to bother  the reader we will embed the proof for the case 
of manifold connected stars (Part \ref{pro:delnodpm}) into the
proof for plain $(d-1)$-connected stars (Part \ref{pro:delnod1}).
This will be done by adding, when needed by Part \ref{pro:delnodpm}, 
the adjectives
(manifold) or (pseudomanifold) between parenthesis. 
The reader should skip this, or read this, 
depending on which proof she  (or he)  wants to read.

Being $\str{\gamma}$ a $(d-1)$-(manifold) connected star there exist, for some $n$ a
(d-1)-(manifold) path of $n+1$ $d$-simplices  {$(\theta^{(i)})_{i=0}^{n}$} in
$\str{\gamma}$ with $\theta^{(0)}=\theta_1$ and $\theta^{(n)}=\theta_2$.
Let us consider the regular (pseudomanifold)  set {${\cal E}_\gamma$} of $n$
\gl\ \inst s of order $(d-1)$ given by:
{${\cal E}_\gamma=
\{{\theta^{(i)}\equatsimp\theta^{(i+1)}}|i=0,\ldots,(n-1)\}$}.
We have to show that {${\theta_1\equatsimp\theta_2}$}
is satisfied by $\equivert^{{\cal E}_\gamma}$.
In fact, for any {$v\in\gamma$}, we have that $v\in\theta^{(i)}$ for 
all $i=0,\dots,n$.  Therefore the set of \verteq\ equations 
associated with  ${{\cal E}_\gamma}$ contains the  $n$ equations
{$v_{\theta^{(i)}}\equivert v_{\theta^{(i+1)}}$} for all $i=0,\ldots,(n-1)$.
Closing with transitivity we have that equation
{$v_{\theta^{(0)}}\equivert v_{\theta^{(n)}}$} must be in
$\equivert^{{\cal E}_\gamma}$.
By construction we have $\theta^{(0)}=\theta_1$ and $\theta^{(n)}=\theta_2$.
Therefore we have that, for any {$v\in\gamma$},
the equation {$v_{\theta_1}\equivert v_{\theta_2}$} is satisfied
by $\equivert^{{\cal E}_\gamma}$.
So the \gl\ \inst\ {${\theta_1\equatsimp\theta_2}$}
is satisfied by $\equivert^{{\cal E}_\gamma}$.
\end{proof}
\end{property}
An easy consequence of the property above is that, in a $d$-complex,
a manifold
\inst\ of order smaller than $(d-1)$ can be replaced by a set of
of manifold \inst s of order $(d-1)$.
This is stated in the following property.
\begin{property}
\label{pro:manid1}
Let $\AComp$ be a $d$-complex and
let {$\theta_1\equatsimp\theta_2$} be a manifold \gl\ \inst of order strictly
smaller than $(d-1)$.
Then, there exist a set  of manifold \gl\ \inst s 
{${\cal E}_\gamma$}, of 
order $(d-1)$, such that {${\theta_1\equatsimp\theta_2}$}
is satified by $\equivert^{{\cal E}_\gamma}$.
\begin{proof}
Let us denote with $\gamma$ the common simplex  i.e.
$\gamma=\theta_1\cap\theta_2$.  By hypothesis $\gamma$ is a manifold 
simplex. Therefore $\lk{\gamma}$, being combinatorially equivalent to
a  $h$-sphere or a  $h$-ball, is a connected  $h$-manifold
(with $h$ the dimension of $\lk{\gamma}$).
By Property \ref{pro:manid1conn} we have that $\lk{\gamma}$ is
$(h-1)$-manifold connected and therefore  
$\str{\gamma}$ is $(d-1)$-manifold connected. 
By applying the previous Property
\ref{pro:delnond} Part \ref{pro:delnodpm} we get the thesis.
\end{proof}
\end{property}

\section{\Cdec}
Going on with our analysis of complexes generated by sets of
manifold \inst s we first consider {\em non closed} sets of manifold
\inst s.
We first note that {\em non closed} 
pseudomanifold sets of \inst s  
can generate complexes that are neither manifold, nor {\em pseudomanifold}.
In the following we will give an example of a non-pseudomanifold 
$3$-complex that 
can be generated by a non-closed  sets of manifold \gl\ \inst.
In general, for $d\geq 3$, 
it is possible to find a (non closed) set of manifold \insts\ that
generate a $d$-complex that is not even a pseudomanifold.
On the other hand, we have seen in Example \ref{ex:moebnonmani} that,
for $d\geq 3$,  there are examples of closed sets of manifold  \insts\
that generate a non-manifold complex.

At this point, one might wonder whether or not sets of manifold
instructions can define any meaningful class of complexes.
The answer to this question is positive.
In fact, in the following,
we will show that two different classes  of non-manifold complexes,
called \Qm\ and \Cdec s,
are  created by sets  of manifold \gl\ \inst s, depending on whether the set of \insts\ is closed or not.
The first class is the class of {\em \Cdec s} complexes.
\Cdec s are generated by non-closed sets of manifold \inst s.
This class of non-manifold complexes was introduced for the first time 
in \cite{Def02b} and.  
it is important to the sequel of this thesis. Indeed since 
\Cdec s characterize the results of our decomposition of non-manifold 
complexes.

\begin{figure}[h]
{
\begin{center}
\framebox{
\parbox[c][0.39\textwidth]{0.45\textwidth}{
\psfig{file=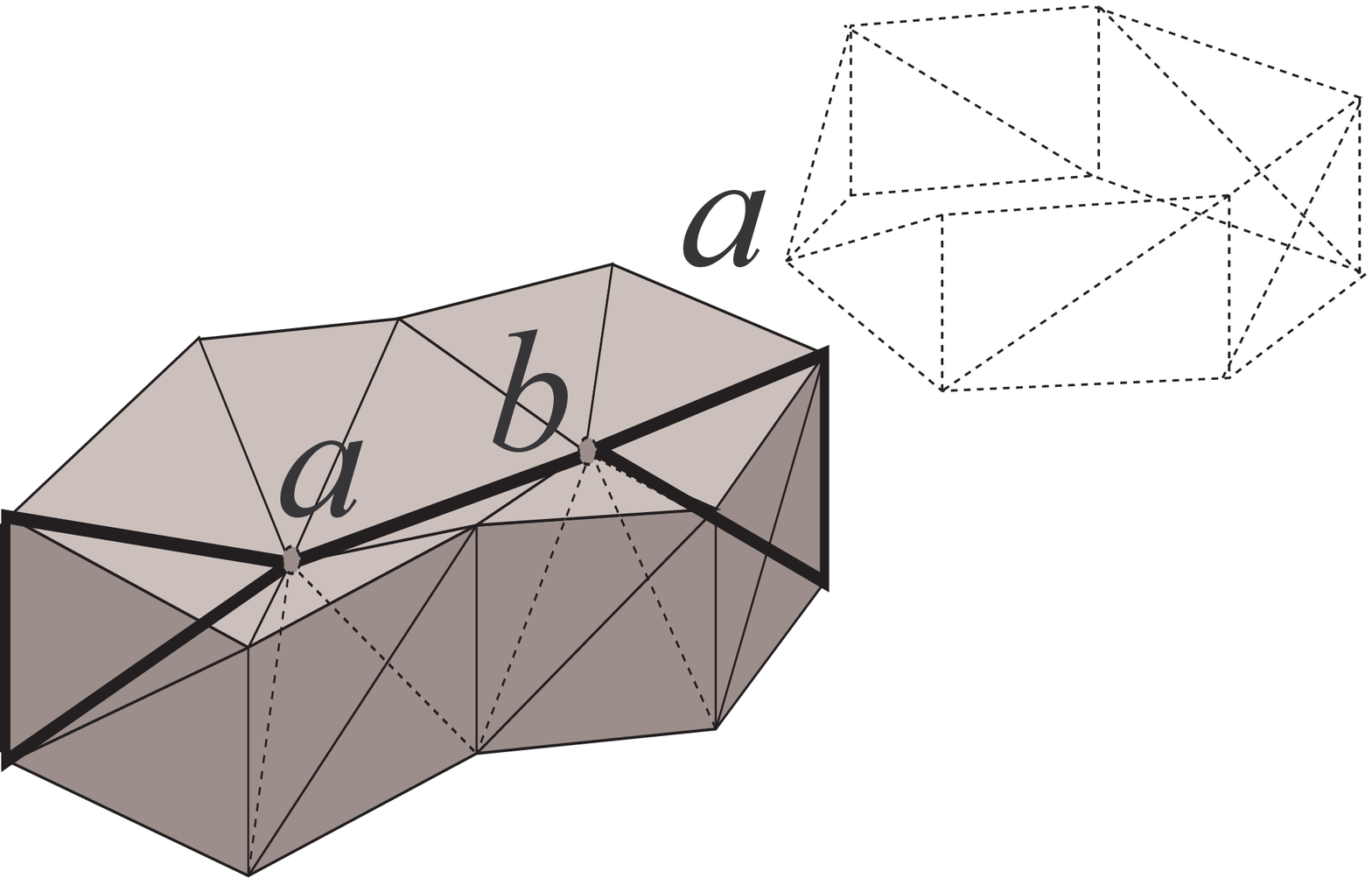,width=0.45\textwidth}
\begin{center}(a)\end{center}
}
}
\framebox{
\parbox[c][0.39\textwidth]{0.30\textwidth}{
\psfig{file=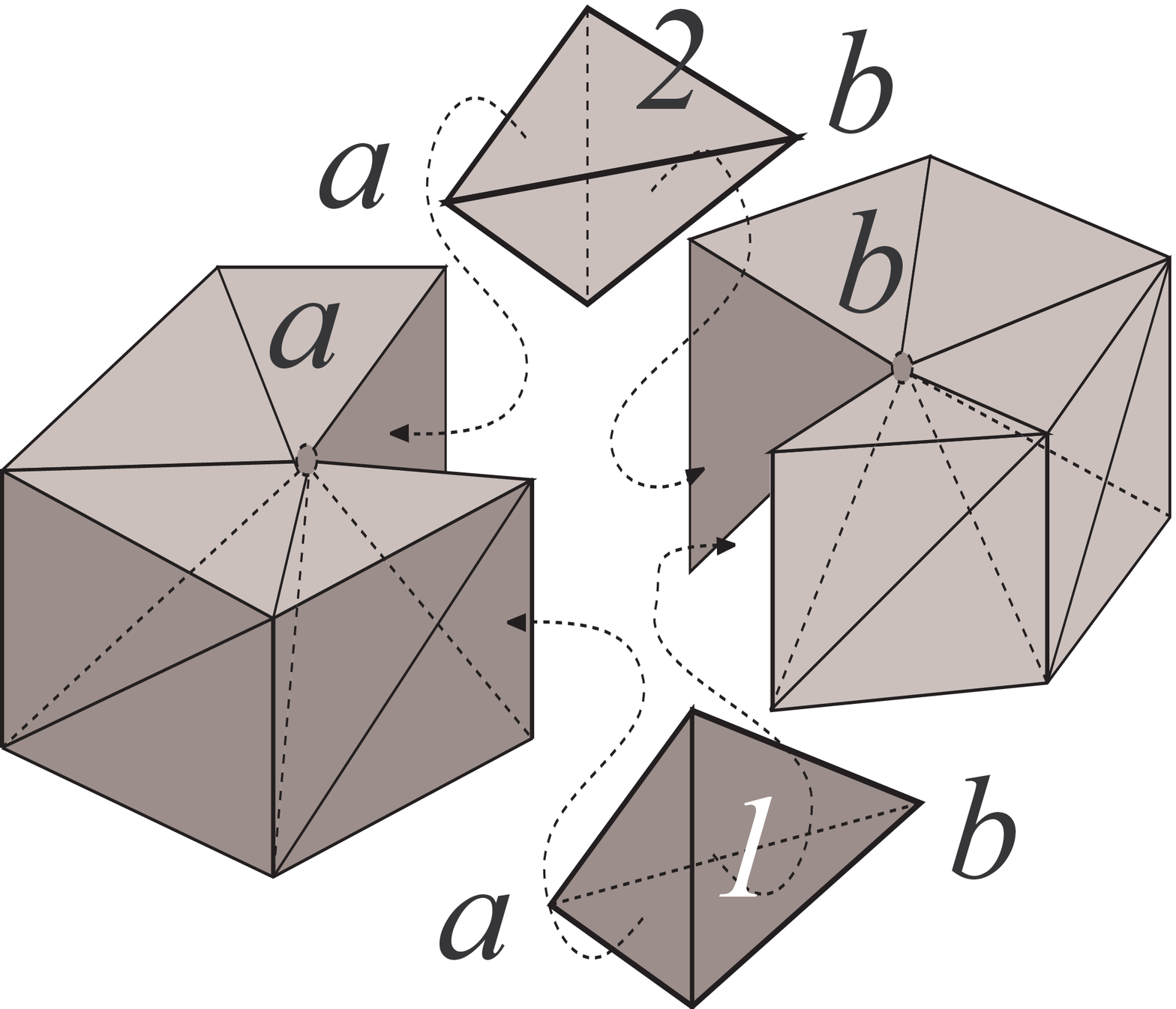,width=0.3\textwidth}
\begin{center}(b)\end{center}
}
}
\end{center}
}
\caption{An example of an \cdec\ complex}
\label{fig:cnondec}
\end{figure}
\begin{example}
In Figure \ref{fig:cnondec}a we report a  examples of an \cdec\ complex.
Note that the complex is {\em punched}
at $a$ and $b$. To show this we marked in black a cross section.
The complex of  Figure \ref{fig:cnondec}  is an \cdec\  $3$-complex.
This is not a manifold. In fact, the two central thick Vertices $a$ and $b$
have a link that is not combinatorially equivalent neither to a sphere
nor to a disk.
The link of vertex $b$ is the dashed surface on the left of Figure  \ref{fig:cnondec}a.
Figure \ref{fig:cnondec}b shows that this complex can
be created \gl\ tetrahedra at manifold triangles.
Thus, this complex is generated by a (non-closed) set  of manifold \gl\ 
\inst s of order $2$. 
These \inst s  induce the  non-manifold \gl\ \inst\ 
{${\theta_1\equatsimp\theta_2}$} of order $1$ (note the two labels 1 and 2 in part   b).
\end{example}

Thus in this section we introduce \cdec s as the class of complexes generated
by (non-closed) sets of  manifold \gl\ \inst s.
Next we will characterize \cdec s through local topological properties.  
In the next section we will show that {\em closed} sets of manifold \gl\ 
\inst s generate the known class of \Qm\ introduced by Lienhardt
\cite{Lie94}.

We start this section introducing  an example of a non-pseudomanifold 
$3$-complex that can be generated by a non-closed pseudomanifold 
sets of \gl\ \inst.
This correspond to the rather counter intuitive fact
there exist non-pseudomanifold $3$-complexes  (although not embeddable in
$\real^3$) that can be generated by 
glueing together tetrahedra at triangles putting glue on
triangles where just two tetrahedra glue at time. 
In other words non-pseudomanifold adjacency can be induced
using the (usual) manifold glue (i.e. manifold adjacencies) on
triangles.
\begin{example}{\bf (A non-pseudomanifold $3$-complex generated by glueing at manifold triangles)}
\label{app:example}
Here, we present an example of a  3-complex that is not a
pseudomanifold and yet it 
can be generated by a non-closed pseudomanifold sets of \inst.
This example is rather complex since it does not admit a geometric
embedding in 3D space.
Therefore, we describe it as an assembly of pieces that may be built
through a pseudomanifold set of gluing $2$-instructions.
The general idea is that, while we explicitly glue tetrahedra at
manifold triangles, some gluing at non-manifold triangles may be
implicitly induced among other faces of such tetrahedra.
We first build a pseudomanifold complex that has a cavity that can be
filled only through a non-pseudomanifold complex made of three tetrahedra incident
at a common triangle.
Then, we fill this cavity by gluing new tetrahedra on the cavity boundary.
Although each tetrahedron introduced to fill the cavity is glued at three
manifold triangles, non-manifold adjacency are induced among such new tetrahedra, and the
final complex is necessarily non-pseudomanifold.

\begin{figure}[h]
\centerline{\fbox{\psfig{file=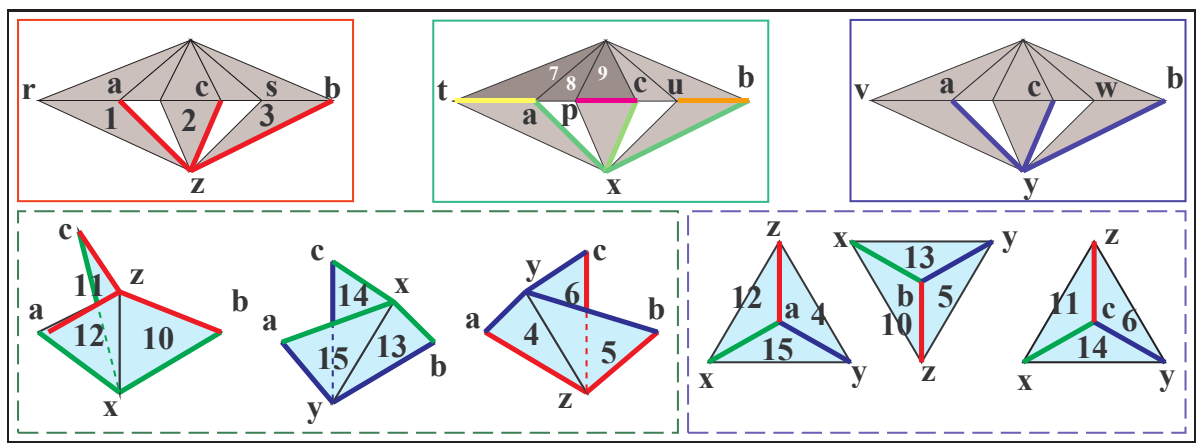,width=0.95\textwidth}}}
\caption{A non-pseudomanifold $3$-complex
generated by a non-closed pseudomanifold set of $2$-\inst s}
\label{fig:tienonreg}
\end{figure}

We start with the 2-complex formed from the three pieces in the first row of
Figure \ref{fig:tienonreg}.
Such pieces form a connected component since they share three vertices
$a$, $b$ and $c$.
Each piece contains eight triangles.
Then we build the following three cones: from $x$ to the complex on the
left (framed in red); from $y$ to the complex in the
middle (framed in green);
and from $z$ to the complex on the
right (framed in blue).
Such cones do not share tetrahedra or triangles because the three
$2$-complexes in the top row do not share either a triangle or an edge.
Therefore, it is easy to see that the resulting complex is a pseudomanifold. The three cones introduce twenty-four tetrahedra. 
The three cones will share some edges, namely those
connecting vertices $x$, $y$ and $z$ to vertices $a$, $b$ and $c$
(thick red, green and blue edges).
Due to this fact,
the boundaries of the three cones form a closed cavity,
which is bounded by
the nine incident triangles numbered
$4$,$5$,$6$,$10$,$11$,$12$,$13$,$14$ and $15$.
On the left side of the second row in Figure \ref{fig:tienonreg}
line)
we report these nine triangles organized into three ``T''s.
Thick red, green and blue edges in the first and second row are shared by
the three cones and are on the boundary of this cavity.

Note that additional thick  colored  lines and some triangle
numbering (that now might seem unnecessary) are  placed here for later reference.
Also the order in which triangles are numbered, that might seem quite
arbitrary,  is relevant for the second part of this example.

On the right side of the second row in Figure \ref{fig:tienonreg}
we report the same nine triangles organized into three fans of triangles
around vertices $a$, $b$ and $c$.
 From this last presentation, we can see that the three ''T''s form the boundary
of a complex made of three tetrahedra $axyz$, $bxyz$ and $cxyz$.
Note that the complex we have built so far does not contain these three tetrahedra,
while it contains all their faces, except $xyz$.
So, we may add such three tetrahedra to the complex through manifold glueing instructions.
The resulting complex is non-pseudomanifold since triangle $xyz$ has three incident tetrahedra.
The total number of tetrahedra for this complex is twenty-seven.
Note that such incidences at $xyz$ were never specified, but implicitly induced, by glueing
instructions.

Up to now we have simply detailed the shape of our non-pseudomanifold 
$3$-complex for this example.
The next step is to show that such a complex can be generated  by 
a set on manifold equations of order $2$. 
More intuitively the question is:  can we build this complex
by stitching tetrahedra
at triangles where only two tetrahedra meet.

This fact can be verified intuitively  
considering the complex on the left of  Figure \ref{fig:tienonreglk}. 
On the left, \begin{figure}[h]
\fbox{\psfig{file=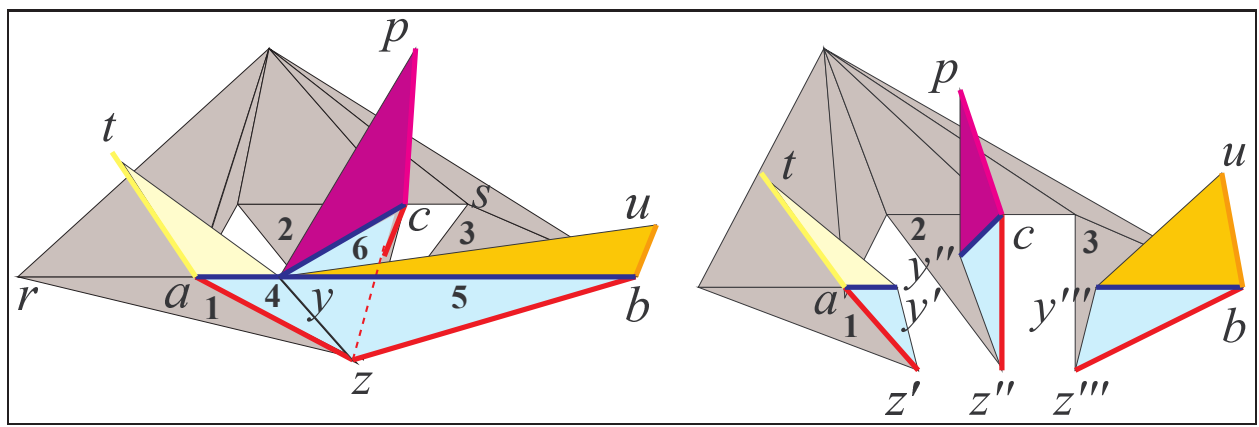,width=\textwidth}}
\caption{How to stitch together $\lk{x}$ (on the left) for the complex of
Figure \ref{fig:tienonreg}}
\label{fig:tienonreglk}
\end{figure}
we have the $2$-complex that is the link of $x$. Pale blue triangles
$4$, $5$ and $6$ comes from the three tetahedra added in the cavity. 
The three colored triangles (yellow,violet and orange)
comes from the cone from $y$ to three lines in the 
complex in the middle of the upper row in Figure \ref{fig:tienonreg}.
The triangle  in yellow,violet and orange comes from the cone from 
$y$ to, respectively, the thick yellow,violet and orange edges 
in the $2$-complex in the middle of upper row of Figure \ref{fig:tienonreg}.
Note that, for symmetry, $\lk{y}$ and $\lk{z}$ must be isomorphic to 
this complex.

Let us consider the $3$-complex that result from the cone from $x$ 
to $\lk{x}$ (i.e. $\clstar{x}$). 
This subcomplex cannot be built 
by stitching together tetrahedra at manifold $2$-faces. 
However the largest complex we can obtain 
by stitching together tetrahedra at manifold $2$-faces. 
is the decomposition obtained as the cone from $x$ to the $2$-complex
on the right of Figure \ref{fig:tienonreglk}.
This takes seven  \inst s to put in place the gray triangles. 
Having obtained this complex 
we start stitching other tetrahedra from $\clstar{y}$ onto it.
Note that six tetrahedra from $\clstar{y}$ are already in place. 
They are  the cones from $x$ to the six colored triangles
in Figure \ref{fig:tienonreglk}. Next we  stitch the three tetrahedra, corresponding
to the cone from $y$  to the darker triangles $7$, $8$, $9$ in  the 
top row of Figure \ref{fig:tienonreg}.
In this way  we add four simplex equations. First and last simplex 
equations are between tetrahedra sharing $tay^\prime$ and $pcy^\second$.
Others connect the three added tetrahedra.
These four simplex equations induce the vertex equation 
$y^\prime\equivert y^\second$.

Similarly we consider the four tetrahedra 
that are cones from $y$ to triangles $10$, $11$, $12$ and $13$. 
Stitching them  from $pcy^\second$ to $uby^\third$ 
we induce $y^\second\equivert y^\third$.
Hence by stitching tetrahedra at  manifold triangles we build all 
$\clstar{y}$ and have $y^\prime$, $y^\second$ and $y^\third$ collapsing into
a unique vertex for $y$. 

For symmetry, it is easy to see that, we can consider the triangles  $1$, $2$
and $3$ and stitch tetrahedra from $\clstar{z}$ to induce \verteq\ equations
$z^\prime\equivert z^\second$ and $z^\second\equivert z^\third$.
The whole process stitch together  tetrahedra $axy^\prime z^\prime$,
$cxy^\second z^\second$ and   $bxy^\third z^\third$ withuot using
the non-manifold simplex equations involving  
$axyz$, $bxyz$ and   $cxyz$. This takes twenty-one instructions for the three cones. Finally we introduce nine  \inst s to fill the cavity for a total of thirty  \inst s to glue the twenty-seven tetrahedra.

The reader that remains skeptical about this construction could refer to Appendix \ref{sec:prolog} where we present a Prolog program for this example. 
This program  details the list ${\cal E}$ of thirty  \inst s. Next, the program  starts from the totally exploded complex $\topAComp$ made up of twenty-seven disjoint  tetrahedra and execute the list of thirty  \inst s ${\cal E}$. 
At each \inst\ adjacencies are updated. Finally, when all \inst\ are executed,  the program  checks that, in the final complex, $\topAComp/{\cal E}$ all tetrahedra join  at manifold triangular faces except for a non pseudomanifold triangular face not considered by any \inst\
 in ${\cal E}$ .
\end{example}
Finally note that taking the cone of from an external vertex to the complex of Example \ref{app:example} we can create an example of a 4-complex that is not a
pseudomanifold and yet it 
can be generated by a non-closed pseudomanifold sets of \inst. Then taking again the cone from a fresh new vertex we can create the same sort of example for $d=5$ and so on for all $d\ge 3$.

Even if non-closed set of $(d-1)$ \inst s can generate quite wierd
complexes 
it is possible to characterize the set of 
of regular $d$-complex that can be generated by a non closed  set of 
$(d-1)$-\inst s. 
This characterization is given by the following property that will lead us
to the definiton of the class of \cdec s complexes.
\begin{property}
\label{pro:qpmeq}
Let $\AComp^\prime$ be a complex in the \dec\ lattice for a $d$-complex $\AComp$.
The following facts hold:
\begin{enumerate}
\item
\label{pro:qpmd1eq}
If the star of every vertex in $\AComp^\prime$ is $(d-1)$-connected
then there exist a set of regular $(d-1)$-\inst s {${\cal D}$}
such that $\AComp^\prime=\topAComp/{\cal D}$.
\item
\label{pro:peqd1qpm}
If the set {${\cal D}$} is a set of $(d-1)$-\inst s 
then, in  the  generated complex $\topAComp/{\cal D}$,
the star of every vertex is $(d-1)$-connected. 
\item
\label{pro:qpmpmeq}
If the star of every vertex in ${\AComp^\prime}$ is 
{$(d-1)$-manifold-connected} 
then there exist a pseudomanifold set of regular 
$(d-1)$-\inst s {${\cal D}$}, that are manifold w.r.t. $\AComp$, and
such that $\AComp^\prime=\topAComp/{\cal D}$.
\item 
\label{pro:peqmqpm}
If the set {${\cal D}$} is a set of $(d-1)$-\inst s  that are 
manifold w.r.t. $\AComp$ then, in the  generated decomposition
of $\AComp^\prime=\topAComp/{\cal D}$,
the star of every vertex is $(d-1)$-manifold-connected. 
\end{enumerate}
\end{property}
\begin{proof}
Let us consider a complex ${\AComp^\prime}=\topAComp/{\cal E}$ 
such that the star of every vertex is $(d-1)$-connected.
We will show that we can delete in {${\cal E}$} every \gl\ \inst\  
provided that we add a certain set of $(d-1)$-\inst s. 
This will prove Part \ref{pro:qpmd1eq}.
We will also show that this added set can be a set of
manifold \inst s provided that 
the star of every vertex in  ${\AComp^\prime}$ is $(d-1)$-manifold-connected
w.r.t. {${\AComp^\prime}/\Rtop$}. 
This will prove Part \ref{pro:qpmpmeq}.
Not to bother  the reader we will embed the proof for the case 
of manifold connected stars (Part \ref{pro:qpmpmeq}) into the
proof for plain $(d-1)$-connected stars (Part \ref{pro:qpmd1eq}).
This will be done, as already done  in proof of Property   \ref{pro:delnond},
by adding in some places the adjective
(manifold) between parenthesis. 
The reader should skip this
depending on which proof  she (or he) is interested in.

Let $\theta_1\equatsimp\theta_2$ be the \inst\ to be deleted and 
let be $\gamma=\theta_1\cap\theta_2$ the common simplex.
Let be $v$ a generic vertex in $\gamma$.
For this generic vertex $v$ we will provide a set of
$(d-1)$ (manifold) \gl\ \inst s {${\cal D}_v$} 
that satisfy the \verteq\ equation 
$v_{\theta_1}\equivert v_{\theta_2}$ (one of those added by
$\theta_1\equatsimp\theta_2$).
To build {${\cal D}_v$},
we start by  noticing  that both $\theta_1$ and $\theta_2$ belongs to 
$\str{v}$. 
The star of $v$ is $(d-1)$-connected in the decomposition,
{${\AComp^\prime}$}. However, pasting ${\AComp^\prime}$ with the relation
$\Rtop$  we do not impair $(d-1)$-connectivity. 
Indeed from {${\AComp^\prime}$} to {${\AComp^\prime}/\Rtop$} 
there an \asm\ that preserve simplex dimension thus preserving
$(d-1)$-paths.   Therefore the star of any
vertex is $(d-1)$-connected  
in  ${{\AComp^\prime}/\Rtop}=\topAComp/\Rtop\simpeq\AComp$.  
So the star $\xstr{\AComp}{v}$ must be $(d-1)$-(manifold)-connected.

Now we are considering $\xstr{\AComp}{v}$. By hypothesis 
this star is $(d-1)$-(manifold)-connected. Then we can find a 
$(d-1)$-(manifold) path {$(\theta^{(i)})_{i=1}^{n}$} in $\str{v}$
made up  of $n$  $d$-simplices $\theta^{(i)}$.
From this path  let us build the set  of $n-1$
(manifold) \inst s. 
{${\cal D}_v=
\{{\theta^{(i)}\equatsimp\theta^{(i+1)}}|i=1,\ldots,(n-1)\}$}.
Proceeding as in 
the proof of Property \ref{pro:delnond} we will find 
that the set of \verteq\ equations 
associated with  {${\cal D}_v$} contains the  $n-1$ equations
{$v_{\theta^{(i)}}\equivert v_{\theta^{(i+1)}}$} with $i=1,\ldots,(n-1)$.
Closing with transitivity we have that equation
{$v_{\theta^{(1)}}\equivert v_{\theta^{(n)}}$} must be in
$\equivert^{{\cal D}_v}$.
Being the start and the end of the path respectively
$\theta_1$ and $\theta_2$
we have that  equivalence $\equivert^{{\cal D}_v}$ 
satisfy $v_{\theta_1}\equivert v_{\theta_2}$.
Summing the sets of the form {${{\cal D}_v}$}, for all $v\in\theta$,
we will obtain a set of $(d-1)$ (manifold) \inst s  
{${{\cal D}_\theta}=\cup_{v\in\theta}{{{\cal D}_v}}$} 
such that,  for all $v\in\theta$, the \verteq\ equation
{$v_{\theta_1}\equivert v_{\theta_2}$} must be in
$\equivert^{{\cal D}_\theta}$.
Therefore $\theta_1\equatsimp\theta_2$ 
is satisfied by $\equivert^{{\cal D}_\theta}$.

Iterating this process we can delete from  {${\cal E}$}
all \inst s  of order smaller that $(d-1)$.
This completes the proof 
in the case of plain $(d-1)$ connected stars (Part \ref{pro:qpmd1eq}).
If we are in the case of manifold connected stars we can use this process
also to delete non manifold \gl\ \inst s of order $(d-1)$.
If the star  of every vertex  is $(d-1)$-(manifold)-connected 
then the added \inst s, using the previous construction.
can be manifold \inst s.
Note that,  in this second  case, by Property \ref{pro:nopair}, 
the resulting set of $(d-1)$ 
\inst s, being a set of manifold \inst s, is also a pseudomanifold
set of \inst s.
This completes the proof of Part \ref{pro:qpmpmeq}.  

Parts  \ref{pro:peqd1qpm} and \ref{pro:peqmqpm} can be proven by induction 
the number $|{\cal D}|$ of  \gl\ \inst s in {${\cal D}$}.
As for the first two parts we merge the proofs of these last two parts 
by adding the adjective (manifold) between parenthesis.  
This  inductive  proof is technically  possible if we prove 
the following, stronger, result.
\begin{lemma}
\label{pro:techlemma}
In each 
$(d-1)$-(manifold)-connected star of a vertex
$\xstr{\topAComp/{{\cal D}}}{v}$, for two given
$d$-simplices in the vertex star: {${\theta^\top_{1}}/{\cal D}$} and 
{${\theta^\top_{n}}/{\cal D}$}, 
the $(d-1)$-(manifold)-path
{$({\theta^\top_{i}}/{\cal D})_{i=1}^{n}$} can be selected such that
the set of \gl\ \inst s 
$\{{\theta_{i}\equatsimp\theta_{i+1}}|i=1,\ldots,(n-1)\}$
is within ${\cal D}$.
\end{lemma}

We start with the inductive basis i.e. we assume $|{\cal D}|=0$.
If $|{\cal D}|=0$
we have that {${\cal D}=\emptyset$} and  the generated complex
is $\topAComp/\emptyset\simpeq\topAComp$. 
This complex, being a collection 
of disjoint simplices, satisfy the thesis.

Now let us consider the inductive step.
Let be ${\cal D}={{\cal D}^\prime}\cup\{\theta_1\equatsimp\theta_2\}$ and
let us assume that the generated complex {$\topAComp/{{\cal D}^\prime}$}
satisfy the thesis. We have to prove that the thesis remains true adding
the last \gl\ \inst\ $\theta_1\equatsimp\theta_2$.
We recall that adding a \gl\ \inst\ 
$\theta_1\equatsimp\theta_2$ we actually  add a set of
associated \verteq\ equations
of the form $v_{\theta_1}\equivert v_{\theta_2}$, one for any
$v\in\theta_1\cap\theta_2$.
These \verteq\ equations act on the complex {$\topAComp/{{\cal D}^\prime}$}.
We have to check that Vertices
that are affected do still have a star that is $(d-1)$-(manifold)-connected.  
Let be {$\theta_1\equatsimp\theta_2$} the \inst\  to be added and
let be $v_{\theta_1}\equivert v_{\theta_2}$ an associated 
\verteq\ equations not already satisfied by  {$\equivert^{{\cal D}^\prime}$}.
If such a new equation do not exist the addition of
{$\theta_1\equatsimp\theta_2$} do not change  
$\equivert^{{\cal D}^\prime}$ and therefore 
{$\topAComp/{{\cal D}^\prime}=\topAComp/{{\cal D}}$} and we are done.

On the other hand, let us assume that there exist an associated 
\verteq\ equations {$v_{\theta_1}\equivert v_{\theta_2}$} not already
satisfied by  $\equivert^{{\cal D}^\prime}$.
Adding the \gl\ \inst\ $\theta_1\equatsimp\theta_2$,
we add the \verteq\ equation $v_{\theta_1}\equivert v_{\theta_2}$,
and we cause  Vertices  ${v_{\theta_1}/{{\cal D}^\prime}}$ and 
$v_{\theta_2}/{{\cal D}^\prime}$ to
stitch together into the common vertex.
Let us call $u$ this common vertex in ${\topAComp/{{\cal D}}}$, i.e. $u=v_{\theta_1}/{\cal D}=v_{\theta_2}/{\cal D}$.

For any vertex $v$ we have to prove that {$\xstr{\topAComp/{{\cal D}}}{v}$}
is $(d-1)$-(manifold)-connected via paths that correspond to \inst s in
${\cal D}$.
Let us first assume that $v$ is not affected by $\theta_1\equatsimp\theta_2$
and let's prove that {$\xstr{\topAComp/{{\cal D}}}{v}$}
is $(d-1)$-(manifold)-connected via paths that correspond to \inst s
that are already in ${\cal D^\prime}$.
By inductive hypothesis we have that
the star   
$\xstr{\topAComp/{{\cal D}^\prime}}{v/{{\cal D}^\prime}}$ 
is (d-1)-(manifold)-connected.

Furthermore we recall the fact that, by inductive hypothesis, 
$\xstr{\topAComp/{{\cal D}^\prime}}{v/{{\cal D}^\prime}}$  is  connected through paths corresponding
to \inst s  in  ${{\cal D}^\prime}$.
We can say that all these $(d-1)$-paths remains in $(d-1)$-paths in
{$\topAComp/{{\cal D}}$} because (by Property \ref{pro:alldec}) 
we have an dimension preserving
\asm\ between {${\topAComp/{{\cal D}^\prime}}$} and {$\topAComp/{{\cal D}}$}.
These remains paths corresponding
to \inst s in  ${{\cal D}^\prime}$

({\em in the case of proof for  Part \ref{pro:peqmqpm} note that these \inst s are 
manifold \inst s w.r.t.  {${\topAComp/\Rtop}$}. Therefore the
$(d-1)$-simplex between the two top simplices in a manifold \gl\ \inst\
by inductive hypothesis must be manifold in
{${\topAComp/{{\cal D}^\prime}}$} and must
remain manifold in {${\topAComp/\Rtop}$} and thus it must be
manifold in {${\topAComp/{{\cal D}}}$}}, see Remark \ref{rem:mani}
later in this proof for details on this).

Therefore the star   
$\xstr{\topAComp/{{\cal D}}}{v/{{\cal D}}}$ 
remains (d-1)-(manifold)-connected whenever $v$ is not affected by $\theta_1\equatsimp\theta_2$.

Now let us return to the case in which,
by adding the \gl\ \inst\ $\theta_1\equatsimp\theta_2$,
we add the \verteq\ equation $v_{\theta_1}\equivert v_{\theta_2}$,
and we cause  Vertices  ${v_{\theta_1}/{{\cal D}^\prime}}$ and 
$v_{\theta_2}/{{\cal D}^\prime}$ to
stitch together into the common vertex $u=v_{\theta_1}/{\cal D}=v_{\theta_2}/{\cal D}$.

By inductive hypothesis we have that
the stars 
$\xstr{\topAComp/{{\cal D}^\prime}}{v_{\theta_1}/{{\cal D}^\prime}}$ 
and
$\xstr{\topAComp/{{\cal D}^\prime}}{v_{\theta_2}/{{\cal D}^\prime}}$ 
are (d-1)-(manifold)-connected.
{
Furthermore we recall the fact that they are connected via paths corresponding
to \inst s  in  ${{\cal D}^\prime}$.
Reasoning as before we can say that all these $(d-1)$-paths remains 
$(d-1)$-paths in
{$\topAComp/{{\cal D}}$} and they 
correspond to \inst s in  ${{\cal D}^\prime}$
({\em in the case of proof for  Part \ref{pro:peqmqpm} recall 
that $(d-1)$-simplex between the two top simplices in this 
manifold \gl\ \inst\
and remains manifold in  {${\topAComp/{{\cal D}}}$},
see Remark \ref{rem:mani}
later in this proof}).
}
Therefore adding ${\theta_1\equatsimp\theta_2}$ we will map 
$(d-1)$-(manifold)-paths 
within 
$\xstr{\topAComp/{{\cal D}^\prime}}{v_{\theta_1}/{{\cal D}^\prime}}$ 
and
$\xstr{\topAComp/{{\cal D}^\prime}}{v_{\theta_2}/{{\cal D}^\prime}}$ 
into $(d-1)$-(manifold)-paths in {$\xstr{\topAComp/{{\cal D}}}{u}$}.

Next we will show that top simplices in {$\xstr{\topAComp/{{\cal D}}}{u}$} 
are $(d-1)$-(manifold)-connected. 
This will be done by considering the fact  that
{${\theta^\top_1/{\cal D}}\in
\xstr{\topAComp/{{\cal D}}}{v_{\theta_1}/{{\cal D}}}$}  and
{${\theta^\top_2/{\cal D}}\in
\xstr{\topAComp/{{\cal D}}}{v_{\theta_2}/{{\cal D}}}$} must share a 
{\em manifold} $(d-1)$-simplex $\gamma_{12}$
in {$\topAComp/{{\cal D}}$}.
Paths going between 
$\xstr{\topAComp/{{\cal D}^\prime}}{v_{\theta_1}/{{\cal D}^\prime}}$ 
and 
$\xstr{\topAComp/{{\cal D}^\prime}}{v_{\theta_2}/{{\cal D}^\prime}}$ 
can always pass through the  manifold ''gate'' $\gamma_{12}$ and
make  {$\xstr{\topAComp/{{\cal D}}}{u}$}  
a unique $(d-1)$-(manifold)-connected star.
.

With this idea in mind we proceed as follows.
We note that, since we added the $(d-1)$-(manifold)-\inst\ 
$\theta_1\equatsimp\theta_2$, the two
$(d-1)$-simplices 
{$\theta^\top_1/{\cal D}$} and {${\theta^\top_2/{\cal D}}$}
will share, in {$\topAComp/{{\cal D}}$},
a $(d-1)$-(manifold)-simplex. Let us denote with  $\gamma_{12}$ this 
simplex in {$\topAComp/{{\cal D}}$}.

\begin{remark}\label{rem:mani}
({In the hypothesis of Part \ref{pro:peqmqpm} 
we can prove that $\gamma_{12}$ is a manifold simplex in
{$\topAComp/{{\cal D}}$} with the following steps.
First note that
\inst\ {$\theta_1\equatsimp\theta_2$} is
manifold w.r.t. {$\AComp$}. 
Next we note that  there is a dimension preserving \asm\
that maps {$\topAComp/{{\cal D}}$} 
into {$\topAComp/\Rtop\simpeq\AComp$}. 
This must be also a bijection between top simplices.
Therefore no more $d$-simplices will share 
$\gamma_{12}$ in 
{$\topAComp/{{\cal D}}$} than those sharing {$\gamma_{12}/\Rtop$}
in {$\topAComp/\Rtop\simpeq\AComp$}.
Since $\gamma_{12}$ is the intersection of
{$\theta^\top_1/{\cal D}$} and {${\theta^\top_2/{\cal D}}$}
the simplex {$\gamma_{12}/\Rtop$} correspond to $\gamma=\theta_1\cap\theta_2$ 
w.r.t.  the isomorphism that maps {$\topAComp/\Rtop$} to {$\AComp$}.
Since just two $d$-simplices  are sharing $\gamma$
in {$\AComp$} two simplices are sharing {$\gamma_{12}/\Rtop$} in {$\topAComp/\Rtop\simpeq\AComp$} and no more
than two $d$-simplices share {$\gamma_{12}$}
in {$\topAComp/{{\cal D}}$} and so
{${\theta^\top_1/{\cal D}}$} and {${\theta_2/{\cal D}}$} are manifold adjacent in 
{$\topAComp/{{\cal D}}$}}).
\end{remark}

We have proven that the two stars
{$\xstr{\topAComp/{{\cal D}}}{v_{\theta_1}/{{\cal D}}}$}  and
{$\xstr{\topAComp/{{\cal D}}}{v_{\theta_2}/{{\cal D}}}$} are $(d-1)$-(manifold)-connected through \inst s  in ${\cal D}^\prime$.
Since
{${\theta^\top_1/{\cal D}}\in
\xstr{\topAComp/{{\cal D}}}{v_{\theta_1}/{{\cal D}}}$}  and
{${\theta^\top_2/{\cal D}}\in
\xstr{\topAComp/{{\cal D}}}{v_{\theta_2}/{{\cal D}}}$},
paths going from one star to the other can always pass through
the ''gate'' $\gamma_{12}$. 
({\em In the hypothesis of Part \ref{pro:peqmqpm} 
we have proven that this is a manifold
$(d-1)$-simplex in  {$\topAComp/{{\cal D}}$}.})
Thus  top simplices in {$\xstr{\topAComp/{{\cal D}}}{u}$} 
must form a unique $(d-1)$-(manifold)-connected star.

Paths going from one star to the other can always pass through
the ''gate'' $\gamma_{12}$ in between
{${\theta^\top_1/{\cal D}}$} and {${\theta^\top_2/{\cal D}}$} and satisfy
the additional inductive hypothesis we have introduced.
Infact in this case paths ''use'' the \gl\ \inst\
{$\theta_1\equatsimp\theta_2$} in ${\cal D}$. This complete the proof of the inductive step.

\end{proof}

The above property supports the definition of \cdec s through local
topological properties.  Next we will give a property that
gives an alternative  characterization in term of \gl\ \inst s.
\begin{definition}[\Cdec]
\label{def:iqm}
An  \emd{\cdec} $d$-complex is 
a regular $d$-complex  where the star of every vertex   is
$(d-1)$-manifold-connected.
\end{definition}

\Cdec s are complexes that are generated by sets of manifold \gl\ \inst s.
\begin{property}
\label{pro:eqiqm}
Let $\AComp^\prime$ be a complex in the \dec\ lattice for a 
$d$-complex $\AComp$.
The complex $\AComp^\prime$ is an \cdec\ \dec\ \iff\
there exist a set {${\cal D}$} of manifold $(d-1)$-\inst s w.r.t. $\AComp$  
such that $\AComp^\prime=\topAComp/{\cal D}$.
\end{property}
\begin{proof}
By Property \ref{pro:qpmeq} Part  \ref{pro:peqmqpm} we have that
a set of $(d-1)$-manifold \inst s  w.r.t. $\AComp$ generates 
a regular $d$-complex  where the star of every vertex  
$(d-1)$-manifold-connected. Thus, by Definition \ref{def:iqm} we
have. that the complex $\topAComp/{\cal D}$ is an \cdec.

Conversely,
by Property \ref{pro:qpmeq} Part  \ref{pro:qpmpmeq}, we have that
any \cdec\  decomposition of a $d$-complex $\AComp$ 
can be generated by a pseudomanifold set of regular 
$(d-1)$-\inst s {${\cal D}$}, that are manifold w.r.t. $\AComp$.
\end{proof}
Note that, by Property \ref{pro:manid1}, 
in a set of manifold \inst s  {${\cal M}$}, we can
purge \inst s  of order smaller than $(d-1)$.
Therefore,
if  ${\cal M}$ is a set of manifold \inst s  w.r.t $\AComp$, then
the complex {$\topAComp/{\cal M}$} is an \cdec\ decomposition of
the $d$-complex $\AComp$.

It is easy to see that manifolds are \cdec s. \Cdec in dimension two are manifolds. In 
dimension three \cdec s are neither manifolds nor pseudomanifolds.
These facts are summarized in the following property
\begin{property}
\label{pro:iqmrel}
The following relations holds between manifolds, pseudomanifolds and 
\cdec s:
\begin{enumerate}
\item
\label{pro:mdcdec}
The class of manifold complexes is a subclass of \cdec\ complexes. 
\item
\label{pro:pmnocdec}
The class of pseudomanifold 3-complexes is neither  a subclass nor
a superclass of \cdec\ 3-complexes. 
\item
\label{pro:d1conn}
\Cdec s are $(d-1)$-manifold connected.
\item
\label{pro:2mani}
The class of $2$-manifolds coincide with the class \cdec\ . 
$2$-complexes.
\end{enumerate}
\end{property}
\begin{proof}
To prove part \ref{pro:mdcdec} we note that a combinatorial manifold is an \iqm\ since,
by Definition \ref{def:combman}, the link of combinatorial $d$-manifold is
either  a $(d-1)$-sphere or a $(d-1)$-ball.
In both cases the link will be $(d-2)$-manifold-connected.
Thus, the closed star of a vertex $v$, being the cone from $v$ to its link,
will be  $(d-1)$-manifold-connected. Thus, by Definition \ref{def:iqm} we 
have that a combinatorial $d$-manifold is an \cdec.
 
Since \cdec s are generated by sets of $(d-1)$-manifold \inst s we have
that by, Property \ref{pro:conn}, each connected component in an \cdec\ is 
$(d-1)$-connected. Putting together paths in each $(d-1)$-manifold
connected vertex star it is easy to
see that \cdec\ are $(d-1)$-manifold-connected. 
This proves part \ref{pro:d1conn}.

To prove part \ref{pro:2mani} we recall that
the link of every vertex in a \cdec\ is a $(d-1)$-manifold-connected
complex.
Therefore, for $d=2$, we have that the link of each vertex is 
$1$-manifold-connected. This means that the link of a vertex in an
\cdec\ $2$-complex
must be a graph that is either a chain or a cycle.
By Definition \ref{def:combman}, the link of combinatorial $1$-manifold must
be combinatorially equivalent 
either to a $1$-sphere  or a $1$-ball. 
This proves that the set of \cdec\ $2$-complexes coincides 
with the set of  $2$-manifolds and is a proper subset of $2$-pseudomanifolds.

To prove part \ref{pro:pmnocdec} we recall that
Example \ref{app:example} shows that \cdec\ $3$-complexes
are not a  subset of $3$-pseudomanifolds.  From
part \ref{pro:2mani} we have that the set of \cdec\ $2$-complexes
are a proper subset of $2$-pseudomanifolds. 
\end{proof}

If we just take decompositions generated by $(d-1)$ manifold \gl\ \inst s,
we obtain a decompositions where ''nearly'' all singularities are
filtered out. Nevertheless this process preserve a good deal of the
connectivity in the original complex. Informally we can say that 
''good'' connectivity is preserved.
These ideas will be developed in the next chapter to define the ''best''
non singular decomposition for a complex.
The following  property, 
that looks quite technical now, will be fundamental to 
characterize the structure of the  ''best'' decomposition for
a given complex.
Informally we can introduce  this property by considering that an \cdec\ $d$-complex can be defined by a rather small set of top $(d-1)$ gluing \inst s. The next propperty states that considering all possible 
top $(d-1)$ gluing \inst s and filtering out singular gluing \inst s  we still preserve ''good'' connectivity. 

\begin{property}
\label{pro:qpmeqfil}
Let be {$\topAComp/{\cal D}$} a decomposition for a $d$-complex $\AComp$.
The following facts hold:
\begin{enumerate}
\item
\label{pro:eqd1qpm}
If the set {${\cal D}$} is the set of all possible $(d-1)$ \gl\
\inst s  then, the two
top simplices $\theta^\top_a/{\cal D}$ and $\theta^\top_b/{\cal D}$ are 
$(d-1)$-connected
in  {$\topAComp/{\cal D}$} if and only if 
$\theta_a$ and $\theta_b$ are  $(d-1)$-connected in $\AComp$.
\item 
\label{pro:eqmqpm}
If the set {${\cal D}$} is the set of all possible $(d-1)$ \gl\
\inst s that are manifold w.r.t. $\AComp$ then, the two
top simplices $\theta^\top_a/{\cal D}$ and $\theta^\top_b/{\cal D}$ are 
$(d-1)$-manifold-connected
in  {$\topAComp/{\cal D}$} if and only if 
$\theta_a$ and $\theta_b$ are  $(d-1)$-manifold-connected in $\AComp$.
\end{enumerate}
\begin{proof}
We will prove Part \ref{pro:eqd1qpm} and Part \ref{pro:eqmqpm}.
merging the two proofs as the in proof of Property   \ref{pro:delnond}.
Proof of Part \ref{pro:eqmqpm} is obtained 
by considering  the adjective
(manifold), that will be  placed  between parenthesis.
The reader should skip or not this
depending on which proof is interested in ({\em note that the proof for Part \ref{pro:eqd1qpm} spans the next paragraph and the third, then all the rest of  the proof is for Part \ref{pro:eqmqpm} and is not all italicized or reported in parenthesis}).

We will first prove that if
$\theta_a$ and $\theta_b$ are  two top simplices 
$(d-1)$-(manifold)-connected in $\AComp$.
then $\theta^\top_a/{\cal D}$ and $\theta^\top_b/{\cal D}$ are 
$(d-1)$-(manifold)-connected in {$\topAComp/{\cal D}$}.
Now let be $\theta_a$ and $\theta_b$ two top simplices in  {$\AComp$}.
These are $(d-1)$-(manifold) connected if and only if there exist a    
$(d-1)$-(manifold) path {$(\theta_{i})_{i=1}^{n}$}  made up  of $n$  
$d$-simplices $\theta_{i}$ with $\theta_a=\theta_1$ 
and $\theta_b=\theta_2$.  
From this path  let us build the set  of $n-1$
\gl\ (manifold) \inst s  
{${\cal D}_v=
\{{\theta_{i}\equatsimp\theta_{i+1}}|i=1,\ldots,(n-1)\}$}.
These must be in ${\cal D}$ because this contains all
(manifold) $(d-1)$-\inst s  for $\AComp$. 
Being 
{$\topAComp/{\cal D}$} a decomposition of $\AComp$
there is a dimension preserving \asm\ between {$\topAComp/{\cal D}$}
and {$\topAComp/{\Rtop}$} that is a bijection
between top simplices. 
For this reason
${\theta_{i}}/{\cal D}$ and ${\theta_{i+1}}/{\cal D}$ 
must share a $(d-1)$ complex in {$\topAComp/{\cal D}$}.
This completes the first half (i.e. $(d-1)$-connected
in $\AComp$  implies   $(d-1)$-connected in {$\topAComp/{\cal D}$} ) of the   proof for Part \ref{pro:eqd1qpm}.

({\em To complete the proof for Part \ref{pro:eqmqpm} we have to note 
that, being {$\topAComp/{\cal D}$} a decomposition, the number of 
$d$-complexes incident to  
${\theta_{i}}/{\cal D}\cap{\theta_{i+1}}/{\cal D}$, can only 
be smaller that those incident to
$\theta_1\cap\theta_2$. 
Therefore, in the hypothesis of  Part \ref{pro:eqmqpm}, we have
that ${\theta_{i}}/{\cal D}$ and ${\theta_{i+1}}/{\cal D}$ 
must share a $(d-1)$ manifold complex in {$\topAComp/{\cal D}$}.
Therefore  {$(\theta_{i})_{i=1}^{n}$} is a $(d-1)$-manifold path.
This complete the first half ot the proof for Part \ref{pro:eqmqpm}})

Conversely, now,
we will prove that if
$\theta_a/{\cal D}$ and $\theta_b/{\cal D}$ are two $d$-simplices that are 
$(d-1)$-(manifold)-connected in {$\topAComp/{\cal D}$}
then $\theta_a$ and $\theta_b$ must be  
$(d-1)$-(manifold)-connected in $\AComp$.
Let be $\theta_a/{\cal D}$ and $\theta_b/{\cal D}$ two top simplices
that are {\em just} $(d-1)$-connected
in  {$\topAComp/{\cal D}$}.
Being 
{$\topAComp/{\cal D}$} a decomposition of $\AComp$
there is a dimension preserving \asm\ between {$\topAComp/{\cal D}$}
and {$\topAComp/{\Rtop}$} that is a bijection
between top simplices. 
Indeed any 
$(d-1)$-path in  {$\topAComp/{\cal D}$} is preserved in
in  {$\AComp/\Rtop\simpeq\AComp$}.  
This completes the second half of the proof for Part \ref{pro:eqd1qpm}.

{\em Next, to complete the proof for Part  \ref{pro:eqmqpm}, we have 
to prove that manifold connectivity in  {${\topAComp/{\cal D}}$} is preserved
in  $\AComp/\Rtop\simpeq\AComp$.  
Let be {$({\theta^\top_{i}}/{\cal D})_{i=1}^{n}$} a $(d-1)$-manifold path
in {$\topAComp/{\cal D}$} in between  
$\theta^\top_a/{\cal D}$ and $\theta^\top_b/{\cal D}$.
Let be ${\theta^\top_{k}}/{\cal D}$ and ${\theta^\top_{k+1}}/{\cal D}$ 
two manifold-adjacent $d$-simplices in this $(d-1)$-manifold-path.
Actually it might happen that
${\theta_{k}}$ and ${\theta_{k+1}}$ are not manifold adjacent in
$\AComp$. In this case we show that we can find a  
$(d-1)$-manifold path in between   
${\theta_{k}}$ and ${\theta_{k+1}}$ in $\AComp$.}
{
\begin{figure}[h]
\begin{center}
\framebox{
\parbox[c][0.35\textwidth]{0.20\textwidth}{
\psfig{file=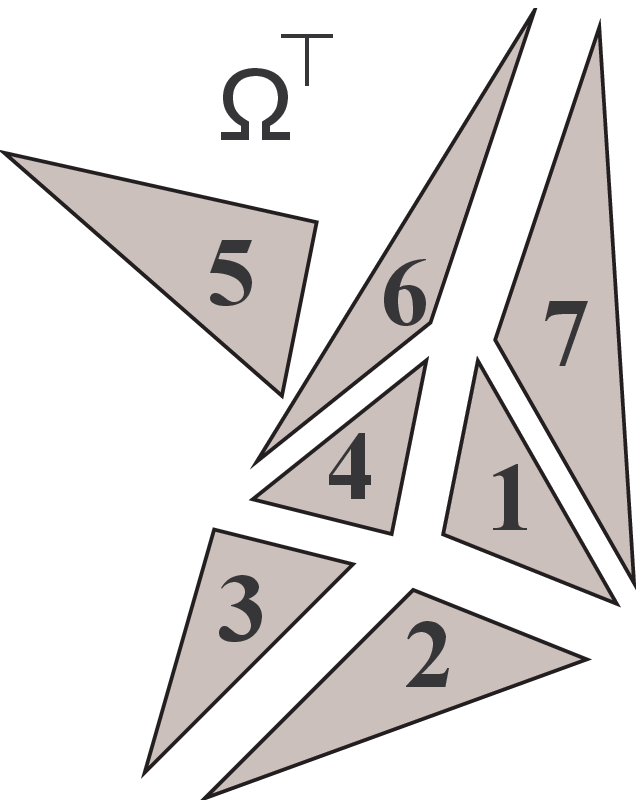,width=0.20\textwidth}
\begin{center}(a)\end{center}
}
}
\framebox{
\parbox[c][0.35\textwidth]{0.20\textwidth}{
\psfig{file=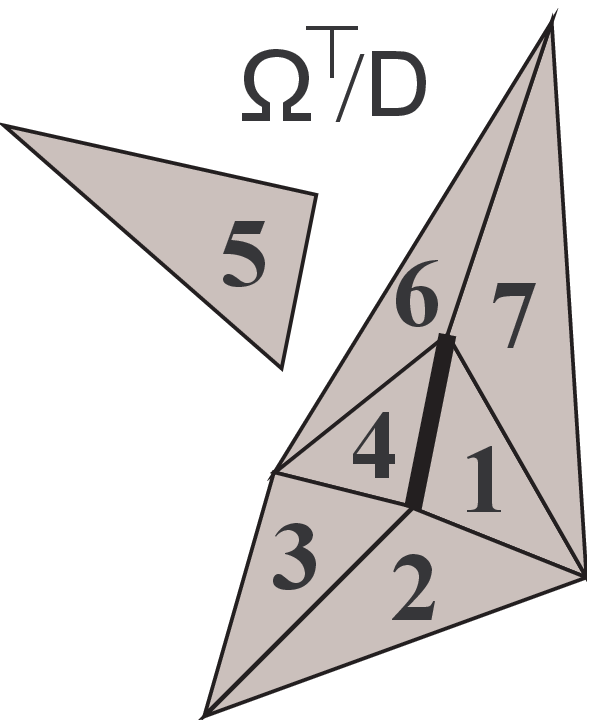,width=0.20\textwidth}
\begin{center}(b)\end{center}
}
}
\framebox{
\parbox[c][0.35\textwidth]{0.15\textwidth}{
\psfig{file=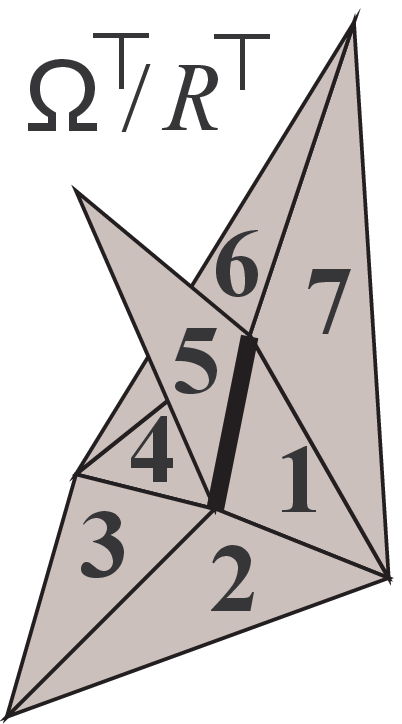,width=0.15\textwidth}
\begin{center}(c)\end{center}
}
}
\end{center}
\caption{A decomposition {$\topAComp/{\cal D}$} (b) where a manifold path 
$(\theta_1,\theta_4)$ is no longer a  manifold path in 
{$\topAComp/\Rtop$} (c) }
\label{fig:nolongm}
\end{figure}
}

{\em An example of this situation is shown in Figure \ref{fig:nolongm}.
From the totally exploded version in {$\topAComp$}
in Figure \ref{fig:nolongm}a
we obtain the decomposition
{$\topAComp/{\cal D}$} in Figure \ref{fig:nolongm}b with the
set of all  manifold  \gl\ \inst s ${\cal D}=\{
{\theta_1\equatsimp\theta_2},
{\theta_2\equatsimp\theta_3},
{\theta_3\equatsimp\theta_4},
{\theta_4\equatsimp\theta_6},
{\theta_6\equatsimp\theta_7},
{\theta_7\equatsimp\theta_1}
\}$.
In this decomposition
the path $(\theta_1,\theta_4)$ is a manifold path and is no longer
a manifold path in {$\topAComp/{\Rtop}$}. However we can turn around one of
the non manifold vertices and obtain back a manifold path, between
$\theta_1$ and $\theta_4$ e.g. with the path 
$(\theta_1,\theta_2,\theta_3,\theta_4)$. Such a creation of a new path  
can be done in general and we will use this fact to prove the thesis.}

So the rest of this is to prove Part \ref{pro:eqmqpm}. By Property \ref{pro:qpmd1eq} Part \ref{pro:peqmqpm}, 
since ${\cal D}$ is a  set of 
$(d-1)$ manifold \inst s, the star of every vertex in 
{$\topAComp/{\cal D}$} must be $(d-1)$-manifold connected.
Let be $v$ a vertex in 
${\theta^\top_{k}}/{\cal D} \cap {\theta^\top_{k+1}}/{\cal D}$. 
The star {$\xstr{\topAComp/{{\cal D}}}{v}$}
must be $(d-1)$-manifold connected. 
Therefore there will be a  $(d-1)$-manifol-path in 
{$\xstr{\topAComp/{{\cal D}}}{v}$} in between  the two top
simplices 
${\theta^\top_{k}}/{\cal D}$ and ${\theta^\top_{k+1}}/{\cal D}$. 
Now we recall that in the proof of Property \ref{pro:peqmqpm}
Parts  \ref{pro:peqd1qpm} and \ref{pro:peqmqpm}
we have proven the following Lemma (see Lemma \ref{pro:techlemma}):

{\em  in each 
$(d-1)$-(manifold)-connected star of a vertex
{$\xstr{\topAComp/{{\cal D}}}{v}$}, for two given
$d$-complexes in the vertex star: {${\theta^\top_{1}}/{\cal D}$} and 
{${\theta^\top_{n}}/{\cal D}$}, 
the $(d-1)$-(manifold)-path
{$({\theta^\top_{i}}/{\cal D})_{i=1}^{n}$} can be selected such that
the set of \gl\ \inst s $\{{\theta_{i}\equatsimp\theta_{i+1}}|i=1,\ldots,(n-1)\}$
is within ${\cal D}$}.

We apply this result and find a 
$(d-1)$-manifold-path in 
{$\xstr{\topAComp/{{\cal D}}}{v}$} in between 
${\theta^\top_{k}}/{\cal D}$ and ${\theta^\top_{k+1}}/{\cal D}$ such that 
the set of corresponding \inst s are in ${\cal D}$. For this reason 
these \inst s must be manifold w.r.t $\AComp$. This set of \inst s
trace a $(d-1)$-manifold path in between   
${\theta_{k}}$ and ${\theta_{k+1}}$ in $\AComp$.
This happens even if 
${\theta_{k}}$ and ${\theta_{k+1}}$, are not manifold
adjacent in $\AComp$. 

Resuming we started with  a 
$(d-1)$-(manifold) path {$({\theta_{i}}/{\cal D})_{i=1}^{n}$}
in {$\topAComp/{\cal D}$}. 
We noted that it can happen that, for some $1\le k\le n$,
the two top $d$-simplices ${\theta_{k}}$ and ${\theta_{k+1}}$ 
are not manifold adjacent in
$\topAComp/\Rtop$. In this case we have shown that we can find a  
$(d-1)$-manifold path in between   
${\theta_{i}}$ and ${\theta_{i+1}}$ in $\AComp$.
Therefore $(d-1)$-manifold connectivity is preserved, although possibly
through alternative paths, passing from {$\topAComp/{\cal D}$} to {$\AComp/\Rtop\simpeq\AComp$}.
This complete the second half ot the proof for Part \ref{pro:eqmqpm}
\end{proof}
\end{property}

\section{\emi\Qm s}\Qm s}
\label{sec:quasi}
The problem of characterizing complexes generated by a {\em closed}
set of manifold \inst s will bring in a known family of complexes 
called {\em \qm}. The informal idea behind this class is that,
from the combinatorial point of view the definition of manifold might seem
rather arbitrary.
In fact there is no reason to privilege the sphere as the canonical form 
for the link of a vertex. 
If we just want links to be, somehow,  {\em regular} we can accept toroidal
links or even we can accept the projective plane as a link.
This idea leads us to the definition of  {\em \qm}. 
This class  was introduced by Lienhardt \cite{Lie94} 
as the class of complexes modeled by n-G-maps.
In this framework we introduce quasi-manifolds by studying  
complexes generated by a {\em $(d-1)$-closed} set of manifold \inst s.
This will give a new characterization of \qm s in term of local topologic
properties.

We will say that a set {${\cal D}$} made up of \inst s of order $(d-1)$ 
is {\em $(d-1)$-closed} if all the \inst s of order $(d-1)$ in 
{$\closeq{\cal D}$} are already in  {${\cal D}$}.
Note that if a  $(d-1)$-closed set of $(d-1)$ \inst s 
{${\cal D}$} generates a 
$d$-complex $\AComp$ then we can find in {${\cal D}$}
all $(d-1)$ \inst s that are satisfied by the equivalence associated to 
$\AComp$. 
The following property gives a characterization of complexes generated by 
a $(d-1)$-closed set of $(d-1)$ manifold \inst s.

\begin{property}
\label{pro:quasi}
A  $d$-complex {$\AComp$} can be generated 
by a $(d-1)$-closed  set {${{\cal D}}$} of 
$(d-1)$-manifold \inst s w.r.t.  {$\AComp$},
if and only if, the complex $\AComp$ is a $d$-pseudomanifold
where the star of every vertex is  $(d-1)$-connected.
\begin{proof}
Let us assume that $\AComp$ can be generated 
by the  $(d-1)$-closed set of $(d-1)$-manifold \inst s {${\cal D}$}
and let us prove that $\AComp$ is a pseudomanifold.
We have that $\AComp$ can also be generated by
the closed set {$\closeq{\cal D}$}. Since all $(d-1)$ 
\inst s in {$\closeq{\cal D}$} are those in {${\cal D}$} and 
since they are all manifold then the set {$\closeq{\cal D}$} is a 
closed pseudomanifold set of \inst s. Therefore, by Property 
\ref{pro:eqpseudo},
the complex generated by {$\closeq{\cal D}$} is a pseudomanifold.
Next we want to prove that in the generated complex {$\topAComp/{\cal D}$}
the star of
a vertex $v$ is $(d-1)$-manifold-connected.
Infact we have that {${{\cal D}}$} is a set of $(d-1)$ manifold \inst s,
therefore, by Property \ref{pro:qpmeq} Part \ref{pro:peqmqpm}, 
the star of every vertex
in the generated complex {$\topAComp/{\cal D}$} is $(d-1)$-connected.

Conversely, for a given $d$-pseudomanifold
$\AComp$ where the star of every vertex is  $(d-1)$-connected.
we have to build a $(d-1)$-closed  set {${{\cal D}}$} of 
$(d-1)$ \inst s that must be manifold  w.r.t.  {$\AComp$} and s.t. 
{$\topAComp/{{\cal D}}\simpeq\AComp$}.
By Property \ref{pro:qpmpmeq} Part \ref{pro:peqmqpm} we  have that
(an isomorphic copy of) the $d$-complex {$\AComp$} can be generated 
by a  set {${{\cal D}_0}$} of 
$(d-1)$-manifold \inst s w.r.t.  {$\AComp$} 
(i.e. {$\AComp\simpeq\topAComp/{{\cal D}_0}$}).
Let be {${{\cal D}}$} the set obtained adding  to {${{\cal D}_0}$} the other 
$(d-1)$-\inst s that are in  {$\closeq{{\cal D}_0}$}. 
Adding the  other  $(d-1)$-\inst s that are in  {$\closeq{{\cal D}_0}$} 
the generated complex do not change.
Thus {$\topAComp/{{\cal D}_0}=\topAComp/{{\cal D}}\simpeq\AComp$}.
By Property \ref{pro:eqpseudo}, since {$\AComp\simpeq\AComp/{\closeq{{\cal D}_0}}$} is 
a pseudomanifold, the set of \gl\ \inst s 
{$\closeq{{\cal D}_0}$} must be  a pseudomanifold set of \gl\
\inst s.
By Property \ref{pro:cpmm} all $(d-1)$ \inst s in 
{$\closeq{{\cal D}_0}$} are manifold w.r.t.  $\AComp$.
Therefore  all $(d-1)$-\inst s in the $(d-1)$-closed set ${\cal D}$ 
are manifold w.r.t. $\AComp$ and 
{$\topAComp/{{\cal D}}\simpeq\AComp$}.
This completes the proof.
\end{proof}
\end{property}
The above property  supports a non constructive definition of 
\qm. In fact by the above Property \ref{pro:quasi} 
the following definition is equivalent to that given by Lienhardt in
(see  the definition of {\em Numbered simplicial quasi-manifolds} in  \cite{Lie94} Pg. 7 and the discussion
in Section \ref{sec:ngmap}).
\begin{definition}[\emi{\Qm}]
	\label{def:quasi}
A  $d$-quasi-manifold is a  $d$-pseudomanifold where the star of 
each vertex is $(d-1)$-connected.
\end{definition}
From the above definition and from  Definition \ref{def:iqm} 
it is easy to see that quasi-manifolds are a subset of  \cdec. 
By Property \ref{pro:iqmrel} it is easy to see that 
$2$-quasi-manifolds coincide with \cdec\ $2$-complexes that in turn
coincide with $2$-manifolds.
For $d\ge 3$ $d$-quasi-manifolds are a proper superset of $d$-manifolds
and a proper subset of $d$-pseudomanifolds.
For $d\ge 3$ $d$-quasi-manifolds are a proper subset of \cdec\ $d$-complexes.
This proper inclusion is given by the fact that,
for $d\ge 3$, there are \cdec\ $d$-complexes that are not
pseudomanifolds (See Example \ref{app:example}).

\chapter{Standard Decomposition}
\label{ch:stdec}
\section{Introduction}
In Chapter \ref{cc:toposgi} we have developed a classification of
complexes generated by a set of \gl\ \inst s ${\cal E}$.
This classification gives some topological  properties
for the decomposition {$\topAComp/{{\cal E}}$}
on the ground of properties of instructions in {${\cal E}$}.
The careful reader may argue that this can be of limited interest
speaking about decomposition because 
(in Example \ref{ex:nonlatt} of Chapter \ref{sec:classify} 
we have seen that) {\em not all decompositions for $\AComp$ can 
be generated by sets of \gl\ \inst}.

This remark is perfectly legal here because we are not interested in
properties of {$\topAComp/{{\cal E}}$} on its own. 
We are interested in  {$\topAComp/{{\cal E}}$} {\em as a decomposition
of}\ \ $\AComp$. This issue, 
in the first part of this chapter. However note that the results in Chapter
\ref{cc:toposgi} might still have some interest if considered on their
own.
In fact, whenever {$\topAComp/{{\cal E}}\simpeq\AComp$},
the \tsimeq\ \inst s in 
${\cal E}$ might be regarded as 
the set of {\em primitive operations} modeling $\AComp$.
From this point of view results in  Chapter \ref{cc:toposgi} gives 
the topological properties of the resulting complex  on the ground of 
syntactical properties of the set of  \inst s ${\cal E}$.

\section{The \Dec\ lattice and \gl\ \inst s}
Now let us revert to the decomposition problem. 
We have seen that \gl\ \inst s  cannot generate all 
decompositions for $\AComp$. In other words
sets of \gl\ \inst s  are not sufficient to label every
path from  $\AComp^\top$ down to a  decomposition
{$\topAComp/{\equivert}$}.
However, we can prove 
that sets of \gl\ \inst s are sufficient to label {\em every
path} from a  decomposition {$\topAComp/{\equivert}$}
down to $\topAComp/\Rtop$.
Therefore sets of \gl\ \inst s gives a set of transformations
quite meaningful in this context. In fact, here,
we focus our attention  on the way in which we go from a
decomposition {$\topAComp/{\equivert}$}.
down to the original complex {$\AComp\simpeq\topAComp/\Rtop$}.
Next Lemma shows that, for such a focus, we can restrict our attention to
paths that can be labeled by sets of \gl\ \inst s. 

In particular, we will prove that, for every pair of 
\eqt\ simplices $\nu_1$, $\nu_2$, in a certain
decomposition {$\topAComp/{\equivert}$}, we can find a \gl\ 
\inst\  {$g=\theta_1\equatsimp\theta_2$} that {\em completely stitch} 
together $\nu_1$ and $\nu_2$. By {\em completely stitch} we mean 
that adding  {$\theta_1\equatsimp\theta_2$} to $\equivert$ we jump into
a decomposition $\topAComp/{\equivert^\prime}$ where $\nu_1/{\equivert^\prime}$ and $\nu_2/{\equivert^\prime}$ becomes the same simplex.
Furthermore, if at least one of the two \eqt\ simplices 
$\nu_i$ is a top \eqt\ simplex in 
{$\topAComp/{\equivert}$}, then the image of both $\nu_1/\Rtop$  and
$\nu_1/\Rtop$ 
in {$\topAComp/\Rtop\simpeq\AComp$} is exactly the common simplex 
{$\gamma^\bot=\theta_1^\top/\Rtop\cap\theta_2^\top/\Rtop$}
(i.e.  the isomoprhic image in  {$\topAComp/\Rtop\simpeq\AComp$} 
of the common simplex $\gamma={\theta_1\cap\theta_2}$
''glued'' by {$\theta_1\equatsimp\theta_2$}).

This fact is  formally expressed by the following Lemma.
\begin{lemma}
\label{lemma:couple}
In a decomposition $\topAComp/\equivert$ there exist
a pair of distinct \eqt\ simplices 
$\nu_1$, $\nu_2$
if and only if there  exist a \gl\ \inst\
$\theta_1\equatsimp\theta_2$ that is not satisfied by $\equivert$
and such that the common image of 
$\nu_1$ and $\nu_2$
in $\topAComp/\Rtop$, (i.e. $\nu_1/\Rtop=\nu_2/\Rtop$) 
lies within the simplex $\gamma^\bot\in\topAComp/\Rtop$ 
corresponding to the common face 
{$\gamma=\theta_1\cap\theta_2$} in $\AComp$. 
In symbols it must hold
$\nu_1/\Rtop=\nu_2/\Rtop\le\gamma^\bot$ 
If either $\nu_1$ or $\nu_2$
is a top \eqt\ simplex the their common image in  $\topAComp/\Rtop$
is exactly $\gamma^\bot$.
\end{lemma}
\begin{proof}
\begin{figure}[h]
\fbox{\psfig{file=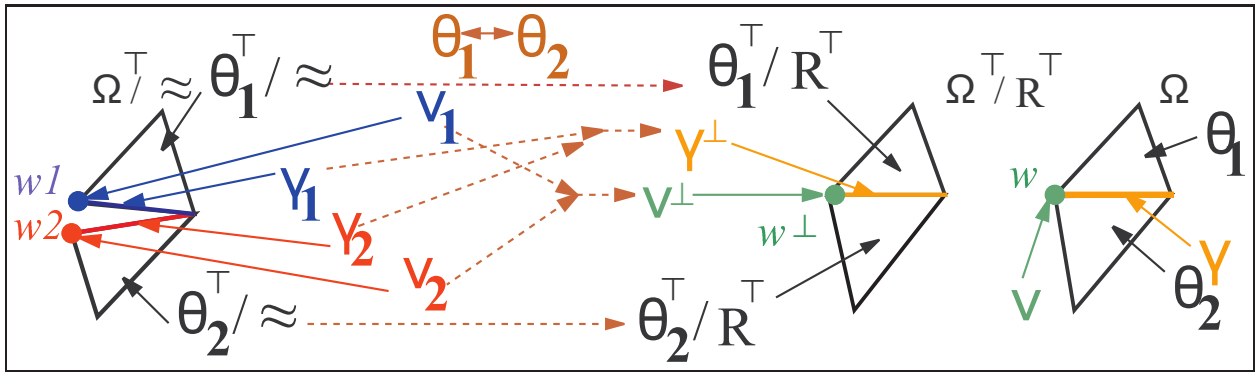,width=\textwidth}}
\caption{Proof of Lemma \ref{lemma:couple}. Dashed arrows denote 
the \asm\ induced by a path in the \quot\ lattice 
(see Property \ref{pro:equiquot}). 
All other arrows are used to name objects in the figure.
Note that for simplicity, in this particular case, we used equating vertices $w_1$ and $w_2$ i.e. we take $w_1=\nu_1$ $w_2=\nu_2$}
\label{fig:equisti}
\end{figure}
A situation coherent with the hypothesis of this Lemma is depicted, just
for reference, in Figure \ref{fig:equisti}.
We first start from the two distinct \eqt\ simplices 
$\nu_1$ and $\nu_2$ and find a \gl\ \inst\
{$\theta_1\equatsimp\theta_2$} that is not satisfied by $\equivert$ with the properties given in the thesis.

Let 
$\theta^\top_1/{\equivert}$ and $\theta^\top_2/{\equivert}$ be 
two top simplices in $\topAComp/\equivert$  that are
cofaces of the two \eqt\ distinct simplices
$\nu_1$ and $\nu_2$ (in certain complexes several choices are possible for $\theta^\top_i/{\equivert}$ ).
Furthermore, we take indices so that 
$\nu_i\le(\theta_i^\top/{\equivert})$ for $i=1,2$.
The top simplices 
$\theta^\top_1/{\equivert}$ and $\theta^\top_2/{\equivert}$ 
must be distinct (i.e. cannot be $\theta^\top_1/{\equivert}§=\theta^\top_2/{\equivert}$) because  going from a decomposition to another, 
in particular going from $\topAComp/{\equivert}$ down to {$\topAComp/\Rtop$},
we can not merge two distinct  faces $\nu_1$ and $\nu_2$
within the same top simplex. 
In fact there is always a dimension preserving map between 
two decompositions that is a  bijection between top simplices
(See Definition \ref{def:decomposition}). Informally, in simple words,  this is about assembling not collapsing.
The
intersection {$\gamma=\theta_1\cap\theta_2$} cannot be empty.
In fact the isomorphic copy of $\gamma$ in 
$\topAComp/\Rtop$, denoted by $\gamma^\bot$ 
must contain $\nu_1/\Rtop\cap\nu_2/\Rtop$.
To show this note that
$\gamma^\bot=\theta_1^\top/\Rtop\cap\theta_2^\top/\Rtop$
and, using $\nu_i\le(\theta_i^\top/{\equivert})$ for $i=1,2$ with
identites \ref{pro:lattid}, it is easy to see that.
$\theta_1^\top/\Rtop\cap\theta_2^\top/\Rtop$ 
must contain $\nu_1/\Rtop\cap\nu_2/\Rtop$.
This proves a part of the thesis.
In fact the common image of two \eqt\ simplices
in {$\topAComp/\Rtop$},
i.e. the simplex  $\nu_1/\Rtop=\nu_2/\Rtop$, must be within 
$\gamma^\bot=\theta_1^\top/\Rtop\cap\theta_2^\top/\Rtop$.

We have proven that the intersection 
{$\gamma=\theta_1\cap\theta_2$} cannot be empty and therefore
{$\theta_1\equatsimp\theta_2$} is a \gl\ \inst\ for $\AComp$ that stitches together $\nu_1$ and $\nu_2$.
We have to prove that this \inst\ is not satified by $\equivert$.
To this aim we note that
$\nu_1$ and $\nu_2$,
that are distinct in $\topAComp/{\equivert}$ and thus they
must differ for at least two distinct Vertices 
$w_1$ and $w_2$ in $\topAComp/{\equivert}$.
Let us assume that we take indices so that 
$w_i\in\nu_i\le\theta_i^\top/{\equivert}$ for 
$i=1,2$.  
These two vertices must map into
a common vertex $w^\bot=w_1/\Rtop=w_2/\Rtop$.
The common image of two \eqt\ simplices
in {$\topAComp/\Rtop$} must be within $\gamma^\bot$
and therefore the vertex $w^\bot$ must be in $\gamma^\bot$.
Let $w$ in $\AComp$ be the vertex corresponding to $w^\bot$
in {$\topAComp/\Rtop$}. We have that $w\in\gamma$ because
$w^\bot$ is in $\gamma^\bot$.
We will show that 
$\theta_1\equatsimp\theta_2$ is not satisfied by $\equivert$
by showing that $w_{\{\theta_1\}}\not\equivert w_{\{\theta_2\}}$.
To this aim we note that 
$w_i\in\theta_i^\top/{\equivert}$ for $i=1,2$.
Furthermore it must be 
$w_{\{\theta_i\}}/{\equivert}\in\theta_i^\top/{\equivert}$ for $i=1,2$.
Since $w^\bot=w_i/\Rtop=w_{\{\theta_i\}}/\Rtop$ it is impossible
that $w_i\neq w_{\{\theta_i\}}$ otherwise top simplex
$\theta_i^\top/{\equivert}$ will decrease its dimension passing
from $\topAComp/{\equivert}$ to
$\theta_i^\top/\Rtop$ in $\topAComp/\Rtop$. This is not possible
since $\topAComp/{\equivert}$ and $\topAComp/\Rtop$ are two decomposition and
between top simplices in a decomposition there is always a dimension preserving
bijection.
Thus $w_i=w_{\{\theta_i\}}/{\equivert}$ 
Since we know that $w_1\neq w_2$ we have that 
$w_{\{\theta_1\}}/{\equivert}\neq w_{\{\theta_2\}}/{\equivert}$
and this proves
$w_{\{\theta_1\}}\not\equivert w_{\{\theta_2\}}$.
This proves 
$\theta_1\equatsimp\theta_2$ is not satisfied by $\equivert$.
 
Finally w.l.o.g. let us assume that  
$\nu_1$ is a top \eqt\ simplex we have to show that
$\nu_1/\Rtop=\nu_2/\Rtop=\gamma^\bot$.
Let us pose $\nu^\bot=\nu_1/\Rtop=\nu_2/\Rtop$.  
We have that $\gamma^\bot$ is  equal to
$\theta_1^\top/\Rtop\cap\theta_2^\top/\Rtop$
and $\theta_1^\top/\Rtop\cap\theta_2^\top/\Rtop$ 
must contain $\nu_1/\Rtop\cap\nu_2/\Rtop=\nu^\bot$.
Thus we have $\nu^\bot\le\gamma^\bot$
Let, for $i=1,2$, $\gamma_i$ the face of $\theta_i^\top/\Rtop$ s.t.
$\gamma_i/\Rtop=\gamma^\bot$ this must exist since
$\gamma^\bot\subset\theta_i^\top/\Rtop$.
Simplex $\gamma_i$ is an \eqt\ simplex and being
$\nu^\bot\le\gamma^\bot$ by remark to Definition \ref{def:abssimap} this implies
$\nu_i\le\gamma_i$ for $i=1,2$. Being $\nu_1$ a top \eqt\ simplex this 
cannot be a proper face of an \eqt\ simplex, Thus we must have
$\nu_1=\gamma_1$  and thus 
$\nu_1/\Rtop=\nu_2/\Rtop=\gamma^\bot$.

Conversely whenever the equation
$\theta_1\equatsimp\theta_2$ is not satisfied by $\equivert$ there
must be a \verteq\ equation such that
$w_{\{\theta_1\}}\not\equivert w_{\{\theta_2\}}$.
Thus $w_{\{\theta_1\}}/{\equivert}\ne w_{\{\theta_2\}}/{\equivert}$
and yet $w_{\{\theta_1\}}/\Rtop=w_{\{\theta_2\}}/\Rtop$ must be
in $\gamma^\bot$.
Thus taking $\nu_i=w_{\{\theta_i\}}/{\equivert}$
for $i=1,2$ we get the thesis.
\end{proof}
In the situation of Lemma \ref{lemma:couple} we will say that 
the \gl\ \inst\ $\theta_1\equatsimp\theta_2$ is {\em associated} 
to the \eqt\ simplices $\nu_1$ and $\nu_2$.
Note that for a couple of \eqt\ simplices there are several different  \gl\ \inst\ that can be associated.  Note, also, that if one of two  \eqt\ simplices $\nu_1$ and $\nu_2$ is a top \eqt\ simplex not necessarily the other is
a top \eqt\ simplex, too.
{
\begin{figure}[h]
\begin{center}
\fbox{\psfig{file=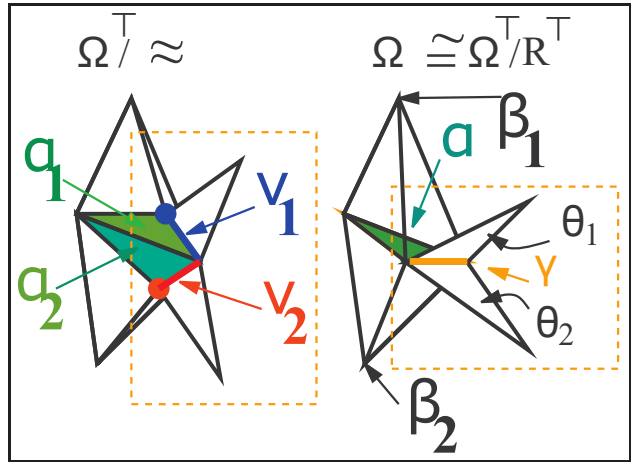,width=0.49\textwidth}}
\caption{An example that shows that the conditions of 
Lemma \ref{lemma:couple} are not sufficient to imply the existience of
top \eqt\  simplices.}   
\end{center}
\label{fig:proofeqtop}
\end{figure}
\begin{example}
Note that the fact that 
$\nu_1/\Rtop=\theta_1^\top/\Rtop\cap\theta_2^\top/\Rtop$ do not
implies that $\nu_1$ is a top \eqt\  simplex.
Consider the situation shown in Figure \ref{fig:proofeqtop}.
In the complex $\AComp$ there is a pair of top 3-simplices
$\beta_1$ and $\beta_2$ such that 
$\beta_1\equatsimp\beta_2$ is a \gl\ \inst\ 
not satisfied in the \dec\  $\topAComp/\equivert$.
In this situation the common simplex
$\alpha=\beta_1\cap\beta_2$ 
splits in $\topAComp/\equivert$ into
two simplices {$\alpha_1$} and {$\alpha_2$}. Each $\alpha_i$, 
is coface of $\nu_i$.
Such a situation can be found in the decomposition on the
left of Figure \ref{fig:proofeqtop}.

In the $3$-complex $\topAComp/\equivert$ 
we have that the stitching simplices $\nu_i$ 
are non top stitching simplices  and 
is coface of the top stitching  triangles 
$\alpha_i$.
Note that the pair of simplices 
$\alpha_2$ and $\nu_1$ and the \gl\ \inst\  $\beta_2\equatsimp\theta_1$
satisfy the hypothesis of Lemma  \ref{lemma:couple} and 
only $\alpha_2$ is the top \eqt\ simplex between $\alpha_2$ and $\nu_1$. 

Finally
note that the pair of top simplices 
$\nu_1$ and $\nu_2$ and the \gl\ \inst\ $\theta_1\equatsimp\theta_2$
satisfy the hypothesis of Lemma  \ref{lemma:couple}. In this situation
the common image  of 
$\nu_1$ and $\nu_2$ in $\topAComp/\Rtop$
is isomorphic to $\gamma=\theta_1\cap\theta_2$. 
Thus all the condition in the hypothesis
and in the thesis of Lemma \ref{lemma:couple} are verified  and yet neither
$\nu_1$ nor $\nu_2$ are top \eqt\ simplices.
\end{example}
}

In order to introduce the central theorem in this thesis
(Theorem \ref{theo:xlat}) we present a second
lemma that gives more details on the relation between  manifoldness of 
the stitching simplex and manifoldness of the associated \tsimeq\
\inst\ $\theta_1\equatsimp\theta_2$.
Indeed, by Lemma \ref{lemma:couple},
for each \eqt\ simplex $\gamma^\prime$ in  $\topAComp/\equivert$ 
there exist at least  one associated \gl\ \inst\
$\theta_1\equatsimp\theta_2$  such that
this equation is not satisfied by $\equivert$ and 
the pasted version of $\gamma^\prime$  is contained in 
$\gamma^\bot=\theta_1^\top/\Rtop\cap\theta_2^\top/\Rtop$.
Note that there are several different  \gl\ \inst\ with this properties that can be associated. 
\begin{lemma}
\label{lemma:couplenm}
Let be  $\nu^\prime$ an \eqt\ simplex in  $\topAComp/\equivert$ 
then there exist at least one associated \gl\ \inst\  {$\theta_1\equatsimp\theta_2$} such that  the  following facts holds:.
{
\begin{enumerate}
\item \label{lemma:stitmani}
if $\nu^\prime$  is a 
manifold \eqt\ simplex the associated \gl\ \inst\ must be manifold.
\item \label{lemma:stitnmani}
if $\nu^\prime$  is a top non manifold \eqt\  simplex
then the associated \gl\ \inst\ must be non-manifold.
\item \label{lemma:couplemani}
For each manifold \gl\ \inst\ {$\theta_1\equatsimp\theta_2$}  
not satisfied by $\equivert$ there exist
a manifold \eqt\  simplex $\nu^\prime$ in $\topAComp/{\equivert}$ 
such that {$\theta_1\equatsimp\theta_2$} can be associated to $\nu^\prime$.
\item \label{lemma:couplenmani}
For each non manifold \gl\ \inst\  {$\theta_1\equatsimp\theta_2$}
not satisfied by $\equivert$ there exist
a non manifold \eqt\  simplex $\nu^\prime$ in $\topAComp/{\equivert}$ 
such that {$\theta_1\equatsimp\theta_2$} can be associated to $\nu^\prime$.
\end{enumerate}
}
\end{lemma}
\begin{proof}
In the proof of Lemma \ref{lemma:couple} we have seen
(Refer to Figure  \ref{fig:equisti} and assume, for instance,
$\nu^\prime=\nu_1$)
that the pasted version of an \eqt\ simplex $\nu^\prime$ is within
(an isomorphic copy of) $\gamma=\theta_1\cap\theta_2$ in $\topAComp/\Rtop$.  
The pasted version is exactly $\gamma^\bot$ if $\nu^\prime$ 
is a top \eqt\ simplex. This proves part \ref{lemma:stitnmani}.
If $\nu^\prime$ is not top then its pasted version is a face of 
$\gamma^\bot$. We prove Part \ref{lemma:stitmani}
since it can be proved (see Remark \ref{rem:manicof})  that
a coface of a manifold simplex is a manifold simplex too.

To prove the remaining two parts we note that we can take for $\nu^\prime$
an \eqt\ simplex (e.g. $\gamma_1$ in Figure \ref{fig:equisti}) whose pasted
version will be $\gamma^\bot$. This will proof Part
\ref{lemma:couplemani} and Part \ref{lemma:couplenmani}
\end{proof}
Note that in general the coface of a manifold simplex is a manifold simplex 
while the face of a non-manifold simplex is a non-manifold simplex
(see Property \ref{pro:maninonmani} Part \ref{pro:3top}).
On the other hand
the face of a manifold simplex need not to be a manifold simplex 
and also the coface of a non-manifold simplex need not to be a 
non-manifold simplex.

{
\begin{example}
As a comment to results in Lemma \ref{lemma:couplenm}, we present some 
counterexamples to claims obtained by slight (and wrong) variatons  of
some of  the properties in the above Lemma. 
The counterexamples are shown in the 
three Figures  \ref{fig:counter}a, \ref{fig:counter}b, \ref{fig:counter}c.
In each figure we represent on the left the 
complex $\topAComp/\equivert$. In this complex we label 
with $1$ and $2$, the pair of top
simplices corresponding to the top simplices
$\theta_1$ and $\theta_2$ mentioned in Lemma \ref{lemma:couplenm}.
In the complex on the left of each figure the red dot 
represents the stitching simplex 
$\nu^\prime$.
On the right of each figure we present 
the final complex {$\topAComp/{\equivert+\{\theta_1\equatsimp\theta_2\}}$}.
In the complex on the right of each figure the red dot 
stands for the pasted version of $\nu^\prime$ in 
{$\topAComp/{\equivert+\{\theta_1\equatsimp\theta_2\}}$}.
Similarly labels $1$ and $2$ are used for the pasted version of 
simplices $\theta_1$ and $\theta_2$. 
With this conventions the three figures  show that all the claims 
below are false.
\begin{figure}[h]
\fbox{\psfig{file=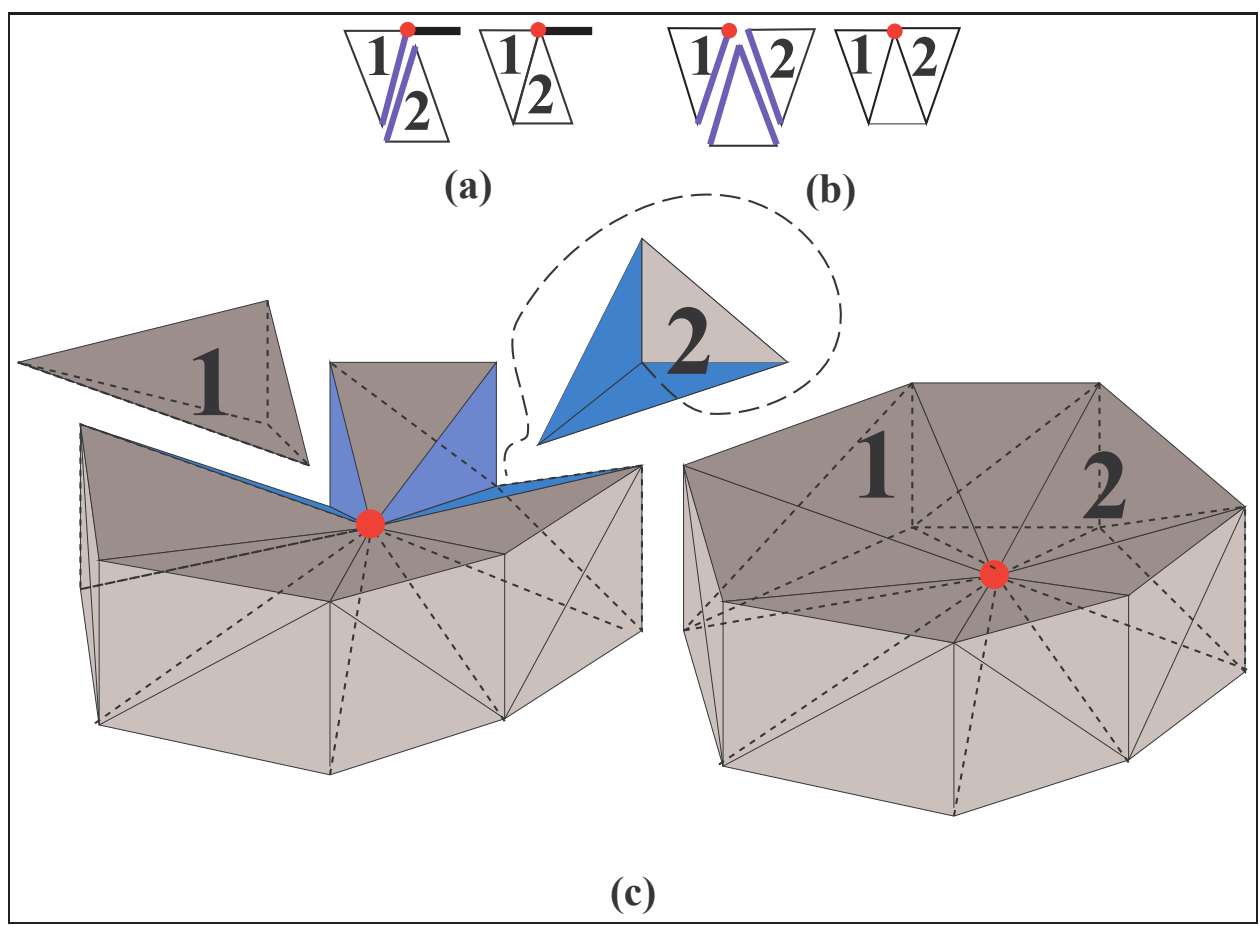,width=\textwidth}}
\caption{Counter examples to some claims about manifold and non manifold
stitching simplices}
\label{fig:counter}
\end{figure}
{
\begin{description}
\item[Wrong variation of Part \ref{lemma:stitnmani} of Lemma \ref{lemma:couplenm}]
if $\nu^\prime$  is a ({\em forget top}) non manifold stitching simplex
then the associated \gl\ \inst\ must be non-manifold.

{\em Indeed in Figure \ref{fig:counter}a
the red dot is a non top,  non manifold, stitching simplex.
The red dot  is a face of the blue edge that is 
a manifold  stitching simplex. The claim is
wrong because in the complex on the left we can only associate the red dot to the
manifold \gl\ \inst\  {$\theta_1\equatsimp\theta_2$}}
\item[Wrong variaton of Part  \ref{lemma:couplemani} of Lemma \ref{lemma:couplenm}]
For each manifold \gl\ \inst\ {$\theta_1\equatsimp\theta_2$}  
not satisfied by $\equivert$ there exist
a top ({\em added top}) manifold stitching simplex 
$\nu^\prime$ in $\topAComp/{\equivert}$
associated to {$\theta_1\equatsimp\theta_2$}.

{\em Indeed in Figure \ref{fig:counter}b the
top simplices in the \gl\ \inst\ {$\theta_1\equatsimp\theta_2$} 
intersect at the red dot 
that is a non top  manifold stitching simplex. 
Top manifold stitching simplices are the blue edges}
\item[Wrong variaton of Part \ref{lemma:couplenmani} of Lemma \ref{lemma:couplenm} ]
For each non manifold \gl\ \inst\   {$\theta_1\equatsimp\theta_2$}
not satisfied by $\equivert$ there exist
a ({\em add top}) top non manifold stitching simplex $\nu^\prime$ in $\topAComp/{\equivert}$ 
associated to {$\theta_1\equatsimp\theta_2$}

{\em Indeed in Figure \ref{fig:counter}c
the \gl\ \inst\ {$\theta_1\equatsimp\theta_2$}
is a non manifold \gl\ \inst.
The red dot is the unique non manifold stitching simplex.
Furthermore the red dot is the unique stitching simplex to which {$\theta_1\equatsimp\theta_2$} can be associated.
Some other stitching simplices are the triangles in blue. The 3-simplex numbered 2 is represented turned upside down to show these manifold stitching 2-faces.
You have to turn it following the dashed line before stitching it. 
So the red dot  is a non top stitching simplex.
The red dot is an example of a non manifold stitching simplex
that is face of a top manifold stitching simplex}
\end{description}
}
\end{example}
}

{
\section{\Cano\ decomposition}
\label{sec:decomp}
Usually one perceives
pasted non manifold top stitching simplices
as ''joints'' and  seems reasonable to expect to build a decomposition
by splitting the complex at non manifold joints.
In this section we formalize  this concept of ''reasonable'' decomposition
and show, in Theorem \ref{theo:xlat}, that the set of such decompositions
admits a least upper bound that is made up of \cdec\ complexes. 
A consequence of this result will be
that, in general, it is not possible
to decompose $d$-complexes 
breaking the complex only at non manifold simplices and expect to
obtain manifold connected components.
Indeed neither  pseudomanifolds connected components can be  assured.
This is true for all $d\ge 3$. On the other hand  a decomposition 
into manifold components exists for non-manifold surfaces.

Actually  it is not possible even to 
decide if a certain $d$-complex admit a decomposition into $d$-manifold
components for all $d$.
This fact is an easy consequence of the non recognizability of 
$d$-manifolds for $d\ge 6$ (see Theorem \ref{teo:halting})
\begin{property}
	\label{pro:nondecdec}
It is not possible, for $d\ge 6$, to build an algorithm
that takes as input
an abstract simplicial $d$-complex $\AComp$ and decompose  
it \iff\ $\AComp$ is non-manifold.
\end{property}
\begin{proof}
Such  an algorithm  can be used to recognize manifolds
simply checking if its output is equal or isomorphic to its input. 
Thus this algorithm do not exist for $d\ge 6$.
\end{proof}
}
{
\subsection{Essential  decompositions} 
An assumption
underlying this  work is that we are interested in
decompositions that splits the original complex only at non manifold simplices.
We will call this kind of decomposition an {\em \natu} decompositions.
Since we split at non manifold joint it seemed  plausible that
decomposition components for \natu\ decompositions must be manifold.
This is not always the case. Actually there are complexes for which an
\natu\ decomposition with manifold components do not exist.
We start the discussion of these problems with the definition of
\natu\ decomposition.
\begin{definition}[Essential Decomposition]
A decomposition {$\AComp^\prime$} will be called an \natu\
decomposition for $\AComp$ if and only if all top \eqt\ simplices
in {$\AComp^\prime$} are non-manifold.
\end{definition}
We recall that  $\gamma^\prime$ is a non manifold 
stitching simplex iff its pasted version is a non manifold simplex.
in $\AComp$

\begin{figure}[h]
\begin{minipage}{0.15\textwidth}
\fbox{\psfig{file=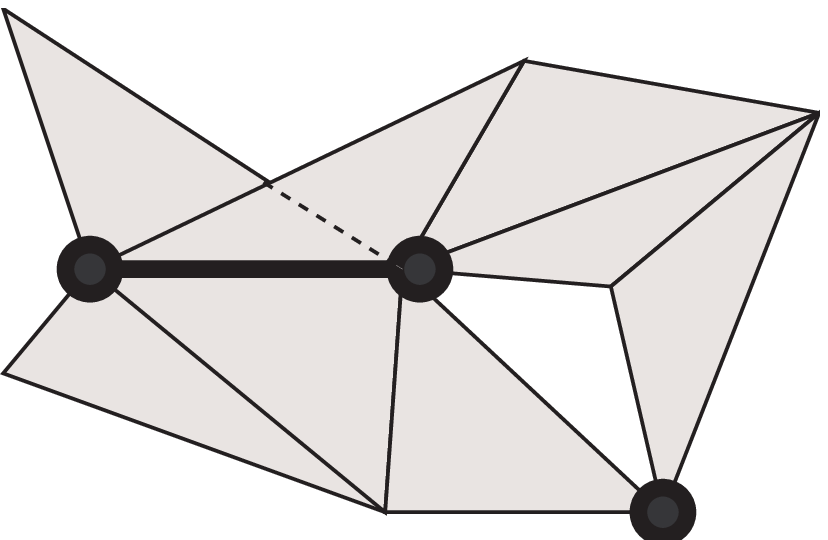,width=\textwidth}}
\begin{center}(a) \end{center}
\end{minipage}
\ \ \ \ \ 
\begin{minipage}{0.15\textwidth}
\fbox{\psfig{file=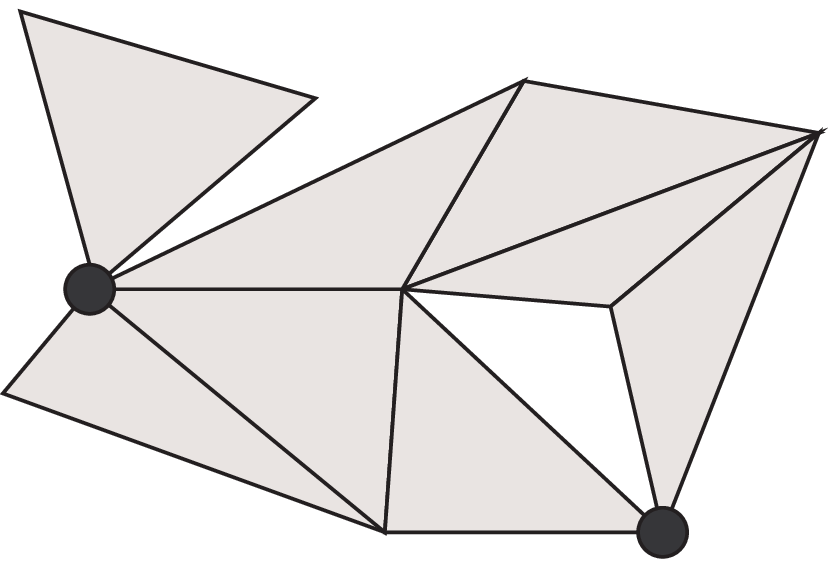,width=\textwidth}}
\begin{center}(b) \end{center}
\end{minipage}
\ \ \ \ \ 
\begin{minipage}{0.15\textwidth}
\fbox{\psfig{file=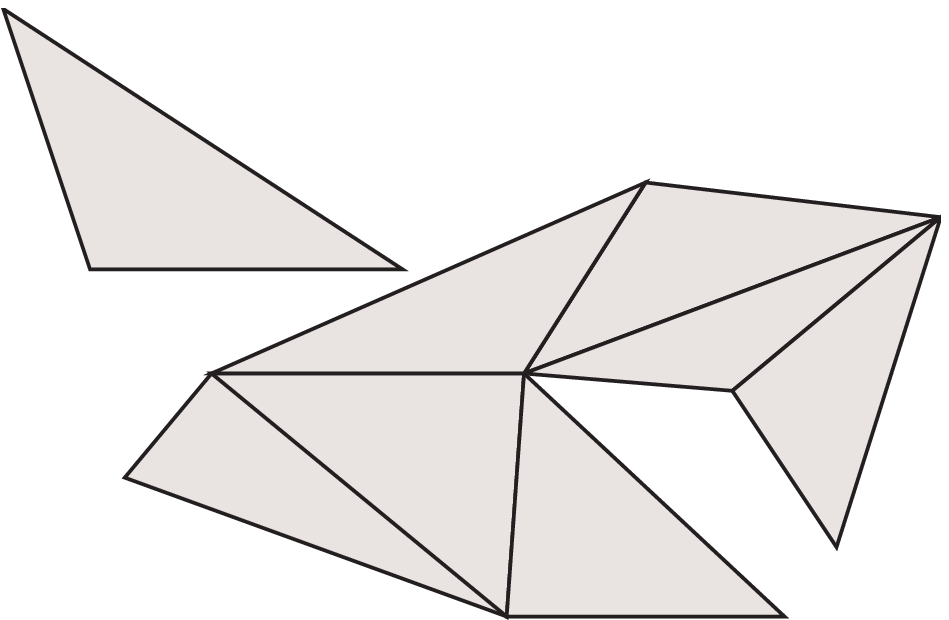,width=\textwidth}}
\begin{center}(c) \end{center}
\end{minipage}
\ \ \ \ \ 
\begin{minipage}{0.15\textwidth}
\fbox{\psfig{file=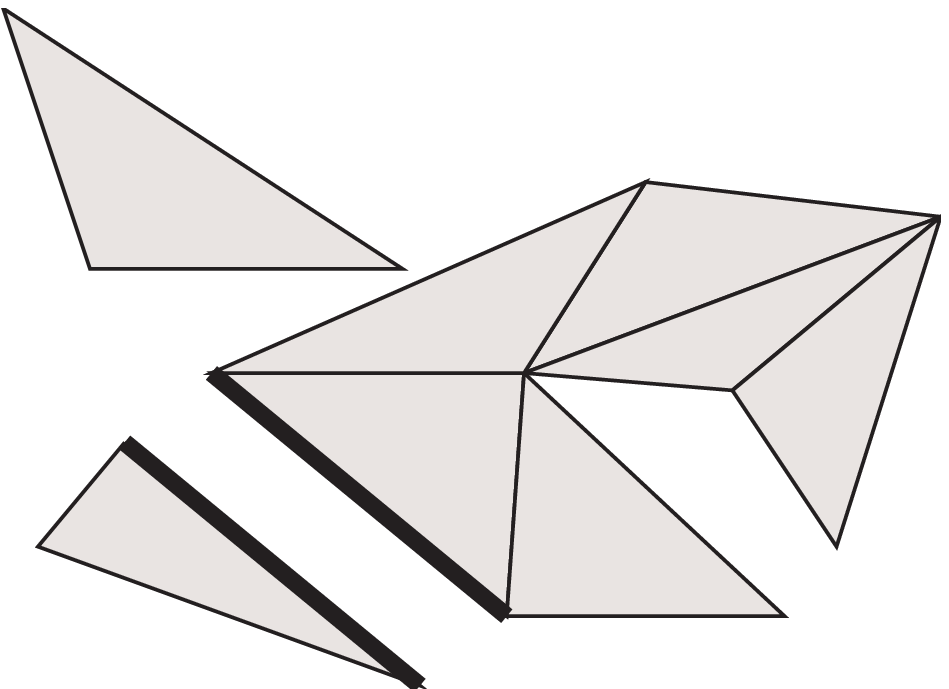,width=\textwidth}}
\begin{center}(d) \end{center}
\end{minipage}
\ \ \ \ \ 
\begin{minipage}{0.15\textwidth}
\fbox{\psfig{file=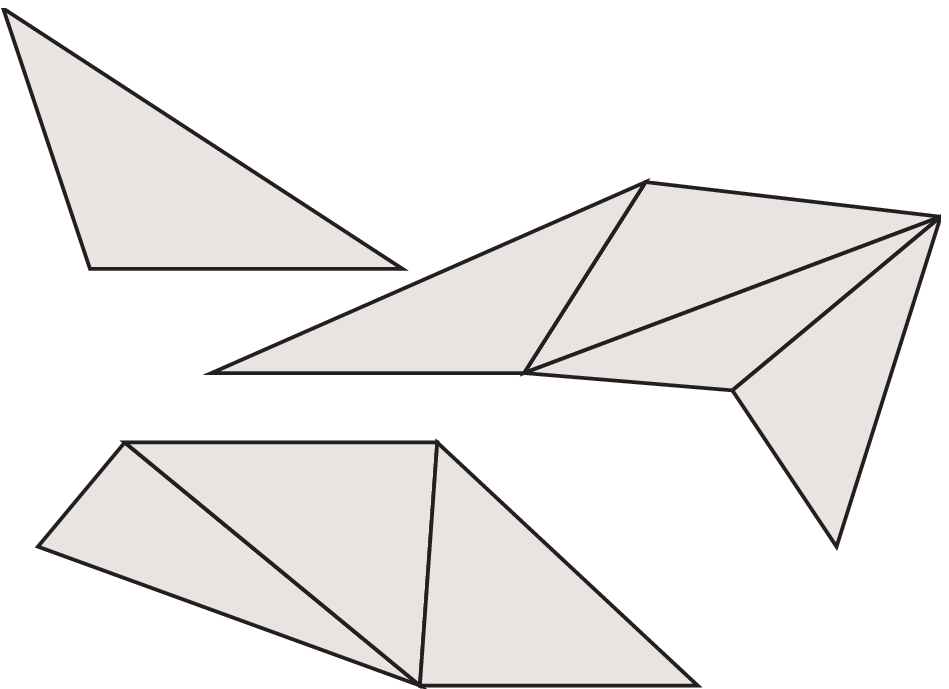,width=\textwidth}}
\begin{center}(e) \end{center}
\end{minipage}
\caption{Four decomposition of the complex (a) 
Decompositions in (b), (c) and (e) are \natu, decomposition in (d) is not}   
\label{fig:natude}
\end{figure}
 
\begin{example}
In Figure \ref{fig:natude} we present three decompositions of the complex
in (a).
Thick dots and the thick edge in figures
(a) and (b) are non manifold simplices. Decomposition in Figure (b)
is \natu\ but is still a  non manifold complex. Decomposition in Figure 
(c) and (e)
are \natu\ decomposition  and connected
components are   a manifold complex. 
Decomposition in Figure (d) is a manifold complex
but is not a \natu\ decomposition because we split along the
thick black edge that is manifold.
\end{example}

A nice property of  \natu\ decompositions is  that they allow
us to define a {\em \cano} decomposition among \natu\ ones.
This decomposition, in some sense,
is the ''most general'' decomposition among those that are \natu.
This is expressed by the following theorem that is the main result
of this thesis.
\begin{theorem}
\label{theo:xlat}
Among 
decompositions for $\AComp$
there exist a unique (up to isomorphism) \natu\
decomposition complex  $\canon{\AComp}$
that is bigger in the \Dec\ Lattice than any other \natu\ decomposition complex.
We will call $\canon{\AComp}$ the \cano\ decomposition for
$\AComp$.
\end{theorem}
\begin{proof}
We first recall that
we only need to consider \dec s in the lattice of
decompositions (defined in \ref{dec:lat-2019}).
Indeed, this lattice, by Property  \ref{pro:declatt},
contains an isomorphic copy of any decomposition of $\AComp$. 

We prove the existence of a \cano\ decomposition by explicilty 
building the \cano\ decomposition $\canon{\AComp}$ for $\AComp$.
Let be ${\cal M}$ the set of \gl\ \insts\ obtained taking all manifold \gl\ 
\insts\ We will show that $\canon{\AComp}=\topAComp/{\cal M}$.

First of all we show that $\topAComp/{\cal M}$ is \natu.
Let us consider a top \eqt\ simplex in $\topAComp/{\cal M}$ and show that it is non-manifold.
By  Lemma \ref{lemma:couple}
we have that for every top \eqt\ simplex $\gamma^\prime$ in
$\topAComp/{\cal M}$ we can find a \gl\ \inst\  $g=\theta_1\equatsimp\theta_2$ such that $g\notin{\cal M}$ and such
that
the pasted version of $\gamma^\prime$ is exactly 
$\gamma^\prime/\Rtop=\gamma^\bot=\theta_1^\top/\Rtop\cap\theta_2^\top/\Rtop$.
Since $g\notin{\cal M}$ we have that $g$ is a non manifold \gl\ \inst\ and
so, by definition \ref{def:manifoldinst}
is $\gamma^\bot$. Hence any generic top \eqt\ simplex $\gamma^\prime$
in  {$\topAComp/{\cal M}$} is non-manifold and therefore  
{$\topAComp/{\cal M}$} is \natu.

Second we show that for any 
\natu\ decomposition
$\topAComp/{\equivert}$ we have
$\topAComp/{\equivert}\quotle\topAComp/{\cal M}$.
We will prove this by proving that for every \inst\ 
$g\in{\cal M}$ we have that $g$ is satisfied by ${\equivert}$.
Therefore we will have that ${\equivert^{\cal M}}{\subset}{\equivert}$ and therefore
$\topAComp/{\equivert}\quotle\AComp^{\cal M}$.
So let be $\topAComp/{\equivert}$ an essential decomposition and
let us assume that there exist a manifold \gl\ \inst s 
$g=(\theta_1\equatsimp \theta_2)\in{\cal M}$
such that $g$ is not satisfied by ${\equivert}$. 
We can  derive a contradiction from this assumption.
By Lemma \ref{lemma:couplenm} part \ref{lemma:couplemani}
there must be a manifold \eqt\ simplex $\gamma^\prime$ such that
$\gamma^\prime/\Rtop=\theta_1^\top/\Rtop\cap\theta_2^\top/\Rtop$.
By definition any \eqt\ simplex is always a face of a top \eqt\
simplex. Therefore $\gamma^\prime$ is face of a top \eqt\ simplex.
$\gamma^{\second}$ in $\topAComp/{\equivert}$ 
(i.e. $\gamma^\prime\le\gamma^\second$).
Thus, by Equation \ref{eg:min} in Property \ref{pro:lattid} we have 
$\gamma^\prime/\Rtop\le\gamma^\second/\Rtop$ and thus
$\gamma^\prime/\Rtop$ must be face of a $\gamma^\second/\Rtop$.
Being  $\topAComp/\equivert$ \natu\ we have that $\gamma^\second$ is
a non-manifold \eqt\ simplex and therefore
$\gamma^\second/\Rtop$ is a non-manifold simplex is $\topAComp/\Rtop$.
Faces of a non manifold simplex are non manifold and so must be
$\gamma^\prime/\Rtop$. This  results in a contradiction since
$\gamma^\prime$ is a manifold stitching simplex.
Hence for all
$g\in{\cal M}$ we have that $g$ is satisfied by ${\equivert}$.
Therefore, ${\equivert^{\cal M}}{\subset}{\equivert}$ and therefore
$\topAComp/{\equivert}\quotle\AComp^{\cal M}$.

This proves that is the l.u.b. of the set of \natu\ decompositions.
Uniqueness then comes  from the fact that the set of all decomposition is
a lattice.
\end{proof}
In the example of 
Figure \ref{fig:natude} the 
decomposition in (d) is the \cano\ decomposition for the 
complex in (a).

Theorem \ref{theo:xlat} proves 
that the \cano\ decomposition 
is the least upper bound among \natu\ decomposition. 
Indeed, it might happen that more interesting decomposition
exist ''below'' the \cano\ decomposition. 
For instance, decomposing particular surfaces,
we can have decompositions with larger
manifold connected components.
For instance in Figure \ref{fig:natude} we have that the complex in
figure (d) is the \cano\ decomposition of the complex in figure (a).
Yet the complex of figure (b) is a decomposition obtained by further
stitching of the \cano\ decomposition  with just two manifold
connected components instead of the three in figure (d).
However the reduction of these connected components is, somehow, arbitrary.
Indeed, in general, if there exist an
\natu\ decomposition $\nabla^\prime$ 
strictly smaller than ${\canon{\AComp}}$ 
(i.e. $\nabla^\prime\dc{\canon{\AComp}}$) there must be 
another \natu\ decomposition $\nabla^\second$ that cannot
be compared against $\nabla^\prime$ and such  that
$\nabla^\second\dc{\canon{\AComp}}$. In this sense the two 
\natu\ decomposition $\nabla^\prime$ and  $\nabla^\second$ are
two {\em arbitrary} options 
(i.e. neither $\nabla^\prime\dc\nabla^\second$ nor
$\nabla^\second\dc\nabla^\prime$). 
This is expressed by the following Property
\begin{property}
Let $\nabla^\prime$ be a \natu\ \dec\ of $\AComp$ 
in the \dec\ lattice such that 
${\topAComp/\Rtop}<\nabla^\prime<\canon{\AComp}$.  
In this situation, in the \dec\ lattice there always exist 
another \natu\ \dec\ $\nabla^\second\neq\nabla^\prime$ such that the following diagram
holds in the \dec\ lattice:
{
	\[\begindc{\commdiag}[500]
\obj(1,1)[b]{$\mbox{\large $\canon{\AComp}$}$}

	\obj(0,0)[a1]{$\mbox{\large $\nabla^\prime$}$}
\obj(2,0)[c1]{$\mbox{\large $\nabla^\second$}$}

\obj(1,-1)[b2]{$\mbox{\large $\topAComp/\Rtop$}$}

	\mor{b}{a1}{}
	\mor{b}{c1}{}
	\mor{a1}{b2}{}
	\mor{c1}{b2}{}
	\enddc\]
	In the diagram at right below, all arrows have the (def
}
i.e. 
${\topAComp/\Rtop}<\nabla^\second<\canon{\AComp}$
\end{property}
\begin{proof}
We can write 
$\canon{\AComp}$ as $\topAComp/E$ 
for some set of \verteq\ equations $E$. 
We can write $\nabla^\prime$ as $\topAComp/(E\sumeq E^\prime)$ 
for some additional set of \verteq\ equations $E^\prime$ and
we can write ${\topAComp/\Rtop}$ as $\topAComp/(E\sumeq E^\prime\sumeq E^\second)$ 
for some additional set of \verteq\ equations $E^\second$. Since the 
. 
Similarly 
(see
section  \ref{sec:pin} in \latt).
By semimodularity we can draw the diamond:
	\[\begindc{\commdiag}[500]
\obj(1,1)[b]{$\mbox{\large $E\sumeq E^\prime\sumeq E^\second$}$}

\obj(0,0)[a1]{$\mbox{\large $E^\prime$} $}
\obj(2,0)[c1]{$\mbox{\large $E^\second$}$}

\obj(1,-1)[b2]{$\mbox{\large $E$}$}

\mor{b}{a1}{}
\mor{b}{c1}{}
\mor{a1}{b2}{}
\mor{c1}{b2}{}
\enddc\]

We obtain the
diamond in the thesis taking all the \quot s of $\topAComp$ w.r.t.
the sets of \verteq\ equations: $E$, $E^\prime$, $E^\second$, 
$E\sumeq E^\prime\sumeq E^\second$ and
naming.  $\nabla^\second$ the \quot\ $\topAComp/E^\second$ 
\end{proof}

Thus, the decomposition ${\canon{\AComp}}$  is the less
decomposed complex obtained cutting only at non-manifold simplices.
The  \cano\ \dec\ is an \natu\ \dec\ and
all other 
\natu\ decompositions, in different ways, are less decomposed than
${\canon{\AComp}}$.

We note that the proof of Theorem \ref{theo:xlat} 
is a constructive proof since it gives a procedure to build the \cano\
decomposition {$\canon{\AComp}$}. This allows to give some properties of the complex  {$\canon{\AComp}$}.
The first fact about standard decomposition is that this decomposition tears apart features with 
mixed dimensionality. This is  due to the fact that manifold \gl\ \insts\ 
must be regular. This is expressed by the following property:
\begin{property}
The connected components of the \cano\ decomposition {$\canon{\AComp}$}
are regular
\end{property}
\begin{proof}
In the proof of Theorem \ref{theo:xlat}
We have seen that {$\canon{\AComp}=\topAComp/{\cal M}$} being $M$ a set of 
manifold \gl\ \inst. By Property \ref{pro:maniregu} we have that 
instructions in ${\cal M}$ must be all regular. 
Hence, we can apply Property \ref{pro:regueqregu} and state that
connected components of {$\topAComp/{\cal M}=\canon{\AComp}$} must be 
regular.
\end{proof}

A second, deeper, characterization of the complex $\canon{\AComp}$
comes from the fact  that, to build this complex, we used 
manifold instructions. The consequence of this fact is that the 
resulting complex has connected components that are  \cdec\ complexes.
This is stated in the following property

\begin{property}
\label{pro:iqmdecomp}
The connected components of the \cano\ decomposition {$\canon{\AComp}$}
are \cdec\ complexes.

\end{property}
\begin{proof}
In the proof of Theorem \ref{theo:xlat}
we have seen that {$\canon{\AComp}=\topAComp/{\cal M}$} being $M$ a set of 
manifold \gl\ \inst s. We have seen in Property
\ref{pro:maniregu} that all manifold \inst s must be regular.
Therefore we can assign a dimension to \inst s in ${\cal M}$.
Let be ${\cal M}_h$ the set of \inst s
of dimension $h$ in  ${\cal M}$.
By Property \ref{pro:partregular} we have that connected components of 
dimension $h$ are all those within the regular $h$-complex 
{$\topAComp_h/{{\cal M}_h}$}, where
$\topAComp_h$ is the subcomplex of top simplices of $\topAComp$ of dimension
$h$.
Since \inst s within {${{\cal M}_h}$} are all manifold by
repeated application of Property  \ref{pro:manid1} we can replace
all equations of order smaller than $(h-1)$ with some others
manifold \inst s of order $(h-1)$.   
The added instructions can generate more identification than the original instruction,but this is not important since we are interested in showing  that
this process will produce a set of $(h-1)$-manifold \inst s 
{${{\cal M}^\prime_h}$} such that 
${\topAComp_h/{{\cal M}^\prime_h}}={\topAComp_h/{{\cal M}_h}}$.
In the end this shows that the connected components of order $h$ can be
generated by a set of $(h-1)$-manifold \inst s. This implies,
by Property \ref{pro:qpmeq} Part \ref{pro:eqmqpm} that
the link of every vertex in {${\topAComp_h/{{\cal M}_h}}$} is $(h-1)$-manifold
connected. Therefore each connected component within 
{${\topAComp_h/{{\cal M}_h}}$} is an \cdec.

\end{proof}
We note that there are examples of 3-complexes $\AComp$ for which
the decomposition  $\canon{\AComp}$ is non manifold.

The complex on the right of Figure \ref{fig:counter} part (c)  shows that it is impossible in general to find a
decomposition of certain non manifolds $d$-complex
by splitting
the complex only at  pasted non manifold top stitching simplices.
In fact the $3$-complex on the right of Figure \ref{fig:counter}c  is an \cdec\  complex.
All simplices but the central   (thick red) vertex are manifold simplices.
The central vertex is not a manifold vertex because its link is not 
homeomorphic to a sphere or a triangle. 
Nevertheless the star of the central vertex  
is, obviously, a $2$-manifold-connected complex.

Therefore this complex  is an 
\cdec\ $2$-complex and its \cano\ decomposition
is the complex itself. This decomposition is therefore 
a non-manifold complex. This is not too counter-intuitive
since we really do not have a good reason to decide how to 
break this complex into manifold pieces. Probably intuition suggests
to {\em inflate} this complex to make it more regular. 
In fact the \cano\ \dec\
of the surface that is the boundary of this complex split the central
vertex in two and yields a decomposition homeomorphic to a sphere.
We recall, from Example \ref{app:example},
that there are  \cdec\ $3$-complexes that are not pseudomanifolds and 
therefore there are \cano\ decompositions for $3$-complexes 
that are not $3$-pseudomanifolds.
Finally we note that  being the set of \cdec\ $2$-complexes equivalent
to $2$-manifolds 
(see Part \ref{pro:2mani} of  Property \ref{pro:iqmrel}) 
it is always possible to decompose $2$-complexes into $2$-manifolds.
}

\section{Computing {$\canon{\AComp}$}}
A byproduct of the findings in the  proof of Property \ref{pro:iqmdecomp}
is the following
property that gives a first procedure to build the \cano\ decomposition.
\begin{property}
\label{pro:firstalgo}
The \cano\ decomposition {$\canon{\AComp}$} for a $d$-complex
$\AComp$ is generated by the set obtained taking, for all $1\le h\le d$,
all \tsimeq\ pairs {${\Pi_\theta^h}=\{\theta_1,\theta_2\}$} 
that satisfy the following constraints:
\begin{itemize}
\item
{${\Pi_\theta^h}$} is  regular of dimension $h$ 
(i.e. such that $\ord{(\theta_1)}=\ord{(\theta_2)}=h$);
\item
{${\Pi_\theta^h}$} is of  order $(h-1)$ 
(i.e.   $\ord{(\theta_1\cap\theta_2)}=h-1$);
\item
{${\Pi_\theta^h}$} is such that the two simplices $\theta_1$ and $\theta_2$ 
gives the star of their intersection (i.e.
 $\str{(\theta_1\cap\theta_2)}=\{\theta_1,\theta_2\}$ ).
\end{itemize}
The set ${{\cal M}_h}$ of all pairs of dimension $h$ that satisfy the above 
requirements will generate the regular connected components of dimension $h$.
If $\topAComp_h$ is the set of top simplices of dimension $h$ in $\topAComp$ we 
have that the $h$-complex {$\topAComp_h/{{\cal M}_h}$} is made up of 
all connected components of dimension $h$ in {$\canon{\AComp}$}.
\begin{proof}
In the proof of Property \ref{pro:iqmdecomp} we have that each connected
component of dimension $h$ is generated by the set 
of regular $(h-1)$-manifold instructions of dimension $h$ 
(denoted by ${{\cal M}_h}$).  Furthermore in the proof of that property 
we have noticed that the set of connected  $h$-components is given by
the complex   {$\topAComp_h/{{\cal M}_h}$}.
Since these are manifold instructions only $h$ simplices 
must be incident to $\theta=\theta_1\cap\theta_2$.
Since $\ord{(\theta)}=(h-1)$  only
two $h$-simplices must be incident to $\theta$.
These two $h$-simplices must be $\theta_1$ and $\theta_2$. 
\end{proof}
\end{property}

Property \ref{pro:firstalgo} gives a computable procedure to select a 
limited number of instructions. By applying these instructions 
repeatedly to simplices in $\AComp^\top$ we can effectively compute 
the \cano\ decomposition {$\canon{\AComp}$} by a top down process.
If $n$ is the number of top simplices in the $d$-complex $\AComp$ 
the construction of  {$\canon{\AComp}$} can be done with a time 
complexity of $\lesscomp{(nd)^2}$.
Infact we simply have to  intersect every top simplex with all others
to build an incidence relation between top simplices. 
Each intersection can be done in $\lesscomp{d^2}$ and therefore
all intersections can be computed in $\lesscomp{(nd)^2}$ time.
For each top $h$-simplex we allocate record with $2^{(h+1)}-1$ 
entries to store possible  incidence at a certain face.   
When this incidence
relation is built, a linear scan of the $n$ records in the 
relation will inspect
all incidence relation for each simplex and select those that  must 
be preserved in {$\canon{\AComp}$}.  Since the possible different
incidence pairs are $\lesscomp{n^2}$ this process takes at most
$\lesscomp{n^2}$. This accounts for an overall time complexity of
$\lesscomp{(nd)^2}$ and an overall space complexity of 
$\lesscomp{n2^d}$ to store the incidence relation.

This top down construction, from $\AComp^\top$ down to
{$\canon{\AComp}$} is only partially satisfactory. 
Indeed, for several
pragmatic reasons, it is interesting to develop a decomposition
procedure that splits the original complex and produces the 
decompostion bottom up by a progressive refinement at different
points. Furthermore we want to develop a decomposition procedure
that splits a certain vertex $v$ using
local topological information round the vertex $v$.

\section{Computing $\canon{\AComp}$ by local editing}
\label{sec:algo}

In this Section, we present a decomposition algorithm that builds
the \cano\ decomposition {$\canon{\AComp}$} by splitting
$\AComp$ at vertices that
violate the condition  for \iqm ness.
The algorithm works iteratively on the vertices of the input complex and recursively on
its dimension.

The algorithm for computing  {$\canon{\AComp}$}
is given by the pseudocode in the Algorithm \ref{algo:decomp}.
This algorithm defines a recursive procedure ${\rm DECOMPOSE}(\AComp,d)$
that returns  the connected components of the
\cano\ decomposition for a $d$-complex $\AComp$.
In the design of this algorithm  
we assume that Vertices are coded as distinct  positive integers 
and that each top simplex receive an index that is a positive integer, too.  
We assume that the complex $\AComp$ is presented in input as a $TV$ map
between top simplex and Vertices,
We assume to have a function ${\rm CONNECTED\_COMPONENTS}(\AComp_c)$
that takes the $TV$ representation of a complex $\AComp_c$ and 
returns a set of $TV$ maps one  for each connected component of $\AComp_c$.  
Note that we assume that this function need not to change the coding
for vertices and top simplices.
Similarly we commit not to change this coding when providing the output
of   ${\rm DECOMPOSE}(\AComp,d)$.
We assume that a standard implementation for sets is used by  
${\rm CONNECTED\_COMPONENTS}(\AComp_c)$ to provide its output.
As a consequence sets operations are freely used with a rather
abstract stlye in the design of  the algorithm.

This function starts by initializing a variable $\AComp_c$ with a copy of
$\AComp$. The $\AComp_c$ holds the current decomposition of
$\AComp$ and the algorithm  splits $\AComp_c$ until it contains {$\canon{\AComp}$}.
The algorithm considers each vertex $v$ of $\AComp$ and computes recursively
the decomposition of the link of $v$ in $\AComp$
(not the link in $\AComp_c$).
Based on such a decomposition, the algorithm decides whether and how $\AComp_c$ should be split at $v$.
Recursion stops for $d=0$  since the decomposition of either a $0$-complex, or of an empty complex trivially coincides with the complex itself.

\begin{algo}{algo:decomp}{\small Computes the connected components in {$\canon{\AComp}$} for the $d$-complex $\AComp$}
\begin{algorithmic}[1]
\STATE {\bf function} DECOMPOSE($\AComp$,$d$) \STATE $\AComp_c\leftarrow\AComp$
\IF{{$d>0$} {\bf and} {$\AComp\neq\emptyset$}}
\FORALL{vertices $v$ of $\AComp$}
\STATE $LK\leftarrow\xlk{\AComp}{v}$
\COMMENT{$LK$ is the link of $v$ in $\AComp$}
\STATE $h\leftarrow\ord(LK)$
\COMMENT{$h$ is the dimension of $LK$}
\STATE $L\leftarrow{\rm DECOMPOSE}(LK,h)$
\COMMENT{compute the components of  {${\canon{LK}}$}}
\IF[$v$ must  split]{($h>0$ {\bf and} $|L|>1$) {\bf or} ($h=0$ {\bf and}
$|L|>2$)}
\FORALL[split $v$ in $\AComp_c$]{$\Psi\in L$}
 \STATE Create {$v_{\Psi}$}
\COMMENT{create a new copy {$v_{\Psi}$} for $v$}
\STATE{Replace {$v$} with {$v_{\Psi}$} in top 
simplices of $\xstr{\AComp_c}{v}$  
incident to a simplex in $\Psi$}
\ENDFOR\COMMENT{the decomposition of vertex $v$ has been completed}
\ENDIF
\ENDFOR
\ENDIF
\STATE {\bf return} CONNECTED\_COMPONENTS($\AComp_c$)
\COMMENT{returns connected components of $\AComp_c$}
\end{algorithmic}
\end{algo}

The general idea at the basis of this algorithm
is to test vertices of $\AComp$ for the local property that
characterizes \iqm s
(i.e. {the star of each vertex $v$ of an \iqm\ must be manifold-connected})
and to split the complex in case a vertex violates such a property.
The manifold connection of the star of a vertex is ensured \siff\
its link (which has a lower dimension) is manifold-connected.
This is true because adding  vertex $v$ to all simplices in the link
we obtain all and alone the simplices in the closed vertex star.
Therefore, we want to decompose the link of $v$ into manifold-connected components.
This process induces only all those
splits that are necessary to obtain the \cano\ decomposition.
Note that the recursive algorithm actually decomposes the link of a vertex into \iqm\ components
rather than manifold-connected components.
On the basis of Part \ref{pro:eqmqpm} in Property \ref{pro:qpmeqfil}  
this result is equivalent, for our purposes,
because the partition of top simplices among connected components is the same in the two cases.
\begin{property}[Correctness of Algorithm \ref{algo:decomp}]
Let $\AComp$ be a TV map of an  abstract simplicia $d$-complex.
The algorithm \ref{algo:decomp} terminates and upon completion
the function  ${\rm DECOMPOSE}(\AComp,d)$ returns a 
set of TV maps each representing a connected component of 
$\canon{\AComp}$. 
The function  ${\rm DECOMPOSE}(\AComp,d)$ 
do not change the assignment of indexes given in input. 
\end{property}
\begin{proof}
Termination can be proved easily  by induction on the dimension of the input
$d$-complex $\AComp$. For $d=0$  procedure stops immediately.
On the other hand, if $d>0$, 
the computation performs a finite number of cycles,
one for each vertex $v$ in $\AComp$, calling 
recursively itself to compute the decomposition of  link $\lk{v}$. 
Since each
link is at most of dimension $(d-1)$, by inductive hypothesis, we can
assume that  the decomposition of each link completes. Therefore the overall
process must complete, too.

The correctness trivially holds for all $d$ whenever $\AComp=\emptyset$
(i.e. $\AComp$ is a map with no entry). 
Whenever $\AComp\neq\emptyset$
the proof can be done by induction on the dimension $d$ of the complex
$\AComp$. If $d=0$ we have $\canon{\AComp}=\AComp$ and the property 
trivially holds, too.

If $d>0$ we assume that the property holds for $d-1$ and show that it holds
for $d$. Thus, by inductive hypothesis we have that the call to
${\rm DECOMPOSE}(LK,h)$ returns the  set of
connected components in the \cano\ decomposition of the link
{$\xlk{\AComp}{v}$}. By the result in  
Part \ref{pro:eqmqpm} in Property \ref{pro:qpmeqfil}   we have that 
this is equivalent to the partition into manifold connected components 
of set of top simplices in {$\xlk{\AComp}{v}$}.
By Property \ref{pro:iqmdecomp} the components of $\canon{\AComp}$
must be \iqm\ an therefore by Definition \ref{def:iqm} the link
of each vertex in $\canon{\AComp}$ is  manifold-connected. 
Thus for each vertex $v$ in $\AComp$ the partition of top 
simplices in {$\xlk{\AComp}{v}$} induced by the 
decomposition $\canon{\AComp}$ is a (possibly identical)
further decomposed version  of the partition 
computed by the algorithm \ref{algo:decomp}.
In fact at line 11 we take the cone from the new  vertex copy  
$v_\Psi$ to each component of the partition
into manifold connected components of  {$\xlk{\AComp}{v}$}.
Thus $\canon{\AComp}$ must be a decomposition of the output of 
${\rm DECOMPOSE}(\AComp,d)$ (up to isomorphism).

On the other hand the partition of top 
simplices in {$\xlk{\AComp}{v}$} used by the  algorithm is 
the partition in $\canon{\xlk{\AComp}{v}}$. This complex in
general is a decomposition of the complex ${\xlk{\canon{\AComp}}{v}}$. 
This second complex induces a partition of top 
simplices in {$\xlk{\AComp}{v}$} that is the partition induced by the 
decomposition $\canon{\AComp}$. Thus this proves 
that the partition of top 
simplices in {$\xlk{\AComp}{v}$} induced by the 
decomposition  computed by the algorithm \ref{algo:decomp}.
is  a (possibly identical) partition of the decomposition 
induced by $\canon{\AComp}$.
Thus the output of ${\rm DECOMPOSE}(\AComp,d)$.
must be  a decomposition of $\canon{\AComp}$.

We have proven both that $\canon{\AComp}$ must be a decomposition of 
${\rm DECOMPOSE}(\AComp,d)$
and that the output of ${\rm DECOMPOSE}(\AComp,d)$.
must be  a decomposition of $\canon{\AComp}$.
This proves that ${\rm DECOMPOSE}(\AComp,d)$ computes an 
isomorphic copy of $\canon{\AComp}$ and this completes the proof .
\end{proof}

{
At an intuitive level we can say that we can obtain
the decomposition
{$\canon{\AComp}$} gluing top simplices in $\topAComp$ using manifold
\gl\ \inst s ${\cal M}$.
Indeed the algorithm splits a vertex $v$ in $\AComp$ 
into a certain number of vertex copies $v_{\Psi}$ to be used in {$\canon{\AComp}$}. Then the stars
of all vertex copies {$v_{\Psi}$} 
(i.e. {$\xclstar{\canon{\AComp}}{v_{\Psi}}$})
can be {\em partially}
formed from the totally exploded version of {$\xclstar{\AComp}{v}$}
applying all manifold \gl\ \inst s  of
the form {$\theta_1\equatsimp\theta_2$} that mention a pair of top
simplices in  {$\xclstar{\AComp}{v}$}.

From the definition of \cano\ decomposition
this means that  we glue from the totally exploded version of
${\xclstar{\AComp}{v}}$
down  towards  {$\canon{\xclstar{\AComp}{v}}$} establishing all manifold joints
between top simplices in ${\xclstar{\AComp}{v}}$.
Thus the manifold connected components in the stars
{${\xclstar{\canon{\AComp}}{v_{\Psi}}}$} 
and in $\canon{\xclstar{\AComp}{v}}$ must be
formed by the same top simplices.

This do not means that the star {$\canon{\xclstar{\AComp}{v}}$}
and the collection of stars {${\xclstar{\canon{\AComp}}{v_{\Psi}}}$} 
need to be the same (isomorphic) complex.
Actually  the links of {$v_{\Psi}$} in the \cano\ decomposition  can be
less decomposed than $\canon{\xlk{\AComp}{v}}$
See for instance the example of  Figure 
\begin{figure}[h]
\psfig{file=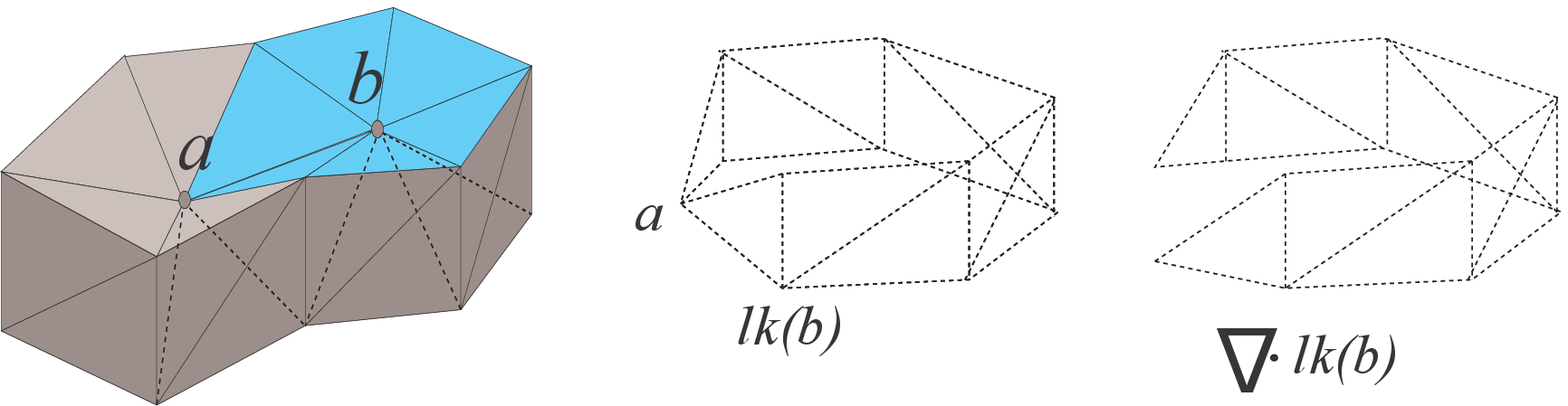,width=\textwidth}
\caption{An  example of a complex where {${\xclstar{\canon{\AComp}}{b}}$} is less decomposed than $\canon{\xlk{\AComp}{b}}$ }
\label{fig:exadeco}
\end{figure}

In Figure \ref{fig:exadeco} we report  an \iqm\ complex.
This complex is not decomposed be the standard decomposition. Let see why. 
The link of $b$ in this complex
is the dashed surface in the middle  of this figure.
The $\canon{\xlk{\AComp}{b}}$ is the dashed surface on the right of this figure. It still have one connected component.
Therefore we do not split b. 
The  ${\xclstar{\canon{\AComp}}{b}}$
is the complex in pale blue and is less decomposed than
$\canon{\xlk{\AComp}{b}}$.
The situation for $a$ is similar.

In spite of this example, connecting fresh vertex copies (i.e. ${v_{\Psi}}$
to the connected components of
$\canon{\xlk{\AComp}{v}}$ we establish the right number of vertex copies
for $v$ and connect them properly to the right component in the link.
${\xlk{\canon{\AComp}}{v_{\Psi}}}$.
Doing this for all vertices (as the main cycle of the algorithm does)
we get {$\canon{\AComp}$}.
}

\begin{figure}[h]
\begin{center}
\framebox{
\parbox[c][0.23\textwidth]{0.14\textwidth}{
\begin{center}
\psfig{file=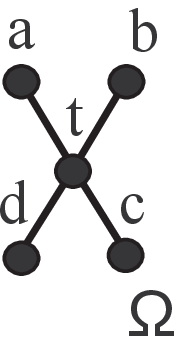,width=0.10\textwidth}
\vfill
(a)\end{center}
}
}
\
\framebox{
\parbox[c][0.23\textwidth]{0.14\textwidth}{
\begin{center}
\psfig{file=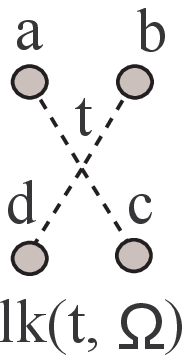,width=0.10\textwidth}
\vfill
(b)\end{center}
}
}
\
\framebox{
\parbox[c][0.23\textwidth]{0.14\textwidth}{
\begin{center}
\psfig{file=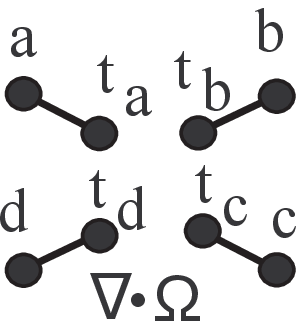,width=0.14\textwidth}
\vfill
(c)\end{center}
}
}
\end{center}
\caption{An  example of the decomposition process for the $1$-complex in (a)}
\label{fig:algoex}
\end{figure}

\begin{example}
We illustrate how the algorithm works on some examples in $1$ and $2$ dimensions.
Consider first the 1-complex on
Figure \ref{fig:algoex}a.
In this complex, the link of vertices $a$, $b$, $c$ and  $d$ is  always
the vertex $t$ and, thus, the algorithm does not split these four vertices.
The link of vertex $t$ is composed of the four gray vertices
 $a$, $b$, $c$ and  $d$
depicted in Figure \ref{fig:algoex}b.
This
is a $0$-complex and is left unchanged, but it consists of
four connected components.
The algorithm then decides to introduce four copies of $t$,
one copy for each connected component in $\xlk{\AComp}{t}$.
Vertex $t_a$ is
introduced for the connected component made of vertex $a$.
Next $t$
is replaced with $t_a$ in simplex $ta$ yielding in the new edge $t_aa$.
Similar moves lead to the introduction of the three
edges $t_bb$, $t_cc$ and $t_dd$.
This generates  the complex of Figure \ref{fig:algoex}c
that is the \cano\ decomposition of the starting complex $\AComp$.

\begin{figure}[h]
\begin{center}
\psfig{file=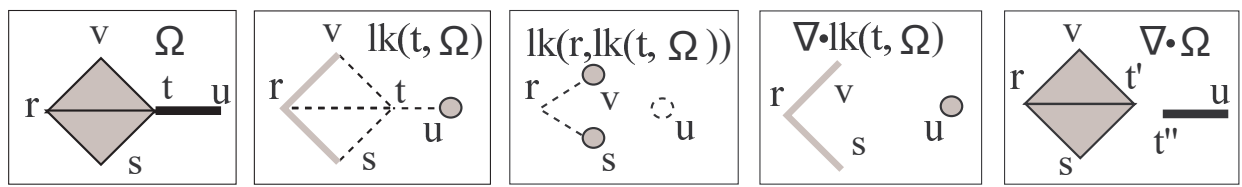,width=\textwidth}
\end{center}
\caption{
Examples of the decomposition process for the $2$-complex of Figure
\ref{fig:lattconf}. 
}
\label{fig:algoex3}
\end{figure}

The decomposition of the
$2$-complex (used in the running
examples of
Figures \ref{fig:lattconf}, \ref{fig:topdef} and many others.
is depicted in Figure \ref{fig:algoex3}.
From left to right, we summarize the step of the algorithm for vertex $t$,
which is the only vertex that induces a split.
First the link {$\xlk{\AComp}{t}$} is computed, next the decomposition
of
the complex  {$\xlk{\AComp}{t}$}
is attempted.
We have that
the link of $u$ in the complex ${\xlk{\AComp}{t}}$
is the empty set
(i.e. {$\xlk{\xlk{\AComp}{t}}{u}=\emptyset$}).
Thus, vertex $u$ is left
unchanged in {$\canon{\xlk{\AComp}{t}}$}.
The link
of $r$ in the complex ${\xlk{\AComp}{t}}$
is a $0$-complex containing the two vertices $u$ an $v$.
Thus, the algorithm does not split vertex $r$ in {$\xlk{\AComp}{t}$}.
Therefore, we obtain that {$\canon{\xlk{\AComp}{t}}={\xlk{\AComp}{t}}$}.
The complex {$\canon{\xlk{\AComp}{t}}$} it consists of two connected components.
Therefore, the algorithm splits $t$ into two copies $t^\prime$ and $t^\second$,
and this yields the decomposition
in the rightmost frame.
Next, for vertex $r$, we must consider
{$\canon{\xlk{\AComp}{r}}$} (not shown in Figure \ref{fig:algoex3}).
It is easy to see that {$\canon{\xlk{\AComp}{r}}={\xlk{\AComp}{r}}$}.
The complex  ${\xlk{\AComp}{r}}$ has just one connected component and thus the
algorithm do not split $r$.
The same happens for the other vertices (i.e. $s$,$u$ and $v$) that
do not split.
Thus, the final $\canon{\AComp}$ is the one depicted in the  rightmost frame.
\end{example}

It is easy to prove  that the Algorithm  \ref{algo:decomp}
has a time complexity that is slightly superlinear in the
size of output.
\begin{property}
\label{pro:decocomp}
The computation of\ \ {${\rm DECOMPOSE}(\AComp,d)$}  can be done in
{$\lesscomp{\factorial{d}\cdot (NT\log{NT}})$}
where $NT$ is the  number of top simplices in the $d$-complex $\AComp$.
\end{property}
\begin{proof}
It can be seen that
all operations, but the computation of connected components
and the decomposition of the link $LK$ can be done in
{$\lesscomp{d\cdot NT\log{NT}}$} .
We start this analysis proving this fact. 
To prove this we recall that we have assumed that vertices and top simplices 
are coded as integers and that the complex $\AComp$ is presented in input as
a $TV$ map between (indexes for) top simplex and $(d+1)$ array of vertices.
The array will be padded with default values for non maximal top simplices.
With these assumptions, it is easy to see that
all operations, but the computation of connected components
and the decomposition of $LK$ can be done in
{$\lesscomp{d\cdot NT\log{NT}}$}. 

We assume a standard implementation for the $TV$ map, for
instance through  a binary search tree. 
It is known that insertion and deletion in such $TV$ map can be done in
logarithmic time vs. the size of the map \cite{Ara89}. 
Thus, the copy operation $1$ can be done with $NT$ reads and $NT$ writes
into a map and this costs $\ordcomp{NT\log{NT}}$ for each read we copy a $(d+1)$ array of Vertices and all these copies  costs $\ordcomp{(d+1)t}$. Thus operation
$1$ takes $\lesscomp{d\cdot NT\log{NT}}$.

To deal with operations $4$,$5$ and $6$ we assume that,
in an initialization phase,
omitted in the abstract version of the algorithm,
with time complexity  of order {$\ordcomp{NV\log{NV}}$}, 
where $NV$ is the number of vertices in $\canon{\AComp}$,
we can generate a $VT$ map  where, for each vertex $v$, we store the simplex indexes
for top simplices in $\xstr{\AComp}{v}$.
Being $NV\le (d+1)NT$ we have that
{$\ordcomp{NV\log{NV}}$} is $\lesscomp{d\cdot NT\log{NT}}$.

Then, for a given vertex $v$, we can build the $TV$ map for the
representation for $\lk{v}$. This can be done with just one access to the  $VT$ map (costing $\ordcomp{\log{NV}}$ and $\lesscomp{log{NT}}$) and $NT_v$ access to the $TV$ being $NT_v$ the number of top simplices in $\lk{v}$ each costing $\ordcomp{\log{NT}}$.
For each  access we copy up to $(d+1)$-Vertices and find the real dimension of the link.
Thus for vertex $v$ for $5$ and $6$ we spend  $\lesscomp{log{NT}}+\ordcomp{NT_v\log{NT}}+\ordcomp{NT_v(d+1)}$
that is $\lesscomp{NT_v\log{NT}}$.
Thus we can provide a $TV$ representation for all links in $\lesscomp{NT\log{NT}}$. Thus we can extract in $4$ one after another
Vertices in $\AComp$ in $\ordcomp{1}$ and all steps $5$ and $6$ takes $\lesscomp{NT\log{NT})}$.

We note that size of a set and sequential access to all its elements can be done in $\ordcomp{1}$ and thus we perform $8$  in
constant time once for each vertex for an overall cost of $\ordcomp{NV}$ and
$\lesscomp{d\cdot NT}$.  

Sequential access to the element in $L$ and the creation of the new vertex can be done in constant time and all the repetitions of operations in $9$ and $10$ accounts for a time that is $\ordcomp{NC}$ where $NC$ is the number of 
vertex copies introduced by the decomposition process. 
Obviously we cannot have  more vertex copies than the number of  Vertices in 
the totally
exploded version of $\AComp$ and thus $NC\le(d+1)NT$.
This proves that also $9$ and $10$ can be done in $\lesscomp{d\cdot NT}$.

Finally step $11$ recalls for the editing of the $TV$ map encoding
$\AComp_c$.
Reasoning as for step $5$ we have that, for a given vertex $v$, we can edit the $TV$ map for the representation for $\xstr{\AComp_c}{v}$. Sequential access to the elements in the domain of the map $\Psi$
will give the indexes of the $NT_{v_{\Psi}}$ top simplices to be
modified, being $NT_{v_{\Psi}}$ the number of top simplices in $\Psi$.
For each simplex we pay $NT_{v_{\Psi}}$ access to the $TV$  each costing $\ordcomp{\log{NT}}$. For each a access we compare up to $(d+1)$-Vertices and find the entries to be modified.
If $NT_v$ is the number of top simplices in $\xstr{\AComp_c}{v}$
for step $11$ we spend  $\ordcomp{NT_v\log{NT}}+\ordcomp{NT_v(d+1)}$
that is $\lesscomp{NT_v\log{NT}}$ to process a single vertex.
Since summing all $NT_v$ for all Vertices yields at most $(d+1)NT$
we have that we can edit the $TV$ representation for all vertices
in $\lesscomp{d\cdot NT\log{NT}}$. 

This proves that all operations, but $7$ and $16$ can be done in
{$\lesscomp{d\cdot NT\log{NT}}$}. 

The subdivision of a complex into connected components
(i.e., the ${\rm CONNECTED\_COMPONENTS}(\AComp_c)$ call in line $16$)
can be performed as the computation of connected components in a graph with
$(d+1)NT$ arcs and $NT+NV'$ nodes, where $NV'$ is the number of
vertices in $\canon{\AComp}$ (note that $NV'\le(d+1)NT$).
This is known (see \cite{Mel84}) to take
{$\ordcomp{d\cdot NT+NV'}$} and thus less than {$\lesscomp{d\cdot NT}$}.

Thus, if we denote with  {$T^d(NT)$} the order of time complexity for the
computation of {${\rm DECOMPOSE}(\AComp,d)$} we have that
\[T^d(NT)=\lesscomp{d\cdot NT\log{NT}}+
\sum_{v\in V}{T^{(d-1)}(NT_{\lk{v}})}\]
where:  $V$ is the set of vertices in $\AComp$,
and $NT_{\lk{v}}$ is the number of top simplices in $\lk{v}$.
We can rewrite this recurrence using the trial solution
{$T^d(NT)=\lesscomp{\factorial{d}\cdot NT\log{NT}}$}.
With some standard algebra and using the fact that
$\sum_{v\in V}{NT_{\lk{v}}}\le (d+1)NT$ we have the thesis.
\end{proof}

 \chapter{Non-Manifold Modeling Through Decomposition}
\label{ch:nmmdl}
\section{Introduction}
In this chapter, we consider the problem of describing a non-manifold 
object through its decomposition.
Based on the theoretical framework from Chapter \ref{ch:stdec}
we design a two-level data structure 
for non-manifold simplicial complexes in arbitrary dimension. 
In such a structure, each \dec\  component is represented through a 
topological data structure, while the connectivity relation among 
components is represented in a second layer 
describing how to stitch components at non-manifold joints.
The resulting  data structure is scalable since all information about
non-manifold features for an object are represented in the upper-layer 
which becomes void in the manifold case. Moreover, the data 
structure is space-efficient and allows navigating in a non-manifold
$d$-complex. In this chapter, we show that we are able to answer queries 
on adjacency and incidence relations efficiently. Please note that from now on we will switch from $d$ to $h$ for the letter used to give the dimension of the complex. This is to stress that, in the applications we have in mind, the dimension is something known and fixed. Even if we have developed many dimension independent results (i.e. for any $h$) still we do not expect to see applications that uses these results and one day are used for $h=2$ (i.e. surfaces) and the next day turn to handle volumetric data i.e. $h=3$. On the other hand, previous material was application independent.

This chapter is organized as a self contained unit. Therefore, in Section
\ref{sec:dsbck}, we introduce some background notions that are used just
in this chapter.  In particular there is a quite lengthy section on face number relations that will be used to assess complexity of our data structure. The reader not interested in technical details of the proofs is advised to skip this.
Next, in Section \ref{sec:dsdup}, we introduce
basic supporting data structures, like: lists, sets etc., that will be used
for the development of our algorithms and for the definition of
the two layer data structure. 

After these two background sections,
in Section \ref{sec:ewrep}, we will introduce the data structure
that will be used  to encode the decomposition components. 
This data structure is defined and analyzed in this section. 
In particular we give algorithms to build and traverse
a decomposition component and evaluate time and space requirements for
these tasks.

Next, in Section \ref{sec:layertwo}
we introduce the second layer data structure.
In particular, in Section
\ref{sec:globewds}, we will introduce a data structure to encode,
in a unique framework, all the components of the decomposition.
$\canon{\AComp}$.

In Sections \ref{sec:sigma} and  \ref{sec:snmnm} we introduce data structures used to encode
non-manifold features. This is the second layer of our data structure for
non-manifolds.   We give algorithms to build this second layer and  
we evaluate the time needed to build this data structure from the 
output of the decomposition algorithm. In the second part of this section we 
develop algorithms to  extract topological relations in the original complex
$\AComp$.

In section \ref{sec:summ} we resume the definitions for the two layer
data structure and present global formulas that gives time and space 
requirements of our two layer data structure.
Next we compare these 
requirements against time and space requirements of
most relevant proposals for non-manifold modeling.

\section{Background}
\label{sec:dsbck}
In this section we report some background notions that are necessary
for the development of this chapter. In particular in sub-section
\ref{sec:toporel}, we introduce basic  notations for
topological relations. In sub-section \ref{sec:facenum}
we report known results on the 
number of faces in particular simplicial complexes.

\subsection{Topological Relations}
\label{sec:toporel}
In the following we will evaluate the effectiveness of our two layers 
representation by considering the complexity for retrieval of basic topological 
relations between simplices. 
Therefore we first introduce  basic  notations for
topological relations we want to compute.
Next we will introduce some standard naming used in modeling for
certain topological relations that are usually called the TV,VT and TT
relations.

\subsubsection{The $\trel{nm}$ relation}
For a given $n$-simplex $\gamma\in\AComp$ in a $h$-complex $\AComp$, for
any $n<m\le h$, we will define the retrieval function 
$\trel{nm}(\gamma)$ as the intersection between the star of  $\gamma$ and the
the set of simplices of order $m$  
of $\AComp$  
i.e. $\trel{nm}(\gamma)=\star{\gamma}\cap \AComp^{[m]}$
(we recall that $\AComp^{[m]}$ denotes 
the set of simplices of $\AComp$ of order $m$). 
We can extend the definition of $\trel{nm}(\gamma)$ to the case $n>m$ as
$\trel{nm}(\gamma)=2^\gamma\cap \AComp^{[m]}$ where $2^\gamma$ denotes the
set of  parts of $\gamma$.
Finally  we will define for $n>0$ the set $\trel{nn}(\gamma)$ as  the set of 
$n$-simplices in $\AComp$ that are  $(n-1)$-adjacent to  $\gamma$.
For $n=0$ we will define $\trel{00}(v)$ as the set of vertices $w$ s.t.
$\{v,w\}$ is a simplex of $\AComp$.
Both sets {$\trel{nn}$} and {$\trel{00}$} can be defined using
sets {$\trel{nm}$} for $n<m$ being:
\begin{eqnarray}
\label{eq:s00}
\trel{00}(v)&=&\cup_{e\in \trel{01}(v)}\{e-\{v\}\}\\
\label{eq:snn}
\trel{nn}(\gamma)&=&\cup_{v\in\gamma}{\trel{(n-1)n}(\gamma-\{v\})}
\end{eqnarray}
When needed we will emphasize the fact that the face relation 
is relative to a certain complex $\AComp$ by writing 
$\trel{nm}(\gamma)$ as $\trel{nm}(\gamma,\AComp)$. 
We note that the set union  in formula \ref{eq:snn}  is a union
between disjoint sets. Thus, the computation of $\trel{nn}(\gamma)$ 
requires the computation of 
{$\trel{(n-1)n}(\gamma-\{v\})$} for all {$v\in\gamma$}.
Similarly, all edges of the form $\{e-\{v\}\}$ in formula \ref{eq:s00}
are distinct. Thus, the computation of 
{$\trel{00}(v)$} reduces to the computation of {$\trel{01}(v)$}.
In the following we will assume that $\trel{nn}$, for $n\ge 0$, is computed
using formulas \ref{eq:s00} and \ref{eq:snn}.
Therefore, we do not exhibit a specific 
algorithm for the computation of $\trel{nn}$.
According to this assumption the time complexity for the computation 
of {$\trel{00}(v)$} will be the time complexity for
the computation of {$\trel{01}(v)$}.
Similarly, the time complexity for the computation 
of {$\trel{nn}(\gamma)$} will be the sum of the
time complexity for the computation of all 
{$\trel{(n-1)n}(\gamma-\{v\})$} for all {$v\in\gamma$}.

Finally we note that
for  $n>m$ the computation of   $\trel{nm}(\gamma)$ is obvious since 
it  reduces to the generation of all subsets of $\gamma$ with $m$ elements. 
Thus in the following we will only consider the computation of
$\trel{nm}$ for $n\le m$.
For $n\neq m$ the set $\trel{nm}(\gamma)$ is the set of $m$-simplices that 
are in a  incidence {\em relation} with  $\gamma$ 
(see Section \ref{par:linkdef}). 
Therefore the set   $\trel{nm}$ is usually called a {\em relation}.
In particular 
relation $\trel{nm}$ is called a {\em boundary relation} if $n>m$
a {\em co-boundary relation} if $n<m$ and an
{\em adjacency relation} if $n=m$
Boundary and coboundary relations toghether are called
{\em incidence relations}.
As already noted above, in the following we will only consider 
those relation we called coboundary relations.

{
\subsubsection{The TV. VT and TT relations}
\label{sec:tvtt}
All topological relations between simplices in an \asc\ are
captured by the set of $\trel{nm}$ relations. Nevertheless, 
a few alternative relations can be defined,
In particular we will consider TV, VT and TT relations.
The TV relation, probably, is the most elementary representation for an
\asc.

For an abstract simplicial complex $\AComp$  
the most obvious representation
is given by the triple $(V,\Theta,\sigma_0)$
where: $V$ is the set of vertices in $\AComp$; $\Theta$ is the set of 
top simplices in $\AComp$ (considered here as atomic objects) and 
$\sigma_0$ is the function
$\funct{\sigma_0}{\Theta}{2^V}$, with $2^V$ being the set of parts of $V$.

The function $\sigma_0$ is usually referred 
to as the TV-relation. Similarly the triple $(V,\Theta,\sigma_0)$
will be referred to as the $TV$ representation. 
A TV representation $(V,\Theta,\sigma_0)$ defines the 
simplicial complex with vertices in $V$ given by the set of simplices 
$\cup_{\theta\in \Theta}\{\sigma_0(\theta)\}$.   
Two TV representations are
equivalent if there are  bijections between respective $V$s and $\Theta$s that
commutes with $\sigma_0$. It is easy to see that equivalent TV 
representations defines isomorphic simplicial complexes.
The function $\sigma_0$ can be defined in term of the previously defined
topologic
relations as $\sigma_0(t)=\trel{h0}(t)$, where $h$ is the dimension of $t$.

A dual representation is the VT representation
given by the triple $(V,\Theta,\sigma_t)$ where
$\sigma_t$ is the function
$\funct{\sigma_t}{V}{2^{\Theta}}$ such that
$\sigma_t(v)$
is  given by the set of top simplices (of any dimension) incident in $v$.
Thus we have $\sigma_t(v)=\cup_{0\le h\le d}\trel{0h}(v)$.
It is easy to see that $\sigma_0(t)=\{v|t\in\sigma_t(v)\}$.
Since we can derive a TV representation from a VT representation, we have
that the VT representation is  unique up to isomorphism and is
non-ambiguous, too.

For a  regular  abstract simplicial $h$-complex $\AComp$  we 
can define another topological relation we called the  {\em TT relation}.
We define the TT relation $\Adj$ as the relation $\Adj\subset \Theta^2$
defined by the condition: $\theta_1 \Adj \theta_2$ iff $\theta_1$ and 
$\theta_2$ are {\em adjacent}. 
Note that relation $\Adj$ is always symmetric.
For a given regular $d$-complex $\AComp$ its TT relation is uniquely
defined. Unfortunately given a TT relation ${\Adj}$ there is not a 
unique regular  $d$-complex whose  TT relation is ${\Adj}$. 
This is true for any $d\ge 2$.

}

\subsection{Face number relations}
\label{sec:facenum}
We define 
the $m$-th \emas{face}{number}
for an \asc\ $\AComp$  (denoted by $f_m(\AComp)$ or $f_m$) as the  
number of $m$-simplices in $\AComp$ i.e. $f_m(\AComp)={|\AComp^{[m]}|}$. 
(see Section \ref{sec:asc} for the definition of {$\AComp^{[m]}$}).
In this section we report some results for face numbers in arbitrary dimension. These are probably not relevant per se but they will be used to assess complexity of algorithms in the forthcoming sections. 
Many theorems gives relations between face numbers being the 
Kruskal-Katona theorem the most known result in this field 
(see \cite{Bil97} for a survey).
We start recalling some simple relations for face numbers in manifold
surfaces.

\subsubsection{Manifold Surfaces}
\label{sec:facenummani}
For a closed manifold surface $\AComp$ we have 
\begin{equation}
\label{eq:pmsurf}
\frac{3}{2}f_2(\AComp)=f_{1}(\AComp)
\end{equation}
while, for a connected manifold surface with boundary, we have that:
\begin{equation}
\label{eq:msurf}
{3}f_2(\AComp)\le 2f_{1}(\AComp)-3
\end{equation}

\subsubsection{Simplicial $h$-complexes embeddable in $\real^h$}
\label{sec:facenumimbed}
Next we will show that 
there are linear inequalities between face numbers $f_h$, $f_{h-1}$ and
$f_{h-2}$ in the star of a vertex of an
$h$-complex that can be 
geometrically embedded into $\real^h$ as a compact
geometric simplicial complex.

\begin{property}
\label{pro:pseudo}
Let $\AComp$ be an  $h$-complex geometrically 
embeddable into $\real^h$ 
then $(h+1)f_h=2f_{h-1}$ for closed complexes and 
$(h+1)f_h\le 2f_{h-1}-(h+1)$ for complexes with non empty boundary.
\end{property}
\begin{proof}
To prove this property  
we proceed as follows.
Let $\CComp\in\carrier{\AComp}$ be an embedding of $\AComp$.
We first show that at most two geometric  $h$-simplices in $\CComp$
can  share a geometric $(h-1)$-face. 
The proof can be done by contradiction assuming that 
three geometric $h$-simplices $\theta_1$, $\theta_2$ and $\theta_3$
share an $(h-1)$-simplex $\gamma$. 
In this case the supporting hyperplane for $\gamma$ divides $\real^h$ in two
disjoint half-spaces. If we assume that $\theta_1$ and $\theta_2$ fall in the
same half-space then these two geometric simplices must share interior points 
and this is not allowed in a geometric simplicial complex.   
Thus we have $(h+1)f_h=2f_{h-1}$ for closed complexes and 
$(h+1)f_h\le 2f_{h-1}-(h+1)$ for complexes with non empty boundary. This is because the smallest boundary has (h+1) (h-1)-simplices. 
This completes the proof.
\end{proof}
Next we show that, in general, for $h\ge 3$ there is a linear inequality 
between face numbers $f_h(\xstr{\AComp}{\gamma})$ and 
$f_{h-2}(\xstr{\AComp}{\gamma})$ in the {\em open} star of a 
$(h-3)$-simplex $\gamma$.
This holds for whenever the simplicial complex
$\AComp$ can be imbedded into $\real^h$ as compact simplicial complex

\begin{property}
\label{pro:euler}
For $h\ge 3$ let $\AComp$ be a regular  $h$-complex geometrically 
embeddable into $\real^h$ as a compact geometric simplicial complex 
and  let $\gamma$ be an $(h-3)$-simplex in $\AComp$.
In this situation the link of  {$\xlk{\AComp}{\gamma}$} is a 
triangulation embeddable in the $2$-sphere and,  
for the open star of {$\gamma$} (i.e. $\xstr{\AComp}{\gamma}$), we have
$f_h(\xstr{\AComp}{\gamma})\le 2(f_{h-2}(\xstr{\AComp}{\gamma})-2)$.
\end{property}
\begin{proof}
We first note that, in a regular $h$-complex, the link of an $(h-3)$-simplex
is a regular $2$-complex. We simply have to prove
that we can imbed {$\xlk{\AComp}{\gamma}$} in a $2$-sphere.
Let $\CComp$ be a geometric simplicial complex in $\real^h$ that is the 
compact geometric  realization of $\AComp$.
This exists by hypothesis.
It is easy to see that a  geometric  realization of the closed star
${\xclstar{\AComp}{\gamma}}$ can be a  compact geometric simplicial 
subcomplex of $\CComp$. Let us call $S$ this subcomplex.
Obviously $S$ is embeddable in $\real^h$ and its
boundary $\bnd{S}$ is a geometric 
realization for {$\xlk{\AComp}{\gamma}$}.

With this situation in mind we first treat the case for $h=3$. 
In this case the geometric realization of the  $(h-3)$-simplex  $\gamma$
must be a single geometric vertex. Let  $P_0$ be this point. 
It is easy to see that we can find a radius $r$ such that the
complex $S$ fall outside the 2-sphere $\Sigma$ 
of radius $r$ centered in $P_0$. 
(i.e. $\norm{P-P_0}>r$ for all $P\in \bnd{S}$).
It is easy to see that we can map geometric vertex $P\in\bnd{S}$
onto the $2$-sphere $\Sigma$ through the continuous mapping
$\send{P}{P_0+(P-P_0)/{\norm{P-P_0}}}$. This mapping,
restricted to $\bnd{S}$,
is invertible with continuous inverse. This proves that $\Sigma$ and $\bnd{S}$
are homeomorphic and thus for $h=3$ we have proved that we can 
imbed {$\xlk{\AComp}{\gamma}$} in a $2$-sphere.

For $h>3$ let {$\{P_0,\ldots,P_{h-3}\}$} the geometric simplex that is the 
geometric realization of $\gamma$ in $\real^h$. Geometric
Vertices {$\{P_0,\ldots,P_{h-3}\}$} must be affinely independent.
This implies that  in the euclidean vector space $\real^h$
(see \cite{Lip74} Chapter 4 for basic definitions in Linear Algebra) 
the $(h-3)$ vectors
$\{(P_i-P_0)|i=1,\ldots,(h-3)\}$ are linearly independent and they generate a 
$(h-3)$-dimensional subspace of $\real^h$. Let {$W^{\gamma}$} be this
subspace and let {$W^{\gamma}_\bot$} the $3$-dimensional space such that
the direct sum ${W^{\gamma}}\oplus{W^{\gamma}_\bot}$ gives $\real^h$
(see for instance Ex. 5.49 \cite{Lip74}).
Next we consider the \asm\ $g$ that maps all and alone the Vertices of 
$\gamma$ into a single vertex $v$. This \asm\ induces a  
geometric simplicial map defined as
$|g|=[\send{P_i}{P_0}|i=1,h-3]$ between $\CComp$ and the geometric realization
of $g(\AComp)$.
In particular we have that  $|g|(S)$ 
must be the geometric realization of $g({\xclstar{\AComp}{\gamma}})$.
Since $g(\gamma)=\{v\}$ we have that $g({\xclstar{\AComp}{\gamma}})$ is 
the cone from $v$ to {$\xlk{\AComp}{\gamma}$}.
Next we note that $|g|({W^{\gamma}})=\{0\}$  and 
$|g|({W^{\gamma}_\bot})={W^{\gamma}_\bot}$.
Thus we have that $|g|(\real^h)={W^{\gamma}_\bot}$ and
therefore ${W^{\gamma}_\bot}$ must include  $|g|(S)$. 
Thus we have a  geometric realization for 
the cone from $v$ to {$\xlk{\AComp}{\gamma}$} in the 
euclidean  subspace ${W^{\gamma}_\bot}$.
This is a $3$-dimensional subspace of $\real^h$.
Thus we have a  geometric realization for 
the cone from $v$ to {$\xlk{\AComp}{\gamma}$} in $\real^3$. 
Reasoning as in the case for $h=3$ we prove that
{$\xlk{\AComp}{\gamma}$} can be embedded in a $2$-sphere.
This proves the first part of the thesis.

The second part comes easily from the Euler formula. We have just proven 
that the  link of each $(h-3)$-simplex $\gamma$ is a triangulation that
can be imbedded in a $2$-sphere. If the link is homeomorphic to the $2$-sphere
we have
{$f_2-f_1+f_0=2$} and $3f_2=2f_1$. This yields $f_2=2(f_0-2)$.
When the link is not homeomorphic to the $2$-sphere we can write
{$f^{k}_2-f^{k}_1+f^{k}_0=1$} and  {$\frac{3}{2}f^{k}_2\le f^{k}_{1}-1$}
being $f^{k}_i$ the face numbers for the $k$-th connected component of the 
link.
This yields {$f^{k}_2\le 2(f^{k}_0-2)$} in the $k$-connected connected 
component of the link. Summing over the $c$ different connected components
of the link we get {$f_2\le 2(f_0-2c)$}.
Thus we can write {$f_2\le 2(f_0-2c)\le 2(f_0-2)$}.
.
Now we note that to a $j$-face in  {$\xlk{\AComp}{\gamma}$} correspond a 
$h-2+j$ face in  the open star {$\xstr{\AComp}{\gamma}$}
and thus we rewrite the previous formula for {$\xlk{\AComp}{\gamma}$} as 
$f_h(\xstr{\AComp}{\gamma})
\le 2(f_{h-2}(\xstr{\AComp}{\gamma})-2)$.
This completes the proof.
\end{proof}
Note that summing all over the $(h-3)$ simplices in a manifold $h$-\asc\  $\AComp$   
we have the following property:
\begin{property}
\label{pro:mainfface}
Let $\AComp$ a closed manifold $h$-complex imbeddable in $\real^h$ in this situation
$$
{\binomial{h+1}{h-2}}f_h(\AComp)-
{\binomial{h}{h-2}}f_{h-1}(\AComp)+
{\binomial{h-1}{h-2}}f_{h-2}(\AComp)=
2f_{h-3}
$$
\end{property}
\begin{proof}
In a closed $h$-manifold complex $\AComp$ we have  every link  
$\xlk{\AComp}{\gamma}$ of an $(h-3)$-simplex $\gamma$ is homeomorphic to a sphere and by Euler formula applied to that link we have
{$f_2-f_1+f_0=2$}.  
Now there are, respectively,  a 2-simplex or a 1-simplex or a 0-simplex in $\xlk{\AComp}{\gamma}$ for, respectively, every
top $h$-simplex or  $(h-1)$-simplex or $(h-2)$-simplex in $\xstr{\AComp}{\gamma}$, because $\gamma$ is a $(h-3)$-simplex. Thus summing the relation
{$f_2-f_1+f_0=2$} 
all over the links of all $(h-3)$ simplices in a manifold \asc\  $\AComp$ 
we mention each
top simplex in $\AComp$ once for each $h-3$ face it has. Thus summing {$f_2-f_1+f_0=2$} all over the complex we obtain something like 
{$\ldots=2f_{h-3}$}.
Let us see what to put instead of dots.
On the other hand an 
$h$-simplex $\theta_h$ is mentioned in this sum as many times as $\theta_h$ enters in a star of some of its $(h-3)$-faces.
Thus we have to count the number of $(h-3)$-faces of $\theta_h$ and put this to multiply $f_{h}$. The number of  $h-3$ faces of an $h$-simplex is given by a choice od $h-2$ vertices among its $h+1$ vertices. So an $h$ face is mentioned ${\binomial{h+1}{h-2}}$ times. Thus summing {$f_2-f_1+f_0=2$} all over the complex we obtain something like 
{${\binomial{h+1}{h-2}}f_h+\ldots=2f_{h-3}$}.
Similarly an
$(h-1)$-simplex $\theta_{h-1}$ is mentioned in this sum as many times as $\theta_{h-1}$ enters in a star of some of its $(h-3)$-faces.
Thus we have to count the number of $(h-3)$-faces of $\theta_{h-1}$ and put this to multiply $f_{h-1}$. The number of  $h-3$ faces of an ${h-1}$-simplex is given by a choice of $h-2$ vertices among its $h$ vertices. So an $h-1$ face is mentioned ${\binomial{h}{h-2}}$ times. Thus summing {$f_2-f_1+f_0=2$} all over the complex we obtain something like 
{${\binomial{h+1}{h-2}}f_h-{\binomial{h}{h-2}}f_{h-1}\ldots=2f_{h-3}$}
Similarly 
we mention each $(h-2)$-simplex ${\binomial{h-1}{h-2}}$ times.
Thus we obtain the formula of the thesis.
\end{proof}

\subsubsection{Non-linear relations between $f_4$ and $f_0$ 
for stars in $4$-balls}
\label{sec:fourdim}
In the following we will be interested in
evaluating the complexity of traversal of our 
two  layer data structure. This require some understanding of how face numbers are related in generic $h$-complexes. These notions will be specialized to initial quasi-manifolds. 
The results shows that, in higher dimension, complexity must  be, at least, quadratic because the complexity of the structure of the complex could be quadratic, or more.

In the previous section we have found linear relations 
between face numbers $f_h$, $f_{h-1}$ and
$f_{h-2}$ in the star of a vertex of an
$h$-complex that can be 
geometrically embedded into $\real^h$.
For $4$-complexes this ensure a linear relation among
$f_4$, $f_3$ and $f_2$.
We note that, for $h=4$, the Properties \ref{pro:pseudo}
and \ref{pro:euler} do not ensure a linear inequality between
the face numbers $f_4$ and $f_1$. Similarly 
no linear relation could be found, in general,  between
face numbers $f_4$ and $f_0$.
in the star of a vertex of a $4$-complex. 
In this sub-section we will study  what happens to face numbers
in stars of $4$-complexes while,
in the following section, we will  investigate face number relations for
complexes that cannot be embedded in $\real^h$.

In particular, we will exhibit a manifold $4$-complex $\AComp$ with a vertex $v$ s.t. $f_4(\xclstar{\AComp}{v})=f_0(f_0-3)/2$ 
with $f_0=f_0(\xclstar{\AComp}{v})$. Therefore complexity of this relation could be quadratic.

In fact
it is possible to build a triangulation of the  $4$-ball where there is
at least a  vertex $w$ for which 
$f_4(\xclstar{\AComp}{w})=f_0(f_0-3)/2$ 
where $f_0=f_0(\xclstar{\AComp}{w})$ is the number of vertices in
$\xclstar{\AComp}{w}$ that are distinct form $w$.
To build such a complex we just have to take as $4$-complex the cone 
from an arbitrary vertex $w$ to the boundary of the {\em Cyclic Polytope} 
{$C_4(f_0)$}.
To show this fact we introduce \ems{polytopes}.
\begin{definition}[Polytope and Cyclic Polytope] \label{def:polyt}
A $k$-\emd{polytope}  with $n$ Vertices  (denoted as $\Pi(n)$) 
is the convex-hull of a set of $n$ points in $\real^k$. 
{The \emd{cyclic} $k$-polytope with $n$-points 
	(denoted by {$C_k(n)$}) is a polytope in $\real^k$ that has the property of 
	having the maximum face number $f_i$ for all $0\le i\lt k$.
}
\end{definition}
Following the conventions
used in this thesis we denote with $f_0$ the number of 
points $n$ and therefore  we will use $\Pi(f_0)$ instead of $\Pi(n)$.

The cyclic $k$-polytope has the maximum face number $f_i$ for all $0\le i<k$.  
Any other $k$-polytope has the face number $f_i$ smaller than the 
corresponding face number of the cyclic polytope.
The so called 
{\em Upper Bound Theorem}
by McMullen \cite{MS71} gives a sets of formulas to compute these upper bounds for face numbers.
(See \cite{Sta96} for  an extension of this result to spheres and recently to {\em homology manifolds} \cite{Nov00} and \cite{Bil97} 
Pg. 298-300).

The explicit formula for the computation of all the face numbers 
$f_i$ of the cyclic polytope vertices is quite complex  
(See \cite{Gru67} Sections 4.7.3. and 9.6.1).
We just report the specialization of this  formula for  $i=d-1$:
\begin{equation}
\label{eq:fd1}
f_{d-1}=
\binomial{f_0-\ceiling{d/2}}{\floor{d/2}}+
\binomial{f_0-\floor{d/2}-1}{\ceiling{d/2}-1}
\end{equation}
In particular for $C_4(f_0)$
we have  $f_3=f_0(f_0-3)/2$ (simply put $d=4$ in equation \ref{eq:fd1}). Since {$C_4(f_0)$}
is a polytope we have that  {$\bnd{C_4(f_0)}$} is homeomorphic to the 
$3$-sphere.
Thus, for all the classes of $3$-complexes that includes the $3$-sphere
(e.g. manifolds, pseudomanifold, \iqm, etc.) we can say that there cannot be, 
in general, a linear relation between $f_0$ and $f_3$.

\begin{property}
\label{pro:nonopt}
There exist a triangulation $TB^4$ of the $4$-ball imbeddable in 
$\real^4$ and a vertex $w$ such that 
$f_4(\xclstar{TB^4}{w})=f_0(f_0-3)/2$ with $f_0$ the number of 
vertices in $\xclstar{TB^4}{w}$ distinct from $w$.
\end{property}
\begin{proof}
We take as $TB^4$ the cone from $w$ to the boundary of the cyclic 
$4$-polytope {$C_4(f_0)$}.
Thus  taking the cone from $w$ to a tetrahedron in the boundary of
{$C_4(f_0)$} we obtain a $4$-simplex in $\xclstar{TB^4}{w}$.
Thus, $f_4(\xclstar{TB^4}{w})$ is the number of tetrahedra in the boundary 
of {$C_4(f_0)$}, This number is  given by
$f_3=f_0(f_0-3)/2$ where $f_0$ is the number of
vertices in $\xclstar{TB^4}{w}$ distinct from $w$.

Now we just have to prove that what we called $TB^4$ is 
a triangulation of the $4$-ball.
To take this cone we just have to take  $w$ not in  {$C_4(f_0)$}.
Since all $3$-spheres are combinatorially equivalent \cite{Moi52} 
we can also say that {$\bnd{C_4(f_0)}$} is a triangulation of the
$3$-sphere embedded in $\real^4$ that is a combinatorial manifold.
Thus, by Property \ref{pro:ballcone},
the cone $TB^4$ from $w$ to {$\bnd{C_4(f_0)}$} is a triangulation of
the $4$-ball as a combinatorial manifold.
If we choose the geometric realization of $w$ within the the geometric realization  of {$C_4(f_0)$} it is easy to see that the
cone $TB^4$ admits a geometric realization in $\real^4$.
\end{proof}
The above Property generalize to $d$-balls showing that the number  of
top simplices in a vertex star can become quite large.
\begin{property}
\label{pro:s01}
There exist a triangulation $TB^d$ of the $d$-ball embeddable in $\real^d$
such that, for a certain vertex $w$ in $TB^d$,
$f_{d}(\xclstar{TB^d}{w})=\ordcomp{f_0^{\floor{d/2}}}$ 
with $f_0$ is the number of vertices in $\xclstar{TB^d}{w}$.
\end{property}
\begin{proof}
Using the explicit formula for face numbers in cyclic polytopes 
${C_{d}(f_0)}$  (see Formula \ref{eq:fd1})
and the approximation suggested by the Stirling
formula \[\lim_{n\rightarrow\infty}{\frac{\factorial{n}}{n^ne^{-n}\sqrt{2\pi\ n}}}=1\] (see, for instance, \cite{KorKor68} \S 21.4-2)
one easily  shows that
the number of top simplices in the cyclic polytope
${C_{d}(f_0)}$  is given by 
{$f_{d-1}=\ordcomp{f_0^{\floor{d/2}}}$} (see \ref{sec:complexity} for the definition of the complexity order $\ordcomp{n}$).
Therefore,  with arguments similar  to those used in the
proof of Property  \ref{pro:nonopt}, one builds 
$TB^d$ by selecting a vertex $w$ within the geometric realization of 
{${C_{d}(f_0)}$} in $\real^d$ and then takes the cone from $w$ to 
{$\bnd{C_{d}(f_0)}$}. The polytope ${C_{d}(f_0)}$ has a boundary that is 
a combinatorial $(d-1)$-sphere. Thus, the resulting cone is a triangulation  
of the combinatorial $d$-ball embeddable in $\real^d$.
Considering the star of vertex $w$ we have
$f_{d}(\xclstar{TB^d}{w})=\ordcomp{f_0^{\floor{d/2}}}$
with $f_0$ the number of vertices in $\xclstar{TB^d}{w}$.
In fact a $d$-simplex in this star correspond to a $(d-1)$-simplex in 
the cyclic polytope {${C_{d}(f_0)}$}.
\end{proof}

Following the ideas in the proofs above 
it is easy to build a $d$-ball $TB^d$, for $d$ sufficiently high, in
which there exist an $n$-simplex $\gamma$ for which   
$f_d(\xclstar{TB^d}{\gamma})=f_3=f_0(f_0-3)/2$ with $f_0$ the number
of vertices in $\xclstar{TB^d}{\gamma}$ that are not in $\gamma$ .
The property below gives a formal statement of this fact.
\begin{property}
\label{pro:nmineff}
For any natural $n\ge 0$,  there exist a combinatorial $d$-ball $TB^d$ that
can be embedded in $\real^d$, for $d=4+n$, 
and an $n$-simplex $\gamma$ such that 
$f_d(\xclstar{TB^d}{\gamma})=f_3=f_0(f_0-3)/2$ with $f_0$ the number
of vertices in $\xclstar{TB^d}{\gamma}$ that are not in $\gamma$.
\end{property}
\begin{proof}
To build such a triangulation of $B^d$ we consider objects in 
the euclidean space $\real^d$ for $d=n+4$.
We first take the geometric realization of
the cyclic polytope  {${{C_{4}(f_0)}}$} in $\real^d$.
Next we take a geometric realization of the  standard $n$-simplex $\gamma$ in
$\real^d$
that fits in the interior of the geometric realization of the cyclic 
polytope  {${{C_{4}(f_0)}}$}.
Then, the complex $TB^d$  is obtained as  the cone from $\gamma$ to 
{${\bnd{C_{4}(f_0)}}$}. We have that  {${\bnd{C_{4}(f_0)}}$} is 
combinatorially equivalent to the $3$-sphere and thus the
cone from the $n$-simplex $\gamma$ to the 
boundary {${\bnd{C_{4}(f_0)}}$} is combinatorially equivalent to
the $d$-ball with $d=n+3+1$ (See Property \ref{pro:ballcone}).
The complex $TB^d$ is  realizable in $\real^{d}$ and is a triangulation
of the $d$-ball as a combinatorial manifold. 
\end{proof}
Thus, also the number of top simplices in an $n$-simplex star 
might become extremely large in a $d$-complex in higer dimension.
The  above property refers to the case $d=n+4$.
However, combining the proof above and that of Property \ref{pro:s01} 
it is easy to build an $h$-complex $TB^h$, for $h$ sufficiently high, in
which there exist a $n$-simplex $\gamma$ for which   
$f_{d}(\xclstar{TB^d}{\gamma})=\ordcomp{f_0^{\floor{d/2}}}$
with $f_0$ is the number of vertices in $\xclstar{TB^d}{\gamma}$.

\subsubsection{Face numbers in stars of non embeddable $3$-manifolds}
\label{sec:nonimbed}
{In the second part of this section we will study the influence
of non embeddability on face numbers for $3$-complexes.
We will find examples of $3$-complexes non-embeddable in $\real^3$
that  contains a vertex $v$ for which there is not a 
linear relation between $f_3(\xclstar{\AComp}{v})$ and
$f_0(\xclstar{\AComp}{v})$.}

The above properties shows that a 
quite large number of $(d-1)$-faces can fit in the
boundary of a cyclic $d$-polytope as $d$ grows. 
This enables us to build $4$-balls where there is a
vertex whose star has a number of top simplices that is non-linear w.r.t.
the number of incident vertices.
Therefore one might suspect that our requirement on 
embeddability in $\real^h$ might be unnecessary when $h$ is low and that
it is not possible to have $3$-complex where the number of tetrahedra in a
vertex star in  non-linear w.r.t. the  number of vertices in the 
star its-self.
This is not the case. In fact similar examples 
pops up already for in $3$-complexes if we drop the requirement
on embeddability in $\real^3$. Indeed if we allow (rather exotic) 
$3$-complexes where a vertex star might have a toroidal surface boundary 
we obtain quite large face numbers with quite few vertices. 
These complexes are neither $3$-balls not $3$-manifolds but they fall in the 
class of \iqm\ $3$-complexes.

This can be shown by providing a 
family of manifold surfaces $\Sigma(g)$, with genus  $g$,
for which do not exist a linear relation between 
$f_0$ and $f_2$. 
If such a family exists, then, with an argument similar to 
that used in the previous property, we can show that, 
in the cone from $w$ to $\Sigma(g)$, 
the number of tetrahedra in the
star of $w$ is  non-linear w.r.t. the  number of vertices in this star.

Indeed the family of {\em minimal triangulations} of 2-manifolds 
provides the required family $\Sigma(g)$.
\begin{property}
\label{pro:mintri}
There exist countably many \iqm\ $3$-complexes
$B(g)$, for any natural $g>0$, not embeddable in $\real^3$, 
s.t. there exist a  vertex $w$ for  which
$f_3(\xclstar{B(g)}{w})=\ordcomp{{f_0}^{2}}$ with $f_0$ the number
of vertices in $\xclstar{B(g)}{w}$ that are distinct from  $w$.
In particular we have:
${{(f_0^2-3f_0+8)}/3}<f_3(\xclstar{B(g)}{w})\le {{f_{0}(f_{0}-1)}/3}$.
\end{property}
\begin{proof}
It  can be shown 
(see \cite{Jun80} Pg. 122) that in a class ${\cal M}$ of
topologically equivalent $2$-manifolds 
there exist an \asc\ with a minimal 
number of triangles given by 
\begin{equation}
\label{eq:mintri}
f_2=2\ceiling{\frac{7+\sqrt{49-24\chi(\cal M)}}{2}}-2\chi(\cal M)
\end{equation}
where $\chi(\cal M)$ is the 
common Euler characteristic for the class of homeomorphic surfaces ${\cal M}$.
This holds for all $2$-manifolds minimal triangulation with the exception 
of the double torus $T_2$, of the Klein's bottle and of
the non orientable surface of genus $3$ whose minimal triangulation 
respectively takes 24, 16 and 20 triangles.

We now prove that, in this family of minimal triangulations of a
$2$-manifold, the face number  $f_2$ is $\ordcomp{f_0^2}$.
This can be proven using the definition of 
$\chi=f_2-f_1+f_0$ and the fact that, for closed $2$-manifolds,
we have that $(3/2)f_2=f_{1}$,
and thus $f_0-(1/2)f_2=\chi$.
By rewriting $\chi$ in Equation  \ref{eq:mintri}
we get
$f_0=\ceiling{\frac{7+\sqrt{49-24\chi}}{2}}$
and therefore must be $f_0\ge {\frac{7+\sqrt{49-24\chi}}{2}}>f_0-1$.  
Solving $f_0\ge {\frac{7+\sqrt{49-24\chi}}{2}}$ with $\chi=f_0-(1/2)f_2$ we obtain  $f_2\le{f_0(f_0-1)}/3$.
Solving ${\frac{7+\sqrt{49-24\chi}}{2}}>f_0-1$
we obtain $f_2>(f_0^2-3f_0+8)/3$
thus we get
\begin{equation}
\label{eq:f2}
{{(f_0^2-3f_0+8)}/3}< f_2\le {{f_0(f_0-1)}/3}
\end{equation}

Thus, to complete the proof we just have to take the minimal triangulation
$\Sigma(g)$
of the surface of genus $g$. 
We recall that for a closed orientable $2$-manifold ${\cal M}$ 
the genus $g$ is given by 
$\chi({\cal M})=2-2g$ while for 
closed non-orientable $2$-manifold we have $\chi({\cal M})=2-g$.
In general we can write these two relations using 
the third  Betti number {$\beta_2$}.
We recall that $\beta_2$ is $0$ for non orientable
surfaces and $1$ for orientable surfaces.
With this definitions we can write $\chi({\cal M})=2-(1+\beta_2)g$.
The number of vertices in
$\Sigma(g)$ can be obtained by reworking Equation  \ref{eq:mintri}
and  using $\chi({\cal M})=2-(1+\beta_2)g$ we get 
$$f_0=\ceiling{\frac{7+\sqrt{1+24(1+\beta_2)g}}{2}}$$
Note that for $f_0>4$ we must have $g>0$.
Taking the $3$-complex $B(g)$ as the cone from a new vertex $w$ to 
$\Sigma(g)$ and reasoning as in the proof of Property \ref{pro:nonopt}
we show that in $B(g)$ we have 
${{(f_0^2-3f_0+8)}/3}<f_3(\xclstar{B(g)}{w})\le {{f_{0}(f_{0}-1)}/3}$
being $f_0$ the number
of vertices in $\xclstar{B(g)}{w}$ that are distinct from  $w$.
In particular we have:
$f_3(\xclstar{B(g)}{w})=\ordcomp{{f_0}^{2}}$.

To complete this proof we just have to show that {$B(g)$} is an \iqm\ and that
it cannot be embedded in $\real^3$.
It is easy  to see that {$\xlk{B(g)}{v}$} is homeomorphic to the  $2$-ball
for $v\neq w$ and for $w$ we know  {$\xlk{B(g)}{w}=\Sigma(g)$}.
Thus, for any vertex $v$ in {$B(g)$} we have that
{$\xlk{B(g)}{v}$} is a $2$-manifold. 
Thus the link of each vertex is $2$-manifold-connected and, therefore, 
the star of each vertex is $3$-manifold-connected.  
Thus {$B(g)$} must be an \iqm.

Finally non embeddability is proven by contradiction.
Let us assume that $B(g)$ is embeddable in $\real^3$. By Property 
\ref{pro:euler} we have that {$\xlk{B(g)}{w}=\Sigma(g)$} 
can be embedded in a $2$-sphere. Thus we can take a pole within a triangle
of $\Sigma(g)$ and project, by stereographic projection, the
$1$-skeleton of {$\Sigma(g)$} as a planar graph.
This graph has a number of faces that is one less the number of
triangles of {$\Sigma(g)$} (i.e. $f_2-1$).
Using the Euler formula we get
$1=(f_2-1)-f_1+f_0$. Thus we have that $\chi(\Sigma(g))=2$.
This implies $g=0$, that,  in general, is false. 
\end{proof}

\subsection{Complexity}
\label{sec:complexity}
We recall briefly the standard  notation for complexity. Let $T(n)$ be a 
function  expressing the measure of a certain quantity $X$ 
w.r.t. a certain parameter $n$ (e.g. $T(n)$ could be the time spent
for computation w.r.t. the size of the input $n$  or $T(n)$ could be the number of bits needed 
to store a certain triangulation  w.r.t. the  number of triangles, etc.).
We will say that the quantity $X$ is in $\ordcomp{f(n)}$ \iff\ there
are three strictly positive constants $c_1$, $c_2$ and $n_0$
such that $c_1 f(n)\le T(n)\le c_2 f(n)$ for all $n>n_0$.

Similarly we will say that the quantity $X$ is in $\lesscomp{f(n)}$ 
\iff\ there are two strictly positive constants $c$  and $n_0$
such that $T(n)\le c f(n)$ for all $n>n_0$.

Finally we recall we say that the quantity $X$ is in $\morecomp{f(n)}$ 
\iff\ there are two strictly positive constants $c$  and $n_0$
such that $c f(n)\le T(n)$ for all $n>n_0$.
Note that $X$ is in  $\ordcomp{f(n)}$ \iff\ $X$ is both in $\lesscomp{f(n)}$
and $\morecomp{f(n)}$.

\section{Supporting Data Structures}
\label{sec:dsdup}
\subsection{Introduction}
In  this section we briefly introduce a number of classic data structures.
For each data structure we will specify the primitives  that will be
used in this thesis to access the data structure, 
the sintax used to denote these primitives and the 
time complexity of the standard algorithm used to implement them.
We will assume to have a pseudocode language for data type definitions that
supports some type constructors we will introduce in the following.

\subsection{Lists \label{sec:lists}}
In the following algorithms we assume to have a pseudocode language
that supports data definitions through a set of type constructors.
For lists we assume a type  constructor of the form 
{{\bf list of} $T$}, being  $T$ a generic  type that do not have to
satisfy any particular constraint.
We will denote with $\brk{\mbox{}}$ the empty list. If
$e$ is an object of type $T$ we will denote with $\brk{e}$ the 
list made up of a single element.
We assume to have an implementation for  the usual operations on lists. 
In particular, if $l$ is an object of type {{\bf list of} $T$} 
we will have that $l.{\rm TOP}$ returns the 
first element in the list and $l.{\rm POP}$ returns a pointer to 
the list without the
first element. We will use the notation $\brk{e}+l$ 
to denote the list obtained appending $e$
at the beginning  of the list $l$. 
We assume that all the above operations on lists can be done in constant time. 
For two objects $l_1$ and $l_2$ of  list type  we assume
to have an operator $l_1+l_2$ to concatenate lists. This can be done in
$\ordcomp{|l_1|}$.

We assume  that is possible to have a method to retrieve,
in constant time, one after one,  all elements $v$  in a list $l$. 
Thus, we will use an expression of the form ''$v$ {\bf in} $l$'' as 
control predicate for loops.
When list $l$ is used for such an iteration  we assume that the list
is not modified and we assume that there exists
a method, denoted by $l.CircularNext$,
that can be used to fetch
the element after the one currently
returned for the iteration. If the last element is currently returned for 
iteration then $l.CircularNext$ returns the first element in the list $l$.
Obviously we expect to do this in constant time.
Finally we assume to have an explicit type 
casting operation ${\bf list}(s)$ that convert a set $s$ into the list
of its elements.  
This latter operation can be done in time proportional to the size of the 
output.

Lists will be used for local variables in algorithms  and are not used within
the topological data structure. Thus the space requirements for this data
structure is not analyzed.

\subsection{Bit and Bit Vectors \label{sec:bitv}}
Next we assume to have the option to define vectors with a huge
number of bits. In fact we will use bit vectors with one bit for 
each top simplex in the \asc\ under analysis.
We assume that a fresh copy of the bit vector $\Theta$ is allocated by the 
declaration  \mbox{{\bf var } $\Theta$: {\bf array} $[Min,\ldots,Max]$
{\bf of  bit}}. This 
creates {$Max-Min+1$} bits initialized to zero.  
We assume to have three operations on bits.
The three operations are:
the operation ${\rm SET}$, that sets the bit;
the operation ${\rm RESET}$, that resets the bit and
the operation ${\rm TEST}$ that returns a boolean value 
that will be {\bf true}
\iff\ the bit is set. 
If $B$ is a bit we also use $B$ as a variable that can hold two values $0$
and $1$. Obviously we will say that the bit is set when it contains 1. 
Finally we assume that we can use bit vectors wherever a positive integer
is needed. In this case we assume an implicit cast from
bit vectors to positive integers.
Bits and bit vectors will be used for temporary marking and are not 
used within
the topological data structure. Thus the space requirements for this data
structure are not analyzed.

\subsection{Sets} 
\label{sec:set}
We assume that in our language for  data type declaration 
we can write: 
{$\mbox{\bf set of } Domain$} being $Domain$ a data type whose 
values must form a totally ordered set.
We do not assume anything else  about sets elements. Thus the time
complexity for sets and maps operations is given, in this section,
by giving the number of 
comparisons between two elements in the $Domain$ set. 
To get the real time complexity for a particular set operation we 
must multiply the number of comparisons  by the time complexity
for a single comparison between two elements in $Domain$.

With this assumption about ordering of the type $Domain$ 
we have that sets  can be implemented with binary search trees (BST)
maintained as an AVL tree (see \cite{Mou42093} Pg. 15-17 for a careful
presentation of 
AVL tree primitives. See also \cite{Wei97} Ch. 4).
Sets will be used for local variables in algorithm and are not used within
the topologic data structure. Thus the space requirements for these
AVL trees  is not taken into account here.

We mention the fact that the set data structure is offered as an 
''off the shelf'' component in the STL (Standard Template Library)
\cite{Pla00,STL} that is part of the  C++ Standard Library and in the 
commercial library LEDA \cite{Nah90}.
What will be the implementation technique used in these packages
we have checked that the time complexity listed in this section 
are guaranteed by these implementations. In particular we report the
time complexity guaranteed by the STL library and add 
the complexity provided by LEDA when this is different.

If $v$ is an object in set element $Domain$ and $S_1$ and $S_2$ are 
object of type {$\mbox{\bf set of } Domain$} we assume that are 
legal expressions of set  type $S_1\cup S_2$ and $S_1\cap S_2$.
Both union and intersection can be done  in STL with  
{$\lesscomp{|S_1|+|S_1|}$}
comparisons between sets elements.
In LEDA union takes  {$\lesscomp{|S_2|\log{|S_1|+|S_1|}}$} and
intersection takes {$\lesscomp{|S_1|\log{|S_1|+|S_1|}}$} comparisons.

Set membership, denoted by $v\in S$, can be tested in $\lesscomp{\log{|S|}}$
comparisons.   
We can compute the subset of element of $S_1$ that are not
in $S_2$, denoted by $S_1-S_2$, with {$\lesscomp{|S_2|+|S_1|}$} comparisons. 
This complexity is offered by STL while LEDA uses a different algorithm that
takes {$\lesscomp{|S_2|\log{|S_1|+|S_1|}}$} comparisons.

We can test in STL whether set $S_1$ is included in $S_2$, 
a test denoted by $S_1\subset S_2$, with {$\lesscomp{|S_1|+|S_2|}$}
comparisons.
However we note that this can be done by testing all elements in $S_1$ and
this takes {$\lesscomp{|S_1|\log{|S_2|}}$}.
We can test whether a set $S$ is empty,  a test
denoted by $S=\emptyset$, in constant time.
We have that 
all elements in a set can be provided in constant time and thus we assume
that we can control loops with expressions of the form 
\mbox{{\bf forall} $t\in S$}, with no extra time overhead. 
 
Finally we recall that we already assumed that is possible to
cast an object of set type into a list.  Similarly we assume that
it is possible to cast an array of $n$ distinct elements $T$ into a 
set denoted by ${\bf set}(T)$. This can be done in $\ordcomp{n\log{n}}$.

\subsection{Maps} 
\label{sec:map}
The {\em map} data type offer a natural implementation for functions.
In fact, maps, from an abstract point of view, are simply sets of
couples, such that there are not two couples with the same first element. 
Implementation for maps can be obtained from implementation for 
sets simply adding extra space to hold the second element of the
couple. For this reason we assume that maps can be implemented as 
a BST maintained as an AVL tree.
This choice is adopted whenever maps are used for local variables within
our algorithms.

However, we anticipate that maps are used in our topological data structure.
In particular, maps are used in the second layer of our two layer data 
structure
that encodes information about non-manifoldness. We will build these maps
using a AVL tree and  convert
these maps into a more compact representation
at the end of the construction phase.
This compact version is  obtained
by transforming the AVL tree into a left complete binary tree
or an {\em  heap} 
(see \cite{Mou25198} Pg. 41-48 and \cite{Cormen90} Ch. 7-8 for an introduction 
to Heaps).
It is well known that an heap can be stored into an array. 
This conversion can be done in
linear time vs. the size of the  domain of the map.
The heap data structure still guarantees logarithmic time access and
supports compact storage of these maps. In fact the space needed to encode
the map as an heap is exactly the space needed to hold all the couples in 
the map.

We assume to have, in our pseudocode, the option to define maps with 
a type constructor of the form
$\mbox{\bf  map of } Domain\, \mbox{\bf  into  } Codomain$.
The type $Domain$ must adhere the same requirements we detailed for sets.
The type $Codomain$ need not to satisfy any particular requirement.
Maps, being sets of couples, inherits all the primitives that we have presented
for sets. In particular, in a map $M$, if we filter out 
the second element, we obtain  the set denoted by $domain(M)$.
This set is available, in constant time, as a set to control loops. 

Maps adds an handy way to store and retrieve elements indexed by elements in the  set $Domain$. If $M$ is a map and $d$ 
is an element in the set $domain(M)$ we will assume that {$M[d]$} denotes
the element in $Codomain$ such that $(d,M[d])\in M$. The element {$M[d]$}
can be obtained with $\ordcomp{\log{|domain(M)|}}$ comparisons.
Similarly, we assume that, in our pseudocode, 
we can associate an element $e$ in $Codomain$ to an element $d$  in
the $Domain$  by an assignment of the form $M[d]\leftarrow e$.
The execution of this assignment, excluding the  time necessary to
retrieve $e$, requires $\ordcomp{\log{|domain(M)|}}$ comparisons.

Finally we note that maps that are present in the topological 
data structure are used only for retrieval and, once created, 
are not modified at all.
Thus, the choice of an heap  implementation for these maps is feasible.
If mesh editing is required then a dynamic data structure is needed and we will have to use an AVL tree
for the implementation of these maps. 
Obviously, this will result in heavier memory requirements. 
However we note that these requirements applies only to  maps that
will be used in the second layer dealing with the non-manifold
structure of the complex. We recall that one of our main assumptions is that
this structure must have a limited extension.

We mention again  the fact that the map data structure, implemented 
with BST  is offered as an 
''off the shelf'' component in STL from the C++ Standard Library.
What LEDA calls maps  is implemented using
hashing and thus it is not suitable for our analysis.
Howeve the LEDA data type called {\em dictionary} offer the right, tree
based, data structure for maps.
What will be the implementation technique used in these packages
we have checked that operations listed here are offered by these
packages with  the time complexity listed in this section.

\subsection{Hashed Sets}
We will assume to have the option to define  sets implemented  with 
an hash table. Hash tables offers insertion and set membership test in
average constant time. Such a set will be defined with 
a  data type declaration of the form 
{$\mbox{{\bf var } H[n] {\bf hashed set of } Domain}$}. 
Where $n$ denotes the size of the hash set.
Usually this size must be set to ten times the number of 
elements that we will ever insert in the set $H$. 
This guarantees have constant time access to the hashed set 
(See for instance \cite{Knu73} Section 6.4).
We assume that the hashed set is initialized in $\lesscomp{n}$.
We assume that the  {\em hashing function} is
provided by the package that implements the hashed set data type. 
This assumption is not arbitrary since we always use hashed sets 
whose elements are made up of a collection of integer indexes.
It will not be too difficult to provide an hashing function for such an
element.

If $H$ is an hashed set and $t$ is an object in $Domain$ we will denote with
$t\in H$ the membership test and with $H\cup \{t\}$ the hashed set obtained
inserting the element $t$ in $H$. 
Hashed sets are offered  both by STL and  LEDA. More precisely
LEDA offers the more classic data type called {\em Hashing array}.

\section{A data structure for \cdec s}
\label{sec:ewrep}
In this work we present a two layer data structure to encode a 
non-manifold complex according to its decomposition into components
that are \cdec s $h$-complexes.
In an upper layer we encode
data necessary to stitch \iqm\ components together.
In a lower layer we use
an extension of the {\em Winged Representation} \cite{PaoAl93} to encode
each   \iqm\ component. We called this extension the {\em \Cdecrep}
Representation (EWR).
The \cdecrep\ representation is designed to be extremely compact and yet  
supports the retrieval of 
topological relations $\trel{nm}(\gamma)$, for any $n<m$.
In this section we first present the \cdecrep\ representation
and evaluate space requirements for this data structure.
Next we evaluate time requirement to extract all topological
relations  in an \iqm\ complex encoded with this representation.

\subsection{The Winged Representation}
We have seen that the TT representation do not 
offer all the information necessary to define completely an \asc.
However, in order to represent an \iqm, it is especially interesting to 
retain this relation in order to navigate easily the \iqm\ complex.
Indeed the TT relation supports an easy navigation of \iqm s because
in this family of $h$-complexes the star of each vertex is $(h-1)$-manifold
connected.
The most natural option in order to build a representation 
that contains the TT relation is simply to add the TT relation 
to the TV representation.  
This idea is at the basis of 
of the Winged  Representation \cite{PaoAl93}.

The original Winged Representation 
for a regular $h$-complex $\AComp$ is a pair $(\AComp^{[h]},\Adj)$ where
$\AComp^{[h]}$ is the subset of the $h$-skeleton $\AComp^h$ made up of
all  $h$-simplices in $\AComp$ and $\Adj$ is an
{\em adjacency function} that associates each $h$-simplex with the 
$(h+1)$-tuple of $h$-simplices that are adjacent to it.
For simplices that are not adjacent exactly to $(h+1)$ $h$-simplices,
the special symbol $\bot$ is used to mean ''no adjacency''.
The domain of the winged  representation is the subclass of regular 
complexes such that a $(h-1)$-face is adjacent  to, at most, two $h$-simplices.

We extended the Winged Representation, using this scheme, beyond
its intended domain to represent \cdec\ complexes.
For this reason we found quite reasonable to give a new name to
this representation to distinguish this from the original one.
We choose  to call this representation the   {\em \Gw} Representation. 
\begin{definition}
\label{def:gw}
The \emad{\Gw}{Representation} 
for an \cdec\ $h$-complex $\AComp$ is the four tuple 
{$\{V,\Theta,\sigma_0,\Adj\}$}
where:
\begin{itemize}
\item $V$ is the set of vertices in $\AComp$;
\item  $\Theta$ is the set of top simplices in $\AComp$;
\item $\sigma_0$ is a mapping $\funct{\sigma_0}{\Theta}{2^V}$ s.t.
the triple $(V,\Theta,\sigma_0)$ is the TV representation for
$\AComp$;
\item the relation $\Adj\subset \Theta^2$ is 
a subset of the TT relation for $\AComp$ such that $\theta_1\Adj \theta_2$ \siff\
the simplex shared by top simplices (indexed by) $\theta_1$ and $\theta_2$ is 
a manifold simplex in $\AComp$.
\end{itemize}
\end{definition}
We note that the \gw\ representation  coincide with the winged
representation but it is {\em generalized} in the sense that it is used 
to represent complexes that are not in the original modeling domain for
the winged data structure.

\subsection{The \Cdecrep\ Representation and
Data Structure}
In order to implement fast computation of the $\trel{nm}$ relations we propose
to further enrich the \Gw\ Representation {$\{V,\Theta,\sigma_0,\Adj\}$}
with a map {$\funct{\sigma_{VT}}{V}{\Theta}$} 
that returns, for each vertex $v$, a top simplex in $\AComp$ incident to $v$. 
We will call {$\sigma_{VT}$} the $\mbox{\rm VT}^\star$ relation.

This mapping is not exactly
defined  here and we just assume that it will satisfy the requirement
that $v\in\sigma_{VT}(v)$. 
We will later specify more in  detail the \VTS\ relation showing an 
optimization that
supports the implicit encoding of this relation using no space at all.
The \gw\ representation, extended with the {$\mbox{\rm VT}^\star$} relation, 
will be called the \Cdecrep\ Representation.
The \cdecrep\ representation is the  representation we used to encode \cdec\
components in our two layer data structure. 
\begin{definition}
\label{def:cdecrep}
The \emad{\Cdecrep}{Representation} is a couple {$EWS=(W,{\sigma_{VT}})$} where 
$W={\{V,\Theta,\sigma_0,\Adj\}}$
is a \gw\ representation  and {$\funct{\sigma_{VT}}{V}{\Theta}$} is a
function such that $v\in\sigma_{VT}(v)$. 
This function is called the  \VTS\ relation of the representation $EWS$.
\end{definition}

In the next section we will define an implementation of the data structure for the 
\cdecrep\ representation. We will  provide an algorithm for the construction of this
data structure and evaluate the complexity of the construction procedure.

\subsubsection{Data Structure Implementation}
Let $\AComp$ be an $h$-complex with Vertices 
in $V$ and top simplices in $\Theta$.
We assume that
Vertices in $V$ will be represented by a set of
integer values (denoted by the identifier \Vertex) that we will
take as   $\Vertex=[\MinV,\ldots,\MaxV]$ with 
$\MaxV-\MinV+1=NV=|V|$.

Similarly, top simplices will be represented by a set of
integer values (denoted by the identifier \TopSimplex) 
that, in this implementation, we will take as
$\TopSimplex=[\MinT,\ldots,\MaxT]$ with 
$\MaxT-\MinT+1=NT=|\Theta|$.   
We assume to have two special values outside the range $\TopSimplex$: one 
to represent the symbol $\bot$ and another to represent the special 
value $\trix$ that will be  used in the following.

In the following, where this is not ambiguous, we will freely use the terms 
''vertex'' to mean ''vertex index''.
Vertex indexes will be denoted by lowercase letters such as $v$, $w$, $o$
Similarly we will use the term  ''simplex'' to mean ''simplex index''.
Simplex indexes will be denoted by lowercase letters such as $t$.
Sometimes we will use the term ''simplex'' to denote a set of
vertex indexes. These objects will be usually denoted by lowercase
greek letters such as $\psi$,$\gamma$ or $\theta$.

Indexes will be organized in the following data structure to 
represent 
the TV,TT and {$\mbox{\rm VT}^\star$} relations.
\begin{data}{data:tttv}{TT and TV and $\mbox{\rm VT}^\star$ Data Structure}
{
\begin{verbatim}
type
 Vertex = [MinV..MaxV];
 TopSimplex = [MinT..MaxT];
var
 TV:array[MinT..MaxT,0..h] of Vertex;
 TT:array[MinT..MaxT,0..h] of TopSimplex;
 VT*:array[MinV..MaxV] of TopSimplex;
\end{verbatim}
}
\end{data}
We will call this data structure the
\emad{\Cdecrep}{Data Structure}.
Note that this data structure takes 
$NT(h+1)\log{NV}+(NT(h+1)+NV)\log{NT}$ bits to encode 
an $h$-dimensional component of the decomposition $\canon{\AComp}$.
{
	\begin{figure}[h]
		\begin{center}
			\psfig{file=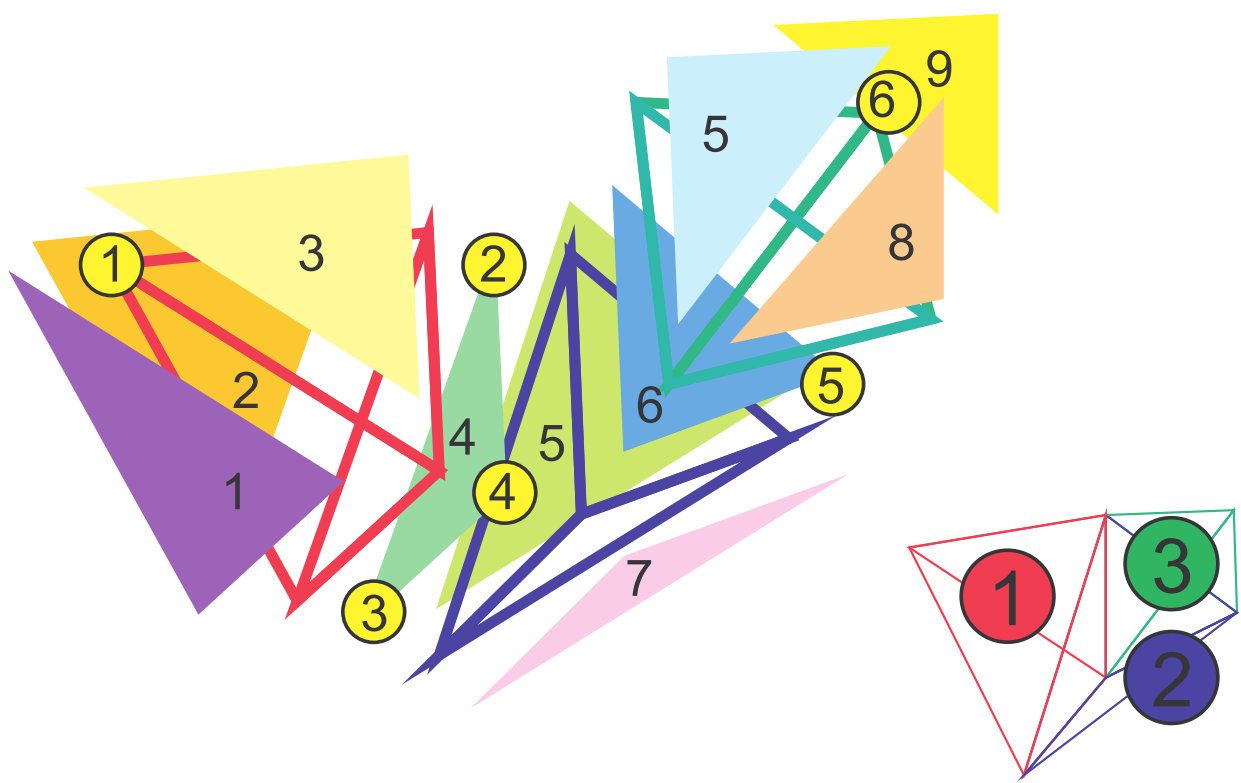,width=\textwidth}
		\end{center}
		\caption{A simple 3-complex used for a running example of the  implementation of the 
			\Cdecrep\  data structure.
		}
		\label{fig:datastrex}
	\end{figure}
	\begin{figure}[h]
		\begin{center}
			\begin{tabular}{|c||c|c|c|c|}\hline
				\multicolumn{5}{|c|}{TV[1..3,0..3]} \\ \hline
				&0&1&2&3    \\ \hline \hline
				1&1&2&3&4    \\ \hline
				2&2&3&4&5   \\ \hline
				3&2&4&5&6    \\ \hline
			\end{tabular}
		\mbox{ }
			\begin{tabular}{|c||c|}\hline
			\multicolumn{2}{|c|}{VT*[1..6]} \\ \hline
1&1    \\ \hline
			2&1   \\ \hline
			3&1    \\ \hline
			4&1    \\ \hline
			5&2    \\ \hline
			6&3    \\ \hline
		\end{tabular}
	\mbox{ }
	\begin{tabular}{|c||c|c|c|c|}\hline
		\multicolumn{5}{|c|}{TT[1..3,0..3]} \\ \hline
		&0&1&2&3    \\ \hline
		1&2&$\bot$&$\bot$&$\bot$    \\ \hline
		2&$\bot$&3&$\bot$&1  \\ \hline
		3&$\bot$&$\bot$&$\bot$&2    \\ \hline

	\end{tabular}
		
				\caption{An example of  TV,TT and {$\mbox{\rm VT}^\star$} relations for the 3-complex of Figure \ref{fig:datastrex}} 
		\label{table:TVTTVT}		
		\end{center}
	\end{figure}
}  
{
Note that sets of integers $\Vertex$ and $\TopSimplex$
need to be a set of consecutive integers.
We will develop algorithms in this section using the
identifier $\Vertex$ for the range $[\MinV,\ldots,\MaxV]$ and 
the identifier $\TopSimplex$ for the range $[\MinT,\ldots,\MaxT]$.
This will be done to stress the fact that our results are still
correct even if we take, for the sets $\Vertex$ and $\TopSimplex$, 
two sets of non contiguous, integers. 
This will be useful in Section \ref{sec:globewds}
when we will merge all the TT, TV, and  {$\mbox{\rm VT}^\star$} 
arrays, one for each connected component, into two larger arrays TT' and TV'  
for all connected components of the original non-manifold complex. 
For optimization reasons, that will be apparent in Appendix \ref{sec:opt}, 
in this merge will be useful 
to spread the range of indexes for TT into a non contiguous set of
indexes for TT'. A similar scattering will be considered for indexes
of TV over the array TV'.  
}
\subsubsection{TV and VT* Construction}
We assume that, at the end of the decomposition process, we are left with 
a TV map $\funct{\sigma_{TV}}{\TopSimplex}{2^{\Vertex}}$ for each 
connected component of $\canon{\AComp}$.
We recall that we  assume an implementation for these  map
that supports the sequential
access to all elements in the domain of {$\sigma_{TV}$} in constant
time.
Therefore, with a linear scan of the map {$\sigma_{TV}$}, we can fill
the arrays TV and VT* in $\ordcomp{NT}$. 

Therefore the construction of TV and \VTS\ relations do not pose particular
problems.
Some more details are needed to show
how to fill the  TT array.
\begin{example}
	In Figure \ref{table:TVTTVT} we report the result of filling TV and VT* arrays for the 3-complex in Figure \ref{fig:datastrex}. 
\end{example}
\subsubsection{TT Construction}
The  TT array, that encodes the TT relation, is 
is filled in two steps. First we build  a map 
{$\funct{\sigma_{\Adj}}{\AComp^{[h-1]}}{{\bf list of }\,\,\TopSimplex}$}.
Upon completion of this first step,
the domain of the map {${\sigma_{\Adj}}$} will be
the set of $(h-1)$-simplices 
in  {$\canon{\AComp}$}.
For each $(h-1)$-simplex $\psi$ the map ${\sigma_{\Adj}}[\psi]$
will give the {\em list} of indexes for the top $h$-simplices that are 
cofaces of $\psi$. 

In a second step we fill the TT array in a way such that
{TT[$t$,$k$]} will be the top simplex adjacent to simplex $t$ and such that
the vertex {TV[$t$,$k$]}
is not in the top simplex  {TT[$t$,$k$]}.
Thus {TT[$t$,$k$]} points the top simplex adjacent to $t$ and incident to its 
$(h-1)$-face that do not contain vertex {TV[$t$,$k$]}.

Such a top simplex do not exist whenever the 
$(h-1)$-face of $t$ that do not contain the vertex {TV[$t$,$k$]} 
is on the boundary.
In this case we put $TT[$t$,$k$]=\bot$.
\begin{example}
	In Figure \ref{table:TVTTVT} we report what must contain the TT array for the 3-complex in Figure \ref{fig:datastrex}. 
\end{example}
Whenever the complex is non-pseudomanifold  we will find situations 
where two or more top simplices satisfy the above condition for $t$. In this 
situation we treat this condition in a way similar to a boundary condition and
put $TT[$t$,$k$]=\trix$
This is the main difference between this  structure  and the classical
Winged Representation.
An alternative approach to this construction, that 
organize top simplices
adjacent to a non-pseudomanifold $(h-1)$-simplex into a cycle using will
be presented in Section \ref{sec:trix}.

{
To sketch the construction algorithm we need to use sets of
vertex indexes
and maps from sets of vertex indexes to sets of top simplices.
To comply with requirements of section \ref{sec:set} we assume that
sets of vertex indexes can be ordered in the following way.
We will say that the set of  vertices $s_1$ is smaller than $s_2$ \siff\
the ordered sequence of vertices in $s_1$ is lexicographically smaller 
than the ordered sequence of  vertices in $s_2$. Thus for instance 
set $s_1=\{3,2,1\}$ is smaller than $s_2=\{1,4,2\}$.
In fact the ordered sequence of elements in $s_1$ is 
$\brk{1,2,3}$ while the ordered sequence of elements in $s_2$ is 
$\brk{1,2,4}$. Lexicographic comparison shows that
$\brk{1,2,3}$ is smaller than $\brk{1,2,4}$. 
}

Finally we assume to have a function {OPPOSITE(TV,$t$,$\psi$)} that
returns the position in  the array TV[$t$] of the vertex in  
the top $h$-simplex $t$ that is not in its $(h-1)$-face  $\psi$ 
(i.e. the vertex in $t$ opposite to $\psi$). In other
words the function OPPOSITE must satisfy the assertion:
TV[$t$,{OPPOSITE(TV,$t$,$\psi$)}]$\not\in\psi$.
With this assumptions we lay the algorithm to fill the TT array as follows:
\begin{algo}{alg:tfill}{{Fill the TT array}}
\begin{algorithmic}
\STATE{{\bf var} $\sigma_{\Adj}$: {\bf map set of }{\tt Vertex }{\bf into list of }{\tt TopSimplex}}; 
\STATE $\sigma_{\Adj}\leftarrow (\forall x)[\send{x}{\brk{\mbox{}}}]$ 
\COMMENT{initially the map $\sigma_{\Adj}$ always  returns the empty list}
\FORALL{$t\in\TopSimplex$} \STATE $\gamma\leftarrow$ {\bf set}(TV[$t$])
\FORALL{$v\in\gamma$} 
\STATE ${{\sigma_{\Adj}}[\gamma-\{v\}]}\leftarrow
\brk{t}+{{\sigma_{\Adj}}[\gamma-\{v\}]}$
\COMMENT{add $t$ to the previous value of 
${\sigma_{\Adj}}[\gamma-\{v\}]$}
\ENDFOR
\ENDFOR
\FORALL{$\psi$ in the domain on $\sigma_{\Adj}$}
\STATE $star_{\psi}\leftarrow \sigma_{\Adj}[\psi]$
\COMMENT{$star_{\psi}$ is the list of cofaces of $(h-1)$-simplex $\psi$}
\IF[$\psi$ is on the boundary]{$star_{\psi}=\brk{t}$}
\STATE TT[$t$,OPPOSITE(TV,$t$,$\psi$)]=$\bot$ 
\COMMENT{$\bot$ stands for no adjacency}
\ELSIF[$\psi$ is a manifold $(h-1)$-simplex]{$star_{\psi}=\brk{t,t^\prime}$}
\STATE TT[$t$,OPPOSITE(TV,$t$,$\psi$)]=$t^\prime$
\STATE TT[$t^\prime$,OPPOSITE(TV,$t^\prime$,$\psi$)]=$t$
\ELSE[$\psi$ is a non-manifold $(h-1)$-simplex]
\STATE TT[$t$,OPPOSITE(TV,$t$,$\psi$)]=$\trix$
\COMMENT{$\trix$ stands for non-manifold adjacency}
\ENDIF
\ENDFOR
\end{algorithmic}
\end{algo}
\begin{example}
	As an example we report the Algorithm \ref{alg:tfill} commented with assertions that applies when filling TT array for the 3-complex in Figure \ref{fig:datastrex}. Loops and instructions within loops are commented with an assertion that holds only the first time the loop is executed. With this convention this example shows how are filled two entries, i.e. TT[1,0]=2 and TT[2,3]=1 in the table TT of Figure \ref{table:TVTTVT}. The annotated algorithm also  shows some actual values for this tiny complex. 
\begin{algorithmic}
		\STATE{{\bf var} $\sigma_{\Adj}$: {\bf map set of }{\tt 1..6 }{\bf into list of }{\tt 1..3}};  
		\STATE $\sigma_{\Adj}\leftarrow (\forall x)[\send{x}{\brk{\mbox{}}}]$ 
		\FORALL[{$t=1$}]{$t\in{\tt 1..3}$}
		\STATE $\gamma\leftarrow$ {\bf set}(TV[$t$])
		\COMMENT{$\gamma=\{1,2,3,4\}$}
		\FORALL[{$v=1$}]{$v\in\gamma$} 
		\STATE ${{\sigma_{\Adj}}[\gamma-\{v\}]}\leftarrow
		\brk{t}+{{\sigma_{\Adj}}[\gamma-\{v\}]}$
		\COMMENT{${{\sigma_{\Adj}}[\{2,3,4\}]=\brk{1}}$}
		\ENDFOR
		\ENDFOR
		\FORALL[{$\psi=\{2,3,4\}$}]{$\psi$ in the domain on $\sigma_{\Adj}$}
		\STATE $star_{\psi}\leftarrow \sigma_{\Adj}[\psi]$
		\COMMENT{$star_{\psi}=\brk{1,2}$}
		\IF[$\psi=\{2,3,4\}$ is not on the boundary,   skip]{$star_{\psi}=\brk{t}$}
		\STATE TT[$t$,OPPOSITE(TV,$t$,$\psi$)]=$\bot$ 
		\COMMENT{$\bot$ stands for no adjacency}
		\ELSIF[$\brk{t,t^\prime}=\brk{1,2}$]{$star_{\psi}=\brk{t,t^\prime}$}
		\STATE TT[$t$,OPPOSITE(TV,$t$,$\psi$)]=$t^\prime$
		\COMMENT{TT[1,0]=2 since OPPOSITE(TV,1,$\{2,3,4\}$)=0}
		\STATE TT[$t^\prime$,OPPOSITE(TV,$t^\prime$,$\psi$)]=$t$
		\COMMENT{TT[2,3]=1 since OPPOSITE(TV,2,$\{2,3,4\}$)=3}
		\ELSE[$\psi$ is a non-manifold $(h-1)$-simplex]
		\STATE TT[$t$,OPPOSITE(TV,$t$,$\psi$)]=$\trix$
		\COMMENT{$\trix$ stands for non-manifold adjacency}
		\ENDIF
		\ENDFOR
	\end{algorithmic}
\end{example}

If we postulate a standard implementation for maps and sets, 
then the following fact holds:
\begin{property}
\label{pro:ttbuildcomp}
Let the  TV array the encoding of a TV relation for a regular $h$-complex
$\AComp$ with $NT$ $h$-simplices.
Then the Algorithm \ref{alg:tfill} computes the TT relation for
$\AComp$ and completes in $\lesscomp{h^2 NT\log{NT}}$
\end{property}
\begin{proof}
To evaluate the time complexity of Algorithm \ref{alg:tfill} we  assume 
that we can use a standard implementation 
for lists, sets and maps as postulated in Section \ref{sec:dsdup}. 
{
When we use maps that have sets as keys 
(as in ${\sigma_{\Adj}}[\gamma-\{v\}]$) we have to consider the fact
that each key must be lexicographically compared. 
Since sets are kept ordered into BSTs the comparison between two sets
of $h$ elements can be done in $\lesscomp{h}$.

With our assumption about sets we have
that all the array-to-set conversions 
performed  by calls to {\bf set}(TV[$t$]) 
are done in $\lesscomp{h\log{h}\cdot NT}$.
}
The construction of {$\sigma_{\Adj}$} is done updating this
map at most {$(h+1)NT$} times. For each update we retrieve the list
${{\sigma_{\Adj}}[\gamma-\{v\}]}$ and add an element.
Since the list can be updated in  constant time the whole construction of
{$\sigma_{\Adj}$} can be done with $\lesscomp{h\cdot NT\log{NT}}$
comparisons between arguments of the form $(\gamma-\{v\})$.
Each comparison is in $\lesscomp{h}$. Therefore the overall time
complexity for 
the construction of {$\sigma_{\Adj}$} must be $\lesscomp{h^2 NT\log{NT}}$.
Upon termination of this phase the map ${\sigma_{\Adj}}[\psi]$ gives
the $h$-simplices incident at the  $(h-1)$-simplex $\psi$.

Next we fill the array TT that has, at most, {$(h+1)NT$} entries.
Actually we consider only $(h-1)$-simplices  that have at most
two incident $h$-simplices.
To fill each entry we have to compute \mbox{OPPOSITE(TV,$t$,$\psi$)}.
This gives the correct TT entry to store into and proves the correctness 
of algorithm.
The computation of OPPOSITE 
can be done testing for all elements in TV[$t$] if they are in
the set $\psi$.
This takes at most  $h\log{h}$. Therefore we can fill TT in
$\lesscomp{h^2NT\log{h}}$ and, assuming that $h<NT$,  we can say that
the whole construction can be done in $\lesscomp{h^2 NT\log{NT}}$
\end{proof}

\subsubsection{TT Navigation and retrieval of $\trel{0h}(v)$ in \Iqm s}
The TT array encodes the TT relation, therefore for every top $h$-simplex 
indexed by  $t$, we have that:
\begin{enumerate}
\item for $k=0\ldots h$ the top simplex recorded in  TT[$t$,$k$] is adjacent with $t$; 
\item the simplices indexed by $t$ and 
TT[$t$,$k$] shares a $(h-1)$-simplex that do not contains the vertex 
indexed by TV[$t$,$k$]. 
\end{enumerate}
These two facts that comes directly from the definition of the TT relation,
easlily implies the following property
\begin{property}
\label{pro:condtt} Let $v$ be a vertex of simplex $t$ such that  {TV[$t$,$h$]$=v$}
and let {$t^\prime=$TT[$t$,$k$]} for some $k\neq h$.
Then the top simplices $t$ and $t^\prime$ are adjacent and they both
belongs to  the star of $v$.
\end{property}

\begin{example}
	As an example we consider  the 3-complex in Figure \ref{fig:datastrex} and the table TT of Figure \ref{table:TVTTVT}. Let's take  $t=2$ and $v=5$. With this choice we have that TV[2,3]=5 so we take $h=3$ and $k=0,1,2$.
	Among TT[2,0], TT[2,1] and TT[2,2] only for $k=1$ TT[2,1] returns a tetrahedron that is tetrahedron 3 that actually shares a triangle with tetrahedron 2.
	Tetrahedra 2 and 3 belongs to the star of vertex 5. The vertex TV[2,1] is vertex 3 that is neither in the star of 5 nor in the triangle shared by tetrahedra 2 and 3.  
\end{example}
	
Considering the above property and the fact that our complex is an \iqm\ we
have that  the set of top simplices  
in $\star{v}$  can be collected  simply using the TT relation.
In fact we can travel completely the star of a vertex by adjacency because,
in an \iqm, the star of each vertex must be $(h-1)$-manifold-connected.

Therefore, we are ready to devise an algorithm for
computation of  $\trel{0h}(v)$ in an $h$-complex:

In the following pseudocode we will use lists as defined in Section
\ref{sec:lists} and a bit vector of $\Theta$ of $NT$ bits, i,e.,
one bit for each  top simplex in the complex  (see Section
\ref{sec:bitv} for the definiton of bit vectors). 
\begin{center}
\mbox{$\Theta$ : {\bf array} $[\MinT,\ldots,\MaxT]$ {\bf of bit}}.
\end{center}
This bit vector will be
used to mark top simplices visited by our algorithm.
We assume that when the application is started this bit vector is
allocated  

With these ancillary data structures it easy to devise an algorithm for
the computation of  $\trel{0h}(v)$ in a  generic $h$-complex:
\begin{algo}{algo:s0h}{Computation of $\trel{0h}(v)$ in an \cdec\ $h$-complex}
\begin{algorithmic} 
\STATE{{\bf Function} $\trel{0h}(v:{\tt Vertex})$ {\bf returns list of} {\tt TopSimplex}}
\STATE $N\leftarrow S\leftarrow {\bf list}(\sigma_{VT^\star}(v))$ 
\WHILE{$N\neq\brk{\mbox{}}$} 
\STATE $t\leftarrow N.{\rm TOP}$; $N\leftarrow N.{\rm POP}$; $\Theta[t].{\rm SET}$
\FOR{$k=0$ to $h$}
\IF{TV[$t$,$k$]$\neq v$} 
\STATE $t^\prime\leftarrow$ TT[$t$,$k$] 
\IF{{\bf not} $\Theta[t^\prime].{\rm TEST}$ {\bf and} $t^\prime \neq\bot$ {\bf and} $t^\prime \neq\trix$} 
\STATE $N\leftarrow \brk{t^\prime}+N$; $S\leftarrow \brk{t^\prime}+S$
\ENDIF
\ENDIF
\ENDFOR
\ENDWHILE
\FORALL{t {\bf in} S}
\STATE $\Theta[t].{\rm CLEAR}$
\ENDFOR
\STATE {\bf return}  $S$
\end{algorithmic} 
\end{algo}
Following the informal idea given in the introduction to this algorithm 
we can prove the following property stating algorithm total correctness.
\begin{property}
\label{pro:corrSoh}
Let $\AComp$ be an \cdec\ $h$-complex.
For any vertex $v \in \AComp$ the above algorithm for $\trel{0h}(\{v\})$ 
terminates.
Upon termination  in the 
variable $S$ we find  the list of elements in the set $\trel{0h}(\{v\})$.
\end{property}
\begin{proof}
By Property \ref{pro:condtt} we have that all simplices added to $S$
belongs to the star of $v$.
To prove termination we note that this computation cannot loop indefinitely because of the 
following facts:
\begin{itemize}
\item a simplex is added to the list $N$ \siff\ it is added to $S$;
\item a simplex already in $S$ is not added to $N$;
\item the algorithm cannot loop for more than $h+1$ steps  without
deleting an element in $N$. 
\end{itemize}
So to loop indefinitely the algorithm must add continuously elements in
$N$ and therefore an indefinite number of new elements must come into 
$S$. This is impossible because $S$ is a subset of the star of $v$.

Obviously the computation for $\trel{0h}(v)$
returns a subset of the star of $v$. We note that during the computation of $\trel{0h}$ if a simplex $t$ is 
inserted in $S$ then all  the  simplices that are manifold adjacent
w.r.t. $t$  will be inserted in $S$, if not already in. 
By transitivity all simplices in the star of $v$ that are 
$(h-1)$-manifold-connected 
with those in  $S$ will eventually  fit into $S$. 
Being  within an \cdec\ $h$-complex the star of $v$ is $(h-1)$-manifold-connected by Property \ref{pro:iqmrel} Part \ref{pro:d1conn}. Therefore upon termination 
$S$ must contain the star of $v$.\end{proof}

The above algorithm computes $\trel{0h}(v)$ in optimal time i.e. in
$\ordcomp{|\trel{0h}(v)|}$. 
This is stated in the following property. 
\begin{property}
\label{pro:compl}
The computation of $\trel{0h}(v)$ takes
$\ordcomp{h|\trel{0h}(v)|}$. 
\end{property}
{For each top simplex inserted in $N$ we perform $(h+1)$ access to the TV
table and next performs $h$ times the test  $\Theta[t^\prime].{\rm TEST}$.
Assuming a standard implementation for the bit vector we can perform 
this test in constant time
for an overall complexity of $\ordcomp{h}$ comparisons for each
top simplex added to  $N$. All other  operations take a constant time
and are performed once for
each element in the output 
All simplices inserted in $N$ are recorded in $S$ and therefore the total
number of elements added to $N$ is given by the size of the output of the set
$\trel{0h}={\trel{0h}(v)}$.
Therefore the algorithm perform a loop for each
element  added in the output.
Thus we have a total complexity of $\ordcomp{h|\trel{0h}|}$.
}
\subsubsection{\Iqm\ Navigation and the computation of $\trel{nm}(\gamma)$}
\label{sec:nmdo}
In the previous section we presented an algorithm for the computation of 
${\trel{0h}}$
In this section we present a couple of algorithms needed for the computation of 
and of ${\trel{nm}}$ for $n<m$. 
Next we evaluate the time complexity of these algorithms.
{
First we need to develop a function \mbox{FaceOf($m,\beta$,CoTop)}
that returns the list of $m$-cofaces of $\beta$ that are
$m$-faces of  $h$-simplices in  CoTop.
The algorithm is developed assuming  that 
all $h$-simplices in  CoTop are cofaces of $\beta$.
In other words the function FaceOf is defined by the equation:
$$
{\rm FaceOf}(m,\beta,CoTop)=
\{\gamma|\ord(\gamma)=m\, \mbox{\bf and}\, 
(\exists \tau\in CoTop)(\beta\le\gamma\le\tau)\} 
$$
}
This auxiliary function can be computed in $\lesscomp{|CoTop|}$
using the following algorithm:
\begin{algo}{algo:faceof}{Computation of \mbox{{\rm FaceOf}($m,\beta$,CoTop)}}  
\begin{algorithmic} 
\STATE{FaceOf($m$, $\beta$: \mbox{\bf set of } {\tt  Vertex},
CoTop: \mbox{\bf list of } (\mbox{\bf set of } {\tt  Vertex})) 
{\bf returns list of} ({\bf set of} {\tt Vertex})}
{\mbox{{\bf var } Inserted[$10\cdot\binomial{h+1}{m+1}|CoTop|$]: {\bf hashed set of } (\mbox{\bf set of } {\tt  Vertex})}}. 
\STATE Result$\leftarrow \brk{\mbox{}}$
\STATE $n\leftarrow |\beta|$
\FORALL{$\tau\in  CoTop$ }
\STATE $\psi\leftarrow \tau-\beta$
\FORALL{$\{v_0,\ldots,v_{m-n}\}\subset\psi$}
\STATE Face $\leftarrow  \beta \cup \{v_0,\ldots,v_{m-n}\}$
\IF[Face not already in]{Face $\not\in$ Inserted}
\STATE Result $\leftarrow \brk{{\rm Face}}$ + Result;
\STATE Inserted $\leftarrow$  Inserted $\cup \{{\rm Face}\}$;
\ENDIF
\ENDFOR
\ENDFOR
\end{algorithmic} 
\end{algo}
The correctness of the above algorithm is obvious. Its  complexitiy is 
discussed by the following property:
\begin{property}
The algorithm \ref{algo:faceof} is linear in the size of the input  CoTop
\end{property}
\begin{proof} 
All operations needed to generate Face can be done in $\lesscomp{h\log{h}}$,
being $h$ the dimension of the complex. We have taken in the implementation the variable  "Inserted" as an  hash set.
The size of this hash table is taken  
ten times  {$\binomial{h+1}{m+1}|CoTop|$}. 
The number {$\binomial{h+1}{m+1}|CoTop|$} represents an
upper bound for the number of $m$-faces of simplices  in CoTop.
Due to this space provision for variable "Inserted",
marking and testing in the hash set "Inserted" can be considered to  be done in constant time.

The test \mbox{Face $\not\in$ Inserted} is performed  at most
{$\binomial{h+1}{m+1}|CoTop|$}. Thus the overall execution  time 
must be in $\lesscomp{{h\log{h}}{\binomial{h+1}{m+1}|CoTop|}}$.
\end{proof} 
Using the auxiliary function FaceOf, the $\trel{nm}(\gamma)$  
relation can be computed following the algorithm below:
\begin{algo}{algo:snm}{Computation of $\trel{nm}(\gamma)$ in a \cdec\ $h$-complex}  
\begin{algorithmic} 
\STATE{{\bf Function} $\trel{nm}(\gamma: \mbox{\bf set of } {\tt  Vertex})$ 
{\bf returns list of} ({\bf set of} {\tt Vertex})}
\STATE {\bf select} $v$ {\bf in} $\gamma$
\STATE $\trel{0h}\leftarrow \trel{0h}(v)$;  $Top\leftarrow \brk{\mbox{}}$
\FORALL{$t\in \trel{0h}$}
\STATE{$\theta\leftarrow$ {\bf set}(TV[$t$])};
\IF{$\gamma\subset\theta$}
\STATE $Top\leftarrow \brk{\theta}+Top$
\ENDIF
\ENDFOR
\STATE {\bf return} FaceOf($m$,$\gamma$,$Top$)  \end{algorithmic} 
\end{algo}
We assume that   the statement {\bf select} $v$ {\bf in} $S$
randomly select an element in $S$.
It is easy to see that the algorithm above computes
the  total function $\trel{nm}(\gamma)$  for every
$\gamma\in\AComp$. The time complexity of this computation in
general is not optimal as it shows the following property.

\begin{property}
\label{pro:complnm}
The computation of $\trel{nm}(\gamma)$ can be done in
$\lesscomp{h|\trel{0h}(v)|}$.
\end{property}
\begin{proof}
The computation of $\trel{nm}(\gamma)$ accounts for an initial computation
of {$\trel{0h}(v)$} for some $v\in\gamma$. This, by property \ref{pro:compl},
takes $\ordcomp{h|\trel{0h}(v)|}$. 
We will show that this dominates the complexity of all other operations.
Having computed {$\trel{0h}(v)$} we are left with a set of 
{$|\trel{0h}(v)|$} top $h$-simplices with $h>n+m$. For each one of these top 
simplices we check if $\gamma$ is a subset of $\theta$. 
This inclusion  check can be done in
$\ordcomp{n\log{(h)}}$ for a total complexity
of $\ordcomp{n\log{(h)}|\trel{0h}(v)|}$. 
From the subset of top simplices left we generate all $m$-faces that
include $\gamma$. This is done with the call
to \mbox{FaceOf($m$,$\gamma$,$Top$)} that takes $\lesscomp{|Top|}$
Since elements in $Top$ are a subset of the elements in
{$|\trel{0h}(v)|$} thus this operation must be in
$\lesscomp{h|\trel{0h}(v)|}$. This completes the proof.
\end{proof}

We notice that the expressions for the complexity of $\trel{nm}(v)$ are only 
partially satisfactory since they do not give an expression that relates 
complexity of the computation with the size of the output 
Indeed the retrieval of 
topologic relations within this  data structure can 
require the exploration of all top simplices around a vertex. 
thus the complexity for {${\trel{nm}(\gamma)}$} depends on  
$\ordcomp{\trel{0h}(v)}$ for some $v\in\gamma$.

Obviously for $n>0$ the extraction of {${\trel{nm}(\gamma)}$} can be grossly 
inefficient (See Property \ref{pro:nmineff} in the next section) but this is a direct consequence of the fact that 
$n$-simplices do not receive a explicit representation in this data structure.
In section \ref{sec:trie} we will show that we can enrich this data
structure with an indexing structure and obtain optimal extraction of
{${\trel{nm}(\gamma)}$} for $n>0$.

However,  in this section we raise the question 
whether this data structure  is acceptable at least for the extraction
of topological relations among the entities that
are explicitly modeled (i.e. vertices).
More precisely we want to show when the data structure can 
extract in optimal time $\trel{0m}$ for $0<m<h$. 
In the next subsection we will show that for certain class of complexes 
the idea of running around a vertex $v$ to find all top simplices
do not impair the optimality of the extraction process, even if we
are just interested in $m$-simplices.

 \subsection{Performance for $S_{0m}$} \label{sec:somopt}
The algorithm for the  computation of $S_{nm}$ is only partially
satisfactory  because its complexity, in general, do not depends  
on the size of the output. 
On the other hand, as we have shown, our data structure, in general,
supports the optimal  computation of {${S_{0h}(v)}$} in an 
$h$-complex. In fact  {${S_{0h}(v)}$} can be recovered in 
$\lesscomp{h|{S_{0h}(v)}|}$.
Since in this data structure vertices receive explicit 
representation we expect that also all the $S_{0m}$ relations, 
for $0<m<h$ can be extracted in optimal time. 

In this section, we will show that optimal extraction of
$S_{0m}$, for $0<m<h$, is possible
for \cdec\ surfaces and for 
\cdec\ tetrahedralizations that are embeddable in $\real^3$ as a 
compact geometric simplicial complex.
In particular we will show that in manifold surfaces
$S_{01}(v)$ is computable in
$\lesscomp{|{S_{01}(v)}|}$.
Similarly we will show that
in a \cdec\ $3$-complex embeddable in $\real^3$
$S_{01}(v)$ is computable in
$\lesscomp{|{S_{01}(v)}|}$ and
$S_{02}(v)$ is computable in
$\lesscomp{|{S_{02}(v)}|}$.
To this aim we will use the relation between {\em face numbers} we 
introduced in Section \ref{sec:facenum}.

\subsubsection{Manifold Surfaces}
We know that all \cdec\ surfaces are manifold
surfaces and, in turn, 
manifold surfaces are
pseudomanifolds. Thus we can apply linear inequalties for face numbers 
we have introduced in Section \ref{sec:facenummani}. In particular we will
use the fact that  $\frac{3}{2}f_2\le f_{1}$. 

We have that 
$S_{01}(v)$ is 
computable in
$\lesscomp{|{S_{02}(v)}|}$.
Being ${|S_{02}(v)|}=f_2(\str{v})$ and being $\frac{3}{2}f_2\le f_{1}$ (see Equation \ref{eq:msurf} and apply some algebra) we 
have that $S_{01}(v)$ is computable in
$\lesscomp{|{S_{01}(v)}|}$.
This proves that, at least for \cdec\ surfaces, the extraction algorithm is 
optimal for the computation of $S_{01}(v)$, too.

\subsubsection{Simplicial $h$-complexes imbeddable in $\real^h$}
Next we will show that our data structure supports optimal 
extraction of topological relations also for 
tetrahedralizations that can be embedded in $\real^3$. 
To this aim, in this section, we will use
the linear inequalities between face numbers $f_h$, $f_{h-1}$ and
$f_{h-2}$ we have developed in Section \ref{sec:facenumimbed}. 
From this we will derive that, for a  complex imbeddable in $\real^h$,
the number of elements in $S_{0h}(v)$ is 
$\lesscomp{|S_{0m}(v)|}$ for all $(h-3)\le m\le (h-1)$. 
From this it is easy to prove that the proposed algorithm 
is optimal for the extraction of $S_{0m}$ for $(h-2)\le m\le h$ whenever
the given $h$-complex is embeddable in $\real^h$. 
This will prove that our extraction algorithm is optimal for
the computation of {$S_{01}$} and {$S_{02}$} 
for \cdec\ tetrahedralizations embeddable in $\real^3$.
In particular this
proves that our algorithm is optimal for the extraction of all 
vertex based topological
relations for \cdec\ tetrahedralizations embeddable in $\real^3$.
This is expressed by the following property.
\begin{property}[Optimality for $S_{0(h-1)}$ $S_{0(h-2)}$]
\label{pro:optext}
The computation of $S_{0m}$ for $(h-2)\le m\le h$ for an \cdec\ $h$-complex
$\AComp$ can be done in $\lesscomp{|S_{0m}|}$ 
whenever the given \asc\ is embeddable in $\real^h$.
\end{property}
\begin{proof}
We already know that computation of $S_{0m}$ for an \cdec\ $h$-complex
can be done in {$\lesscomp{|S_{0h}|}$}.
Thus properties
\ref{pro:pseudo} and \ref{pro:euler} shows that $|S_{0h}|$ is both 
{$\lesscomp{|S_{0(h-1)}|}$} (by Property \ref{pro:pseudo}) and 
{$\lesscomp{|S_{0(h-2)}|}$} (by Property \ref{pro:euler}).
This proves that the computation of $S_{0m}$ 
can be done in $\lesscomp{|S_{0m}|}$
for all $(h-2)\le m\le h$. 
\end{proof}

\subsubsection{Non-optimal Extraction in $4$-manifolds}
Although we have assessed the optimality of vertex based extractions for 
tetrahedralizations in $\real^3$.
A natural question is what happens for $4$-complexes. 
By property \ref{pro:optext} we know that if the $4$-complex is
an \iqm\ embeddable
in $\real^4$ our data structure  supports optimal extraction of
relations $S_{04}$,  $S_{03}$ and  $S_{02}$, while the
optimal extraction of $S_{01}$ is not assured. 
In this section we will prove that our algorithm
is not always optimal for the extraction of $S_{01}$ in a  $4$-complexes. 
Next we will show that non optimal extraction exist in $3$-complexes
not embeddable in $\real^3$.
To this aim we will use the results of Section \ref{sec:fourdim}

For the $4$-dimensional case
it is possible to build a triangulation of the  $4$-ball where there is
at least a  vertex $w$ for which  the extraction of $S_{01}(w)$ takes 
much more than $|S_{01}(w)|$.
To build such a complex we just have to take as $4$-simplex the
triangulation of the $4$-ball mentioned in Property  \ref{pro:nonopt}
obtained as the cone from an arbitrary vertex $w$ to the boundary of 
the {\em Cyclic Polytope} in $\real^4$.

\begin{property}
\label{pro:nonoptext}
There exist a triangulation $TB^4$ of the $4$-ball imbeddable in 
$\real^4$ and a vertex $w$ such that the extraction of $S_{01}(w)$
following algorithm \ref{algo:snm} takes 
$\morecomp{|S_{01}(w)|^2}$.
\end{property}
\begin{proof}
As $4$-complex we take the 
triangulation $TB^4$ of the $4$-ball mentioned in Property  \ref{pro:nonopt}.
This ensure that there is a vertex $w$ for which $f_4(\xclstar{TB^4}{w})=f_0(f_0-3)/2$
 with $f_0$ the number of 
vertices in $\xclstar{TB^4}{w}$.
Next we consider the computation of $S_{01}(w)$ and
recall that the computation of $S_{01}(w)$ implies the
computation of {$S_{04}(w)$}. Thus we have that
the computation of $S_{01}(w)$ is  $\morecomp{{|S_{04}(w)|}}$. 
For the $4$-complex $TB^4$ we have that
${|S_{04}(w)|}=f_3({\bnd{C_4(f_0)}})=f_0(f_0-3)/2$ where
$f_0$ must be  ${|S_{01}(w)|}$. Thus, for this particular $4$-complex, the
computation of {$S_{01}(w)$} takes $\morecomp{|S_{01}(w)|^2}$.
Thus our algorithm fails to be optimal for a $4$-complexes
embeddable in $\real^4$.
\end{proof}
The above Property generalize to $h$-balls showing that the extraction of
$S_{01}(w)$ is non-optimal and can become quite inefficient as $h$ grows.
\begin{property}
\label{pro:s01ext}
There exist a triangulation $TB^h$ of the $h$-ball embeddable in $\real^h$
such that, for a certain vertex $w$ in $TB^h$,
the extraction of $S_{01}(w)$, performed by algorithm
\ref{algo:snm} is non-optimal and is
$\morecomp{|S_{01}(w)|^{\floor{h/2}}}$.
\end{property}
\begin{proof}
By Property \ref{pro:s01ext} we know that there exist a triangulation 
$TB^h$ of the $h$-ball embeddable in $\real^h$
such that, for a certain vertex $w$ in $TB^h$,
$f_{h}(\xclstar{TB^h}{w})=\ordcomp{f_0^{\floor{h/2}}}$ 
with $f_0$ is the number of vertices in $\xclstar{TB^h}{w}$.
Next we consider the computation of $S_{01}(w)$ and
recall that the computation of $S_{01}(w)$ implies the
computation of {$S_{0h}(w)$}. Thus we have that
the computation of $S_{01}(w)$ is  $\morecomp{{|S_{0h}(w)|}}$. 
For the complex $TB^h$ we have that
${|S_{0h}(w)|}=f_h(\xclstar{TB^h}{w})=\ordcomp{f_0^{\floor{h/2}}}$
with $f_0$ is the number of vertices in $\xclstar{TB^h}{w}$.
Therefore $|S_{01}(w)|$
is $\morecomp{|S_{01}(w)|^{\floor{h/2}}}$.
\end{proof}

Following Property \ref{pro:nmineff} 
it is easy to build an $h$-ball $TB^h$, for $h$ sufficiently high, in
which there exist an $n$-simplex $\gamma$ for which   
{$S_{n(n+1)}(\gamma)$} is 
$\morecomp{|S_{n(n+1)}(\gamma)|^2}$.
The property below, as formal statement of this fact, gives a concrete 
reference for our claim for non optimality of $S_{nm}$ we gave in
Section \ref{sec:nmdo}.
\begin{property}
\label{pro:nmineffext}
For any natural $n>0$,  there exist a combinatorial $h$-ball $TB^h$ that
can be embedded in $\real^h$, for $h=4+n$, 
and an $n$-simplex $\gamma$ such that 
the computation of $S_{n(n+1)}(\gamma)$ is 
$\ordcomp{|S_{n(n+1)}(\gamma)|^{2}}$.
\end{property}
\begin{proof}
By Property \ref{pro:nmineff}, for $h=4+n$,  
there exist a combinatorial $h$-ball $TB^h$ that
can be embedded in $\real^h$, 
$f_h(\xclstar{TB^h}{\gamma})=f_0(f_0-3)/2$ with $f_0$ the number
of vertices in $\xclstar{TB^h}{\gamma}$ that are not in $\gamma$.

By Property \ref{pro:compl} we have  that 
the computation of $S_{n(n+1)}(\gamma)$ can be done in
{$\ordcomp{|S_{0h}(v)|}$} being $v$ a vertex in $\gamma$. 
By the way we have built $TB^h$ 
{$|S_{0h}(v)|=f_{h}(\xclstar{TB^h}{\gamma})=f_0(f_0-3)/2$}
and therefore, with some easy algebra, we show that the computation
of  $S_{n(n+1)}(\gamma)$ is 
$\ordcomp{f_0^{2}}$ .i.e.
$\ordcomp{|S_{n(n+1)}(\gamma)|^{2}}$.
\end{proof}
Thus, also the computation of {$S_{nm}(\gamma)$} in an $h$-complex
might become grossly inefficient for $m=n+1$.
Finally, combining the proof above and that of Property \ref{pro:s01} 
it is easy to build an $h$-complex $B^h$, for $h$ sufficiently high, in
which there exist a $n$-simplex $\gamma$ for which   
{$S_{n(n+1)}(\gamma)$} is 
$\ordcomp{|S_{n(n+1)}(\gamma)|^{\floor{h/2}}}$.

\subsubsection{Non-optimal Extraction in non embeddable $3$-manifolds}
The above property shows that the algorithm \ref{algo:snm} is non optimal
in dimension  higher than three.
Lack of optimality is also present in relation extraction for
$3$-complexes if we drop the requirement
on embeddability in $\real^3$. This is an easy consequence
of properties developed in Section \ref{sec:nonimbed}.
\begin{property}
\label{pro:mintriext}
There exist countably many \iqm\ $3$-complexes
$B(g)$, for any natural $g>0$, not embeddable in $\real^3$, 
such that there exist a  vertex $w$ for 
which $|S_{03}(w)|= \ordcomp{|S_{01}(w)|^2}$.
In particular we have
${{(S_{01}^2-3S_{01}+8)}/3}<S_{03}\le {{S_{01}(S_{01}-1)}/3}$.
\end{property}
\begin{proof}
The proof builds easily upon the  result in Property \ref{pro:mintri}
Taking the family of $3$-complexes $B(g)$ mentioned in this property and
the associated vertex $w$
and reasoning as in the proof of Property \ref{pro:nonopt}
we show that in $B(g)$ we have 
$|S_{03}(w)|=\ordcomp{{|S_{01}(w)|}^2}$.
\end{proof}

 \subsection{\Cdecrep\ Data Structure for non-pseudomanifolds}
\label{sec:trix}
In this section we describe a further extension  for
the \cdecrep\ data structure (EWDS) that will be crucial in Section 
\ref{sec:globewds}
to travel the star of a generic $n$-simplex $\gamma$ in an \iqm. 
In fact, in an \iqm\ $h$-complex, the star of a vertex must be 
$(h-1)$-manifold connected. This condition ensures
the correctness of the extraction algorithms \ref{algo:s0h} and
\ref{algo:snm}. Whenever we pretend to travel the star
of an arbitrary $n$-simplex, for $n>0$, we cannot assume this
star to be manifold-connected.
Thus, in this section, we will present a revised version of the 
\cdecrep\ representation that takes into account non-pseudomanifold
adjacencies. This will allow to navigate the star of $\gamma$ whenever this
is $(h-1)$-connected even if it is not $(h-1)$-manifold connected.
Next in section \ref{sec:snmnm} 
we will introduce the $\splitmap$ relation that will allow to jump
from one $(h-1)$-connected component into another $(h-1)$-connected 
component in the star of a given simplex $\gamma$. These two extension will
allow to retrieve in optimal time all top simplices in the star of a
arbitrary simplex.
This extension to the EWDS exploits the fact that some features in this data 
structure were left unspecified or unused. 
In particular we will use  the indexes of the TT relation holding 
the symbol $\trix$ to express non-pseudomanifold adjacencies.

We recall that an \iqm\ $d$-complex in general is not a $d$-pseudomanifold.
In Example \ref{app:example} we have shown that this is the case for $d\ge 3$.
However we can ensure that the star of each vertex, in an \iqm\ $h$-complex,
is manifold-connected and this supports the correctness of the algorithms
developed so far. Thus a first thing to note is that using the $\trix$
simbol for other purposes do not impair neither the correctness nor 
the complexity of the given algorithms that remain the same.

The idea at the basis of this extension is that we can 
use indexes previously set to $\trix$ to connect
the three or more $h$-simplexes sharing the same  $(h-1)$-simplex, in a
non-pseudomanifold  \iqm.
This extension do not require to modify the algorithm for the 
extraction of $S_{0m}$. We simply have to  modify the way in which 
the TT relation is built.
Therefore we adopt the assumption of Section \ref{sec:lists} 
about lists. In particular we recall that is possible
to have a method to retrieve,
in constant time, one after one,  all elements in a list. 
When list $l$ is used to control an iteration  we recall that $l.CircularNext$
can be used to fetch, in the list $l$, the element after the one currently
returned for the iteration,  w.r.t. the circular order induced by the list 
$l$. 

With these assumptions it easy to see that the following algorthm do 
the job and  fills the TT array without using the symbol $\trix$.
In the following we will assume that this algorithm is used instead of 
Algorithm \ref{alg:tfill} to fill the TT array in the EWDS we use.
\begin{algo}{algo:nmtt}{Fill the TT array for non-pseudomanifolds} 
\begin{algorithmic}
\STATE{{\bf var} $\sigma_{\Adj}$: {\bf map set of }{\tt Vertex }{\bf into list of }{\tt TopSimplex}}; 
\STATE $\sigma_{\Adj}\leftarrow (\forall t)[\send{t}{\brk{\mbox{}}}]$ 
\COMMENT{initially the map $\sigma_{\Adj}$ always  returns the empty list}
\FORALL{$t\in\TopSimplex$} \STATE $\gamma\leftarrow$ {\bf set}(TV[$t$])
\FORALL{$v\in\gamma$} 
\STATE ${{\sigma_{\Adj}}[\gamma-\{v\}]}\leftarrow \brk{t}+{{\sigma_{\Adj}}[\gamma-\{v\}]}$
\COMMENT{add simplex $t$ to the list in  ${\sigma_{\Adj}}[\gamma-\{v\}]$}
\ENDFOR
\ENDFOR
\FORALL{$\psi$ in the domain on $\sigma_{\Adj}$}
\STATE $l\leftarrow \sigma_{\Adj}[\psi]$
\IF{$l=\brk{t}$}
\STATE TT[$t$,OPPOSITE(TV,$t$,$\psi$)]=$\bot$ 
\COMMENT{$\bot$ stands for no adjacency}
\ELSE \FORALL{ $t$ in $l$ }
\STATE TT[$t$,OPPOSITE(TV,$t$,$\psi$)]=$l$.CircularNext
\ENDFOR
\ENDIF
\ENDFOR
\end{algorithmic}
\end{algo}
This algorithm has the same requirements and performance of 
the Algorithm \ref{alg:tfill} from which it is derived.
In fact all added operations can be done in constant time and each
added operation sets an entry that the  previous algorithm was
filling with $\trix$. 
A $\trix$ entry is used, in the previous algorithm, when there are  
three or more top simplices $t_1,t_2, \ldots, t_n$
that are jointly adjacent.
In the new TT relation,  
now we have a sequence of $n>2$ 
indexes $j_1,j_2,\ldots,j_n$ such that, whenever in 
the previous algorithm we had TT[$t_{k}$,$j_k$]=$\trix$ now we have:
$t_{k+1}=TT[t_{k},j_k]$ for $1\ge k\ge n$ with $t_{n+1}$=$t_1$.

\begin{remark}\label{rem:cdecrep}
This extension require a slight modification in the definition of the 
\gw\ representation that is the main component of the 
\cdecrep\ representation. 
We recall that, in definition \ref{def:gw}, we required  
that the relation $\Adj\subset \Theta^2$ must be 
a subset of the TT relation for $\AComp$ such that 
$\theta_1\Adj \theta_2$ \siff\
the simplex shared by top simplexes (indexed by) $\theta_1$ and $\theta_2$ is 
a manifold simplex in $\AComp$.
Now we extend this definition by removing this requirement and simply 
ask that the relation $\Adj\subset \Theta^2$ must be 
a subset of the TT relation for $\AComp$.
In the following we will assume that the \cdecrep\ representation satisfy
this, more general, definition.
\end{remark} 
 
\section{The \emi{Non-manifold Layer}}
\label{sec:layertwo}
In the previous section  we have considered the problem of representing 
a single \iqm\ component through the \Cdecrep\ Representation.
This representation is the basis for the lower layer of our two layer data 
structure.
In this section we introduce the \NMWDS\ representation and
give a rationale for this representation.
Next we define a data structure to store this representation
and  evaluate its space requirements.
We will give procedures to
build this data structures using the output of the decomposition process.
Finally  we will show
that the two layers together supports the extraction of all topological
relations in $\lesscomp{n\log{n}}$ for  a $h$-complex for $h\le 3$,
being $n$ the size of the output. 

In general, for $h>3$ for an $h$-complex embeddable in $\real^h$,
the two layer data structure supports the 
extraction of $S_{0m}$ for $(h-2)\le m\le h$ and the
extraction  of $S_{nm}$ for $(h-3)\le n<m\le h$ in $\lesscomp{n\log{n}}$ .
On the other hand, as already shown in Property \ref{pro:s01ext}, for instance,  for $4$-complexes the extraction of relation $S_{01}$ is 
non-optimal.  

\subsection{ The \NMWDS\ Representation}
To give the definition of the \NMWDS\ we need some preliminary notation.
If $\AComp$ is an \asc\ with Vertices in $V$ and top simplices in $\Theta$
we will use the notation {$\canon{V}$} and $\canon{\Theta}$ to denote,
respectively, the set of vertices and top simplices in $\canon{\AComp}$.
We will denote with $\NRA$ the subset 
of non-manifold $n$-simplices $\gamma$ in $\AComp$ having one these 
two problem:
\begin{itemize}
\item the simplex $\gamma$ is a splitting simplex w.r.t. the decomposition $\canon{\AComp}$;
\item the simplex $\gamma$ is not splitting simplex and its star in $\canon{\AComp}$ is not $(h-1)$-connected.
\end{itemize}
Note that the second issue can occur only for an $n$-simplex $\gamma$ for $n>0$. Indeed for $n=0$ we know that the star of a vertex in an $h$-component in $\canon{\AComp}$ is always 
$(h-1)$-connected being $\canon{\AComp}$ an \iqm\ (see Definition \ref{def:iqm} and Property \ref{pro:iqmdecomp}).
We will denote with {$\canon{\NRA}$} the set of all simplex copies
for all simplices in $\NRA$.
Note that non-splitting simplices in $\NRA$ remains in  {$\canon{\NRA}$}, too.
In Figure \ref{fig:tienonstar} there is an example of an \iqm\  3-complex. In this non-manifold complex the star of the orange central edge is made up of the two colored tetrahedra. Thus the orange edge has a  star that is not a 2-connected complex. Therefore, the set $\NRA$ for this complex is the singleton containing the orange edge. 
	\begin{figure}
	\begin{center}
		\psfig{file=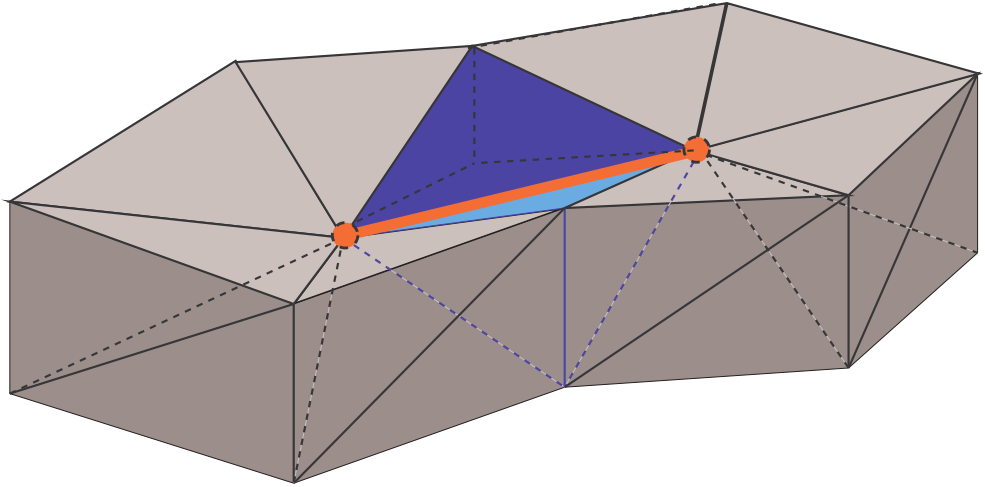,width=0.8\textwidth}
	\end{center}
	\caption{A  3-complex in which the central orange edge has a star that is not 2-connected.
	}
	\label{fig:tienonstar}
\end{figure}
Note that in general, in the following,
we will use primed symbols (e.g. $\gamma^\prime$)
for elements in $\canon{\AComp}$ and non primed symbols for the
corresponding element in $\AComp$.
Whenever the operator $\nabla$ is used in our notation we will define
things so that the element coming from the not decomposed complex must
stay {\em after} $\nabla$, as in $\canon{\AComp}$, while the
elements from $\AComp$ are usually placed {\em before} $\nabla$.
With this idea in mind we can introduce the pair of relations
$\splitsimp0 $ and $\splitstar0 $ that represents the
{\em upper layer} of our two layer representation.

The first relation combines a splitting simplex $\gamma\in\AComp$
with all its simplex copies $\gamma^\prime\in\canon{\AComp}$.
We will write $\xsplitsimp{\gamma}{\gamma^\prime}$ to mean that
$\gamma^\prime$ is a simplex copy for $\gamma$.

The second relation $\splitstar0 $ is a subset of the
restriction of the face relation to the set
${(\canon{\Theta})\times(\canon{\NRA})}$. This simply means that whenever
$\xsplitstar{\gamma^\prime}{\theta^\prime}$ we must have that
${\gamma^\prime}\in \canon{\NRA}$ and
${\theta^\prime}\in \canon{\Theta}$ and $\theta^\prime\ge\gamma^\prime$.
This sub relation is not  uniquely identified but it must satisfy
the following requirements:
\begin{itemize}
	\item
	first we ask 
	$\xsplitstar{\gamma^\prime}{\theta^\prime}$ for some top
	simplex $\theta^\prime\in\canon{\Theta}$
	\iff\ $\xclstar{\canon{\AComp}}{\gamma^\prime}$
	has at least two $(h-1)$-connected components;
	\item second, we ask that
	for each $(h-1)$-connected component in
	the $\xclstar{\canon{\AComp}}{\gamma^\prime}$
	there must be one and only one top simplex $\theta^\prime\in\canon{\AComp}$
	such that $\xsplitstar{\gamma^\prime}{\theta^\prime}$;
\end{itemize}

With the above notations we can define the \NMWDS\  Representation as follows:
\begin{definition}
\label{def:nmwds}
If $\AComp$ is an \asc\ with Vertices in $V$ and top simplices in $\Theta$
then the \emad{\NMWDS}{Representation} for $\AComp$ is a triple 
{$NMWS=(EWS,\splitsimp0 ,\splitstar0 )$} where:
\begin{itemize}
\item $EWS$ is an \Cdecrep\ Representation for $\canon{\AComp}$;
\item the relation $\splitsimp0 \subset(\canon{\AComp})\times\AComp$
is such that$\xsplitsimp{\gamma}{\gamma^\prime}$ 
\iff\ $\gamma$ is a splitting simplex and $\gamma^\prime$ is one of its
simplex copies;
\item the relation  $\splitstar0 \subset(\canon{\Theta})\times(\canon{\NRA})$ 
is such that the following conditions are satisfied:
\begin{enumerate}
\item for each $(h-1)$-connected component in
the star  $\xstr{\canon{\AComp}}{\gamma^\prime}$ 
there exist a top simplex $\theta^\prime\in\canon{\AComp}$ 
for which $\xsplitstar{\gamma^\prime}{\theta^\prime}$;  

\item if $\xsplitstar{\gamma^\prime}{\theta_1^\prime}$ and
$\xsplitstar{\gamma^\prime}{\theta_2^\prime}$ then $\theta_1$ and
$\theta_2$ must belong to two  distinct $(h-1)$-connected component in
the star  $\xstr{\canon{\AComp}}{\gamma^\prime}$ 
\end{enumerate}
\end{itemize}
\end{definition}
We will refer to the couple of relations $\splitsimp0 $ and $\splitstar0 $
as the {\em upper layer} of our two layer representation.
On the other hand the $EWS$ \Cdecrep\ Representation  will be referred as the
{\em lower layer} of the  \NMWDS\ Representation

To define a data structure for this representation we assume that is possible
to represent the complex $\canon{\AComp}$ with a global \Cdecrep\ 
Data Structure obtained by the merge of all data structures for 
each connected component in $\canon{\AComp}$.
We assume that exists a data type named EWDS for such a data structures.
In particular
if $NV'$ and $NT'$ are respectively the number of vertices and top 
simplices in {$\canon{\AComp}$}. 
Obviously $NT=NT'$ but we choose to use two different symbols to have a uniform notation.  We assume that the declaration
${\bf var} EWS:EWDS(NV',NT')$ expands to the declaration of
a valid data structure to encode {$\canon{\AComp}$}.
The data structure for EWDS is developed  from data structure in \ref{data:tttv}.
The EWDS data structure and its optimization will be discussed in the
following Section \ref{sec:globewds}.

With these assumptions the data structure encoding the \NMWDS\ is the following:{
\begin{data}{data:nmwds}{\NMWDS\ Data Structure for  {$\AComp$}} 
{
\begin{tabbing}
{\bf type}\=\\
 \>Vertex = [1..NV];{$\{${\em range of indexes for vertices of\mbox{}  $\AComp$} $\}$}\\
 \>TopSimplex = [1..NT];{$\{${\em range of indexes for top simplices of\mbox{}  $\AComp$} $\}$}\\
\>Simplex = {\bf set of} Vertex;{$\{${\em a generic simplex of\mbox{} $\AComp$ represented as a set of vertex indexes} $\}$}\\ 
 \>Vertex' = [1..NV'];{$\{${\em range of indexes for vertices of\mbox{}  $\canon{\AComp}$} $\}$}\\
 \>TopSimplex' = [1..NT'];{$\{${\em range of indexes for top simplices of\mbox{}  $\canon{\AComp}$} $\}$}\\
\>Simplex' = {\bf set of} Vertex';{$\{${\em a generic simplex of\mbox{} $\canon{\AComp}$ represented as a set of vertex indexes} $\}$}\\ 
{\bf var}\=\\
\>EWS\=: EWDS(NV',NT');\\
\>\>{\em $\{$ A Global \Cdecrep\ Data Structure for $\canon{\AComp}$ 
(see Section \ref{sec:globewds})$\}$}\\
\>$\sigma$\=:
{\bf map of} Vertex' {\bf into} Vertex;
$\sigma^{-1}$\=: {\bf map of} Vertex {\bf into set of} Vertex';$\{${\em See Section \ref{sec:sigma}}$\}$\\
\>\>{$\{${\em $\sigma[v^\prime]$ maps a vertex copy $v^\prime$ into the corresponding splitting vertex in ${\AComp}$} $\}$}\\
\>\>{$\{${\em $\sigma^{-1}[v]$ maps a splitting vertex $v$ into the
set of its corresponding vertex copies in $\canon{\AComp}$} $\}$}\\
\>$\splitmap$\=:
{\bf map of} Simplex {\bf into} ({\bf map of} Simplex' {\bf into set of} TopSimplex');\\
\>\>{\em $\{\theta\in\splitmap[\gamma][\gamma^\prime]$ \iff\
($\gamma=\gamma^\prime \mbox{ \bf or } \xsplitsimp{\gamma}{\gamma^\prime}) \mbox{ \bf and } 
\xsplitstar{\gamma^\prime}{\theta^\prime}\}$ (see Section \ref{sec:splitmap})}  
\end{tabbing}
}
\end{data}
}
As reported before, note that in this data structure definition, we have used primed identifiers (e.g. Vertex') 
to denote elements that refers to the decomposition $\canon{\AComp}$.
Similarly will use primed letters, e.g. $v^\prime$, 
$\gamma^\prime$ to denote elements in the decomposition $\canon{\AComp}$.
Plain letters and identifiers (e.g. Vertex) are used for elements that 
refers to $\AComp$.
{
	\begin{figure}
		\begin{center}
			\psfig{file=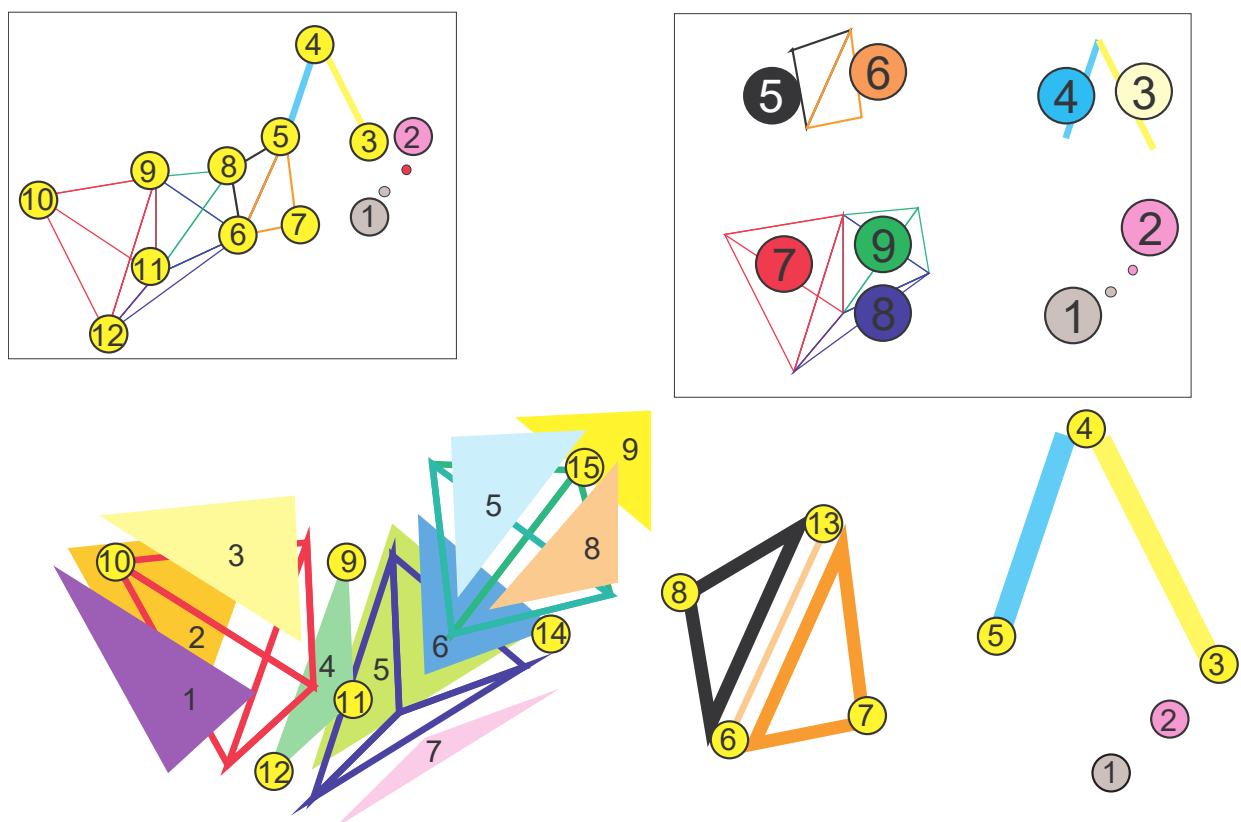,width=\textwidth}
		\end{center}
		\caption{A  3-complex $\AComp$ used for a running example of the  implementation of the 
		\emad{\NMWDS}{Representation}.
		On the top left we have the un-decomposed complex $\AComp$. Note that we do not stitch to the definition On the top right the decomposed complex $\canon{\AComp}$ with top simplices numbering. In larger bottom figure vertices numbers are given in yellow balls. Note that, due to the addition of dimensionally mixed parts, the vertices and simplices numbers are different from those in Figure \ref{fig:datastrex}. All other numbers are for $(h-1)$ simplices in the $h$-complexes in the decomposition.
		}
		\label{fig:datastrexbig}
	\end{figure}

}
\begin{example}
With reference to Figure \ref{fig:datastrexbig} we can begin a chain of running examples where we apply the concepts so far defined. Up to now we can say that for this complex the first part of the data structure rewrites as
\begin{tabbing}
	{\bf type}\=\\
		\>TopSimplex = [1..9];{$\{${\em range of indexes for vertices of\mbox{}  $\AComp$} $\}$}\\
	\>Vertex = [1..12];{$\{${\em range of indexes for vertices of\mbox{}  $\AComp$} $\}$}\\
	\>Vertex' = [1..15];{$\{${\em range of indexes for vertices of\mbox{}  $\canon{\AComp}$} $\}$}\\
	\>TopSimplex' = [1..9];{$\{${\em range of indexes for top simplices of\mbox{}  $\canon{\AComp}$} $\}$}
\end{tabbing}
being NV=12, NT=9,NV'=15 and NT=NT'=9.
\label{ex:datastrexbig}
\end{example}
\subsection{A Global Data Structure for $\canon{\AComp}$}
\label{sec:globewds}
We can build, quite easily, a single \Cdecrep\ data structure
EWDS(NV',NT') that
contains the representation for all the \iqm\ components of $\canon{\AComp}$.
Let $\AComp$ be a $d$-complex with Vertices in $V$ and top simplices
in $\Theta$ (note that $d$ is used for the dimension of the complex ${\AComp}$ and $h$ for the dimension of each component in $\canon{\AComp}$).
A global data structure for $\canon{\AComp}$ is the following:
{
\begin{data}{data:globtt}{Global \Cdecrep\ Data Structure for  {$\canon{\AComp}$}} 
{
\begin{tabbing}
{\bf type}\=\\
\> Vertex' = [1..NV'];\\
\> TopSimplex' = [1..NT'];\\
\> INDEX =[1..SIZE];\\
{\bf var}\=\\
\> TV':array[\mbox{}][\mbox{}][\mbox{}] of Vertex';\\
\> TT':array[\mbox{}][\mbox{}][\mbox{}] of TopSimplex';\\
\> VT*':array[1..NV'] of TopSimplex';\\
\> TBase,TBaseAddr:array[0..D] of INDEX;\\
\end{tabbing}
}
\end{data}
}
Where $NT^\prime$ and $NV^\prime$ are, respectively the number of
top simplices and of vertices in the decomposition $\canon{\AComp}$.
{
To build a global data structure for $\canon{\AComp}$
we simply have to  take the data structures for each components in
$\canon{\AComp}$ (see the Data Structure \ref{data:tttv}) and 
assume that the ranges of indexes  $\Vertex=[\MinV,\ldots,\MaxV]$ and 
$\TopSimplex=[\MinT,\ldots,\MaxT]$ for different components 
are a partition of the
the  larger ranges, 
$[1,\ldots,NV^\prime]$ and $[1,\ldots,NT^\prime]$.

We can assume that space for the TT, TV and TV* arrays for each 
decomposition components is allocated in non overlapping sub-areas
within the areas for the corresponding \TTP, \TVP and TV*' arrays.
We assume that allocation   starts from components of dimension $0$ and goes 
up.
That is to say elements in lower dimensional components receive lower indexes.

Note that within the \TTP\ and \TVP\ arrays coexist slices of different
length. This require some ancillary data structures for indexing.
To this goal ve provided  the $\TBase$ and the $\TBaseAddr$ arrays.
We discuss in the next lines details of this allocation mechanism.
With the notation
{\tt TV':array[][][] of Vertex';} and
{\tt  TT':array[][][] of TopSimplex';} we denote arrays whose three
ranges are not known at compile time. These array will be allocated 
dynamically to the size given by 
$\SIZE=\sum_{0\le h\le d}{|\Theta^{[h]}|(h+1)}$,
where with  {$\Theta^{[h]}$} we denote the set of top $h$-simplices.
In the following discussion we assume that we can use each of these two 
arrays as large unidimensional arrays $\TTP[1,\ldots,\SIZE]$ and 
$\TVP[1,\ldots,\SIZE]$.

The occupation of the Global \Cdecrep\ Data Structure is given by the
following property:
\begin{property}
\label{pro:ewdsocc}
Let $\AComp$ a $d$-complex with top simplices in $\Theta$ and vertices in
$V$.
Let $NS$ and $NC$ be respectively the number of splitting vertices
and the number all vertices that are the vertex copies of the $NS$ splitting vertices  \cano\ decomposition
$\canon{\AComp}$. In this situation    
the \TTP, \TVP, VT*' arrays in the Global \Cdecrep\ Data Structure takes 
$$\SIZE\cdot(\log{NT'}+\log{NV'})+NV'\log{NT^\prime}\, \mbox{\rm bits}$$ where:
$\SIZE=\sum_{0\le h\le d}{|\Theta^{[h]}|(h+1)}$,
with  {$\Theta^{[h]}$} the set of top $h$-simplices in $\AComp$.
$NT'=\sum_{0\le h\le d}{|\Theta^{[h]}|}$ is the number of top simplices
in $\AComp$ and $NV'=|V|-NS+NC$.
\end{property}
{ 
\begin{proof}
It is easy to see that the number $NV'$ of vertices in $\canon{\AComp}$ is
given by  $NV'=|V|-NS+NC$.
On the other hand the number of top  simplices is NT' both in $\AComp$ and in
$\canon{\AComp}$.
The \TTP, \TVP, VT*' arrays in the global data structure takes 
$\SIZE\cdot(\log{NT'}+\log{NV'})+NV'\log{NT^\prime}$ bits to store $\canon{\AComp}$.
This formula comes easily from thee fact that 
 $\log{NV'}$ bits are used to code a pointer to a vertex for each  top simplex
in the TV' array.
Thus an $h$ top simplex takes $(h+1)\log{NV'}$ bits and all $h$ top simplices take
 $|\Theta^{[h]}|(h+1)\log{NV'}$. Summing for all $h$ we obtain  $\SIZE\cdot(\log{NV'})$ bits occupation for \TVP.
 Then $\log{NT'}$ bits are necessary to code a reference to a top simplex in
 the TT' table, so
 summing all over the TT' table we obtain an occupation of $\SIZE\cdot(\log{NT'})$ bits. 
 Then
Finally the VT*' table has NV' entries each storing a pointer to a top simplex taking $\log {NT^\prime}$ bits for an overall sum of  $NV'\log{NT^\prime}$.
This completes the proof.
\end{proof}
An obvious possible space optimization can be implemented by coding these
$\log{NT'}$ and $\log{NV'}$ 
pointers as displacements within the ranges of 
$\Vertex=[\MinV,\ldots,\MaxV]$ and $\TopSimplex=[\MinT,\ldots,\MaxT]$. 
}

In the array element $\TBase[h]$ we store the index assigned to 
the first top $h$-simplex.
In the array element {$\TBaseAddr[h]$} we store the base address to access
data for $h$-simplices  in \TTP\ and \TVP\ arrays.
More precisely {$\TBaseAddr[h]$} stores  the index of the last 
element of the last slice of $h$ elements within the \TTP\ and \TVP\ arrays. 
{
The array $\TBase$ is filled when we build the global data structure 
for $\canon{\AComp}$ by counting the number of top $h$-simplices inserted.
We will put  $\TBase[h]=\TBase[h+1]$  if no top
$h$-simplex  exist. 
When the array $\TBase[h]$ is filled the array $\TBaseAddr[h]$ is
computed using the recurrence $\TBaseAddr[0]=1$ and
$$\TBaseAddr[h+1]=\TBaseAddr[h]+(h+1)(\TBase[h+1]-\TBase[h])\label{eq:tbaseaddrfill}$$
}
For a top $h$-simplex that received as index $t^\prime$ 
we will retrieve the TT slice of length $h+1$ starting beyond 
{\TTP[\TBaseAddr[$h$]+$(t^\prime-\TBase[h])(h+1)$]}.
Similarly 
we will retrieve the TV slice of length $h+1$ starting beyond 
{\TVP[\TBaseAddr[$h$]+$(t^\prime-\TBase[h])(h+1)$]}.
In the following we will forget these access details and use the shortcuts:
\begin{equation}
\label{eq:ttcomp}
\TTP[h,t^\prime,k]=\\{\TTP[\TBaseAddr[h]+(t^\prime-\TBase[h])(h+1)+k-1]} \mbox{ with } 1\le k \le h+1
\end{equation} 
\begin{equation}
\label{eq:tvcomp}
\TVP[h,t^\prime,k]=\\{\TVP[\TBaseAddr[h]+(t^\prime-\TBase[h])(h+1)+k-1]} \mbox{ with } 1\le k \le h+1
\end{equation}
Note that this is just a convenient notation and must be implemented
directly (for instance using macros in C).  
In fact such an array, where the range of the second and third index 
depends upon the value of the first index are not supported by
standard programming languages.

{
We note that the dimension of a top simplex $t^\prime$ can be computed 
in $\ordcomp{d}$ by finding the index $h$ for which 
$\TBase[h]\le t^\prime<\TBase[h+1]$.
Thus in the following we will use the notation $\ord(t^\prime,\canon{\AComp})$
to denote the result of this computation, i.e. the dimension of
the simplex (indexd by) $t^\prime$.
In general 
we will denote with {$\ord(v^\prime,\canon{\AComp})$}
the dimension of the component of $\canon{\AComp}$ to which $v^\prime$
belongs to.  
The function $\ord(v^\prime,\canon{\AComp})$ can be computed in $\ordcomp{d}$ with using
the \VTS\ relation with
the formula 
$\ord(v^\prime,\canon{\AComp})=\ord(\sigma_{VT^\star}(v^\prime),\canon{\AComp})$.
Similarly for a simplex $\gamma^\prime\in\canon{\AComp}$ we will define  
{$\ord(\gamma^\prime,\canon{\AComp})$}  
 as 	  
{$\ord(\gamma^\prime,\canon{\AComp})=\ord(v^\prime,\canon{\AComp})$}  
for some $v^\prime\in\gamma^\prime$.
When this is not ambiguous we will use simply {$\ord(\gamma^\prime)$}
as a shortcut for {$\ord(\gamma^\prime,\canon{\AComp})$}
}
}
\begin{example}
\label{ex:datastrexbig0}
We continue our running example started at Example \ref{ex:datastrexbig}. W.r.t. Figure \ref{fig:datastrexbig} we can say that, in the situation of Property \ref{pro:ewdsocc} we have $d=3$, NS=3 and NC=6. The top simplices are of order 0,1,2,3 and we have $\Theta^{[0]}=2$, $\Theta^{[1]}=2$, $\Theta^{[2]}=2$ and  $\Theta^{[3]}=3$. Therefore SIZE must be 24. The arrays  \TBase and \TBaseAddr\ are filled as shown in Figure \ref{fig:bigtv}.
To play with them we can try to develop the expression in Formula \ref{eq:tvcomp}
\begin{eqnarray*}
\TVP[2,5,3]&=&{\TVP[\TBaseAddr[2]+(5-\TBase[2])(2+1)+3-1]}\\
&=&
\TVP[7+(5-5)3+3-1]=\TVP[9]=8
\end{eqnarray*} i.e. the third vertex of 2-simplex 5 is 8.
Next we report in Figure \ref{table:tbase} the complete filling of \TVP\ and \TTP.
\end{example}

\begin{figure}[h]
	\begin{center}
		\begin{tabular}{|c||c|}\hline
			\multicolumn{2}{|c|}{\TBase[0..3]} \\ \hline
			0&1    \\ \hline 
			1&3    \\ \hline
			2&5   \\ \hline
			3&7    \\ \hline
		\end{tabular}
		\mbox{ }
	\begin{tabular}{|c||c|}\hline
		\multicolumn{2}{|c|}{\TBaseAddr[0..3]} \\ \hline
		0&1    \\ \hline 
		1&3    \\ \hline
		2&7   \\ \hline
		3&13    \\ \hline
	\end{tabular}\mbox{ }\\
\vspace*{0.3cm }	
	\begin{tabular}{|c|c||c|c|c|c|c|c|}\hline
		Simplex&@&\multicolumn{4}{|c|}{\TVP[1..24]}&{Comment} \\ \hline
		1&1&1&\multicolumn{3}{|c|}{}&{\TBaseAddr[0]=1}    \\ \hline 
		2&2&2&\multicolumn{3}{|c|}{} &  \\ \hline
		3&3:4&3&4&\multicolumn{2}{|c|}{}&{\TBaseAddr[1]=3}    \\ \hline
		4&5:6&4&5&\multicolumn{2}{|c|}{}&    \\ \hline
		5&7:9&6&13&8&&{\TBaseAddr[2]=7}    \\ \hline
		6&10:12&6&7&13&&{}    \\ \hline
		7&13:16&10&9&12&11&{\TBaseAddr[3]=13}    \\ \hline
		8&17:20&9&12&11&14&{}    \\ \hline
		9&21:24&9&11&14&15&{SIZE=24}    \\ \hline
	\end{tabular}
\end{center}
\label{fig:bigtv}
	\caption{An example of arrays \TBase,  \TBaseAddr, \TVP  for the 3-complex of Figure \ref{fig:datastrexbig}. Note that in these tables the values in contiguous cells in the  array \TVP, whenever needed, are grouped horizontally in slices. Thus $x:y$ on the @ column corresponds to $y-x+1$ cells on the right, showing values stored at $x,x+1,\ldots,y$. Vector VT*' is represented horizontally simply to save space. } 	
\end{figure}	
\mbox{ }
\begin{figure}
	\begin{center}
	\begin{tabular}{|c|c|c|c|c|c|c|c|c|c|c|c|c|c|c|}\hline
\multicolumn{15}{|c|}{VT*'[1..15]} \\ \hline\hline
	                 1&2&3&4&5&6&7&8&9&10&11&12&13&14&15   \\ 
	                 \hline
	                 1&2&3&3&4&6&6&5&7&7&7&7&5&8&9
	    \\ \hline
\end{tabular}
	
\mbox{ }\\
\vspace*{0.3cm }	
	\begin{tabular}{|c|c||c|c|c|c|c|c|}\hline
	Simplex&@&\multicolumn{4}{|c|}{\TTP[1..24]}&{Comment} \\ \hline
	1&1&$\bot$&\multicolumn{3}{|c|}{}&{\TBaseAddr[0]=1}    \\ \hline 
	2&2&$\bot$&\multicolumn{3}{|c|}{} &  \\ \hline
	3&3:4&4&$\bot$&\multicolumn{2}{|c|}{}&{\TBaseAddr[1]=3}    \\ \hline
	4&5:6&$\bot$&3&\multicolumn{2}{|c|}{}&    \\ \hline
	5&7:9&$\bot$&$\bot$&6&&{\TBaseAddr[2]=7}    \\ \hline
	6&10:12&$\bot$&5&$\bot$&&{}    \\ \hline
	7&13:16&8&$\bot$&$\bot$&$\bot$&{\TBaseAddr[3]=13}    \\ \hline
	8&17:20&$\bot$&9&$\bot$&8&{}    \\ \hline
	9&21:24&$\bot$&$\bot$&$\bot$&8&{SIZE=24}    \\ \hline
\end{tabular}		
		\caption{An example of arrays \TTP and VT*' for the 3-complex of Figure \ref{fig:datastrexbig}. Vector VT*' is represented horizontally simply to save space. } 
		\label{table:tbase}		
	\end{center}
\label{fig:bigtt}
\end{figure}
\begin{example}
	\label{ex:datastrexbig1}
	We continue our running example  from  Example 	\ref{ex:datastrexbig0}. W.r.t. Figure \ref{fig:datastrexbig} the arrays  VT*' and \TTP\ are filled as shown in Figure \ref{fig:bigtt}.
	To play with them we can try to develop the expression in Formula \ref{eq:ttcomp}
		\begin{eqnarray*}
\TTP[2,5,3]&=&{\TTP[\TBaseAddr[2]+(5-\TBase[2])(2+1)+3-1]}\\
&=&\TTP[7+(5-5)3+2]=\TTP[9]=6
		\end{eqnarray*}
i.e. the top 2-simplex adjacent to top 2-simplex 5 opposite to \TVP[2,5,3] is 6. Indeed, from Example \ref{ex:datastrexbig1} we have 
	 \TVP[2,5,3]=8 and looking at Figure  \ref{fig:datastrexbig} we can see that the triangle adjacent to 5 opposite to 8 is 6.
\end{example}
In the next section, 
we will discuss a possible space optimization  
that brings the size
of our global data structure below the usual reference limit, 
attained by  the Winged edge, of six pointers per triangle. 

\subsection{Splitting Vertices: Maps $\sigma$ and $\sigma^{-1}$ as a  $\split0 $ restriction  }
\label{sec:sigma}
In this section we start to discuss the relation $\split0 $ in the \NMWDS\ 
Representation (see Definition \ref{data:nmwds}).
In particular we will introduce maps $\sigma$ and $\sigma^{-1}$.
These two maps are mentioned 
in the \NMWDS\ Data Structure \ref{data:nmwds}.
We will detail them here and define them as the
restriction of relation {$\xsplit{\gamma}{\gamma^\prime}$} to pair of vertices. We first introduce some ideas and then state them formally giving a proof in Property \ref{pro:sigmaspace}.

To this aim we recall that, upon termination of the decomposition process, we are left with a simplicial complex 
$\canon{\AComp}$ that represents the decomposition (see Algorithm \ref{algo:decomp}).
The decomposition Algorithm \ref{algo:decomp} can be patched to return some ancillary data structures. We will introduce here the patches and discuss the (negligible) impact on time complexity.

Indeed, upon completion of the decomposition process, 
we need to have a  map $\funct{\sigma}{{\canon{V}}}{V}$ 
that gives, for each vertex copy $v^\prime$ in $\canon{\AComp}$ the corresponding splitting vertex in  $\AComp$. 
We also assume that the decomposition process 
gives a set of Vertices $V_{\NRA}$ with potential problems. 
The set $V_{\NRA}$ and its construction will be described more precisely 
in Section \ref{sec:snmnm}.  

We model the splitting process by the  $\split0  $ relation
in the  upper layer of our representation. The relation $\split0 $ is  a
non symmetric relation $\split0 \subset (\canon{V})\times V$.
When restricted to Vertices the relation $\split0 $ 
relates each splitting vertex $v$ in $\AComp$, 
with the set of its vertex copies in $\canon{\AComp}$. 
In other words {$\xsplit{v}{v^\prime}$} \siff\ $v$ is a splitting vertex and
$v^\prime$ is one of its vertex copies.
More in general $\split0 $ is a  relation between pairs of simplices,
i.e.    $\split0 \subset (\canon{\AComp})\times\AComp$.
We will write $\xsplit{\gamma}{\gamma^\prime}$
whenever $\gamma^\prime$ is a splitting simplex for $\gamma$.

In the \NMWDS\ Data Structure \ref{data:nmwds}
we represent separately the restriction of  $\split0 $ to set of vertices, i.e. the restriction of  
$\split0 $ to  $(\canon{V})\times V$. This portion of the relation
$\split0 $ is represented with the two maps $\sigma^{-1}$ and 
${\sigma}$.
The map $\sigma$ can be constructed with a minor modification of the Algorithm \ref{algo:decomp}  and the other  map
$\funct{\sigma^{-1}}{\canon{V}}{V}$ is defined by the condition 
$v^\prime\in\sigma^{-1}[\sigma[v^\prime]]$.

We will extend $\sigma^{-1}$ and $\sigma$ to all Vertices in $\AComp$ and $\canon{\AComp}$
by assuming $\sigma^{-1}(v)=\{v\}$ whenever
$v$ is not a splitting vertex.
Similarly we will pose  $\sigma(v^\prime)=v^\prime$ whenever $v^\prime$ is not a  
vertex copy.
Given a simplex copy $\gamma^\prime$ in the 
decomposed complex we will denote with $\sigma(\gamma^\prime)$ 
its translation into the original complex 
(i.e. $\cup_{v^\prime\in\gamma^\prime}{\sigma({v^\prime})}$).
Similarly given a set of simplices $\Gamma\subset\canon{\AComp}$ we
define $\sigma(\Gamma)=\{\sigma(\gamma)|\gamma\in\Gamma\}$. These extensions are used for convenience in proofs and statements but it is not necessary to implement them, now. Actually only $\sigma$ and $\sigma^{-1}$ are detailed here.

The requirements for construction and storage of $\sigma$ and $\sigma^{-1}$ are given by
the following  property.
\begin{property} 
\label{pro:sigmaspace}
Let $NS$ be the number of splitting Vertices in $\AComp$ and
let  $NC$ be the total number of vertex copies introduced by the 
decomposition process. 
In this situation the following facts holds:
\begin{enumerate}
\item 
\label{pro:sigmaspace1}
the construction of the map $\sigma$ takes $\ordcomp{NC\log{NC}}$;
\item 

the construction of the map $\sigma^{-1}$ takes $\ordcomp{NC\log{NC}}$; 
\item 
the map $\sigma^{-1}$ takes up 
to $NS\log{NS}+NC\log{NC}$ bits to be encoded.
\item 
the map $\sigma$ takes up 
$NC(\log{NC}+\log{NS})$ bits to be encoded.
\end{enumerate}
\end{property}

\begin{proof}
We note, as detailed in \ref{sec:map}, that a standard implementation for maps can attain a logarithmic
access time. 

Looking at Algorithm \ref{algo:decomp} it is easy to see that step 10 can be modified to build 
the $\sigma$ map with time complexity $\ordcomp{NC\log{NC}}$.
Indeed each vertex copy deserve just one attempt to be stored in  $\sigma$. This storage is done in logarithmic time. Hence  the map $\sigma$ can be constructed in  $\ordcomp{NC\log{NC}}$. This proves part  \ref{pro:sigmaspace1}. 
The original Algorithm \ref{algo:decomp} time complexity was {$\lesscomp{\factorial{d}\cdot (NT\log{NT}})$}.
Where $NT$ is the  number of top simplices in the $d$-complex $\AComp$ (see \ref{pro:decocomp}). Being $NC\le (d+1)\cdot NT$, we can say that this patch to line 10 of Algorithm \ref{algo:decomp} do not change its time complexity.

For proving the other parts we note that a 
standard implementations of maps usually supports sequential access to
the elements of the domain of the map in constant time.
Each entry in $\sigma^{-1}$  is a set of vertex copies and
again we assume a standard implementation for these sets with a logarithmic
access time. 
Thus, scanning the domain of the  map $\sigma$, 
it is easy to build the map $\sigma^{-1}$ in $\ordcomp{NC\log{NC}}$. 
This is done by the following fragment of code.
\par
\begin{algorithmic}
\FORALL{$v^\prime\in\domain(\sigma)$}
\STATE{${\sigma^{-1}[\sigma[v^\prime]]}\leftarrow
 {\sigma^{-1}[\sigma[v^\prime]]}\cup\{v^\prime\}$}
\ENDFOR
\end{algorithmic}
This is done in $\lesscomp{NC\log{NC}}$.
In fact, it is easy to see,
that this loop is executed once for each vertex copy
$v^\prime$ and thus the loop gets executed $NC$ times.
At each iteration we retrieve, modify and store back the entry
${\sigma^{-1}[\sigma[v^\prime]]}$. Modification is done in constant time and
store and retrieval takes $\lesscomp{\log{NS}+\log{NC}}$, that is
$\lesscomp{\log{NC}}$.

To analyze space requirements for these map we note that maps can be
implemented with {\em binary search trees} (BST)  
that in turn can be implemented using {\em heaps} i.e.
maintaining the BST a  complete binary tree. 
Thus it can be represented with an array with
just one index for each node in the tree. 
Thus, if $NS$ be the number of splitting Vertices in $\AComp$ then the
map $\sigma^{-1}$ will take $NS\log{NS}$ bits for the indexing structure for this map.
Similarly sets in map entries can be implemented as complete BST.
Then we have that all sets takes less than
$NC\log{NC}$ bits for all the sets of vertex copies. 
With these assumptions we have that $\sigma^{-1}$ takes up to
$NS\log{NS}+NC\log{NC}$ bits to be encoded.
No sets are needed for $\sigma$, being $\sigma(v^\prime)$ a single vertex in $V$.
Thus, the space requirement for  the map $\sigma$ will be
$NC(\log{NC}+\log{NS})$.
\end{proof}
Finally we note that our data structure  the \NMWDS\ 
representation (see Definition \ref{data:nmwds}) implicitly requires that vertex index used for a vertex $v$ in $\AComp$ is used again in $\canon{\AComp}$ even if $v$ is a splitting vertex.
So we assume that, when generating an index for a new vertex copy $v^\prime$,  when splitting  $v$, we store in $\sigma$ the same index for $v$ for the first copy and a brand new index only at the second try.
With this last assumption we can draw a small example for the $\sigma$ and $\sigma^{-1}$ maps.
\begin{example}
 	\label{ex:datastrexbig2}
 	We continue our running example  from  Example 	\ref{ex:datastrexbig1}. W.r.t. Figure \ref{fig:datastrexbig} we have that the two maps $\sigma$ and $\sigma^{-1}$ must be.
\begin{center}
	 	\begin{tabular}{|c||c|}\hline
		$v$&$\sigma^{-1}$
		\\ \hline \hline
		5&$\{5,13\}$    \\ \hline
		6&$\{6,14\}$   \\ \hline
		8&$\{8,15\}$    \\ \hline
	\end{tabular}
\mbox{ }  
	 	\begin{tabular}{|c||c|}\hline
	$v^\prime$&$\sigma$
	\\ \hline \hline
		5&$5$    \\ \hline
	6&$6$   \\ \hline
	8&$8$    \\ \hline
		13&$5$    \\ \hline
	14&$6$   \\ \hline
	15&$8$    \\ \hline
\end{tabular}  
\end{center} 	

\end{example}
This completes the description and analysis of the two maps $\sigma$ and $\sigma^{-1}$
that are the first 
two elements in the upper layer of our representation.
Using the data structures $\sigma$ and $\sigma^{-1}$ we can devise algorithms
to extract, in optimal time, the topological relations $S_{0m}(\canon{\AComp},v^\prime)$, for
$0<m\le h$.

\subsubsection{Extraction of $S_{0m}$ in the Non-manifold Complex}
The two maps $\sigma$ and $\sigma^{-1}$ are sufficient to perform 
vertex based queries and navigation of a non-manifold complex $\AComp$
using the \Cdecrep\ Representation of the decomposition $\canon{\AComp}$.
In fact starting from a vertex $v$ in $\AComp$ we can find all the 
topological relations $S_{0m}(\AComp,v)$ in $\AComp$ using 
the topological relations ${S_{0m}}$ in $\canon{\AComp}$
following formula:
\begin{equation}\label{eq:nmnav1}
{S_{0m}(\AComp,v)}=\bigcup_{v^\prime\in\sigma^{-1}[v]}{\sigma(S_{0m}(\canon{\AComp},v^\prime))}
\end{equation}
We must turn this formula into an algorithm with some care.
The problem that requires some care is related to non
regularity in the $\AComp$  complex. Indeed  we must not perform the 
computation of $S_{0m}(\canon{\AComp},v^\prime)$ for those vertex copies 
$v^\prime$ that fall in an \iqm\  of dimension strictly smaller than $m$.

With this caveat it is easy to translate  formula
\ref{eq:nmnav1}  into an algorithm 
and it is easy to see that the computation of {$S_{0m}(\AComp,v)$} 
can be done in optimal
time only if {$S_{0m}(\canon{\AComp},v^\prime)$} can be done in optimal time.
However this condition might not be sufficient. We will see that this is a
sufficient condition for $3$-complexes.
In general, the algorithm for the computation of ${S_{0m}(\AComp,v)}$ is the following:
\begin{algo}{algo:somnav}{Computation of {$S_{0m}(\AComp,v)$}}
\begin{algorithmic}
\STATE{{\bf Function} $S_{0m}(v: {\tt  Vertex})$ 
{\bf returns } {\bf set of} {\tt Vertex}}
\IF[$v$ is a splitting vertex]{$v\in\domain(\sigma)$}
\STATE $VertexCopies\leftarrow\sigma^{-1}[v]$;
\ELSE
\STATE $VertexCopies\leftarrow\{v\}$;
\ENDIF
\STATE $Result\leftarrow\emptyset$;
\FORALL{$v^\prime\in VertexCopies$ {\bf and }{$\ord{(v^\prime,\canon{\AComp})}\ge m$}}
\STATE{$Result\leftarrow Result\cup \sigma(S_{0m}(\canon{\AComp},v^\prime))$}
\COMMENT{refer to Algorithm \ref{algo:snm}
for {$S_{0m}(\canon{\AComp},v^\prime)$}}
\ENDFOR
\STATE {\bf return} $Result$;
\end{algorithmic}
\end{algo}
According to the forthcoming results of Property \ref{pro:nmopt} we have the
following property. 
\begin{property}
\label{pro:somopt}
Let  $\AComp$ be a  $d$-dimensional \asc\  and let $v$ be a vertex 
in $\AComp$.
In this situation the relation {$S_{0m}({\AComp},v)$} in $\ordcomp{n\log{n}}$,
being $n$ the size of the output in the following cases:
\begin{itemize}
\item for $d=2$ and $1\le m\le 2$;
\item for $d=3$ and $1\le m\le 3$ whenever $\AComp$ is embeddable in $\real^3$;
\end{itemize}
\end{property}
\begin{proof}
This property is a particular case of Property \ref{pro:nmopt}. The latter
can be proven independently of this one, thus we reference to the proof
of Property \ref{pro:nmopt} for a proof of this property.
Note that Property \ref{pro:nmopt} deals about the computation of
 ${S_{0m}({\AComp},v|\theta^\prime)}$ i.e. the computation of
{${S_{0m}({\AComp},v)}$} provided that a top simplex $\theta^\prime$ 
incident to $v$ in $\AComp$ is known. This applies to  this case since
we can write  {${S_{0m}({\AComp},v)}$}  as
{${S_{0m}({\AComp},v)}=S_{0m}({\AComp},v|VT*'[v])$}.
\end{proof}
Optimal extraction is possible for $S_{01}$ and  $S_{02}$ in  
$2$-complexes and for $S_{01}$, $S_{02}$
and $S_{03}$ in $3$-complexes embeddable in $\real^3$ as compact geometric
complexes. 
The extraction algorithm fail to be optimal for $4$-complexes
For an example of non optimality of $S_{01}$ in a $4$-complex 
see Property \ref{pro:s01ext}.

\subsection{Computation of $S_{nm}(\AComp,\gamma)$: the $\splitsimp0 $ and $\splitstar0 $  Relations and $\splitmap $}
\label{sec:snmnm}
{
\subsubsection{Introduction}
\label{sec:summ}
In the previous section we have given a rationale for the introduction of
$\sigma$ and $\sigma^{-1}$ by showing that these relations support the
extraction of $S_{0m}$ in optimal time at least for $2$ and $3$-complexes.
We note here that these two maps are {\em necessary} 
to develop such an extraction.
The rationale for the introduction of $\splitsimp0 $, $\splitstar0 $ and
$\splitmap $ is slightly different. We introduce the relations 
$\splitsimp0 $, $\splitstar0 $ in the \NMWDS\ representation to support
the computation of {$S_{nm}(\AComp,\gamma)$} in optimal time. These
relations {\em are not necessary} to perform this task. 
This is a consequence of the fact that the \Cdecrep\ Representation match
exactly the decomposition procedure and thus all information 
about the non-manifoldness that has been removed by the decomposition can be encoded in the 
$\sigma$ map. Furthermore this result is dimension independent.

In fact the computation of {$S_{nm}(\AComp,\gamma)$} can be performed as
follows.
If $\gamma=\{v_1,\ldots,v_n\}$ is an $n$-simplex in $\AComp$ it is easy to see that,
for all $n\le m$:
\begin{equation}
\label{eq:stitch}
S_{nm}(\AComp,\{v_1,\ldots,v_n\})=
\bigcup_{v_i^\prime\in\sigma^{-1}(v_i)}
\sigma({S_{nm}(\canon{\AComp},\{v^\prime_1,\ldots,v^\prime_n\})})
\end{equation}
It is easy to turn equation \ref{eq:stitch} into an algorithm.
In fact the algorithm we have proposed  for (the computation of) 
{$S_{nm}(\canon(\AComp),\gamma)$} (see Algorithm \ref{algo:snm})
gives a correct result (i.e. the empty set) even if the  supplied argument
$\{v^\prime_1,\ldots,v^\prime_n\}$  is a set of $n$ Vertices that 
is not a simplex in $\canon{\AComp}$.
}

Therefore the \Cdecrep\ 
Data Structure extended with maps $\sigma$ and $\sigma^{-1}$
contains sufficient information to extract the
topological relations $S_{nm}$.
However this cannot be done in optimal time, 
In fact the extraction of {$S_{nm}(\canon{\AComp},\gamma)$},
performed following  Algorithm \ref{algo:snm}
can be grossly inefficient. 
In fact our algorithms
always fetch a fan of top simplices around {\em a
	vertex } whatever will be the topological relation requested.
In higher dimension and we have shown that, in some cases, for
$m=n+1$, Algorithm \ref{algo:snm}
yields a processing time for {$S_{n(n+1)}(\gamma)$} that is polynomial
i.e.  
${S_{n(n+1)}(\canon{\AComp},\gamma)}$ is
 $\ordcomp{|S_{n(n+1)}(\canon{\AComp},\gamma)|^{\floor{h/2}}{\log{|S_{n(n+1)}(\canon{\AComp},\gamma)|}}}$.
(see the remark at the end of Property 
\ref{pro:nmineffext}).

In this section we will analyze and solve this problem introducing
the $\splitsimp0 $ and $\splitstar0 $ relations.

\subsubsection{The $S_{nh}(\AComp,\gamma|\theta)$ topological relation}
A processing time of 
$\ordcomp{n^{\floor{h/2}}{\log{n}}}$ for the computation of
${S_{n(n+1)}(\canon{\AComp},\gamma)}$
is not acceptable since the TT relation contains all information
that are necessary to perform this computation in optimal time. 
We feel that this problem comes from two facts. 

A first, quite obvious, fact is that both in the original 
Winged Representation and in the \Cdecrep\ Representation in
each $h$-dimensional component the $n$-simplices, for
$0<n<h$, do not receive a direct representation. In fact we represent 
$n$-simplices as set of $n$ Vertices. For $n=0$ and $n=h$ we assign 
an explicit representation to $0$-simplices in the range $[1\ldots NV']$ and 
we assign an explicit representation to $h$-simplices in the range 
$[1\ldots NT']$. 
Representing explicitly $n$-simplices can be quite heavy and, furthermore,
there are  many situations in which this is not needed. 
In fact, there are situations in which
we start from  a top $h$-simplex $\theta$ and want to find all {\em top} 
$h$-simplices in the star $\star{\gamma}$ being $\gamma$ a proper face of 
$\theta$. In this case we will say that we want to compute the topologic
relation {$S_{nh}(\AComp,\gamma|\theta)$} 
(read $S_{nh}(\AComp,\gamma)$ {\em given} $\theta$). 
Next we take a retrieved  $h$-simplex 
$\theta^\prime\in S_{nh}(\AComp,\gamma|\theta)$, 
select a face $\gamma^\prime\le\theta^\prime$ and
repeat this kind of query. 
In this  case we do not need to introduce an explicit modeling
entity for $n$-simplices. In fact,
as we will show,  our data structure can support
the computation of {$S_{nh}(\AComp,\gamma|\theta)$} in optimal time.
So, in this section, we will discuss the $\splitsimp0 $ and $\splitstar0 $
relations that are introduced to support the 
computation of {$S_{nh}(\AComp,\gamma|\theta)$} in optimal time.
In Section \ref{sec:trie} we will devise an auxiliary data structure,
the $\FT{n}$ map,  that introduces an explicit representation 
for $n$-simplices thus supporting the computation of {$S_{nh}(\AComp,\gamma)$} 
in optimal time.

For sake of clarity we choose to leave the $\FT{n}$  out of the
\NMWDS\ Representation. In fact, the problem of giving an explicit 
representation to intermediate objects need not to be
confused with the problem of representing non-manifoldness. 
Another benefit of the \NMWDS\ is that these two aspects can be
clearly identified and treated with separate mechanism.

To support the optimal computation of {$S_{nh}(\AComp,\gamma|\theta)$ we need to
take care of a second issue.
In general, in an  \iqm\ complex, 
the star $\xstr{\AComp}{{\gamma}}$ of a generic $n$-simplex $\gamma$, 
for $n>0$, it  is not $(h-1)$-connected. This might happen because 
$\gamma$  splits into several copies $\gamma^\prime_i$.
However it is also possible to find complexes $\AComp$ for which 
$\AComp=\canon{\AComp}$ and yet there exist 
an $n$-simplex $\gamma\in\AComp$ that have non-$(h-1)$-connected star. 
We have shown \cite{Def02b} that this problem
do not depends on the particular decomposition process.
In fact there are 
$h$-complexes, for $h\ge 3$, that can not be decomposed into
a complex where all simplices has $(h-1)$-connected star.
A complex where all the stars of
all simplices are $(h-1)$-connected is called a {\em \ra}
complex. A simplex whose star is $(h-1)$-connected will be called a 
\ra\ simplex.  
In Figure \ref{fig:tienonstar} we have shown an example of an \iqm\ 3-complex with a 2-simplex that is not \ra.

We note that the problem related to non-$(h-1)$-connectedness of
a $n$-simplex stars is present \iff\ the complex $\AComp$
is a non-manifold complex. 
For this reason we do not want to solve this 
second problem adding a mechanism that
adds extra memory when the complex $\AComp$ is regularly adjacent
or even manifold.
The introduction of the relations $\splitsimp0 $ and $\splitstar0 $
solves this problem in this direction and supports, with a small
increase in memory requirements, 
the optimal computaton of {$S_{nh}(\gamma|\theta)$}

\subsubsection{The $\splitsimp0 $ and $\splitstar0 $ Relations} 
Thus we  want to overcome these two problems in order to obtain a data 
structure that supports the optimal extraction of   {$S_{nm}$} by 
adding an explicit representation just for the subset of
{non-manifold $n$-simplices $\gamma$ in $\AComp$
having one these two problem:
\begin{itemize}
\item the simplex $\gamma$ is a splitting simplex w.r.t. the decomposition $\canon{\AComp}$;
\item the simplex $\gamma$ is not splitting simplex and its star in $\canon{\AComp}$ is not $(h-1)$-connected.
\end{itemize}
The set of simplices with one of these problems will be denoted by $\NRA$.
Similarly we will denote with $\canon{\NRA}$ the set of all simplex copies
for all splitting simplices in $\NRA$ plus all non-splitting simplices
in $\NRA$ 
(i.e. $\canon{\NRA}=\{\gamma^\prime|\sigma(\gamma^\prime)\in\NRA\}$). 
}
We first note that it might happen that 
the simplex $\gamma$ is a splitting simplex and still 
for one or more of its simplex
copies $\gamma^\prime$ the complex $\xstr{\canon{\AComp}}{\gamma^\prime}$ is not
$(h-1)$-connected.
Second note that we simply ask for $(h-1)$-connectivity and not for
$(h-1)$-manifold-connectivity. This is due to the possibilities offered
by the optimization described in Section \ref{sec:trix}. In fact using
the relation \TTP, optimized deleting symbols $\trix$,  
we can retrieve all the top  simplices that are 
adjacent to a given simplex even if we traverse non-manifold 
$(h-1)$-simplices.

With the above discussion in mind, we are ready to devise completelty 
the two relations in the upper layer of our representation.
We have already presented the relation $\splitsimp0 $ in Section
\ref{sec:sigma}
and we have encoded its restriction to $0$-simplices with the
maps $\sigma$ and $\sigma^{-1}$. We choose to encode separately
this restriction because, as shown by equation \ref{eq:stitch}, these
two maps are sufficient to extract all topological relations.

The second relation that is present in the \NMWDS\ Representation, 
denoted by $\splitstar0 $, is the relation
$\splitstar0 \subset\canon{\Theta}\times\canon{\NRA}$ such that the following
conditions are satisfied:
\begin{itemize}
\item for each $(h-1)$-connected component in
the star  $\xstr{\canon{\AComp}}{\gamma^\prime}$ 
there exist a top simplex $\theta^\prime\in\canon{\AComp}$ 
for which $\xsplitstar{\gamma^\prime}{\theta^\prime}$;  
\item if $\xsplitstar{\gamma^\prime}{\theta_1^\prime}$ and
$\xsplitstar{\gamma^\prime}{\theta_2^\prime}$ then $\theta_1$ and
$\theta_2$ must belong to two  distinct $(h-1)$-connected component in
the star  $\xstr{\canon{\AComp}}{\gamma^\prime}$ 
\end{itemize}

\subsubsection{The $\splitmap$ Map} 
\label{sec:splitmap}
The remaining part of relation $\splitsimp0 $ and the   $\splitstar0 $ 
relation will be jointly implemented in the \NMWDS\ Data Structure
by the map $\splitmap$.
Putting together the two things, somehow, make things a little obscure. The remaining part of the relation $\splitsimp0 $ is the subset of the 
relation $\split0 \subset \canon{\AComp}\times\AComp$ obtained deleting 
the couples in $\canon{V}\times V$  from the relation $\splitsimp0 $. 
We will have that $\xsplitsimp{\gamma}{\gamma^\prime}$ 
\iff\ $\gamma$ is a splitting $n$-simplex and $\gamma^\prime$ is one of its
 simplex copies.  

The two relations  $\splitsimp0 $ and the   $\splitstar0 $ are jointly coded into $\splitmap$ by taking  map $\splitmap$ as a map that sends each simplex $\gamma\in\NRA$ into 
a particular map. 
In other words the element returned by  {$\splitmap[\gamma]$}
is a map $\funct{\splitmap[\gamma]}{(\canon{\NRA})}{2^{\canon{\Theta}}}$.
Sometimes, however, we will use $\splitmap[\gamma,\gamma^\prime]$ for $\splitmap[\gamma][\gamma^\prime]$.
The most handy definition of $\splitmap$ is given by the procedure below

\begin{algorithmic}
	\STATE{ {\bf Function} $\splitmap(\gamma,\gamma^\prime)$
		{\bf returns } {\bf subset of} {${\canon{\Theta}}$}}
	\IF[$\xsplitsimp{\gamma}{\gamma^\prime}$]{$\gamma$ is a splitting simplex}
\STATE {\bf return}$\{\theta_1\ldots\theta_n\}$
\COMMENT{ Each $\theta_i$	 from a  distinct $(h-1)$-connected component in
	$\xstr{\canon{\AComp}}{\gamma^\prime}$}
	\ELSE
	\IF[must be $\gamma=\gamma^\prime$]{$\gamma$ is not a splitting simplex}
	\STATE {\bf return}$\{\theta_1\ldots\theta_n\}$
	\COMMENT{ Each $\theta_i$	 from a  distinct $(h-1)$-connected component in
		  $\xstr{\canon{\AComp}}{\gamma^\prime}$}
	\ENDIF
	\ENDIF
\end{algorithmic}
Looking at $\splitmap$ as a map that returns a map we can say that
the domain of the map $\splitmap[\gamma]$ 
(denoted by {$\domain(\splitmap[\gamma])$} is the set of simplex copies 
for $\gamma$. This domain is a sigleton for all and alone simplices in
$\NRA$ that are not splitting simplices. 
Thus we have:
\begin{equation}
  \xsplitsimp{\gamma}{\gamma^\prime}\Leftrightarrow
(\,{|\domain(\splitmap(\gamma))|>1}\, \mbox{\bf   and } 
{\gamma^\prime\in \domain(\splitmap(\gamma))\,)}.
\end{equation}
More in general, the map $\splitmap(\gamma)$ is defined completely by the condition:
\begin{equation}
\theta\in\splitmap[\gamma][\gamma^\prime] \Leftrightarrow
((\,\gamma=\gamma^\prime \mbox{\bf or   } \xsplitsimp{\gamma}{\gamma^\prime}\,)
\mbox{\bf  and  } 
\xsplitstar{\gamma^\prime}{\theta^\prime}\,).  
\end{equation}
The implementation of this map can be done using a well
known data structure for dictionaries called {\em trie} and
is described in Section \ref{sec:trie} together with the implementation 
of similar data structures that will be introduced later.
\begin{example}
	\label{ex:datastrexbig3}
	We continue our running example  from  Example 	\ref{ex:datastrexbig2}. W.r.t. Figure \ref{fig:datastrexbig} we have that $\splitmap[\gamma]$
	is defined only for $\gamma=	\{6,8\}$ and
	$\domain(\splitmap[\{6,8\}])=\{\{6,8\},\{14,15\}\}$
	\begin{center}
		\begin{tabular}{|c||c|}\hline
			${\gamma^\prime}$&$\splitmap[\{6,8\}][{\gamma^\prime}]$
			\\ \hline \hline
			$\{6,8\}$&$\{5\}$    \\ \hline
			$\{14,15\}$&$\{9\}$   \\ \hline
		\end{tabular}
	\end{center} 
\end{example}
\subsubsection{Construction of the map $\splitmap$}
In this section we assume a standard implementation for this map and 
describe an algorithm to fill the $\splitmap$ map.
Unless otherwise stated, in the description of this algorithm, we will
use the greek primed letters, e.g. 
$\gamma^\prime$ and $\theta^\prime$, 
to denote   simplices
in $\canon{\AComp}$ that are represented as sets of indexes.
We will use latin letters, e.g.  $t^\prime$ or $v^\prime$ 
to denote  top  simplices and vertices
in $\canon{\AComp}$ that are represented as sets of indexes.

This algorithm is described in two parts. First we develop a recursive 
procedure {${\rm TravelStar}(\gamma^\prime,t^\prime)$}
that travels the star of $\gamma^\prime$ in $\canon{\AComp}$ by adjacency, 
using the $\Adj$ relation (i.e. the \TTP\ array). 
The visit performed by {${\rm TravelStar}(\gamma^\prime,t^\prime)$}
starts from top simplex  $\theta^\prime$ 
and marks all  visited top simplices.
Marking of top simplices in $\canon{\AComp}$ 
is performed setting a bit into an the array of bits 
\begin{verbatim}
FT_FLAGS: array [1..NT'][1..2**(d+1)-2] of bits
\end{verbatim}
All bits in this array are initially reset to zero.
We have a bit for each face of each top simplex.
We assume to have function {\INDEX($t^\prime$,$\gamma^\prime$)} 
that takes a
top simplex index $t^\prime$ and a $n$-simplex $\gamma$ and returns a 
bit pattern of length $(h+1)$ with $n$ ''1'' such that
{\INDEX($t^\prime$,$\gamma^\prime$)[$i$]=''1''} \iff\ {\TVP[$h$,$t^\prime$,$i$]$\in\gamma^\prime$}.
We will use this index to locate the right flag to be set.
In particular  we will have that 
the bit {FT\_FLAGS[$t^\prime$][\INDEX($t^\prime$,$\gamma^\prime$)]}
will be set to ''1'' to indicate that the algorithm, exploring 
the star of $\gamma^\prime$, has taken into account the top 
simplex $\tau\in\xstr{\canon{\AComp}}{\gamma^\prime}$
indexed by $t^\prime$.
For this reason we extend the FT\_FLAGS array from 
$1$ to $2^{d+1}-2$. We leave out of this range
$0$, corresponding to the empty face, and $2^{(d+1)}-1$, corresponding to 
the top $d$-simplex itsself.

With these assumptions we have that the algorithm for 
{${\rm TravelStar}(\gamma^\prime,t^\prime)$} is the following.
\begin{algo}{algo:travelstar}{{${\rm TravelStar}(\gamma^\prime,t^\prime)$} travels
{$\xstr{\canon{\AComp}}{\gamma^\prime}$} 
by adjacency and marks visited top simplices in global variable $FT\_FLAG$}
\begin{algorithmic}
\STATE{{\bf Procedure} {{\rm TravelStar}($\gamma^\prime$: {\bf set of }{\tt Vertex'}, $t^\prime$: {\tt TopSimplex'})}  }
\STATE $FLAG\leftarrow \INDEX(t^\prime,\gamma^\prime)$;
\IF[simplex $t^\prime$ not visited yet when going round $\gamma^\prime$]
    {{\bf not} {$FT\_FLAGS[t^\prime][FLAG]$}}
\STATE{{$FT\_FLAGS[t^\prime][FLAG]\leftarrow 1$}}
\COMMENT{we are going to visit it so mark it}
\STATE{$h\leftarrow \ord(t^\prime,\canon{\AComp})$};
\FORALL[ {FLAG[$i$]=0  \siff\ $\TVP[h,t^\prime,i]\not\in\gamma^\prime$} ]
{$1\le i\le h+1$ {\bf and} FLAG[$i$]=0}
\STATE {${\rm TravelStar}(\gamma,\TTP[h,t^\prime,i])$};
\ENDFOR
\ENDIF
\end{algorithmic}
\end{algo}

It is easy to see that the simplices marked during the
execution of {${\rm TravelStar}(\gamma^\prime,t^\prime)$} are all and 
alone the top simplices in $(h-1)$-connected component of
{$\xstr{\canon{\AComp}}{\gamma^\prime}$} that contains $t^\prime$.
The time complexity of {${\rm TravelStar}(\gamma^\prime,\theta^\prime)$}
is linear w.r.t. the number of
top simplices in this $(h-1)$-connected component.

With this procedure at hand it is quite easy to
give an algorithm that builds the map $\splitmap$.
We recall that the map $\splitmap$ uses simplices as indexes and 
returns a map between simplices and top simplex indexes.  
Thus  the assignment 
$\splitmap[\gamma][\gamma^\prime]\leftarrow t^\prime$ 
is perfectly legal
and associate in the map $\splitmap[\gamma]$ the top simplex $t^\prime$ to the key $\gamma^\prime$.
We also need an inverse to the function $\INDEX$. We will denote
this inverse with {\SIMPLEX($t^\prime$,$Idx$)}. The function \SIMPLEX\ takes 
an $h$-simplex index  $t^\prime$, a bit pattern $Idx$ with $n$ ''1'',
and returns the  $n$-simplex $\gamma^\prime$ 
such that  $t^\prime$ is a face of $\gamma^\prime$ and
{INDEX($t^\prime$,$\gamma^\prime$)=$Idx$}.
The two functions \INDEX\ and \SIMPLEX\ can easily be implemented using the 
\TVP\ array. 
The two functions must jointly satisfy the equations:
$${\INDEX(t^\prime,{\SIMPLEX(t^\prime,Idx)})=Idx}\,  \mbox{\bf  and  }\,
{\SIMPLEX(t^\prime,{\INDEX(t^\prime,\gamma^\prime)})=\gamma^\prime}
$$

We also need to use an auxiliary function CopiesOf($v$). This 
function must return the set $\sigma^{-1}(v)$ if $v$ is a splitting vertex 
and returns the singleton $\{v\}$ otherwise.
Finally we assume that  we can delete a key 
in the map $\splitmap$ with a call to the method
$\splitmap$.RemoveKey($\gamma^\prime$).
{The  algorithm for the construction of $\splitmap$ will consider all simplices incident to a vertex. It would be nice to limit the set of vertices considered to those with potential problems. 
	This is the set of 
	Vertices in $\AComp$, denoted by $V_{\NRA}$ such that $v\in V_{\NRA}$ \iff\
	one of the following two conditions occurs:
	\begin{itemize}
		\item the vertex $v$ is a splitting vertex in $\AComp$ and there exist a vertex copy
		$v^\prime$ of $v$ incident to a non-\ra\ simplex in $\canon{\AComp}$.
		\item  the vertex $v$ is not a splitting vertex and yet 
		$v$ is incident to a non-\ra\ simplex in $\canon{\AComp}$.
	\end{itemize}
	The set $V_{\NRA}$ is a superset of Vertices incident to a non-\ra\ simplex.
	On the other hand  the set  {$V_{\NRA}$} is
	always a subset of the set of non-manifold vertices.
In the following we assume that some initialization loads the set  {$V_{\NRA}$} possibly setting $V_{\NRA}=V$ } 
With these assumptions we are ready to devise the algorithm that builds the 
map $\splitmap$.
\begin{algo}{algo:splitmap}{Builds the map $\splitmap$ }
\begin{algorithmic}
\FORALL[consider all Vertices $v$ in $\AComp$ with potential problems]{$v\in V_{\NRA}$}
\FORALL[consider all vertex copies of $v$]{$v^\prime\in$ CopiesOf($v$)}
\STATE $h\leftarrow{\ord(v^\prime,\canon{\AComp})}$
\FORALL[for all $t^\prime$ incident to $v^\prime$
 (see Algorithm \ref{algo:s0h} for {$S_{0h}(\canon{\AComp},v^\prime)$})]
{$t^\prime\in {S_{0h}(\canon{\AComp},v^\prime)}$}
\FOR[loop for all $\gamma^\prime\subset t^\prime$]{$Idx=1$ to $2^{d+1}-2$}
\IF{{\bf not} {$FT\_FLAGS[t^\prime][Idx]$}}
\STATE $\gamma^\prime\leftarrow  SIMPLEX(t^\prime,Idx)$
\COMMENT{$\gamma^\prime$ not visited yet}
\STATE $\splitmap[\sigma(\gamma^\prime)][\gamma^\prime]\leftarrow t^\prime$;
\COMMENT{(\dag 1) record $t^\prime$ then mark and forget all other top simplices...}
\STATE {${\rm TravelStar}(\gamma^\prime,t^\prime)$}
\COMMENT{.... in the $(h-1)$-connected component for $t^\prime$ in $\xstr{\canon{\AComp}}{\gamma^\prime}$}
\ENDIF \ENDFOR
\ENDFOR
\ENDFOR
\ENDFOR
\FORALL{$\gamma\in\domain(\splitmap)$}
\FORALL{$\gamma^\prime\in\domain(\splitmap[\gamma])$}
\IF[$\gamma$ is not a splitting simplex]{$|\domain(\splitmap[\gamma])|<2$}
\IF[$\gamma$ has 1 $(d-1)$-connected component]{$|\splitmap[\gamma][\gamma^\prime]|=1$ }
\STATE $\splitmap$.RemoveKey($\gamma$);
\ENDIF
\ENDIF
\ENDFOR
\ENDFOR
\end{algorithmic}
\end{algo}
The time complexity of this algorithm is given by the following property:
\begin{property}
\label{pro:splitmap}
The Algorithm \ref{algo:splitmap} computes the map $\splitmap$ in
$\lesscomp{NSP\cdot 2^d\log{NSP}}$ being $NSP$ 
the number of top simplices in $\AComp$ incident to a non-manifold vertex in $V_{\NRA}$.
\end{property}
\begin{proof}
In fact the set $V_{\NRA}$ is a subset of non-manifold Vertices in $\AComp$.
Therefore the set of top simplices incident to a vertex in   $V_{\NRA}$ is
smaller than $NSP$.
The Algorithm  \ref{algo:splitmap}
visits each face of each top simplex incident to a vertex in $V_{\NRA}$
once. These faces are less than $NSP\cdot 2^d$. 
For each visit operations that are performed are dominated by
the map insertion at line (\dag 1). This map insertion take less that
$\log{|S|}$ being $|S|$ the number of elements in the map when we execute
line (\dag 1). Thus this operation takes less than $\log({NSP\cdot 2^d})$.
The term $2^d$ comes from the fact that a top simplex in the 
$d$-complex $\AComp$ has at most $2^d$ faces.
Summing over all the insertion we obtain the upper bound
$\lesscomp{NSP\cdot 2^d\log{NSP}}$.
\end{proof}
The complexity of $\lesscomp{NSP\cdot 2^d\log{NSP}}$ 
represent a drastic reduction in complexity w.r.t. a
global analysis.
A reduction is possible if load in $V_{\NRA}$ a small set.

\subsubsection{Computation of {$S_{nh}(\canon{\AComp},\gamma^\prime|\theta^\prime)$}} 
In this section we develop algorithms to extract the topological relation
{$S_{nh}(\canon{\AComp},\gamma^\prime|\theta^\prime)$} being
$h$ the dimension of the component of $\canon{\AComp}$ 
containing simplex $\gamma^\prime$.
The computation of the relation 
{$S_{nh}(\canon{\AComp},\gamma^\prime|\theta^\prime)$} 
can be done in optimal time using the $\splitmap$ map. 
In this section we will exhibit an algorithm to compute the
function that returns  the set of 
indexes for top simplices in  
{$S_{nh}(\canon{\AComp},\gamma^\prime|\theta^\prime)$}.
We will denote this function with 
{$S_{nh}[\canon{\AComp}](\gamma^\prime,t^\prime)$} 
In this algorithm we assume that $t^\prime$ is the index of a top
$h$-simplex  $\theta^\prime$ incident to $\gamma^\prime$. 
We recall that {$S_{nh}(\canon{\AComp},\gamma^\prime|\theta^\prime)$}
must contain just all  $h$-simplices that are incident to  
$\gamma^\prime$ in $\canon{\AComp}$. Since we are assuming that
$h=\ord(\theta^\prime,\canon{\AComp})$ we have that all $h$-simplices
in {$S_{nh}(\canon{\AComp},\gamma^\prime|\theta^\prime)$} must be 
top $h$-simplices.  Nevertheless, if $\gamma^\prime$ is a splitting 
simplex there can be non top $h$-simplices 
that are incident to another simplex copy of  
$\sigma(\gamma^\prime)$ in another component of
$\canon{\AComp}$. Since they are in another component they
are not included in 
{$S_{nh}(\canon{\AComp},\gamma^\prime|\theta^\prime)$}.
The algorithm that computes 
{$S_{nh}[\canon{\AComp}](\gamma^\prime,t^\prime)$} 
is the following
\begin{algo}{algo:Snmnm}{Computation of {$S_{nh}[\canon{\AComp}](\gamma^\prime,t^\prime)$}}
{
\begin{algorithmic} 
\STATE{{\bf Function} $S_{nh}[\canon{\AComp}]$($\gamma^\prime$: {\bf set of }{\tt Vertex'}, $t^\prime$: {\tt TopSimplex'}) {\bf returns set of} {\tt TopSimplex'}}
\STATE $\gamma\leftarrow\sigma(\gamma^\prime);$
\STATE{$h\leftarrow\ord(t^\prime,\canon{\AComp})$;}
\IF{$\gamma\in\domain{(\splitmap)}$}
\STATE $S\leftarrow{\splitmap[\gamma][\gamma^\prime]};$
\ELSE[$\gamma$ is not a splitting simplex and $\xstr{\canon{\AComp}}{\gamma^\prime}$ is $(h-1)$-connected] 
\STATE $S\leftarrow{\{t^\prime\}};$
\ENDIF
\COMMENT{ (\dag 1) an element in $S$  for each $(h-1)$-connected component in $\xstr{\canon{\AComp}}{\gamma^\prime}$} 
\STATE $N\leftarrow S$ 
\COMMENT{$t^\prime\in N\subset{\xstr{\canon{\AComp}}{\gamma^\prime}}$ \siff\ 
$t^\prime$ adjacent to a simplex in $S$ and $t^\prime$ not visited}
\FORALL {$t^\prime\in N$}  
\FOR[{search for a new $t^\second$ incident to $\gamma^\prime$ and
adjacent to $t^\prime$}]{$k=1$ to $(h+1)$}
\IF{\TVP[$h$,$t^\prime$,$k$]$\not\in\gamma^\prime$}
\STATE $t^\second\leftarrow$ \TTP[$h$,$t^\prime$,$k$] 
\IF[{found a new $t^\second$ incident to $\gamma^\prime$ and
adjacent to $t^\prime$}]{$t^\second \notin S$ {\bf and} $t^\second \neq\bot$} 
\STATE $N\leftarrow N\cup\{t^\second\}$
\STATE $S\leftarrow S\cup\{t^\second\}$
\ENDIF
\ENDIF
\ENDFOR
\STATE $N\leftarrow N-\{t^\prime\}$
\COMMENT{all top simplices adjacent to $t^\prime$ has been visited}
\ENDFOR
\STATE {\bf return}  $S$
\end{algorithmic} 
}
\end{algo}
The following property gives correcteness and complexity of the above 
algorithm.
\begin{property}
\label{pro:corrSnh}
Let $\AComp$ be a $d$-complex and let $\gamma^\prime$ be a $n$-simplex
in $\canon{\AComp}$.
For any top $h$-simplex 
$\theta^\prime\in\xstr{\gamma^\prime}{\canon{\AComp}}$ 
we have that the above algorithm for 
{$S_{nh}(\canon{\AComp},\gamma^\prime|\theta^\prime)$} terminates.
Upon termination  in the 
variable $S$ we find  the set of 
top $h$-simplices in {$S_{nh}(\canon{\AComp},\gamma^\prime)$}.
This computation can be done in $\ordcomp{nt\log{nt}}$ where $nt$ is the number
of top $h$-simplices in {$S_{nh}(\canon{\AComp},\gamma^\prime)$}.
\end{property}
\begin{proof} The proof of the correctness of this algorithm
is nearly the same  as in Property
\ref{pro:corrSoh} and will not be developed in detail. The only
relevant difference with the proof of Property \ref{pro:corrSoh} 
is that after the execution of the if-then-else (\dag 1) 
we have in $S$ an $h$-simplex for each $(h-1)$-connected component of 
$\xstr{\gamma^\prime}{\canon{\AComp}}$. 
Similarly the proof of optimal time complexity follows the proof of
property \ref{pro:compl}
\end{proof}

\subsubsection{Computation of $S_{nm}(\AComp,\gamma|\theta)$}  
If $\theta$ is a top simplex {\em of any dimension} incident to
the $n$-simplex $\gamma$, in a $d$-complex $\AComp$, then, 
for any $n<m\le d$ we can easily compute $S_{nm}(\AComp,\gamma|\theta)$
using the map $\splitmap$ and Algorithm  \ref{algo:Snmnm}.
We recall that {$S_{nm}({\AComp},\gamma|\theta)$}
must contain all  $m$-simplices that are incident to  
$\gamma$ in ${\AComp}$.
The computation of {$S_{nm}({\AComp},\gamma|\theta)$}
can be done with the Algorithm \ref{algo:Snm} that computes the function
{$S_{nm}[\AComp](\gamma,t)$}. 
In this algorithm we assume that $t$ is the index of the top simplex
$\theta$ incident to $\gamma$. 
We note that the decomposition algorithm do not introduce new top
simplices thus we have that the identity is the conversion
function between types {\tt TopSimplex} and {\tt TopSimplex'}.
Thus we can assume to have a valid type cast between these two types
such that ${\tt TopSimplex}(t^\prime)=t$ and 
${\tt TopSimplex'}(t)=t^\prime$.

We assume to have function FaceOf($m,\beta$,Top)
that returns the set of $m$-cofaces of $\beta$ that are
$m$-faces of simplices in Top.
In other words the function FaceOf is defined by the equation:
$$
{\rm FaceOf}(m,\beta,Top)=
\{\gamma|\ord(\gamma)=m\, \mbox{\bf and}\, 
(\exists \tau\in Top)(\beta\le\gamma\le\tau)\} 
$$
In order to  develop Algorithm \ref{algo:Snm} we present
the function {$\sigma_n^{-1}(\gamma,t)$} that
returns, for a non-splitting $n$-simplex $\gamma$, the $n$-simplex 
$\gamma^\prime\in\canon{\AComp}$ such that $\sigma(\gamma^\prime)=\gamma$.
The index $t$ is given as an hint and is a top simplex incident to
$\gamma$ in $\AComp$.
The function {$\sigma_n^{-1}(\gamma,t)$} is  computed  by the 
following fragment of code:
\begin{algo}{algo:sinvn}{Computation of {$\sigma_n^{-1}(\gamma,t)$}}
\begin{algorithmic} 
\STATE{{\bf Function} $\sigma_n^{-1}(\gamma$: {\bf set of }{\tt Vertex}, $t$: {\tt TopSimplex}) {\bf returns set of} {\tt Vertex'}}
\STATE $t^\prime\leftarrow {\tt TopSimplex'}(t)$;
\COMMENT{cast $t$ into the index type for the data structure for $\canon{\AComp}$}
\STATE $h\leftarrow \ord(t,\AComp)$;
\STATE $\theta^\prime\leftarrow SetOf(\TVP[h,t^\prime])$;
\COMMENT{(\dag 1) convert index $t^\prime$ into a set of Vertices in $\canon{\AComp}$}
\STATE $\gamma^\prime\leftarrow \emptyset$;
\COMMENT{accumulate in $\gamma^\prime$ the simplex such that $\sigma(\gamma^\prime)=\gamma$}
\FORALL[(\dag 2) check all verices of $\theta^\prime$]{$v^\prime\in\theta^\prime$}
\IF{$\sigma(v^\prime)\in\gamma$}
\STATE{$\gamma^\prime\leftarrow\gamma^\prime\cup\{v^\prime\}$}
\ENDIF
\ENDFOR
\STATE {\bf return} $\gamma^\prime$;
\end{algorithmic} 
\end{algo}
The correctness of this algorithm is given by the following property:
\begin{property}
Let $\gamma\in\AComp$ be a non-splitting $n$-simplex incident to the 
top simplex $\theta\in\AComp$ and let $t$ be the index for $\theta$.
Let (EWS,$\sigma$,$\sigma^{-1}$,$\splitmap$) an \NMWDS\ Data Structure  
for $\AComp$. 
Let  $NC$ be the total number of vertex copies introduced by the
\cano\ decomposition $\canon{\AComp}$. 

In this situation there exist a unique $n$-simplex
$\gamma^\prime\in\canon{\AComp}$ such that 
$\gamma=\sigma(\gamma^\prime)$ and   the Algorithm \ref{algo:sinvn} 
returns in $\lesscomp{h(\log{h}+\log{NC})}$ 
the simplex $\gamma^\prime$ (with $h=\ord(t,\AComp)$). 
\end{property}
\begin{proof}
By hypothesis we have {$\gamma\le\theta$}. There is a top simplex
$\theta^\prime\in\canon{\AComp}$ such that
{$\theta=\sigma(\theta^\prime)$}. Therefore
{$\gamma\le\sigma(\theta^\prime)$} and thus there is a simplex 
$\gamma^\prime\le\theta^\prime$ such that 
{$\gamma=\sigma(\gamma^\prime)\le\sigma(\theta^\prime)$}.
Being $\gamma$ a non splitting simplex there can not be two distinct 
simplices $\gamma^\prime$ and $\gamma^\second$ such that 
{$\gamma=\sigma(\gamma^\prime)=\sigma(\gamma^\second)$}.
Therefore such a $\gamma^\prime$ is unique.

Line (\dag 1) in Algorithm \ref{algo:sinvn}  find a $\theta^\prime$
such that {$\theta=\sigma(\theta^\prime)$}. The loop (\dag 2) 
checks all verices of $\theta^\prime$ and builds a simplex $\gamma^\prime\le\theta^\prime$ such that {$\sigma(\gamma^\prime)\le\gamma$}.
Eventually we will reach the condition
{$\gamma=\sigma(\gamma^\prime)\le\sigma(\theta^\prime)$}.

To reach this condition we perform  the body of loop (\dag 2) at most
$h$ times. The body of the loop contains set operations 
(element insertion and set membership) that takes $\lesscomp{\log{h}}$
The application of map $\sigma$ to a simplex takes $\lesscomp{h\log{NC}}$.
Summing the two terms we obtain the thesis.
\end{proof}
With these auxiliary functions we can give the algorithm for the computation of {$S_{nm}[\AComp](\gamma,t)$}:
\begin{algo}{algo:Snm}{Computation of {$S_{nm}[\AComp](\gamma,t)$}}
{
\begin{algorithmic} 
\STATE{{\bf Function} $S_{nm}[\AComp](\gamma$: {\bf set of }{\tt Vertex}, $t$: {\tt TopSimplex}) {\bf returns set of}({\bf set of} {\tt Vertex})}
\STATE $t^\prime\leftarrow {\tt TopSimplex'}(t)$;
\IF[{$\gamma$ do not split and its star is \ra}]{$\gamma\not\in\domain{(\splitmap)}$}
\STATE $\gamma^\prime\leftarrow{\sigma_n^{-1}(\gamma,t)}$;
\STATE {$h\leftarrow\ord(\gamma^\prime,\canon{\AComp})$}
\IF{$h\ge m$}
\STATE $Top\leftarrow{S_{nh}[\canon{\AComp}](\gamma^\prime,t^\prime)}$;
\COMMENT{(\dag 1)}
\ENDIF
\ELSE 
\STATE {$Top\leftarrow\emptyset$;}
\FORALL[{$\domain(\splitmap[\gamma])$ is the set of simplex copies of $\gamma$}]
{$\gamma^\prime\in\domain({\splitmap[\gamma]})$}
\STATE {$h\leftarrow\ord(\gamma^\prime,\canon{\AComp})$}
\IF{$h\ge m$}
\FORALL[{a $t^\prime$ for each $(h-1)$-connected part in $\xstr{\canon{\AComp}}{\gamma^\prime}$}]
{$t^\prime\in \splitmap[\gamma][\gamma^\prime]$}
\STATE $Top\leftarrow Top\cup{S_{nh}[\canon{\AComp}](\gamma^\prime,t^\prime)}$;
\COMMENT{(\dag 2) see Algorithm \ref{algo:Snmnm} for {$S_{nh}[\canon{\AComp}](\gamma^\prime,t^\prime)$}}
\ENDFOR[{all $(h-1)$-connected component in $\xstr{\canon{\AComp}}{\gamma^\prime}$ visited}]
\ENDIF
\ENDFOR[all simplex copies of $\gamma$ considered]
\STATE {\bf return} FaceOf($m,\gamma$,$\sigma(Top)$);
\COMMENT{(\dag 3)}
\ENDIF
\end{algorithmic} 
}
\end{algo}
It is easy to see that the above algorithm computes  
{$S_{nm}(\AComp,\gamma)$}.
\begin{property}
If $t$ is the index of a top simplex {$\theta$}, 
incident to the $n$-simplex $\gamma$
in $\AComp$, then Algorithm \ref{algo:Snm}, upon
termination computes, returns the set  {$S_{nm}(\AComp,\gamma)$}
\end{property}
\begin{proof}
By Property \ref{pro:corrSnh} we have that 
$S_{nh}[\canon{\AComp}](\gamma^\prime,t^\prime)$
computes all top $h$-simplices incident to the simplex copy $\gamma^\prime$.
Control ensure that the algorithm computes this function for
all the $\gamma^\prime$ that are simplex copies of $\gamma$ with
$\ord(\gamma^\prime,\canon{\AComp})\ge m$.
The result of this computation are all disjoint and are accumulated
into the variable $Top$ at lines (\dag 1) and (\dag 2).
Taking $\sigma{(Top)}$ we have
all top $h$-simplices incident to $\gamma$ in $\AComp$ for $h\ge m$.
Taking the $m$-faces with  ${\rm FaceOf}(m,\gamma,\sigma(Top))$ (see line (\dag 3))
we generate, from the set  $\sigma{(Top)}$ of
all top $h$-simplices incident to $\gamma$, the set of all
$m$-faces $\beta$ such that 
$(\exists \tau\in \sigma(Top))(\gamma\le\beta\le\tau)$
This proves the correctness of Algorithm \ref{algo:Snm}
\end{proof}
The complexity of this computation is not always satisfactory but,
under some reasonable conditions the above algorithm is acceptable.
In particular the above algorithm supports the optimal extraction 
of $S_{12}(\gamma|\theta)$ and $S_{13}(\gamma|\theta)$ 
in a $3$-complex embeddable in $\real^3$.
This fact is expressed in the following property.
\begin{property}
\label{pro:nmopt}
The computation of {$S_{nm}(\AComp,\gamma|\theta)$} in a $d$-complex
$\AComp$ for $(d-3)\le n<m\le d$ 
can be done in $\lesscomp{|S_{nm}|\log{|S_{nm}|}}$ 
whenever the given complex is embeddable in $\real^d$.
\end{property}
\begin{proof}
We have to split this proof in several cases according to different $n$ and
$m$. There are six cases, three as $n=d-3$ and $m\in\{d-2,d-1,d\}$, two as $n=d-2$ and 
$m\in\{d-1,d\}$ and one for $n=d-1$ and $m=d$. In the body of this proof we will use $S_{nm}$ as
a shortcut for {$S_{nm}(\AComp,\gamma|\theta)$}.
Similarly we will use {$T_{nh}(\AComp,\gamma)$} or $T_{nh}$ to denote the set
of top $h$-simplices incident at $\gamma$. Since the \cano\ decomposition
neither creates nor deletes any top simplex in $\AComp$ we have that
the set {$T_{nh}(\AComp,\gamma)$} is accumulated in the variable
$Top$ during the computation of $S_{nm}$. Thus, by Property
\ref{pro:corrSnh} Algorithm \ref{algo:Snm}
performs in $\ordcomp{nt\log{nt}}$ with 
$nt=\sum_{h\le m}{{|\T_{nh}(\AComp,\gamma)|}}$.

For $m=d$ we have that the algorithm takes all top simplices  incident 
to each simplex copy of $\gamma$ and insert them in $Top$. 
For each simplex copy the incident top $d$-simplices are retrieved 
in optimal ($n\log{n}$) time as proven in Property \ref{pro:corrSnh}. 
Thus  the  thesis is proven for the three cases with  $m=d$ and for all $(d-3)\le n\le (d-1)$. 
For $n=(d-1)$ we must only consider the case $m=d$ and therefore the 
thesis remains proven for all $m$ when $n=(d-1)$. Now let us increase $n=(d-2)$ and add the two cases $m=(d-1)$ and $m=d$

Following the scheme  used 
in the proof of Property  \ref{pro:euler}, it is easy to see
that for any $n$-simplex
$\gamma$ in a $d$-complex $\AComp$ embeddable in $\real^d$ then   
the cone from a new vertex $w$ to {$\xlk{\AComp}{\gamma}$} is embeddable 
in $\real^{d-n}$ and the link {$\xlk{\AComp}{\gamma}$} is embeddable  in
the $(d-n-1)$-sphere. being $\gamma$ an $n$-simplex in $\AComp$.

For $n=(d-2)$ we can have $m=d$ and $m=(d-1)$. The case for $m=d$ has already been proved.
For $m=(d-1)$, for the computation of {$S_{(d-2)(d-1)}$} we consider all 
top $d$-simplices and all top $(d-1)$-simplices.
We have that the algorithm performs in $\lesscomp{nt\log{nt}}$ with
$nt=|T_{(d-2)(d)}|+|T_{(d-2)(d-1)}|$.
Top $(d-1)$-simplices in ${T_{(d-2)(d-1)}}$  are inserted
directly in the output and therefore {$|{T_{(d-2)(d-1)}}|\le|S_{(d-2)(d-1)}|$}.
To complete the  case for $m=(d-1)$ we have to show that 
{$|T_{(d-2)d}|$} is ${\lesscomp{|S_{(d-2)(d-1)}|}}$.
To this aim we note that
we can project the geometric realization of $\xlk{\AComp}{\gamma}$ onto a 
$1$-sphere $\Gamma$.
This projection is a bijection that sends the set {$T_{(d-2)d}$} 
to a sets of $f_1$ non-overlapping  arcs  in $\Gamma$. The set 
{$S_{(d-2)(d-1)}$} will project to the set of $f_0$  
endpoints of these arcs. Clearly must be $f_1\le 2f_0$. Therefore 
{$|T_{(d-2)d}|\le 2|S_{(d-2)(d-1)}|$} and thus it remains
proven that
{$|T_{(d-2)d}|$} is ${\lesscomp{|S_{(d-2)(d-1)}|}}$.
Therefore $|T_{(d-2)(d-1)}|+|T_{(d-2)(d-1)}|$ is ${\lesscomp{|S_{(d-2)(d-1)}|}}$.
Since the computation of {$S_{(d-2)(d-1)}$} can be done in
$\lesscomp{nt\log{nt}}$ with $nt=|T_{(d-2)d}|+|T_{(d-2)(d-1)}|$
we have that the computation of {$S_{(d-2)(d-1)}$} can be done in
$\lesscomp{nt\log{nt}}$ with $nt=|S_{(d-2)(d-1)}|$.
This completes the case $m=(d-1)$ and $n=(d-2)$ and therefore for $n=(d-2)$ all cases has been proved.

For $n=(d-3)$ regardless of $m$ we have that we can project the link of $\gamma$ onto 
a $2$-sphere $\Sigma$.
Using Properties
\ref{pro:pseudo} and \ref{pro:euler}, we obtain that the
number  of top $d$-simplices in the star of $\gamma$
(i.e. {$|T_{(d-3)d}|$}) is both
{$\lesscomp{|S_{(d-3)(d-1)}|}$} (by Property \ref{pro:pseudo}) and 
{$\lesscomp{|S_{(d-3)(d-2)}|}$} (by Property \ref{pro:euler}).

With this idea, for $n=(d-3)$, we have to show a proof for $m=(d-2)$ and $m=(d-1)$ being $m=d$ already proved in the beginning. For the case $m=(d-1)$ i.e. for the computation of 
{$S_{(d-3)(d-1)}$}
we have that the algorithm performs in $\lesscomp{nt\log{nt}}$ with
$nt=|T_{(d-3)(d-1)}|+|T_{(d-3)d}|$.
We have that $|T_{(d-3)d}|$ is 
{$\lesscomp{|S_{(d-3)(d-1)}|}$} (by Property \ref{pro:pseudo}).
Similarly  top $(d-1)$-simplices in $T_{(d-3)(d-1)}$  are inserted
directly in the output and therefore {$|T_{(d-3)(d-1)}|\le|S_{(d-3)(d-1)}|$},
Therefore 
$nt=|T_{(d-3)(d-1)}|+|T_{(d-3)d}|$. is {$\lesscomp{|S_{(d-3)(d-1)}|}$}
and  thus the computation of {$S_{(d-3)(d-1)}$} can be done in
$\lesscomp{nt\log{nt}}$ with $nt=|S_{(d-3)(d-1)}|$.
This completes the case $m=(d-1)$.

For $m=(d-2)$ i.e. for the computation of {$S_{(d-3)(d-2)}$} the
we have that the algorithm performs in $\lesscomp{nt\log{nt}}$ with
$nt=|T_{(d-3)(d-2)}|+|T_{(d-3)(d-1)}|+|T_{(d-3)d}|$.
We have that $|T_{(d-3)d}|$ is, by Property \ref{pro:euler}, 
{$\lesscomp{|S_{(d-3)(d-2)}|}$}.
Similarly  top $(d-1)$-simplices in $T_{(d-3)(d-2)}$  are inserted
directly in the output and therefore {$|T_{(d-3)(d-2)}|\le|S_{(d-3)(d-2)}|$},
To end this proof we have to prove that {$|T_{(d-3)(d-1)}|$} is
$\lesscomp{|F_{(d-3)(d-2)}|}$ where {$F_{(d-3)(d-2)}$} is the set of
$(d-2)$-faces  of simplices in {$T_{(d-3)(d-1)}$}.
When we project the link of $\gamma$ onto the $2$-sphere $\Sigma$ the top
$(d-1)$-faces in {$|T_{(d-3)(d-1)}|$} project to an arc on $\Sigma$ 
and each $(d-2)$-face in {$T_{(d-3)(d-2)}$} project
to an arc endpoint. 
Between arcs $e$ and vertices $v$ in a graph on a sphere holds the relation  
$e\le 3v-5$ thus
{$|T_{(d-3)(d-1)}|$} is $\lesscomp{|F_{(d-3)(d-2)}|}$.
Being {$F_{(d-3)(d-2)}\subset S_{(d-3)(d-2)}$} 
we have {$|F_{(d-3)(d-2)}|\le|S_{(d-3)(d-2)}|$}.
In conclusion $n=|T_{(d-3)(d-2)}|+|T_{(d-3)(d-1)}|+|T_{(d-3)d}|$.
is {$\lesscomp{|S_{(d-3)(d-2)}|}$}.
Thus we can say that the algorithm perform the 
extraction of {$S_{(d-3)(d-2)}$} in
 {$\lesscomp{|S_{(d-3)(d-2)}|\log}{|S_{(d-3)(d-2)}|}$}.
This completes the proof.
\end{proof}
The above property shows that within $3$-complexes 
we are able to compute  {$S_{12}(\AComp,\gamma|\theta)$} and {$S_{13}(\AComp,\gamma|\theta)$}
in optimal time.
Similarly for $4$-complexes 
we are able to compute  {$S_{12}$},{$S_{13}$},{$S_{14}$},{$S_{23}$}
and {$S_{24}$} in optimal time.
Thus for $3$-complexes embeddable in $\real^3$ 
all topological relations can be computed in optimal
time if we can provide a top simplex within the set to be computed.
For $4$-complexes, even for those embeddable in $\real^4$  this
algorithm fails to be optimal for
$S_{01}$ (see \ref{pro:s01ext}).

\subsection{The relation $S_{nm}(\AComp,\gamma)$}
\label{sec:trie}
The computation of the relation {$S_{nm}(\AComp,\gamma)$} reduces
to  the computation of {$S_{nm}(\AComp,\gamma|\theta)$} if we can 
provide a top simplex $\theta$ incident to $\gamma$.
To satisfy  this requirement one will have to introduce some 
sort of indexing for $n$-simplices and associate a top simplex
with each $n$-simplex. Thus we can extend  the upper layer of our 
representation with an optional function $\FT{n}$ that,
for each $n$-simplex $\gamma$ gives a top 
simplex in the star of $\gamma$, i.e. 
$\FT{n}(\gamma)$ is a top simplex such that 
$\FT{n}(\gamma)\in\xstr{\AComp}{\gamma}$.

This relation represent a possible option for the explicit modeling of 
$n$-simplices  in $\AComp$. The choice of this kind of modeling for
$n$-simplices is actually quite compact.
In fact, in general, modeling  explicitly $n$-simplices means to
introduce some sort of association between $n$-simplices and some other
entity in the model.
Whatever will be the class of elements 
$n$-simplices are associated with, we will have to add at least an array 
storing a pointer for each $n$-simplex. This requires something 
in between  {$f_n(\AComp)\log{NV}$}   and {$f_n(\AComp)\log{NT}$}
bits for this array.
(recall that $f_n(\AComp)$ is the number of $n$ simplices in $\AComp$).  
We believe that is impossible, in general, to obtain optimal
extraction of {$S_{nm}(\gamma)$} without this extra price.
A comparison with existing data structures for {\em manifold}
tetrahedralizations that support the optimal extraction of $S_{12}$ and 
$S_{13}$ is shown in the conclusions and confirms this claim.
However extra memory requirements must be
as close as possible to {$f_n(\AComp)\log{NV}$}.

We note that due to our prior decomposition procees,
we have confined non-manifoldness into maps $\sigma$,  $\sigma^{-1}$
and $\splitmap$.
In this section we will show that it is possible to compute in optimal
time {$S_{nm}$} extending the \Cdecrep\ Representation with an
auxiliary relation (denoted by $\FT{n}$) whose data structure takes less
than  $f_n\log{f_n}+f_n\log{NT}$ bits. 
Note that here and in the following we use $f_n$ for the
face number of the original, un-decomposed complex $\AComp$ i.e.
$f_n=f_n(\AComp)=|\AComp^{[n]}|$.

Furthermore we will show that we can compress all auxiliary data 
structures for $\FT{k}$ relations, for $k\le n$, in the the data structure
for $\FT{n}$ and encode them using  less 
than  $f_n\log{f_n}+\sum_{0<k\le n}f_k\log{NT}$
bits.
In particular for $n=d-1$ we have that we can code all what is needed to
extract in optimal time all topological relations using less than
than  $f_{d-1}\log{f_{d-1}}+\sum_{0<k\le {d-1}}f_k\log{NT}$.

\subsubsection{\emi{Dictionaries}Dictionaries for $n$-simplices}
We now consider the design of an efficient data structure to encode
the mapping $\FT{m}$   and the relation 
$\splitmap$. These data structures are similar since they are 
maps whose keys are simplices represented as set of vertex indexes.  
To develop a compact and efficient implementation  for these
abstract data structures  we consider the set of $m$-simplices in $\AComp$ 
as a set of {\em words} of length $m+1$ over 
an alphabet given by the set of vertex indexes given by 
$Vertex=[1.\ldots,NV]$.
Thus the design of a data structure for these maps
reduces to the probelm of designing a {\em dictionary}.

For a  dictionary  the {\em trie} \cite{Fre60} is the  classic data
structure. In the following we will introduce tries and specialize them to 
this particular task. The result will be a data structure that can encode
collectively relations $\FT{m}$ for $0\le m<d$ using 
exactly $|\AComp^{d-1}|\log{|\AComp^{d-1}|}$ bits for the indexing structure
where ${|\AComp^{d-1}|}$ is the number of (all) simplices in the 
$(d-1)$-skeleton of $\AComp$.
This trie will support access to $\FT{m}(\gamma)$ in 
{$\lesscomp{\log{|\AComp^[m]|}}$} (i.e. {$\lesscomp{\log{f_m(\AComp)}}$}).  
Similar, logarithmic access time holds for the $\splitmap$ data structure.

\subsubsection{Tries}
In particular a sub-section is reserved to present {\em tries}.
A trie is a data structure used to encode dictionaries. 
In the context of this thesis tries are used as an indexing structure whose
keys are simplices described by the ordered sequence of simplex vertices.

Maps are used in this thesis to implement functions of the
form $\sigma[\gamma]$ being $\gamma$ an $m$-simplex represented
by a set of vertex indexes.
Obviously a binary search tree (BST) whose nodes holds sets of vertices is the simplest 
option to implement the map  $\sigma[\gamma]$.
In this case the access to the map $\sigma$ implies lexicographic 
comparison between two sets representing two $d$-simplices in a $d$-complex.
Since each sets is kept within a BST the comparison between
two simplices is in $\lesscomp{d}$
and operations on the map $\sigma$ takes  $\lesscomp{dn\log{n}}$, where
$n=|domain(\sigma)|$ is the number of elements in the map.

To develop a compact and efficient implementation  for these
maps we consider the set of $m$-simplices in the map domain
as a set of {\em words} of length $m+1$ over 
an alphabet given by the set of vertex indexes.
Thus the problem of implementing a map $\sigma[\gamma]$.
for simplices is quite similar to the problem of implementing a
{\em dictionary}.

The {\em trie} data structure is the classical solution to implement 
dictionaries.
A \ems{trie} \cite{Fre60} is a tree-like abstract data type  
for storing {\em words} in which there is one node for every common 
prefix. The name {\em trie} comes from re{\em trie}val.
We will briefly describe tries and a possible implementation.

Let $V$ be an ordered set called the alphabet, we will call $w$ a {\em word} if $w$ is a  list of elements in $V$.
If $W$ is a set of  finite words $\brk{v_1\ldots v_n}$ we will denote with $W/v$ the set of words  obtained deleting the prefix  $v$
from the words of $W$ that starts with $v$.
If no word in $W$ starts with $v$ the set $W/v$ is an empty set.
In particular if $W$ contains the word $\brk{v}$, then
$W/v$ contains the {\em empty word} $\brk{\mbox{}}$.
With this notation we can associate a tree to a set of words with $W$. This tree will be 
denoted by $\tr{W}$. This tree is  
defined inductively 
as follows:
\begin{definition}[Trie]
	\label{def:trie}
	Given a finite set $V$, called {\em alphabet}, and a set 
	$W\subset V^{\star}$ of sequences of symbols in $V$, called a
	\ems{word} set, we will define 
	the \emd{trie} associate to $W$, denoted by $\tr{W}$, as the tree 
	inductively defined as  follows. 
	\begin{enumerate}
		\item If $W=\emptyset$, $\tr{W}$ is a tree with just one node. We will say that this node consumes the empty word.
		\item If $W=\{\omega\}$, i.e. $W$ contains just one word, then
		$\tr{{\{\omega\}}}$ is a tree with just one node labeled with word $\omega$.
		We will say that this node consumes the word $\omega$.
		\item In all other cases  $\tr{W}$ is the tree where the root has 
		a subtree {$\tr{W/v}$} for each (vertex) $v\in V$ such that $W/v$ is non empty.
		In this case we will say that the root of the subtree {$\tr{W/v}$} is a node 
		labeled with $v$ and that the transition from the root to this node consumes 
		$v$.
	\end{enumerate}
\end{definition}
The fundamental property of a trie is that a word $\omega$ is in $W$ \siff\
in the trie $\tr{W}$ we can go
from the root to some leaf $l$ by consuming all and alone the symbols in $\omega$
in the order in which they appear in $\omega$.
In this situation we will say that the word $\omega$ indexes the leaf $l$.
In this sense the trie $\tr{W}$ is the Deterministic  Finite Automaton (DFA) that recognize all and alone the 
words in $W$.
A trie is an indexing data structure and we will say that
a word indexes a particular leaf of the tree $\tr{W}$ \siff\ $\omega\in W$.
{
	\begin{figure}
	\begin{center}
	\framebox{
		\parbox[0.45\textwidth]{0.45\textwidth}{
			\psfig{file=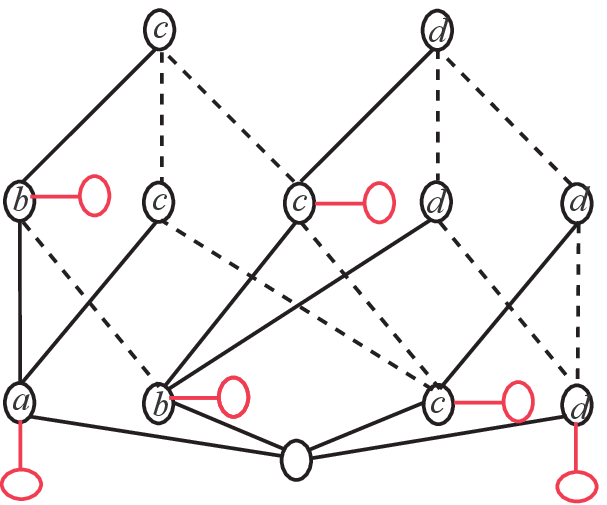,width=0.45\textwidth}
			\begin{center}(a)\end{center}
		}
	}
	\framebox{
		\parbox[0.45\textwidth]{0.45\textwidth}{
			\psfig{file=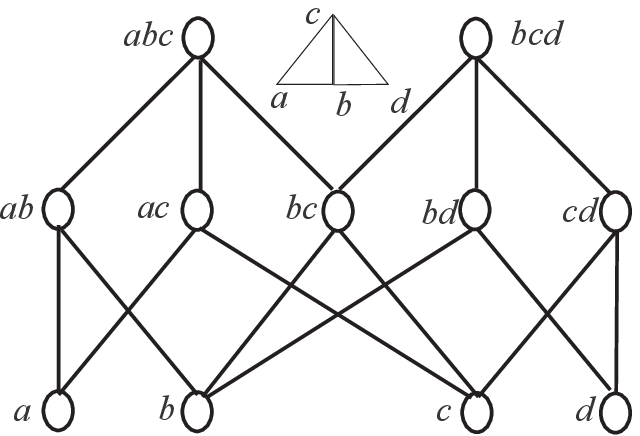,width=0.45\textwidth}
			\begin{center}(b)\end{center}
		}
	}
\end{center}
\caption{Hasse diagrams and tries: 
	the Hasse diagram for the lattice of faces of the two triangles $abc$, $bcd$ from Figure 
	\ref{fig:triglatt} (b) and the corresponding trie (a) (see Example \ref{ex:trie})} 
\label{fig:trigtrie}
\end{figure}
}
\begin{example}
	\label{ex:trie}
Consider for instance a set of words $W$ over the vocabulary $V=\{a, b, c, d\}$. Let be $W=\{a, ab, ac, abc, b, bc, bd, bcd, c, cd, d\}$.  
The corresponding
trie $\tr{W}$ is shown in Figure \ref{fig:trigtrie}a
The red nodes consume an empty word and the dashed lines must be ignored and are there for future reference.
\end{example}
Tries can be implemented using a map for each node in the tree $\tr{W}$.
Each map can be implemented as a
{\em binary search trees} (BST). This kind of data structure was
proposed in \cite{Ben97,Cla64} under the name of \emas{ternary}{search tree}.
In a  $\tr{W}$, implemented with a ternary search tree,
we can decide whether $\omega\in W$ or access information indexed by $\omega$ or insert a key $\omega$
with $\ordcomp{\log{|W|}}$  comparisons (see Theorem 5 in \cite{Ben97}).

Each BST 
in a trie node can be heapified and represented with an array with no
extra space. This array representation for each heap can be packed within
a single array.
For each map we propose to store a heapified BST. This could take at least,
 $\log{|V|}$ bits for each symbol to be consumed and 
  $\log{N}$ bits  to store and index for the next node to go. Here $N$ is the number of nodes in the tree.
  It can be proven that this minimal 
   occupation can be attained (See references in section \ref{sec:map}) With this convention the lenght of the array representation for all BST
is easily available.
Each node in the trie is reached from just one path and therefore we can implement
the trie $\tr{W}$  using at most $\log{|V|}+\log{N}$ bits for each node
in the tree $\tr{W}$.
We note that this space is used to build  the indexing data structure for
the tree $\tr{W}$ and we still have to add the space for the information
associated to each word in  $W$.

\subsubsection{Tries and Complexes}
It is easy to associate a set of words, $W(\AComp)$ 
to an \asc\ $\AComp$ with Vertices 
in $V'$. We just have to take for each simplex 
$\gamma=\{v_j|j=1,\ldots,n\}$ the  word $\omega(\gamma)=
\brk{v_{p_1},\dots,v_{p_n}}$ being $p_i$ the permutation of ${1\ldots n}$.
such that $v_{p_i}>v_{p_{i+1}}$ in the (now ordered) set $V$.
Thus we can associate a set of words $W(\AComp)$ to an \asc\ $\AComp$ given 
by $W(\AComp)=\{\omega(\gamma)|\gamma\in\AComp\}$. 
Similarly we can 
associate the trie  {$\tr{W(\AComp)}$} to the \asc\ $\AComp$. 

We note there is a tight relation between the \asc\ $\AComp$ and the 
associated trie {$\tr{W(\AComp)}$}. In fact it can be proved
that the Hasse diagram for the 
lattice associated with $\AComp$ contains 
(an isomorphic copy of) the forest of trees obtained from the trie 
{$\tr{W(\AComp)}$}
deleting the root and all leaves consuming an empty word. For this reason we will reserve a special notation to this forest and
we will denote it as $\tr{\AComp}$.

It can be proved that  the forest $\tr{\AComp}$ spans completely
the Hasse diagram, i.e. $\tr{\AComp}$  contains all Vertices in the
Hasse diagram.
The proof of this property builds upon
the fact the forest obtained deleting the root from the trie  subtree {$\tr{W(\AComp)/v}$} is a forest that 
is (isomorphic to) a subgraph within the Hasse diagram for $\lk{v}$.
This forest spans completely the Hasse diagram for $\lk{v}$, too. 
\begin{example}
	Consider for instance the complex $\AComp$ made up of two adjacent
	triangles $abc$ and $bcd$ in Figure \ref{fig:trigtrie}b.
	The corresponding set of  words  $W(\AComp)$ is
	$\{a, ab, ac, abc, b, bc, bd, bcd, c, cd, d\}$ and the corresponding
	$\tr{\AComp}$ is shown in Figure \ref{fig:trigtrie}a. 
	In Figure \ref{fig:trigtrie}b we have the Hasse diagram for the lattice of the set of faces
	in an \asc\ $\AComp$ ordered by the face relation
	(see Definition \ref{def:asc}). This is well known poset called
	the \emas{face}{lattice} (see Example \ref{ex:hasse}). Looking at the two figures is easy to see that the {$\tr{W(\AComp)}$} is contained in the face lattice, just delete the red nodes, the root and add dashed lines.
\end{example}
This happens in general for every \asc.
This claims is proven in the following property
\begin{property}
\label{pro:latt}
Let $\AComp$ an \asc with Vertices in $V$ with $|V|>1$.
An isomorphic copy of  a forest  contained in {$\tr{\AComp}$} is contained
in the lattice 
induced by the poset of simplices of  $\AComp$ ordered by the face relation.
This forest spans completely all elements of the poset for $\AComp$.
For each vertex $v$ in $\AComp$
we have that the forest  obtained deleting the root and leaves from the
 {$\tr{W(\AComp)/v}$} 
is isomorphic to a forest within the lattice for $\lk{v}$.
This forest spans completely all elements of the poset for $\lk{v}$.
\end{property}
\begin{proof}
The proof comes easily by induction of the number $|V|$ of Vertices in 
$\AComp$. 
If $|V|=1$ we have $\AComp={\{\{a\}\}}$ and both $\tr{\AComp}$  and the Hasse diagram for the face lattice are isomorphic, with just one node.

For the case $|V|>1$ we first note that 
the face lattice associated to $\AComp$ is a {\em geometric lattice}
(see \cite{Bir67} \S IV.11) whose minimal elements are 
vertices and whose maximal elements are top simplices.
Directly from the definition of link we can say that the lattice
for {$\xlk{\AComp}{v}$} is isomorphic to a sublattice of $\AComp$ that 
contains the set of simplices $\gamma$ such that  $v\in\gamma$.
Thus, for each forest $F$ contained in the lattice for 
{$\xlk{\AComp}{v}$}, whose roots are minimal elements in
{$\xlk{\AComp}{v}$}, we can find,  in the lattice for $\AComp$ a tree rooted
at $v$ and whose elements of depth $1$ are roots of the forest $F$.

Next we note that {${W(\AComp)/v}$} is  exactly
{${W(\xlk{\AComp}{v})}$} and thus by inductive hypothesis
deleting the root from {$\tr{W(\AComp)/v}$} and empty leaves we obtain a forest
within the Hasse diagram for {$\xlk{\AComp}{v}$} that spans all 
nodes in {$\xlk{\AComp}{v}$}.
Thus {$\tr{W(\AComp)/v}$} is a tree within the lattice for $\AComp$.
By the trie definition, deleting the root in {$\tr{W(\AComp)}$},
we obtain a forest whose trees are the trees {$\tr{W(\AComp)/v}$},
one  for each vertex $v$.
Thus an isomorphic copy of the forest {$\tr{\AComp}$} is contained
in the lattice $\AComp$. 
We note that for all $v\in V$ the set of words  ${W(\AComp)/v}$ is non-empty 
therefore we span all nodes in the lattice for $\AComp$.
\end{proof}
We note that different forests {$\tr{\AComp}$} are generated changing the 
ordering of $V$. However a  particular forest, 
in general, is a proper subgraph of the 
Hasse diagram of the lattice $\AComp$ simply because some edges are missing.
In general the tree {$\tr{\AComp}$} contains all and alone the paths in the 
lattice  for simplex $\gamma$.

\subsubsection{Trie implementation}
The implementation of this trie provide easily a data structure for the 
{\em collective} implementation of maps {$\FT{m}$} for $0\le m<d$.
We just have to link information for {$\FT{m}(\gamma)$} to the leaf \null 
node indexed by $\omega(\gamma)$. Similarly the map $\splitmap$ and
each map indexed by $\splitmap(\gamma)$ can be implemented using a trie.

Tries can be implemented using {\em binary search trees} (BST) 
that in turn can be  maintained as a complete binary trees 
and implemented by an {\em heap} using arrays using just one 
index for each node in the tree. 
If we assume this fact we can prove the following.

\begin{property}
\label{pro:trievtn}
\mbox{}
\begin{enumerate}
\item\label{pro:constr}
The construction of a collective trie for
{$\FT{m}$} for $0\le m<d$ can be done with a time complexity of 
{$\lesscomp{|\AComp^{d-1}|\log{|\AComp^{d-1}|}}$}.
where $|\AComp^{d-1}|$ is the number of simplices in the $d-1$ skeleton 
$\AComp^{d-1}$.
\item\label{pro:acc}
The the access time  to $\FT{m}(\gamma)$ is  
{$\lesscomp{\log{f_m}}$}.
\end{enumerate}
\end{property}
\begin{proof}
To prove \ref{pro:constr} we proceed as follows.
The construction of {$\FT{m}$} proceeds by inserting one after one
all the words $\omega(\gamma)$ one for each $m$ simplex $\gamma$. 
Each insertion takes $m+1$ access to $m+1$ BST down the trie tree. 
Each access can be done in {$\ordcomp{\log{s_k}}$} where $s_k$ is the number of sons of the
$k$-th node in  the trie path we are following for insertion.
The total access time is  {$\ordcomp{\log{s_1 s_2\ldots s_{m+1}}}$}.
Obviously the product ${s_1 s_2\ldots s_{m+1}}$ is smaller than the number of
leaves  in the trie for {$\FT{m}$}. The number of leaves in the trie for
{$\FT{m}$} must be smaller than the face number $f_m$ of $m$-simplices 
in $\AComp$.
Therefore the construction of the trie for  {$\FT{m}$} takes 
{$\lesscomp{f_m\log{f_m}}$} and the construction of a collective trie for
{$\FT{m}$} for $0\le m<d$ takes {$\lesscomp{|\AComp^{d-1}|\log{|\AComp^{d-1}|}}$}.
where $|\AComp^{d-1}|$ is the total number of all simplices (not only top simplices) in the $d-1$ skelton 
$\AComp^{d-1}$.

Similar arguments can be used to prove Part \ref{pro:acc}.
In fact the mechanism used to insert $\gamma$ is used also to access 
{$\FT{m}(\gamma)$}.
Therefore the access to {$\FT{m}(\gamma)$} takes  
{$\lesscomp{\log{f_m}}$} even if a single trie is used to implement 
collectively all the $\FT{m}$ for $0\le m<d$. 
\end{proof}

{
To analyze space requirements for this trie we assume that the BST  
is implemented using heaps. This choice do not change access time that
remains logarithmic. With this implementation each BST in each trie node
is maintained as complete binary tree. 
Thus it can be represented with an array with
just one index for each node in the tree. 
}
Thus the number of indexes needed is just the number of nodes in the trie
minus one.
With this assumption we can prove the following
\begin{property}
	\label{pro:vmtsize}
The construction of a collective trie for
{$\FT{m}$} for $0\le m<d$ takes less than
{$|\AComp^{d-1}|(\log{|\AComp^{d-1}|}+\log{|V|}+\log{NT})$} bits.
\end{property}
\begin{proof}
Now, due to the relation between  the trie $\tr{\AComp}$ and the lattice
for $\AComp$, expressed by Property \ref{pro:latt}, we 
can say that the number of nodes in {$\tr{\AComp}$} is 
exactly  $|\AComp^{d-1}|$. 
Therefore each index will take {$\log{|\AComp^{d-1}|}$} bits and the whole 
indexing structure will take {$|\AComp^{d-1}|(\log{|\AComp^{d-1}|}+\log{|V|})$}.

Next we need to evaluate the space needed to reference a top simplex
for each node in the trie for {$\tr{\AComp}$}.
We have that each node $n$ in the trie corresponds to a simplex $\gamma(n)$
and must point a top simplex in $\xstr{\AComp}{\gamma}$.  
Actually, we can reduce the number of these references to top simplices.
We need only to associate top simplices to 
leaves of the trie tree. Infact when we reach $n(\gamma)$
we can still travel the tree to a leaf successor of $n$ and
find a top simplex $\theta$ referenced by that leaf.
This successor will be associated to a coface of $\gamma$ and therefore 
the top simplex $\theta$ must be  incident to $\gamma$, too.
Note that the leaves associated to $(d-1)$-simplices
are not all the leaves of the tree {$\tr{\AComp}$} and there are  
leaves that are associated to non-top simplices
in $\AComp$. 
Thus these references will be coded with less than {$|\AComp^{d-1}|\log{NT}$}
Summing the two terms we obtain the thesis
\end{proof}

Finally note that we do not need to insert in the trie
{$\tr{\AComp}$} the  simplices that are already in the domain of $\splitmap$.
These simplices will never be used as keys to access $\FT{m}$.
Thus we can save in the implementation of the trie for  {$\tr{\AComp}$} 
the space needed for the indexing structure of $\splitmap$.
The following property evaluates the space needed to code the
elements in the codomain of the map  $\splitmap$. 
\begin{property}
\label{pos:sigmaspl}
All the elements of the form {$\splitmap[\gamma]$} can be encoded using
less  than {$\phi(d\cdot\log{d\cdot\phi}+\log{NT'})$} bits where {$\phi=(2^{d+1}-(d+3))NSP$}
and with $NSP$ 
we denote with $NSP$ the number of top simplices in the $d$-complex
that are incident at least to a non-manifold simplex.
\end{property}
\begin{proof}
We need to count the number of references to top simplices
in the maps of the form $\splitmap[\gamma]$. 
The non-manifold simplices and the simplex copies in $\canon{\AComp}$
maps to  a subset of the non-manifold simplices in $\AComp$.
Therefore simplices referenced by the map $\splitmap$ are simplices that 
are incident to a non-manifold simplex in $\AComp$.
There cannot be, in the $\splitmap$ map, more references to a certain 
top simplex $\theta$ than the number of its $m$-faces,  for $0<m<d$.
In fact, if for the same simplex $\gamma^\prime$  in $\canon{\AComp}$
we have non empty entries in $\splitmap[\gamma][\gamma^\prime]$
and $\splitmap[\beta][\gamma^\prime]$ must be $\gamma=\beta$.
The overall number of $m$-faces for for $0<m<d$ is exactly 
{$(2^{d+1}-(d+3))$}. Thus there are at most  {$\phi=(2^{d+1}-(d+3))NSP$} 
references to top simplices in $\splitmap$. 
Considering that each reference to a top simplex takes $\log{NT'}$
bits we need $\phi\log{NT'}$ for all these references.
Assuming the 
worst case of a tree with a distinct chain of maximum length
for each reference we have
$d\phi$ nodes and $d\log{d\phi}$ bits for each reference. 
This leads to  $\phi d\log{d\phi}$ bits for the whole indexing
structure of this trie.
Summing the two terms we get the thesis.
\end{proof}
We note that this is not a rough estimate of the space requested 
by this structure. It is quite easy to find a non-manifold for which 
the \cano\ decomposition generates exactly {$\phi=(2^{d+1}-(d+3))NSP$}
simplex copies. We present this situation in the next example.
 {
	\begin{figure}
		\begin{center}
			{
					\psfig{file=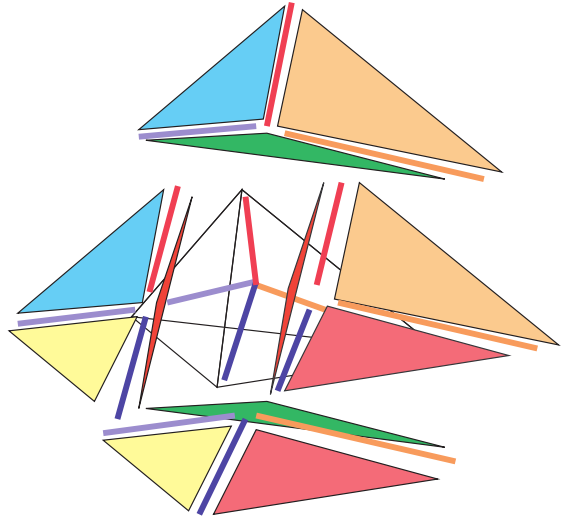,width=0.45\textwidth}
			}
		\caption{A 2-complex whose standard decomposition is totally exploded, see Example 
			\ref{ex:totalexp}} 
		\label{fig:totalexp}
\end{center}
	\end{figure}
}
\begin{example}
	\label{ex:totalexp}
The simplest example is given  for $d=2$ by the 
$2$-skeleton of the tetrahedralization obtained by starring the 
standard $3$-simplex at some internal point.
See Figure \ref{fig:totalexp}. The four transparent traingles defines the initial tetrahedron.
This starring will split the $3$-simplex into a $3$-complex
with  $4$ tetrahedra. The $2$-skeleton of this complex has 
$10$ triangles. In Figure each new triangle is represented thrice with the same color. They are six. The triangles of the original tetrahedron are transparent and represented only once.  The  $2$-skeleton has $10$ edges.  Each edge is a non-manifold edge. Indeed all  edges are adjacent  to three triangles. We represented internal edges with thick colored lines, four copies for each. Edges of the original tetrahedron are thin and black. Each is represented once.
Thus each triangle is incident to a non-manifold edge.
Thus for this complex we have $NSP=10$ and $d=2$ and therefore $\phi=30$.
The \cano\ decomposition of this complex is its totally exploded version 
with $10$ distinct triangles. 
Thus we have $30$ $1$-simplex copies.
\end{example}
\section{Summary and Discussion}
In this section we recall the proposed representation and the 
associated date structure. We summarize space and time performance and 
compare space requirements with existing solutions. 

The \NMWDS\ Representation (see Definition \ref{def:nmwds}) and 
the associated Data Structure (see Data Structure \ref{data:nmwds})
NMWDS=(EWS,$\sigma$,$\sigma^{-1}$,$\splitmap$) can encode a $d$-complex 
$\AComp$  encoding its \cano\ decomposition $\canon{\AComp}$.
This representation is a two layer representation. In the lower layer we
encode the connected components of the complex $\canon{\AComp}$ using the \Cdecrep\ Representation ($EWS$ )
(see Definition \ref{def:cdecrep}). The associated Data Structure
(see Data Structure \ref{data:tttv}) is extended to a Global \Cdecrep\
Data Structure (see Data Structure \ref{data:globtt}) 
to accommodate, in the single data structure $EWS$, the decomposition
$\canon{\AComp}$. This data structure is then optimized to the
the more compressed Implicit Data Structure \ref{data:globimptt}.

On the other hand, the upper layer is made up of the two relations
$\splitsimp0 $ and $\splitstar0 $
implemented by the three maps: $\sigma$,$\sigma^{-1}$
(see Section \ref{sec:sigma}) and $\splitmap$
(see Section \ref{sec:splitmap}).

\subsection{Space Requirements}
If the original complex has $NV$ vertices and $NT$ top simplices and
if we denote with $NS$ the number of splitting vertices and with 
$NC$ the  number of vertex copies introduced by the decomposition process
we have that the encoding of $\canon{\AComp}$ takes
(see Property \ref{pro:ewdsocc})
$$\SIZE(\log{NT'}+\log{NV'})+NV'\log{NT^\prime}\, \mbox{\rm bits}$$ where:
$\SIZE=\sum_{0\le h\le d}{|\Theta^{[h]}|(h+1)}$,
being  {$\Theta^{[h]}$} the set of top $h$-simplices in $\AComp$.
and $NV'=NV-NS+NC$.
This representation can be compressed saving 
(see Property \ref{pro:optspace}) $NV'(\log{NV'}+\log{NT'})$ bits.

The upper layer, made up of the maps: $\sigma$,$\sigma^{-1}$ and $\splitmap$
takes: 
{
\begin{enumerate}
\item 
up to $NS\log{NS}+NC\log{NC}$ bits to encode the map $\sigma^{-1}$  (see Property \ref{pro:sigmaspace})
\item 
$NC(\log{NC}+\log{NS})$ bits to be encode the map $\sigma$ (see Property \ref{pro:sigmaspace})
\item
the map {$\splitmap$} can be encoded using
less  than {$\phi(d\log{d\phi}+\log{NT'})$} bits where 
{$\phi=(2^{d+1}-(d+3))NSP$} and where with $NSP$ we
we denote the number of top simplices in the $d$-complex
that are incident at least to a non-manifold simplex.(see Property \ref{pos:sigmaspl})
\end{enumerate}
}
\subsection{Information and Non-Manifoldness}
Note that  the three parameters  $NS$, $NSP$, $NC$ gives a measure of three
different aspects of the non manifoldness in the complex $\AComp$:
$NS$ gives an idea how how many singularities are present. $NSP$ gives an idea
of which portion  of the original complex is incident to non-manifold
situations.
$NC$ gives  an idea of the severity of the non-manifoldness in terms of
the complexity of the actions needed to split the complex into \cdec\ 
components.
The results above shows that, for a complex with a certain number $NT'$ of
top simplices, the information needed to encode its singularities is an 
increasing function of these three parameters.
Thus, this gives some theoretical foundation to a classification of 
the complexity of the non-manifold structure of a complex in term of 
the three parameters $NC$, $NS$ and $NSP$, using the information given by
the formula:
\begin{equation}
\label{eq:info}
\widehat{H(\AComp)}=NS\log{NS}+NC\log{NC}+\phi(d\log{d\phi}+\log{NT'}) 
\end{equation}
This formula is obtained summing the space needed by the map $\sigma$ and by
the map $\splitmap$. We omit the map $\sigma^{-1}$ being clearly redundant 
w.r.t. $\sigma$. The analysis of our two layer data structure shows that
the quantity $\widehat{H(\AComp)}$ in formula \ref{eq:info} 
is surely an upper bound for the information  
associated with the non-manifold aspects in a complex.
This quantity, or a normalized version of it, can be used for 
the classification of non-manifold structures according to the complexity
of their non-manifold structure.

.
\subsection{Time Requirements for the Construction of the \NMWDS\ Data Structure}
We assumed that the map $\sigma$ 
is a by-product of the output of  the decomposition 
algorithm that takes $\lesscomp{NT\log{NT}}$ (see \ref{pro:decocomp}). 
From this output we can build the  
NMWDS=(EWS,$\sigma$,$\sigma^{-1}$,$\splitmap$) data structure with the 
following time requirements:
\begin{itemize}
\item 
The EWS data structure  can be built in  $\lesscomp{d^2 NT\log{NT}}$
(see Property \ref{pro:ttbuildcomp}) using  Algorithm \ref{alg:tfill} and 
can be optimized using Algorithm \ref{algo:nmtt} and
Algorithms from \ref{algo:opt} to \ref{algo:optren}. This
optimization takes  $\lesscomp{dNT'}$ (see Property \ref{pro:optspace});
\item
The map $\sigma$ comes from the decomposition process and the
map $\sigma^{-1}$ can be build out of $\sigma$ in $\ordcomp{NC\log{NC}}$
(See Property \ref{pro:sigmaspace});
\item
The map $\splitmap$ can be computed using Algorithms \ref{algo:travelstar}
and \ref{algo:splitmap} in  
$\lesscomp{NSP\cdot 2^d\log{NSP}}$ 
(See Property \ref{pro:splitmap})
\end{itemize}
The complexity of $\lesscomp{NSP\cdot 2^d\log{NSP}}$ 
represent a drastic reduction in complexity w.r.t. a
global analysis.
This reduction is possible due to the prior decomposition of the original
complex $\AComp$ that allows, when building $\splitmap$, to restrict the search to the neighborhood of a subset of non-manifold vertices.

\subsection{Extraction of Topological relations $S_{nm}(\AComp,\gamma)$}
The \NMWDS\ Data Structure supports the computation
of the topological relations $S_{nm}(\AComp,\gamma)$ using only the
$\sigma$ and $\sigma^{-1}$ maps 
(see Formula \ref{eq:stitch} in Section \ref{sec:snmnm}).
In particular, using Algorithm \ref{algo:somnav} we can compute
{$S_{0m}(\AComp,\gamma)$} in $\lesscomp{n\log{n}}$ being $n$ the size of the
output. This holds for $2$-complexes and for $3$-complexes embeddable
in $\real^3$ (See Property \ref{pro:somopt}).
However, in general the
complexity of the associated computation can become  polynomial w.r.t. 
the dimension of the output 
(see Properties \ref{pro:s01}, \ref{pro:nmineff} and \ref{pro:mintri}).
The computation of $S_{nm}(\AComp,\gamma)$ in $\lesscomp{n\log{n}}$
can be supported adding the $\splitmap$ relation. 
In this case, using Algorithm \ref{algo:Snm}, we can compute 
{$S_{nm}(\AComp,\gamma|\theta)$} in $\lesscomp{n\log{n}}$ being
$n$ the size of the output and assuming that at
least a top simplex $\theta$ incident to $\gamma$ is given at the beginning
of  the computation (see Property \ref{pro:nmopt}).

Finally the task of providing a top simplex incident to a given $n$-simplex
$\gamma$ is  carried through by an auxiliary data structure that we called the
$\FT{m}$ amp.   This data structure represents the way in which we introduce,
in the framework of the \NMWDS\ Representation, an explicit modeling
for $m$-simplices for $1\le m\le d-1$.

The basic facts about the map $\FT{m}$ are the following:
\begin{itemize}
\item
The construction of a collective trie for
{$\FT{m}$} for $0\le m<d$ takes less than
{$|\AComp^{d-1}|(\log{|\AComp^{d-1}|}+\log{|V|}+\log{NT})$} bits (see Property 	\ref{pro:vmtsize})
being $\AComp^{d-1}$ the $(d-1)$ skeleton of $\AComp$.
Hence $|\AComp^{d-1}|=\sum_{0\le k\le(d-1)}f_k\log{f_k}$
\item
The construction of a collective data structure for
{$\FT{m}$} for $0\le m<d$ can be done with a time complexity of 
{$\lesscomp{|\AComp^{d-1}|\log{|\AComp^{d-1}|}}$}.
where $|\AComp^{d-1}|$ is the number of simplices in the $d-1$ skelton 
(See Property \ref{pro:trievtn} Part \ref{pro:constr})
\item 
The the access time  to $\FT{m}[\gamma]$ is  
{$\lesscomp{\log{f_m}}$}
(See Property \ref{pro:trievtn} Part \ref{pro:acc}) 
\end{itemize}

\subsection{Comparison with  existing approaches} 
For the comparison between manifold and non-manifold approaches we refer to
the empirical study in \cite{LeeLee01} where classic 
non-manifold modeling  schemes are compared with classical manifold 
modeling approaches such as the Winged Edge \cite{Bau72} and the Half Edge 
\cite{Man83}.
This empirical analysis was conducted on the ground of statistical
data from \cite{Wil88} for solid models represented by
cellular complexes. The results shown in Figure \ref{table:ha} (but row 4) are  reported in \cite{LeeLee01}. They  give some ratios for 
known modeling approaches for non-manifold surfaces:
{
\begin{figure}
\caption{Space requirements for different data structures for surfaces normalized vs. space requirements for the Half Edge. Data (but the fourth row) are
from \cite{LeeLee01}} 
\label{table:ha}
\begin{center}
\begin{tabular}{|c|l|c|}\hline\hline
\multicolumn{2}{|c|}{Modeling Data Structure} & Ratio to HE  \\ \hline
\multicolumn{2}{|c|}{Half Edge} & 1  \\ \hline
1& Winged Edge & 0.80   \\ \hline
2& Partial Edge & 1.7   \\ \hline
3& Radial Edge & 3.5   \\ \hline
4& \ewds & 0.92   \\ \hline
\end{tabular}
\end{center}
\end{figure}
}  
The fourth row is theoretically computed assuming a manifold
triangulated closed surface of genus $0$.
This ratio comes from the fact that the \NMWDS\ Data Structure reduces to the \Cdecrep\ Data Structure for
manifold objects and we have seen that an optimized Global \Cdecrep\ Data
Structure uses 5.5 pointers for each triangle (see remark at the 
end of Property \ref{pro:optspace}). Thus we have that for the \Cdecrep\
Data Structure the ratio to the Half Edge is $5.5/6=0.92$ since the
Half Edge uses six  pointers for each triangle.

If we add an explicit representation for edges we must add $f_1$ pointers
for edges plus $f_0+f_1$ pointers for the indexing structure of $\FT{1}$
yielding additional 3.5 pointers for each triangle. In this case the
ratio to the Half Edge rise to  $9/6=1.5$. However note that neither 
the Half Edge nor the Winged Edge offer an explicit representation for 
vertices and faces   since  they are based on explicit
modeling for edges. On the other hand, the \Cdecrep\ Data Structure
is based on the explicit representation of triangles and vertices.

We next compare the \Cdecrep\ with approaches for manifold volumetric cellular
representations \cite{DobLas87,LopTav97}. In the Handle-Face
Data Structure \cite{LopTav97} we have 36
pointers for each tetrahedron for a occupancy of $36f_3$
while in the Facet-Edge Data Structure \cite{DobLas87} 
we have a data structure requiring $12f_2$ pointers.
These two approaches compares with 
$\SIZE(\log{NT'}+\log{NV'})-NV'\log{NV^\prime}$ bits where $\SIZE=4f_3$.
If we assume that $\log{NT'}$ and $\log{NV'}$ are replaced by the size
of a pointer we got an occupancy of $8f_3-f_0$ pointers.
Adding an explicit representation for edges and triangles  we have to
add  the maps $\FT{1}$ and  $\FT{2}$.
Thus we need $f_0+2f_1+2f_2$ pointers for a total
of $o=8f_3+2f_1+2f_2$ pointers. Since we are comparing the occupancy for
manifold models we have from Property
\ref{pro:pseudo} $4f_3\le 2f_2$ and using the fact that in a 
tetrahedralization $f_0\le 4f_3$, $f_1\le 6f_3$ and $f_2\le 4f_3$ we have that
$o\le 28f_3$.
This shows that the ratio with the Handle-Face is above $36/28=1.28$
using also Property \ref{pro:mainfface}
$4f_3-3f_2+2f_1=2f_0$ it is easy to see that 
$o=4f_3+5f_2+2f_0\le 12f_3+5f_2\le 11f_2$.
This shows that the ratio with the Facet-Edge is above $12/11=1.09$.

A comparison is possible also with models based on incidence graph.
These \cite{Edel87,Woo85}  encodes completely the lattice associated with
the complex $\AComp$ with using two pointers to represent each edge 
in the Hasse Diagram for $\AComp$. This requires, for a
tetrahedralization $2(4f_3+3f_2+2f_1)$ pointers.  relations. 
A reduced scheme can reach the occupancy of at most $8f_3+3f_2+3f_1+f_0$.
This is above the occupancy of the \NMWDS\ (i.e. $o=8f_3+2f_1+2f_2$)
of a difference of $f_2+f_1+f_0$. It is easy to see that the ratio vs.
this reduced version of incidence graph is  above $1.07$.
{
Finally in 
Brissons's Cell Tuple \cite{Bri93} and in
Lienhardt's  n-G-maps \cite{Lie91} the representation is $\ordcomp{df_d(d+1)!}$
while the \Cdecrep, enriched with $\FT{d-1}$, occupies $\lesscomp{f_d2^d}$
thus the ratio between the two is $\lesscomp{2^d/d(d+1)!}$ that
asymptotically tends to zero. For $d=3$ We have a ratio w.r.t. the Cell Tuple
that is above $3.43$ and the ratio   w.r.t the n-G-maps is above $2,57$.
}
The following table summarize this analysis and gives the relation between the
\Cdecrep\ Data structure and other modeling approaches compared over the 
domain of tetrahedralizations of $3$-manifolds.
{
\begin{figure}
\caption{Space requirements for different data structure for $d$-manifolds
for $d\ge 3$ normalized v.s. space requirements for \NMWDS\ Data Structure} 
\label{table:tetra}
\begin{center}
\begin{tabular}{|c|l|c|}\hline\hline
\multicolumn{2}{|c|}{Modeling Data Structure} & Ratio to \ewds  \\ \hline
\multicolumn{2}{|c|}{\ewds} & 1  \\ \hline
1& Handle-Face & $>1.28$   \\ \hline
2& Facet-Edge & $>1.09$   \\ \hline
3& Incidence Graph & $>1.07$   \\ \hline
4& Cell Tuple & 3.43   \\ \hline
5& n-G-map & 2.57   \\ \hline
\end{tabular}
\end{center}
\end{figure}
}
Obviously this comparison do not take into account the fact that
these approaches can model cellular complexes and thus the comparison might
be not so  bright in all those cases in which  if the applicative domain 
do not impose a simplicial representation.
Nevertheless our scheme is compact enough to encode manifold surfaces and
manifold tetrahedralizations, providing explicit representation for
all $n$-simplices being always more compact than classic approaches in
the field. 

When we apply this approach  to model non-manifold complexes the 
space requirements of our data structure grows linearly with the 
structural complexity of non-manifoldness in the modeled complex.  
This data structure is useful as long as the structural complexity of
non-manifoldness is not too deep. Otherwise the size of auxiliary 
structures used to model non-manifoldness might be too important and probaly 
a data structure conceived to model non-manifoldness everywhere is a 
better modeling choice. To give a rough estimate of the tradeoff between 
this approach and the other approaches we restrict out attention to
regular non-manifold complexes.
We rework the formula that gives the occupancy of  the upper layer i.e.
the sum of the three terms
$$
NS\log{NS}+NC\log{NC};\, 
NC(\log{NC}+\log{NS});\,
{\phi(d\log{d\phi}+\log{NT'})} 
$$
where 
{$\phi=(2^{d+1}-(d+3))NSP$} and where with $NSP$ we
we denote the number of top simplices in the $d$-complex
that are incident at least to a non-manifold simplex.(see Property \ref{pos:sigmaspl});
with $NS$ we denote the number of splitting vertices and with 
$NC$ the  number of vertex copies introduced by the decomposition process.
We can assume that all logarithmic terms in formulas above are 
replaced by the size of a pointer. Then we use the fact
that $NS<NSP$ and $NC<NSP$ to see that less than $(\phi(d+1)+4)NSP$
pointers are necessary to encode the upper layer.
For $d=2$ we have to add  at most
$13$ pointers for each triangle incident to a non
manifold vertex. For $d=3$ 40 more pointers are needed.

Using data from table in Figure \ref{table:ha} and restricting 
our attention to regular 2-complexes
we can predict that 
21 pointers per triangle are needed by the Radial Edge and 10.2 pointers
per triangle are needed by the Partial Edge Data Structure. 
We ask for 5.5 pointers for triangle to encode the decomposition 
and 9 pointers  for triangle to encode the decomposition modeling
explicitly all simplices. w.r.t. the Radial Edge we use from
15.5 to 12 pointers less for each triangle  and this gain is 
balanced when $13NSP=12f_2$ thus our representation is a viable substitute
for the Radial Edge when the ratio $NSP/f_2$ is below $12/13=0.92$.

Thus as long as the number of non-manifold triangles adjacent to a 
non-manifold simplex is less than the 92\% we have some profit in using
our data structure instead of the Radial Edge. 
Furthermore our solution  is always preferable if we do not need to model
all simplices and thus we do not introduce $\FT{2}$.

The same comparison for the Partial Edge yields an interval from 13\% to  40\%.
Namely if $NSP/f_2$ is below 13\% we can save space  using our
data structure instead of the Partial Entity Data Structure.
If we do not need the $\FT{2}$ relation we can use our data structure 
whenever $NSP/f_2$ is below 40\%.

For regular $3$-complexes the most relevant modeling
option to compare with  is the Incidence Graph. The Incidence Graph data
structure requires $8f_3+6f_2+4f_1$ relations to be stored. 
Since the lower layer takes
$o=8f_3+2f_1+2f_2$ pointers we remain with a gain of $4f_2+2f_1$ 
pointers that are used by the upper layer. Obviously in this case we
consider a data structure containing $\FT{1}$ and $\FT{2}$ because
all simplices are explicitly modeled by the Incidence Graph.
The upper layer takes less than $40NSP$ pointers and this equates the 
gain of $4f_2+2f_1$ when $NSP/f_3$ is the limit ratio 
$NSP/f_3=(4f_2+2f_1)/40f_3$.
If we assume that we are dealing with complexes imbeddable in $\real^3$
we have $4f_3\le 2f_2$ and
thus the limit ratio is always bigger than  
$8f_3+2f_1/40f_3\ge 1/5$.
Thus in this case our data structure  is a more  efficient option
than Incidence Graph whenever $NSP/f_3$ is below 20\%.
 \chapter{Conclusions}
\label{ch:concl}
In this short chapter we recall briefly several proposals from the state-of-the-art in Chapter \ref{ch:related} and show how the results in this thesis represent a (possibly) relevant option w.r.t. existing solutions.
\section{The decomposition problem} 
In the first part of this thesis we studied the problem of decomposing a generic non--­manifold simplicial 
d-­complex into more regular components \cite{Def02b}. We called these components \iqm. 
This problem was 
considered in the context of Combinatorial Topology and the results obtained are dimension independent. This first part provided a better understanding of the combinatorial structure of non-manifolds and 
represents a contribution in the direction of topology-based geometric modeling \cite{Lie97}. 
The problem of decomposing non-manifolds has been already studied in geometric modeling. However, the few proposed solutions \cite{Des92,FaRa92,RosCad99} are limited to the problem of decomposing surfaces. In 
particular, in \cite{FaRa92} the problem of decomposing two-dimensional boundary representations of r-sets 
(i.e., uniformly two-dimensional objects) is studied. In \cite{Des92} this former decomposition of the boundary 
of r-sets is assumed and extended Euler operators are introduced to build such a model. Finally, in \cite{RosCad99} is presented an algorithm, called Matchmaker, that decomposes the boundary of r-sets (or even a 
2-complex) into a set of manifold surfaces. This latter approach, although restricted to non-manifold 
2-complexes, attempts to minimize the number of duplications introduced by the decomposition process. In \cite{Gui98,Gui99} we also find the idea of cutting a two-dimensional non-manifold complex into 
manifold pieces. This decomposition is used within a geometric compression algorithm that uses a 
two level representation: in a first level manifold components are encoded separately, the second level 
encodes stitching instructions for components. 
From the very beginning of our study we have noticed that the decomposition of a complex could 
be studied with no reference to the geometry of the complex and, thus, we adopted the classical 
framework of the Combinatorial Topology \cite{Gla70,Hud69}. Indeed, a strong relation between Combinatorial 
Topology and Topology based Geometric Modeling is clearly pointed out in \cite{Lie97}. A major benefit of this (purely combinatorial) approach is the possibility of treating the general problem of decomposition in any dimension. A second benefit is the possibility of 
providing proofs that do not require geometric intuition. 

\section{Decomposition: a formal definition} 
In this framework we first tackled the problem of giving a non-operational definition of the notion 
of decomposition. A naive statement of the decomposition problem might require to search for a 
decomposition algorithm that decompose a complex into maximal manifold connected components. 
By requiring maximal components, we mean that we  do not accept 
decomposition where there are components that can be merged into a bigger 
manifold subcomplex. This 
requirement about maximal components seems quite ''natural'' since, otherwise, the problem becomes 
quite trivial and the collection of all top simplices in the original complex, each considered as separate 
component, would be a solution to this problem. 
A first result (that it is quite easy to prove) was that the decomposition problem, with such a requirement about maximal manifold components, is,
in general, unsolvable. 
Classical results in combinatorial topology \cite{Mar58,Vol74} easily imply 
that, for $d\ge 6$, there is not a decomposition algorithm 
that splits a generic d-complex into maximal manifold parts.
Such a decomposition problem is actually 
equivalent to recognizability of d-manifolds. This problem is settled for $d=4$ \cite{Tho94}, it is still an open 
problem for $d=5$ and is known to be unsolvable for $d\ge 6$ \cite{Vol74}. 
Moreover, already for surfaces, 
several possible decompositions exists. The non-uniquiness of decomposition is usually neglected in existing approaches (with 
the notable exception of \cite{RosCad99}). 

We have defined a decomposition using a specific class of abstract simplicial maps and using the notion of combinatorial manifold. Intuitively we can say that we have chosen to 
define a decomposition as the result of a process that is allowed to cut a complex just along singular 
(non-manifold) features. 
We have found that this statement, although apparently ''natural'', has some important consequences. First of all, there are 3-complexes that cannot be decomposed into manifold parts. It 
is interesting to note that some of those unbreakable tetrahedralizations fell into the class of non-manifolds defined by Lienhardt and called quasi-manifold \cite{Lie94}. However we have found examples of 
3--complexes that cannot be decomposed into pseudomanifolds. Equivalently we can find examples of 
non-pseudomanifold tetrahedralizations that cannot be decomposed (according to the decomposition 
notion given above) into pseudomanifold parts. This means that there are certain tetrahedralizations 
that cannot decomposed into {\em pseudomanifold} components by simply cutting them at singular features. 
An example of such a tetrahedralization is given in the thesis. 

A possible solution is to break the unbreakable complex along non singular features. This is 
clearly not desirable since several arbitrary cuts are possible. There are examples of unbreakable 
tetrahedralizations where different cuts, along non singular features, will yield alternative non homeomorphic  decompositions. 

This problem of choosing among several possible decomposition is not addressed too often in existing literature. 
The best answer to the possible impasse  in choosing is in Matchmaker \cite{RosCad99}. Matchmaker is a decomposition algorithm 
for surfaces that searches the space of possible decomposition by looking for an optimal solution 
according to some a priori criterion.  Another finding in this thesis is that there exist always  a most general decomposition and all the others proposed solutions are options taking some more stitching for  components that are separated in the most general solution.
\section{The Standard Decomposition}
Indeed, our goal was to see if it could be possible to define a non-arbitrary non a priori decomposition. We assumed that a 
non arbitrary decomposition must be, somehow, more general than other decomposition. Intuitively, 
we look for a decomposition obtained by further cuts in other arbitrary decomposition. We search 
the decomposition space for a general solution by cutting
arbitrary decomposition. 
In this search, we forbid 
to cut along features that were manifold (i.e. non-singular) in the original complex. In other words, 
we ask to cut as much as possible whenever we cut at a singular features and see if a most general 
decomposition exist. 
One of the main results of this first theoretical part is that such a decomposition exists for a generic d-complex. Furthermore, we have proven that, for any d, this decomposition is unique up to isomorphism. We called such a decomposition the {\em Standard Decomposition}. In spite of the fact that a 
decomposition into manifolds is not computable, we have found that the standard decomposition can be 
computed. If $t$ is the number of top simplices in the standard 
decomposition, this computation can be 
done at least in (nearly) linear time (i.e.  $\ordcomp{t\log{t}}$   ). 
We have developed an algorithm that transforms a 
complex into its standard decomposition by a sequence of local operations modifying just simplices 
which are incident at a vertex. Each local operation is computed using just local information about 
the star of the vertex (i.e., the set of simplices incident to a vertex). 
\section{The Decomposition Lattice and Initial-Quasi-Manifolds}
To provide a framework for the results of this first part of the thesis  
we have defined an ordering among complexes and built a poset. This poset orders all 
the possible modifications for a complex. Possible modifications we considered are those obtained 
by a sequence of vertex pair stitching (a vertex pair stitching being the transformation that collapses 
together two vertices). This poset turns out to be a {\em lattice} with a top and a bottom element. The top 
element is the complex made up of a set of isolated simplices and the bottom element is the complex 
made up of a single point. Every complex in this lattice is associated with an equivalence relation 
between vertices. One can move on the lattice adding equations that stitch together just two vertices 
at time. 
In this framework the standard decomposition becomes simply the least upper bound for a specific 
set of decompositions, obtained by cutting a complex at singularities. 

Although this lattice originates from the particular goal of studying decompositions we think that 
this construction can be useful in general. This framework supports the study of topological properties of complexes by tracing syntactical properties of sets of equations between vertices. Using 
equations, we have analyzed the topological property of the connected components in standard decomposition. We singled out the class of complexes that are possible connected components in a 
standard decomposition. We called these complexes {\em initial-quasi-manifolds}. It is an easy consequence of this theoretical development that initial-quasi-manifold are all and alone the unbreakable 
complexes. Thus, the standard decomposition of an initial-quasi-manifold complex is the complex 
itself. 

We have proven that initial-quasi-manifold can be defined in terms of local properties of the 
star of each vertex. In the star of an initial-quasi-manifold two top d-simplices must be connected 
with a path of d-simplices, each linked to the other via a (d-1)-manifold (non singular) joint. It 
can be proven that initial-quasi-manifold d-complexes are a proper superset of d-manifolds 
for $d\ge 3$. They coincide with manifolds for $d=2$. Furthermore, it is easy to see 
that initial-quasi-manifolds are a decidable set of d-complexes for any d. 
As we already mentioned, there are (initial-quasi-manifold) tetrahedralizations that are unbreakable and that are not pseudomanifold. 
However, such a tetrahedralization cannot be embedded in $\real^3$ . We like 
to mention that a nice benefit of this theoretical framework has been the possibility of checking a 
proof of the correctness of this example, that we cannot visualize, 
by manipulating equations via a Prolog program. 

Furthermore, we felt that the methodology and notations developed for this part can be used to give a set of algebraic tools that can be used to study other application domains in Computer Graphics such as:  Simplification, Compression, Multi-resolution and Feature based classification. According to this strong feeling we found interesting to develop this study within a formal settlement as an example of how algebraic techniques and combinatorics can be used for this task.

\section{The Non-Manifold Data Structure}  
The second part of this thesis dealt with a two layered data structure conceived to model the 
decomposed non-manifold. This data structure models separately the structure of the decomposition 
and each connected component of the decomposition. 
The decomposition structure, at an abstract level, is 
modeled via an hypergraph \cite{DeFMagMorPup00}. Each component of the decomposition is encoded using an extended 
version of the Winged Representation \cite{PaoAl93}. This approach offered a compact, dimension independent, 
data structure for non-manifolds that can be used whenever the modeled object has few non-manifold 
singularities. Algorithms used to build and navigate this data structure were presented. We have shown that they have optimal time  performance in the usual domain of 2-complexes and 3-complexes.
We also analyzed the space requirements of the two layer data structure and 
discussed a possible approach  for the computation of all topological relations.

Most common data structures for non-manifolds \cite{Wei86,Gur90,LeeLee01} are quite space-consuming since they 
assume that non-manifoldness can occur very often in the model. The resulting data structures are 
designed to accommodate a singularity everywhere in the modeled object. As a result, storage costs 
do not scale with the number of non-manifold singularities. In other words, the performance of such 
non-manifold data structure is quite poor when used to code a manifold. On the other hand, very 
efficient data structures for subdivided 2-manifolds \cite{Bau72,Man83,LopTav97,GuiSto85,DobLas87} do exist. Furthermore, most 
modeling proposals for non--manifolds are limited to 2-complexes and the few data structures for 
dimension independent modeling do not pay special attention to singularities. Some of them code just 
manifolds \cite{Bri93}, or a special subset of non-manifolds \cite{Lie94}. Others are quite general since they simply 
code incidence relations between cells \cite{Edel87} or a subset of these incidence relations \cite{Woo85,DefSob02}. Models 
based on incidence relations can be quite satisfactory, supporting even a full set of Boolean operations 
\cite{RosCon90}. However. in general, models encoding incidence 
relations  between cells can be quite space consuming 
if simple cells (e.g. triangles) are 
used. 
On the contrary, models based on incidence relations can be quite compact 
provided that one decompose the object into few components and a 
single cell is used for each component. Obviously, to 
have a compact model, a compact coding scheme  for each component
must exist. Thus 
we studied the problem of representing a simplicial d-complex using our the decomposition into  initial-quasi-manifold  
connected components. 

\subsection{The Two-Layer Data Structure}
Having established a sound notion of decomposition, we have considered the design of a dimension 
independent layered data structure that exploits this decomposition. 
In this direction we have been influenced by the idea behind SGC \cite{RosCon90} where complexes are modeled with a cellular complex with quite complex cells. In an SGC the cellular complex, represented 
via an incidence graph, can be regarded as an upper layer that ties together quite complex cells that 
can, for instance, be any open manifold merged in $\real^3$. 

For the data structure used to code initial-quasi-manifolds components, we see some relations with ideas in some works in the field of dimension independent 
modeling. In particular we were influenced by ideas of Lienhardt on nG-maps \cite{Lie91,LieElt93} and somehow by Brissons's cell-tuples \cite{Bri93}. All these approaches provide a dimension independent machinery 
to represent, respectively, either a certain subclass of non-manifold complexes or plain d-manifolds. 
Similarly we adopted a uniform, dimension independent, scheme to represent initial-quasi-manifold. 
This scheme is an extended version of the Winged Representation \cite{PaoAl93} and the modeled class of complexes (i.e. initial-quasi-manifold) is very close to Lienhardt' s quasimanifold. 

On the other hand the 
concept behind data structure we propose is, somehow, complementary to the Radial-Edge structure 
\cite{Wei86}. 
In our two-layer data structure the upper level is used to encode the structure of the decomposition. This is done by encoding an hypergraph \cite{DeFMagMorPup00}. 
From the results about the decomposition process   we have that, in the decomposition process, 
just singular (i.e. non-manifold) simplices are duplicated across different 
components. 
Furthermore there is no need to represent explicitly sub-faces of a duplicated singular simplex. 
Hence, in this hypergraph, nodes contain the representation of initial-quasi-manifold components and hyperarcs need to encode just top singular simplices. In each hyperarc we store the description of non-manifold vertices for the corresponding top singular simplex. On the other hand, 
the lower level is devoted to the 
coding of (initial-quasi-manifold) decomposition components. 

The coding of (initial-quasi-manifold) decomposition components 
is accomplished via an unconstrained usage of the the Winged Representation \cite{PaoAl93}. The Winged Representation considers complexes built stitching d-simplices at (d-1)-faces. This offers exactly the 
right level of abstraction to treat initial-quasi-manifold components. We point out that this scheme 
is used here beyond its intended domain in order to treat initial-quasi-manifolds that can be even 
non-pseudomanifolds. 

To encode the hypergraph, some additional information must be added to the 
coding of singular vertices in each initial-quasi-­manifold. We stress that this additional data need only 
to be provided for non-manifold vertices. Whenever the number of non-manifold vertices is low, this 
additional information can be stored and retrieved easily using hashing techniques. 

We note that completeness of this data structure comes from the definition and the existence 
of standardized decompositions. Similarly uniqueness of representation comes from our result on 
uniqueness of the standardized decompositions. 

Finally, in the last part of the thesis, the two layer data structure is developed in detail showing how we 
can construct this data structure using the results of the decomposition algorithm. The complexity of 
the proposed solution is analyzed. In particular, we detail space requirements and give the complexity 
required to build the data structure and to extract all topological relations between simplices of different dimension. The proposed two layer data structure proves to be fast and compact with respect to 
existing approaches. 

Another interesting result in the second part of this thesis comes from a trivial application of deep results from Combinatorics. Data structures that stores only (d-1)-adjacency between top d-simplices are bound to travel all (d-1) simplices around a vertex to find all d-simplices incident to a certain vertex. This choice bounds extraction to be non optimal in higher dimension.  
The idea was to see what happens of this search scheme  when going in higher dimension. We were interested in that because we adopted such a scheme to encode \iqm\ components for our decomposition. For $d\ge 4$ classical results in combinatorial topology (i.e. the so called 
{\em Upper Bound Theorem}
by McMullen \cite{MS71}) implies that extraction time of some topological relations could be non-optimal. More precisely (see Property \ref{pro:s01} and \ref{pro:s01ext}) complexity is 
polynomial. For instance, finding all edges round a vertex is  $\ordcomp{e^{\floor{d/2}}}$ where $e$ is the number of edges in the output. \appendix
\chapter{Posets, Lattices and the Partition Lattice}
\label{sec:thlatt}
\section{Introduction}
In this thesis it is crucial to compare decompositions as generated
by equivalences among vertices.
The goal of this appendix is to introduce the right mathematical framework to
order equivalences.
Equivalences, ordered by set inclusion, form a partially ordered set that is
a well known object in Lattice Theory.
This lattice is called the Partition Lattice and is usually denoted by $\Pi_n$.
In this appendix we summarize basic notions from Lattice Theory  necessary
to present the properties of the Partition Lattice.
These properties, often recalled in this thesis by a claim to intuitive arguments, 
are used extensively. In this section intuition is left
apart and we show that
there is a  mathematical framework behind this  claim to intuition.

The last section of this appendix develops, is more specific to this thesis 
and specialize this notion to the aims of our work. The reader that might find
hard to read the first three sections might browse directly to the last 
section to have an idea of what is this all about.
The material reported here is not meant to
be a self-contained presentation of this subject. Thus, some useful
results are reported without proof. However precise reference to 
textbooks is given for each result. 
The interested reader may refer to  \cite{Bir67,Coh81,Gra68} for a
complete treatment.

\section{Partially Ordered Sets (Poset)}
\label{sec:poset}
A (binary) \ems{relation} $R$ on a set $X$  
is any subset of the cartesian product $X\times X$. We will write
$xRy$ to mean $(x,y)\in R$.  
A relation $R$ is a \emas{reflexive}{relation} \iff\ $xRx$ for all $x\in X$.
The relation $\Delta_X=\{(x,x)|x\in X\}$, called the \ems{diagonal} in
$X$. The diagonal is the smallest reflexive relation on $X$. 
A relation $\lt$ is a \emas{transitive}{relation} \iff\ for all
$x$, $y$ and $z$ such that $x\lt y$ and $y\lt  z$ we have that that $x\lt z$.
A transitive and reflexive  binary relation $x\le y$ is called 
a \ems{preorder} (see  \cite{Coh81} Pg. 10).
A relation $R$ is a \emas{symmetric}{relation} \iff\ for all
$x$, $y$ we have that  $xRy$ implies  $yRx$.
A relation $R$ is a \emas{antisymmetric}{relation} \iff\ for all
$x$, $y$  we have that $xRy$ and $yRx$ implies that $x\equiv y$ 
(i.e. $x$ and $y$ are the same element in $X$).
An antisymmetric preorder $x\leq y$ is called an \ems{ordering} or a
\emas{partial}{order} on $X$. For a given ordering $\leq$ we will denote 
with $x\lt y$  the relation obtained deleting from $\leq$ all elements in the
diagonal $\Delta_X$ (i.e. $x\lt y$ \siff\ $x\leq y$ and $x\neq y$).
If $\leq$ is an ordering on $X$ the set  $X$ is called a partially
ordered set and the pair {$P=\ualg{X}{\leq}$} is called a
\emas{partially ordered}{set} or  \ems{poset}.
With some abuse of notation we will write $x\in P$ to mean $x\in X$.
{
Finite posets can be represented by particular diagrams called 
\emas{Hasse}{diagrams}. To define Hasse diagrams we need to introduce some
more definitions. 
\begin{definition}[Immediate Superior (see \cite{Bir67} Pg. 4)]
We will say that in a poset $P=\ualg{X}{\leq}$ $x$ \emd{covers} $y$
whenever $x\gt y$ and, for no $z\in X$, $x\gt z\gt y$.
In this situation $x$ is called the  \emad{immediate}{superior} of $y$ in $P$.
\end{definition}
Similarly $y$ is called the  \emas{immediate}{inferior} of $x$ in $P$.
Using the covering relation one can represent a poset $P$ with a directed 
graph $G(P)$ such that $(x,y)$ is an edge in $G(P)$ \iff\ $x$ covers $y$.
It can be proved that the graph $G(P)$ is an acyclic graph (see Lemma 2 Pg. 2 \cite{Bir67})
If we take the unoriented graph associated  with $G(P)$ and we draw it
with straght arcs placing $x$ above $y$ 
whenever $x$ covers $y$ we have what is
called   define the \emas{Hasse}{diagram} for  $P$.
}
\begin{example}
\label{ex:hasse}
As an example of a poset and of its Hasse diagram consider the set of faces
in an \asc\ $\AComp$ ordered by the face relation
(see Definition \ref{def:asc}). This is well known poset called
the \emas{face}{lattice} (see \cite{Zie97} Pg. 247).
Figure \ref{fig:triglatt}a shows the Hasse diagram for the face lattice
for the triangle $abc$. Elements of this poset are all the faces of
the triangle. Thus we have the three vertices, $a$, $b$ and $c$ and  
the three edges $ab$, $bc$, $ca$. The top element is the triangle itsself.
In Figure \ref{fig:triglatt}b we have the Hasse diagram for the complex
made up of the two triangles $abc$ and $cbd$.
\begin{figure}
\begin{center}
\framebox{
\parbox[c][0.45\textwidth]{0.30\textwidth}{
\psfig{file=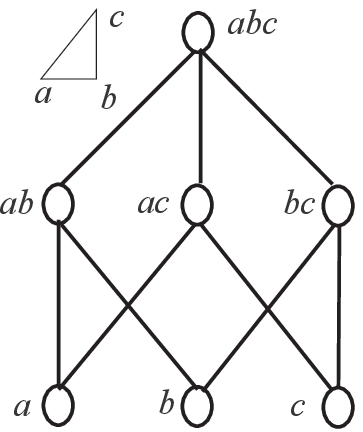,width=0.30\textwidth}
\begin{center}a\end{center}
}
}
\framebox{
\parbox[c][0.45\textwidth]{0.45\textwidth}{
\psfig{file=abcdok.eps,width=0.45\textwidth}
\begin{center}b\end{center}
}
}
\end{center}
\caption{Two Hasse diagrams: the Hasse diagram of the face lattice
for the triangle $abc$ (a) and the Hasse diagram for the two triangles $abc$, $bcd$} 
\label{fig:triglatt}
\end{figure}
A classification of polytopes 
(see proof of Property  \ref{pro:nonopt} for a definition of polytope)
can be given abstracting completely from geometry (see \cite{Zie097} Pg. 129 and \cite{BLS93}).
\end{example}

A partially ordered set where, for all $x,y\in X$,
either $x\leq y$ or $y\leq x$ is called a \emas{totally ordered}{set} or
a \emas{linearly ordered}{set} or a \ems{chain}.
Two elements  $x,y\in X$ are called \ems{incomparable} \iff\ 
neither $x\leq y$ nor $y\leq x$.  They are called \ems{comparable}
otherwise.
It is easy to see  that (Theorem 1 \cite{Bir67}) a subset of a poset (chain)
is still a poset (chain) w.r.t. the same ordering relation.
Given two comparable elements $x\leq y$ 
the \emas{closed}{interval} {$[x,y]$} is the set of
all elements  $z$ such that $x\leq z\leq y$.  
It is easy to see that {$[x,y]$} is a poset, too.

We will write $x \gteq y$ and $x \gt y$ to mean, respectively, $y\leq x$ and 
$y\lt x$.
It is easy to see (Theorem 2 \cite{Bir67})
that $x \gteq y$ defines an ordering relation. This  is called the 
\ems{converse} of relation $\leq$. Hence the pair  
$\dual{P}=\ualg{X}{\gteq}$ is a poset   that is called the 
\ems{dual}{poset} of $P$.
It is easy to see that $\dual{\dual{P}}$ is exaclty the poset $P$.
When dealing with dual posets one can prove just properties just for $P$ 
and use the \emas{duality}{principle} (see \cite{Gra68} Pg. 10) to show 
properties for $\dual{P}$.
The duality principle says that if a statement $S$ holds in $P$ 
then the dual statement $\dual{S}$ obtained exchanging $\gteq$ with $\leq$
must  hold in $\dual{P}$.
The Hasse  diagram of the dual poset $\dual{P}$ is obtained turning the
diagram for $P$ upside down.

An element $o$  in  a poset {$P=\ualg{X}{\leq}$} is  called an 
\emas{lower}{bound}  for a set $H\subset X$
\iff\ $o\leq x$ for all $x\in H$.
Similarly an element $i$  is  called a 
\emas{upper}{bound}  for a set $H\subset X$
\iff\ $x\leq i$ for all $x\in H$.
A lower bound for the whole ordered set $X$, if  exists, is unique and  
is called the \emas{least}{element} of the poset $P$.
Similarly, an upper bound for the whole ordered set $X$, if it exists,
is unique and  is called the \emas{greatest}{element} of the poset $P$.
Least and greatest elements for a poset will be 
denoted respectively with symbols $\bot$ (bottom) and $\top$ (top).

The set of lower bounds (upper bounds) of a poset is a poset, too.
The least element in the poset of upper bounds for a subset $H$ is called
the \emas{least}{upper bound} or \ems{l.u.b.} of $H$.
Similarly, the greatest element in the poset of lower bounds for a subset 
$H$ is called the \emas{greatest}{lower bound} or \ems{g.l.b.} of $H$.
Lub and glb for a set H, when they exist, are unique and will be 
denoted respectively with $\lub H$  and $\glb H$.
The g.l.b. $\glb H$ is also called the \ems{meet} or the {\em product} of
the elements in $H$.
The l.u.b. $\lub H$ is also called the \ems{join} or the {\em sum} of
the elements in $H$.
The names {\em sum} and {\em product} refer to the usual interpretation 
for $\glb$ and $\lub$ in the Boolean algebra that, on the other hand, is a particular lattice.
If the  set $H$ is made up of two elements $a$ and $b$ we will write the 
l.u.b. and g.l.b. respectively as $x\lub y$ and $x\glb y$.
It is easy to prove (see \cite{Bir67} Theorem 4) 
that the l.u.b. and the g.l.b. of a chain always exist.
They are called, respectively, the {\em first} and the {\em last} elements
of the chain.  The chain is said to  lay {\em between} its first and its
last elements. 

A function $\funct{\theta}{P}{Q}$ from a poset $P$ to a poset $Q$ is called 
\emas{order}{preserving} or \ems{isotone} \iff\ $x\leq y$ implies that
$\theta(x)\leq \theta(y)$.
An isotone function with isotone inverse is called an 
\emas{poset}{isomorphism} (see \cite{Bir67} Pg. 3). 
Two posets are called \ems{isomorphic} (in symbols $P\poeq Q$) \iff\ 
there exists an isomorphism between them. 
Two finite posets are isomorphic \iff\ they have the same Hasse diagram.  
A function $\funct{\theta}{P}{Q}$ from a poset $P$ to a poset $Q$ is called 
\ems{antitone} \iff\ $x\leq y$ implies that
$\theta(y)\leq \theta(x)$.
An antitone function with antitone inverse is called an \emas{dual}{isomorphism} (see \cite{Bir67} Pg. 3). 
Two posets are called \ems{anti-isomorphic} \iff\ 
there exists a dual isomorphism between them. 
Obviously a poset and its dual are anti-isomorphic. 
Two finite posets are 	anti-isomorphic if they have nearly the same
Hasse  diagram apart from the fact that one is turned upside down w.r.t.
the other. 
A poset that is anti-isomorphic with itself is called a \emad{self-dual}{poset}.
The Hasse diagram of a self-dual poset remains the same turning it
upside down.

It can be proved  (see Theorem 5 \cite{Bir67}) that a finite chain of
$n$ elements is always isomorphic to the chain of integers $\{1,\ldots,n\}$
(i.e. the ordinal ${\bf n}$) in the poset $\ualg{\natural}{\le}$. With this 
result in mind we can define the \ems{length} $l(C)$ of a chain $C$ as
$l(C)=|C|-1$. Next we define the length  $l(P)$ of $P$ as the ordinal 
that is the l.u.b. of integer legths of its finite chains. 
The poset $P$ is said to be of \emas{finite}{length} whenever the
ordinal $l(P)$ is finite. In particular every finite poset is of finite length.
If a poset $P$ is of finite length and it has a least element $\bot$ we 
define the \ems{height} or \ems{dimension} $h[x]$ of an element $x\in P$ 
as the l.u.b. of the lengths of the chains between $\bot$ and $x$.
Clearly we have that $h[x]=1$ \iff\ $x$ covers $\bot$. 
Such an element $x$ will be called a \ems{point} or an \ems{atom}.
A poset where all elements have a finite dimension $h[x]$ is called a finite  
dimensional poset.
Dimension $h$  can give a way to find the distance between 
comparable elements in a  poset. To show this we need to introduce some 
others definitions.
\begin{definition}  
A poset $P$ is \ems{graded} by a function $\funct{g}{P}{\integer}$ \iff\ $g$
satisfy the following conditions.
\begin{enumerate}
\item $x\lt y$ implies that $g[x]\lt g[y]$;
\item if $x$ covers $y$ then $g[x]=g[y]+1$
\end{enumerate}
\end{definition}
In a graded poset all maximal chains between two points have the same finite
length ({\em Jordan-Dedekind chain condition}) conversely,  
a poset with a least element where all chains are finite  
is graded by $h[x]$ \iff\ it holds the Jordan-Dedekind chain condition.
In this situation the length of any maximal chain between $a$ and $b$ is given
by $|h[a]-h[b]|$. 

\section{The Partition Poset}
\label{sec:partposet}
As an example of a graded poset let us consider the poset of equivalence relations on a finite set $V$ of $n$ elements. To study this
poset 
we need some more definitions. We start by recalling the definition of
equivalence relation. A reflexive,  symmetric and 
transitive relation ${\equivert}\subset(V\times V)$ 
is called an equivalence relation. 
Equivalence relations ordered by set inclusion are a finite poset.
It is easy to see that the least element for this poset is the diagonal
(or identity relation) $\Delta_V$, while the greatest element is $V\times V$.
For a given equivalence relation $R$ we define the \emas{equivalence}{class}
of $x$, denoted by $[x]$ as the set $[x]=\{y|x\equivert y\}$. 
The set of all equivalence classes of all elements in  $V$ form a partition
(see Definition  \ref{def:covering})
of $V$ called the \emas{quotient}{set} of $V$ by ${\equivert}$ 
(usually denoted by $V/{\equivert}$).

Conversely, given a partition  $\Pi$ we can easily define a relation 
$(V/\Pi)$ by saying that $x(V/\Pi)y$ \iff\ $x$ and $y$ belongs to the
same block in the partition $\Pi$. It is easy to see  that 
$(V/\Pi)$ is an equivalence relation. We have seen that set of partitions 
of $V$ is a poset ordered by the refinement relation 
(see Definition  \ref{def:covering}). 
In this poset the least element is the most refined partition i.e.
the partition made up  of singletons. The greatest element is the
coarsest partition. This is the 
singleton partition i.e. the partition made up just one set: the set $V$. 
It can be proved (see Theorem 1 Pg. 3 \cite{Gra68}) that the functions
$\send{{\equivert}}{V/{\equivert}}$ and $\send{\Pi}{V/\Pi}$ are both isotone and that
$V/(V/\Pi)=\Pi$ and $V/(V/{\equivert})={\equivert}$. Thus the poset of equivalence relations and
the poset of partitions for a set $V$ are isomorphic.
Furthermore it is easy to
see that the isomorphism class of these two posets do not depend on
$V$ but just on $|V|=n$.  
Hence this poset is simply denoted  by $\Pi_n$ and is called the 
\emas{partition}{poset}. 
However, in this thesis, we will always use the term
{\em partition poset} to denote the poset of equivalence relations. 
It is easy to see (cfr. Example 9 Pg. 15 \cite{Bir67})
that, for any pair of equivalence relation 
{${\equivert_1}$} and {${\equivert_2}$} we have that
{${\equivert_1}\glb {\equivert_2}=({\equivert_1}\cap{\equivert_2})$}.
It can be proved (see \cite{Gra68} Pg. 19) that
{${\equivert_1}\lub {\equivert_2}$} is the smallest transitive
relations containing both ${\equivert_1}$ and ${\equivert_2}$
(i.e. the transitive closure of ${\equivert_1}\cup{\equivert_2}$).

The poset $\Pi_n$ is graded by $h[\equivert]=n-|V/{\equivert}|$ and 
we have that $\equivert_1$ covers $\equivert_2$ \iff\ $V/{\equivert_2}$
is the coarsening of the partition $V/{\equivert_2}$ obtained uniting
two distinct equivalence classes $[u]_2$ and $[v]_2$ 
(see \cite{Bir67} Pg. 15).
This is equivalent to say that ${\equivert_1}$ is the
smallest equivalence containing both ${\equivert_2}$ and a 
pair $(u,v)$ such that $u{\not\equivert_2}v$.
In particular the points of $\Pi_n$ are equivalence relations obtained
extending the  diagonal with a couple of distinct elements (u,v).

\section{Lattice}
\label{sec:lattice}
Lattices are introduced in this thesis basically to study the structure of 
the quotient poset $\Pi_n$. 
In fact the quotient poset belongs to a particular class of lattices 
called {\em geometric lattices}.
It is interesting to note that partition lattices are a paradigm for
lattices. In fact a deep result of lattice theory \cite{Whi46} shows that any 
lattice is isomorphic to a sublattice of the partition 
lattice $\Pi_{\infty}$ for $\natural$.
\begin{definition}
A \emd{lattice} is a poset $L$ where any pair of elements $x$ and $y$ in $L$ 
have a g.l.b.  or \ems{meet}, denoted by $x\glb y$ and a l.u.b. or 
\ems{join} denoted by $x\lub y$.
\end{definition}
Lattices are a particular subclass of the class of posets. 
Therefore they inherit all attributes of this class.  Therefore, in the 
following, we will freely talk about, for instance, finitely dimensional
lattices to denote a lattice whose associated poset is finitely dimensional. 
A lattice is \emas{complete}{lattice}  when each of its subsets
$S$ has a l.u.b. (denoted by $\lub S$)
and a g.l.b. (denoted by $\glb S$).
It is easy to see (see Lemma 1 Pg. 8 \cite{Bir67}) that the two operations
$\glb$ and $\lub$ are both commutative and associative. 
Furthermore they are both  \emas{idempotent}{operations}  i.e. for all $x\in L$
$x\glb x=x\lub x=x$.
Finally it holds  the \emas{absorption}{law}
$x\glb(x\lub y)=x\lub(x\glb y)=x$.
Lattices can also be defined with no reference to posets as {\em an algebraic structures $L$ with two 
operations that are commutative associative idempotent and that jointly
satisfy the absorption law}. It can be proved (Theorem 8 Pg. 10 \cite{Bir67})
that, in this case the algebraic structure becomes a poset ordered by the 
relation $x\geq y=(x\glb y=x)$ or, equivalently, by the relation $x\geq y=(x\lub y=y)$.

A set {${\cal X}$} of parts of a certain set $X$ has the \emas{closure}{property} 
\iff\ {${\cal X}$} contains $X$ and is closed under intersection. 
In this case, given a set of $S\subset X$ we define the \ems{closure}
of $S$ (denoted by $\closure{S}$) as the smallest subset in
{${\cal X}$} that contains $S$. It can be proved 
(see \cite{Bir67} Theorem 2 Pg. 112) that any set with the 
closure property is a complete lattice ordered by set inclusion where
the two lattice operations are given by $U\glb V=U\cap V$ and
$U\lub V=\closure{(U\cup V)}$.

A subset {$L^\prime$} of a lattice $L$ need not to be a lattice or, if it 
is a lattice, need
not to be a lattice  w.r.t. the  lattice operations of $L$. When this 
happens the subset {$L^\prime$} is called a \ems{sublattice} of $L$.
This happens \iff\ when the subset $L^\prime$ is closed under 
operations in $L$. 
Closed intervals of a lattice are sublattices.

The partition  lattice belongs to a particular  class of lattices called
{\em geometric} lattices.                      
A \emas{geometric}{lattice} is a finitely dimensional semimodular
point lattice.
Thus, to introduce the  partition lattice we need to give the basic 
definitions about point lattices and semimodular lattices.
\begin{definition}[Point Lattice (\cite{Bir67} Ch. IV)]
\label{def:pointlatt}
Let be $L$ a lattice which is a poset of finite length with a least 
element. The  lattice  $L$  is a \emad{point}{lattice} \iff\ every element 
in $L$ is the join $\lub p_i$ of a set of points $p_i$.  
\end{definition}
A lattice that can be organized as a repetition of small ''squares'' is
called semimodular
\begin{definition}[Semimodular Lattice]
\label{def:semi}
A lattice is semimodular \iff\ whenever two elements has a 
common immediate inferior they also have a common immediate superior
\end{definition}
A \emas{geometric}{lattice} is a finitely dimensional semimodular
point lattice.
If the geometric lattice $L$ is of finite length we can give some
interesting properties of the height function $h$ for $L$.
Any semimodular lattice of finite length is graded by the height
function $h[x]$ (Corollary to Th. 14 Pg. 40 \cite{Bir67}).
In a geometric lattice of finite length every element can be expressed as the
join of a finite set of points. Several sets can define the same element.
Hovever  there are sets of points that, in some  sense, are minimal.
These are sets of points that can be ordered into a sequence of 
{\em independent} points.
\begin{definition}[Independent points]
\label{def:indip}
In a geometric lattice of finite length a sequence of points
$p_1,\ldots,p_r$ is   \emd{independent} \iff\ \[
(p_1\lub\ldots\lub p_k)\glb p_{k+1}=\bot\, {\rm for}\, k=1.\ldots.r-1\] 
\end{definition}
If a sequence of points is independent then any permutation of this
sequence is independent, too. Therefore we will talk about 
independent set of points. It is easy ot see that any subset of
a set of  independent 
points is a set of independent points, too.
It can be proven (see \cite{Bir67} IV\S 4) that geometric lattices 
are generated by sets of independent points.
Indeed  a  
generic set of points $Q$ always contains a subset of independent points.
If $Q$ is finite, then all independent subsets of $Q$ have the same number 
of elements. 
This number is called the \ems{rank} of $Q$ (denoted by $r(Q)$).  
In a geometric lattice of finite length each element $x$ is the
join of $h[x]$ independent points.

Geometric lattices are \emas{relatively}{complemented} that to say 
we can go up  from an element $x$ ({$x=p_1\lub\ldots\lub p_k$} with $k=h[x]$)
to its immediate superior adding  a point to the join of independent 
$h[x]$  points $p_i$  that gives $x$.
This is expressed by the following property
\begin{property}
Let  $a$ and $b$ two elements in a geometric lattice of finite length.
Let $a\leq b$ and let 
{$a=p_1\lub\ldots\lub p_k$} with $k=h[a]$, then we can  add
$h[b]-h[a]$ independent points to the set {$\{p_1,\ldots,p_k\}$} 
to get a set of 
$h[b]$ independent points whose join is  $b$. 
\end{property}
\begin{proof}
By Lemma Theorem 6 Pg. 88 in \cite{Bir67} we have that a geometric 
lattice of finite length is
relatively complemented and thus, by definition of
relatively complemented (see Def. L7B Pg. 88 \cite{Bir67}),
we have that there  exist $x$ and $y$ such that $a=x\glb y$ and $b=x\lub y$.
For $y$ we can find $h$ independent points {$\{q_1,\ldots,q_h\}$}
whose join gives $y$. Thus $b$ is the l.u.b of {$\{p_1,\ldots,p_k,q_1,\ldots,q_h\}$}.
We can extract a set of $h[b]$ independent points 
out of {$\{p_1,\ldots,p_k,q_1,\ldots,q_h\}$} putting all the $p_i$ first.
This completes the proof.
\end{proof}
As an example of geometric lattice of finite length we reviev the
lattice $\Pi_n$.

\section{The Partition Lattice $\Pi_n$}
\label{sec:pin}
It is easy to see that the partition poset $\Pi_n$ with the operations 
$\glb$ and $\lub$ defined in section \ref{sec:partposet} is a lattice.
Indeed it can be prove that $\Pi_n$ is a geometric lattice of 
finite length. 

Due to the extensive usage of this lattice in this thesis we introduce 
some ad hoc terminology that makes statements about $\Pi_n$ more intuitive
in this context. In this last paragraph of this appendix we summarize the links between the general lattice theory and the use we have done of it dealing with the decomposition process.
The interested reader might have at hand  examples \ref{ex:latticequo1}, 
\ref{ex:latticequo2}, \ref{ex:equats} and \ref{ex:lattconf} 
to see where we have applied, in the thesis,
definitions and properties given in this appendix.

First we note that points in $\Pi_n$ are the equivalence relation that add
to the diagonal just a pair of couples, let us say $(u,v)$ and $(v,u)$.
We will denote such a point by the \emas{\verteq}{equation} $u\equivert v$ or 
{$v\equivert u$}.
{
We will say that the equivalence ${\equivert}\subset V\times V$ 
\ems{satisfies} the equation $u\equivert v$  if and only if 
$(u,v)\in{\equivert}$.
We will write $u\not\equivert v$ if and only if $(u,v)\not\in{\equivert}$.
In this case we will say that the equivalence $\equivert$ do not
satisfies the equation $u\equivert v$.
}
Furthermore, we will use the operator $+$ instead of $\lub$ and the operator
$\cdot$ instead of $\glb$ to denote lattice operations in the $\Pi_n$ lattice.
Indeed we recall the the g.l.b. of two relations is simply their intersection.

Being $\Pi_n$ a point lattice, an equivalence $\equivert$ is usually given 
(or {\em generated})
by the join of a set of equations (i.e. points) of the form 
$E=\{u_i\equivert v_i|i=1,\ldots k\}$. The equivalence
given by such a set of equations $E$, denoted
by $\equivert^E$, Note that $\equivert^E$ is 
the smallest equivalence $\equivert^E$
such that $\equivert^E$ satisfies all the equations in $E$.
(i.e. $u_i\equivert^E v_i$ for $i=1,\ldots,k$).
We extend this notation to the empty set by taking the identity relation
for {$\equivert^{\emptyset}$}. 

If  $E$ is a  set of equations that defines a
\verteq\ equivalence, denoted by $\equivert^E$,
we will use $E$ as a shortcut for $\equivert^E$ in all the expressions
where this is not ambiguous. In particular we will
write $\equivert\sumeq \;E$ and
$\equivert\proeq\; E$ as a shortcut
for $\equivert\sumeq \equivert^E$  and $\equivert\proeq \equivert^E$.

For each equivalence $\equivert$ there is always a set of equations $E$
such that $\equivert=\;\equivert^E$.
However note that the same equivalence can be defined by different set of
equations. The notion of independent points translates in $\Pi_n$ in the
familiar concept of logic independence of equations. Indeed it is easy to see
a set of \verteq\ equations $E=\{e_i|1\le i\le n\}$
is a set of $n$  {\em independent} equations
if and only if, for all  $1\le k\le n$,
we have that $\{e_k\}$ is not satisfied by the equivalence
generated by  the subset $E_k=\{e_i|1\le i<k\}$.  
Since any permutation of independent points is a sequence of independent
points we have that for any subset $E^\prime\subset E$, 
all equations in $E-E^\prime$ are not satisfied by the equivalence
generated by $E^\prime$.
In general, whenever all equations in a set $E$  is not satisfied
by an equivalence $\equivert$ we will say that $E$ is independent w.r.t.
$\equivert$.

Note that sets of equations are basically synctactic objects that are
quite different from the corresponding  relation.
Some analogy remains, indeed 
if $E_1\subset E_2$ then $\equivert^{E_1}\subset\;\equivert^{E_2}$ and 
$\equivert^{E_1}\sumeq\equivert^{E_2}=\equivert^{E_1\cup E_2}$.
However a similar property
do not exist for  intersection of sets of \verteq\ equations.
Infact, in general,
the equivalence
$\equivert^{E_1}\proeq\equivert^{E_2}$ is not necessarily  
$\equivert^{E_1\cap E_2}$;
Take for instance 
$E_1=\{u\equivert v,v\equivert w\}$ and $E_2=\{u\equivert w\}$.
we have that the
$E_1\cap E_2=\emptyset$ while
$\equivert^{E_1}\proeq\equivert^{E_2}=\{(u,w),(w,u)\}$.
Hence the poset of sets of equations,
ordered by set inclusion, is not isomorphic to the poset of
$\Pi_n$.

Similarly, in  general $\equivert^E{\supset}\equivert^{E^\prime}$
do not implies an inclusion between sets $E$ and ${E^\prime}$, 
not even if  $E$ and ${E^\prime}$ are sets of independent equations.
This is due to the fact that the mapping that sends a (possibly
independent) set of  equations $E$ into the equivalence $\equivert^E$
is not injective.

The height function $h[x]$ in $\Pi_n$ admits a nice interpretation in term 
of equations. The height $h[E]$ 
(recall that $E$ here is a shorthand for $\equivert^E$) is such that $h[E]\le |E|$ and
equality holds \iff\  $E$ is a set of independent equations 
(see Equations (2) Pg. 81 and (13) Pg. 86 in \cite{Bir67}).
In particular passing from an element $E$ ot its immediate superior $E^\prime$
we just have to add an equation $e$ independent w.r.t. $E$.
In this case we will say that equation $e$ is a label for the $1$-chain from
$E$ to $E^\prime$.
Labels for longer chains are obtained collecting equations on $1$-chains.
Note that different equations can label the same chain of length $1$. 
In general if $E\le E^{\prime}$ we must add $h[E^\prime]-h[E]$
independent  equations  to $E$ 
to get an independent set of equations that generates 
{$\equivert^{E^\prime}$}.
Note that in general we will not get {\em exactly} the set of equations
$E^\prime$. We just obtain a set of independent equations that contains 
$E$ and generates the equivalence {$\equivert^{E^\prime}$}.
If added equations are of the form $u\equivert v$ then the couple
$(u,v)$ must be in $\equivert^{E^\prime}-\equivert^{E}$. 

Semimodularity is another property of the geometric lattices that is useful
to derive  properties of the partition lattice.
Let be $E_a$ and $E_b$ two elements in $\Pi_n$ with a common immediate
superior ${E_a\sumeq E_b}$ and a common immediate inferior $E_a\proeq E_b$.
There must be two equations $e_a$ and $e_b$ such that  the following diagram
is a portion of the partition lattice.
{
	\[\begindc{\commdiag}[500]
\obj(1,1)[b]{$\mbox{\large $E_a\sumeq E_b$}$}

\obj(0,0)[a1]{$\mbox{\large $E_a$}$}
\obj(2,0)[c1]{$\mbox{\large $E_b$} $}

\obj(1,-1)[b2]{${\mbox{\large $E_a\proeq E_b$}}$}

\mor{b}{a1}{$e_a$}[-1,0]
\mor{b}{c1}{$e_b$}
\mor{a1}{b2}{$e_b$}[-1,0]
\mor{c1}{b2}{$e_a$}
\enddc\]
}
{In general, iterating the construction of the above diagram to a larger 
portion of the lattice one can prove that, 
given two elements $E_a$ and $E_b$ 
with an  upper  bound $E_{x}$ and a lower bound $E_{y}$ 
we can use the same 
label $E_{xa}$ for the chains from $E_{x}$ to $E_{a}$ and from 
$E_{b}$ to $E_{y}$.
Similarly 
we can use the same 
label $E_{xb}$
for the chains from $E_{x}$ to $E_{b}$ and from $E_{a}$ to $E_{y}$.
The situation is depicted in the following diagram:
	\[\begindc{\commdiag}[500]
\obj(1,1)[b]{$E_{x}$}

\obj(0,0)[a1]{$E_{a}$}
\obj(2,0)[c1]{$E_{b}$}

\obj(1,-1)[b2]{$E_{y}$}

\mor{b}{a1}{$E_{xa}$}[-1,0]
\mor{b}{c1}{$E_{xb}$}
\mor{a1}{b2}{$E_{xb}$}[-1,0]
\mor{c1}{b2}{$E_{xa}$}
\enddc\]

 \chapter{A Space optimization for the \Cdecrep\ Data Structure}
\label{sec:opt}
We first note that
in the original EWDS we can choose the 
numbering of Vertices and top simplices independently.
We will  impose here a relation between these two numbering and 
develop an {\em \ids} for the EWDS that saves,
the space occupied by $2NV'$ indexes in the \ewds.

In this optimization we basically develop an algorithm that outputs 
two maps $\funct{f_{VV}}{Vertex'}{Vertex'}$ and $\funct{f_{TT}}{Simplex'}{Simplex'}$ that represent a coherent
change in the numbering of, respectively, 
Vertices and top simplices in a given global \Cdecrep. Applying this renumbering to the global \Cdecrep\ we will obtain a new \Cdecrep\ Data Structure in which the \VTS\ relation and the NV' elements in the array $\TVP$ are known.
Therefore this new data structure can be coded
saving $NV'(\log{NT'}+\log{NV'})$ bits w.r.t. the original data structure. 
This saving comes from the $NV'$ simplex indexes needed to code the 
\VTS\ relation and from saving $NV'$ vertex indexes for $NV'$ elements in $\TVP$.

The re-numbering $f_{VV}$ and $f_{TT}$ are built
visiting the \iqm\ complex $\canon{\AComp}$ and using the fact that an 
\iqm\ $h$-complex is $(h-1)$-manifold-connected. Thus starting from an 
arbitrary $h$-simplex,  using adjacency encoded by the TT relation,
we visit and re-number, one after another, all $h$-simplices and all vertices 
in a connected component in  $\canon{\AComp}$.
Whenever, by adjacency, we reach a non-yet-visited $h$-simplex $\theta$ 
we give it a new index \iff\  there is a Vertices $v$ 
of $\theta$ that do not have a new index yet.
In this case both $v$ and $\theta$  receive  two new indexes {$v_{t^\prime}$} 
and $t^\prime$ that remains liked by a known relation.
Note that is possible to visit $\canon{\AComp}$ using $(d-1)$-adjacency,
adding at most one  vertex $v$ for every new top $d$-simplex $\theta$ visited.
This vertex $v$ is still  unnumbered and also $\theta$ is unnumbered. The two gets two indexes  that are linked so that there is no need to store
information  in the \VTS\ and, for that vertex, in the \TVPP\ relation. 
On the other hand there were top simplices that are not linked in this scheme and receive a, somehow, more free index. The relation from these latter top simplices to its vertices needs to be fully stored in the \TVP.  
Obviously the first top simplex from which we start to visit of a certain 
connected component has all its $(h+1)$ Vertices unnumbered and all its vertices need not to be stored in \VTS\ and \TVPP. Going further into detail we will see that one of these 
vertices will receive an index with smallest number and the other $h$ receive an higher index when all other vertices are numbered.

{
We will develop now the renumbering algorithm.
We first assume that the encoding for functions $f_{VV}$ and $f_{TT}$ must 
be stored into two arrays FTT and FVV.
{
We assume that arrays FVV and FTT are initialized with a {\bf null} value.

We assume to have an array of boolean flags called VISITED 
that is initialized to {\bf false}.
The flag VISITED[t] is turned to {\bf true} when the simplex $t$ has
been considered for renumbering.

Finally, we assume that  the decomposition algorithm fills an array of positive integers, we called $\CC$,
by providing in $\CC[h]\ge 0$ the number of 
connected components of dimension $h$ are present in $\canon{\AComp}$.

As a byproduct of the optimization process we will also fill an array $\VBase$
that gives, in $\VBase[h]$, the base index for Vertices in $h$-dimensional
components. Thus $\ord(v,\canon{\AComp})=h$ \iff\ $v\in [\VBase[h],\ldots,\VBase[h+1]-1]$. 

The array $\VBase$ is the analogous of the array $\TBase$ in Data Structure
\ref{data:globtt} 

We will use the shortcuts \TVP[$h$,$t$,$v$] and \TTP[$h$,$t$,$k$] as defined in equations \ref{eq:tvcomp}  and \ref{eq:ttcomp}.  
With these assumption we have that the data structure needed for the 
optimization is the following
{
\begin{data}{data:opt}{Data Structure used by the optimization}
{
\begin{verbatim}
var
 FVV:array[1..NV'] of Vertex';
 FTT:array[1..NT'] of TopSimplex';
 VISITED:array[1..NT'] of BOOLEAN;
 VBase:array[0..D] of Vertex';
 CC:array[0..D] of POSITIVE;
\end{verbatim}
}
\end{data}
}

Here and in the following we assume that
NewVertex() is a function with memory that returns contiguous
increasing vertex indexes starting from seed that is
given by a call to NewVertex(seed).
A similar specification is assumed for function NewSimplex.
We assume that we can use two special index values, denoted by {\bf dontcopy}
and {\bf reserve}
to mark entries in the \TVP\ array. 
We use {\bf reserve} to say that
a  FVV entry FVV[$v$] must not be considered for further remembering.
The new number for $v$ will be assigned later.
We use {\bf dontcopy} to say that
a \TVP\ entry must not be copied in the optimized \TVP\ array.
With this assumptions we can give the following algorithm to build
FVV and FTT.  (note that in  comments appended to statement we sometimes 
insert tags (\dag 1), (\dag 2) etc.  to reference tagged lines later).
The correctness of this algorithm should not obvious at once for the reader. We will see why this algorithm works, after its step-wise design, in the proof of Property \ref{pro:optspace}.
{
\begin{algo}{algo:opt}{Fill FTT and FVV arrays and VBase}
\begin{algorithmic}
\STATE NewVertex(1);
\FOR[loop through dimensions]{$h=0$ {\bf to} $d$}
\STATE $\VBase[h]\leftarrow NewVertex()$
\COMMENT{$v\in [\VBase[h],\ldots,\VBase[h+1]-1]$ if $\ord(v,\canon{\AComp})=h$}
\STATE $TIdx\leftarrow\TBase[h]$
\COMMENT{$TIdx$ is the next simplex index in {$[\TBase[h],\ldots,{(\TBase[h]+\CC[h]-1)}]$}}
\STATE NewSimplex(\TBase[h]+\CC[h]);
\COMMENT{now NewSimplex returns indexes $\ge{\TBase[h]+\CC[h]}$}
\STATE $VIdx\leftarrow\VBase[h]$
\COMMENT{$VIdx$ is the next vertex index in {$[\VBase[h],\ldots,{(\VBase[h]+\CC[h]-1)}]$}}
\STATE NewVertex(\VBase[h]+\CC[h]);
\COMMENT{now NewVertex returns indexes $\ge{\VBase[h]+\CC[h]}$}
\FOR[find a $t$ top $h$-simplex in a not visited component]{$t=\TBase[h]$ {\bf to} $\TBase[h+1]-1$}
\IF[from $t$ start visiting a new component]{VISITED[$t$]={\bf false}}
\STATE $t_{new}\leftarrow TIdx$; $TIdx\leftarrow TIdx+1$;
\COMMENT{$t_{new}\in{[\TBase[h],\ldots,{(\TBase[h]+\CC[h]-1)}]}$}
\STATE $v_{new}\leftarrow VIdx$; $VIdx\leftarrow VIdx+1$;
\COMMENT{$v_{new}\in{[\VBase[h],\ldots,{(\VBase[h]+\CC[h]-1)}]}$}
\STATE FTT[{{t}}]$\leftarrow t_{new}$ 
\COMMENT{rename top $h$-simplex at the start of the new visit}
\STATE FVV[\TVP[$h$,$t$,$(h+1)$]]$\leftarrow v_{new}$
\COMMENT{(\dag 1) rename the  vertex of $t$ at $\TVP[h,t,(h+1)]$}
\STATE $\TVP[h,t,h+1]\leftarrow{\bf dontcopy}$;
\COMMENT{do not copy $\TVP[h,t,h+1]$ in optmized TV}
{
\FOR[(\dag 2) reserve other Vertices at $t$ for later renumbering]{$j=1$ {\bf to} $h$}
\STATE FVV[\TVP[$h$,$t$,$j$]]$\leftarrow$ {\bf reserved};
\COMMENT{reserve vertex in $t$ and at $\TVP[h,t,j]$}
\ENDFOR
}
\STATE {AdjacentRenumber($h$,$t$)};
\COMMENT{renumber the $h$-simplices $(h-1)$-connected to $t$}
\ENDIF
\ENDFOR[all $h$-simplices explored once but some not renumbered yet]
\FOR[loop again through top  $h$-simplices]{$t=\TBase[h]$ {\bf to} $\TBase[h+1]-1$}
\IF[(\dag 2) not renumbered yet]{FTT[$t$]={\bf null}}
\STATE FTT[$t$]$\leftarrow$NewSimplex();
\ENDIF
\ENDFOR[all $h$-simplices renumbered]
\FORALL{{$\TBase[h]\le t<\TBase[h+1]$} {\bf and} {FTT[$t$]$<\TBase[h]+\CC[h]$}}
{
\FOR[(\dag 3) from $t$ we started a visit, so its $h$ vertices were {\bf reserved}]{$j=1$ {\bf to} $h$}
\STATE FVV[\TVP[$h$,$t$,$j$]]$\leftarrow$  NewVertex();
\COMMENT{rename the  vertex in $t$ and at $\TVP[h,t,j]$}
\STATE $\TVP[h,t,j]\leftarrow{\bf dontcopy}$;
\COMMENT{do not copy $\TVP[h,t,j]$ in optmized \TVP}
\ENDFOR
}
\ENDFOR[All $\CC[h]$ $h$-components settled]
\ENDFOR
\end{algorithmic} 
\end{algo}
}
The above algorithm uses 
the recursive procedure {AdjacentRenumber($h$,$t$)} 
that visit a $(h-1)$-connected $h$-components and builds the renaming for 
simplices in it.
This procedure uses the functions NewVertex() and NewSimplex()
to get new indexes for, respectively, Vertices  and simplices.
Thus, due to initialization in the previous algorithm, during  AdjacentRenumber execution, 
NewSimplex() returns indexes $NewSimplex()\ge{\TBase[h]+\CC[h]}$ and
NewVertex() returns indexes $NewVertex()\ge{\VBase[h]+\CC[h]}$.

Thus the new indexes assigned by the procedure AdjacentRenumber are distinct
from those assigned by the previous algorithm.
These latter stand in 
{$[\TBase[h],\ldots,{(\TBase[h]+\CC[h]-1)}]$} for simplices and in
{$[\VBase[h],\ldots,{(\VBase[h]+\CC[h]-1)}]$} for Vertices. 

After that, for other vertices that were initially {\bf reserved}, an index is assigned. These are $\CC[h]\cdot h$. Before this latter phase ({\bf reserved} vertices are assigned in the last loop), the above algorithm assigns in parallel and pairwise a code to a new vertex and to a new top $h$-simplex. In the last two loops other indexes are freely assigned to top simplices and to {\bf reserved} vertices. These vertices must be assigned in this latter phase so that
in the previous phase vertices and top simplices are numbered pairwise.
This is the rationale for  (the rather obscure) {\bf reserved} vertices.

Finally we note that in this algorithm we assume the
availability of the  procedures EXX($x$,$y$) and OPPOSITE. 
The procedure EXX($x$,$y$)  exchange the
contents of the array elements at $x$ and $y$.
For the definition of the procedure  OPPOSITE($\TVP[h],t^\prime,\phi$)
see the remark before Algorithm \ref{alg:tfill}. 
Note that here we  use the shortcut \TVP[$h$] as the array such that
 \TVP[$h$][$t$,$v$]=\TVP[$h$,$t$,$v$] the latter being  defined as in equations \ref{eq:tvcomp}  and \ref{eq:ttcomp}.  
 With these assumptions we can define the following algorithm:
{
\begin{algo}{algo:optrec}{Recursive Procedure AdjacentRenumber($h$,$t$)}
\begin{algorithmic}
\STATE {\bf Procedure} {AdjacentRenumber($h$,$t$)} 
\FOR[consider all $h$-simplices adjacent to $t$]{$i$=1 {\bf to} $h+1$}
\STATE $t^\prime\leftarrow$ \TTP[$h$,$t$,$i$] 
\IF{$t^\prime \neq\bot$ {\bf and} $t^\prime \neq\trix$} 
\IF[$t^\prime$ not visited yet]{VISITED[$t^\prime$]={\bf false}}
\STATE {VISITED[$t^\prime$]={\bf true}};
\COMMENT{mark $t^\prime$ as visited}
\STATE $\phi$= SetOf(\TVP[$h$,$t^\prime$]) $\cap$ SetOf(\TVP[$h$,$t$]);
\COMMENT{$\phi$ is the $(h-1)$-face between $t$ and $t^\prime$} 
\STATE $k_{t^\prime}\leftarrow$ OPPOSITE($\TVP[h],t^\prime,\phi$); 
\STATE $v_{t^\prime}\leftarrow$ $\TVP[h,t^\prime,k_{t^\prime}]$;
\COMMENT{vertex {$v_{t^\prime}$} is the vertex of $t^\prime$ that is not in $t$}
\IF[{$v_{t^\prime}$} not visited yet]{FVV[{$v_{t^\prime}$}]={\bf null}}
\STATE FVV[{$v_{t^\prime}$}]$\leftarrow$ NewVertex();
\COMMENT{generate and assign a new index to {$v_{t^\prime}$}}
\STATE FTT[{{$t^\prime$}}]$\leftarrow$ NewSimplex(); 
\COMMENT{in touch generation of a new index for {${t^\prime}$}}
\STATE {EXX($\TVP[h,t^\prime,k_{t^\prime}]$,$\TVP[h,t^\prime,h+1]$)};
\COMMENT{(\dag 1)Place new vertex at $(h+1)$ position}
\STATE EXX($\TTP[h,t^\prime,k_{t^\prime}]$,$\TTP[h,t^\prime,h+1]$);
\COMMENT{(\dag 2)adjust \TTP according to the EXX above}
\STATE $\TVP[h,t^\prime,h+1]\leftarrow{\bf dontcopy}$;
\COMMENT{forget $\TVP[h,t^\prime,h+1]$ in optmized \TVP}
\ENDIF
\STATE {AdjacentRenumber($h$,$t^\prime$);}
\COMMENT{renumber the $h$-simplices $(h-1)$-connected to $t^\prime$}
\ENDIF
\ENDIF
\ENDFOR
\end{algorithmic} 
\end{algo}
}
}

\begin{figure}[h]
	\begin{center}
		\begin{tabular}{|c||c|}\hline
			\multicolumn{2}{|c|}{\TBase[0..3]} \\ \hline
			0&1    \\ \hline 
			1&3    \\ \hline
			2&5   \\ \hline
			3&7    \\ \hline
		\end{tabular}
		\mbox{ }
	\begin{tabular}{|c||c|}\hline
		\multicolumn{2}{|c|}{\CC[0..3]} \\ \hline
		0&2    \\ \hline 
		1&1    \\ \hline
		2&1   \\ \hline
		3&1     \\ \hline
	\end{tabular}
	\mbox{ }
		\vspace*{0.3cm }	
		\begin{tabular}{|c|c||c|c|c|c|c|}\hline
			Simplex&@&\multicolumn{4}{|c|}{\TVP[1..24]}&{Comment} \\ \hline
			1&1&1&\multicolumn{3}{|c|}{}&{\TBaseAddr[0]=1}    \\ \hline 
			2&2&2&\multicolumn{3}{|c|}{} & {} \\ \hline
			3&3:4&3&4&\multicolumn{2}{|c|}{}&{\TBaseAddr[1]=3}    \\ \hline
			4&5:6&4&5&\multicolumn{2}{|c|}{}& {}   \\ \hline
			5&7:9&6&13&8&&{\TBaseAddr[2]=7}    \\ \hline
			6&10:12&6&7&13&&{}    \\ \hline
			7&13:16&10&9&12&11&{\TBaseAddr[3]=13}    \\ \hline
			8&17:20&9&12&11&14&{}    \\ \hline
			9&21:24&9&11&14&15&{SIZE=24}    \\ \hline
		\end{tabular}
	\end{center}
	\label{fig:optbigtv1}
	\caption{An example of arrays \TBase, \CC\ and \TVP\  for the 3-complex of Figure \ref{fig:datastrexbig}. We will show in Example \ref{ex:datastrexbigapp} the \FVV\ and \FTT\ maps that are computed for this complex. } 	
\end{figure}	
\begin{example}
	\label{ex:datastrexbigapp}
	We continue our running example  from  Example 	\ref{ex:datastrexbig1}. W.r.t. Figure \ref{fig:datastrexbig} the array \TVP\ is filled as shown in Figure \ref{fig:optbigtv1}.
	We can imagine to run Algorithm \ref{algo:opt} on this complex. 
	
	The reader can verify that the resulting renaming \FTT\ is the identity while \FVV\ is given in Figure	\ref{fig:opbigtv}. 
	To give a short comment to the construction of \FVV\ we can say
	that, for $h=2$, when assigning $\FVV[8]$ the entries for  $\FVV[6]$ and $\FVV[13]$ are {\bf reserved}.
	When assigning $\FVV[7]$ the locations in \TVP\ containing 7 and 13 are swapped.
	For $h=3$, 
	when assigning $\FVV[11]$ the entries for  $\FVV[9]$,$\FVV[10]$ and $\FVV[12]$ are {\bf reserved}.
	When assigning $\FVV[14]$ the location in \TVP\ containing 14 must be swapped with its-self via EXX. Same for $\FVV[15]$.
\end{example}
\begin{figure}[ht]
	\begin{center}
				\begin{tabular}{|c||c|}\hline
			\multicolumn{2}{|c|}{\VBase[0..3]} \\ \hline
			0&1    \\ \hline 
			1&3    \\ \hline
			2&6   \\ \hline
			3&10    \\ \hline
		\end{tabular}
		\mbox{ }
		\begin{tabular}{|c|c|c|c|c|c|c|c|c|c|c|c|c|c|c|}\hline
\multicolumn{15}{|c|}{FVV[1..15]} \\ \hline
			1&2&3&4&5&6&7&8&9&10&11&12&13&14&15   \\ 
			\hline
			1&2&5&3&4&8&7&6&14&13&10&15&9&11&12
			\\ \hline
		\end{tabular}
		\caption{Arrays \VBase\ and \FVV\  for the 3-complex of Figure \ref{fig:datastrexbig} (see Example \ref{ex:datastrexbigapp} ). } 
	\end{center}
	\label{fig:opbigtv}
\end{figure}

The renumbering computed by the previous algorithm is applied producing 
two new arrays \TTPP\ and \TVPP\ out of the old arrays \TTP\ and \TVP.
The array \TTPP\ has the same size and the same addressing method of
array \TTP. The array \TVPP\ is smaller than the array \TVP\ and has different
addressing methods.
More precisely we define the Implicit Global \Cdecrep\ Data Structure to be
the concrete data structure below: 
{
\begin{data}{data:globimptt}{Global Implicit \Cdecrep\ Data Structure for  {$\canon{\AComp}$}} 
{
\begin{verbatim}
type
 Vertex' = [1..NV'];
 TopSimplex' = [1..NT'];
var
 TV'':array[][][] of Vertex';
 TT'':array[][][] of TopSimplex';
 TBase,TBaseAddr,TVAddr,VBase:array[0..D] of TopSimplex';
\end{verbatim}
}
\end{data}
}
Arrays \TBase, \TBaseAddr, are the same arrays that are present 
in the non-implicit data structure \ref{data:globtt}.
The array \VBase\ is filled by the Algorithm \ref{algo:opt}.
With these assumptions we can  give the algorithm to build 
the arrays \TTPP\ and \TVPP. The  algorithm for \TTPP\ is easy to design, simply rename vertices while copying \TTP\ into \TTPP.
 The  algorithm for \TVPP\ renames vertices while copying \TVP\ into \TVPP\ but does not copy information for top simplices that has been assigned an index pairwise with a vertex. These will be recovered using a fixed scheme devised in forthcoming Algorithms \ref{algo:opttvp} and \ref{algo:opttsp}.
{
	\begin{algo}{algo:optren}{Renumbering of \TTP and \TVP arrays}
		\begin{algorithmic} 
			\FOR{$h=0$ {\bf to} $d$}
			\FOR{$\TBase[h]\le t<\TBase[h+1]$}
			\FOR{$k=1$ {\bf to} $h+1$}
			\STATE $\TTPP[h,FTT[t],k]=\FTT[\TTP[h,t,k]]$
			\IF{$\TVP[h,t,k]\neq{\bf dontcopy}$}
			\STATE $\TVPP[h,FTT[t],k]=\FVV[\TVP[h,t,k]]$
			\ENDIF
			\ENDFOR[simplex $t$ completed]
			\ENDFOR[dimension $h$ completed]
			\ENDFOR[\TTP relation renumbered into \TTPP]
		\end{algorithmic} 
	\end{algo}
	Applying this renumbering we can save NV' entries in the coding of the 
	$\TVP$ relation.
}
It is easy to see that the renumbering gives a consistent \TTPP\ relation. More difficult is to see that \TVPP\ can support the computation of ${\TVP[h,t,k]}$. To show this we must add some remarks.

The array \TAddr\ holds the base index at which are stored, in \TVPP, the
entries for $h$-dimensional top simplices for the optimized version of \TVP.
The array $\TAddr[h]$ is initialized using a recursive formula  we will introduce later, when are more intuitive the reasons for this formula.  

In the following discussion we assume that we can use the 
array \TVPP\ as a large mono-dimensional array
$\TVPP[1,\ldots,\SIZE-NV']$. 
Thus, in the following, we will see how to compute ${\TVP[h,t,k]}$.

For a certain class of top $h$-simplices $t$ the vertices for $k=h+1$ are implicitly coded. The simplices $t$ with {$\TBase[h]\lt t\le\TBase[h]+\CC[h]$} are the $\CC[h]$ simplices from which we started to visit one of the $h$-dimensional connected component of the decomposition. Then 
${\TVP[h,t,h+1]}$ is $\VBase[h]+(t-\TBase[h])$. This happens for $\CC[h]$
vertices assigned first by the algorithm starting from $\VBase[h]$.

Next for some $t\ge TBase[h]+\CC[h]$ we know that the algorithm, for all the vertices in the $h$ dimensional connected component, assign first a linked index for $t$ and one of its vertices. 
Again for those ${\TVP[h,t,h+1]}=\VBase[h]+(t-\TBase[h])$.
This happens for $t$ between $t\ge TBase[h]+\CC[h]$ and
$t\lt\TBase[h]+\VBase[h+1]-\VBase[h]-\CC[h]h$. 

The number of top simplices $t$ that have an index linked to a vertex is not the number of vertices, i.e. $\VBase[h+1]-\VBase[h]$ but something less i.e. $\VBase[h+1]-\VBase[h]-\CC[h]h$. Not all vertices (that are $\VBase[h+1]-\VBase[h]$) are linked to a top simplex because
 $\CC[h]\cdot h$ vertices are {\bf reserved} and are assigned altogether in the end. So only for $t$ s.t.  $TBase[h]+\CC[h]\le t\lt\TBase[h]+\VBase[h+1]-\VBase[h]-\CC[h]h$ must be
${\TVP[h,t,k]}=\VBase[h+1]-\CC[h]h+(t-\TBase[h])h+k-1$.

We have given formula to find ${\TVP[h,t,k]}$ with no need to store information in \TVPP. In all other cases we find ${\TVP[h,t,k]}$ stored in \TVPP.
This happens for  all top simplices $t$ that do not enter any optimization. From the reasoning above it is reasonable to define 
$$\IIBND[h]=\TBase[h]+\VBase[h+1]-\VBase[h]-\CC[h]h$$ as a limit for $t$.
When this must be computed for $h=d$ take $\VBase[h+1]=NV'+1$.

 For $t\ge \IIBND[h]$ no optimization takes place and for all $1\le k\le h+1$ and ${\TVPP[h,t,k]}$ must be retrieved from \TVPP.  The problem is to find where is it.  These data are stored in the \TVPP\ at the end of the section for optimized top $h$-simplices. This is because in the renumbering of algorithm 
\ref{algo:opt} (\dag 2) these simplices are numbered for last and receive higher numbers. What comes before are optimized top simplices. The $\CC[h]$ top simplices, from where we start the visit, do not store anything and $(h+1)$ vertices receive this best optimization. Then for each one of the other vertices, assigned pairwise with a top simplex, we store $h$ vertices. The latter are  $(\VBase[h+1]-\VBase[h]-\CC[h](h+1))$.   Therefore the optimized section occupies $(\VBase[h+1]-\VBase[h]-\CC[h](h+1))h$ and un-optimized storage of the \TVPP\ starts at 
$$\IITAddr[h]=\TAddr[h]+(\VBase[h+1]-\VBase[h]-\CC[h](h+1))h$$   Each un-optimized top simplex has an index $t\ge\IIBND[h]$  and stores $(h+1)$
entries.  Therefore the $t$ simplex beyond optimized simplices stores at 
$\IITAddr[h]+(t-\IIBND[h])(h+1)$ and must be
$${\TVP[h,t,k]}=\TVPP[\IITAddr[h]+(t-\IIBND[h])(h+1)+k-1]$$

We have in the \TVPP\ $h$ vertices for  $1\le k\le h$  for the top simplices $t$ s.t. ${\TBase[h]+\CC[h]\le t\lt  \IIBND[h]}$. These are stored starting from $\TAddr[h]$ and each simplex stores $h$ vertices so,
vertex $k$ is at $\TAddr[h]+(t-\TBase[h]-\CC[h])h+k-1$ because $\CC[h]$ top simplices do not store a vertex.
Therefore for those we have $\TVP[h,t,k]=\TVPP[\TAddr[h]+(t-\TBase[h]-\CC[h])h+k-1]$

Now is more apparent how to find a recursion for $\TAddr[h]$. This could be:
$\TAddr[0]=1$ and
\begin{equation}
\label{eq:taddr}
\TAddr[h+1]=
\TAddr[h]+(\TBase[h+1]-\TBase[h])(h+1)-(\VBase[h+1]-\VBase[h])
\end{equation}
We omit the code for this initialization.  In the equation \ref{eq:taddr} the expression $(\TBase[h+1]-\TBase[h])(h+1)$ gives the number of cells to store vertices for $h$-simplices without optimization. The expression $(\VBase[h+1]-\VBase[h])$ is the number of entries in \TVPP\ saved by the optimization that, therefore, are subtracted.

Putting all things together we have that $\TVP[h,t,k]$ is given by the following algorithm:
{
\begin{algo}{algo:opttvp}{Computation of the $\TVP[h,t,k]$ relation}
\begin{algorithmic}
\IF{$\TBase[h]\le t\lt\TBase[h]+\CC[h]$}
\IF{$k=h+1$}
\STATE {\bf return} $\VBase[h]+(t-\TBase[h])$;
\COMMENT{(\dag 1)}
\ELSE
\STATE {\bf return} $\VBase[h+1]-\CC[h]h+(t-\TBase[h])h+k-1$;
\COMMENT{(\dag 2)}
\ENDIF
\ELSIF{$\TBase[h]+\CC[h]\le t\lt  \IIBND[h]$}
\IF{$k=h+1$}
\STATE {\bf return} $\VBase[h]+(t-\TBase[h])$;
\COMMENT{(\dag 3)}
\ELSE
\STATE {\bf return} $\TVPP[\TAddr[h]+(t-\TBase[h]-\CC[h])h+k-1]$;
\COMMENT{(\dag 4)}
\ENDIF
\ELSE
\STATE {\bf return} $\TVPP[\IITAddr[h]+(t-\IIBND[h])(h+1)+k-1]$;
\COMMENT{(\dag 5) $t\ge  \IIBND[h]$}
\ENDIF
\end{algorithmic}
\end{algo}
}
\begin{example}
	\label{ex:datastrexbigapp1}
	We continue our running example  from  Example 	\ref{ex:datastrexbigapp}. W.r.t. Figure \ref{fig:datastrexbig} the  Figure \ref{fig:optvp}  shows the result of the computation for $\TVP[h,t,k]$,\VBase, \TBase\ and \IIBND. We also reported the computation for 
	\TAddr\ following recurrence \ref{eq:taddr}. The last column gives 
	\IITAddr. Note that $\IITAddr[h]=\TAddr[h+1]$. Recalling that $\IITAddr[h]$ is the start of the un-optimized area in the \TVPP\ and that $\TAddr[h+1]$ is the start of storage for $h+1$ components in the \TVPP, it is easy to understand  that encoding for this complex optimize all entries of the \TVP\ relation. This is easy to see in Figure \ref{fig:optvp}. Only nine entries need to be stored in \TVPP. They are reported in boldface in Figure \ref{fig:optvp}. Figure \ref{fig:opbigbnd} shows the actual \TVPP.
\end{example}
\begin{figure}
\begin{center}
	\begin{tabular}{|c|c|c||c|c|c|c|c|c|c|c|c|}\hline
		h&t&k&\multicolumn{4}{|c|}{$\TVP[h,t,k]$}&
		\rotatebox[origin=c]{270}{ \VBase\ }
		&\rotatebox[origin=c]{270}{ \TBase\ }
		&\rotatebox[origin=c]{270}{ \IIBND\  }
		&\rotatebox[origin=c]{270}{ \TAddr\  }
		&\rotatebox[origin=c]{270}{ \IITAddr  }
		\\ \hline
		0&1&1&1&\multicolumn{3}{|c|}{}&{1}&1&3&1&1    \\ \hline 
		0&2&1&2&\multicolumn{3}{|c|}{}&{}& &&&\\ \hline
		1&3&1:2&5&3&\multicolumn{2}{|c|}{}&{3}&{3}&5&1&2    \\ \hline
		1&4&1:2&\bf 3&4&\multicolumn{2}{|c|}{}&{}& & & & \\ \hline
		2&5&1:3&8&9&6&&{6}&5  &7 &2 & 4 \\ \hline
		2&6&1:3&\bf 8&\bf 7&9&{}&  & & & &\\ \hline
		3&7&1:4&13&14&15&10&{10} &7 &10 & 4 &10 \\ \hline
		3&8&1:4&\bf 14&\bf 15&\bf 10&11&{}   & & & & \\ \hline
		3&9&1:4&\bf 14&\bf 10&\bf 11&12&{} & & & &\\ \hline
	\end{tabular}
\end{center}
\caption{The $\TVP[h,t,k]$ for the 3-complex of Figure \ref{fig:datastrexbig} after applying the renaming \FVV\ in Example \ref{ex:datastrexbigapp} and Figure \ref{fig:opbigtv}	in boldface indexes that are actually stored in \TVPP. } 
\label{fig:optvp}
\end{figure}

\begin{figure}
	\begin{center}
		\begin{tabular}{|c|c|c|c|c|c|c|c|c|}\hline
\multicolumn{9}{|c|}{\TVPP[1..9]} \\ \hline\hline
			1&2&3&4&5&6&7&8&9  \\ 
			\hline
			3&8&7&14&15&10&14&10&11
			\\ \hline
		\end{tabular}
		\caption{The array \TVPP\ for the 3-complex of Figure \ref{fig:datastrexbig}. Array \TVPP\ is represented horizontally simply to save space. } 
	\end{center}
	\label{fig:opbigbnd}
\end{figure}

{Similarly the relation {$\sigma_{VT^\star}(h,v^\prime)$} 
need not to be stored and can  be computed as follows.
All vertices except the {\bf reserved} ones are assigned an index pairwise w.r.t. a top simplex. Thus when $\VBase[h]$ is assigned to a vertex $\TBase[h]$ is assigned to a top simplex $t$. In general vertex $v^\prime$ is paired to simplex  $t^\prime=\TBase[h]+(v^\prime-\VBase[h])$.

Last $\CC[h]h$ vertices are {\bf reserved} and then are assigned a code
$v^\prime$ that is given for $1\le k\le h$ by the equation
 $v^\prime=\VBase[h+1]-\CC[h]h+(t^\prime-\TBase[h])h+k-1$
(See Algorithm \ref{algo:opttvp} line \dag 1).
Therefore, with some algebra we can write
$v^\prime-\VBase[h+1]+\CC[h]h-k+1=(t^\prime-\TBase[h])h$
and therefore $t^\prime=\frac{v^\prime-\VBase[h+1]+\CC[h]h-k+1}{h}+\TBase[h]$
The formula gives the correct index for the top simplex of $v^\prime$ for $k=1$ for the vertex with the smallest index  $v^\prime$. Then for the other $h-1$ indexes the contribution of $1-k$ is a decrement
by a fractional difference from $\frac{0}{h}$ to $\frac{1-h}{h}$.
If we delete $1-k$ from the formula $v^\prime$ gives an unbalanced increment that never exceeds  $\frac{1-h}{h}$.
Taking the floor of this expression i.e. $\floor{\frac{v^\prime-\VBase[h+1]+\CC[h]h}{h}}+\TBase[h]$ we get the 
correct value for $t^\prime$. Putting together all these ideas we can define the following algorithm for the $VT^\star$  relation, for which nothing needs to be stored.
\begin{algo}{algo:opttsp}{Computation of the {$\sigma_{VT^\star}(h,v^\prime)$} relation}
\begin{algorithmic}
\IF{$\VBase[h]\le v^\prime\lt \VBase[h+1]-\CC[h]h$} 
\STATE {\bf return} $\TBase[h]+(v^\prime-\VBase[h])$;
\COMMENT{(\dag 2)}
\ELSE
\STATE {\bf return} $\TBase[h]+\floor{\frac{v^\prime-\VBase[h+1]+\CC[h]h}{h}}$.
\COMMENT{(\dag 1)}
\ENDIF
\end{algorithmic}
\end{algo}
}

\begin{example}
	\label{ex:datastrexbigapp2}
	We end our running example and continue the analysis of   example 	\ref{ex:datastrexbigapp1}.  We have summarized in Figure \ref{fig:vtopt} what is needed for the computation of  $\sigma_{VT^\star}(h,v^\prime)$. The result of the computation of this relation is displayed in the last column of Figure \ref{fig:vtopt} together with the reference to a line of code in Algorithm \ref{algo:opttsp}. The interested reader can verify that the reported line is the one actually executed to find the reported value.
\end{example}
\begin{figure}
	\begin{center}
		\begin{tabular}{|c|c||c|c|c|c|}\hline
			h&$v^\prime$&
			\rotatebox[origin=c]{270}{ \VBase\ }
			&\rotatebox[origin=c]{270}{ \TBase\ }
			&\rotatebox[origin=c]{270}{ \CC\  }
			&{$\sigma_{VT^\star}(h,v^\prime)$  }
			\\ \hline
			0&1&1&1&2&1 (\dag 2) \\ \hline 
			0&2&&&&2 (\dag 2)\\ \hline
			1&3&3&3&1&3 (\dag 2)\\ \hline
			1&4&&&&4 (\dag 2)\\ \hline
			1&5&&&&3 (\dag 1)\\ \hline
			2&6&6&5&1&5 (\dag 2)\\ \hline
			2&7&&&&6 (\dag 2)\\ \hline
			2&8&&&&5 (\dag 1)\\ \hline
			2&9&&&&5 (\dag 1)\\ \hline
			3&10&10&7&1&7 (\dag 2)\\ \hline
			3&11&&&&8 (\dag 2)\\ \hline
			3&12&&&&9 (\dag 2)\\ \hline
			3&13&&&&7 (\dag 1)\\ \hline
			3&14&&&&7 (\dag 1)\\ \hline
			3&15&&&&7 (\dag 1)\\ \hline
		\end{tabular}
	\end{center}
	\caption{The $\sigma_{VT^\star}(h,v^\prime)$ for the 3-complex of Example \ref{ex:datastrexbigapp2}  } 
	\label{fig:vtopt}
\end{figure}

The above set of algorithms gives a procedure to strip away 2NV' indexes from
the global \Cdecrep\ Data Structure. The correctness of the above procedure 
is ensured by the following property. The proof of this property simply
recalls all the ideas we reported before each algorithm. 
\begin{property}
\label{pro:optspace}
Let {$\canon{\AComp}$} the decomposition of complex represented by a
{Global \Cdecrep\ Data Structure} (i.e. the data structure \ref{data:globtt})
and let FTT, FVV and \VBase\  the arrays of the Data Structure 
\ref{data:opt} computed by the Algorithm \ref{algo:opt}.
Let \TTPP and \TVPP and \TAddr\ the arrays in the Implicit Data Structure 
\ref{data:globimptt} filled by algorithms \ref{algo:optren}. 
In this situation:
\begin{enumerate}
\item\label{pro:corr}
the array \TTPP\ from data structure \ref{data:globimptt} and the algorithms 
\ref{algo:opttsp} and \ref{algo:opttvp} gives respectively the relations 
\TTP, \TVP\ and {$\sigma_{VT^\star}(h,v^\prime)$} of a 
coherent {Global \Cdecrep\ Data Structure} for the complex {$\canon{\AComp}$};
\item\label{pro:size}
the Implicit Data Structure \ref{data:globimptt}
saves  $NV'(\log{NV'}+\log{NT'})$ bits
over the {Global \Cdecrep\ Data Structure}
\item \label{pro:time}
the Algorithm \ref{algo:opt} optimizes the {Global \Cdecrep\ Data Structure} in  $\lesscomp{hNT'}$
\end{enumerate}
\end{property}
\begin{proof}[Proof of Part \ref{pro:corr}]
We prove Part \ref{pro:corr}
by showing first that the renaming computed 
by Algorithm \ref{algo:opt}, when applied by Algorithm  \ref{algo:optren},
assigns new indexes such that some values of the relation   
$\TVP$ and the whole relation {$\sigma_{VT^\star}$} can
be computed by Algorithms \ref{algo:opttvp} and \ref{algo:opttsp}
and need not to be stored.
Next we show that the array $\TVPP$ stores non computed values
for relation $\TVP$  and 
Algorithm \ref{algo:opttvp} fetches them correctly.

The Algorithm \ref{algo:opt} visit the {Global \Cdecrep\ Data Structure}
by adjacency and assigns new indexes to top simplices and Vertices.
The new simplex index for $t$ is stored in $\FTT[t]$.
The new vertex index for $v$ is stored in $\FVV[v]$.
We start the visit at some top simplex and visit a whole $(h-1)$-connected
component by a call to {recursive procedure AdjacentRenumber($h$,$t$)}
(see Algorithm \ref{algo:optrec}).
When all Vertices in all $h$-components have received a new index 
we have also given a new index to a subset of the top $h$-simplices 
in $\AComp$.
We note that not all  $h$-simplices might have been renumbered.
in this phase. Top $h$-simplices missing a new index  
must be renumbered using indexes beyond those used so far.
This is correctly handled by the second loop in Algorithm  \ref{algo:opt}.

It is easy to see that  traversing by adjacency each
$h$-component, for $0\le h\le d$, eventually, all $NV'$ 
Vertices in $\canon{\AComp}$ receive a new index. 
New vertex indexes are assigned using  scheme that allow not to
store $NV'$ entries in $\TVPP$ w.r.t. $\TVP$.
To explain how indexes are assigned we recall that
the element $\TBase[h]$ contains the base index for the $h$-simplices
(i.e. the first $h$-simplex is at $\TBase[h]$, see remark after  property \ref{pro:ewdsocc}). 
In this renumbering  we will use the $\CC[h]$ array. Therefore we recall
that the element $\CC[h]$ contains the number of $h$-dimensional
components in the decomposition $\canon{\AComp}$.
We start to explain how vertex indexes are assigned by considering
all formulas 
in Algorithms \ref{algo:opttvp} and \ref{algo:opttsp}.
that do not reference the array $\TVPP$.

\paragraph{Defaults for Vertices in Algorithms \ref{algo:opttvp} and \ref{algo:opttsp}}
Simplex indexes from $\TBase[h]$ to $\TBase[h]+\CC[h]-1$
are reserved for the $\CC[h]$ top $h$-simplices from which the
Algorihm \ref{algo:opt} started the visit of
one of the $\CC[h]$ connected components. 
These simplices are indexed by $t$
with {$\TBase[h]\le t \lt \TBase[h]+\CC[h]$}.
The vertex in $\TVP[h,t,h+1]$ will receive, in Algorithm \ref{algo:opt} line
(\dag 1),
 index {$\VBase[h]+t-\TBase[h]$} and Vertices at
$\TVP[h,t,k]$ for $1\le k\le h$ will be reserved and will receive the  last $\CC[h]h$ 
indexes  in the vertex range i.e.
\begin{equation}
\label{eq:lastind}
\TVP[h,t,k]=
\VBase[h+1]-\CC[h]h+(t-\TBase[h])h+k-1 
\end{equation}
for $1\le k\le h$ and {$\TBase[h]\le t \lt \TBase[h]+\CC[h]$}.
These Vertices are first {\em reserved} for later renumbering in
loop (\dag 2) in Algorithm \ref{algo:opt}.
The actual renumbering takes place  in loop (\dag 3) in Algorithm \ref{algo:opt}. 
This explains the formulas used in Algorithms \ref{algo:opttvp} (\dag 2)
for the relation $\TVP$ for {$\TBase[h]\le t \lt \TBase[h]+\CC[h]$}. 
The inversion of the formula \ref{eq:lastind} w.r.t. $t$ with
$v^\prime=\TVP[h,t,k]$  gives the
formula used in the first else statement of Algorithm \ref{algo:opttsp} at (\dag 2).

There is also a known bijection between the index for a vertex $v$ and 
the index for an $h$-simplex $t$ incident to $v$. 
This is given by {$v=\VBase[h]+t-\TBase[h]$}.
This bijection is established by the procedure AdjacentRenumber in \ref{algo:optrec}.
This bijection holds for {${\TBase[h]+\CC[h]\le t\lt  \IIBND[h]}$} where
$\IIBND[h]$ is the index of the last top $h$-simplex associated with a 
vertex. Note that there might be more simplices than Vertices thus 
$\IIBND[h]=\TBase[h]+\VBase[h+1]-\VBase[h]-\CC[h]h-1$.
For $t$ in this range it is useless to store the index for $v$ in  
$\TVP[h,t,j]$ for some $j$. 
We can assume that the known vertex  $v$ in bijection with $t$ is  present at
the $(h+1)$-position in the array. If this is not the case this situation is
enforced by the EXX exchanges  (\dag 1 and \dag2) in Algorithm \ref{algo:optrec}.
This bijection, enforced between $t$ and  $\TVP[h,t,h+1]$, explains the 
formulas in (\dag 3)  Algorithm \ref{algo:opttvp} 
for  {${\TBase[h]+\CC[h]\le t\lt  \IIBND[h]}$}.
The inversion of the bijection formula {$v=\VBase[h]+t-\TBase[h]$}
w.r.t. $v$  gives
formulas in (\dag 3)  Algorithm \ref{algo:opttsp}. 

In fact the encoding of the \VTS\ relation 
becomes useless, being the top simplex indexed by 
$t=\TBase[h]+(v-\VBase[h])$ incident to $v$.
This formula (\dag 2) in Algorithm \ref{algo:opttsp} is obtainded by 
inversion of the formula $v^\prime=\TVP[h,t,h+1]=\VBase[h]+(t-\TBase[h])$ 
w.r.t. The original formula as valid  for
${{\TBase[h]+\CC[h]}\le t\lt  \IIBND[h]}$.  
Therefore formula (\dag 2) in Algorithm \ref{algo:opttsp} 
is valid for
{$\VBase[h]+CC[h]\le v^\prime \lt \VBase[h+1]-\CC[h]h$}.
To see this simply take $t=\TBase[h]+(v-\VBase[h])$ in ${{\TBase[h]+\CC[h]}\le t\lt  \IIBND[h]}$.  

{
For ${\TBase[h]\le t\lt{\TBase[h]+\CC[h]-1}]}$ we have found that vertex 
$v=\VBase[h]+(t-\TBase[h])$ is incident to simplex $t$.
Therefore for $\VBase[h]\le v \lt \VBase[h+1]-\CC[h]h$ we have that 
$v$ is incident to $\TBase[h]+(v-\VBase[h])$. This explain the computation of formula (\dag 2) in Algorithm \ref{algo:opttsp}. 

Finally, for formula (\dag 1) in Algorithm \ref{algo:opttsp}
we can take as part of this proof the remarks in the last paragraph before  Algorithm \ref{algo:opttsp}. 
} 
This completes the
checking of all formulas for defaults  in 
Algorithms  \ref{algo:opttvp} and \ref{algo:opttsp}

{
\paragraph{Allocation of the \TVPP\ array}
Next we explain how non default values are 
stored in array \TVPP.
We first assume   that this fragment of code is executed to fill
an auxiliary array we called the \TVXP\ array:
{
\begin{algorithmic} 
\FOR{$h=0$ {\bf to} $d$}
\FOR{$\TBase[h]\le t<\TBase[h+1]$}
\FOR{$k=1$ {\bf to} $h+1$}
\STATE $\TVXP[h,FTT[t],k]=\FVV[\TVP[h,t,k]]$
\ENDFOR[simplex $t$ completed]
\ENDFOR[dimension $h$ completed]
\ENDFOR
\end{algorithmic} 
}

This fragment of code generates the table \TVXP\ that is the result of a
plain renaming of table \TVP\ with substitutions in FTT and FVV. 
Therefore using \TVXP\ instead of \TVP\ and \TTPP\ instead of \TTP\ we still
have a coherent \Cdecrep\ Data Structure.
In the rest of this proof we will show that 
that several areas in \TVXP\ bears no information at all
and can be deleted. From this deletion we obtain the array \TVPP\ that
is the table used in the implicit representation. 
We have just  shown in the previous paragraph of this proof that Algorithms \ref{algo:opttvp} and 
\ref{algo:opttsp} complies with this deletion and,  for some ranges of
indexes, they compute correctly default values for the TV and \VTS\ relations.
Finally note that
the array \TVXP\ is relevant just for the purpose of this proof.
.

Upon renaming of the array \TVP\ we have some peculiar situation in
the array \TVXP.
The slice of the  \TVXP\ array used by dimension $h$ is subdivided in
three areas.
A first area is made up of  the $\CC[h]$ entries indexed by $t$ such that 
$${\TBase[h]\le t\lt\TBase[h]+\CC[h]}.$$
These entries are no longer useful.
In fact, for each $t\in[\TBase[h],\ldots, \TBase[h]+\CC[h]-1]$  
the entry {$\TVXP[h,t,h+1]$} is assigned, 
by Algorithm \ref{algo:opt} (\dag 1),
to the first $\CC[h]$ values starting for $\VBase[h]$, i.e.  
$$t\in[\TBase[h],\ldots, \TBase[h]+\CC[h]-1]\Rightarrow 
{\TVXP[h,t,h+1]}=\VBase[h]+(t-\TBase[h]).$$
This situation is exploited also in the computation of \ref{algo:opttvp} (\dag 1).
The last ${\TVXP[h,t,k]}$ entries for $1\le k \le h$ are assigned,
by Algorithm \ref{algo:opt} (\dag 3),
to the last $h\cdot \CC[h]$ values in the range $[\VBase[h],\ldots,\VBase[h+1]-1]$.
Thus, in this case   $${\TVXP[h,t,k]}=\VBase[h+1]-\CC[h]h+(t-\TBase[h])h+k-1.$$
This situation is exploited also in the computation of \ref{algo:opttvp} (\dag 2).

A second area is starts at
${\TBase[h]+\CC[h]}$ and extends to all simplices that
receive an index in relation with a vertex index.
Thus, this area must extend to index 
$$\IIBND[h]=\TBase[h]+\VBase[h+1]-\VBase[h]-\CC[h]h$$
In this area, for $t\in[\TBase[h]+\CC[h],\ldots,\IIBND[h]-1$  
each ${\TVXP[h,t,h+1]}$ entry is assigned,
by Algorithm \ref{algo:optrec} (\dag 1) and (\dag 2),
to default values i.e. ${\TVXP[h,t,h+1]}=\VBase[h]+(t-\TBase[h])$.
This situation is exploited also in the computation of \ref{algo:opttvp} (\dag 3).

These two  areas described above are not transferred the array \TVPP\
by the renaming generated by  Algorithm \ref{algo:opt}.
This algorithm takes care to mark the \TVP\ entries that must not
be transferred to \TVXP\ writing into the appropriate entires of \TVP\
the special symbol {\bf dontcopy}.
The formulas in Algorithms
\ref{algo:opttvp} (\dag 1),(\dag 2) and (\dag 3) takes care to generate known
values for these omitted entries.

On the other hand, the entries {$\TVXP[h,t,k]$} outside these two areas 
must be stored and no default value is available from them.
We have that these values are
stored in an array \TVPP\ starting from $\TAddr[h]$.
For each top simplex $t\in[\TBase[h]+\CC[h],\ldots,\IIBND[h]-1]$ we save 
one element for each entry in the TV table. Thus we just have $h$ elements 
in \TVPP\ for each top $h$-simplex $t$ in this range. 
Thus, in this case the value ${\TVP[h,t,k]}$
is stored  at 
\begin{equation}
\label{eq:atopt}
\TVPP[\TAddr[h]+(t-\TBase[h]-\CC[h])h+k-1]...
\end{equation}

This formula explains line  (\dag 4) in Algorithm \ref{algo:opttvp}.

Finally for $t$  from $\IIBND[h]$ to $\TBase[h+1]-1$ no optimization
is possible and we have entries with $(h+1)$ elements. 
These entries are stored in the \TVPP\ array beyond
index
\begin{eqnarray*}
\lefteqn{\TAddr[h]+(\IIBND[h]-\TBase[h]-\CC[h])h+h=}\\ 
&&\TAddr[h]+(\VBase[h+1]-\VBase[h]-\CC[h](h+1))-1. 
\end{eqnarray*}
To see this simply take formula \ref{eq:atopt} and put $t=\IIBND[h]-1$ and
$k=h$. In this way we find where the last element of the previous area is stored. We add one and we get where the next area must start. We denote this location  with 
$$\IITAddr[h]=\TAddr[h]+(\VBase[h+1]-\VBase[h]-\CC[h](h+1))h$$ 
Thus in this case ${\TVP[h,t,k]}$
is at $$\TVPP[\IITAddr[h]+(t-\IIBND[h])(h+1)+k-1]$$
This formula explains line  (\dag 5) in Algorithm \ref{algo:opttvp}.

We note that the expression for  $\TAddr[h+1]$ comes from the expression between square brackets in:
$$\TVPP[\IITAddr[h]+(t-\IIBND[h])(h+1)+k-1]$$ taking  $k=h+1$ and $t=\TBase[h+1]-1$ and adding one.
Considering
$$\IITAddr[h]=\TAddr[h]+(\VBase[h+1]-\VBase[h]-\CC[h](h+1))h$$
$$\IIBND[h]=\TBase[h]+\VBase[h+1]-\VBase[h]-\CC[h]h$$
This gives the recurrence \ref{eq:taddr}
$$\TAddr[h+1]=
\TAddr[h]+(\TBase[h+1]-\TBase[h])(h+1)-(\VBase[h+1]-\VBase[h])$$
}
This completes the proof of the mutual correctness of all the algorithms 
involved in this optimization.
\end{proof}
{
\begin{proof}[Proof of Part \ref{pro:size}]
We note  that we do not save space encoding the \TTP\ table.
In fact,  even with this optimization, the encoding of an \iqm\ with $NT'$ top
simplices takes the same space 
for the \TTP\ table i.e.  $NT'(h+1)\log{NT'}$ bits.
On the other hand the \TVP\ table scale down, with the optimization,
from $NT'(h+1)\log{NV'}$.  to $NV'h\log{NV'}+(NT'-NV')(h+1)\log{NV'}$ 
saving at least $NV'\log{NV'}$ bits.
Finally the VT* array is no longer necessary and this
saves $NV'\log{NT'}$ bits. 
Summing the terms for the tow savings  we obtain a reduction of, 
$NV'(\log{NT'}+\log{NV'})$ bits. 
This proves part \ref{pro:size}.
\end{proof}
}

\begin{proof}[Proof of Part \ref{pro:time}]
To prove part \ref{pro:time} we note that 
the overall computations of
{AdjacentRenumber($h$,$t$)} and of Algorithm \ref{algo:opt} 
visits each top simplex once. 
During each top $h$-simplex  visit up to  $(h+1)$ vertexes are
checked to see if they received a new index and
up to $(h+1)$ adjacent top simplices are checked to see
if they have already been visited or not. Thus, the
overall computation of all the calls to 
{AdjacentRenumber($h$,$t$)} can be done in $\lesscomp{hNT'}$.
It is easy to see that the initialization of FVV and FTT 
can be done in constant time and thus all the computations for 
FTT and FVV can be done $\lesscomp{hNT'}$. 
All the operations necessary to apply the renumbering FTT can be 
performed in $\ordcomp{hNT'}$.
Similarly, all the operations necessary to apply the renumbering FVV can be 
performed in $\ordcomp{hNT'}$ since $NV'$ is $\lesscomp{NT'}$.
\end{proof}

 \chapter{A Prolog application}
\label{sec:prolog}
\section{Introduction}
In this section we present the approach we have used to verify that the complex of Example \ref{app:example} could be 
 generated by a pseudomanifold sets of \inst s.
 This approach is based on a small software package we have implemented in Prolog.
The package runs under SWI-Prolog \cite{Wie03b} and  is available from GitHub \cite{git}. 

Several forms of  automated reasoning has been used in Combinatorial Topology and Combinatorics to prove remarkable properties. The most notable example being the recent version of the proof of the four-color theorem.
Such kind of proofs usually involves more powerful theorem proving systems. Nevertheless Prolog seemed powerful enough for the task described here. 

Our package simply builds and maintain a TV relation (see Sections \ref{sec:wingedpao} and \ref{sec:tvtt})
for a simplicial complex $\AComp$. Simplexes can be added by calls to {\tt addSimplex}. Basic topological checks for this complex are implemented in our package. The package contains definitions for the predicates {\tt incident, adjacent, orderOf, dm1connectedComponents, notPseudoManifolds, boundary}.
A short presentation of these predicates is the goal of Section \ref{sec:complex}. The package is of some interest when considering uniformly dimensional complexes i.e. a complex where all top simplices have the same dimension.

Next, in Section \ref{sec:complex}  we introduce facilities dealing with \gl\ \inst s (see Section \ref{sec:gluinst}).  The Prolog package (via predicate {\tt buildTotallyExploded}) takes care to build a separate copy of the totally exploded version $\topAComp$ (see \ref{def:top} ) of the complex $\AComp$. This second complex is stored via a VT relation (see \ref{sec:tvtt}). 

So,  Section \ref{sec:complex} continues an example that uses the primitives {\tt doVertexEquation},\\ {\tt doGluingInstruction} and  {\tt doPseudoManifoldGluingInstruction}}. The example transforms the totally exploded version $\topAComp$ into a
complex {$\topAComp/{\cal E}$} obtained applying a set of \gl\ \inst s ${\cal E}$.
The package takes care to maintain the VT relation for this. Thus we keep in memory the TV relation for the complex $\AComp$ and the VT relation for the modified version of $\topAComp$, going into {$\topAComp/{\cal E}$}, as \gl\ \inst s in ${\cal E}$ are applied. 

Facilities are provided to turn the VT relation into a TV relation (predicate {\tt  vtToTv}).
The user can delete the current version of  $\AComp$ and store in the TV the result of  {\tt  vtToTv} in  order to analyze the topological properties of the complex {$\topAComp/{\cal E}$}.

In Section \ref{sec:expro} we will present the outcomes of running some Prolog programs to prove properties of the complex in Example  \ref{app:example}. The proof offered by the Prolog program confirms the intuitive ideas introduced in Example \ref{app:example}. 
Finally we report in Sections  \ref{sec:iqm} and \ref{sec:utilities}
the API description of this package generated by SWI-Prolog.
 
We note that Prolog is used here as an {\em 
assistant} to proof that there exist a non-pseudomanifold tetrahedralization that can be built stitching together tetrahedra at manifold triangles.
What is described here is, by no means, an argument for the assertion that automated reasoning can find 
 the complex of the  Example \ref{app:example} on its own. 
\section{Primitives for Initial Quasi Manifolds}
\label{sec:complex}
With reference to the API in section \ref{sec:iqm} we can give a small example of what can be done with this Prolog package. We must first clear the TV relation, this is done with a call to {\tt resetComplex} next several calls to {\tt  addSimplex} are needed to build the complex.  For instance consider the complex of Figure \ref{fig:topdef1} (a) that is 
a modified version of Figure \ref{fig:topdef}.
Indeed  we stripped away the 1-dimensional edge.
We recall that this package, although easily extensible, is conceived for uniformly dimensional simplicial complexes.
The complex is built by the following program fragment.
\begin{lstlisting}[language=Prolog,firstline=1, lastline=7]
:-consult("iqm.pl").
test1:-
	resetComplex,
	addSimplex(1,[j,k,q]),
	addSimplex(2,[j,k,l]),
	addSimplex(3,[j,k,m]),
	addSimplex(4,[j,l,n]),
\end{lstlisting}
Next we run a set of test calls, one  for each primitive that checks
different topological properties of this complex. The proposed test code is:
\begin{lstlisting}[language=Prolog,firstline=8,lastline=15]
	incident(1,[j,k]),
	not(incident(1,[l,n])),
	adjacent(1,2,X),write("1 and 2 adjacent at:"),write(X),nl,
	orderOf([j,k],N),write("N="),write(N),nl,
	dm1connectedComponents(C),
	 write("1-connected component:"),write(C),nl,
      	listNonPseudoManifolds, 
	boundary(B),write("boundary"),write(B),nl,
\end{lstlisting}
The output is:
\lstset{language=Prolog}
\begin{lstlisting}
?- test1.
1 and 2 adjacent at:[j,k]
N=3
1-connected component:[[1,2,3,4]]

Non PseudoManifold Top d-Simplexes:
[j,k,l]
[j,k,m]
[j,k,q]
Non Manifold d-1-Simplexes:
[j,k] of order 3
boundary[[k,q],[j,q],[k,m],[j,m],[k,l],[l,n],[j,n]]
true.
\end{lstlisting}
Next we start to use the stitchin/decomposition facilities, therefore,
an appropriate code segment could be:
\begin{lstlisting}[language=Prolog,firstline=16]
	resetDecomposition,buildTotallyExploded,
	doGluingInstruction(1,2),
	dumpDecomp,
	doVertexEquation(2,3,k),
	doPseudoManifoldGluingInstruction(2,4),
	dumpDecomp,
	splitVertex(V),write("splitted "),write(V),nl,
	doVertexEquation(2,3,j),
	not(splitVertex(_)).

\end{lstlisting}
the resulting output is following:
\begin{lstlisting}
simplex 1=[ q-[1] j-[1,2] k-[1,2] ]
simplex 2=[ l-[2] j-[1,2] k-[1,2] ]
simplex 3=[ j-[3] k-[3] m-[3] ]
simplex 4=[ j-[4] l-[4] n-[4] ]

simplex 1=[ q-[1] k-[1,2,3] j-[1,2,4] ]
simplex 2=[ k-[1,2,3] j-[1,2,4] l-[2,4] ]
simplex 3=[ j-[3] m-[3] k-[1,2,3] ]
simplex 4=[ n-[4] j-[1,2,4] l-[2,4] ]
splitted j
true 
\end{lstlisting} 
The first output is from {\tt  dumpDecomp} corresponds  to  the situation in Figure \ref{fig:topdef1} (b). The second output from {\tt  dumpDecomp} corresponds  to  the situation in Figure \ref{fig:topdef1} (c).

\begin{figure}
		\begin{minipage}{0.32\textwidth}
			\mbox{\psfig{file=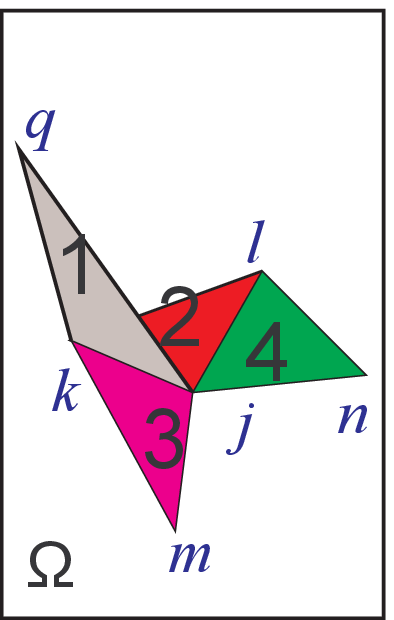,width=\textwidth}}
			\begin{center}(a)\end{center}
		\end{minipage}
		\begin{minipage}{0.33\textwidth}
			\mbox{\psfig{file=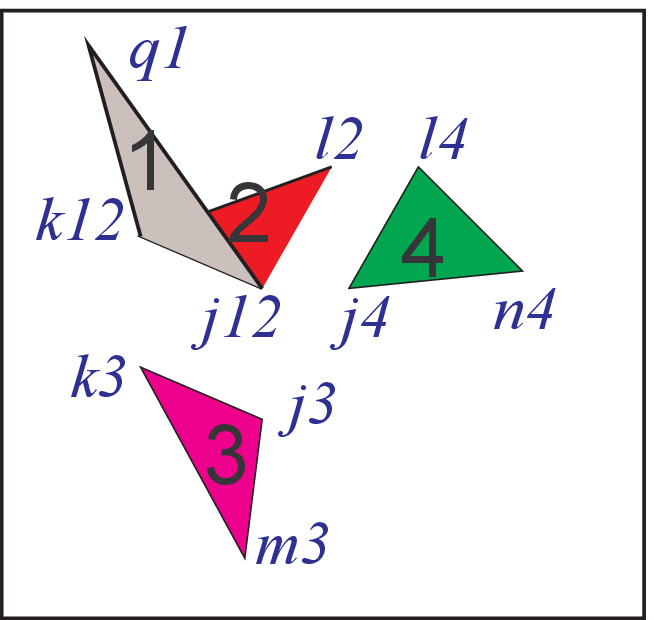,width=\textwidth}}
			\begin{center}(b)\end{center}
		\end{minipage}
			\begin{minipage}{0.33\textwidth}
		\mbox{\psfig{file=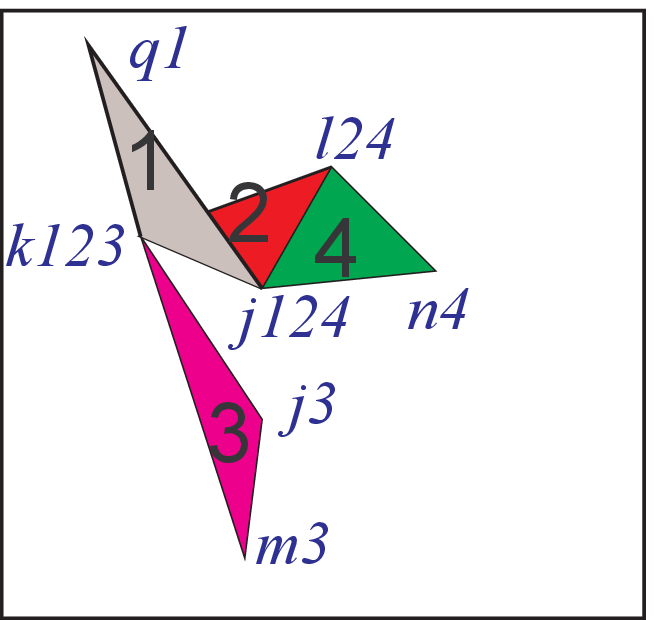,width=\textwidth}}
		\begin{center}(c)\end{center}
	\end{minipage}
		\caption{Example application of Prolog primitives}
		\label{fig:topdef1}
	\end{figure}
In the next section we will present some code for the verification of the non-pseudomanifoldness of the complex of Example \ref{app:example}.
\section{An initial quasi manifold, non pseudomanifold, 3-complex}
\label{sec:expro}
In this section we present the Prolog program we have used to verify that the complex $\AComp$  of Example \ref{app:example}  could be 
generated by a pseudomanifold sets of \inst s (see \ref{def:pminst}).
First we present the instruction to build the complex $\AComp$ and verify that it is not a pseudomanifold. Next we will generate $\topAComp$, the totally decomposed version of $\AComp$, and show  a set of  pseudomanifold gluing \inst s that turns $\topAComp$ into 
an isomorphic copy of $\AComp$.

It is useful to have a look to the complex of Example \ref{app:example}, so we reported it in Figure \ref{fig:tienonreg-all}.
\begin{figure}[ht]
	\centerline{\mbox{\psfig{file=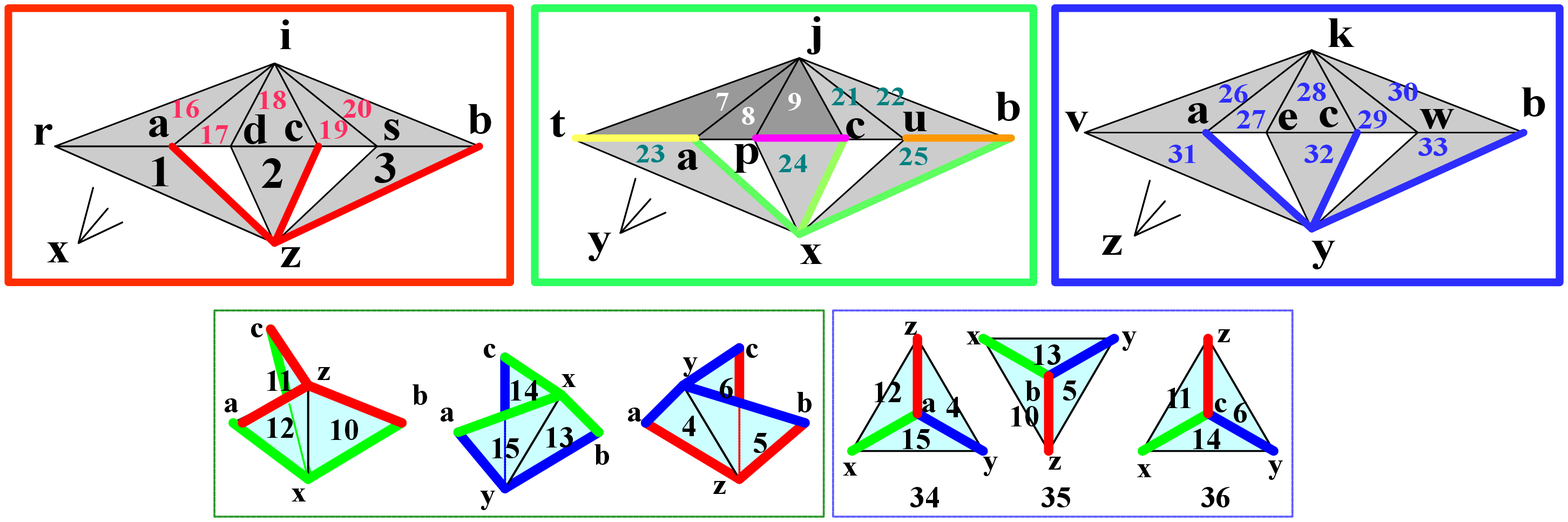,width=0.95\textwidth}}}
	\caption{A non-pseudomanifold $3$-complex
		generated by a non-closed pseudomanifold set of $2$-\inst s}
	\label{fig:tienonreg-all}
\end{figure}
The complex $\AComp$ is the same of Figure \ref{fig:tienonreg}.
We kept the same numbering for top simplices (i.e. tetrahedra). Note that we used indexes 4, 5, 6, 10, 11, 12, 13, 14 and 15 to number 2-faces in the context of discussion of Example \ref{app:example}. For this reason, here, the 27 tetrahedra of this complex receive non contiguous  indexes in the range from 1 to 36. This is not a problem due to the Prolog flexibility. Similarly vertexes are coded by non contiguous single character atoms keeping
the assignments already made in the discussion of Example \ref{app:example}. 

Looking at this figure and consulting Example  \ref{app:example} one can see that we are asked to build the \ems{cone}  from $x$ to the triangles in the red frame (see \ref{par:linkdef} for te definiton of cone). This task is done by this fragment of Prolog code:
\begin{lstlisting}[language=Prolog,firstline=18,lastline=26]
buildRedFrame:-
  addSimplex(1,[x,r,z,a]),
  addSimplex(2,[x,d,z,c]),
  addSimplex(3,[x,s,z,b]),
  addSimplex(16,[x,r,a,i]),
  addSimplex(17,[x,i,a,d]),
  addSimplex(18,[x,i,c,d]),
  addSimplex(19,[x,i,c,s]),
  addSimplex(20,[x,i,s,b]).
\end{lstlisting}
Similar fragments of codes takes care to build the partial assembly of the complex $\AComp$  depicted as the  cones in the green and blue frames of Figure \ref{fig:tienonreg-all}. 
The needed calls to  {\tt addSimplex} are easy to find using the numbering  of Figure \ref{fig:tienonreg-all} and they are not detailed here. The interested reader can find them in the source code available for download from GitHub \cite{git}. So to build all the complex, first we have to complete the operations for all the cones, from x, y and z. The scheme for these cones is  detailed in the three colored frames on top of Figure \ref{fig:tienonreg-all}. Then we have to add
the three tetrahedra 34,35 and 36 using the numbering  on the right bottom of Figure
\ref{fig:tienonreg-all}.
The fragment of Prolog code for all this is the following:
\begin{lstlisting}[language=Prolog,firstline=10,lastline=17]
buildComplex:-
	buildFrame,
	% add the three central simplices
	addSimplex(34,[x,y,z,a]),
  	addSimplex(35,[x,y,z,b]),
  	addSimplex(36,[x,y,z,c]).
% Builds the complex of the example without 3 central simplices.
buildFrame:-buildRedFrame,buildGreenFrame,buildBlueFrame.
\end{lstlisting}
Now that the complex is loaded in memory, storing its TV relation, we can check if it is manifold or not. 
Actually, we already know that triangle xyz is common to tetrahedra 34,35,36 so it is interesting to ask Prolog to execute this fragment of 
code:
\begin{lstlisting}[language=Prolog,firstline=45,lastline=48]
test1:-
  resetComplex,
  buildComplex,
  listNonPseudoManifolds,  
\end{lstlisting}
and the output is:
\begin{lstlisting}
?- test1.

Non PseudoManifold Top d-Simplexes:
[a,x,y,z]
[b,x,y,z]
[c,x,y,z]
Non Manifold d-1-Simplexes:
[x,y,z] of order 3
true.
\end{lstlisting}
that confirms our intuition. Actually, even within this trivial 
framework, the benefits of using Prolog is evident. We cannot say that the discussion of Example \ref{app:example} is a proof of the 
claims we stated there. This program checks all possible triangles for non-pseudomanifoldness and returns a more reliable answer.

The next step is to show that we have an assembly to build this complex  $\AComp$  from its totally exploded version $\topAComp$
using only {\em manifold glue} i.e. pseudomanifold \gl\ \inst s (see \ref{def:pminst}).

Again we split the description of this set of  \gl\ \inst s in three subsets needed to build the cones in red, green and blue frames of Figure  \ref{fig:tienonreg-all}. It is easy to see that to build the complex in the red frame  from the totally exploded version  $\topAComp$ we need this fragment of Prolog code: 
\begin{lstlisting}[language=Prolog,firstline=218,lastline=225]
stitchTheRedFrame:-
  doPseudoManifoldGluingInstruction(1,16),
  doPseudoManifoldGluingInstruction(16,17),
  doPseudoManifoldGluingInstruction(17,18),
  doPseudoManifoldGluingInstruction(18,2),
  doPseudoManifoldGluingInstruction(18,19),
  doPseudoManifoldGluingInstruction(19,20),
  doPseudoManifoldGluingInstruction(20,3).
\end{lstlisting}
Using {\tt doPseudoManifoldGluingInstruction} we are sure that, if construction do not fail,  we are using  \gl\ \inst s that are
pseudomanifold \gl\ \inst s. Again, the code for the green and blue frames is similar and it is not shown here. So, in the end, we will stitch together the three cones and then stitch simplices 34,35 and 36 with the rest of the complex. 
This fragment of code will do the job:
\begin{lstlisting}[language=Prolog,firstline=203,lastline=215]
stitchTheComplex:-
  stitchTheRedFrame,
  doPseudoManifoldGluingInstruction(1,34),
  doPseudoManifoldGluingInstruction(23,34),
  stitchTheGreenFrame,
  doPseudoManifoldGluingInstruction(31,34),
  stitchTheBlueFrame,
  doPseudoManifoldGluingInstruction(3,35),
  doPseudoManifoldGluingInstruction(25,35),
  doPseudoManifoldGluingInstruction(33,35),
  doPseudoManifoldGluingInstruction(2,36),
  doPseudoManifoldGluingInstruction(24,36),
  doPseudoManifoldGluingInstruction(32,36).
\end{lstlisting}
We note that we do not use  \gl\ \inst s that stitch two of the three tetrahedra 34,35 and 36.
It will be the form of the surrounding complex that will constrain the three tetrahedra to share triangle xyz, that's where the trick is! At least this is what intuition suggest. The  \gl\ \inst s in the code above are arranged to show  that the complex is a  \emas{shelllable}{complex} i.e. during the construction of the complex $\AComp$ the \gl\ \inst s takes care to grow a 2-connected tetrahedralization adding one new tetrahedra at every step. The new tetrahedron is introduced  stitching one or more triangles between the newly added tetrahedron and the growing 3-complex.
Obviously the complex $\AComp$ cannot be embedded into a three dimensional space so a formal proof of the above claims is quite appropriate. 

We have a couple of tools for this. One is to
use {\tt splitVertex} to see if, for each vertex $v$,  all \ema{vertex}{copies}  (see \ref{pro:uniqver}) $v_{\theta}$
created in  $\topAComp$ (see \ref{def:top}) are collapsed into a single vertex. The predicate {\tt splitVertex(V)} is the appropriate tool since it succeeds if V is a \ema{splittig}{vertex} 
(see remarks after Property \ref{pro:uniqver}).
If no splitting vertex is found this means that the \gl\ \inst s collapses all
the vertex copies $v_{\theta}$ in  $\topAComp$ into a single vertex (isomorphic to) $v$. 
A second option is to use brute force and check that, having applied all \gl\ \inst s, the TV and VT relations models two isomorphic simplicial complexes.  The predicate {\tt vtIsoTv} can do this for us. We prepared a fragment of code to stitch all together and perform the two tests.
\begin{lstlisting}[language=Prolog,firstline=49,lastline=53]
  resetDecomposition,
  buildTotallyExploded,
  stitchTheComplex,
  not(splitVertex(_)),
  vtIsoTv.
\end{lstlisting}
Running this we do not get any output but {\tt true.} This confirms that there are no splitting vertices  and that the two complexes are isomorphic. 

Please note that the success of {\tt vtIsoTv} is equivalent to the success of {\tt not(splitVertex(V))} because this package do not models stitching in general.  Only \gl\ \inst s that can form a decomposition of  $\AComp$ from $\topAComp$ are supported.
If we ask to do some \gl\ \inst\ that is not  
in the lattice of decompositions the relative call to {\tt doGluingInstruction} fails.
Therefore, what is modeled in the TV is always a complex in the lattice of decompositions (see \ref{sec:declatt}) between  $\topAComp$ and  $\AComp$.  The package presented here needs to  be properly extended if one needs a tool for modeling simplicial complexes. 

On the other hand, this limitation helps us a lot. Indeed it is not know if checking isomorphism for graphs is polynomial or even NP. Still, our trivial algorithm  behind {\tt vtIsoTv} checks isomorphism of
the current decomposition  w.r.t. the complex modeled by the TV in linear time w.r.t. the number of vertices. 

Finally we can use this package to understand how this example works. To this aim we stitch all the tetrahedra but 34,35,36 
and then dump splitting vertices and non pseudomanifold triangles (if any). The fragment of code for this is the following:
\begin{lstlisting}[language=Prolog,firstline=172,lastline=178]
  resetComplex,
  buildComplex,
  resetDecomposition,
  buildTotallyExploded,
  stitchTheFrame, % just build the three cones.
  dumpSplitVertex, 
  listNonPseudoManifolds,
\end{lstlisting}
where {\tt stitchTheFrame} is:
\begin{lstlisting}[language=Prolog,firstline=216,lastline=217]
stitchTheFrame:-
     stitchTheRedFrame,stitchTheGreenFrame,stitchTheBlueFrame.
\end{lstlisting}
The output from {\tt dumpSplitVertex} is quite lengthy because the complex without 34,35,36 is quite open. 
To give an idea of the situation we report  in Figure \ref{fig:tienonreg-frame} the portion of the output that is relevant for vertex x. 
Arrows in figure link vertex copies of x to their location in the (not completely closed) cones.
\begin{figure}[ht]
	\centerline{\mbox{\psfig{file=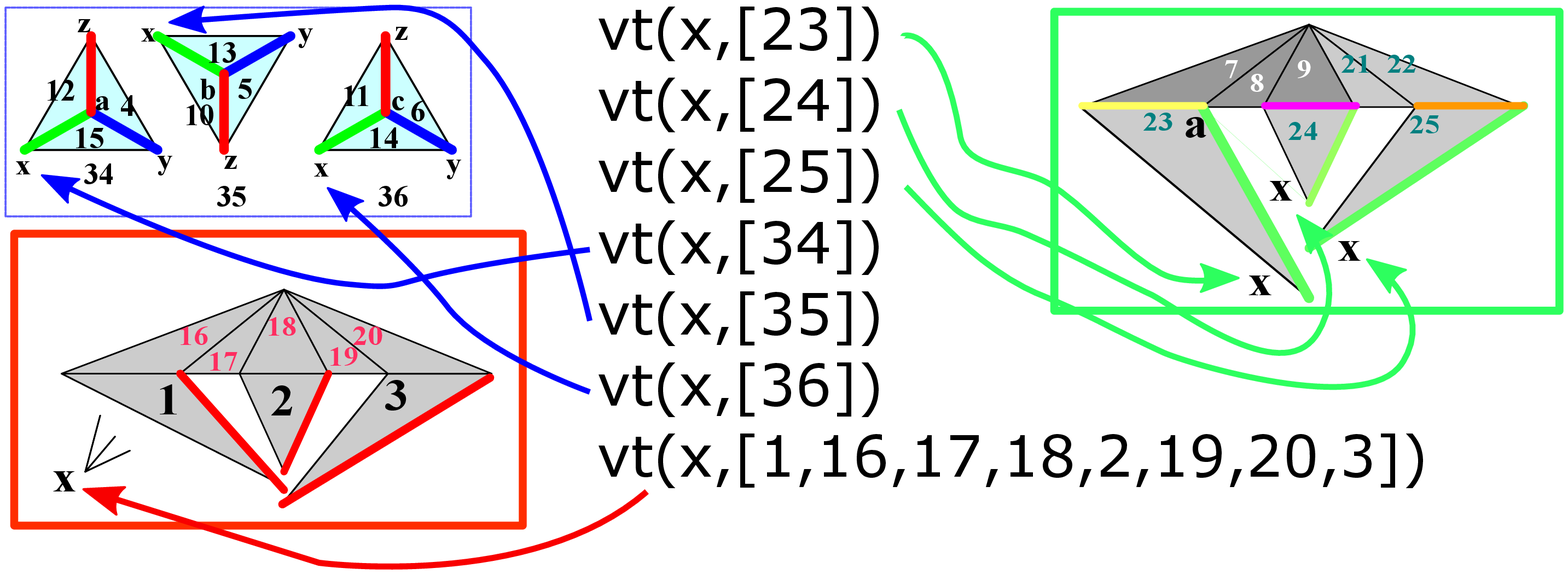,width=0.95\textwidth}}}
	\caption{the splitting vertices resulting from an incomplete
		stitching}
	\label{fig:tienonreg-frame}
\end{figure}
On the other hand the call to  {\tt listNonPseudomanifolds} confirms that there are no non-pseudomanifold triangles. So we can expect to have three distinct 2-connected components, each bounded by a separate manifold surface. These hypothesis can be checked by the following code fragment:
\begin{lstlisting}[language=Prolog,firstline=183,lastline=193]
  dm1connectedComponents(C), % complex has 6 2-connected components
  % Three are isolated tetrahedra.
  print_term(C,[]),
  boundary(Bnd), % it has a boundary surface that has 3 1-connected
  % components each of 16 triangles plus three tetrahedra.
  resetComplex,
  addComplex(bnd,Bnd), 
  % now the TV holds the boundary
  dm1connectedComponents(CBnd), % boundary is made up of 6 surfaces
  print_term(CBnd,[]),
  listNonPseudoManifolds. % no pseudomanifolds in the boundary
\end{lstlisting}
output from the call to {\tt print_term(C,[])} confirms that the complex is made up of three connected components plus the 
three tetrahedra 34,35 and 36.
\begin{lstlisting}
[ 
[34],
[35],
[36],
[1,16,17,18,2,19,20,3],
[23,7,8,9,21,24,22,25],
[31,26,27,28,29,32,30,33]
]
\end{lstlisting}
The first three components are the three isolated tetrahedra.
The fourth component  is the cone from x to the complex in the red frame in Figure \ref{fig:tienonreg-frame}. The fifth component is the cone from y to the complex in the green frame.
The sixth cone is not shown in   Figure \ref{fig:tienonreg-frame}a and is in the last line of output. The remaining (omitted) output
confirms that the boundary of this complex is made up of six 1-connected  surfaces without any non pseudomanifold edge.
The remaining sections present the API of this package.
\clearpage{}

\section{iqm.pl: iqm library for Initial Quasi Manifolds}

\label{sec:iqm}

\begin{tags}
    \pldoctag{author}
Franco Morando
    \pldoctag{license}
GPL
\end{tags}

This program takes a simplicial complex defined via a TV relation and
handles its totally decomposed version giving the possibility to
stitch together top simplexes either using vertex equations or simplex
gluing instructions. The totally decomposed complex is represented
via VT relation\vspace{0.7cm}

\begin{description}
    \predicate[nondet]{tv}{2}{?T:atom, ?V:list}
is a dynamic predicate used to encode the TV for complex to be decomposed.
If vt(foo,[a,b,c]) is stored then triangle foo with vertices a,b and c exist.
User are advised to use carefully this predicate possibly ignoring it.

\begin{arguments}
\arg{T} & an atom that is the index of some top d-simplex. \\
\arg{V} & a set of atoms that are indexes of vertices. \\
\end{arguments}

    \predicate[nondet]{vt}{2}{?V:atom, ?T:list}
is a dynamic predicate used to encode the VT for a decomposition of
the complex stored in the TV.
If tv(a,[foo,bar]) is stored then top simplexes foo and bar share
vertex a. To be consistent with the rest of the package no two
entries like tv(a,[foo,bar]) and tv(a,[some,thing]) must exist.
User are advised to use carefully this predicate possibly ignoring it.

\begin{arguments}
\arg{V} & an atom that is the index of some vertex. \\
\arg{T} & a set of atoms that are indexes of top simplexes. \\
\end{arguments}

    \predicate[det]{resetComplex}{0}{}
deletes the TV relation, complex is erased.

    \predicate[det]{vtToTv}{1}{-Res:list}
This procedure extracts in \arg{Res} a TV from the VT representation of
the complex obtained by stitching the totally decomposed complex.
Indeed the stitched complex is recorded in this package
in a VT using asserts to vt(V,L).
This procedure leaves dynamic predicates \predref{tv}{2} and \predref{vt}{2} unchanged.

\begin{arguments}
\arg{Res} & a list of terms of the form vt($<$some vertex$>$,[s1,...,s1])
representing a VT relation. \\
\end{arguments}

    \predicate[det]{vtIsoTv}{0}{}
succeeds if tv and vt are isomorphic.
We assume that VT is created by a call to buildTotallyExploded and
several calls to doPseudoManifoldGluingInstruction or to doGluingInstruction.
We recall that the effect of these two calls is the same but
doPseudoManifoldGluingInstruction checks that stitching simplex is a pseudomanifold
one. If not it fails.

    \predicate[det]{addSimplex}{2}{+Gamma:atom, +SetGamma:list}
adds a simplex to this complex
Top simplexes are encoded by atoms (could be integers)

\begin{arguments}
\arg{Gamma} & an atom that is used as an index for the top simplex to be added. \\
\arg{SetGamma} & a set of atoms used to encode vertices for this top simplex.  \\
\end{arguments}

    \predicate[det]{addComplex}{2}{+U:atom, +SimplexList:list}
adds a complex made up of a list of simplexes, e.g. [[a,b,c],[b,c,d]]
for the rectangle abdc.
Simplex indexes are atoms U1, U2 etc. that are created randomly
using \arg{U} as prefix.

\begin{arguments}
\arg{U} & an atom that is used as a prefix for top simplex indexes. \\
\arg{SimplexList} & a set of sets of vertices.
Vertices must be encoded by atoms. e.g. [[a,b,c],[b,c,d]]. \\
\end{arguments}

    \predicate[det]{eulerX}{1}{+X:int}
returns the Euler characteristics of the closed 2-complex in TV.
Works only if in TV is stored a closed 2-complex. Otherwise
results are meaningless.

\begin{arguments}
\arg{X} & the Euler characteristics of surface.  \\
\end{arguments}

    \predicate[det]{link}{2}{+V:atom, -S:list}
returns the set of all maximal simplices in TV in the link of \arg{V}.

\begin{arguments}
\arg{V} & a vertex \\
\arg{S} & a set of lists each being a maximal simplex. \\
\end{arguments}

    \predicate[det]{star}{2}{+V:atom, -S:list}
returns the set of all top simplices in TV in the star of \arg{V}.

\begin{arguments}
\arg{V} & a vertex \\
\arg{S} & a set of atoms each being a top simplex index. \\
\end{arguments}

    \predicate[det]{skeleton}{2}{-S:list, +D:int}
returns the set of all \arg{D} simplices in TV.

\begin{arguments}
\arg{S} & a set of sets each being a simplex. \\
\arg{D} & dimension of the simplexes to be returned \arg{D}=0 for points \\
\end{arguments}

    \predicate[nondet]{incident}{2}{?Theta:atom, +S:list}
succeeds iff \arg{Theta} is incident to simplex \arg{S}.

\begin{arguments}
\arg{Theta} & an atom that is the index of some top simplex. \\
\arg{S} & a set of vertices. \\
\end{arguments}

    \predicate[det]{orderOf}{2}{+S:list, -N:int}
counts in \arg{N} the number of top simplexes that are incident to
the simplex given by the set of vertices in \arg{S}.

\begin{arguments}
\arg{S} & a set of vertices. \\
\arg{N} & a positive integer that gives the number of top simplexes that are incident to \arg{S}. \\
\end{arguments}

    \predicate[nondet]{adjacent}{3}{?Theta1:atom, ?Theta2:atom, ?SetTheta:list}
succeeds iff \arg{Theta1} and \arg{Theta2} are adjacent via the set of vertexes
in \arg{SetTheta}. If \arg{Theta1} is equal to \arg{Theta2} it fails.
To succeed \arg{Theta1} and \arg{Theta2} must be two top d-simplexes and they must share
the d-1 face in \arg{SetTheta}.

\begin{arguments}
\arg{Theta1} & an atom that is the index of some top simplex. \\
\arg{Theta2} & an atom not equal to \arg{Theta1} that is the index of some top simplex. \\
\arg{SetTheta} & a set of vertices. \\
\end{arguments}

    \predicate[nondet]{adjacent}{2}{?Theta:atom, ?Theta2:atom}
succeeds iff Theta1 and \arg{Theta2} are adjacent.
If Theta1 is equal to \arg{Theta2} it fails.
Theta1 and \arg{Theta2} must be two d-simplexes and they must share
a d-1 face.

\begin{arguments}
\arg{Theta1} & an atom that is the index of some top simplex. \\
\arg{Theta2} & an atom not equal to Theta1 that is the index of some top simplex. \\
\end{arguments}

    \predicate[det]{dm1connectedComponents}{1}{-C:list}
always succeeds and returns a set of sets. Each set contains atoms that
are indexes of top simplices. Top simplicies in this set are uniformly
dimensional. All top d-simplices in such sets are d-1 connected.

\begin{arguments}
\arg{C} & a set of sets of top simplexes.  \\
\end{arguments}

    \predicate[nondet]{nonPseudoManifold}{1}{?Theta1:atom}
succeed if \arg{Theta1} is a top d-simplex d-1-adjacent to three or more
top d simplexes.

\begin{arguments}
\arg{Theta1} & an atom that is the index of some top simplex. \\
\end{arguments}

    \predicate[nondet]{nonPseudoManifold}{2}{?Theta1:atom, ?SetTheta:list}
succeeds iff the top d-simplex \arg{Theta1} is d-1 adjacent via the set
of vertexes in \arg{SetTheta} to three or more top d-simplexes.

\begin{arguments}
\arg{Theta1} & an atom that is the index of some top d-simplex. \\
\arg{SetTheta} & a set of d-1 vertices. \\
\end{arguments}

    \predicate[nondet]{nonPseudoManifoldPair}{2}{?Theta1:atom, ?Theta2:atom}
succeed if two top d-simplexes \arg{Theta1} and \arg{Theta2} meet at a
non-manifold d-1-face.

\begin{arguments}
\arg{Theta1} & an atom that is the index of some top simplex. \\
\arg{Theta2} & an atom not equal to \arg{Theta1} that is the index of some top simplex. \\
\end{arguments}

    \predicate[det]{printSimplex}{1}{+S:atom}
is a printing utility that prints the TV relation for the top simplex \arg{S}.

\begin{arguments}
\arg{S} & an atom that is the index of some top simplex. \\
\end{arguments}

    \predicate[det]{listNonPseudoManifolds}{0}{}
is a listing utility that lists the set of vertices for all top
d-simplexes involved in a non-manifold d-1-adjacency for some d.
The utility lists also all d-1 simplexes involved in a non-manifold
adjacency of top d-simplexes for some d.

    \predicate[det]{listAdjacents}{0}{}
is a listing utility that lists the set of triples
[Theta1,Theta2,SetTheta] for all top d-simplexes Theta1,Theta2
involved in a d-1-adjacency via the set of vertices in SetTheta
for some d.

    \predicate[det]{boundaryface}{1}{+F:list}
succeeds if \arg{F} is d-1 face of a top d-simplex and \arg{F} is on the
boundary.

\begin{arguments}
\arg{F} & a set of d-1 vertices. \\
\end{arguments}

    \predicate[det]{boundary}{1}{-Bnd:list}
returns the list of boundary simplexes.

\begin{arguments}
\arg{Bnd} & a set of sets of vertices. \\
\end{arguments}

    \predicate[det]{dumpTv}{0}{}
lists dynamic predicate tv(T,[v1,...]) that encodes
the TV for the original complex.

    \predicate[det]{resetDecomposition}{0}{}
resets the stored decomposition by deleting the VT.

    \predicate[det]{buildTotallyExploded}{0}{}
The complex given by the TV relation is turned into a totally
exploded version where all top simplexes are distinct components
and vertex v is split into n-copies being n the number of top
simplices incident to v in the TV complex.
The result is the creation of a VT relation storing vt(foo,[bar])
one for every vertex foo and for every top simplex bar.
Vertices of the form vt(foo,[bar]) are called "vertex copies" of
the "splitting vertex" foo.
As stitcing operation takes place relations like vt(foo,[bar]) and
vt(foo,[biz]) might disappear and will be substituted by vt(foo,[bar,biz]).

    \predicate[det]{doVertexEquation}{3}{+Theta1:atom, +Theta2:atom, +V:atom}
Vertexes $v_{\sigma_1}$ and $v_{\sigma_2}$ with $\theta_1\in\sigma_1$ and
$\theta_2\in\sigma_2$ are identified. Note that the resulting
complex, modeled by the VT relation, is a decomposition of the
complex modeled by the TV relation.

\begin{arguments}
\arg{Theta1} & an atom that is the index of some top simplex. \\
\arg{Theta2} & an atom, distinct from \arg{Theta1}, that is the index of some top simplex. \\
\arg{V} & a vertex that must be a vertex of \arg{Theta1} and \arg{Theta2} in the TV relation. \\
\end{arguments}

    \predicate[det]{doGluingInstruction}{2}{+Theta1:atom, +Theta2:atom}
merges vertexes according to simplex gluing instruction \arg{Theta1} $<$\Sifthen{} \arg{Theta2}.

\begin{arguments}
\arg{Theta1} & an atom that is the index of some top simplex. \\
\arg{Theta2} & an atom, distinct from \arg{Theta1}, that is the index of some top simplex. \\
\end{arguments}

    \predicate[det]{doPseudoManifoldGluingInstruction}{2}{+Theta1:atom, +Theta2:atom}
merges vertexes according to simplex gluing instruction \arg{Theta1} $<$\Sifthen{} \arg{Theta2}.
Before executing checks that \arg{Theta1} and \arg{Theta2} are pseudomanifold adjacent in the TV.

\begin{arguments}
\arg{Theta1} & an atom that is the index of some top simplex. \\
\arg{Theta2} & an atom, distinct from \arg{Theta1}, that is the index of some top simplex. \\
\end{arguments}

    \predicate[det]{dumpDecomp}{0}{}
dumps each simplex for the current decomposition. The dump lists
each vertex with the corresponding VT record.

    \predicate[nondet]{splitVertex}{1}{?V:atom}
is a topological check on the decomposition that
succeeds if the vertex \arg{V} has more than one vertex copy.

    \predicate[det]{dumpSplitVertex}{0}{}
dumps the VT for all splitting vertexes.
\end{description}

\clearpage{}
\clearpage{}

\section{utilities.pl: utilities for application}

\label{sec:utilities}

\begin{tags}
    \pldoctag{author}
Franco Morando
    \pldoctag{license}
GPL
\end{tags}

\vspace{0.7cm}

\begin{description}
    \predicate[det]{write_ln}{1}{+X}
Write \arg{X} on a single line

\begin{arguments}
\arg{X} & the item to be written. \\
\end{arguments}

    \predicate[nondet]{asublist}{3}{?Sub:list, +L:list, +N:int}
upon backtrack returns all ordered sublist of \arg{L} of length \arg{N}.

\begin{arguments}
\arg{Sub} & sublist returned. \\
\arg{L} & the complete list. \\
\arg{N} & length of the list to be returned.  \\
\end{arguments}

    \predicate[det]{disjoint}{2}{+C1:list, +C2:list}
succeeds if \arg{C1} and \arg{C2} are disjoint.

\begin{arguments}
\arg{C1} & and \arg{C2} two lists to be considered as sets \\
\end{arguments}

    \predicate[det]{closurePartition}{3}{:Rel, +L:list, -Partition:list}
returns the partition of \arg{L} given by the quotient \arg{L}/\arg{Rel}*.
The binary relation \arg{Rel} must be
symmetric and \arg{Rel}* is the transitive closure of \arg{Rel}.

\begin{arguments}
\arg{Rel} & a binary relation. \\
\arg{L} & a set of elements as a list. \\
\arg{Partition} & a list of sets one for each set in the partition.
of \arg{L} defined by the transitive closure of \arg{Rel}. \\
\end{arguments}
\end{description}

\clearpage{}
 \bibliographystyle{plainurl}
\bibliography{library}
\printindex
\end{document}